%% file: spinor.tex
\documentclass[final,1p,times]{elsarticle}

\usepackage{graphicx}
\usepackage{amssymb}
\usepackage{amsthm}
\usepackage{amsmath,bm,lscape}
\usepackage{color}
\usepackage{ulem}
\usepackage[mathscr]{eucal}
\usepackage{ogonek}

\newcommand{\elemag}[1]{\mathfrak{#1}}
\newcommand{\mani}[1]{\mathscr{#1}}

\journal{Physics Report}

\begin{document}

\begin{frontmatter}


\title{Spinor Bose-Einstein condensates}

\author[label1]{Yuki Kawaguchi \footnote{Current address: Department of applied physics and quantum-phase electronics center, University of Tokyo, Yayoi  2-11-16, Bunkyo-ku, Tokyo 113-0033, Japan}}
\ead{yuki@cat.phys.s.u-tokyo.ac.jp}
\author[label1,label2]{Masahito Ueda}
\ead{ueda@phys.s.u-tokyo.ac.jp}

\address[label1]{Department of Physics, University of Tokyo, Hongo
 7-3-1, Bunkyo-ku, Tokyo 113-0033, Japan}
\address[label2]{ERATO Macroscopic Quantum Control Project, JST, Tokyo 113-8656, Japan}

\begin{abstract}
An overview on the physics of spinor and dipolar Bose-Einstein condensates (BECs) is given.
Mean-field ground states, Bogoliubov spectra, and many-body ground and excited states of spinor BECs are discussed.
Properties of spin-polarized dipolar BECs and those of spinor-dipolar BECs are reviewed.
Some of the unique features of the vortices in spinor BECs such as fractional vortices and non-Abelian vortices are delineated.
The symmetry of the order parameter is classified using group theory, and various topological excitations are investigated based on homotopy theory. 
Some of the more recent developments in a spinor BEC are discussed.
\end{abstract}


\end{frontmatter}


\tableofcontents

\begin{table}
\begin{center}
\begin{tabular}{ll} \hline
symbol                                    & definition  \\ \hline \hline
$\hat{A}_\mathcal{FM}({\bm r},{\bm r}')$  & annihilation operator for atomic pairs in $|\mathcal{F},\mathcal{M}\rangle$ state\\
$A$                                       & pair amplitude of the spin-singlet pair, $\langle \hat{A}_{00}\rangle_0$  \\
${\bm A}, \Phi$                           & geometric gauge potential\\
$\elemag{A}$                              & electromagnetic four potential\\
$a_\mathcal{F}$                           & {\it s}-wave scattering length of the total spin-$\mathcal{F}$ channel\\
${\bm B}$                                 & external magnetic field\\
$c_0,c_1,c_2,\cdots$                      & interaction coefficients for the short-range interaction\\
$c_{\rm dd}$                              & interaction coefficient for the dipole-dipole interaction\\
$\hat{\bm d}$                             & unit vector indicating the nematic director \\
$\mathcal{F},\mathcal{M}$                 & total spin of two colliding atoms and its projection  \\
$f, m$                                    &  atomic spin and its projection \\
${\bf f}=({\rm f}_x,{\rm f}_y, {\rm f}_z)$&  vector of $f\times f$ spin matrices or spin operators \\
$\hat{\bm F}({\bm r})=\sum_{mm'}\hat{\psi}^\dagger_m {\bf f}_{mm'}\hat{\psi}_{m'} $ & spin operator  \\
${\bm F}({\bm r})=\sum_{mm'}{\psi}^\ast_m {\bf f}_{mm'}{\psi}_{m'} $                & spin expectation value \\
${\bm f}({\bm r})=\sum_{mm'}\zeta_m^\ast {\bf f}_{mm'}\zeta_{m'}$                   & spin expectation value per atom \\
$j,k$                                     & indices for $x$, $y$, and $z$ in the coordinate space\\
$M$                                       & atomic mass  \\
$M_z$                                     & total longitudinal magnetization of the condensate \\
$m,m',m_1,m_2,\cdots$                     & indices for the magnetic sublevel \\
$\hat{\mathcal{N}}_{\nu\nu'}$             & nematic tensor operator \\ 
$\mathcal{N}_{\nu\nu'}$                   & expectation value of the nematic tensor operator \\ 
$n({\bm r}) = \sum_{m}{\psi}^\ast_m {\psi}_m $ & number density \\
$p$                                       & linear Zeeman energy\\
$q$                                       & quadratic Zeeman energy \\
$Q_{\nu\nu'}$                             & kernel of the dipole-dipole interaction \\
$\mani{R}$                                & order-parameter manifold \\
$\hat{\bm s}={\bm F}/|{\bm F}|$           & unit vector spin along the spin expectation value  \\
$\mathcal{T}$                             & Time-reversal operator \\
${\bm v}^{\rm mass}$                      & superfluid velocity\\
${\bm v}_\nu^{\rm spin}$                  & superfluid velocity for the $\nu=x,y,z$ component of spin\\
$\alpha,\beta,\gamma$                     & Euler angles describing the SO(3) spin rotation\\
$\hat{\psi}_m({\bm r})$                   & field operator  \\
$\psi_m({\bm r})$                         & order parameter \\
$\zeta_m\equiv \psi_m/\sqrt{n}$           & normalized spinor order parameter  \\
$\mu$                                     & chemical potential \\
$\nu,\nu',\nu_1,\nu_2,\cdots$             & indices for $x$, $y$, and $z$ in the spin space\\
\hline
\end{tabular}
\caption{Table of symbols}
\end{center}
\end{table}

\begin{table}
\begin{center}
\begin{tabular}{ll}\hline
abbreviation  & definition  \\ \hline \hline
AF  & antiferromagnetic   \\
BA  & broken axisymmetry  \\
BEC & Bose-Einstein condensate  \\
BKT & Berezinskii-Kosterlitz-Thouless\\
BN  & biaxial nematic     \\
C   & cyclic  \\
CSV & chiral spin vortex  \\
DDI & dipole-dipole interaction  \\
DW  & density wave      \\
EdH & Einstein-de Haas \\
F   & ferromagnetic     \\
FL  & flower            \\
GPE  & Gross-Pitaevskii equation \\
MI  & Mott insulator    \\
P   & polar  \\
PCV & polar-core vortex  \\
SF  & superfluid  \\
SO  & spin-orbit \\
SS  & supersolid  \\
UN  & uniaxial nematic \\
\hline
\end{tabular}
\caption{Table of abbreviations}
\end{center}
\end{table}

\clearpage

\input introduction.tex

\input generaltheory.tex

\input meanfield.tex

\input experiments.tex

\input bogoliubov.tex
\input dipole.tex

\input hydro.tex
\input vortices.tex

\input symmetry.tex

\input topology.tex
\input manybody.tex

\input summary.tex

\section*{Acknowledgements}

We acknowledge the fruitful collaborations 
with
P. Blair Blakie,
Michikazu Kobayashi,
Shingo Kobayashi,
Kazue Kudo,
Muneto Nitta,
Nguyen Thanh Phuc,
Hiroki Saito,
Satoshi Tojo,
Shun Uchino, and
Zhifang Xu.
MU acknowledges the Aspen Center for Physics, where part of this work was carried out.
This work was supported by Grants-in-Aid for Scientific Research (Nos. 22340114, 22740265), 
a Grant-in-Aid for scientific Research on Innovative Areas ``Topological Quantum Phenomena'' (No. 22103005), 
a Global COE Program ``the Physical Science Frontier'', and the Photon Frontier Network Program of MEXT of Japan.
YK acknowledge the financial support from Inoue Foundation.



\bibliographystyle{elsarticle-num-names}
\bibliography{reference}

\end{document}

%% file: introduction.tex
\section{Introduction}

Gaseous Bose-Einstein condensate (BEC) was first created using atoms in a single spin state of rubidium 87 ($^{87}$Rb)~\cite{Anderson1995} and later in single spin states of sodium 23 ($^{23}$Na)~\cite{Davis1995} and lithium 7 ($^7$Li)~\cite{Bradley1997}. In these systems, only those atoms in a week-field seeking state were magnetically trapped, and therefore, their spin degrees of freedom were frozen. 
A spinor Bose-Einstein condensate (BEC), namely a BEC with spin internal degrees of freedom, was first realized in a gas of spin-1 $^{23}$Na atoms confined in an optical dipole trap in 1998~\cite{Stamper-Kurn1998}, opening up a new research arena of ultracold atomic systems.
Confined in an optical trap, the direction of atomic spins can change due to the interparticle interaction. Consequently, the order parameter of a spin-$f$ BEC has $2f+1$ components that can vary over space and time, producing a very rich variety of spin textures. 
In contradistinction to a spinor BEC, a BEC of atoms in a single spin state is referred to as a scalar BEC.
This article provides an overview of the physics of spinor BECs.

Bose-Einstein condensation is a genuinely quantum-mechanical phase
transition in that it occurs without the help of interaction. However,
in the case of spinor BECs, several phases are possible below the
transition temperature $T_{\rm BEC}$, and which phase is realized at what temperature does
depend on the nature of the interaction. With the spin and gauge degrees of freedom, the full symmetry of the system above $T_{\rm BEC}$ is ${\rm SO}(3)\times {\rm U}(1)$. Below the transition temperature, the symmetry is spontaneously broken in many different ways: The number of ground-state phases in the absence of an external magnetic field is two for the spin-1 case~\cite{Ohmi1998,Ho1998}, three for the spin-2 case~\cite{Chibanu2000,Ueda2002}, and  eleven for the spin-3 case~\cite{Diener2006,Kawaguchi2011}. The number of phases further increases under the external magnetic field. Such a wealth of possible phases make a spinor BEC a fascinating area of research for quantum gases.
Another striking consequence of the spin degrees of freedom is the spin dynamics: The initial population balance between each magnetic sublevel can change via the spin-exchange collision.
For example, in a spin-1 spinor BEC, two atoms in a magnetic sublevel $m=0$ can coherently and reversibly scatter into a pair of atoms in $m=+1$ and $m=-1$ states, and vice versa.
So far, spin-exchange collision has been observed in systems of 
spin-1 $^{87}$Rb~\cite{Chang2004}, 
spin-1 $^{23}$Na~\cite{Stenger1998,Black2007},
spin-2 $^{87}$Rb~\cite{Chang2004,Schmaljohann2004,Kuwamoto2004}, 
and spin-3 chromium 52 ($^{52}$Cr)~\cite{Pasquiou2011b}.
The basic properties of the mean-field spinor condensates, such as ground-state phase diagrams and spin dynamics, are reviewed in Sec.~\ref{sec:meanfield}, and the experimental achievements are summarized in Sec.~\ref{sec:experiments}.

The BEC responds to an external perturbation in a very unique manner. 
Even if the perturbation is weak, the response may be nonperturbative; for example, the phonon velocity depends on the scattering length in a nonanalytic manner. 
The low-lying excitations of spinor BECs are described by the Bogoliubov theory, as discussed in Sec.~\ref{sec:Bogoliubov}.
A spinor BEC can be experimentally prepared in an unstable stationary state.
The dynamics starting from such an unstable state can be understood in terms of a growth of unstable Bogoliubov modes or the dynamical instability which may be triggered by quantum fluctuations.
In Sec.~\ref{sec:Bogoliubov} we discuss how topological defects are nucleated after a quantum phase transition in the context of the dynamical instability.

The magnetic moment of the atom causes the magnetic dipole-dipole interaction (DDI). The long-range and anisotropic nature of DDI leads to new phenomena even in a spin-polarized BEC~\cite{Baranov2008,Lahaye2009}. Although the strength of the magnetic DDI is the smallest among all the energy scales involved, it plays a pivotal role in a spinor BEC in producing the spin texture---the spatial variation of the spin direction. In particular, the magnetic DDI couples spin and orbital angular momenta under an ultralow magnetic field (below 10~$\mu$G), which spontaneously generates superfluid flow of atoms.
Properties of spin-polarized and spinor dipolar BECs are discussed in Sec.~\ref{sec:DipolarBEC}.

The dynamics of spinor BECs may be understood more intuitively if the equations of motion are expressed in terms of physical quantities such as superfliud velocity, spin superfluid velocity, magnetization, and nematic directors, as discussed in Sec.~\ref{sec:hydro}.
We will see that the space and time dependences of spin configurations naturally generate a geometric gauge field.

One of the hallmarks of superfluidity manifests itself in its response to an external rotation. In a scalar BEC, the system hosts vortices that are characterized by the quantum of circulation, $\kappa=h/M$, where $h$ is the Planck constant and $M$ is the mass of the atom. The origin of this quantization is the single-valuedness of the order parameter. As mentioned above, however, in spinor BECs, the gauge degree of freedom is coupled to the spin degrees of freedom. This spin-gauge coupling gives rise to some of the unique features of the spinor BEC. For example, the fundamental unit of circulation can be a rational fraction of $\kappa$, and when two vortices collide, they may not reconnect unlike the case of the scalar BEC but form a rung vortex that bridges the two vortices. Vortices of spinor BECs are discussed in Sec.~\ref{sec:Vortices}.

In Sec.~\ref{sec:symmetry}, we discuss symmetry properties of mean-field ground states.
The ground-state phases in spinor BECs differ from each other in gauge, spin-rotation, or combined symmetries.
For example, the ferromagnetic phase has the SO(2) spin-gauge coupled rotational symmetry, whereas the spin-2 cyclic phase has the symmetry of tetrahedron.
We will show that ground-state order parameters can be found from the symmetry consideration without minimizing the mean-field energy.

Once we know the symmetry property of the condensed phase under consideration, the homotopy theory  tells us possible types of topological excitations.
In general, the direction of the spin can vary rather flexibly over space and time. Nonetheless, the global configuration of the spin texture must satisfy certain topological constraints.
This is because the order parameter in each phase belongs to a particular order-parameter manifold whose symmetry leads to a conserved quantity called a topological charge.
Such a topological constraint determines the nature of topological excitations such as line defects, point defects, Skyrmions, and knots. 
Elements of homotopy theory with applications to spinor BECs are reviewed in Sec.~\ref{sec:topology}.

In the last part of this review, we discuss the ground states and spin dynamics beyond the mean-field theory by exactly diagonalizing the many-body Hamiltonian.
In some parameter regions, the many-body spin correlations dramatically alter the ground states of spinor BECs and fragmented ground states arise so as to recover the SO(3) spin rotational symmetry as discussed in Sec.~\ref{sec:ManyBodyTheory}.

In view of these ongoing developments, it now seems appropriate to consolidate the knowledge that has been accumulated over the past decade. 
In this paper, we provide an overview of the basics and recent developments on the physics of spinor and dipolar BECs.
Some topics and related problems which we do not cover the main text, such as finite-temperature effects, low-dimensional systems, optical lattice, and spin-orbit coupling are overviewed in Sec.~\ref{sec:Summary}.

One of the major topics that is not treated in this review is fictitious spin systems such as a binary mixture of hyperfine spin states~\cite{Hall1998a,Hall1998b}.
Since the intra- and interspecies interactions of this mixture are almost identical, this system has the SU(2) symmetry and is regarded as a pseudo-spin-1/2 system.
They have some intrinsic interest such as interlaced vortex lattices and vortex molecules caused by the Josephson coupling. A comprehensive review of this subject is given in Ref.~\cite{Kasamatsu2005}.

This paper is organized as follows.
Section~\ref{sec:GeneralTheory} describes the fundamental Hamiltonian of the spinor BEC.
Section~\ref{sec:meanfield} develops the mean-field theory of spinor condensates and discusses the ground-state properties and spin dynamics of the spin-1, 2, and 3 BECs.
Section~\ref{sec:experiments} summarizes the experimental achievements so far.
Section~\ref{sec:Bogoliubov} develops the Bogoliubov theory of the spinor BEC.
Section~\ref{sec:DipolarBEC} provides an overview of the dipolar BEC and the spinor-dipolar BEC.
Section~\ref{sec:hydro} derives the hydrodynamic equations of motion of superfluid velocity and magnetization.
Section~\ref{sec:Vortices} discusses various types of vortices that can be created in spinor BECs.
Section~\ref{sec:symmetry} classifies the ground-state order parameters based on group theory and discuss the symmetry property of each phase.
Section~\ref{sec:topology} examines the topological aspects of spinor BECs. Possible topological excitations such as non-Abelian vortices are investigated using homotopy theory.
Section~\ref{sec:ManyBodyTheory} reviews the many-body aspects of spinor BECs.
Section~\ref{sec:Summary} summarizes the main results of this paper and discusses possible future developments.

%% file: generaltheory.tex
\section{General theory} 
\label{sec:GeneralTheory}

\subsection{Single-particle Hamiltonian}
The fundamental Hamiltonian of a spinor BEC can be constructed quite generally based on the symmetry argument. We consider a system of identical bosons with mass $M$ and spin $f$ that are described by the field operators $\hat{\psi}_m({\bm r})$, where $m=f,f-1,\cdots,-f$ denotes the magnetic quantum number. The field operators are assumed to satisfy the canonical commutation relations
\begin{align}
[\hat{\psi}_m({\bm r}),\hat{\psi}^\dagger_{m'}({\bm r}')]
&=\delta_{mm'}\delta({\bm r}-{\bm r}'), \nonumber \\
[\hat{\psi}_m({\bm r}),\hat{\psi}_{m'}({\bm r}')]
&= [\hat{\psi}^\dagger_m({\bm r}),\hat{\psi}^\dagger_{m'}({\bm r}')] = 0,
\label{fieldCR}
\end{align}
where $\delta_{mm'}$ is the Kronecker delta which takes the value 1 if $m=m'$ and 0 otherwise. 

Here, $f$ is the hyperfine spin of an atom, that is, the composition of the electron spin $s$, the orbital angular momentum $l$, and the nuclear spin $i$,
and it is a good quantum number when an external magnetic field is so low that the Zeeman energy is much smaller than the hyperfine energy splitting.
The angular momenta for various atomic species are summarized in Table~\ref{table:atomic_species}.
As of the writing of this article, spin 1, 2 and 3 spinor BECs have been realized utilizing $^{87}$Rb, $^{23}$Na, and $^{52}$Cr atoms.
\begin{table}[ht]
\begin{center}
\begin{tabular}{lllllll} \hline
atom       &$s$ &$l$ &$j=s+l$& $i$           &$f=j+i$ & $\Delta E_{\rm hf}/h$ (MHz)\\ \hline\hline
$^{1}$H    &1/2 &0   &1/2    & 1/2           &0, 1    &1420\\
$^{7}$Li   &1/2 &0   &1/2    & 3/2           &1, 2    & 804\\
$^{23}$Na  &1/2 &0   &1/2    & 3/2           &1, 2    &1772\\
$^{39}$K   &1/2 &0   &1/2    & 3/2           &1, 2    & 462\\
$^{41}$K   &1/2 &0   &1/2    & 3/2           &1, 2    & 254\\
$^{52}$Cr  &3   &0   &3      & 0             &3       & --\\
$^{85}$Rb  &1/2 &0   &1/2    & 5/2           &2, 3    &3036\\
$^{87}$Rb  &1/2 &0   &1/2    & 3/2           &1, 2    &6835\\
$^{133}$Cs &1/2 &0   &1/2    & 7/2           &3, 4    &9193\\ 
$^{164}$Dy &2   &6   &8      & 0             &8       &--\\   
$^{168}$Er &1   &5   &6      & 0             &6       &--\\ \hline
\end{tabular}
\caption{The angular momenta for electric ground states of various atomic species.
Here $j=s+l$ is the electronic total angular momentum and $f=|j\pm i|$ is the hyperfine spin, where $s$, $l$, and $i$ are the electron spin, the orbital angular momentum, and the nuclear spin, respectively.
The rightmost column shows the hyperfine energy splitting.
For alkali-metal atoms listed in the table (namely, except for $^{52}$Cr, $^{164}$Dy, and $^{168}$Er), the hyperfine state with $f=i-1/2$ lies lower in energy than that with $f=i+1/2$.}
\label{table:atomic_species}
\end{center}
\end{table}

The noninteracting part of the Hamiltonian comprises the kinetic term, trapping potential $U_{\rm trap}({\bm r})$, and linear and quadratic Zeeman terms:
\begin{align}
\hat{H}_0=\int d{\bm r} \sum_{m,m'=-f}^{f}
\hat{\psi}_m^\dagger
\left[
-\frac{\hbar^2\nabla^2}{2M}+U_{\rm trap}({\bm r})-p({\rm f}_z)_{mm'}+q({\rm f}_{z}^2)_{mm'}
\right]
\hat{\psi}_{m'},
\label{H_0}
\end{align}
where ${\rm f}_z$ is the $z$-component of the spin matrix whose matrix elements are given by $({\rm f}_z)_{mm'}=m\delta_{mm'}$, and hence, $({\rm f}_z^2)_{mm'}=m^2\delta_{mm'}$; and $p=-g\mu_{\rm B}B$ is the product of the Land\'e hyperfine $g$-factor $g$, the Bohr magneton $\mu_{\rm B}=e\hbar/2m_{\rm e}$ ($m_{\rm e}$ is the electron mass, and $e>0$ is the elementary charge), and an external magnetic field $B$ that is assumed to be applied in the $z$ direction. 
For atoms with $i=0$, $g$ is the the Land\'e $g$-factor for an electron:
\begin{align}
 g_j=\frac{3}{2}+\frac{s(s+1)-l(l+1)}{2j(j+1)},
\end{align}
whereas for atoms with $i\neq 0$, $g$ is given by
\begin{align}
 g= g_j\frac{f(f+1)+j(j+1)-i(i+1)}{2f(f+1)}.
\end{align}
(See, e.g., \textsection~113 and \textsection~121 of Ref.~\cite{LandauLifshitz_QM}.)
In particular, the Land\'e hyperfine $g$-factor for an alkali-metal atom (i.e., for $s=1/2$, $l=0$, and $j=1/2$) is given by 
\begin{align}
 g=\left\{
\begin{matrix}
-\displaystyle\frac{1}{i+1/2} & \textrm{ for } &f=i+\frac{1}{2},\\[3mm]
\displaystyle\frac{1}{i+1/2} & \textrm{ for } &f=i-\frac{1}{2}.\\
\end{matrix}
\right.
\end{align}

The coefficient $q$ of the quadratic Zeeman term is contributed from an external field ($q_B$) and from a microwave or light field ($q_{\rm MW}$).
The former is calculated by using the second-order perturbation theory as
\begin{align}
q_B=\frac{(g\mu_{\rm B}B)^2}{\Delta E_{\rm hf}},
\end{align}
where $\Delta E_{\rm hf}=E_{\rm m}-E_{\rm i}$ is the hyperfine energy splitting and it is given by the difference between the initial ($E_{\rm i}$) and intermediate ($E_{\rm m}$) energies.
For example, for the case of $^{87}$Rb, $g=-1/2$ and  $E_{\rm hf}\simeq 6.8$~GHz for the $f=1$ hyperfine manifold and  $g=1/2$ and  $E_{\rm hf}\simeq -6.8$~GHz for the $f=2$ hyperfine manifold.
For the case of $^{52}$Cr, $^{164}$Dy, and $^{168}$Er, $q_B=0$ due to the absence of the hyperfine structure.
The value of $q_{\rm MW}$ can be tuned independently of $q_B$ by employing a linearly polarized microwave field which is off-resonant with the other hyperfine state~\cite{Gerbier2006b,Leslie2009}.
For the case of $^{52}$Cr, a similar effect can be realized by using a quasi-resonant light field 
which couples the ground state of $^{52}$Cr ($^7{\rm S}_3$) to different excited P-states ($^7{\rm P}_{2,3,4}$)~\cite{Santos2007}.

\subsection{Interaction Hamiltonian}
\subsubsection{Symmetry considerations and irreducible operators}
Next, we consider the interaction Hamiltonian. 
We discuss the case of the magnetic dipole-dipole interaction later as it does not conserve the total spin of the system, and hence, it requires a separate treatment. Because the system is very dilute, we consider only the binary interaction. Upon the exchange of two identical particles of spin $f$, the many-body wave function changes by the phase factor $(-1)^{2f}$. By the same operation, the spin and orbital parts of the wave function change by $(-1)^{\mathcal{F}+2f}$ and $(-1)^\mathcal{L}$, respectively, where $\mathcal{F}$ is the total spin of the two particles and $\mathcal{L}$ is their relative orbital angular momentum. To be consistent, $(-1)^{2f}$ must be equal to $(-1)^{\mathcal{F}+2f}\times(-1)^\mathcal{L}$; hence, $(-1)^{\mathcal{F}+\mathcal{L}}=1$. Thus, $\mathcal{F}+\mathcal{L}$ must be even, regardless of the statistics of the particles. In the following, we consider only the $s$-wave scattering ($\mathcal{L}=0$), and, therefore, the total spin of two interacting particles must be even. 
Moreover, since the total spin $\mathcal{F}$ is conserved in the binary collision,
the interaction Hamiltonian can be divided according to the spin channel $\mathcal{F}$ as
\begin{align}
\hat{V}=\sum_{\mathcal{F}=0,2,\cdots,2f}\hat{V}^{(\mathcal{F})},
\label{sumV^F}
\end{align}
where $\hat{V}^{(\mathcal{F})}$ is the interaction Hamiltonian between two bosons whose total spin is $\mathcal{F}$. 

The interaction $\hat{V}^{(\mathcal{F})}$ in Eq.~(\ref{sumV^F}) can be constructed from the irreducible operator $\hat{A}_{\mathcal{F}\mathcal{M}}({\bm r},{\bm r}')$ that annihilates a pair of bosons at positions ${\bm r}$ and ${\bm r}'$. The irreducible operator is related to a pair of field operators via the Clebsch-Gordan coefficients $\langle \mathcal{F},\mathcal{M}|f,m;f,m'\rangle$ as follows:
\begin{align}
\hat{A}_{\mathcal{F}\mathcal{M}}({\bm r},{\bm r}')=
\sum_{m,m'=-f}^f
\langle \mathcal{F},\mathcal{M}|f,m;f,m'\rangle
\hat{\psi}_{m}({\bm r})\hat{\psi}_{m'}({\bm r}').
\label{A_FM}
\end{align}
Because the boson field operators $\hat{\psi}_{m}({\bm r})$ and $\hat{\psi}_{m'}({\bm r}')$ commute, $\hat{A}_{\mathcal{F}\mathcal{M}}({\bm r})$ vanishes identically for odd $\mathcal{F}$. In fact, when $\mathcal{F}=1$ and $f=1$, the Clebsch-Gordan coefficient is given by
\begin{align}
\langle 1, \mathcal{M} | 1, m ; 1, m' \rangle
= &\frac{(-1)^{1-m}}{\sqrt{2}} \;
\delta_{m + m', M} 
\nonumber \\
& \times
\left[ \delta_{\mathcal{M},1} (\delta_{m,1} + \delta_{m,0}) + \delta_{\mathcal{M},0} \;
m - \delta_{\mathcal{M}, -1}(\delta_{m,0} + \delta_{m,-1}) \right]\!.
\label{eqIDF16}
\end{align}
Substituting this into Eq.~(\ref{A_FM}), we find that $\hat{A}_{1\mathcal{M}}({\bm r})=0$. Similarly, we can show that $\hat{A}_{\mathcal{F}\mathcal{M}}({\bm r})=0$ if $\mathcal{F}$ is odd. Incidentally, for the case of fermions, we obtain $\hat{A}_{\mathcal{F}\mathcal{M}}({\bm r})=0$ for even $\mathcal{F}$ because the fermion field operators anticommute.

Because the interaction Hamiltonian is a scalar operator, it must take the following form:
\begin{align}
\hat{V}^{(\mathcal{F})}=\frac{1}{2}\int d{\bm r}\int d{\bm r}'
v^{(\mathcal{F})}({\bm r},{\bm r}')
\sum_{\mathcal{M}=-\mathcal{F}}^\mathcal{F}
\hat{A}^\dagger_{\mathcal{F}\mathcal{M}}({\bm r},{\bm r}')\hat{A}_{\mathcal{F}\mathcal{M}}({\bm r},{\bm r}'),
\label{intvF}
\end{align}
where $v^{(\mathcal{F})}({\bm r},{\bm r}')$ is a scalar function that describes the dependence of the interaction on the positions of the particles.
When the system is dilute and the range of interaction is negligible compared with the interparticle spacing, the interaction kernel $v^{(\mathcal{F})}$ may be approximated by a delta function with an effective coupling constant $g_\mathcal{F}$:
\begin{align}
v^{(\mathcal{F})}({\bm r},{\bm r}')=g_\mathcal{F}\delta({\bm r}-{\bm r}'),
\label{eq:interaction_delta}
\end{align}
where $g_\mathcal{F}$ is related to the $s$-wave scattering length of the total spin-$\mathcal{F}$ channel, $a_\mathcal{F}$, as
\begin{align}
g_\mathcal{F}=\frac{4\pi\hbar^2}{M}a_\mathcal{F}.
\label{eq:def_g_F}
\end{align}
The scattering lengths of $^{23}$Na, $^{87}$Rb, and $^{52}$Cr atoms are summarized in Table~\ref{table:scattering_length}.
According to the high resolution photoassociation spectroscopy of spin-1 $^{87}$Rb BEC~\cite{Hamley2009},
$\Delta a\equiv a_2-a_0$ is predicted to greatly change by means of the optical Feshbach resonance,
whose sign determines the magnetism of the system [see Eq.~\eqref{c(f=1)} and Sec.~\ref{sec:MFTspin1}].
However, since the lifetime of the condensate also becomes shorter due to the photoassociation laser light,
it is still challenging to change the magnetic nature of the condensate.
Another possibility to tune the spin-dependent interaction is using the microwave-induced Feshbach resonance proposed in Ref.~\cite{Papoular2010}.
Different from optical Feshbach resonances, microwave Feshbach resonances do not suffer from the associated atomic losses
because all relevant states correspond to the electronic ground level of the atoms.

Using Eqs.~\eqref{A_FM} and \eqref{intvF}--\eqref{eq:def_g_F}, Eq.~\eqref{sumV^F} can also be written as
\begin{align}
 \hat{V}=\frac{1}{2}\int d{\bm r}\sum_{m_1m_2m_1'm_2'}C^{m_1m_2}_{m_1'm_2'} \hat{\psi}_{m_1}^\dagger({\bm r})\hat{\psi}_{m_2}^\dagger({\bm r})\hat{\psi}_{m_2'}({\bm r})\hat{\psi}_{m_1'}({\bm r}),
\label{eq:V_Cmnm'n'}
\end{align}
where
\begin{align}
 C^{m_1m_2}_{m_1'm_2'}\equiv \frac{4\pi\hbar^2}{M} \sum_{\mathcal{F}=0,2,\cdots,2f} a_\mathcal{F}\langle f,m_1;f,m_2|\hat{\mathcal{P}}_\mathcal{F}|f,m_1';f,m_2'\rangle,
\label{eq:def_Cmnm'n'}
\end{align}
with $\hat{\mathcal{P}}_{\mathcal{F}}$ being the projection operator onto a two-body state with the total spin angular momentum $\mathcal{F}$:
\begin{align}
\hat{\mathcal{P}}_{\mathcal{F}}\equiv \sum_{\mathcal{M}=-\mathcal{F}}^\mathcal{F}|\mathcal{F},\mathcal{M}\rangle\langle \mathcal{F},\mathcal{M}|.
\label{eq:def_mathcalP}
\end{align}

\begin{table}
\begin{tabular}{cllllll}\hline
spin & atom   & scattering length (a.u.) & method & reference\\ \hline\hline
   1 & $^{23}$Na & $a_2=52.98\pm 0.40,\ a_0=47.36\pm 0.80$ & MS & Ref.~\cite{Samuelis2000}&\\
     &           & $a_2-a_0=3.5\pm 1.5$ & SD & Ref.~\cite{Stenger1998}\\
     &           & $a_2-a_0=2.47\pm 0.27$ & SMD & Ref.~\cite{Black2007}\\[2mm]
     & $^{87}$Rb & $a_2=100.4\pm 0.1,\ a_0=101.8\pm 0.2$ & MS & Ref.~\cite{Kempen2002}\\
     &           & $a_2-a_0=-1.45\pm 0.32$ & SMD & Ref.~\cite{Chang2005}\\
     &           & $a_2-a_0=-1.07\pm0.09$ & SMD in MI & Ref.~\cite{Widera2006}\\[2mm]
   2 & $^{23}$Na & $a_4=62.51\pm 0.50$ & MS & Ref.~\cite{Samuelis2000}\\[2mm]
     &           & $a_4=64.5\pm 1.3,\ a_2=45.8 \pm 1.1,\ a_0=34.9\pm 1.0$ & -- & Ref.~\cite{Ciobanu2000}\\[2mm]
     & $^{87}$Rb & $a_2-a_0=3.51\pm0.54,\ a_4-a_2=6.95\pm0.35$ & SMD in MI & Ref.~\cite{Widera2006}\\[2mm]
   3 & $^{52}$Cr & $a_6=112\pm 14, \ a_4=58\pm 6, \ a_2=-7\pm 20$ & MS & Ref.~\cite{Werner2005}\\[2mm]
     &           & $a_6=102.5 \pm 0.4$                            & DR + MS & Ref.~\cite{Pasquiou2010} \\ \hline
\end{tabular}
\caption{
Scattering lengths of $^{23}$Na, $^{87}$Rb, and $^{52}$Cr atoms in units of the Bohr radius ($a_{\rm B}=0.0529$~nm), where
MS, SD, SMD (in MI), and DR denote molecular spectroscopy, spin domain, spin-mixing dynamics (in the Mott insulator state), and dipolar relaxation, respectively (see Sec.~\ref{sec:experiments}).
For $^{52}$Cr, the scattering length of total spin 0 channel ($a_0$) is not determined in the experiment but theoretically predicted to be positive and around $30a_{\rm B}$--$50a_{\rm B}$~\cite{Sadeghpour}.}
\label{table:scattering_length}
\end{table}

\subsubsection{Operator relations}
We derive some useful operator relations which relate irreducible operators $\hat{A}_{\mathcal{F}\mathcal{M}}({\bm r},{\bm r}')$
to physical observables such as the total density operator
\begin{align}
\hat{n}({\bm r})\equiv\sum_{m=-f}^f\hat{\psi}_m^\dagger({\bm r})
\hat{\psi}_m({\bm r}),
\end{align}
the singlet-pair operator
\begin{align}
 \hat{A}_{00}({\bm r},{\bm r}') = \frac{1}{\sqrt{2f+1}}\sum_{m=-f}^f (-1)^{f-m} \hat{\psi}_m({\bm r}) \hat{\psi}_{-m}({\bm r}'),
\label{eq:hatA00}
\end{align}
the spin density operator
\begin{align}
\hat{F}_\nu({\bm r})=\sum_{m,m'=-f}^f ({\rm f}_\nu)_{mm'}\hat{\psi}_m^\dagger({\bm r})
\hat{\psi}_{m'}({\bm r}) \ \ \ (\nu=x,y,z),
\label{spindensity}
\end{align}
and the rank-$k$ ($k=2,3,\cdots$) spin nematic tensor operator
\begin{align}
 \hat{\mathcal{N}}^{(k)}_{\nu_1\nu_2\cdots\nu_k}({\bm r}) = \sum_{m,m'=f}^f ({\rm f}_{\nu_1}{\rm f}_{\nu_2}\cdots{\rm f}_{\nu_k})_{mn} \hat{\psi}_m^\dagger({\bm r}) \hat{\psi}_{m'}({\bm r})\ \ \ (\nu_1,\nu_2,\cdots\nu_k=x,y,z),
\label{spinnematic}
\end{align}
where the Clebsch-Gordan coefficient for $\mathcal{F}=\mathcal{M}=0$,
\begin{align}
\langle 0, 0|f,m;f,m'\rangle
=\delta_{m+m',0} \; \frac{(-1)^{f-m}}{\sqrt{2f+1}},
\label{CG}
\end{align}
is used to obtain Eq.~\eqref{eq:hatA00}, and 
$({\rm f}_\nu)_{mm'}$ $(\nu=x,y,z)$ in Eqs.~\eqref{spindensity} and \eqref{spinnematic} are the $(m,m')$-components of
spin matrices ${\rm f}_\nu$:
\begin{align}
 ({\rm f}_x)_{mm'} &= \frac{1}{2}\left[\sqrt{(f-m+1)(f+m)}\delta_{m-1,m'}+\sqrt{(f+m+1)(f-m)}\delta_{m+1,m'}\right],\\
 ({\rm f}_y)_{mm'} &= \frac{1}{2i}\left[\sqrt{(f-m+1)(f+m)}\delta_{m-1,m'}-\sqrt{(f+m+1)(f-m)}\delta_{m+1,m'}\right],\\
 ({\rm f}_z)_{mm'} &= m\delta_{mm'}.
\end{align}

We start from the completeness relation
\begin{align}
\hat{1}=\sum_{\mathcal{F}}\hat{\mathcal{P}}_{\mathcal{F}},
\label{completeness}
\end{align}
where $\hat{\mathcal{P}}_\mathcal{F}$ is defined in Eq.~\eqref{eq:def_mathcalP}. 
By applying $\hat{\psi}_{m_1}^\dagger(\bm r)\hat{\psi}_{m_2}^\dagger(\bm r')\langle f,m_1;f,m_2|$ from the left and $|f,m'_1;f,m'_2\rangle\hat{\psi}_{m'_2}(\bm r')\hat{\psi}_{m'_1}(\bm r)$ from the right of Eq.~\eqref{completeness},
and taking the summation with respect to $m_1, m_2, m'_1$ and $m'_2$,
we obtain
\begin{align}
:\hat{n}({\bm r})\hat{n}({\bm r}'):
=\sum_{\mathcal{F}=0,2,\cdots,2f}\sum_{\mathcal{M}=-\mathcal{F}}^\mathcal{F}
\hat{A}^\dagger_{\mathcal{F}\mathcal{M}}({\bm r},{\bm r}')\hat{A}_{\mathcal{F}\mathcal{M}}({\bm r},{\bm r}'),
\label{totaldensity}
\end{align}
where :~: denotes normal ordering that places annihilation operators to the right of creation operators.
Here, $\mathcal{F}$ in the summation on the right-hand side of Eq.~\eqref{totaldensity} takes only even values since $\hat{A}_{\mathcal{F},\mathcal{M}}({\bm r})=0$ if $\mathcal{F}$ is odd.

Another useful relation can be derived from the composition law of the angular momentum:
\begin{align}
{\bf f}_1\cdot{\bf f}_2
= \frac{1}{2}\left[ ({\bf f}_1+{\bf f}_2)^2-{\bf f}_1^2-{\bf f}_2^2 \right]
=\frac{1}{2}{\bf f}_{\rm tot}^2-f(f+1),
\end{align}
where  ${\bf f}_{\rm tot}={\bf f}_1 + {\bf f}_2$ is the total spin angular momentum vector.
Operating this on the completeness relation (\ref{completeness}), we obtain
\begin{align}
{\bf f}_1\cdot{\bf f}_2
=\sum_{\mathcal{F}}\left[\frac{1}{2}\mathcal{F}(\mathcal{F}+1)-f(f+1)\right]\hat{\mathcal{P}}_\mathcal{F}.
\label{AMR}
\end{align}
By calculating
\begin{align}
\sum_{m_1m_2m'_1m'_2}\hat{\psi}_{m_1}^\dagger(\bm r)\hat{\psi}_{m_2}^\dagger(\bm r')\langle f,m_1;f,m_2|\cdots |f,m'_1;f,m'_2\rangle\hat{\psi}_{m'_2}(\bm r')\hat{\psi}_{m'_1}(\bm r)
\end{align}
for Eq.~\eqref{AMR}, the following operator relation is obtained:
\begin{align}
:\hat{\bm F}({\bm r})\cdot\hat{\bm F}({\bm r}'):
=\sum_{\mathcal{F}=0,2,\cdots, 2f}\left[\frac{1}{2}\mathcal{F}(\mathcal{F}+1)-f(f+1)\right]\sum_{\mathcal{M}=-\mathcal{F}}^\mathcal{F}
\hat{A}^\dagger_{\mathcal{F}\mathcal{M}}({\bm r},{\bm r}')\hat{A}_{\mathcal{F}\mathcal{M}}({\bm r},{\bm r}'),
\label{eq:AMR_spinvector}
\end{align}
where we have used $\langle f,m_1;f,m_2|{\bf f}_1 \cdot {\bf f}_2|f,m'_1;f,m'_2\rangle = \sum_{\nu=x,y,z}({\rm f}_\nu)_{m_1m'_1}\cdot({\rm f}_\nu)_{m_2m'_2}$.
In a similar manner, using the relations
\begin{align}
&({\bf f}_1\cdot{\bf f}_2)^k
=\sum_{\mathcal{F}}\left[\frac{1}{2}\mathcal{F}(\mathcal{F}+1)-f(f+1)\right]^k\hat{\mathcal{P}}_\mathcal{F}\ \ \ (k=1,2,\cdots),
\label{AMR2}\\
&\langle f,m_1;f,m_2|({\bf f}_1\cdot{\bf f}_2)^k|f,m'_1;f,m'_2\rangle \nonumber\\
&= \sum_{\nu_1\nu_2\cdots\nu_k=x,y,z} \langle f,m_1;f,m_2|{\rm f}_{1\nu_1}{\rm f}_{2\nu_1}{\rm f}_{1\nu_2}{\rm f}_{2\nu_2}\cdots {\rm f}_{1\nu_k}{\rm f}_{2\nu_k}|f,m'_1;f,m'_2\rangle \nonumber\\
&= \sum_{\nu_1\nu_2\cdots\nu_k=x,y,z} ({\rm f}_{\nu_1}{\rm f}_{\nu_2}\cdots{\rm f}_{\nu_k})_{m_1m'_1}({\rm f}_{\nu_1}{\rm f}_{\nu_2}\cdots {\rm f}_{\nu_k})_{m_2m'_2},
\end{align}
and Eq.~\eqref{spinnematic},
we obtain
\begin{align}
&\sum_{\nu_1 \nu_2 \cdots\nu_k=x,y,z} :\hat{\mathcal{N}}_{\nu_1\nu_2\cdots\nu_k}^{(k)}({\bm r})\hat{\mathcal{N}}_{\nu_1\nu_2\cdots\nu_k}^{(k)}({\bm r}'):\nonumber\\
&=\sum_{\mathcal{F}=0,2,\cdots, 2f}\left[\frac{1}{2}\mathcal{F}(\mathcal{F}+1)-f(f+1)\right]^k\sum_{\mathcal{M}=-\mathcal{F}}^\mathcal{F}
\hat{A}^\dagger_{\mathcal{F}\mathcal{M}}({\bm r},{\bm r}')\hat{A}_{\mathcal{F}\mathcal{M}}({\bm r},{\bm r}').
\label{eq:AMR_spinnematic}
\end{align}
Equation~\eqref{eq:AMR_spinvector} is the spacial case of Eq.~\eqref{eq:AMR_spinnematic} with $k=1$.

Applying Eqs.~\eqref{totaldensity} and \eqref{eq:AMR_spinvector} to the case of $f=1$, we obtain
\begin{align}
:\hat{n}({\bm r})\cdot\hat{n}({\bm r}'):
&=
\sum_{\mathcal{M}}\hat{A}^\dagger_{2\mathcal{M}}({\bm r},{\bm r}')\hat{A}_{2\mathcal{M}}({\bm r},{\bm r}')
+\hat{A}^\dagger_{00}({\bm r},{\bm r}')\hat{A}_{00}({\bm r},{\bm r}'),\label{eq:spin1identity1}\\
:\hat{\bm F}({\bm r})\cdot\hat{\bm F}({\bm r}'):
&=
\sum_{\mathcal{M}}\hat{A}^\dagger_{2\mathcal{M}}({\bm r},{\bm r}')\hat{A}_{2\mathcal{M}}({\bm r},{\bm r}')
-2\hat{A}^\dagger_{00}({\bm r},{\bm r}')\hat{A}_{00}({\bm r},{\bm r}'),\label{eq:spin1identity2}
\end{align}
which lead to the following operator identity:
\begin{align}
:\hat{\bm F}({\bm r})\cdot\hat{\bm F}({\bm r}'):+3\hat{A}_{00}^\dagger({\bm r},{\bm r}')\hat{A}_{00}({\bm r},{\bm r}')
=:\hat{n}({\bm r})\hat{n}({\bm r}'):.
\label{eq:spin1identity3}
\end{align}
Thus, the spin density operator and the spin singlet-pair operator are not independent of each other but related by Eq.~\eqref{eq:spin1identity3}.

For the case of $f=2$, Eqs.~\eqref{totaldensity} and \eqref{eq:AMR_spinvector} respectively give
\begin{align}
:\hat{n}({\bm r})\cdot\hat{n}({\bm r}'):
&=
\sum_{\mathcal{M}}\hat{A}^\dagger_{4\mathcal{M}}({\bm r},{\bm r}')\hat{A}_{4\mathcal{M}}({\bm r},{\bm r}')
+\sum_{\mathcal{M}}\hat{A}^\dagger_{2\mathcal{M}}({\bm r},{\bm r}')\hat{A}_{2\mathcal{M}}({\bm r},{\bm r}')
+\hat{A}^\dagger_{00}({\bm r},{\bm r}')\hat{A}_{00}({\bm r},{\bm r}'),\label{eq:spin2identity1}\\
:\hat{\bm F}({\bm r})\cdot\hat{\bm F}({\bm r}'):
&=
4\sum_{\mathcal{M}}\hat{A}^\dagger_{4\mathcal{M}}({\bm r},{\bm r}')\hat{A}_{4\mathcal{M}}({\bm r},{\bm r}')
-3\sum_{\mathcal{M}}\hat{A}^\dagger_{2\mathcal{M}}({\bm r},{\bm r}')\hat{A}_{2\mathcal{M}}({\bm r},{\bm r}')
-6\hat{A}^\dagger_{00}({\bm r},{\bm r}')\hat{A}_{00}({\bm r},{\bm r}'). \label{eq:spin2identity2}
\end{align}
It follows from these results that
\begin{align}
:\hat{{\bm F}}({\bm r})\cdot\hat{{\bm F}}({\bm r}'):+
7\sum_{\mathcal{M}=-2}^2
\hat{A}_{2\mathcal{M}}^\dagger({\bm r},{\bm r}')\hat{A}_{2\mathcal{M}}({\bm r},{\bm r}')
+10\hat{A}_{00}^\dagger({\bm r},{\bm r}')\hat{A}_{00}({\bm r},{\bm r}')
=4:\hat{n}({\bm r})\hat{n}({\bm r}'):.
\label{spin2identity3}
\end{align}
Hence, for $f=2$, 
there are three independent operators $:\hat{n}({\bm r})\hat{n}({\bm r}'):$, $\hat{A}_{00}^\dagger({\bm r}, {\bm r}')\hat{A}_{00}({\bm r}, {\bm r}')$, and $:\hat{\bm F}({\bm r})\cdot \hat{\bm F}({\bm r}'):$.

For the case of $f=3$, Eqs.~\eqref{totaldensity}, \eqref{eq:AMR_spinvector}, and \eqref{eq:AMR_spinnematic} with $k=2$, respectively, give
\begin{align}
:\hat{n}({\bm r})\cdot\hat{n}({\bm r}'):
=&
\sum_{\mathcal{M}}\hat{A}^\dagger_{6\mathcal{M}}({\bm r},{\bm r}')\hat{A}_{6\mathcal{M}}({\bm r},{\bm r}')
+\sum_{\mathcal{M}}\hat{A}^\dagger_{4\mathcal{M}}({\bm r},{\bm r}')\hat{A}_{4\mathcal{M}}({\bm r},{\bm r}')\nonumber\\
&+\sum_{\mathcal{M}}\hat{A}^\dagger_{2\mathcal{M}}({\bm r},{\bm r}')\hat{A}_{2\mathcal{M}}({\bm r},{\bm r}')
+\hat{A}^\dagger_{00}({\bm r},{\bm r}')\hat{A}_{00}({\bm r},{\bm r}'),\label{eq:spin3identity1}\\
:\hat{\bm F}({\bm r})\cdot\hat{\bm F}({\bm r}'):
=&
9\sum_{\mathcal{M}}\hat{A}^\dagger_{6\mathcal{M}}({\bm r},{\bm r}')\hat{A}_{6\mathcal{M}}({\bm r},{\bm r}')
-2\sum_{\mathcal{M}}\hat{A}^\dagger_{4\mathcal{M}}({\bm r},{\bm r}')\hat{A}_{4\mathcal{M}}({\bm r},{\bm r}')\nonumber\\
&-9\sum_{\mathcal{M}}\hat{A}^\dagger_{2\mathcal{M}}({\bm r},{\bm r}')\hat{A}_{2\mathcal{M}}({\bm r},{\bm r}')
-12\hat{A}^\dagger_{00}({\bm r},{\bm r}')\hat{A}_{00}({\bm r},{\bm r}'),\label{eq:spin3identity2}\\
\sum_{\nu\nu'}:\hat{\mathcal{N}}_{\nu\nu'}^{(2)}({\bm r})\hat{\mathcal{N}}_{\nu\nu'}^{(2)}({\bm r}'):
=&
81\sum_{\mathcal{M}}\hat{A}^\dagger_{6\mathcal{M}}({\bm r},{\bm r}')\hat{A}_{6\mathcal{M}}({\bm r},{\bm r}')
+4\sum_{\mathcal{M}}\hat{A}^\dagger_{4\mathcal{M}}({\bm r},{\bm r}')\hat{A}_{4\mathcal{M}}({\bm r},{\bm r}')\nonumber\\
&+81\sum_{\mathcal{M}}\hat{A}^\dagger_{2\mathcal{M}}({\bm r},{\bm r}')\hat{A}_{2\mathcal{M}}({\bm r},{\bm r}')
+144\hat{A}^\dagger_{00}({\bm r},{\bm r}')\hat{A}_{00}({\bm r},{\bm r}'). \label{eq:spin3identity3}
\end{align}
Thus, four of the seven operators appearing in Eqs~\eqref{eq:spin3identity1}-\eqref{eq:spin3identity3} 
[namely, $\sum_{\mathcal{M}}\hat{A}_{\mathcal{FM}}^\dagger({\bm r},{\bm r}')\hat{A}_{\mathcal{FM}}({\bm r},{\bm r}')$ ($\mathcal{F}=0,2,4$, and 6), $:\hat{n}({\bm r})\hat{n}({\bm r}'):$, $:\hat{\bm F}({\bm r})\cdot \hat{\bm F}({\bm r}'):$,
and $\sum_{\nu\nu'}:\hat{\mathcal{N}}_{\nu\nu'}^{(2)}({\bm r})\hat{\mathcal{N}}_{\nu\nu'}^{(2)}({\bm r}'):$] are independent of each other.

In a similar manner, in general, 
the operator $\sum_{\mathcal{M}=-\mathcal{F}}^\mathcal{F} \hat{A}_{\mathcal{F},\mathcal{M}}^\dagger({\bm r},{\bm r}')\hat{A}_{\mathcal{F},\mathcal{M}}({\bm r},{\bm r}')$ for spin $f\ge 2$
can be expressed in terms of $f+1$ operators:
the total density operator, the spin-singlet pair operator, the spin density operator, and rank-$k$ ($2\le k\le f-1$) spin nematic tensor operators.
The nematic tensor of rank higher than $f-1$ are written with these $f+1$ operators.

\subsubsection{Interaction Hamiltonian of spin-1 BECs}
We derive the interaction Hamiltonians for the $f=1$ case, which was first derived by Ohmi and Machida~\cite{Ohmi1998} and Ho~\cite{Ho1998}.
For $f=1$, the total spin $\mathcal{F}$ of two colliding bosons must be $0$ or $2$, and the corresponding interaction Hamiltonian (\ref{intvF}) gives
\begin{align}
\hat{V}^{(0)}
&=\frac{g_0}{2} \int d{\bm r} \hat{A}_{00}^\dagger({\bm r}) \hat{A}_{00}({\bm r})
\label{V0}
\\
\hat{V}^{(2)}
&=\frac{g_2}{2} \int d{\bm r} \sum_{\mathcal{M}=-2}^2 
\hat{A}_{2\mathcal{M}}^\dagger({\bm r}) \hat{A}_{2\mathcal{M}}({\bm r})
\nonumber \\
&=\frac{g_2}{2} \int d{\bm r} 
\left[ : \hat{n}^2({\bm r}) : - \hat{A}_{00}^\dagger({\bm r}) \hat{A}_{00}({\bm r}) \right ],
\label{V2}
\end{align}
where $\hat{A}_{\mathcal{F}\mathcal{M}}({\bm r})\equiv \hat{A}_{\mathcal{F}\mathcal{M}}({\bm r},{\bm r})$, and
Eq.~(\ref{eq:spin1identity1}) is used to derive the last equality. Combining Eqs.~(\ref{V0}) and (\ref{V2}), we obtain
\begin{align}
\hat{V}
=\int d{\bm r} 
\left[\frac{g_2}{2} : \hat{n}^2 ({\bm r}) : + \frac{g_0-g_2}{2} 
\hat{A}_{00}^\dagger({\bm r}) \hat{A}_{00}({\bm r}) \right].
\label{V(f=1)}
\end{align}
We use the operator identity~\eqref{eq:spin1identity3} to rewrite Eq.~(\ref{V(f=1)}) as 
\begin{align}
\hat{V}
=\frac{1}{2}\int d{\bm r} 
\left[ c_0 : \hat{n}^2 ({\bm r}) : + c_1 :\hat{\bm F}^2({\bm r}): \right],
\label{V(f=1)2}
\end{align}
where 
\begin{align}
c_0=\frac{g_0+2g_2}{3}, \ \ 
c_1=\frac{g_2-g_0}{3}.
\label{c(f=1)}
\end{align}

\subsubsection{Interaction Hamiltonian of spin-2 BECs}
For $f=2$, $\mathcal{F}$ must be 0, 2, or 4, and the interaction Hamiltonian~\eqref{sumV^F} is given by
\begin{align}
\hat{V} 
&= \hat{V}^{(0)} + \hat{V}^{(2)} + \hat{V}^{(4)}
\nonumber \\
&= \frac{1}{2}\int d{\bm r}
\left[ 
   g_0\hat{A}_{00}^\dagger({\bm r}) \hat{A}_{00}({\bm r})
 + g_2\sum_{\mathcal{M}=-2}^2\hat{A}_{2\mathcal{M}}^\dagger({\bm r}) \hat{A}_{2\mathcal{M}}({\bm r}) 
 + g_4\sum_{\mathcal{M}=-4}^4\hat{A}_{4\mathcal{M}}^\dagger({\bm r}) \hat{A}_{4\mathcal{M}}({\bm r}) 
\right].
\label{V(f=2)'}
\end{align}
We can eliminate the second and third terms in the right-hand side of Eq.~\eqref{V(f=2)'} by using the operator identities~\eqref{eq:spin2identity1} and \eqref{eq:spin2identity2}, obtaining
\begin{align}
\hat{V}
&=\frac{1}{2}\int \!d{\bm r}
\left[
  c_0:\hat{n}^2({\bm r}):
+ c_1:\hat{{\bm F}}^2({\bm r}):
+ c_2\hat{A}_{00}^\dagger({\bm r})\hat{A}_{00}({\bm r})
\right],
\label{V(f=2)}
\end{align}
where 
\begin{align}
c_0=\frac{4g_2+3g_4}{7}, \ \ 
c_1=\frac{g_4-g_2}{7}, \ \ 
c_2=\frac{7g_0-10g_2+3g_4}{7}.
\label{c(f=2)}
\end{align}
We use the same notations $c_0$ and $c_1$ for both spin-1 and spin-2 cases because there is no fear of confusion.

\subsubsection{Interaction Hamiltonian of spin-3 BECs}

In a similar manner, by using the operator identities~\eqref{eq:spin3identity1} and \eqref{eq:spin3identity2},
the interaction Hamiltonian of the spin-3 spinor BEC is given by~\cite{Diener2006,Santos2006}
\begin{align}
\hat{V}
=\frac{1}{2}\int \!d{\bm r}
\left[
  c_0:\hat{n}^2({\bm r}):
+ c_1:\hat{{\bm F}}^2({\bm r}):
+ c_2\hat{A}_{00}^\dagger({\bm r})\hat{A}_{00}({\bm r}) 
+ c_3\sum_{\mathcal{M}=-2}^2\hat{A}^\dagger_{2\mathcal{M}}({\bm r})\hat{A}_{2\mathcal{M}}({\bm r})
\right],
\label{V(f=3)}
\end{align}
where 
\begin{align}
c_0=\frac{9g_4+2g_6}{11}, \ \ 
c_1=\frac{g_6-g_4}{11}, \ \ 
c_2=\frac{11g_0-21g_4+10g_6}{11}, \ \ 
c_3=\frac{11g_2-18g_4+7g_6}{11}.
\label{c(f=3)}
\end{align}
When we use the identity~\eqref{eq:spin3identity3}, 
the interaction Hamiltonian is also written as
\begin{align}
\hat{V}
=\frac{1}{2}\int \!d{\bm r}
\left\{
  \tilde{c}_0:\hat{n}^2({\bm r}):
+ \tilde{c}_1:\hat{{\bm F}}^2({\bm r}):
+ \tilde{c}_2\hat{A}_{00}^\dagger({\bm r})\hat{A}_{00}({\bm r}) 
+ \tilde{c}_3 :{\rm Tr}[\hat{\mathcal{N}}^2({\bm r})]:
\right\},
\label{V(f=3)2}
\end{align}
where $\hat{\mathcal{N}}$ is the symmetrized rank-2 tensor:
\begin{align}
 \hat{\mathcal{N}}_{\nu_1\nu_2} &\equiv \frac{\hat{\mathcal{N}}^{(2)}_{\nu_1\nu_2}+\hat{\mathcal{N}}^{(2)}_{\nu_2\nu_1}}{2}
= \hat{\mathcal{N}}^{(2)}_{\nu_1\nu_2}- \frac{i}{2}\sum_{\nu_3=x,y,z}\epsilon_{\nu_1\nu_2\nu_3}\hat{f}_{\nu_3},
\label{hatNsym}
\end{align}
and the coefficients are given by
\begin{align}
\tilde{c}_0=c_0-\frac{c_3}{7},\ \ 
\tilde{c}_1=c_1-\frac{5c_3}{84},\ \ 
\tilde{c}_2=c_2-\frac{5c_3}{3}, \ \ 
\tilde{c}_3=\frac{c_3}{126}.
\label{c(f=3)2}
\end{align}
Here, the commutation relation of spin matrices, $[{\rm f}_{\nu_1},{\rm f}_{\nu_2}]=i\sum_{\nu_3=x,y,z}\epsilon_{\nu_1\nu_2\nu_3}{\rm f}_{\nu_3}$, is used to obtain the last equality of Eq.~\eqref{hatNsym}.

%% file: meanfield.tex
\section{Mean-field theory of spinor condensates}
\label{sec:meanfield}
\subsection{Number-conserving theory}
The mean-field theory is usually obtained by replacing the field operator with its expectation value $\langle\hat{\psi}_m\rangle$. This recipe, though widely used and technically convenient, has one conceptual difficulty; that is, it breaks the global U(1) gauge invariance, which implies that the number of atoms is not conserved. However, in reality, the number of atoms is strictly conserved, as are the baryon (proton and neutron) and lepton (electron) numbers. In fact, it is possible to construct the mean-field theory without breaking the U(1) gauge symmetry~\cite{Koashi2000,Ueda2002}.

To construct a number-conserving mean-field theory, we first expand the field operator in terms of a complete orthonormal set of basis functions $\{ \varphi_{mi} ({\bm r}) \}$:
\begin{eqnarray}
\hat{\psi}_m ({\bm r})
=\sum_i \hat{a}_{mi} \varphi_{mi} ({\bm r}) \ \ (m=f,f-1,\cdots, -f),
\label{expansion}
\end{eqnarray}
where $\varphi_{mi}({\bm r})$ describes the basis function for the magnetic quantum number $m$ and spatial mode $i$, 
and $\hat{a}_{mi}$'s are the corresponding annihilation operators that satisfy the canonical commutation relations
\begin{eqnarray}
[ \hat{a}_{mi}, \hat{a}_{m'j}^\dagger ]
= \delta_{mm'} \delta_{ij}, \ \ \ 
[ \hat{a}_{mi}, \hat{a}_{m'j} ]
= [ \hat{a}_{mi}^\dagger, \hat{a}_{m'j}^\dagger ]
=0.
\label{eqMFT39}
\end{eqnarray}
The basis functions are assumed to satisfy the orthonormality conditions
\begin{eqnarray}
\int d{\bm r} \; \varphi_{mi}^\ast ({\bm r}) \; \varphi_{mj} ({\bm r}) = \delta_{ij},
\label{orthonormality}
\end{eqnarray}
and the completeness relation
\begin{eqnarray}
\sum_i \varphi_{mi} ({\bm r}) \varphi_{mi}^\ast({\bm r}')
= \delta ({\bm r}-{\bm r}').
\label{completeness2}
\end{eqnarray}
Then, the field operators (\ref{expansion}) satisfy the field commutation relations (\ref{fieldCR}).

In the mean-field approximation, it is assumed that all Bose-condensed bosons occupy a single spatial mode, say $i=0$, and a single spin state specified by a linear superposition of magnetic sublevels. Then, the state vector is given by
\begin{eqnarray}
| \boldsymbol{\zeta} \rangle
= \frac{1}{\sqrt{N!}}
\left( \sum_{m=-f}^f \zeta_m \hat{a}_{m0}^\dagger \right)^N
| {\rm vac} \rangle,
\label{statevector}
\end{eqnarray}
where $| {\rm vac} \rangle$ denotes the particle vacuum and $\zeta_m$'s are assumed to satisfy the following normalization condition:
\begin{eqnarray}
\sum_{m=-f}^f|\zeta_m|^2=1.
\label{normalization}
\end{eqnarray}
It is straightforward to show that
\begin{align}
\langle \hat{\psi}_m ({\bm r}) \rangle_0 
&= \langle \hat{\psi}_m^\dagger ({\bm r}) \rangle_0 =0, 
\label{MFa} \\
\langle \hat{\psi}_m^\dagger ({\bm r}) \hat{\psi}_{m'} ({\bm r}') \rangle_0  
&=\psi_m^\ast ({\bm r}) \psi_{m'} ({\bm r}'), 
\label{MFb} \\
\langle \hat{\psi}_{m_1}^\dagger ({\bm r}) \hat{\psi}_{m_2}^\dagger ({\bm r}') 
\hat{\psi}_{m_2'} ({\bm r}'')  \hat{\psi}_{m_1'} ({\bm r}''') \rangle_0 
&= \left(1-\frac{1}{N}\right)\psi_{m_1}^\ast ({\bm r}) \psi_{m_2}^\ast ({\bm r}') \psi_{m_2'} ({\bm r}'') \psi_{m_1'} ({\bm r}''') ,
\label{MFc}
\end{align}
where $\langle \cdots\cdot \rangle_0 \equiv \langle \boldsymbol{\zeta} | \cdots\cdot | \boldsymbol{\zeta} \rangle$ and
\begin{eqnarray}
\psi_m ({\bm r}) = \sqrt{N} \zeta_m \varphi_{m0} ({\bm r}).
\label{GPWF}
\end{eqnarray}
As shown in Eq.~(\ref{MFa}), the expectation value of the field operator vanishes as it should in a number-conserving theory. Nevertheless, all experimentally observable physical quantities, which are expressed in terms of the correlation function of the field operators such as (\ref{MFb}) and (\ref{MFc}), can have nonzero values. These values agree with those obtained by the U(1) symmetry-breaking approach except for the factor of $1/N$ in Eq.~(\ref{MFc}). 

\subsection{Graphical representation of a mean-field state}
In the spin-$f$ system, the mean-field state can be described with a $(2f+1)$-component order parameter:
\begin{align}
\bm\psi(\bm r)\equiv (\psi_f(\bm r), \psi_{f-1}(\bm r),\dots, \psi_{-f}(\bm r))^{\rm T}.
\end{align}
Before discussing the detail of the mean-field theory,
we introduce two methods to visualize the mean-field spinor state,
which help us identify the symmetry structure of a spin-$f$ condensate.

\subsubsection{Spherical-harmonic representation}
\begin{figure}[ht]
\begin{center}
\resizebox{0.8\hsize}{!}{
\includegraphics{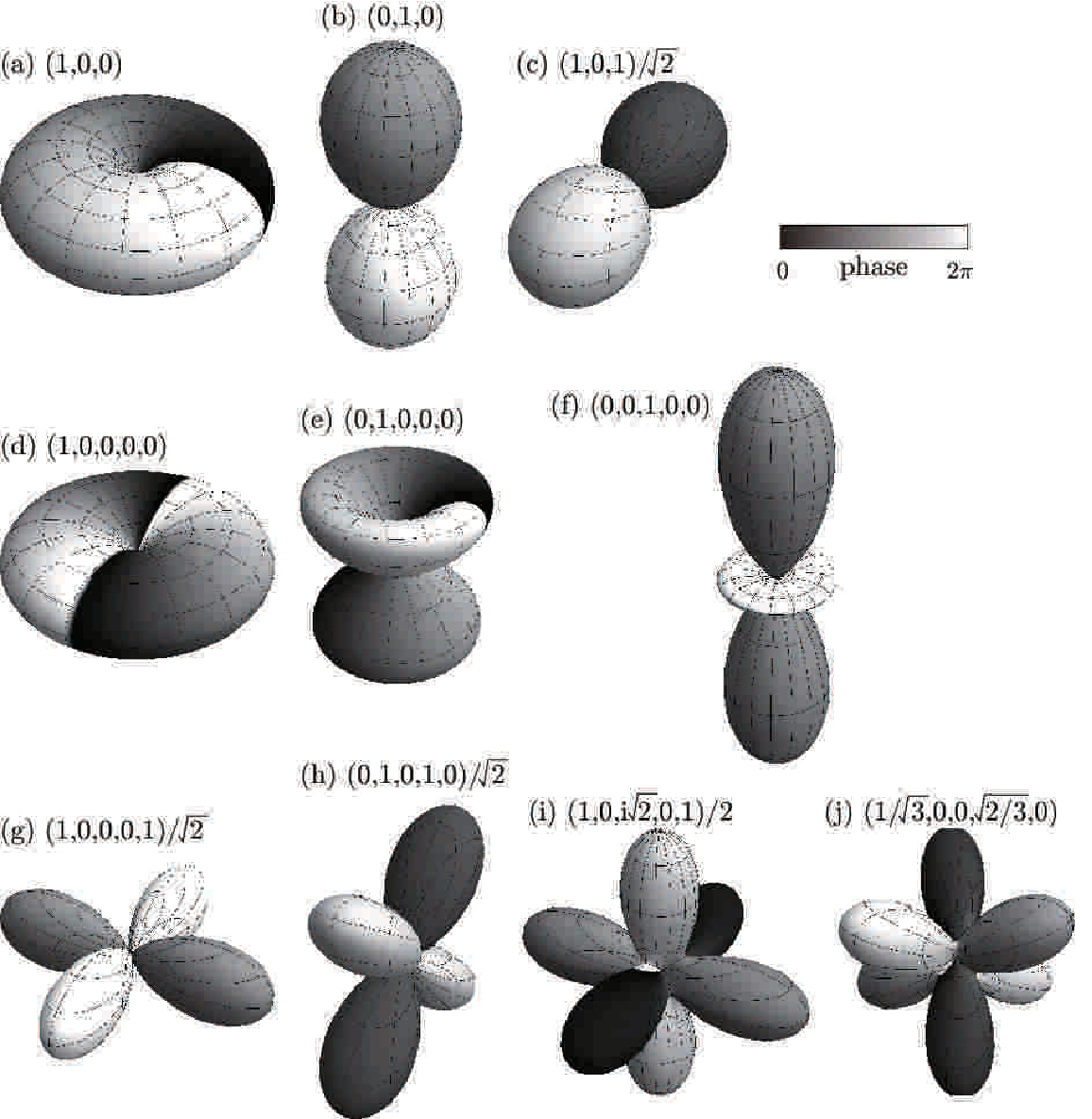}
}
\end{center}
\caption{Spherical-harmonic representation for (a)--(c) $f=1$ and (d)--(j) $f=2$ spinor order parameters.
Shown are the surface plots of $|\Psi(\hat{s})|^2$ defined in Eq.~\eqref{eq:shr_psi}, where the gray scale on the surface represents ${\rm arg}\Psi(\hat{s})$.
}
\label{fig:SHR}
\end{figure}

Because an integer-spin state can be described in terms of the spherical harmonics $Y_f^m(\hat{s})$, where $\hat{s}$ is a unit vector in spin space,
the order parameter for a spin-$f$ system can be expressed in terms of a complex wave function given by
\begin{align}
\Psi(\hat{s}) = \sum_m \psi_m Y_f^m(\hat{s}).
\label{eq:shr_psi}
\end{align}
Figure~\ref{fig:SHR} illustrates examples of spin-1 and spin-2 order parameters,
where the surface plot of $|\Psi(\hat{s})|^2$ is shown with a gray scale representing ${\rm arg}\Psi(\hat{s})$.
The order parameters of Figs.~\ref{fig:SHR} (b) $(0,1,0)$ and (c) $(1,0,1)/\sqrt{2}$ are different, but they transform into each other via a rotation in spin space.
Similar situations hold in Figs.~\ref{fig:SHR} (g) and (h), and Figs.~\ref{fig:SHR} (i) and (j).
On the other hand, $(0,0,1,0,0)$ [Fig.~\ref{fig:SHR} (f)] and $(1,0,0,0,1)/\sqrt{2}$ [Fig.~\ref{fig:SHR} (g)] are different in their symmetries, as is obvious from the figures.

\subsubsection{Majorana representation}
\label{sec:Majorana}
Another method is a geometrical representation invented by Majorana~\cite{Majorana1932,Zhou2001,Barnett2006}.
A state of the spin-$f$ system can be specified by providing a symmetric configuration of $2f$ spin-1/2 systems, except for an overall phase.
Since the state of a spin-1/2 system can be described by the unit-sphere Bloch vector,
the state of a spin-$f$ system can be described by $2f$ vertices on the unit sphere.

\begin{figure}[ht]
\begin{center}
\resizebox{0.8\hsize}{!}{
\includegraphics{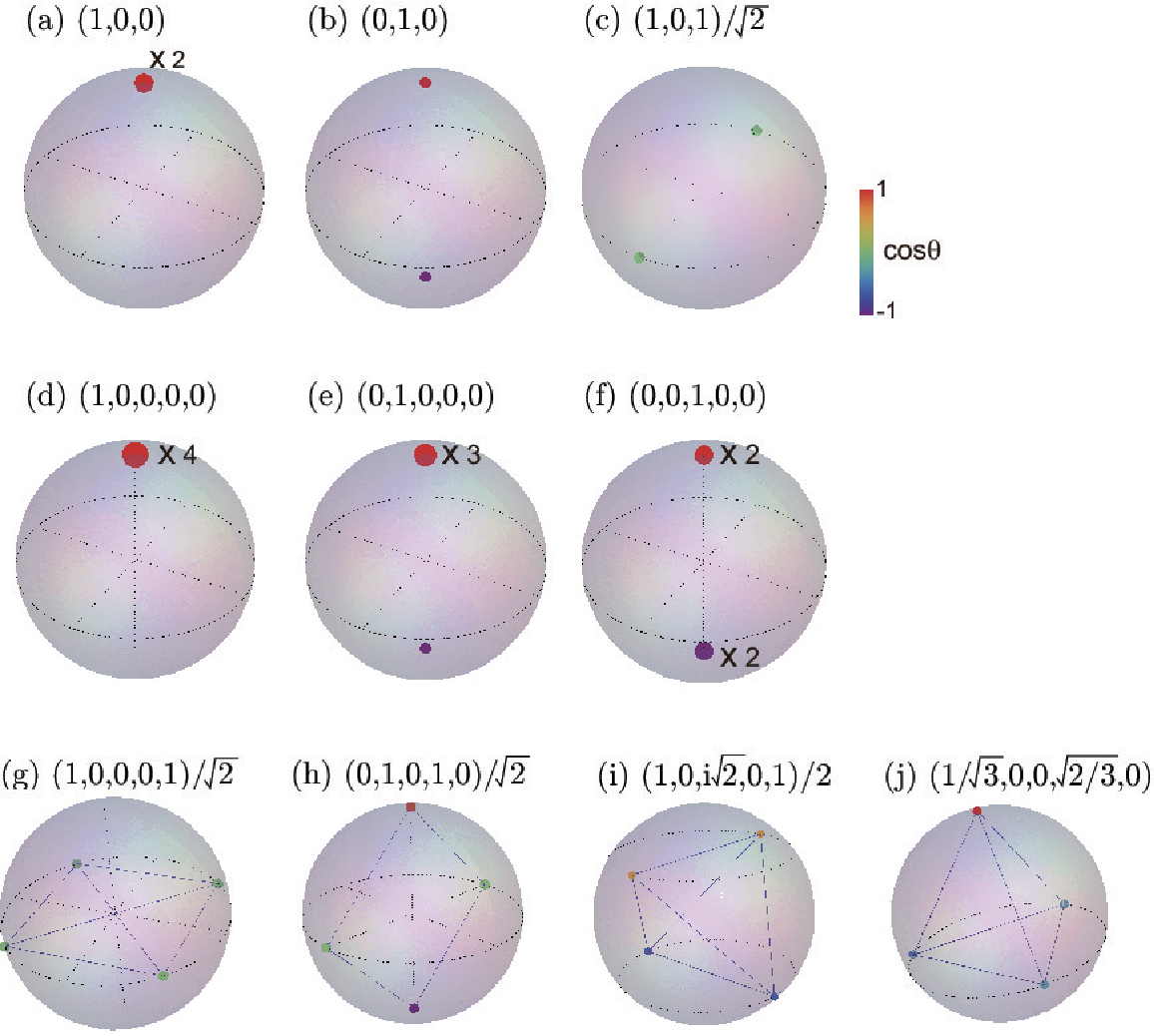}
}
\end{center}
\caption{Majorana representation for (a)--(c) $f=1$  and (d)--(j) $f=2$ spinor order parameters.
The color of the points shows the value of $\cos\theta=(1-|z|^2)/(1+|z|^2)$ according to the color gauge.
When the polynomial $P^{(f)}_\psi(z)$ has an $n$-multiple root, the root is indicated with $\times n$.
The time-reversal symmetry is broken in (a), (d), (e), (i) and (j), since the antipodal map does not leave the configurations of vertices unaltered.
}
\label{fig:MR}
\end{figure}

We consider a polynomial of degree $2f$ for a given order parameter $\bm\psi$:
\begin{align} 
P^{(f)}_{\bm \psi}(z) = \sum_{\alpha=0}^{2F} \sqrt{\begin{pmatrix} 2F \\ \alpha \end{pmatrix}} \psi_{F-\alpha}^* z^{\alpha}.
\end{align}
Then, the $2f$ complex roots of $P^{(f)}_{\bm \psi}(z)=0$ give $2f$ vertices on the unit sphere 
through the stereographic mapping $z=\tan(\theta/2)e^{i\phi}$.
For the cases of spin $f=1, 2$ and 3, the polynomials $P^{(f)}_{\bm\psi}(z)$ are given, respectively, by
\begin{align}
 P^{(1)}_{\bm\psi}(z) =& \psi^*_1 z^2 + \sqrt{2} \psi^*_0 z + \psi^*_{-1},\\
 P^{(2)}_{\bm\psi}(z) =& \psi^*_2 z^4 + 2 \psi^*_1 z^3 + \sqrt{6}\psi^*_0 z^2 + 2\psi^*_{-1} z + \psi^*_{-2},\\
 P^{(3)}_{\bm\psi}(z) =& \psi^*_3 z^6 + \sqrt{6} \psi^*_2 z^5 + \sqrt{15}\psi^*_1 z^4 + \sqrt{20} \psi^*_0 z^3
+ \sqrt{15} \psi^*_{-1} z^2  + \sqrt{6} \psi^*_{-2} z +\psi^*_{-3}.
\end{align}
Note that if $z = z_0$ is a root of $P^{(f)}_{\bm\psi}(z)=0$,
then $z = -1/z_0^*$ is a root of the polynomial for the time-reversed state, $P^{(f)}_{\mathcal{T}\bm\psi}(z)=0$,
which corresponds to the antipole of $z_0$ on the unit sphere. Here, the time-reversal operator $\mathcal{T}$ is defined by
\begin{align}
 (\mathcal{T}\bm \psi)_m = (-1)^m\psi_{-m}^*.
\label{eq:def_mathcalT}
\end{align}
Hence, the time-reversed state is described with the antipoles of vertices of the original state.
The Majorana representations for the same order parameters as in Fig.~\ref{fig:SHR} are shown in Fig.~\ref{fig:MR}.
The symmetry in spin space is clearer in the Majorana representation than the spherical-harmonic representation, whereas the information on the condensate phase is left out in the Majorana representation~\cite{Yip2006b,Barnett2006b}.

\subsection{Mean-field theory of spin-1 BECs}
\label{sec:MFTspin1}

\subsubsection{Gross-Pitaevskii equations}
We use Eqs.~(\ref{MFb}) and (\ref{MFc}) to evaluate the expectation value of the Hamiltonian $\hat{H}=\hat{H}_0+V$ over the state (\ref{statevector}) with $f=1$, where $\hat{H}_0$ and $\hat{V}$ are given in Eqs.~(\ref{H_0}) and (\ref{V(f=1)2}), respectively. By neglecting the terms of the order of $1/N$ in Eq.~(\ref{MFc}), we obtain
\begin{eqnarray}
E[\psi] \equiv \langle \hat{H} \rangle_0 
= \int d{\bm r} 
\Bigg\{ \sum_{m=-1}^1 \psi_m^\ast \left[ -\frac{\hbar^2 \nabla^2}{2M} + U_{\rm trap}({\bm r}) - pm + qm^2 \right] \psi_m  
 +\frac{c_0}{2}n^2 + \frac{c_1}{2} | {\bm F} |^2 \Bigg\},
\label{energy_functional(f=1)}
\end{eqnarray}
where 
\begin{eqnarray}
n({\bm r}) \equiv \langle \hat{n}({\bm r}) \rangle_0 = \sum_{m=-1}^{1} | \psi_m({\bm r}) |^2
\label{density}
\end{eqnarray}
is the particle density and 
${\bm F} = (F_x,F_y,F_z)$ is the spin density vector defined by
\begin{eqnarray}
F_\nu({\bm r}) \equiv \langle \hat{F}_\nu({\bm r}) \rangle_0
= \sum_{m, m'=-1}^{1} \psi_m^\ast({\bm r}) ({\rm f}_\nu)_{mm'} \psi_{m'}({\bm r}) \ \ \ 
(\nu = x, y, z).
\label{spindensity2}
\end{eqnarray}
Because the spin-1 matrices are given in the irreducible representation by
\begin{eqnarray}
{\rm f}_x =\frac{1}{\sqrt{2}}
\begin{pmatrix}
	0 & 1 & 0 \\
	1 & 0 & 1 \\
	0 & 1 & 0 
\end{pmatrix}, \ \ 
{\rm f}_y =\frac{i}{\sqrt{2}}
\begin{pmatrix}
	0 & -1 & 0 \\
	1 & 0 & -1 \\
	0 & 1 & 0
\end{pmatrix},\ \ 
{\rm f}_z= \begin{pmatrix}
	1 & 0 & 0 \\
	0 & 0 & 0 \\
	0 & 0 & -1
\end{pmatrix},
\label{spin1matrices}
\end{eqnarray}
the components of the spin vector ${\bm F}$ are expressed as
\begin{align}
F_x &= \frac{1}{\sqrt{2}} 
\left[ \psi_1^\ast \psi_0 + \psi_0^\ast (\psi_1 + \psi_{-1} ) + \psi_{-1}^\ast \psi_0 \right],
\label{fx} \\
F_y &= \frac{i}{\sqrt{2}}
\left[ -\psi_1^\ast \psi_0 + \psi_0^\ast (\psi_1 - \psi_{-1}) + \psi_{-1}^\ast \psi_0 \right],
\label{fy} \\
F_z &= |\psi_1|^2 - |\psi_{-1}|^2.
\label{fz}
\end{align}
From the energy functional in Eq.~(\ref{energy_functional(f=1)}), we find that $c_0$ must be nonnegative; otherwise, the system collapses. This condition also guarantees that the Bogoliubov excitation energy corresponding to density fluctuations is real, as shown in Sec.~\ref{sec:Bogoliubov}.

The time evolution of the mean field is governed by
\begin{eqnarray}
i\hbar \frac{\partial \psi_m ({\bm r})}{\partial t}
= \frac{\delta E}{\delta \psi_m^\ast ({\bm r})}.
\label{dynamics}
\end{eqnarray}
Substituting Eq.~(\ref{energy_functional(f=1)}) in the right-hand side of Eq.~(\ref{dynamics}) gives
\begin{align}
i\hbar \frac{\partial \psi_m}{\partial t}
=& \left[ -\frac{\hbar^2 \nabla^2}{2M} + U_{\rm trap}({\bm r}) - pm + qm^2 \right] \psi_m
\nonumber \\
& + c_0 n \psi_m + c_1 \sum_{m'=-1}^1 
{\bm F} \cdot {\bf f}_{mm'} \psi_{m'} \ \ (m=1,0,-1),
\label{spin-1GPE}
\end{align}
which are the multi-component Gross-Pitaevskii equations (GPEs) that describe the mean-field dynamics of spin-1 Bose-Einstein condensates~\cite{Ohmi1998,Ho1998}.
In a stationary state, we substitute
\begin{eqnarray}
\psi_m ({\bm r}, t) = \psi_m({\bm r}) \; e^{-i \mu t/\hbar} 
\end{eqnarray}
in Eq.~(\ref{spin-1GPE}), where $\mu$ is the chemical potential, obtaining
\begin{eqnarray}
\left[ -\frac{\hbar^2 \nabla^2}{2M} + U_{\rm trap}({\bm r}) - pm + qm^2 \right]
\psi_m + c_0 n \psi_m + c_1 \sum_{m'=-1}^1 {\bm F} \cdot {\bf f}_{mm'} \psi_{m'} 
 = \mu \psi_m.
\label{stationaryGPE(f=1)}
\end{eqnarray}
Writing down the three components $m=1,0,-1$ explicitly, we obtain
\begin{align}
\left[ -\frac{\hbar^2 \nabla^2}{2M} + U_{\rm trap}({\bm r}) - p+q+c_0n+c_1 F_z  - \mu \right] \psi_1 +\frac{c_1}{\sqrt{2}}  F_-  \psi_0  &=0,
\label{eqMFT55}
\\ 
\frac{c_1}{\sqrt{2}}  F_+  \psi_1 +
\left[ -\frac{\hbar^2 \nabla^2}{2M} + U_{\rm trap}({\bm r}) +c_0n - \mu \right] \psi_0
+\frac{c_1}{\sqrt{2}}  F_-  \psi_{-1}
&=0 ,
\label{eqMFT56}
\\
\frac{c_1}{\sqrt{2}}  F_+  \psi_0 + 
\left[ -\frac{\hbar^2 \nabla^2}{2M} + U_{\rm trap}({\bm r}) + p+q+c_0n-c_1  F_z  -\mu \right]  \psi_{-1} &=0,
\label{eqMFT57}
\end{align}
where $F_\pm  \equiv  F_x  \pm i  F_y$.
By solving the set of equations (\ref{eqMFT55})--(\ref{eqMFT57}), we can investigate the properties of the ground and excited states of spin-1 condensates.

\subsubsection{Ground states in a uniform system}
When we consider a uniform system with a fixed number density $n$,
we rewrite the order parameter with a normalized spinor $\zeta_m$ as
\begin{align}
 \psi_m = \sqrt{n}\zeta_m,
\end{align}
and define the spin expectation value per particle
\begin{align}
 \bm f = \sum_{mm'} \zeta_m^* {\bf f}_{mm'}\zeta_{m'},
\end{align}
which is related to $\bm F$ through $\bm F=n\bm f$.

It is straightforward to identify the ground-state magnetism of a uniform system when the Zeeman terms are negligible ($p=q=0$). From the energy functional (\ref{energy_functional(f=1)}), 
one can see that the ground state is ferromagnetic ($|{\bm f}|=1$) when $c_1<0$ and antiferromagnetic or polar ($|{\bm f}|=0$) when $c_1>0$. 
Here, the order parameter of the antiferromagnetic phase~\cite{Ohmi1998} is given by $\bm\zeta = (1,0,1)^{\rm T}/\sqrt{2}$ and that of the polar state~\cite{Ho1998} is $(0,1,0)^{\rm T}$. 
The former state is obtained by rotating the latter about the $y$ axis by $\pi/2$ [see Figs.~\ref{fig:SHR}(b) and \ref{fig:SHR}(c)], and therefore, these two states are equivalent and degenerate in the absence of an external magnetic field. 

Note that because the system is suspended in a vacuum chamber, the projected spin angular momentum in the direction of the magnetic field is conserved for a long time ($\ge 1$~s)~\cite{Chang2004}.
Because the atomic interaction given by Eq.~\eqref{intvF} conserves the total spin of two colliding atoms,
the spin dynamics obeying the GPEs conserves the total magnetization.
Therefore, when we search for the ground state for a given magnetization $M_z=\int d{\bm r}F_z({\bm r})$, 
we should replace $E$ in Eq.~\eqref{dynamics} with $E-\lambda M_z$, where $\lambda$ is a Lagrange multiplier.
Then, $p$ in Eqs.~\eqref{eqMFT55}--\eqref{eqMFT57} is replaced by $\tilde{p}=p+\lambda$, which is determined as a function of $M_z$.

Here, we solve Eqs.~\eqref{eqMFT55}--\eqref{eqMFT57} 
and investigate the ground-state phase diagram~\cite{Stenger1998,Zhang2003,Murata2007} in a parameter space of $(q, p)$.
We assume a uniform system for a given number density $n$
and solve Eqs.~\eqref{eqMFT55}--\eqref{eqMFT57} by
setting the kinetic energy and $U_{\rm trap}({\bm r})$ to be zero.
The ground state is obtained by comparing the energy per particle which is given by
\begin{align}
\epsilon = \sum_m (-pm+qm^2)|\zeta_m|^2 + \frac{1}{2}c_0 n + \frac{1}{2}c_1 n|{\bm f}|^2.
\label{eq:spin1MFenergy}
\end{align}
Note that we may assume $f_y =0$ without loss of generality, because the system is symmetric about the $z$ axis.
Then, if we choose the overall phase so that $\zeta_0$ is real, $\zeta_{\pm 1}$ satisfy ${\rm Im}(\zeta_1)={\rm Im}(\zeta_{-1})$.
Rewriting $\zeta_{\pm1}$ as $\zeta_{\pm 1}=\zeta_{\pm1}^{\rm Re}+i\delta$, we obtain
\begin{align}
(-p+q+c_1 n f_z -\tilde{\mu}) \zeta_1 + c_1 n (\zeta_1^{\rm Re} + \zeta_{-1}^{\rm Re}) \zeta_0^2 &=0,
\label{f=1GPE1}\\
[ \tilde{\mu} - c_1 n (\zeta_1^{\rm Re} + \zeta_{-1}^{\rm Re})(\zeta_1^{\rm Re}+\zeta_{-1}^{\rm Re}+2i\delta) ] \zeta_0 &=0,
\label{f=1GPE2}\\
c_1 n (\zeta_1^{\rm Re} + \zeta_{-1}^{\rm Re} ) \zeta_0^2 +( p+q-c_1 n f_z  - \tilde{\mu} ) \zeta_{-1} &=0,
\label{f=1GPE3}
\end{align}
where $\tilde{\mu}\equiv\mu-c_0n$. 

From Eq.~(\ref{f=1GPE2}), we have either (i) $\zeta_0=0$ or (ii) $\tilde{\mu} = c_1 n (\zeta_1^{\rm Re} + \zeta_{-1}^{\rm Re})(\zeta_1^{\rm Re}+\zeta_{-1}^{\rm Re}+2i\delta)$.
In case (i), we have three stationary states:
\begin{align}
{\rm I:  }\ &\left(e^{i\chi_1},0,0\right)^{\rm T},\ \ \ \epsilon_{\rm I}=-p+q+\frac{1}{2}c_0 n + \frac{1}{2}c_1 n,\label{eq:spin1_ss_1}\\
{\rm II: }\ &\left(0,0,e^{i\chi_{-1}}\right)^{\rm T},\ \ \ \epsilon_{\rm II}= p+q+\frac{1}{2}c_0 n + \frac{1}{2}c_1 n,\label{eq:spin1_ss_2}\\
{\rm III:}\ &\left(e^{i\chi_1}\sqrt{\frac{1}{2}\left(1+\frac{p}{c_1n}\right)},0,e^{i\chi_{-1}}\sqrt{\frac{1}{2}\left(1-\frac{p}{c_1n}\right)}\right)^{\rm T},
\ \ \ \epsilon_{\rm III}= q+\frac{1}{2}c_0 n - \frac{p^2}{2c_1 n},\label{eq:spin1_ss_3}
\end{align}
where $\chi_{\pm1}$ are arbitrary real numbers.
States I and II are fully polarized, giving $f_z=1$ and $-1$, respectively, and therefore, ferromagnetic,
whereas state III has a longitudinal magnetization that depends on
$p$ such that $f_z=p/(c_1 n)$.

In case (ii), we have $\delta=0$, since $\tilde{\mu}$ should be real.
By solving Eqs.~(\ref{f=1GPE1}) and (\ref{f=1GPE3}), together with the normalization condition $\sum_{m=-1}^1 | \zeta_m |^2 = 1$, the following two stationary states are obtained: one, referred to as a polar state, is given by
\begin{align}
{\rm IV:}\ \left(0,e^{i\chi_0},0\right)^{\rm T},\ \ \ \epsilon_{\rm IV}=\frac{1}{2}c_0 n,\label{eq:spin1_ss_4}
\end{align}
which has no magnetization in any direction; the other is given by~\cite{Murata2007}
\begin{align}
{\rm V:}\ \zeta_{\pm 1} &= e^{i(\chi_0 \mp \chi_z)}\frac{q \pm p}{2q} \sqrt{ \frac{-p^2+q^2+2c_1nq}{2c_1nq}},\label{eq:spin1_ss_5-1}\\
\zeta_0 &= e^{i\chi_0}\sqrt{\frac{(q^2-p^2)(-p^2-q^2+2c_1nq)}{4c_1nq^3}},\label{eq:spin1_ss_5-2}\\
\epsilon_{\rm V}&=\frac{(-p^2+q^2+2qc_1n)^2}{8c_1nq^2}+\frac{1}{2}c_0n. \label{eq:spin1_ss_5-3}
\end{align}
This state exists only when the argument of the square root is nonnegative for every $\zeta_m$.
In Eqs.~\eqref{eq:spin1_ss_4}--\eqref{eq:spin1_ss_5-2}, we have recovered the phases of the order parameter
by performing a gauge transformation $e^{i\chi_0}$ and spin rotation $e^{-i{\rm f}_z\chi_z}$ about the $z$ axis.
By comparing Eqs.~\eqref{eq:spin1_ss_4} and \eqref{eq:spin1_ss_5-3},
we find that  theV state can be the ground state only for $c_1<0$.
The magnetization for the V state tilts against the external magnetic field:
\begin{align}
 f_z &= \frac{p(-p^2+q^2+2qc_1n)}{2c_1n q^2},\ \ \ 
 f_+ = e^{i\chi_z}\frac{\sqrt{(q^2-p^2)\{(p^2-2c_1nq)^2-q^4\}}}{2|c_1|nq^2}.
\label{eq:BA_fz}
\end{align}
The polar angle [${\rm arctan}(|f_+|/f_z)$] is determined by the interaction and the magnetic field, whereas the azimuthal angle ($\chi_z$) is spontaneously chosen in each realization of the system. Because the axial symmetry with respect to the applied magnetic field axis is spontaneously broken, this phase is referred to as the broken-axisymmetry phase~\cite{Murata2007}.

\begin{figure}[ht]
\begin{center}
\resizebox{0.8\hsize}{!}{
\includegraphics{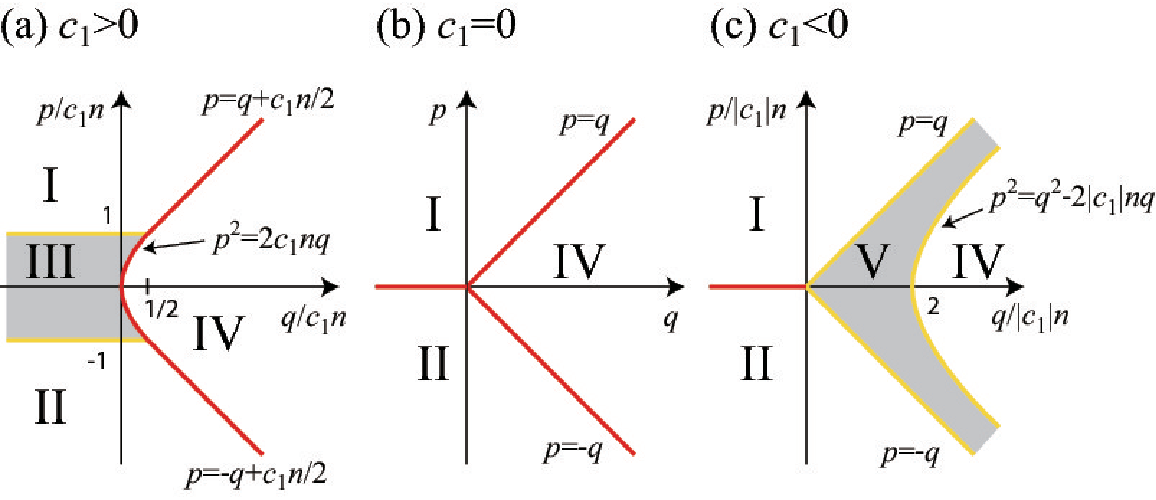}
}
\end{center}
\caption{Phase diagrams of spin-1 Bose-Einstein condensates for (a) $c_1>0$, (b) $c_1=0$, and (c) $c_1<0$.
The SO(2) rotational symmetry about the magnetic field is broken in the shaded region.
The yellow (light-colored) boundaries indicate second-order phase boundaries across which the derivatives of $\epsilon$ with respect to $p$ and $q$ change continuously.
}
\label{fig:spin-1PD}
\end{figure}

Comparing the energies of Eqs.~\eqref{eq:spin1_ss_1}--\eqref{eq:spin1_ss_4} and \eqref{eq:spin1_ss_5-3}, we obtain the phase diagram in a parameter space of $(q,p)$, as shown in Fig.~\ref{fig:spin-1PD}~\cite{Stenger1998}.
When $c_1$ is positive, the spin exchange interaction favors a state with no spontaneous magnetization, i.e., $\bm\zeta=(0,1,0)^{\rm T}$ or $(1,0,1)^{\rm T}/\sqrt{2}$ (or the state which is obtained by rotating these states in spin space).
The former (latter) state is stable for $p=0$ and $q>0\, (q<0)$. When $p$ is nonzero, the condensate tends to be magnetized along the $z$ axis.
When $q<0$ and $p\neq 0$, the magnetization gradually increases as a function of $p$ by changing the population balance in the $m=1$ and $-1$ states. This is the state III, which eventually goes to the ferromagnetic phase (I and II).
When $q$ is positive, however, as $|p|$ increases, $|\zeta_0|^2$ discretely changes from 1 to 0 and the magnetization jumps from zero to a nonzero value.
This is because the order parameter in the form of $(\delta,\sqrt{1-|\delta|^2},0)\,(\delta\in \mathbb{C}, |\delta|\ll 1)$ always leads to the transverse magnetization $|f_+|\sim \sqrt{2}|\delta|$, which is unfavorable for the antiferromagnetic interaction.
On the other hand, when $c_1$ is negative, the system tends to be magnetized. When $q<0$, both linear and quadratic Zeeman interactions are compatible with the ferromagnetic interaction, 
and the direction of the spontaneous magnetization is determined by the linear Zeeman term.
When $q>0$, however, the quadratic Zeeman interaction competes with the ferromagnetic interaction because the quadratic Zeeman term favors the $m=0$ state.
For $p=0$, for example, the condensate is magnetized perpendicular to the external field, and its amplitude monotonically decreases as a function of $q$, until finally the BEC enters the polar (IV) phase.

The phase diagram in a parameter space of $(q,f_z)$ is investigated in Ref.~\cite{Zhang2003}, in which state V appears also for $c_1\ge 0$ due to conservation of the magnetization,
whereas phase separation occurs for $c_1\le 0$ and $q<0$ if the system size is larger than the spin healing length, $\hbar/\sqrt{2M|c_1|n}$.

\begin{table}
{\renewcommand{\arraystretch}{1.5}
\begin{tabular}{@{}llllll@{}} \hline
      &phase& order parameter $\bm\zeta^{\rm T}$     & $f_z$          & $\tilde{\mu}=\mu-c_0n$ & $\tilde{\epsilon}=\epsilon-\frac{1}{2}c_0n$ \\ \hline\hline
(I)   &F    & $(e^{i\chi_1},0,0)$     & $ 1$           & $-p+q+c_1n$            & $-p+q+\frac{1}{2}c_1n$ \\ 
(II)  &F    & $(0,0,e^{i\chi_{-1}})$  & $-1$           & $ p+q+c_1n$            & $ p+q+\frac{1}{2}c_1n$ \\ \hline
(III) &AF   & $(e^{i\chi_1}\sqrt{\frac{1+p/(c_1n)}{2}},0,e^{i\chi_{-1}}\sqrt{\frac{1-p/(c_1n)}{2}})$   
                                                          & $\frac{p}{c_1n}$        & $q$                    & $ q-\frac{p^2}{2c_1n}$ \\ \hline
(IV)  &P    & $(0,e^{i\chi_0},0)$     & $0$            & $0$                    & $0$ \\ \hline
(V)   &BA   & Eqs.~\eqref{eq:spin1_ss_5-1}, \eqref{eq:spin1_ss_5-2} & $\frac{p(-p^2+q^2+2qc_1n)}{2c_1nq^2}$ & $\frac{q}{2}+c_1n-\frac{p^2}{2q}$  & $\frac{(-p^2+q^2+2qc_1n)^2}{8c_1nq^2}$\\ \hline
\end{tabular}}
\caption{Possible ground-state phases of spin-1 Bose-Einstein condensates, where  
F, AF, P, and BA denote ferromagnetic, antiferromagnetic, polar, and broken-axisymmetry phases, respectively.
}
\label{spin1table}
\end{table}

\begin{figure}[ht]
\begin{center}
\resizebox{0.8\hsize}{!}{
\includegraphics{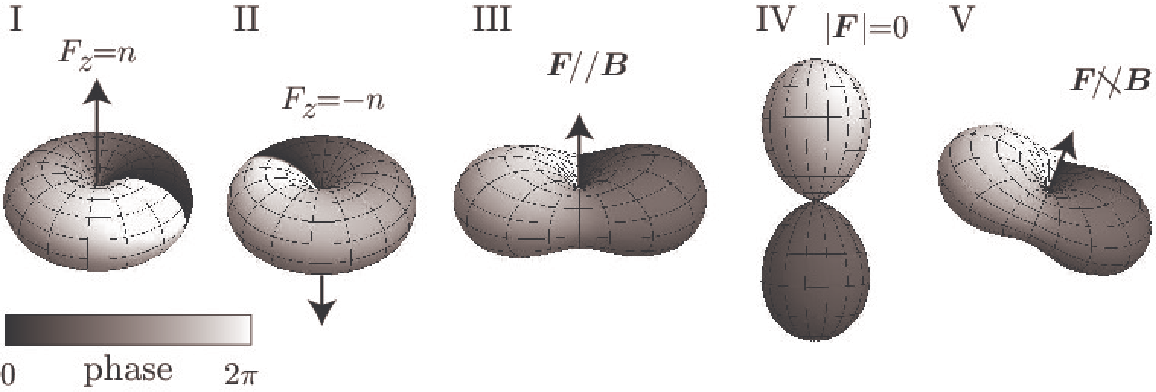}
}
\end{center}
\caption{Spherical-harmonic representation of the order parameters for spin-1 stationary states. 
}
\label{fig:spin1_OP_shape}
\end{figure}

Note that the rotational symmetry about the magnetic-field axis is also broken in state III, despite the fact that the magnetization is parallel to the magnetic field in this phase. 
To understand the underlying physics, we introduce the spin nematic tensor defined as the expectation value of the rank-2 symmetric nematic tensor operator [defined in Eq.~\eqref{hatNsym}]:
\begin{align}
 \mathcal{N}_{\nu\nu'}(\bm r) &\equiv \left\langle \hat{\mathcal{N}}_{\nu\nu'}({\bm r}) \right\rangle_0 
= \sum_{m,m'} \psi_m^*({\bm r})
\left(\frac{{\rm f}_\nu {\rm f}_{\nu'} +
 {\rm f}_{\nu'} {\rm f}_\nu}{2}\right)_{mm'}\psi_{m'}({\bm r})
\ \ \ \ \ (\nu,\nu'=x,y,z).
\label{eq:def_nematic_tensor}
\end{align}
The spin nematic tensor describes an anisotropy of spin fluctuations.
We also define the same quantity per particle as 
\begin{align}
\bar{\mathcal{N}}_{\nu\nu'}&\equiv \frac{\mathcal{N}_{\nu\nu'}}{n} =
  \sum_{m,m'} \zeta_m^*
\left(\frac{{\rm f}_\nu {\rm f}_{\nu'} +
 {\rm f}_{\nu'} {\rm f}_\nu}{2}\right)_{mm'}\zeta_{m'}
\ \ \ \ \ (\nu,\nu'=x,y,z).
\label{eq:def_bar_nematic_tensor}
\end{align}
The nematic tensors for the ferromagnetic (I, II) and polar (IV) phases are independent of $\chi_{\pm1}$ and $\chi_0$
and given by
\begin{align}
\bar{\mathcal{N}}^{\rm I,II} = \begin{pmatrix} \frac{1}{2} & 0 & 0 \\ 0 & \frac{1}{2} & 0 \\ 0 & 0 & 1\end{pmatrix},\ \ \ 
\bar{\mathcal{N}}^{\rm IV} = \begin{pmatrix} 1 & 0 & 0 \\ 0 & 1 & 0 \\ 0 & 0 & 0\end{pmatrix},
\label{eq:nematictensor_FP}
\end{align}
respectively.
Here, $\tilde{\mathcal{N}}_{xx}=\tilde{\mathcal{N}}_{yy}$ indicates that the order parameter has rotational symmetry about the $z$ axis.
On the other hand, the nematic tensor for state III for $\chi_1=\chi_{-1}$ is given by
\begin{align}
\bar{\mathcal{N}}^{\rm III} =  \begin{pmatrix} \frac{1}{2}(1+\alpha) & 0 & 0 \\ 0 & \frac{1}{2}(1-\alpha) & 0 \\ 0 & 0 & 1\end{pmatrix},\ \ \ 
\alpha = \sqrt{1-\left(\frac{p}{c_1n}\right)^2}.
\end{align}
Since $\bar{\mathcal{N}}_{xx}\neq\bar{\mathcal{N}}_{yy}$, the spin fluctuation is anisotropic on the $xy$ plane. 
This is the physical origin of the axisymmetry breaking of state III. 
In fact, it can be shown that off-diagonal elements such as $\mathcal{N}_{xy}$ appear when $\chi_1\neq\chi_{-1}$.
Figure~\ref{fig:spin1_OP_shape} shows the spherical-harmonic representations of the order parameters for states I--V.
From Fig.~\ref{fig:spin1_OP_shape}, one can see that the order parameters for states III and V have the same morphology and that they differ in their orientations with respect to the magnetic field.
Additional symmetry properties are discussed in Sec.~\ref{sec:symmetry}.

\subsubsection{Representation in the Cartesian basis}
\label{sec:spin1_Cartesian}
It is often convenient to represent the spinor order parameter in the Cartesian basis~\cite{Ohmi1998} which is defined as
\begin{align}
{\rm f}_\nu |\nu\rangle = 0\ \ (\nu=x,y,z).
\end{align}
The order parameter in the Cartesian representation transforms as a three-dimensional complex vector in spin space.
To relate the Cartesian basis to the irreducible one $|m\rangle$ which is defined as an eigenstate of ${\rm f}_z$ (${\rm f}_z|m\rangle = |m\rangle$), we note that $|z\rangle =|0\rangle$, where we choose an arbitrary phase factor to be unity. Then, $|x\rangle$ and $|y\rangle$ are obtained by rotating $|z\rangle$ as follows:
\begin{align}
 |x\rangle &= e^{-i {\rm f}_y \pi/2} |z\rangle = \frac{1}{\sqrt{2}}(-|1\rangle + |-1\rangle),\\
 |y\rangle &= e^{ i {\rm f}_x \pi/2} |z\rangle = \frac{i}{\sqrt{2}}(|1\rangle + |-1\rangle),
\end{align}
where we have used the matrix representation of ${\rm f}_\nu$ in the irreducible basis given in Eq.~\eqref{spin1matrices}.
The transformation from the Cartesian basis to the irreducible one is given in the matrix form as
\begin{align}
 |m\rangle = \sum_{\nu=x,y,z}\mathcal{U}_{m\nu}|\nu\rangle,
\label{eq:spin1_orthogonal_transformation}
\end{align}
where
\begin{align}
\mathcal{U}_{m\nu} = \begin{pmatrix}
    -1/\sqrt{2} & -i/\sqrt{2} & 0 \\ 0 & 0 & 1 \\ 1/\sqrt{2} & -i/\sqrt{2} & 0
   \end{pmatrix}.
\label{eq:spin1_orthogonal_U}
\end{align}
Let $\eta_\nu$ be the normalized order parameter in the Cartesian basis and $\zeta_m$ be the corresponding order parameter in the irreducible one. 
Since they both describe the same state, we have $\sum_{\nu=x,y,z} \eta_\nu |\nu\rangle = \sum_{m=0,\pm1} \zeta_m|m\rangle$. Since the two bases are related through Eq.~(\ref{eq:spin1_orthogonal_transformation}), the unitary transformation between two bases is given by
\begin{align}
 \begin{pmatrix} \eta_x \\ \eta_y \\ \eta_z \end{pmatrix} 
 = \mathcal{U}^{\rm T}
 \begin{pmatrix} \zeta_1 \\ \zeta_0 \\ \zeta_{-1} \end{pmatrix}
 = \begin{pmatrix}
    -1/\sqrt{2} & 0 & 1/\sqrt{2} \\ -i/\sqrt{2} & 0 & -i/\sqrt{2} \\ 0 & 1 & 0
   \end{pmatrix}
 \begin{pmatrix} \zeta_1 \\ \zeta_0 \\ \zeta_{-1} \end{pmatrix}.
\label{eq:transform_eta_zeta}
\end{align}
For example, the order parameter for the polar phase is given by $\bm\eta=(0,0,1)^{\rm T}$,
while that for the ferromagnetic phase is $\bm\eta=-(1,i,0)^{\rm T}$.

The spin matrix in the Cartesian basis is given by $({\rm f}_{\nu_1})_{\nu_2\nu_3} = -i\epsilon_{\nu_1\nu_2\nu_3}\ (\nu_1, \nu_2, \nu_3=x, y, z)$. The expectation value of the spin vector is therefore expressed as
\begin{align}
 \bm f({\bm r}) &= \sum_{\nu\nu'}({\bf f})_{\nu\nu'} \eta_\nu^*({\bm r})\eta_{\nu'}({\bm r}) = -i \bm\eta^*({\bm r}) \times \bm\eta({\bm r}).
\label{eq:spin1_orthogonal_F}
\end{align}
Thus, the expectation value vanishes when $\bm\eta^* // \bm\eta$, i.e.,
when the order parameter is real except for an overal phase factor.

\subsection{Mean-field theory of spin-2 BECs} 
\label{sec:MFTspin2}

\subsubsection{Gross-Pitaevskii equations}
The expectation value of the Hamiltonian $\hat{H}=\hat{H}_0+\hat{V}$ over the state in Eq.~(\ref{statevector}) with $f=2$, where $\hat{H}_0$ and $\hat{V}$ are given in Eqs.~(\ref{H_0}) and (\ref{V(f=2)}), is given as
\begin{align}
E [\psi]
&\equiv\langle\hat{H}\rangle_0 \nonumber\\
&=
\int d{\bm r} \Bigg\{  \sum_{m=-2}^2 \psi_m^\ast
\left[ -\frac{\hbar^2 \nabla^2}{2M} + U_{\rm trap}({\bm r}) -pm + qm^2 \right]\psi_m 
+\frac{c_0}{2}n^2 +\frac{c_1}{2} | {\bm F} |^2 +\frac{c_2}{2} |A_{00}|^2 \Bigg\}.
\label{energy_functional(f=2)}
\end{align}
Here, the spin-2 matrices are given by
\begin{eqnarray}
{\rm f}_x &=& \begin{pmatrix}
       0 & 1 & 0 & 0 & 0\\
       1 & 0 & \sqrt{\frac{3}{2}} & 0 & 0\\
       0 & \sqrt{\frac{3}{2}} & 0 & \sqrt{\frac{3}{2}} & 0\\
       0 & 0 & \sqrt{\frac{3}{2}} & 0 & 1\\
       0 & 0 & 0 & 1 & 0 \end{pmatrix},\ \ \ 
{\rm f}_y = \begin{pmatrix}
       0 & -i & 0 & 0 & 0\\
       i & 0 & -i\sqrt{\frac{3}{2}} & 0 & 0\\
       0 & i\sqrt{\frac{3}{2}} & 0 & -i\sqrt{\frac{3}{2}} & 0\\
       0 & 0 & i\sqrt{\frac{3}{2}} & 0 & -i\\
       0 & 0 & 0 & i & 0 \end{pmatrix},\ \nonumber \\ 
{\rm f}_z &=& \begin{pmatrix} 2 & 0 & 0 & 0 & 0 \\ 0 & 1 & 0 & 0 & 0 \\ 0 & 0 & 0 & 0 & 0 \\ 0 & 0 & 0 & -1 & 0 \\ 0 & 0 & 0 & 0 & -2 
\end{pmatrix}.
\label{spin-2Matrices}
\end{eqnarray}
In comparison with the energy functional (\ref{energy_functional(f=1)}) of the spin-1 BEC, the new term $c_2|A|^2/2$ appears in Eq.~(\ref{energy_functional(f=2)}),
where
\begin{eqnarray}
A_{00}({\bm r}) \equiv \langle \hat{A}_{00}({\bm r})\rangle_0 = \frac{1}{\sqrt{5}} \; \left[2 \psi_2({\bm r})\psi_{-2}({\bm r})-2\psi_1({\bm r})\psi_{-1}({\bm r}) +\psi_0^2({\bm r})\right]
\label{eqS2107}
\end{eqnarray}
is the amplitude of the spin-singlet pair.

The time-dependent GPEs for the spin-2 case can be obtained by 
substituting Eq.~(\ref{energy_functional(f=2)}) in Eq.~(\ref{dynamics}):
\begin{align}
i\hbar \frac{\partial \psi_{\pm 2}}{\partial t}
 =& \left[ -\frac{\hbar^2 \nabla^2}{2M} +U_{\rm trap}({\bm r}) \mp 2p + 4q + c_0n \pm 2c_1  F_z -\mu  \right] \psi_{\pm 2} \nonumber \\
  & + c_1  F_\mp  \psi_{\pm 1} + \frac{c_2}{\sqrt{5}} A_{00} \psi_{\mp 2}^\ast,\label{f=2GPE2}\\
i\hbar \frac{\partial \psi_{\pm 1}}{\partial t}
 =& \left[ -\frac{\hbar^2 \nabla^2}{2M} +U_{\rm trap}({\bm r}) \mp p + q + c_0n \pm c_1  F_z -\mu  \right] \psi_{\pm 1} \nonumber \\
  & + c_1 \left( \frac{\sqrt{6}}{2}  F_\mp  \psi_0 +  F_\pm  \psi_{\pm 2} \right) -\frac{c_2}{\sqrt{5}} A_{00} \psi_{\mp 1}^\ast,\label{f=2GPE1}\\
i\hbar \frac{\partial \psi_0}{\partial t}
 =&\left[ -\frac{\hbar^2 \nabla^2}{2M} +U_{\rm trap}({\bm r}) + c_0n -\mu \right] \psi_0 
  + \frac{\sqrt{6}}{2} c_1 \left(  F_+  \psi_1 +  F_-  \psi_{-1} \right) +\frac{c_2}{\sqrt{5}} A_{00} \psi_0^\ast,\label{f=2GPE0}
\end{align}
where
\begin{align}
 F_+ &= F_-^\ast = 2 \left(\psi_2^\ast \psi_1 + \psi_{-1}^\ast \psi_{-2} \right) + \sqrt{6}\left(\psi_1^\ast \psi_0 + \psi_0^\ast \psi_{-1}\right),\label{f_+}\\
 F_z &= 2 \left(|\psi_2|^2 - |\psi_{-2}|^2 \right) + |\psi_1|^2 - |\psi_{-1}|^2.\label{f_z}
\end{align}
The time-independent GPEs can be obtained by substituting $\psi_m({\bm r}, t)=\psi_m ({\bm r}) e^{-i \mu t/\hbar}$ in Eqs.~(\ref{f=2GPE2})--(\ref{f=2GPE0}). 

\subsubsection{Ground states in a uniform system}
\label{sec:MF_spin2_GS}
For the spin-1 case, the spin-singlet amplitude is uniquely related to the magnetization by relation~\eqref{eq:spin1identity3}, and only one of them can change independently in the energy functional. 
However, for the spin-2 case, they can change independently in a certain region of the $(|\bm F|,|A_{00}|)$ space,
giving rise to new ground-state phases. 
This can be best illustrated at zero magnetic field.

Here, we consider a uniform system with a fixed density $n$ and rewrite the order parameter as $\psi_m=\sqrt{n}\zeta_m$, where $\bm\zeta^\dagger \bm \zeta=1$.
The expectation values of the spin and spin-singlet pair amplitude per particle are defined by
\begin{align}
 {\bm f} &\equiv \frac{{\bm F}}{n}=\sum_{mm'}\zeta_m^*({\bf f})_{mm'}\zeta_{m'}, \label{eq:def_f=F/n}\\
 a_{00} &\equiv \frac{A_{00}}{n} = \frac{1}{\sqrt{5}}\left(2\zeta_2\zeta_{-2}-2\zeta_1\zeta_{-1}+\zeta_0^2\right).\label{eq:def_a=A/n}
\end{align}
Then, the ground-state magnetism is determined by the last two terms in Eq.~\eqref{energy_functional(f=2)}, i.e., $c_1n|{\bm f}|^2$ and $c_2n|a_{00}|^2$.

\begin{figure}[ht]
\begin{center}
\resizebox{0.6\hsize}{!}{
\includegraphics{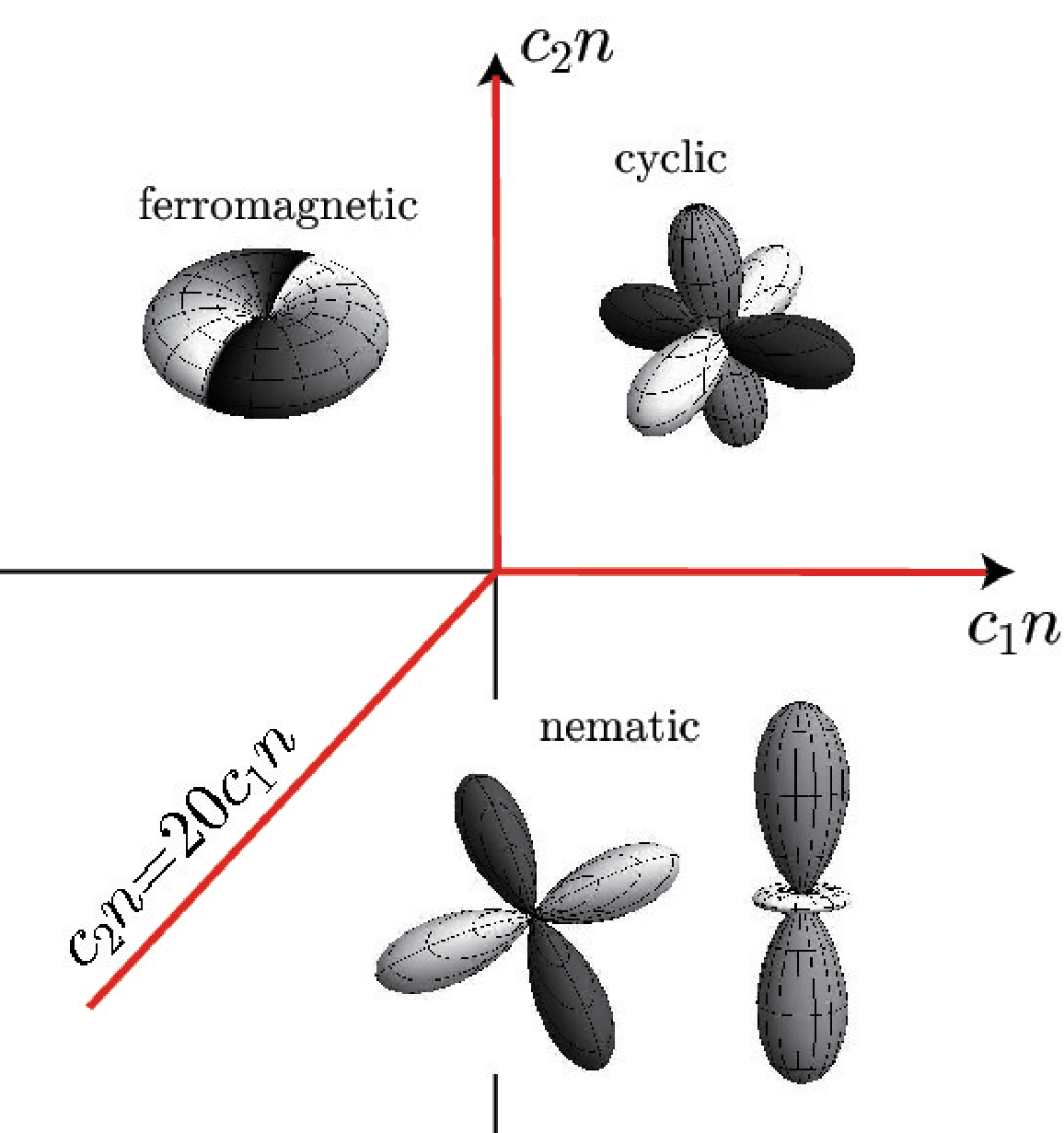}
}
\end{center}
\caption{Phase diagram of spin-2 Bose-Einstein condensates at zero magnetic field.
In each phase, the spherical-harmonic representation of the order parameter is shown.
All phase boundaries are first order.
}
\label{fig:spin-2PD_B=0}
\end{figure}

Clearly, $|\bm f|$ can vary within $0 \le |\bm f| \le 2$ for the $f = 2$ system. 
Note that $|a_{00}|$ is proportional to the inner product of the order parameter and its time reversal: 
$|a_{00}|=(\mathcal{T}\bm\zeta)^\dagger \bm \zeta/\sqrt{5}$,
where the time-reversal operator $\mathcal{T}$ is defined in Eq.~\eqref{eq:def_mathcalT}.
It takes the maximum value of $1/\sqrt{5}$
when the order parameter is invariant under time reversal, while it should vanish for the ferromagnetic state. 
Then, the ground-state magnetism is determined as follows:

\begin{itemize}
\item When $c_1<0$ and $c_2>0$, the energy of the system is lowered as magnetization increases and time-reversal symmetry is broken.
Therefore, the ground state is ferromagnetic, and the order parameter is given by ${\bm\zeta}^{\rm ferro}=(1,0,0,0,0)^{\rm T}$ or its rotated state in the spin space.

\item When $c_1>0$ and $c_2<0$, the energy of the system becomes lowest when magnetization vanishes and the system possesses time-reversal symmetry.
Therefore, the ground state is nematic, and the order parameter is given by
${\bm\zeta}^{\rm uniax}=(0,0,1,0,0)^{\rm T}$ (uniaxial nematic, UN) or ${\bm\zeta}^{\rm biax}=(1,0,0,0,1)^{\rm T}/\sqrt{2}$ (biaxial nematic, BN).
While these states are often referred to as polar or antiferromagnetic~\cite{Koashi2000,Ciobanu2000,Ueda2002}, here, we adopt the liquid-crystal terminology which implies an inversion symmetric nature of the order parameter (see the insets in Fig.~\ref{fig:spin-2PD_B=0}).
The order parameter $(0, 1, 0, 1, 0)^{\rm T}/\sqrt{2}$ also represents the BN state,
which is obtained from ${\bm\zeta}^{\rm biax}$ via a rotation as $i e^{-i{\rm f}_y \pi/2} e^{-i {\rm f}_z \pi/4}{\bm\zeta}^{\rm biax}$,
while $\bm\zeta^{\rm uniax}$ cannot be obtained by a rotation of $\bm\zeta^{\rm biax}$ since the symmetries of the order parameters are different [see Figs.~\ref{fig:SHR}(f), \ref{fig:SHR}(g), and \ref{fig:SHR}(h)].
Note that the symmetries of the UN and BN states differ; however, the two states are degenerate at the mean-field level.
Moreover, the superposition of these two states $(\cos\xi,0,\sqrt{2}\sin\xi,0,\cos\xi)^{\rm T}/\sqrt{2}$, where $\xi$ is an arbitrary real number, is also degenerate.
The same degeneracy is pointed out to appear in a {\it d}-wave superconductor~\cite{Mermin1974}.
The degeneracy between the UN and BN phases has been shown to be lifted by quantum or thermal fluctuations~\cite{Song2007,Turner2007} (see Sec.~\ref{sec:Bogoliubov}).
In the many-body ground state corresponding to the nematic phase, spin-singlet pairs of atoms undergo Bose-Einstein condensation (see Sec.~\ref{sec:ManyBodyTheory}).

The uniaxial and biaxial nematic phases are also referred to as the antiferromagnetic phase in Refs.~\cite{Koashi2000,Ueda2002} and the polar phase in Ref.~\cite{Ciobanu2000}.

\item When both $c_1$ and $c_2$ are negative, the competition between energies associated with spontaneous magnetization ($4c_1n$ for the ferromagnetic phase and 0 for the nematic phase)
and spin-singlet pairing (0 for the ferromagnetic phase and $c_2n/5$ for the nematic phase)
leads to the phase boundary at $c_2n=20 c_1 n$.

\item When both $c_1$ and $c_2$ are positive, 
neither the ferromagnetic nor nematic phase is energetically favorable and frustration arises, resulting in a new phase called the cyclic phase~\cite{Koashi2000,Ciobanu2000,Ueda2002}. In this phase, time-reversal symmetry is broken even though there is no spontaneous magnetization. The order parameter is given by ${\bm\zeta}^{\rm cyclic}=(1,0,i\sqrt{2},0,1)^{\rm T}/2$, which possesses tetrahedral symmetry, as shown in Fig.~\ref{fig:SHR}(i). 
In the many-body state corresponding to the cyclic phase, three bosons form a spin-singlet trimer and the boson trimmers undergo Bose-Einstein condensation (see Sec.~\ref{sec:ManyBodyTheory}).
\end{itemize}

Next, we investigate the ground-state phases in the presence of an external magnetic field.
In this case, the mean-field energy per particle is given by
\begin{align}
\epsilon 
&= \sum_{m=-2}^2 (-pm+qm^2)\,|\zeta_m|^2 + \frac{c_0n}{2}
+ \frac{c_1n}{2} |\bm f|^2 + \frac{c_2n}{2}|a_{00}|^2.
\label{eqS2ad6}
\end{align}
Substituting $\psi_m({\bm r}, t)=\sqrt{n}\zeta_m e^{-i \mu t/\hbar}$ in Eqs.~(\ref{f=2GPE2})--(\ref{f=2GPE0}),
and assuming that the system is uniform (i.e., $U_{\rm trap}({\bm r})=0$), Eqs.~(\ref{f=2GPE2})--(\ref{f=2GPE0}) reduce to
\begin{eqnarray}
&&( 4q+2\tilde{\gamma}_0-\tilde{\mu}) \; \zeta_2 + a \zeta_{-2}^\ast
= -\gamma_- \zeta_1 ,
\label{StGPE1}
\\[2mm]
&& a^\ast \zeta_2 + (4q-2\tilde{\gamma}_0-\tilde{\mu}) \; \zeta_{-2}^\ast
=  -\gamma_- \zeta_{-1}^\ast ,
\label{StGPE2}
\\
&& \gamma_+ \zeta_2 + (q+\tilde{\gamma}_0-\tilde{\mu}) \; \zeta_1 -a \zeta_{-1}^\ast
= -\frac{\sqrt{6}}{2} \; \gamma_- \zeta_0,
\label{StGPE3}
\\
&& -a^\ast \zeta_1 + (q-\tilde{\gamma}_0-\tilde{\mu}) \; \zeta_{-1}^\ast + \gamma_+ \zeta_{-2}^\ast
= -\frac{\sqrt{6}}{2} \; \gamma_- \zeta_0^\ast,
\label{StGPE4}
\\
&& \tilde{\mu} \zeta_0 - a \zeta_0^\ast
=\frac{\sqrt{6}}{2} \; ( r_+ \zeta_1 + \gamma_- \zeta_{-1}),
\label{StGPE5}
\end{eqnarray}
where $ \tilde{\mu}\equiv\mu-c_0n, \gamma_\pm\equiv c_1 n f_\pm, \gamma_0 \equiv c_1n f_z, \tilde{\gamma}_0\equiv\gamma_0-p$, and $a \equiv c_2 n a_{00}/\sqrt{5}$.
We use the degree of gauge transformation (i.e., the global U(1) phase) to make $\zeta_0$ real.
Furthermore, because of the SO(2) rotational symmetry about the direction of the magnetic field (i.e., the $z$ axis), 
we may choose the coordinate system to make $ f_y =0 $ such that $ \gamma_+ = \gamma_- \equiv \gamma$. 
A mean-field ground state can be obtained as a solution of Eqs.~(\ref{StGPE1})--(\ref{StGPE5}) under such a simplification.

Here, we consider the case of $ \gamma=0 $, i.e., the case of zero transverse magnetization. 
This assumption is valid as long as $q<0$, because the quadratic Zeeman energy and ferromagnetic interaction are compatible for $q<0$.
Then, Eqs.~(\ref{StGPE1})--(\ref{StGPE5}) reduce to
\begin{align}
(4q+2\tilde{\gamma}_0 -\tilde{\mu} ) \, \zeta_2 + a \zeta_{-2}^\ast &= 0,
\label{ad1}
\\
a^\ast \zeta_2 + (4q - 2\tilde{\gamma}_0 -\tilde{\mu} ) \, \zeta_{-2}^\ast &=0,
\label{ad2}
\\
(q + \tilde{\gamma}_0 -\tilde{\mu} ) \, \zeta_1 -a \zeta_{-1}^\ast &=0,
\label{ad3}
\\
-a^\ast \zeta_1 + (q - \tilde{\gamma}_0 -\tilde{\mu} )\, \zeta_{-1}^\ast &=0,
\label{ad4}
\\
(\tilde{\mu} - a) \, \zeta_0 &= 0.
\label{ad5}
\end{align}

Because Eqs.~(\ref{ad1})--(\ref{ad5}) are decoupled into three parts, the solutions can be classified according to the determinant of the coefficient matrix of Eqs.~(\ref{ad1}) and (\ref{ad2}),
\begin{eqnarray}
\mathcal{D}_2 \equiv (4q - \tilde{\mu})^2 - 4\tilde{\gamma}_0^2 - |a|^2,
\label{D2}
\end{eqnarray}
and
the determinant of the coefficient matrix of Eqs.~(\ref{ad3}) and (\ref{ad4}),
\begin{eqnarray}
\mathcal{D}_1 \equiv (q-\tilde{\mu})^2 -\tilde{\gamma}_0^2 -|a|^2.
\label{D1}
\end{eqnarray}

\begin{itemize}
\item
If $\mathcal{D}_1\neq0$ and $\mathcal{D}_2\neq0$, then we have $\zeta_1 =\zeta_{-1} = 0$ and $ \zeta_2 = \zeta_{-2} = 0 $, so that the solution is the UN state with the order parameter given by
\begin{eqnarray}
\textrm{UN:\ \ \ } (0,0,1,0,0)^{\rm T}.
\end{eqnarray}
The chemical potential is determined from Eq.~(\ref{ad5}) to be $ \tilde{\mu} = \frac{1}{5}c_2n$ and $f_z =0$. 
The energy per particle is found from Eq.~(\ref{eqS2ad6}) to be $\epsilon = \frac{1}{2}(c_0+\frac{1}{5}c_2)n$.

\item
If $\mathcal{D}_1=0$ and $\mathcal{D}_2 \neq 0$, then we have $\zeta_2=\zeta_{-2}=0$. From the assumption that there is no transverse magnetization,
$\zeta_{\pm 1}\neq 0$ leads to either (i) $\zeta_0=0$ or (ii) $\zeta_1^*=\zeta_{-1}$.

For case (i), 
\begin{align}
\textrm{F}_{1+}:\ \ \ (0,e^{i\chi_1},0,0,0)^{\rm T}
\end{align}
and
\begin{align}
\textrm{F}_{1-}:\ \ \ (0,0,0,e^{i\chi_{-1}},0)^{\rm T}
\end{align}
are clearly the solutions of Eqs.~\eqref{ad3} and \eqref{ad4}
with $f_z=1$, $\tilde{\mu}=-p+q+c_1n$, $\epsilon=-p+q+\frac{1}{2}(c_0+c_1)n$ for F1$_+$, and
$f_z=-1, \tilde{\mu}=p+q+c_1n$, $\epsilon=p+q+\frac{1}{2}(c_0+c_1)n$ for F1$_-$.
A solution with nonzero $\zeta_{\pm 1}$ is written as
\begin{eqnarray}
C_2:\ \ \ \left( 0, e^{i\chi_1} \sqrt{\frac{1+ f_z}{2}}, 0, e^{i\chi_{-1}} \sqrt{\frac{1-f_z}{2}}, 0 \right)^{\rm T}.
\end{eqnarray}
Substituting this in Eqs.~\eqref{ad3}, \eqref{ad4}, and $\mathcal{D}_1=0$, we obtain
$f_z=p/[(c_1n-c_2/5)n]$ and $\tilde{\mu}=q+c_2n/5$.
The energy per particle is obtained as
$\epsilon=q+(c_0+c_2/5)n/2 -p^2/[2(c_1-c_2/5)n]$.
This state exists for $|p|<|c_1-c_2/5|n$
and continuously becomes the BN state at $p=0$.

For case (ii), the order parameter is written in the following form:
\begin{align}
 \left(0, e^{i\chi}\sqrt{\frac{1-\zeta_0^2}{2}},\zeta_0,-e^{-i\chi}\sqrt{\frac{1-\zeta_0^2}{2}},0\right)^{\rm T},
\end{align}
where $\zeta_0$ is a real number from the assumption.
However, by substituting this order parameter in Eqs.~\eqref{ad3}--\eqref{ad5},
we find that this state becomes stationary only when $p=q=0$.
This state is the superposition of the UN and BN states.

\item
If $\mathcal{D}_1\neq 0$ and $\mathcal{D}_2=0$, then we have $\zeta_1=\zeta_{-1}=0$. Moreover, if $\tilde{\mu}\neq a$, we have $ \zeta_0=0$. 
In a manner similar to the previous case, Eqs.~\eqref{ad1} and \eqref{ad2} have three solutions.
Two of them are ferromagnetic and given by
\begin{align}
{\rm F}_{2+}:\ \ \ (e^{i\chi_2},0,0,0,0)^{\rm T}
\end{align}
with  $f_z=2, \tilde{\mu}=-2p+4q+4c_1n$, and $\epsilon=-2p+4q+\frac{1}{2}(c_0+4c_1)n$, and by
\begin{align}
{\rm F}_{2-}:\ \ \ (0,0,0,0,e^{i\chi_{-2}})^{\rm T}
\end{align}
with  $f_z=-2, \tilde{\mu}=2p+4q+4c_1n$, and $\epsilon=2p+4q+(c_0+4c_1)n/2$.
When $|p|<|2c_1-c_2/10|n$, the other solution is given by
\begin{eqnarray}
C_4:\ \ \ \left( e^{i\chi_2} \sqrt{\frac{1+f_z/2}{2}}, 0, 0, 0, e^{i\chi_{-2}}\sqrt{\frac{1-f_z/2}{2}}\right)^{\rm T},
\label{eq:MFTspin2C4}
\end{eqnarray}
where $f_z= p/[(c_1-c_2/20)n],\, \tilde{\mu}=4q+c_2n/5$ and 
$\epsilon=4q+(c_0+\frac{1}{5}c_2)n/2-p^2/[2(c_1-c_2/20)n]$.
This state becomes the BN state at $p=0$.

\item
If $\mathcal{D}_1\neq 0$ and $\mathcal{D}_2=0$ and $\tilde{\mu}=a$, we have  $\zeta_1=\zeta_{-1}=0$; however, $\zeta_{\pm 2}$ and $\zeta_0$ can be nonzero.
They obey $|\zeta_2|^2 + |\zeta_{-2}|^2 + |\zeta_0|^2 = 1 $ and $ |\zeta_2|^2 - |\zeta_{-2}|^2 = f_z/2$, and hence,
\begin{eqnarray}
C_2':\ \ \ \left(e^{i\chi_2}\sqrt{\frac{1 - \zeta_0^2+f_z/2}{2}},0,\zeta_0,0,e^{i\chi_{-2}}\sqrt{\frac{1 - \zeta_0^2-f_z/2}{2}} \right)^{\rm T}.
\label{C2'}
\end{eqnarray}
Because  $a=(c_2n/5)\left[e^{i(\chi_2+\chi_{-2})}\sqrt{(1-\zeta_0)^2-f_z^2/4}+\zeta_0^2\right]=\tilde{\mu}$ is real, we have $\chi_2+\chi_{-2}=0$ or $\pi$. 
Here, $\tilde{\mu}$ and $f_z$ are determined from Eqs.~(\ref{ad1}) and (\ref{ad2}) to be $\tilde{\mu}=2q-\tilde{\gamma}_0^2/(2q)$ and $f_z = (\tilde{\gamma}_0+p)/(c_1 n)$, respectively, 
where $\tilde{\gamma}_0$ is a real solution to the following equation:
\begin{align}
\tilde{\gamma}_0^3 + p\tilde{\gamma}_0^2 + 4q[q+2c_1 n(1-\zeta_0^2)]\tilde{\gamma}_0 + 4pq^2=0,
\label{eq:tilde_gamma_0}
\end{align}
and $\zeta_0$ is determined so as to minimize the energy per particle given by
\begin{align}
\epsilon=4q\left(1-\zeta_0^2\right) + \frac{1}{2}c_0 n
+ \frac{\tilde{\gamma_0}^2-p^2}{2c_1 n} + \frac{c_2 n}{10}\left| \sqrt{(1-\zeta_0^2)^2-\frac{f_z^2}{4}} e^{-i(\chi_2+\chi_{-2})} + \zeta_0^2 \right|^2.
\label{eq:energy_for_c2'phase}
\end{align}
In particular, when $p=0$ and $|q|>2c_1 n (1-\zeta_0^2)$, $\tilde{\gamma}_0$ is determined to be zero, leading to $f_z=0$.
Moreover, when $c_2>0$, Eq.~\eqref{eq:energy_for_c2'phase} is minimized at $\chi_2+\chi_{-2}=\pi$ and $\zeta_0^2=\frac{1}{2}+5q/(c_2n)$, with the result 
$\epsilon=2q+\frac{1}{2}c_0n - 10 q^2/(c_2n)$.
The corresponding order parameter is given by
\begin{align}
D_2':\ \ \ \left( ie^{i\chi}\frac{\sqrt{1-10q/(c_2n)}}{2}, 0, \sqrt{\frac{1+10q/(c_2n)}{2}}, 0, ie^{-i\chi}\frac{\sqrt{1-10q/(c_2n)}}{2}\right)^{\rm T},
\end{align}
where we set $\chi_{\pm2}=\frac{\pi}{2}\pm\chi$.
This state continuously becomes the cyclic state at $q=0$.
On the other hand, when $c_2<0$, Eq.~\eqref{eq:energy_for_c2'phase} is minimized at $\chi_2+\chi_{-2}=0$ and
the state is the UN state for $q>0$ and the BN state for $q<0$.

\item
If $\mathcal{D}_1=\mathcal{D}_2=0$, we solve the simultaneous equations $\mathcal{D}_1=0$ and $\mathcal{D}_2=0$, obtaining
\begin{eqnarray}
\tilde{\mu}
= \frac{5q^2 - \tilde{\gamma}_0^2}{2q}
= \pm\sqrt{4q^2+|a|^2}.
\label{eqS2ad7}
\end{eqnarray}
When $q \neq 0$, $\tilde{\mu} \neq a$, we find from Eq.~(\ref{ad5}) that $\zeta_0=0$. Then, $ \gamma = 2c_1n (\zeta_2^\ast \zeta_1 + \zeta_{-1}^\ast \zeta_{-2} ) =0$ by assumption. We may use Eqs.~(\ref{ad1})--(\ref{ad4}) and (\ref{eqS2ad7}) to show that
\begin{eqnarray}
\gamma
= 2c_1 \zeta_2^\ast \zeta_1 
\! \left( \! 1 \! - \! \frac{\tilde{\gamma}_0 + 3q}{\tilde{\gamma}_0 - 3q} \right)
= 2c_1 \zeta_{-1}^\ast \zeta_{-2} 
\! \left( \! 1 \! - \! \frac{\tilde{\gamma}_0 - 3q}{\tilde{\gamma}_0 + 3q} \right)
\! = \! 0.
\label{eqS2ad8}
\end{eqnarray}
Thus, we must, in general, have $ \zeta_2 \zeta_1 = \zeta_{-1} \zeta_{-2} =0 $. 
To be consistent with Eqs.~(\ref{ad1})--(\ref{ad4}), we find that either $ \zeta_1 = \zeta_{-2} =0 $ or $ \zeta_2 = \zeta_{-1} =0 $ should hold. In the former case, the order parameter is given by 
\begin{eqnarray}
C_{3+}:\ \ \ 
\left( e^{i\chi_2} \sqrt{\frac{1+f_z}{3}}, 0, 0, e^{i\chi_{-1}} \sqrt{\frac{2-f_z}{3}}, 0 \right)^{\rm T}
\end{eqnarray}
with $ f_z = (p-q) / (c_1n)$ and $\epsilon=2q-c_0n/2 - (p-q)^2/(2c_1 n)$. 
In the latter case, the solution is given by 
\begin{eqnarray}
C_{3-}:\ \ \ 
\left( 0, e^{i\chi_1} \sqrt{\frac{2+f_z}{3} }, 0, 0, e^{i\chi_{-2}}\sqrt{\frac{1-f_z}{3}} \right)^{\rm T}
\end{eqnarray}
with $ f_z = (p+q) / (c_1n)$ and $\epsilon=2q-c_0 n/2 - (p+q)^2/(2c_1n)$. 
At zero magnetic field (i.e., $p=q=0$), both of these two states are cyclic states which are related to ${\bm\zeta}^{\rm cyclic}=(1,0,i\sqrt{2},0,1)^{\rm T}/2$ via rotations as 
\begin{align}
\frac{1}{\sqrt{3}}\begin{pmatrix} 1 \\ 0 \\ 0 \\ \sqrt{2} \\ 0 \end{pmatrix} &=
 -i \exp\left(i {\rm f}_z \frac{\pi}{4}\right) \exp\left[ -i \frac{{\rm f}_x - {\rm f}_y}{\sqrt{2}}\arccos\left(\frac{1}{\sqrt{3}}\right)\right]
{\bm\zeta}^{\rm cyclic},\\
\frac{1}{\sqrt{3}}\begin{pmatrix} 0 \\ \sqrt{2} \\ 0 \\ 0 \\ 1 \end{pmatrix} &=
 -i \exp\left(i {\rm f}_z \frac{3\pi}{4}\right) \exp\left[ -i \frac{{\rm f}_x + {\rm f}_y}{\sqrt{2}}\arccos\left(\frac{1}{\sqrt{3}}\right)\right]
{\bm\zeta}^{\rm cyclic}.
\end{align}
These relations are also understood from the spherical-harmonic representation of the order parameters [Figs.~\ref{fig:SHR}(i) and \ref{fig:SHR}(j)].

\end{itemize}

\begin{figure}[htb]
\begin{center}
\resizebox{0.5\hsize}{!}{
\includegraphics{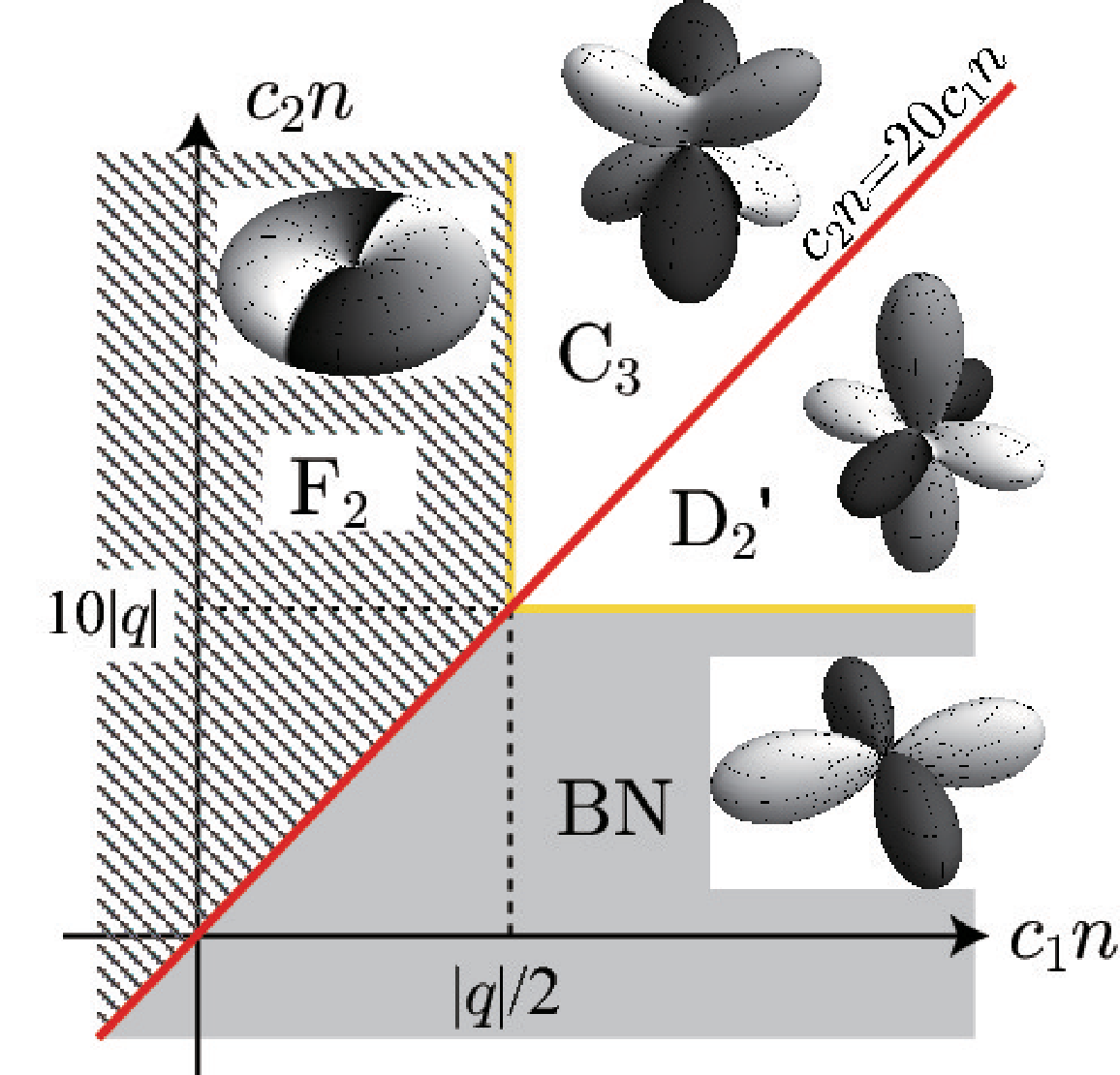}
}
\end{center}
\caption{
Ground-state phase diagram of spin-2 Bose-Einstein condensates for $p=0$ and $q<0$,
where the F$_{2\pm}$ states and the C$_{3\pm}$ states are degenerate in the region of F$_2$ and C$_3$, respectively,
while the order parameter in the region of BN is given by that of $C_4$ state in Table~\ref{table3} with $p=0$ ($f_z=0$).
The C$_3$--F$_2$ and $D_2'$--BN phase boundaries are second order across which the derivatives of the mean-field energy $\epsilon$ with respect to $c_1$ and $c_2$, respectively, change continuously.
The other phase boundaries are first order.
The M state and C state in Ref.~\cite{Saito2005b} correspond to $C_3$ and $D_2'$ in this phase diagram, respectively.
}
\label{fig:spin-2PD}
\end{figure}

The above results are summarized in Table~\ref{table3}.
By comparing the energy of the obtained state, we find the ground-state phase diagram of a spin-2 BEC.
Figure~\ref{fig:spin-2PD} shows the special case of $p=0$ and $q<0$~\cite{Saito2005b}.
By comparing Figs.~\ref{fig:spin-2PD_B=0} and \ref{fig:spin-2PD}, we find that
the quadratic Zeeman energy lifts the degeneracy of UN and BN states
and fixes the cloverleaf shape of the BN state perpendicular to the external field.
Moreover, it shrinks the region of the cyclic phase in the phase diagram and divides it into two phases: C3 and D2$'$.
The orientation of the undirected triad of the cyclic order parameter is fixed in these states and its shape is deformed along the $z$ direction.

\begin{landscape}
\begin{table}[htb]
\begin{center}
{\renewcommand{\arraystretch}{1.5}
\begin{tabular}{@{}llcll@{}} \hline
state & order parameter $\bm\zeta^{\rm T}$ & $f_z$ & $\tilde{\mu}=\mu-c_0n$ & $\tilde{\epsilon}=\epsilon-\frac{1}{2}c_0n$ \\ \hline\hline
F$_{2+}$ & $(e^{i\chi_2},0,0,0,0)$    & $ 2$ & $-2p+4q+4c_1n$ & $-2p+4q          +2c_1n$ \\ 
F$_{2-}$ & $(0,0,0,0,e^{i\chi_{-2}})$ & $-2$ & $ 2p+4q+4c_1n$ & $ 2p+4q          +2c_1n$ \\ \hline
F$_{1+}$ & $(0,e^{i\chi_1},0,0,0)$    & $  1$ & $ -p +q +c_1n$ & $ -p +q+\frac{1}{2}c_1n$ \\ 
F$_{1-}$ & $(0,0,0,e^{i\chi_{-1}},0)$ & $ -1$ & $  p +q +c_1n$ & $  p +q+\frac{1}{2}c_1n$ \\ \hline
UN      & $(0,0,e^{i\chi_0},0,0)$ & $0$ & $\frac{1}{5}c_2n$ & $\frac{1}{10}c_2n$ \\ \hline
$C_4$   & $\left(e^{i\chi_2}\sqrt{\frac{1+f_z/2}{2}}, 0,0,0,e^{i\chi_{-2}}\sqrt{\frac{1-f_z/2}{2}}\right)$ 
	& $\frac{p}{(c_1-c_2/20)n}$ & $4q+\frac{1}{5}c_2n$ & $4q+\frac{1}{10}c_2n-\frac{p^2}{2(c_1-c_2/20)n}$ \\ \hline
$C_{3+}$& $\left(e^{i\chi_2}\sqrt{\frac{1+f_z}{3}}, 0,0,e^{i\chi_{-1}}\sqrt{\frac{2-f_z}{3}},0 \right)$ 
	& $\frac{p-q}{c_1n}$ & $2q$ & $2q-\frac{(p-q)^2}{2c_1n}$ \\ 
$C_{3-}$& $\left( 0,e^{i\chi_1}\sqrt{\frac{2+f_z}{3}},0,0,e^{i\chi_{-2}}\sqrt{\frac{1-f_z}{3}} \right)$
	& $\frac{p+q}{c_1n}$ & $2q$ & $2q-\frac{(p+q)^2}{2c_1n}$ \\ \hline
$C_2$   & $\left( 0, e^{i\chi_1}\sqrt{\frac{1+f_z}{2}}, 0, e^{i\chi_{-1}}\sqrt{\frac{1-f_z}{2}},0  \right)$
	& $\frac{p}{(c_1-c_2/5)n}$ & $ q+\frac{1}{5}c_2n $ & $q+\frac{1}{10}c_2n-\frac{p^2}{2(c_1-c_2/5)n}$ \\ \hline
$C_2'$  & $\left( e^{i\chi_2}\sqrt{\frac{1-\zeta_0^2+f_z/2}{2}},0,e^{i\chi_0}\zeta_0,0,e^{i\chi_{-2}}\sqrt{\frac{1-\zeta_0^2-f_z/2}{2}} \right)$
	& $\frac{p+\tilde{\gamma}_0}{c_1n}$ & $2q-\frac{\tilde{\gamma}_0^2}{2q}$ & $4q(1-\frac{\zeta_0^2}{n}) + \frac{\tilde{\gamma}_0^2-p^2}{2c_1n}$ \\
& & & & $+\frac{c_2}{10n} \left| \sqrt{(n-\zeta_0^2)^2-\frac{F_z^2}{4}}e^{i(\chi_2+\chi_{-2}-2\chi_0)}+\zeta_0^2\right|^2$ \\ \hline
$D_2'$  & $\left( ie^{i(\chi_0+\chi)}\frac{\sqrt{1-10q/(c_2n)}}{2},0,e^{i\chi_0}\sqrt{\frac{1+10q/(c_2n)}{2}},0,ie^{i(\chi_0-\chi)}\frac{\sqrt{1-10q/(c_2n)}}{2} \right)$
	& 0 & $2q$ & $2q-10q^2/(c_2n)$ \\ \hline
\end{tabular}}
\caption{Possible ground states of spin-2 Bose-Einstein condensates. 
Each state is associated with a particular symmetry, as identified in Sec.~\ref{sec:symmetry}.
In the C$_2'$ state, $\tilde{\gamma}_0$ is a real solution of Eq.~\eqref{eq:tilde_gamma_0}, and $\chi_2+\chi_{-2}-2\chi_0=\pi$ for $c_2>0$ and
 $\chi_2+\chi_{-2}-2\chi_0=0$ for $c_2<0$.
At $p=0$ the C$_4$ and C$_2$ states continuously become the BN state,
while the C$_{3\pm}$ states become the cyclic state.
For $c_2>0$, the C$_2'$ state becomes the D$_2'$ state at $p=0$
and the cyclic state at $p=q=0$.
}
\label{table3}
\end{center}
\end{table}
\end{landscape}

\subsubsection{Representation in the Cartesian basis}
\label{sec:spin2_Cartesian}
The order parameter for the spin-2 system is a rank-2 tensor~\cite{Zhou2006} and described with a 5-component spinor in the irreducible basis, while the order parameter is represented by a $3\times 3$ traceless symmetric tensor in the Cartesian basis.
To identify the relationship between these bases,
we first construct a spin-2 state from two spin-1 states:
\begin{align}
|f=2,m\rangle &= \sum_{m_1,m_2=0,\pm1} \langle 1,m_1;1,m_2|2,m\rangle |f=1,m_1;f=1,m_2\rangle.
\end{align}
Using the unitary transformation defined in Eq.~\eqref{eq:spin1_orthogonal_transformation}, we expand the above state in terms of the Cartesian basis $|\nu_1\rangle\otimes|\nu_2\rangle\ (\nu_1,\nu_2=x,y,z)$ as
\begin{align}
|f=2,m\rangle &= \sum_{\nu_1,\nu_2=x,y,z} \sum_{m_1,m_2=0,\pm1} \langle 1,m_1;1,m_2|2,m\rangle \mathcal{U}_{m_1\nu_1}\mathcal{U}_{m_2\nu_2}|\nu_1\rangle\otimes|\nu_2\rangle,
\end{align}
where the matrix elements of $\mathcal{U}$ is given in Eq.~\eqref{eq:spin1_orthogonal_U}.
Then, the order parameter $\bm\chi$ in the Cartesian basis is defined as $\sum_m\zeta_m|f=2,m\rangle = \sum_{\nu_1,\nu_2=x,y,z}\chi_{\nu_1\nu_2}|\nu_1\rangle\otimes|\nu_2\rangle$, resulting in a traceless symmetric tensor:
\begin{align}
 \chi_{\nu_1\nu_2} &= \sum_{m=-2}^2\sum_{m_1,m_2=0,\pm1} \langle 1,m_1;1,m_2|2,m\rangle \mathcal{U}_{m_1\nu_1}\mathcal{U}_{m_2\nu_2} \zeta_m\\
&=\frac{1}{2}\begin{pmatrix}
\zeta_2 - \sqrt{\frac{2}{3}}\zeta_0 + \zeta_{-2} & i(\zeta_2-\zeta_{-2}) & -\zeta_1+\zeta_{-1} \\
i(\zeta_2-\zeta_{-2}) & - \zeta_2 - \sqrt{\frac{2}{3}}\zeta_0 - \zeta_{-2}  & -i(\zeta_1+\zeta_{-1}) \\
-\zeta_1+\zeta_{-1} & -i(\zeta_1+\zeta_{-1}) & 2\sqrt{\frac{2}{3}} \zeta_0
 \end{pmatrix}.
\end{align}
In particular, the order parameter for the cyclic state $\bm\zeta=(1,0,i\sqrt{2},0,1)^{\rm T}/2$ is transformed as
\begin{align}
\bm\chi = \frac{i}{\sqrt{3}}\begin{pmatrix} 
e^{i4\pi/3} & 0 & 0 \\	   
0 & e^{i2\pi/3} & 0 \\	   
0 & 0 & 1 \\	   
\end{pmatrix},
\end{align}
indicating that this state is highly symmetric in the spin space.
Indeed, the cyclic state has the symmetry of the tetrahedron, as shown in the inset of Fig.~\ref{fig:spin-2PD_B=0}, where the (relative) phase of each lobe is 0, $2\pi/3$ and $4\pi/3$ as can be seen from the diagonal elements in the above cyclic order parameter.

\subsection{Mean-field theory of spin-3 BECs} 
\label{sec:MFTspin3}

For the case of $f=3$, we use Eq.~\eqref{V(f=3)} for $\hat{V}$ and 
take the expectation value of the Hamiltonian $\hat{H}=\hat{H}_0+\hat{V}$ over the state in Eq.~(\ref{statevector}):
\begin{align}
E [\psi]
\equiv&\langle\hat{H}\rangle_0 \nonumber\\
=&
\int d{\bm r} \Bigg\{  \sum_{m=-3}^3 \psi_m^\ast
\left[ -\frac{\hbar^2 \nabla^2}{2M} + U_{\rm trap}({\bm r}) -pm + qm^2 \right]\psi_m \nonumber\\
&+\frac{c_0}{2}n^2 +\frac{c_1}{2} | {\bm F} |^2 +\frac{c_2}{2} |A_{00}|^2 +\frac{c_3}{2}\sum_{\mathcal{M}=-2}^2 |A_{2\mathcal{M}}|^2 \Bigg\}.
\label{energy_functional(f=3)}
\end{align}
Here, the spin-3 matrices are given by
\begin{align}
\rm f_x &=
\begin{pmatrix}
0 & \sqrt{3/2} & 0 & 0 & 0 & 0 & 0 \\
\sqrt{3/2} & 0 & \sqrt{5/2} & 0 & 0 & 0 & 0 \\
0 & \sqrt{5/2} & 0 & \sqrt{3} & 0 & 0 & 0 \\
0 & 0 & \sqrt{3} & 0 & \sqrt{3} & 0 & 0 \\
0 & 0 & 0 & \sqrt{3} & 0 & \sqrt{5/2} & 0 \\
0 & 0 & 0 & 0 & \sqrt{5/2} & 0 & \sqrt{3/2} \\
0 & 0 & 0 & 0 & 0 & \sqrt{3/2} & 0 
\end{pmatrix},\\
\rm f_y &=
\begin{pmatrix}
0 & -i\sqrt{3/2} & 0 & 0 & 0 & 0 & 0 \\
i\sqrt{3/2} & 0 & -i\sqrt{5/2} & 0 & 0 & 0 & 0 \\
0 & i\sqrt{5/2} & 0 & -i\sqrt{3} & 0 & 0 & 0 \\
0 & 0 & i\sqrt{3} & 0 & -i\sqrt{3} & 0 & 0 \\
0 & 0 & 0 & i\sqrt{3} & 0 & -i\sqrt{5/2} & 0 \\
0 & 0 & 0 & 0 & i\sqrt{5/2} & 0 & -i\sqrt{3/2} \\
0 & 0 & 0 & 0 & 0 & i\sqrt{3/2} & 0 
\end{pmatrix},\\
\rm f_z &=
\begin{pmatrix}
3 & 0 & 0 & 0 & 0 & 0 & 0 \\
0 & 2 & 0 & 0 & 0 & 0 & 0 \\
0 & 0 & 1 & 0 & 0 & 0 & 0 \\
0 & 0 & 0 & 0 & 0 & 0 & 0 \\
0 & 0 & 0 & 0 &-1 & 0 & 0 \\
0 & 0 & 0 & 0 & 0 &-2 & 0 \\
0 & 0 & 0 & 0 & 0 & 0 & -3
\end{pmatrix},
\end{align}
and the components of the spin density vector $\bm F$ are given by
\begin{align}
F_+&=F_x+iF_y\nonumber\\
   &=\sqrt{6}\psi_{3}^*\psi_{2}+\sqrt{10}\psi_{2}^*\psi_{1}+2\sqrt{3}\psi_{1}^*\psi_{0}+2\sqrt{3}\psi_{0}^*\psi_{-1}+\sqrt{10}\psi_{-1}^*\psi_{-2}+\sqrt{6}\psi_{-2}^*\psi_{-3},\\
F_z&=3|\psi_{3}|^2+2|\psi_{2}|^2+|\psi_{1}|^2-|\psi_{-1}|^2-2|\psi_{-2}|^2-3|\psi_{-3}|^2.
\end{align}
The spin-singlet ($A_{00}$) and spin-quintet ($A_{2\mathcal{M}}$) pair amplitudes are written in terms of the order parameter as follows:
\begin{align}
A_{00} &\equiv \langle \hat{A}_{00}({\bm r})\rangle_0 = \frac{1}{\sqrt{7}} \; \left(2 \psi_3\psi_{-3}-2\psi_2\psi_{-2} +2\psi_1\psi_{-1} -\psi_0^2\right),\\
A_{20} &\equiv \langle \hat{A}_{20}({\bm r})\rangle_0 = \frac{1}{\sqrt{7}} \; \left(\frac{5}{\sqrt{3}}\psi_3\psi_{-3}   - \sqrt{3}\psi_1\psi_{-1}            +\sqrt{\frac{2}{3}}\psi_0^2\right),\\
A_{2\pm1} &\equiv \langle \hat{A}_{2\pm1}({\bm r})\rangle_0 = \frac{1}{\sqrt{7}} \; \left(\frac{5}{\sqrt{3}}\psi_{\pm3}\psi_{\mp2} - \sqrt{5}\psi_{\pm2}\psi_{\mp1}          +\sqrt{\frac{2}{3}}\psi_{\pm1}\psi_0\right),\\
A_{2\pm2} &\equiv \langle \hat{A}_{2\pm2}({\bm r})\rangle_0 = \frac{1}{\sqrt{7}} \; \left(\sqrt{\frac{10}{3}}\psi_{\pm3}\psi_{\mp1} - \sqrt{\frac{20}{3}}\psi_{\pm2}\psi_0 +\sqrt{2}\psi_{\pm1}^2\right).
\end{align}
If we use Eq.~\eqref{V(f=3)2} instead of Eq.~\eqref{V(f=3)}, the energy functional is written as
\begin{align}
E [\psi]
\equiv&\langle\hat{H}\rangle_0 \nonumber\\
=&
\int d{\bm r} \Bigg\{  \sum_{m=-3}^3 \psi_m^\ast
\left[ -\frac{\hbar^2 \nabla^2}{2M} + U_{\rm trap}({\bm r}) -pm + qm^2 \right]\psi_m \nonumber\\
&+\frac{\tilde{c}_0}{2}n^2 +\frac{\tilde{c}_1}{2} | {\bm F} |^2 +\frac{\tilde{c}_2}{2} |A_{00}|^2 +\frac{\tilde{c}_3}{2}{\rm Tr}\mathcal{N}^2 \Bigg\},
\label{energy_functional(f=3)2}
\end{align}
where $\mathcal{N}$ is a $3\times 3$ real symmetric tensor defined by Eq.~\eqref{eq:def_nematic_tensor},
and $\tilde{c}_i$ and $c_i$ $(i=0,1,2,3)$ are related with each other through Eq.~\eqref{c(f=3)2}.
The time-dependent GPEs for the spin-3 case is obtained by substituting Eq.~\eqref{energy_functional(f=3)} in Eq.~(\ref{dynamics}).

\begin{figure}[ht]
\begin{center}
\resizebox{\hsize}{!}{
\includegraphics{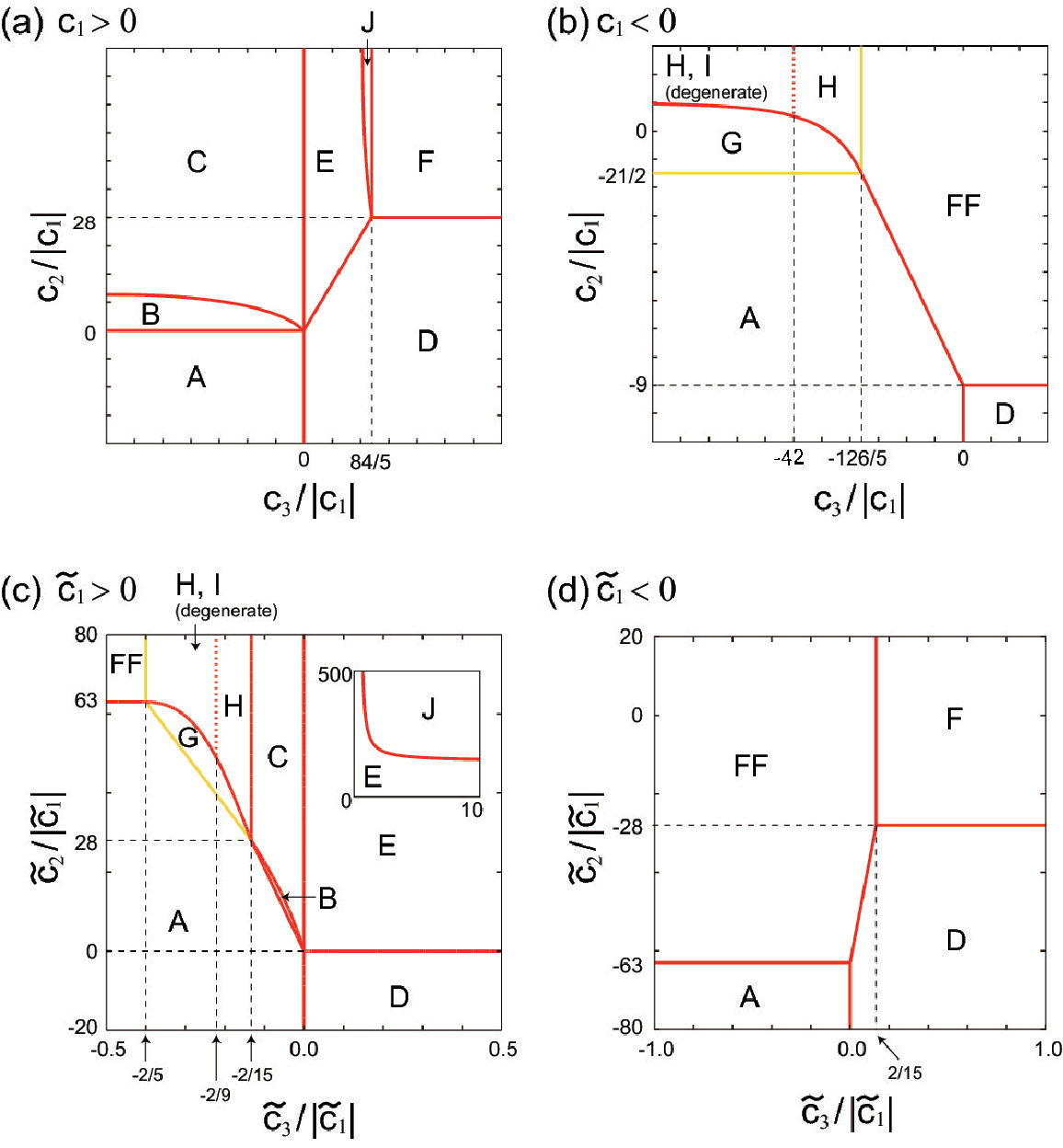}
}
\end{center}
\caption{Ground-state phase diagram of spin-3 Bose-Einstein condensates for $p=q=0$ 
in the parameter space of (a) $(c_3/|c_1|,c_2/|c_1|)$ for $c_1>0$, (b) $(c_3/|c_1|,c_2/|c_1|)$ for $c_1<0$,
(c) $(\tilde{c}_3/|\tilde{c}_1|,\tilde{c}_2/|\tilde{c}_1|)$ for $\tilde{c}_1>0$, and 
(d) $(\tilde{c}_3/|\tilde{c}_1|,\tilde{c}_2/|\tilde{c}_1|)$ for $\tilde{c}_1<0$.
For case (c), phase J exists in the region of large $c_2/|c_1|$ and large $c_3/|c_1|$, as shown in the inset of (c).
The phase boundaries in (a) and (b) are given by B-C: $y=-3x/(21-2x)$, E-J: $y=(1764x-25x^2)/[12(19x-252))]$,
and G-H: $y=7(168+5x)/(5x^2+168x+1764)$, where $x=c_3/|c_1|$ and $y=c_2/|c_1|$.
The phase boundaries in (c) are given by B-C: $\tilde{y}=168\tilde{x}(4-15\tilde{x})/(21\tilde{x}-2)$, G-H: $\tilde{y}=252\tilde{x}(5\tilde{x}-2)/(45\tilde{x}^2+12\tilde{x}+4)$,
and E-J: $\tilde{y}=567\tilde{x}/(4\tilde{x}-2)$,
where $\tilde{x}=\tilde{c}_3/|\tilde{c}_1|$ and $\tilde{y}=\tilde{c}_2/|\tilde{c}_1|$.
In the region indicated by ``H, I degenerate'', the order parameters H and I, which have different symmetries, are degenerate.
The yellow (light-colored) lines in (b) and (c) indicate second-order phase boundaries, while others are first-order.
}
\label{fig:spin-3PD}
\end{figure}

Since the order parameter has seven components, it is very involved to solve the GPEs even for the case of $p=q=0$.
Here, we show only the result for $p=q=0$ in a uniform system.
In this case, the ground state is obtained by exploiting the symmetry property of the system~\cite{Kawaguchi2011},
which will be discussed in Sec.~\ref{sec:symmetry}.
Figures~\ref{fig:spin-3PD}(a) and (b) show the phase diagram for $c_1>0$ and $c_1<0$, respectively, in the parameter space of $(c_3/|c_1|, c_2/|c_1|)$~\cite{Diener2006,Kawaguchi2011},
whereas Figs.~\ref{fig:spin-3PD}(c) and (d) are the same phase diagrams shown in the parameter space of $(\tilde{c}_3/|\tilde{c}_1|, \tilde{c}_2/|\tilde{c}_1|)$~\cite{Makela2007b,Kawaguchi2011}.

\begin{landscape}
\begin{table}
\begin{center}
{\renewcommand{\arraystretch}{1.2}
\begin{tabular}{lllll}\hline
 phase & order parameter $\bm\zeta^{\rm T}$ & $|\bm f|$ & $|a_{00}|$ & ${\rm Tr}\bar{\mathcal{N}}^2$ \\ \hline\hline
 A     & $(1,0,0,0,0,0,1)/\sqrt{2}$  & 0 & $1/\sqrt{7}$ & 171/2\\ \hline
 B     & ${(0,\sqrt{\frac{1-\eta}{4}},0,i\sqrt{\frac{1+\eta}{2}},0,\sqrt{\frac{1-\eta}{4}},0)}$ & 0 & $|\eta|/\sqrt{7}$ & $6(9+2\eta+\eta^2)$\\ 
       & $\eta={-\frac{c_\beta}{3c_\alpha-4c_\beta}}$ \\ \hline
 C     & $(a,0,b,0,c,0,d)$ & \multicolumn{3}{l}{numerically calculated} \\ 
       &  $a,b,c,d\in \mathbb{R},\ a,b,d>0, c<0$ \\ \hline 
 D     & $(0,1,0,0,0,1,0)/\sqrt{2}$  & 0 & $1/\sqrt{7}$ & 48\\ \hline
 E     & ${(\sqrt{\frac{1-\eta}{4}},0,0,\sqrt{\frac{1+\eta}{2}},0,0,\sqrt{\frac{1-\eta}{4}})}$ & 0 & $|\eta|/\sqrt{7}$ & $\frac{9}{8}(43-6\eta+27\eta^2)$ \\ 
       & $\eta={\frac{9c_\beta}{48c_\alpha+c_\beta}}$ \\ \hline
 FF    & $(1,0,0,0,0,0,0)$           & 3 & 0 & 171/2\\ \hline
 F     & $(0,1,0,0,0,0,0)$           & 2 & 0 & 48\\ \hline
 G     & $(0,\sqrt{\frac{1-\eta-\xi}{2}}+\sqrt{\frac{\xi}{2}},0,\sqrt{\eta},0,\sqrt{\frac{1-\eta-\xi}{2}}-\sqrt{\frac{\xi}{2}},0)$ & $4\sqrt{\xi(1-\eta-\xi)}$ & $|2\xi-1|/\sqrt{7}$ & $24[2-4\eta^2+5\eta(1-\xi)]$\\ 
       & \multicolumn{4}{l}{
        $\eta={\frac{(5 c_\beta - 6 c_\gamma) (c_\alpha - 4 c_\gamma)} {8 (c_\alpha c_\beta -4c_\beta  c_\gamma + 3 c_\gamma^2)}}$,
        $\xi={\frac{c_\beta (2c_\alpha - 3 c_\gamma)}{4 (c_\alpha c_\beta -4c_\beta  c_\gamma + 3 c_\gamma^2)} }$} \\ \hline
 H     & ${(\sqrt{\frac{2+\eta}{5}},0,0,0,0,\sqrt{\frac{3-\eta}{5}},0)}$  & $\eta$ & 0 & $\frac{3}{2}(36+4\eta+\eta^2)$\\ 
       & $\eta={\frac{c_\beta}{2(c_\beta-3c_\gamma)}}$                    &  \\ \hline
 I(HH) & ${(0,\sqrt{\frac{1+\eta}{3}},0,0,\sqrt{\frac{2-\eta}{3}},0,0)}$ & $\eta$ & 0 & $\frac{3}{2}(36-4\eta+\eta^2)$ \\ 
       & $\eta={\frac{c_\beta}{2(3c_\gamma-c_\beta)}}$  \\ \hline
 J     & ${(\sqrt{\frac{1+\eta}{4}},0,0,0,\sqrt{\frac{3-\eta}{4}},0,0)}$  & $\eta$ & 0 & $\frac{99}{2}-6\eta+6\eta^2$\\ 
       & $\eta={\frac{2c_\beta}{12c_\gamma-c_\beta}}$  \\ \hline
\end{tabular}}
\caption{Order parameter, magnetization $|\bm f|$, singlet-pair amplitude $|a_{00}|$, and ${\rm Tr}\bar{\mathcal{N}}^2$ for the ground state phases of spin-3 Bose-Einstein condensates at $p=q=0$.
}
\label{table:spin3}
\end{center}
\end{table}
\end{landscape}

The phase diagram for a uniform system with a fixed density is determined by the terms $\tilde{c}_1n |\bm f|^2$, $\tilde{c}_2n|a_{00}|^2$, and $\tilde{c}_3n{\rm Tr}\bar{\mathcal{N}}^2$,
where $\bm f$, $a_{00}$, and $\bar{\mathcal{N}}$ are defined in a manner similar to Eqs.~\eqref{eq:def_f=F/n}, \eqref{eq:def_a=A/n}, and \eqref{eq:def_bar_nematic_tensor}, respectively.
Here, the magnetization and the singlet-pair amplitude can vary between $0\le|\bm f|\le3$ and $0\le|a_{00}|\le 1/\sqrt{7}$, respectively.
As explained in Sec.~\ref{sec:MF_spin2_GS}, $|a_{00}|$ takes its maximum when the order parameter is invariant under time reversal.
As for the nematic tensor, since $\bar{\mathcal{N}}$ is a $3\times 3$ real symmetric tensor, it has three real eigenvalues $\lambda_i \,(i=1,2,3)$.
By definition [Eq.~\eqref{eq:def_bar_nematic_tensor}], ${\rm Tr}\bar{\mathcal{N}}=\langle {\rm f}_x^2+ {\rm f}_y^2 + {\rm f}_z^2\rangle_0=f(f+1)$.
Hence, for $f=3$, we have ${\rm Tr}\bar{\mathcal{N}}=\sum_{i=1}^3\lambda_i=12$.
Under this constraint, ${\rm Tr}\bar{\mathcal{N}}^2=\sum_{i=1}^3\lambda_i^2$ can vary between $48$ and $171/2$;
it takes the minimum at $(\lambda_1,\lambda_2,\lambda_3)=(4,4,4)$ and the maximum at $(\lambda_1,\lambda_2,\lambda_3)=(9,3/2,3/2)$ (note that $0\le\lambda_i\le 9$).
In the obtained phase diagram [Figs.~\ref{fig:spin-3PD} (c) and (d)], A, D, and FF phases arise for both $\tilde{c}_1>0$ and $\tilde{c}_1<0$ and occupy considerably large parameter spaces.
For these three phases, $|\bm f|^2$, $|a_{00}|^2$, and ${\rm Tr}\bar{\mathcal{N}}^2$ take their maxima or minima as summarized in Table~\ref{table:spin3_property}.
Hence, phase A minimizes the mean-field energy for $\tilde{c}_{2}/|\tilde{c}_1|\to -\infty$ and $\tilde{c}_{3}/|\tilde{c}_1|\to -\infty$;
phase D minimizes the mean-field energy for $\tilde{c}_{2}/|\tilde{c}_1|\to -\infty$ and $\tilde{c}_{3}/|\tilde{c}_1|\to +\infty$; 
phase FF minimizes the mean-field energy for $\tilde{c}_{2}/|\tilde{c}_1|\to +\infty$ and $\tilde{c}_{3}/|\tilde{c}_1|\to -\infty$.
Phase FF also becomes the ground state for $\tilde{c}_1<0$ and $|\tilde{c}_{2,3}/\tilde{c}_1|\ll 1$.
These results agree with Figs.~\ref{fig:spin-3PD}(c) and (d).
In the region of $\tilde{c}_{2}/|\tilde{c}_1|\to +\infty$ and $\tilde{c}_{3}/|\tilde{c}_1|\to +\infty$ for $\tilde{c}_1<0$, 
phase F appears since
this phase has the second largest magnetization among the possible configurations of the order parameter and minimizes both $|a_{00}|^2$ and ${\rm Tr}\bar{\mathcal{N}}^2$.
Since there is no order parameter that minimizes all $|\bm f|^2$, $|a_{00}|$, and ${\rm Tr}\bar{\mathcal{N}}^2$ simultaneously,
many phases arise in the phase diagram for $\tilde{c}_1>0$.
\begin{table}[ht]
\begin{center}
{\renewcommand{\arraystretch}{1.2}
\begin{tabular}{llll}\hline
phase & $|{\bm f}|^2$ & $|a_{00}|^2$ & ${\rm Tr}\bar{\mathcal{N}}^2$\\ \hline\hline
A  &min  &max  &max\\
D  &min	 &max  &min\\
FF &max	 &min  &max\\ 
F  &2nd max &min  &min\\ \hline
\end{tabular}}
\caption{Comparison of the terms that appear in the interaction Hamiltonia in Eq.~(\ref{energy_functional(f=3)2}) for A, D, FF, and F phases.}
\label{table:spin3_property}
\end{center}
\end{table}

For $f=3$, the phase diagram for $p\neq 0$ and $q=0$ is investigated in Refs.~\cite{Santos2006,Diener2006},
in which the effect of the magnetic dipole-dipole interactions are taken into account within the single-mode approximation (SMA, see Sec.~\ref{sec:SMA}).
The phase diagram under a fixed magnetization is investigated in Ref.~\cite{Makela2007b},
and that for $p\neq 0$ and $q\neq 0$ is discussed in Ref.~\cite{Santos2007}.
All these references assume that the order parameter is uniform or has a single spatial mode.
He and Yi investigate the phase diagram of a spin-3 system beyond SMA for $p\neq 0$ and predicts 
phase separation in a certain parameter region due to the magnetic dipole-dipole interaction~\cite{He2009}.

\subsection{Spin-mixing dynamics}
\label{sec:SMD}
One of the salient feature that distinguishes a spinor BEC from a binary mixture of BECs is the spin-mixing dynamics.
The initial population balance can change in a spinor BEC via the spin exchange collision, whereas it is fixed in a mixture.
For example, in a spin-1 spinor BEC, two atoms in a magnetic sublevel $m=0$ can coherently and reversibly scatter into a pair of atoms in the $m=+1$ and $m=-1$ states, and vice versa.
In this section, we describe the spin-mixing dynamics by assuming that all spin states share the same spatial dependence. Such an assumption is called the single-mode approximation (SMA).
The spin-mixing dynamics observed in experiments~\cite{Chang2005,Kronjager2005,Black2007} are well described with the SMA.
The spin dynamics beyond the SMA is discussed in Sec.~\ref{sec:Bogoliubov}.

\subsubsection{Single-mode approximation}
\label{sec:SMA}
So far, we have considered the ground-state phase diagram in a uniform system.
In experiments, however, condensates are trapped in an optical trap.
For a trapped system, the single-mode approximation (SMA) is often adopted for the studies of spin dynamics and many-body physics,
where all spin components are assumed to share the same spatial dependence and only the spin components vary in time:
\begin{align}
 \psi_m(\bm r,t) = \sqrt{N}\zeta_m(t) \psi_{\rm SMA}({\bm r})e^{-i\mu t/\hbar}.
\label{eq:SMA_OP}
\end{align}
Here, $\zeta_m(t)$ is a space-independent normalized spinor, and $\psi_{\rm SMA}({\bm r})$ is determined from the following equation:
\begin{align}
\left[-\frac{\hbar^2}{2M}\nabla^2 + U_{\rm trap}(\bm r) + c_0 N |\psi_{\rm SMA}(\bm r)|^2\right]\psi_{\rm SMA}(\bm r) = \mu\psi_{\rm SMA}(\bm r).
\label{eq:SMA_GP}
\end{align}
Under the SMA, the ground-state phase diagrams are the same as those obtained in Secs.~\ref{sec:MFTspin1}--\ref{sec:MFTspin3}, the only change being that the number density $n$ is replaced with $N/V^{\rm eff}$, where $V^{\rm eff}\equiv\left(\int d{\bm r}|\psi_{\rm SMA}|^4\right)^{-1}$ is an effective volume of the system.

The SMA can capture the physics of spinor condensates if 
all spin-dependent interaction energies, $|c_i|N/V^{\rm eff}\ (i=1,2,3,\cdots)$, are much smaller than the spin-independent interaction $c_0N/V^{\rm eff}$ and the kinetic energy $E_{\rm kin}\sim\hbar^2/(2MR^2)$,
where $R$ is the radius of the condensate.
Note, however, that the SMA is not exact even in the ground state, if 
the quadratic Zeeman effect and the spin-conservation law (or the linear Zeeman effect) are not compatible with the magnetism determined by the spin-dependent interactions~\cite{Yi2002}.
This fact can be understood if we consider, for example, a spin-1 ferromagnetic BEC ($c_1<0$) at $q>0$.
In this region, the broken-axisymmetry phase emerges due to the interplay between the quadratic Zeeman effect and ferromagnetic interaction,
in which the polarization of the magnetization ($|\bm f|=|\bm F|/n$) is determined as functions of $p/(|c_1|n)$ and $q/(|c_1|n)$ [see Eq.~\eqref{eq:BA_fz}].
In a trapped system with the SMA, however, the number density varies in space as $n({\bm r})=|\psi_{\rm SMA}(\bm r)|^2$,
and therefore, the polarization also varies in space.
The energy cost for such deformation of $|\bm f(\bm r)|$ is much smaller than $E_{\rm kin}$.
Another example is a spin-1 antiferromagnetic BEC ($c_1>0$) with $q>0$.
When the total magnetization $M_z=\int d{\bm r}F_z({\bm r})$ is fixed, the polarization $|\bm f|=M_z/N$ is constant in the SMA,
while it varies in realistic systems, since the interaction energy can be lowered by decreasing the polarization in the high-density region of the central part in the trap.

We should also emphasize that spin-orbit coupling interactions, such as the magnetic dipole-dipole interaction and the spin-orbit coupling, can generate spatial spin structures even in the ground state (see Secs.~\ref{sec:dipole_spinor} and \ref{sec:Summary}).
Therefore, these interactions must be smaller than $E_{\rm kin}$ in order for the SMA to be valid.

\subsubsection{Spin-mixing dynamics in a spin-1 BEC within the SMA}
Within the SMA, the out-of-equilibrium dynamics of a spin-1 condensate corresponds to that of a nonrigid pendulum,
which can be characterized using semi-classical trajectories in the phase space~\cite{Zhang2005a}.
This is because the system is integrable as a consequence of the conservation of the total number of atoms and the total magnetization~\cite{Romano2004}.
The spin-mixing dynamics is analogous to Josephson oscillations in weakly connected superconductors and represents a certain type of matter-wave four-wave mixing.

\begin{figure}[ht]
\begin{center}
\resizebox{0.9\hsize}{!}{
\includegraphics{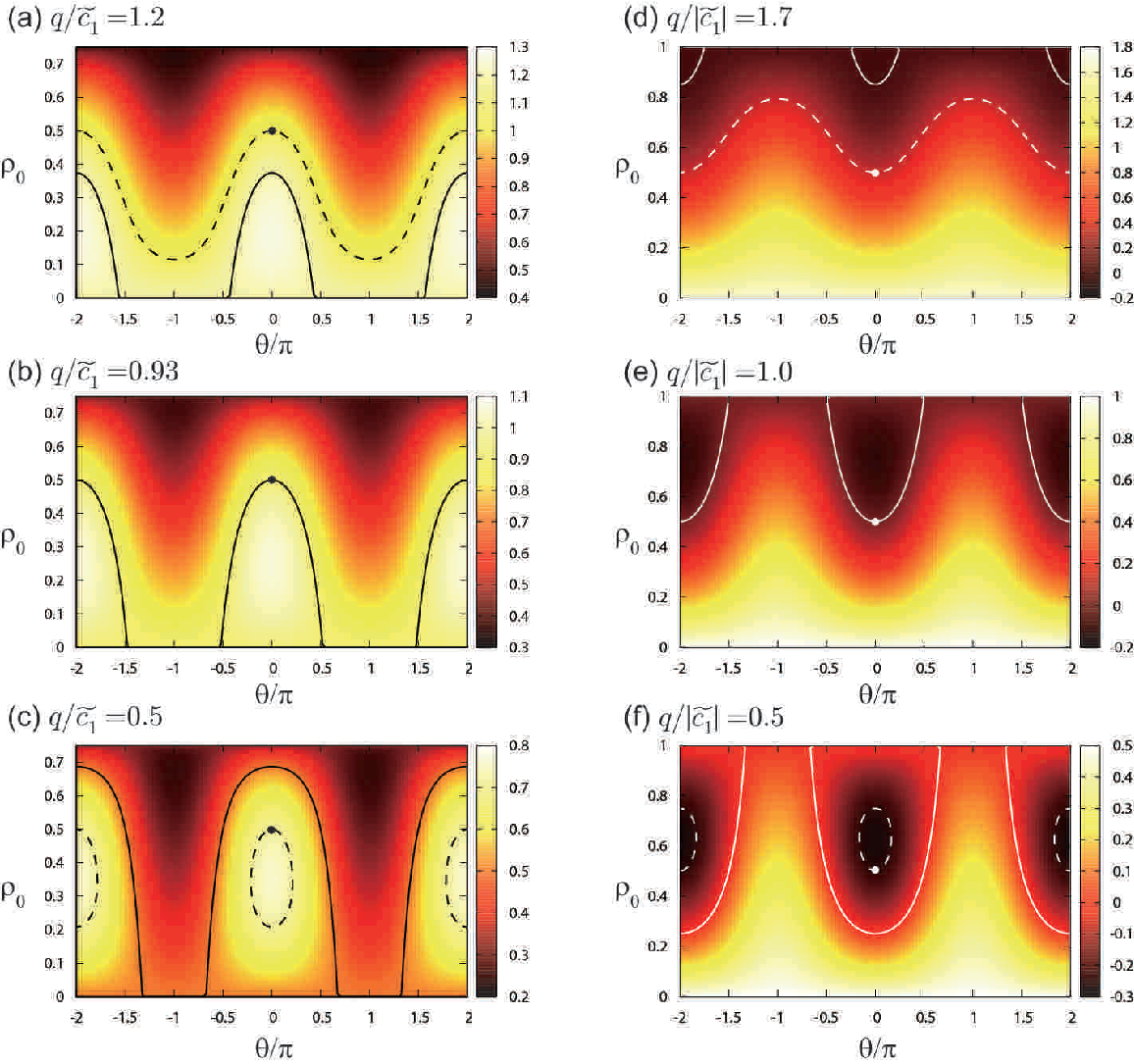}}
\end{center}
\caption{
Contour plot of the mean-field energy~\eqref{eq:SMAdynamics_ene} for (a) $\tilde{c}_1>0$ and $f_z=0.25$, and (b) $\tilde{c}_1<0$ and $f_z=0$.
The color scale shows the energy which is scaled by $|\tilde{c}_1|$.
In each panel, the dashed curve indicates the contour at $E(\theta=0,\rho_0=1/2)$,
and the solid curves show the separatrix given by Eq.~\eqref{eq:SMAdynamics_qc_polar} [(a)--(c)]
 and Eq.~\eqref{eq:SMAdynamics_qc_ferro} [(d)--(f)].
In (b) and (e), these two contours coincide with each other.
}
\label{fig:SMAdynamics}
\end{figure}

Substituting the order parameter~\eqref{eq:SMA_OP} to the spin-1 GPE~\eqref{spin-1GPE}, we obtain the equation of motion of $\zeta_m$:
\begin{align}
 i\hbar\dot{\zeta}_{\pm1} &= (\mp p+q)\zeta_{\pm1} + \tilde{c}_1[(\rho_{\pm 1}+\rho_0-\rho_{\mp1})\zeta_{\pm1}+\zeta_0^2\zeta_{\mp1}^*],\\
 i\hbar\dot{\zeta}_0 &= \tilde{c}_1[(\rho_1+\rho_{-1})\zeta_0+2\zeta_1\zeta_{-1}\zeta_0^*],
\label{eq:MF_SMD_dzeta0dt}
\end{align}
where $\tilde{c}_1\equiv c_1 N/V^{\rm eff}$ and $\rho_m\equiv |\zeta_m|^2$.
With the above set of equations of motion, the total longitudinal magnetization $f_z=|\zeta_1|^2-|\zeta_{-1}|^2$ is conserved, and the $p$ dependence can be eliminated by
transforming the order parameter as
\begin{align}
\zeta_m = \sqrt{\rho_m}e^{-i\theta_m} e^{ipmt/\hbar}.
\end{align}
From the normalization condition, $\sum_m|\zeta_m|^2=1$, and the conservation of the total magnetization, $|\zeta_1|^2-|\zeta_{-1}|^2=f_z$,
the population in the $m=\pm 1$ states are given in terms of $\rho_0$ as $\rho_{\pm1}=(1-\rho_0\pm f_z)/2$.
Then, after some algebraic manipulation, we obtain the following equations of motion for $\rho_0$ and $\theta\equiv \theta_1+\theta_{-1}-2\theta_0$:
\begin{align}
 \hbar\dot{\rho}_0 &= 2\tilde{c}_1 \rho_0 \sqrt{(1-\rho_0)^2-f_z^2}\sin\theta,\label{eq:SMAdynamics_rho}\\
 \hbar\dot{\theta} &= -2q + 2\tilde{c}_1(1-2\rho_0) + 2\tilde{c}_1 \frac{(1-2\rho_0)(1-\rho_0)-f_z^2}{\sqrt{(1-\rho_0)^2-f_z^2}}\cos\theta.
\end{align}
Note that the only the relative phase $\theta$ appears in the equations of motion
because the system has the U(1) gauge symmetry and SO(2) spin rotational symmetry about the $z$ axis.
The above two coupled equations give rise to a classical dynamics of a nonrigid pendulum.
The same set of equations are also obtained from the mean-field energy~\eqref{energy_functional(f=1)} with the SMA
\begin{align}
 E(\theta,\rho_0)=q(1-\rho_0) + \tilde{c}_1 \rho_0\left[ 1-\rho_0 + \sqrt{(1-\rho_0)^2-f_z^2}\cos\theta\right],
\label{eq:SMAdynamics_ene}
\end{align}
by using the relations $\dot{\rho}_0=-(2/\hbar)\partial E/\partial \theta$ and $\dot{\theta}=(2/\hbar)\partial E/\partial \rho_0$.
Figure~\ref{fig:SMAdynamics} shows the contour plot of $E$ in the phase space of $(\theta,\rho_0)$
for (a)--(c) $\tilde{c}_1>0$ and $f_z=0.3$, and (d)--(f) $\tilde{c}_1<0$ and $f_z=0$.
The system traces a constant-energy contour in the phase space.
When we start from the same initial state, say $(\rho_0,\theta)=(0.5,0)$ which is indicated by the solid circle in each panel in Fig.~\ref{fig:SMAdynamics}, the trajectory changes depending on the value of $q$:
from the running-phase [Figs.~\ref{fig:SMAdynamics}(a) and (d)] to periodic-phase [Figs.~\ref{fig:SMAdynamics} (c) and (f)] regime.
The $q$-dependence of the spin-mixing dynamics is also pointed out in Ref.~\cite{Schmaljohann2004b}.

\begin{figure}[ht]
\begin{center}
\resizebox{0.9\hsize}{!}{
\includegraphics{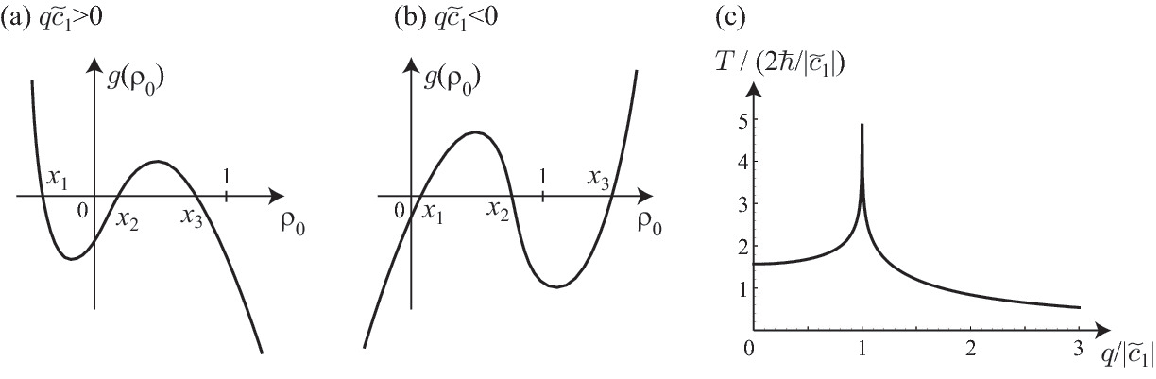}}
\end{center}
\caption{
(a), (b) $\rho_0$ dependence of $g(\rho_0)$ defined in Eq.~\eqref{eq:SMAdynamics_def_g} for (a) $q \tilde{c}_1>0$ and (b) $q\tilde{c}_1<0$.
$\rho_0(t)$ oscillates between $x_2$ and $x_3$ in (a) and between $x_1$ and $x_2$ in (b).
(c) $q$-dependence of the oscillation period $T$ starting from $\rho_0=0.5$, $f_z=0$, and $\theta=0$.
}
\label{fig:SMAdynamics_g}
\end{figure}

Using Eq.~\eqref{eq:SMAdynamics_ene}, Eq.~\eqref{eq:SMAdynamics_rho} can be rewritten as follows:
\begin{align}
 (\dot{\rho}_0)^2 &= g(\rho_0), \label{eq:dot_rho0}\\
g(\rho_0)&\equiv -\frac{4}{\hbar^2} \left\{[q(1-\rho_0)-E][(2\tilde{c}_1\rho_0+q)(1-\rho_0)-E]+(\tilde{c}_1\rho_0f_z)^2\right\} \label{eq:SMAdynamics_def_g}\\
&=-\frac{8\tilde{c}_1q}{\hbar^2}(\rho_0-x_1)(\rho_0-x_2)(\rho_0-x_3),
\end{align}
where $x_{i=1,2,3}\ (x_1\le x_2 \le x_3)$ are the roots of $g(x)=0$, and $E$ is the mean-field energy~\eqref{eq:SMAdynamics_ene} for the initial state.
Note that $g(0)$ and $g(1)$ are both negative, while Eq.~\eqref{eq:dot_rho0} is meaningful only when $0\le \rho_0\le 1$ and $g(\rho_0)\ge0$.
It follows that $x_i$'s satisfy the relation
$x_1\le 0 \le x_2\le \rho_0(t)\le x_3\le 1$ for $\tilde{c}_1 q>0$ and
$0\le x_1 \le \rho_0(t)\le x_2 \le 1 \le x_3$ for $\tilde{c}_1 q<0$ [see Figs.~\ref{fig:SMAdynamics_g}(a) and (b)].
Then, the time evolution of $\rho_0$ can be expressed in terms of the Jacobi elliptic function ${\rm cn}(z,k^2)$ as
\begin{align}
 \rho_0(t) &= x_2+(x_3-x_2){\rm cn}^2\left(\frac{\sqrt{2\tilde{c}_1q(x_3-x_1)}}{\hbar}t+\gamma_+,\frac{x_3-x_2}{x_3-x_1}\right)  \ \  \textrm{for}\ \  \tilde{c}_1q>0,\\
 \rho_0(t) &= x_2-(x_2-x_1){\rm cn}^2\left(\frac{\sqrt{2|\tilde{c}_1q|(x_3-x_1)}}{\hbar}t+\gamma_-,\frac{x_2-x_1}{x_3-x_1}\right)\ \  \textrm{for}\ \  \tilde{c}_1q<0,  
\end{align}
where ${\rm cn}^2(\gamma_+,(x_3-x_2)/(x_3-x_1))=[\rho_0(0)-x_2]/(x_3-x_2)$  and ${\rm cn}^2(\gamma_-,(x_2-x_1)/(x_3-x_1))=[x_2-\rho_0(0)]/(x_2-x_1)$.
Hence, $\rho_0(t)$ oscillates between $x_2$ and $x_3$ for $\tilde{c}_1q>0$ and between $x_1$ and $x_2$ for $\tilde{c}_1q<0$, and 
the oscillation periods are given by
\begin{align}
T&=\sqrt{\frac{2\hbar^2}{\tilde{c}_1q(x_3-x_1)}}K\left(\frac{x_3-x_2}{x_3-x_1}\right)\ \  \textrm{for}\ \  \tilde{c}_1q>0,\\
T&=\sqrt{\frac{2\hbar^2}{|\tilde{c}_1q|(x_3-x_1)}}K\left(\frac{x_2-x_1}{x_3-x_1}\right)  \ \  \textrm{for}\ \  \tilde{c}_1q<0,  
\end{align}
where $K(k^2)$ is the elliptic integral of the first kind.
Figure~\ref{fig:SMAdynamics_g}(c) shows the $q$ dependence of the oscillation period starting from $(\theta(0),\rho_0(0))=(0,1/2)$ and $f_z=0$. In this case, $T$ is analytically obtained as a function of $\tilde{q}=q/|\tilde{c}_1|$ as
$T=(2\hbar/|\tilde{c}_1|) K(\tilde{q}^2)$ for $0<\tilde{q}<1$ and $T=(2\hbar/|\tilde{c}_1|) K(\tilde{q}^{-2})/\tilde{q}$ for $\tilde{q}>1$ for both $\tilde{c}_1>0$ and $\tilde{c}_1<0$,
and there is a resonance peak at $q=\tilde{c}_1$.

For the case of $\tilde{c}_1q>0$, the period diverges when $x_1=x_2=0$, at which $\rho_0(t)$ converges to $0\,(=x_2)$ at $t\to\infty$.
Substituting $\rho_0(t\to\infty)=0$ in Eq.~\eqref{eq:SMAdynamics_ene}, we obtain $E=q$, which is equivalent to
\begin{align}
q = \tilde{c}_1 \left[ 1-\rho_0(0) + \sqrt{[1-\rho_0(0)]^2-f_z^2}\cos\theta(0)\right].
\label{eq:SMAdynamics_qc_polar}
\end{align}
Equation~\eqref{eq:SMAdynamics_qc_polar} gives the separatrix between the running-phase and periodic-phase regimes in the $(\theta,\rho_0)$ space which is indicated by a solid curve in Fig.~\ref{fig:SMAdynamics}(a)--(c).
For a fixed $q$, Eq.~\eqref{eq:SMAdynamics_qc_polar} also gives the critical value of the quadratic Zeeman energy $q_c$ for an initial state $(\theta(0), \rho_0(0))$:
$\theta(t)$ monotonically increases or decreases for $q>q_c$ [Fig.~\ref{fig:SMAdynamics}(a)], while it is periodic for $q<q_c$ [Fig.~\ref{fig:SMAdynamics}(c)].

For the case of $\tilde{c}_1q<0$, the period diverges when $x_2=x_3=1$, at which $\rho_0(t)$ converges to $1\,(=x_2)$ at $t\to\infty$.
This solution can be achieved only when $f_z=0$ since $\rho_0$ is bounded above by $1-f_z$.
When $f_z=0$, by substituting $\rho_0(t\to\infty)=1$ in Eq.~\eqref{eq:SMAdynamics_ene}, we obtain $E=0$, i.e.,
\begin{align}
q = -\tilde{c}_1\rho_0(0) [1+\cos\theta(0)].
\label{eq:SMAdynamics_qc_ferro}
\end{align}
Equation~\eqref{eq:SMAdynamics_qc_ferro} gives the separatrix in the $(\theta,\rho_0)$ plane, as indicated by a solid curve in Fig.~\ref{fig:SMAdynamics}(d)--(f).
Temporal modulation of the spin-exchange interaction is predicted to localize the spin-mixing dynamics~\cite{Zhang2010}.
 
\subsubsection{Spin-mixing dynamics in a spin-2 BEC within the SMA}
For the case of $f=2$, there are two types of spin-dependent interaction.
The $c_1$ term consists of elementary processes such as $m_1+m_2\leftrightarrow(m_1+1)+(m_2-1)$.
The elementary processes in the spin-singlet channel ($c_2$ term) are 
$0+0\leftrightarrow1+(-1)$, $0+0\leftrightarrow2+(-2)$, and $1+(-1)\leftrightarrow2+(-2)$.
The spin-independent interaction does not flip the spin.
Thus the elementary process $0+0\leftrightarrow2+(-2)$ appears only in the $c_2$ term.

In order to focus on the process in the spin-singlet channel, we consider the initial state of the form
\begin{align}
\bm \zeta = \left(e^{i\theta_2}\sqrt{\rho_2},0,e^{i\theta_0}\sqrt{\rho_0},0,e^{i\theta_{-2}}\sqrt{\rho_{-2}}\right).
\label{eq:SMA_spin2_init}
\end{align}
If $\zeta_{\pm1}$ are exactly zero in the initial state as in Eq.~\eqref{eq:SMA_spin2_init}, we find from Eq.~\eqref{f=2GPE1} that $\zeta_{\pm1}(t)$ will remain zero
within the mean-field approximation.
Then, the equations of motion can be solved in the same manner as in the case of $f=1$.
Here, the equations of motion are given by
\begin{align}
\hbar\dot{\rho_0} &=\frac{2\tilde{c}_2}{5}\rho_0\sqrt{(1-\rho_0)^2-f_z^2}\sin\theta,\\
\hbar\dot{\theta} &= -8q-\frac{2\tilde{c}_2}{5}(1-2\rho_0) + \frac{2\tilde{c}_2}{5}\frac{(1-\rho_0)(1-2\rho_0)-f_z^2}{\sqrt{(1-\rho_0)^2-f_z^2}}\cos\theta,
\end{align}
where $\theta\equiv \theta_2+\theta_{-2}-2\theta_0$ and $\tilde{c}_2\equiv c_2N/V^{\rm eff}$.
Note that the sign of $\tilde{c}_2$ is diagnozed from the qualitative behavior of the spin-mixing dynamics~\cite{Saito2005b}.
Suppose that $\theta=0$ and $f_z=0$ in the initial state and $q<-|\tilde{c}_2|/10<0$.
(Note that the quadratic Zeeman energy is negative for spin-2 BECs of $^{87}$Rb and $^{23}$Na atoms in the absence of the off-resonant microwave field.)
Then, $\theta$ monotonically increases as a function of time, and hence, $\rho_0$ increases (decreases) for $\tilde{c}_2>0$ ($\tilde{c}_2<0$) at the beginning of the dynamics.

On the other hand, when $\zeta_{\pm1}\neq 0$ in the initial state,
the spin dynamics via the $c_1$ term becomes dominant in a $^{87}$Rb BEC since $c_1\gg |c_2|$.
For example, when we prepare a BEC in the $m=0$ state with small fluctuations in the other spin states,
the $m=\pm1$ components first grow, and then the $m=\pm2$ components increase via the $c_1$ process~\cite{Chang2004}.

%% file: experiments.tex
\section{Experiments on spinor Bose-Einstein condensates}
\label{sec:experiments}

A spinor condensate was first realized by Stamper-Kurn {\it et al}.~\cite{Stamper-Kurn1998} by using a spin-1 $^{23}$Na condensate confined in an optical dipole trap, and the spin-dependent interaction coefficient $c_1$ was estimated from the domain size~\cite{Stenger1998}.
The ground-state phase diagram in the space of linear ($p$) and quadratic ($q$) Zeeman energies (see Fig.~\ref{fig:spin-1PD}) was theoretically predicted and experimentally verified for $c_1>0$ from an analysis of spin-domain structures subject to a magnetic field gradient~\cite{Stenger1998}. By changing the magnetic field linearly across the condensate, one can determine the $p$ dependence of the phase diagram from a single sample. The quadratic Zeeman coefficient $q$ can be controlled independently by application of a uniform magnetic field~\cite{Stenger1998}.
The phase diagram of a spin-1 BEC of $^{23}$Na atoms at a fixed longitudinal magnetization was experimentally verified in Ref.~\cite{Black2007}.
Miesner {\it et al}. observed a long-lived metastable state in which the condensate was segmentalized into many spin domains~\cite{Miesner1999}, and Stamper-Kurn {\it et al}. investigated the decay toward the ground state via quantum tunneling~\cite{Stamper-Kurn1999}.
Spin-1 spinor condensates of $^{87}$Rb were first created in 2001 with an all-optical scheme~\cite{Barrett2001}.
The equilibrium spin configurations of $^{87}$Rb atoms in the spin-1 manifold were measured for different magnetic fields and found to show ferromagnetic behavior~\cite{Chang2004}.

The stretched state ($|f=2,m=-2\rangle$) of the spin-2 $^{23}$Na condensate was observed to have a lifetime of several seconds~\cite{Gorlitz2003}, but the $m=0$ state has a very short lifetime of the oder of milliseconds due to hyperfine-changing interactions which render two atoms in the $|f=2,m=0\rangle$ states decay into other states such as $|f=1,m=0\rangle$ and $|f=1, m=1\rangle + |f=2,m=-1\rangle$. 
The spin-2 $^{87}$Rb spinor condensate has a relatively long lifetime of the order of a few hundreds of milliseconds~\cite{Chang2004,Schmaljohann2004,Kuwamoto2004} because inelastic collisions are suppressed due to a fortuitous near coincidence of singlet and triplet scattering lengths~\cite{Burke1997}. The spin-dependent two-body loss rates of $^{87}$Rb atoms were investigated in Ref.~\cite{Tojo2009}.

The spin-3 $^{52}$Cr BEC was first realized in an optical dipole trap by Griesmaier {\it et al.}~\cite{Griesmaier2005}. 
Because of the large magnetic dipole-dipole interaction of $^{52}$Cr atoms, only atoms in the lowest Zeeman sublevel can be stable in the presence of an external magnetic field above a few milli gauss; atoms in other magnetic sublevels would decay into lower ones and become thermalized~\cite{Hensler2003, Pasquiou2010}. 
Recently, Pasquiou {\it et al}. successfully created a spin-3 spinor condensate of $^{52}$Cr by active stabilization of magnetic field fluctuations at the level of 100~$\mu$G and lowering an ambient mangetic field below 1~mG~\cite{Pasquiou2011b,Pasquiou2012}.
Because the magnetic dipole-dipole interaction is significant in this atomic species, the spinor $^{52}$Cr condensate also possesses dipolar characters as discussed in Sec.~\ref{sec:DipolarBEC}.

\subsection{Spin-mixing dynamics in a spin-1 BEC}
The spin-mixing dynamics in spin-1 spinor condensates was discussed in Sec.~\ref{sec:SMD} and observed in $^{87}$Rb~\cite{Chang2005,Kronjager2005} and $^{23}$Na~\cite{Black2007,Liu2009a}.
In these experiments, the magnetic field dependence (i.e., $q$-dependence) of the oscillation period was investigated, and the transition between the running-phase and periodic-phase regimes was observed, in good agreement with the theoretical prediction~\cite{Zhang2005a}.
The spin-dependent interaction coefficient $c_1$ can be determined from the period of these oscillations. The results are summarized in Table~\ref{table:scattering_length}.
The total longitudinal magnetization was observed to be conserved during the spin-mixing dynamics~\cite{Chang2004}.
In long-time evolution, however, the spatial domain formation and the damped oscillations~\cite{Chang2005,Kronjager2005} show that the single-mode approximation (SMA) is no longer valid.
The domain-formation mechanism was explained by the dynamical instability of the SMA solution~\cite{Zhang2005b} (see also Secs.~\ref{sec:dynamicalinstability}--\ref{sec:Bog_Domainformation}).
In the presence of energy dissipation, the system changes from the periodic-phase to running-phase regime as the energy decreases, crossing the separatrix in the $(\theta, \rho_0)$ phase space.
Liu {\it et al.} observed this dynamics in a $^{23}$Na BEC and found that the fluctuation in each spin population shows a peak when the system crosses the separatrix~\cite{Liu2009b}.

\subsection{Spin-mixing dynamics in a spin-2 BEC}
The spin-mixing dynamics in spin-2 spinor condensates was investigated using $^{87}$Rb atoms~\cite{Chang2004,Schmaljohann2004,Kuwamoto2004}.
Starting from a condensate in the $|f=2,m=0\rangle$ state, first the $m=\pm1$ components and then the $m=\pm2$ components were observed to grow, which implies that $|c_2|\ll |c_1|$ because the process of $0+0\leftrightarrow 2+(-2)$ is caused only by the $c_2$ term. This is in agreement with the theoretical calculations of the scattering lengths~\cite{Klausen2001}.
Starting from a spin-polarized condensate in the direction perpendicular to an external magnetic field, the periodic oscillations of the $m=0$ population and their magnetic-field dependence were observed~\cite{Kronjager2006,Kronjager2008}.
Since $|c_2|\ll |c_1|$ for $^{87}$Rb, the spin-mixing dynamics is caused mainly by the $c_1$ term. In this case, 
the spin-mixing dynamics can be described analytically in the limit of $q\ll \tilde{c}_1$ and $q\gg \tilde{c}_1$,
where $\tilde{c}_1$ is defined below Eq.~\eqref{eq:MF_SMD_dzeta0dt}.
The observed oscillation period and amplitude found good agreement with the predictions in the SMA~\cite{Kronjager2006,Kronjager2008}.
The same group also observed spontaneous pattern formation of spins in an elongated trap~\cite{Kronjager2010}.
The observed $q$-dependence of the spatial pattern and its growth rate are well described by the Bogoliubov theory~\cite{Kronjager2010}.

\subsection{Spin dynamics in the Mott-insulator state}
\label{sec:exp_MI}
In the Mott-insulator state, a pair of atoms confined in each site of an optical lattice undergo oscillations due to spin-exchange interactions between Zeeman states having the same total magnetization.
Because the number of atoms can be controlled precisely in the Mott-insulator state and probed by observing the oscillating amplitude~\cite{Gerbier2006a}, the oscillations are long-lived and hence allow precise determination of the scattering lengths~\cite{Widera2005,Widera2006}.
The scattering lengths are determined from the magnetic-field dependence of the oscillation frequency, with the results summarized in Table~\ref{table:scattering_length}.

\subsection{Thermalization dynamics}

Spinor condensates provide a unique platform for studying magnetic relaxation and thermalization in superfluid systems. 
Erhard {\it et al}.~\cite{Erhard2004} and Schmaljohann {\it et al}.~\cite{Schmaljohann2004b} observed such thermalization dynamics in a spin-1 $^{87}$Rb BEC.
When atoms reside initially in the $m=1$ and $-1$ states, they can collide into the $m=0$ state with an excess kinetic energy equal to twice the quadratic Zeeman energy $2q$.
Since the spin-dependent interaction ($c_1$) is very small for spin-1 $^{87}$Rb atoms, the spin-mixing dynamics proceeds on the time scale of $\sim 1$~s (for the atomic density in Refs.~\cite{Erhard2004,Schmaljohann2004b}) which is much longer than the time scale for thermalization $\sim 50$~ms, which is proportional to the density of the thermal cloud.
Hence, the spin-flipped atoms quickly thermalize, and the condensate in the $m=0$ state emerges only after the accumulated thermal fraction becomes large enough to undergo Bose-Einstein condensation.
On the other hand, for a spin-2 $^{87}$Rb BEC, the spin dynamics caused by the $c_1$ term ($\sim10$~ms) proceeds faster than the process of thermalization ($\sim 50$~ms), and hence the condensate first appears, followed by kinetic thermalization~\cite{Schmaljohann2004}.
The same group also observed that the spin-mixing dynamics is enhanced at high temperature~\cite{Kronjager2005}.

When a thermal gas of $^{87}$Rb atoms is cooled down and reaches Bose-Einstein condensation,
small magnetic domains were observed to develop~\cite{Vengalattore2010}.
Similar magnetic domains were also observed to form from an initial spin helix at low temperatures,
suggesting the presence of the dipole-dipole interaction~\cite{Vengalattore2008}.
However, the small magnetic domains were found to disappear in the long-time evolution~\cite{Guzman2011}.
The origin of this domain formation and dissolution has not yet been fully understood.

Thermodynamics of a $^{52}$Cr BEC was experimentally studied in Ref.~\cite{Pasquiou2012}.
For the case of a $^{52}$Cr BEC, the longitudinal magnetization is not conserved due to a large magnetic dipole-dipole interaction~\cite{Hensler2003,Pasquiou2010,Pasquiou2011a,Pasquiou2011b}.
Hence, it is possible in this system to investigate thermalization
dynamics which is free from the constraint of the conservation of magnetization.

\subsection{Parametric amplification}
A stationary solution of the GPE is stable within mean-field theory, if it lies at a local maximum of the energy functional.
In reality, however, such a state cannot be stable because thermal, quantum, and extrinsic fluctuations cause dynamical or energetic instabilities of the system.
For example, if we prepare a BEC in the $m=0$ state, atoms in the $m=\pm1$ states will be created via the  process $0+0\leftrightarrow 1+(-1)$ and will grow exponentially if the initial state is dynamically unstable.
This is an example of parametric amplification of vacuum fluctuations for matter waves.

Sadler {\it et al.} used an {\it in-situ} phase-contrast imaging~\cite{Higbie2005} to observe the magnetization dynamics of a spin-1 $^{87}$Rb BEC~\cite{Sadler2006}.
In this experiment, the system was initially prepared in the $m=0$ state, which is the ground state when $q>2|c_1|n$,
and then, the external magnetic field was quenched to $q<2|c_1|n$  so that the initial state becomes dynamically unstable.
They then observed that the populations of $m=\pm1$ states spontaneously grow and the relative phase between each spin component, which corresponds to the direction of the transverse magnetization, fluctuates in space.
In a subsequent paper, they found it consistent to consider that the magnitude and spatial distribution of magnetization fluctuations in the initial state are due to quantum noise~\cite{Leslie2009}.
In a similar experimental situation, Bookjans {\it et al}. observed sub-Poissonian fluctuations in the relative atom numbers in the $m=1$ and $-1$ states~\cite{Bookjans2011a}.

Klempt {\it et al}. investigated the parametric amplification starting from the $m=0$ state of a spin-2 BEC of $^{87}$Rb atoms under a magnetic field.
The initial state is dynamically unstable and the fluctuations in the $m=\pm1$ components are amplified because $q<0$ for the $f=2$ manifold (in the absence of an off-resonant microwave).
They pointed out that compared with a uniform system the trapping potential qualitatively changes the $q$ dependence of the growth rate~\cite{Klempt2009}.
By investigating the $q$ dependence of the growth rate, they confirmed that the quantum noise, rather than the classical one, is amplified in the experiment~\cite{Klempt2010}.
The same group also observed the parametric amplification of fluctuations with a finite orbital angular momentum
in a cylindrically symmetric trap~\cite{Scherer2010}.

\subsection{Spin correlations}

As for beyond-mean-field effects, the spin correlations, that is,
population fluctuations in each spin component, have been studied in experiments:
Super-Poissonian fluctuations have been observed in the spin-mixing dynamics~\cite{Liu2009b}
and in parametric amplification~\cite{Leslie2009, Klempt2010}, while
sub-Poissonian fluctuations in parametric amplification were demonstrated in Ref.~\cite{Bookjans2011a}.
Very recently, quadrature squeezing of the spin and nematic variables below the standard quantum limit was observed~\cite{Hamley2012}.
By observing the time evolution in the number of atoms in each magnetic sublevel after quenching the quadratic Zeeman energy $q$, the first-order quantum phase transition of a spin-1 $^{23}$Na BEC crossing $q=0$ was investigated in Ref.~\cite{Bookjans2011b}.

\subsection{Topological excitations}
Due to the rich variety of phases, spinor condensates are considered to be an ideal testing ground for investigating topological excitations (see Secs.~\ref{sec:Vortices} and \ref{sec:topology}).
The first experimental study on topological excitations in spinor condensates was the creation of vortices with the winding number two and four in spin-1 and 2 $^{23}$Na BEC, respectively~\cite{Leanhardt2002}.
The same scheme has also been applied for a $^{87}$Rb BEC~\cite{Kumakura2006}.
In these experiments, the orbital angular momentum was imprinted by using the Berry phase associated with a spatial spin configuration.
Using this scheme, Leanhardt {\it et al}. created a coreless vortex state, in which each spin component has a different phase winding~\cite{Leanhardt2003}.
A similar structure was created in a spin-2 $^{87}$Rb BEC by transferring the orbital angular momentum from a Laguerre-Gaussian beam to the condensate~\cite{Leslie2009}.
Using two-photon Raman process with Laguerre-Gaussian beams, the coherent superposition of a vortex and antivortex was created in spinor condensates~\cite{Wright2008,Wright2009} in a manner similar to scalar condensates~\cite{Andersen2006}.

%% file: bogoliubov.tex
\section{Bogoliubov theory}
\label{sec:Bogoliubov}

\subsection{Bogoliubov Hamiltonian and its diagonalization}

Quantum and thermal fluctuations as well as external perturbations induce excitations from the mean-field ground state. When the excitations are weak, they can be described by the Bogoliubov theory. We express the field operator $\hat{\psi}_m$ as a sum of its mean-field value $\psi_m$ and the deviation from it, $\delta\hat{\psi}_m$:
\begin{eqnarray}
\hat{\psi}_m=\psi_m+\delta\hat{\psi}_m \ \ (m=f,f-1,\cdots,-f).
\label{decomp}
\end{eqnarray}
Here, the mean-field part can be calculated from the Gross-Pitaevskii theory described in Sec.~\ref{sec:meanfield}. The basic idea of the Bogoliubov theory is to substitute Eq.~(\ref{decomp}) in the second-quantized Hamiltonian and retain the terms up to the second order in $\delta\hat{\psi}_m$. The resulting Hamiltonian can be diagonalized by means of the Bogoliubov transformations. The Bogoliubov transformations are canonical transformations that couple particle-like excitations with hole-like ones via a Bose-Einstein condensate, and they are non-perturbative. The resultant dispersion relation therefore may depend on the coupling constant in a non-analytic manner. 

We consider a spatially uniform system [$U_{\rm trap}({\bm r})=0$]. In this case, it is convenient to analyze the system in momentum space. We expand the field operator in plane waves as
\begin{eqnarray}
\hat{\psi}_m=\frac{1}{\sqrt{\Omega}}\sum_{\bm k}\hat{a}_{{\bm k},m}e^{i{\bm k}\cdot{\bm r}},
\label{eq:FT_psi_m}
\end{eqnarray}
where $\Omega$ is the volume of the system and $\hat{a}_{{\bm k},m}$ is the annihilation operator of a boson with wave vector ${\bm k}$ and magnetic quantum number $m$. 
Then, the noninteracting and interacting parts of the Hamiltonian, which are defined in Eqs.~\eqref{H_0} and \eqref{eq:V_Cmnm'n'}, are written as
\begin{align}
\hat{H}_0&=\sum_{{\bm k}m}(\epsilon_{\bm k}-pm+qm^2)\hat{n}_{{\bm k},m},
\label{H_0Fourier}\\
\hat{V}&=\frac{1}{2\Omega}\sum_{m_1m_2m_1'm_2'}\sum_{{\bm k}_1{\bm k}_2{\bm k}_3{\bm k}_4}C^{m_1m_2}_{m_1'm_2'} \delta_{{\bm k}_1+{\bm k}_2,{\bm k}_3+{\bm k}_4}
\hat{a}_{{\bm k}_1,m_1}^\dagger\hat{a}_{{\bm k}_2,m_2}^\dagger\hat{a}_{{\bm k}_3,m_2'}\hat{a}_{{\bm k}_4,m_1'},
\label{V_fourier}
\end{align}
respectively, where $\epsilon_{\bm k}=\hbar^2{\bm k}^2/2M$, $\hat{n}_{{\bm k},m}=\hat{a}_{{\bm k},m}^\dagger\hat{a}_{{\bm k},m}$,
and $C^{m_1m_2}_{m_1'm_2'}$ is defined in Eq.~\eqref{eq:def_Cmnm'n'}.
For the case of spin-1 and 2, $C^{m_1m_2}_{m_1'm_2'}$ are simply written as follows:
\begin{align}
f=1 :\  C^{m_1m_2}_{m_1'm_2'} &= c_0 \delta_{m_1m_1'}\delta_{m_2m_2'} + c_1 \sum_{\nu=x,y,z} ({\rm f}_\nu)_{m_1m_1'}({\rm f}_\nu)_{m_2m_2'}, \label{eq:C_f=1}\\
f=2 :\  C^{m_1m_2}_{m_1'm_2'} &= c_0 \delta_{m_1m_1'}\delta_{m_2m_2'} + c_1 \sum_{\nu=x,y,z} ({\rm f}_\nu)_{m_1m_1'}({\rm f}_\nu)_{m_2m_2'} + c_2 (P_0)_{m_1m_2}(P_0)_{m_1'm_2'},\label{eq:C_f=2}
\end{align}
where 
\begin{align}
(P_0)_{m_1m_2}\equiv \langle 0,0|f,m_1;f,m_2\rangle = \frac{1}{\sqrt{5}}
\begin{pmatrix}
0 & 0 & 0 & 0 & 1 \\
0 & 0 & 0 &-1 & 0 \\
0 & 0 & 1 & 0 & 0 \\
0 &-1 & 0 & 0 & 0 \\
1 & 0 & 0 & 0 & 0 \\
\end{pmatrix}.
\end{align}
Note that $C^{m_1m_2}_{m_1'm_2'}$ is symmetric under the following permutation of spin indices:
\begin{align}
 C^{m_1m_2}_{m_1'm_2'} = C_{m_1m_2}^{m_1'm_2'} = C^{m_2m_1}_{m_2'm_1'} = C_{m_2m_1}^{m_2'm_1'}.
\label{eq:Cmnm'n'_symmetry}
\end{align}

In the Bogoliubov theory, we assume that the Bose-Einstein condensation occurs in the $\bm k=\bm 0$ state and expand the Hamiltonian up to the second order in $\hat{a}_{{\bm k}\neq{\bm 0},m}$,
obtaining
\begin{align}
\hat{H}_0=&\sum_m(-pm+qm^2)\hat{n}_{\bm 0,m}+\sum_{{\bm k}\neq{\bm 0},m}(\epsilon_{\bm k}-pm+qm^2)\hat{n}_{{\bm k},m},
\label{eq:H_0_00}\\
\hat{V}\simeq&\frac{1}{2\Omega}\sum_{m_1m_2m_1'm_2'}C^{m_1m_2}_{m_1'm_2'} \hat{a}_{{\bm 0},m_1}^\dagger\hat{a}_{{\bm 0},m_2}^\dagger\hat{a}_{{\bm 0},m_2'}\hat{a}_{{\bm 0},m_1'}\nonumber\\
&+\frac{1}{\Omega}\sum_{m_1m_2m_1'm_2'}\sum_{{\bm k}\neq{\bm 0}}C^{m_1m_2}_{m_1'm_2'} \left(
\hat{a}_{{\bm 0},m_1}^\dagger\hat{a}_{{\bm k},m_2}^\dagger\hat{a}_{{\bm 0},m_2'}\hat{a}_{{\bm k},m_1'}+\hat{a}_{{\bm 0},m_1}^\dagger\hat{a}_{{\bm k},m_2}^\dagger\hat{a}_{{\bm k},m_2'}\hat{a}_{{\bm 0},m_1'}
\right) \nonumber\\
&+\frac{1}{2\Omega}\sum_{m_1m_2m_1'm_2'}\sum_{{\bm k}\neq{\bm 0}}C^{m_1m_2}_{m_1'm_2'} \left(
\hat{a}_{{\bm 0},m_1}^\dagger\hat{a}_{{\bm 0},m_2}^\dagger\hat{a}_{-{\bm k},m_2'}\hat{a}_{{\bm k},m_1'}
+\hat{a}_{{\bm k},m_1}^\dagger\hat{a}_{-{\bm k},m_2}^\dagger\hat{a}_{{\bm 0},m_2'}\hat{a}_{{\bm 0},m_1'}\right).
\label{eq:V_00}
\end{align}
We furthermore assume that the spinor part of the condensate is described by the order parameter $\zeta_m$, that is a stationary solution of the GPE for a uniform system,
and make the following replacements:
\begin{align}
\hat{a}_{{\bm 0},m}^\dagger\hat{a}_{{\bm 0},m'} &\to \hat{N}_0\zeta_m^*\zeta_{m'}, \label{eq:Bog_a2zeta}\\
\hat{a}_{{\bm 0},m_1}^\dagger\hat{a}_{{\bm 0},m_2}^\dagger\hat{a}_{{\bm 0},m_2'}\hat{a}_{{\bm 0},m_1'} &\to \hat{N}_0(\hat{N}_0-1)\zeta_{m_1}^*\zeta_{m_2}^*\zeta_{m_2'}\zeta_{m_1'},
\end{align}
where
\begin{align}
\hat{N}_0\equiv\hat{N}-\sum_{{\bm k}\neq{\bm 0},m}\hat{n}_{{\bm k},m}
\label{eq:Bog_norm}
\end{align}
is the number operator for the condensate atoms.
For a system with a fixed number of atoms, $\hat{N}$ can be replaced with a scalar quantity $N$.
By subtracting the last term in Eq.~\eqref{eq:Bog_norm} from $\hat{N}$, we can construct a number-conserving Bogoliubov theory without introducing the chemical potential. 

Then, we obtain the Bogoliubov Hamiltonian given by
\begin{align}
\hat{H}^{\rm B} = 
&E_0
+\sum_{\bm k\neq \bm 0}\frac{D^{\rm corr}}{4\epsilon_{\bm k}}\nonumber\\
&+\sum_{{\bm k}\neq{\bm 0},m}(\epsilon_{\bm k}-pm+qm^2 -\mu)\hat{n}_{{\bm k},m}\nonumber\\
&+\frac{N}{\Omega}\sum_{m_1m_2m_1'm_2'}\sum_{{\bm k}\neq{\bm 0}}\left(C^{m_1m_2'}_{m_1'm_2}+C^{m_1m_2'}_{m_2m_1'}\right)
\zeta_{m_1'}\zeta^*_{m_2'}\hat{a}_{{\bm k},m_1}^\dagger\hat{a}_{{\bm k},m_2}
\nonumber\\
&+\frac{N}{2\Omega}\sum_{m_1m_2m_1'm_2'}\sum_{{\bm k}\neq{\bm 0}}  C^{m_1m_2}_{m_1'm_2'} \left(
\zeta_{m_1}^*\zeta_{m_2}^*\hat{a}_{-{\bm k},m_2'}\hat{a}_{{\bm k},m_1'}
+\zeta_{m_2'}\zeta_{m_1'}\hat{a}_{{\bm k},m_1}^\dagger\hat{a}_{-{\bm k},m_2}^\dagger\right),
\label{eq:Bog_Heff}
\end{align}
where we have used the symmetry property~\eqref{eq:Cmnm'n'_symmetry}, and
$E_0$ and $\mu$ originate from the first terms on the right-hand sides of Eqs.~\eqref{eq:H_0_00} and \eqref{eq:V_00}, and are given by
\begin{align}
E_0 &\equiv N\left[\sum_m(-pm+qm^2)|\zeta_m|^2 + \frac{N-1}{2\Omega}\sum_{m_1m_2m_1'm_2'}C^{m_1m_2}_{m_1'm_2'}\zeta_{m_1}^*\zeta_{m_2}^*\zeta_{m_2'}\zeta_{m_1'}\right],\\
\mu &\equiv \sum_m (-pm+qm^2)|\zeta_m|^2+\frac{2N-1}{2\Omega}\sum_{m_1m_2m_1'm_2'}C^{m_1m_2}_{m_1'm_2'}\zeta_{m_1}^*\zeta_{m_2}^*\zeta_{m_2'}\zeta_{m_1'}.
\label{eq:Bog_defmu}
\end{align}
The second term in Eq.~\eqref{eq:Bog_Heff} arises also from the first term on the right-hand side of Eqs.~\eqref{eq:V_00}.
Note that the third and fourth lines of Eq.~\eqref{eq:Bog_Heff} are of the order of the five-halves power of the interaction strength, because the number of excited atoms is  proportional to the three-halves power of the scattering length [see, e.g., Eq.~\eqref{eq:Nqntm_spin1F}].
On the other hand, the interaction potential that we have defined in Eqs.~\eqref{eq:interaction_delta} and \eqref{eq:def_g_F}
is correct only within the first-order Born approximation.
According to the higher-order perturbation theory (see, e.g., \textsection 130 of Ref.~\cite{LandauLifshitz_QM}),
$g_\mathcal{F}$ in Eq.~\eqref{eq:interaction_delta} should be replaced with
\begin{align}
 \tilde{g}_\mathcal{F} = g_\mathcal{F}+\frac{g_\mathcal{F}^2}{\Omega}\sum_{\bm k\ne\bm 0}\frac{1}{2\epsilon_{\bm k}},
\label{eq:g_F_high}
\end{align}
where $g_\mathcal{F}=4\pi\hbar^2a_{\mathcal{F}}/M$.
The last term in Eq.~\eqref{eq:g_F_high} leads to $D^{\rm corr}$ in Eq.~\eqref{eq:Bog_Heff}, given by
\begin{align}
 D^{\rm corr}=\frac{N(N-1)}{\Omega^2}\tilde{C}^{m_1m_2}_{m_1'm_2'}\zeta_{m_1}^*\zeta_{m_2}^*\zeta_{m_2'}\zeta_{m_1'},
\end{align}
where $\tilde{C}^{m_1m_2}_{m_1'm_2'}$ for $f=1$ and 2 are defined as follows:
\begin{align}
f=1 :\  \tilde{C}^{m_1m_2}_{m_1'm_2'} =& (c_0^2+2c_1^2)\delta_{m_1m_1'}\delta_{m_2m_2'} + c_1(2c_0-c_1) \sum_{\nu=x,y,z} ({\rm f}_\nu)_{m_1m_1'}({\rm f}_\nu)_{m_2m_2'}, \label{eq:tildeC_f=1}\\
f=2 :\  \tilde{C}^{m_1m_2}_{m_1'm_2'} =& (c_0^2+12c_1^2) \delta_{m_1m_1'}\delta_{m_2m_2'} + c_1(2c_0+c_1) \sum_{\nu=x,y,z} ({\rm f}_\nu)_{m_1m_1'}({\rm f}_\nu)_{m_2m_2'} \nonumber\\
 &+ [30c_1^2-12c_1c_2+c_2(2c_0+c_2)] (P_0)_{m_1m_2}(P_0)_{m_1'm_2'}.\label{eq:tildeC_f=2}
\end{align}

Next, we define $(2f+1)\times(2f+1)$ matrices:
\begin{align}
 \bm H^{(0)}_{\bm k} &= (\epsilon_{\bm k}-\mu){\bm 1}-p{\rm f}_z+q{\rm f}_z^2,
\label{eq:def_H^0}\\
 H^{(1)}_{m_1m_2} &= \frac{N}{\Omega}\sum_{m_1'm_2'}\left(C^{m_1m_2'}_{m_1'm_2}+C^{m_1m_2'}_{m_2m_1'}\right)\zeta_{m_1'}\zeta^*_{m_2'},
\label{eq:def_H^1}\\
 H^{(2)}_{m_1m_2} &= \frac{N}{\Omega}\sum_{m_1'm_2'}C^{m_1m_2}_{m_1'm_2'}\zeta_{m_1'}\zeta_{m_2'},
\label{eq:def_H^2}
\end{align}
where ${\bm 1}$ is the identity matrix.
Here, ${\bm H}_{\bm k}^{(0)}$ and ${\bm H}^{(1)}$ are Hermitian matrices, whereas ${\bm H}^{(2)}$ is a symmetric matrix.
Using these matrices, Eq.~\eqref{eq:Bog_Heff} is rewritten in a matrix form as
\begin{align}
\hat{H}^{\rm B} =& E_0 
+ \frac{1}{2}\sum_{\bm k\neq\bm 0}
\begin{pmatrix} \bar{\hat{\bm a}}^\dagger_{\bm k} & \bar{\hat{\bm a}}_{-\bm k} \end{pmatrix}
\begin{pmatrix} \bm H^{(0)}_{\bm k} + \bm H^{(1)} & \bm H^{(2)} \\
[\bm H^{(2)}]^* & [\bm H^{(0)}_{-\bm k} + \bm H^{(1)}]^*\end{pmatrix}
\begin{pmatrix} \hat{\bm a}_{\bm k} \\ \hat{\bm a}^\dagger_{-\bm k} \end{pmatrix}
\nonumber\\
&
-\frac{1}{2}\sum_{\bm k\neq\bm 0} \left\{{\rm Tr}[\bm H_{-\bm k}^{(0)}+\bm H^{(1)}]-\frac{D^{\rm corr}}{2\epsilon_{\bm k}}\right\}.
\label{eq:Bog_Heff_matrixform}
\end{align}
Here, ${\rm Tr}$ denotes the trace of the $(2f+1)\times(2f+1)$ matrix, and $\hat{\bm a}_{\bm k}^\dagger$ and $\hat{\bm a}_{\bm k}$  are the vectors of the creation and annihilation operators:
\begin{align}
&\hat{\bm a}_{\bm k}=\begin{pmatrix} \hat{a}_{{\bm k},f} \\ \hat{a}_{{\bm k},f-1} \\ \vdots \\\hat{a}_{{\bm k},-f}\end{pmatrix},\ \ \ 
\bar{\hat{\bm a}}_{\bm k}=\begin{pmatrix} \hat{a}_{{\bm k},f}, & \hat{a}_{{\bm k},f-1}, & \cdots, & \hat{a}_{{\bm k},-f}\end{pmatrix},\\
&\hat{\bm a}^\dagger_{\bm k}=\begin{pmatrix} \hat{a}^\dagger_{{\bm k},f} \\ \hat{a}^\dagger_{{\bm k},f-1} \\ \vdots \\ \hat{a}^\dagger_{{\bm k},-f}\end{pmatrix},\ \ \ 
\bar{\hat{\bm a}}^\dagger_{\bm k}=\begin{pmatrix} \hat{a}^\dagger_{{\bm k},f}, & \hat{a}^\dagger_{{\bm k},f-1}, & \cdots, & \hat{a}^\dagger_{{\bm k},-f}\end{pmatrix}.
\end{align}
Although ${\bm H}^{(0)}_{\bm k}={\bm H}^{(0)}_{-\bm k}$ from Eq.~\eqref{eq:def_H^0} in the preent situation, 
we keep ${\bm H}^{(0)}_{-\bm k}$ in the second line of Eq.~\eqref{eq:Bog_Heff_matrixform} so that we can apply the following discussions to situations in which the space inversion symmetry is broken (e.g.,  a condensate is in a spin helix state~\cite{Kawaguchi2010}).

We diagonalize Eq.~\eqref{eq:Bog_Heff_matrixform} by means of the Bogoliubov transformation:
\begin{align}
 \hat{a}_{{\bm k},m} &= \sum_{m'=-f}^f(U_{{\bm k},mm'}\hat{b}_{{\bm k},m'}+V_{-{\bm k},mm'}^*\hat{b}_{-{\bm k},m'}^\dagger),
\label{eq:Bog_tramsform_1}\\
 \hat{a}^\dagger_{-{\bm k},m} &= \sum_{m'=-f}^f(U^*_{-{\bm k},mm'}\hat{b}^\dagger_{-{\bm k},m'}+V_{{\bm k},mm'}\hat{b}_{{\bm k},m'}),
\label{eq:Bog_tramsform_2}
\end{align}
which is written in the matrix form as
\begin{align}
 \begin{pmatrix} \hat{\bm a}_{\bm k} \\ \hat{\bm a}^\dagger_{-\bm k} \end{pmatrix}= 
\begin{pmatrix} {\bm U}_{\bm k} & {\bm V}_{-\bm k}^* \\ {\bm V}_{\bm k} & {\bm U}^*_{-\bm k}\end{pmatrix}
 \begin{pmatrix} \hat{\bm b}_{\bm k} \\ \hat{\bm b}^\dagger_{-\bm k} \end{pmatrix}.
\label{eq:Bog_transform_3}
\end{align}
Here, we impose the canonical commutation relations for the Bogoliubov quasi-particle operator $\hat{b}_{{\bm k},m}$.
Note that the index $m$ in $\hat{b}_{{\bm k},m}$ means not the magnetic sublevel but a label that distinguishes spin modes.
Substituting Eqs.~\eqref{eq:Bog_tramsform_1} and \eqref{eq:Bog_tramsform_2} into the canonical commutation relations for $\hat{a}$-particles,
we obtain
\begin{align}
\left({\bm U}_{\bm k}{\bm U}_{\bm k}^\dagger - {\bm V}^*_{-\bm k}{\bm V}^{\rm T}_{-\bm k}\right)_{mm'}&=\delta_{mm'},
\label{eq:uv_relation_1}\\
\left({\bm U}_{\bm k}{\bm V}_{\bm k}^\dagger - {\bm V}^*_{-\bm k}{\bm U}^{\rm T}_{-\bm k}\right)_{mm'}&=0.
\label{eq:uv_relation_2}
\end{align}
These relations can be used to invert Eq.~\eqref{eq:Bog_transform_3} as 
\begin{align}
 \begin{pmatrix} \hat{\bm b}_{\bm k} \\ \hat{\bm b}^\dagger_{-\bm k} \end{pmatrix}= 
\begin{pmatrix} {\bm U}_{\bm k}^\dagger & -{\bm V}_{\bm k}^\dagger \\[1mm] -{\bm V}_{-\bm k}^{\rm T} & {\bm U}^{\rm T}_{-\bm k}\end{pmatrix} 
 \begin{pmatrix} \hat{\bm a}_{\bm k} \\ \hat{\bm a}^\dagger_{-\bm k} \end{pmatrix}.
\label{eq:Bog_transform_4}
\end{align}
Substituting Eq.~\eqref{eq:Bog_transform_4} into the canonical commutation relations for $\hat{b}$-particles,
we obtain
\begin{align}
\left({\bm U}_{\bm k}^\dagger{\bm U}_{\bm k} - {\bm V}^\dagger_{\bm k}{\bm V}_{\bm k}\right)_{mm'}&=\delta_{mm'},
\label{eq:uv_relation_3}\\
\left({\bm U}_{\bm k}^{\rm T}{\bm V}_{-\bm k} - {\bm V}^{\rm T}_{\bm k}{\bm U}_{-\bm k}\right)_{mm'}&=0.
\label{eq:uv_relation_4}
\end{align}

Under the Bogoliubov transformation~\eqref{eq:Bog_transform_3}, the Bogoliubov Hamiltonian~\eqref{eq:Bog_Heff_matrixform} is written in terms of $\hat{b}$-particles as
\begin{align}
\hat{H}^{\rm B} =& E_0 
-\frac{1}{2}\sum_{\bm k\neq\bm 0} \left\{{\rm Tr}[\bm H_{\bm k}^{(0)}+\bm H^{(1)}]-\frac{D^{\rm corr}}{2\epsilon_{\bm k}}\right\}
\nonumber\\
&+\frac{1}{2}\sum_{\bm k\neq\bm 0}
\begin{pmatrix} \bar{\hat{\bm b}}^\dagger_{\bm k} & \bar{\hat{\bm b}}_{-\bm k} \end{pmatrix}
\begin{pmatrix} {\bm U}_{\bm k}^\dagger & {\bm V}_{\bm k}^\dagger \\[1mm] {\bm V}_{-\bm k}^{\rm T} & {\bm U}^{\rm T}_{-\bm k}\end{pmatrix}
\begin{pmatrix} \bm H^{(0)}_{\bm k} + \bm H^{(1)} & \bm H^{(2)} \\
[\bm H^{(2)}]^* & [\bm H^{(0)}_{-\bm k} + \bm H^{(1)}]^*\end{pmatrix}
\begin{pmatrix} {\bm U}_{\bm k} & {\bm V}_{-\bm k}^* \\ {\bm V}_{\bm k} & {\bm U}^*_{-\bm k}\end{pmatrix}
\begin{pmatrix} \hat{\bm b}_{\bm k} \\ \hat{\bm b}^\dagger_{-\bm k} \end{pmatrix}.
\end{align}
Note that, using the orthonormal conditions~\eqref{eq:uv_relation_3} and \eqref{eq:uv_relation_4}, $\hat{H}^{\rm B}$ is diagonalized if ${\bm U}_{\bm k}$ and ${\bm V}_{\bm k}$ satisfy the following equation:
\begin{align}
\begin{pmatrix} \bm H^{(0)}_{\bm k} + \bm H^{(1)} & \bm H^{(2)} \\
[\bm H^{(2)}]^* & [\bm H^{(0)}_{-\bm k} + \bm H^{(1)}]^*\end{pmatrix}
\begin{pmatrix} {\bm U}_{\bm k} \\ {\bm V}_{\bm k} \end{pmatrix}
=
\begin{pmatrix} {\bm U}_{\bm k} {\bm E}_{\bm k} \\ -{\bm V}_{\bm k}{\bm E}_{\bm k} \end{pmatrix},
\label{eq:eigenvalue_matrix_eq}
\end{align}
where ${\bm E}_{\bm k}$ is a $(2f+1)\times (2f+1)$ diagonal matrix.
Expressing ${\bm U}_{\bm k}$ and ${\bm V}_{\bm k}$ in terms of column vectors ${\bm u}_{{\bm k},m}$ and ${\bm v}_{{\bm k},m}$ as
\begin{align}
{\bm U}_{\bm k} &= \begin{pmatrix} {\bm u}_{\bm k,f}, & {\bm u}_{{\bm k},f-1}, & \cdots, & {\bm u}_{{\bm k},-f} \end{pmatrix},\\
{\bm V}_{\bm k} &= \begin{pmatrix} {\bm v}_{\bm k,f}, & {\bm v}_{{\bm k},f-1}, & \cdots, & {\bm v}_{{\bm k},-f} \end{pmatrix},
\end{align}
the eigenvalue matrix~\eqref{eq:eigenvalue_matrix_eq} reduces to
\begin{align}
{\bm \sigma}_z{\bm M}^{\rm B}_{\bm k}\begin{pmatrix} {\bm u}_{{\bm k},m} \\ {\bm v}_{{\bm k},m} \end{pmatrix}
= E_{{\bm k},m}\begin{pmatrix} {\bm u}_{{\bm k},m} \\ {\bm v}_{{\bm k},m} \end{pmatrix},
\label{eq:Bog_eigenvalue_equation}
\end{align}
where $E_{{\bm k},m}$ is the diagonal element of $\bm E_{\bm k}$, and
\begin{align}
\bm M^{\rm B}_{\bm k}&\equiv \begin{pmatrix} \bm H^{(0)}_{\bm k} + \bm H^{(1)} & \bm H^{(2)} \\
[\bm H^{(2)}]^* & [\bm H^{(0)}_{-\bm k} + \bm H^{(1)}]^*\end{pmatrix},\\
{\bm \sigma}_z&=\begin{pmatrix} {\bm 1} & {\bm 0} \\ {\bm 0} & -{\bm 1}\end{pmatrix},
\end{align}
are $(4f+2)\times (4f+2)$ matrices with ${\bm 1}$ and ${\bm 0}$ being the $(2f+1)\times(2f+1)$ identity and zero matrices, respectively.
It follows from Eq.~\eqref{eq:uv_relation_3} that ${\bm u}_{{\bm k},m}$ and ${\bm v}_{{\bm k},m}$ satisfy the orthonormality relation:
\begin{align}
{\bm u}^\dagger_{{\bm k},m}{\bm u}_{{\bm k},n}-{\bm v}^\dagger_{{\bm k},m}{\bm v}_{{\bm k},n}
=\delta_{mn}.
\label{eq:norm_uv}
\end{align}
Note that if $\begin{pmatrix} {\bm u}_{{\bm k},m} \\ {\bm v}_{{\bm k},m} \end{pmatrix}$ is an eigenvector of ${\bm \sigma}_z{\bm M}^{\rm B}_{\bm k}$ with an eigenvalue $E_{\bm k,m}$, 
then $\begin{pmatrix} {\bm v}^*_{{\bm k},m} \\ {\bm u}^*_{{\bm k},m} \end{pmatrix}$ is an eigenvector of ${\bm \sigma}_z{\bm M}^{\rm B}_{-\bm k}$ with an eigenvalue $-E^*_{\bm k,m}$.
Hence, although ${\bm \sigma}_z{\bm M}^{\rm B}_{\bm k}$ has $4f+2$ eigenvectors, only the half of them satisfy the normalization condition.
Note also that the eigenvalues of ${\bm \sigma}_z{\bm M}^{\rm B}_{\bm k}$ are not necessarily real because it is not Hermitian. (${\bm M}^{\rm B}_{\bm k}$ is Hermitian.)
When ${\bm \sigma}_z{\bm M}^{\rm B}_{\bm k}$ has complex eigenvalues, the system is dynamically unstable as discussed in Sec.~\ref{sec:dynamicalinstability}.
When all eigenvalues are real, on the other hand, the Bogoliubov Hamiltonian can be diagonalized with canonical commutation relations, resulting in
\begin{align}
\hat{H}^{\rm B} =& E_0 + \frac{1}{2}\sum_{{\bm k}\ne{\bm 0}}\left\{{\rm Tr}[{\bm E}_{\bm k}-\bm H_{\bm k}^{(0)}-\bm H^{(1)}]+\frac{D^{\rm corr}}{2\epsilon_{\bm k}}\right\}
+\sum_{\bm k\neq\bm 0,m} E_{{\bm k},m}\hat{b}^\dagger_{{\bm k},m}\hat{b}_{{\bm k},m}.
\label{eq:Bog_Heff_diagonalized}
\end{align}
The second term on the right-hand side of Eq.~\eqref{eq:Bog_Heff_diagonalized}
\begin{align}
E^{\rm qntm}\equiv \frac{1}{2}\sum_{{\bm k}\ne{\bm 0}}\left\{{\rm Tr}[{\bm E}_{\bm k}-\bm H_{\bm k}^{(0)}-\bm H^{(1)}]+\frac{D^{\rm corr}}{2\epsilon_{\bm k}} \right\}
\end{align}
gives the zero-point energy.
If $E_{\bm k,m}<0$ for certain $\bm k$ and $m$, the mean-field state ($\zeta_m$) is not the ground state of the system
because the energy of the system can be lowered by creating the corresponding quasi-particles $\hat{b}_{\bm k,m}^\dagger$.
This is called the Landau instability.
In order for the system to be stable, all eigenvalues of ${\bm \sigma}_z{\bm M}^{\rm B}_{\bm k}$ under the normalization condition of Eq.~\eqref{eq:norm_uv} have to be real and semi-positive definite.

The ground state $|{\rm vac}_{\rm B}\rangle$ of the Bogoliubov Hamiltonian is defined as the vacuum of the Bogoliubov quasi-particles:
\begin{align}
 \hat{b}_{{\bm k},m}|{\rm vac}_{\rm B}\rangle = 0.
\end{align}
This Bogoliubov ground state is not the vacuum of the original $\hat{a}$-particles (real particles), but involves pairs of $\hat{a}$-particles that are virtually excited by the interparticle interaction. In fact, the fraction of virtually excited particles is calculated to be
\begin{align}
N^{\rm qntm}&\equiv \sum_{\bm k\neq\bm 0}\langle{\rm vac}_{\rm B}|\bar{\hat{\bm a}}^\dagger_{\bm k}\hat{\bm a}_{\bm k}|{\rm vac}_{\rm B}\rangle \nonumber\\
&=\sum_{\bm k\neq\bm 0}\langle{\rm vac}_{\rm B}|(\bar{\hat{\bm b}}^\dagger_{\bm k}{\bm U}^\dagger_{\bm k}+\bar{\hat{\bm b}}_{-\bm k}{\bm V}^{\rm T}_{-\rm k}) ({\bm U}_{\bm k}\hat{\bm b}_{\bm k}+{\bm V}_{-\bm k}^*\hat{\bm b}^\dagger_{-\bm k})|{\rm vac}_{\rm B}\rangle \nonumber\\
&=\sum_{\bm k\neq\bm 0} {\rm Tr}({\bm V}^\dagger_{\bm k}{\bm V}_{\bm k})\nonumber\\
&=\sum_{\bm k\neq\bm 0,m} |{\bm v}_{\bm k,m}|^2.
\end{align}

Due to the virtual pair excitations,
the properties of the Bogoliubov ground state are very different from those of the mean field. 
We see this point for the case of a scalar BEC with scattering length $a$.
A mean-field ground state is
\begin{eqnarray}
\frac{\left(\hat{a}_{{\bm 0}}^\dagger\right)^N}{\sqrt{N!}}|{\rm vac}\rangle,
\label{MF_FM}
\end{eqnarray}
where $|{\rm vac}\rangle$ is the vacuum of the real particles, i.e., $\hat{a}_{{\bm k}}|{\rm vac}\rangle=0$.
On the other hand, the Bogoliubov ground state is described in terms of $\hat{a}_{\bm k}$ as
\begin{eqnarray}
|{\rm vac}_{\rm B}\rangle = \frac{1}{\sqrt{\mathcal{N}}}\exp\left(\phi_0\hat{a}_{{\bm 0}}^\dagger
+{\sum_{\bm k}}'\alpha_{\bm k}\hat{a}_{{\bm k}}^\dagger
\hat{a}_{-{\bm k}}^\dagger\right)
|{\rm vac}\rangle,
\label{BogGS}
\end{eqnarray}
where 
$ |\phi_0|^2 = N\left[1-(8/3)\sqrt{na^3/\pi}\right]$ with $n$ being the number density,
$\alpha_{\bm k}=v_{\bm k}/u_{\bm k}$ with $u_{\bm k}$ and $v_{\bm k}$ being defined in a manner similar to ${\bm U}_{\bm k}$ and ${\bm V}_{\bm k}$  for the case of spinless particles, and
$\mathcal{N} = e^{|\phi_0|^2}{\prod_{\bm k}}'1/(1-|\alpha_{\bm k}|^2)$.
Here, the prime on the summation and product means that we sum or take a product over a half momentum space excluding ${\bm k}={\bm 0}$, in order to count each pair $({\bm k},-{\bm k})$ only once.

In contrast to the single-particle state in Eq.~(\ref{MF_FM}), the Bogoliubov ground state is a pair-correlated state.
In momentum space, the Bogoliubov ground state exhibits pair correlation having opposite momenta;
in real space, it describes the pairwise repulsive interaction. To show this, let us write down the many-body wave function in the coordinate representation:
\begin{align}
\Psi({\bm r}_1,{\bm r}_2,\cdots,{\bm r}_N)=\langle {\bm r}_1,{\bm r}_2,\cdots,{\bm r}_N|{\rm vac}_{\rm B}\rangle
=\langle{\rm vac}|
\hat{\psi}({\bm r}_1)\cdots\hat{\psi}({\bm r}_N)
|{\rm vac}_{\rm B}\rangle.
\label{eq:Bog_Psi_Nbody}
\end{align}
Substituting Eq.~(\ref{BogGS}) into Eq.~\eqref{eq:Bog_Psi_Nbody} and ignoring the terms of the order of $1/N$, we obtain~\cite{UedaTextbook}
\begin{eqnarray}
\Psi({\bm r}_1,{\bm r}_2,\cdots,{\bm r}_N)\simeq 
\frac{\exp\!\left(-\frac{4}{9}(3\pi-8)N\sqrt{na^3/\pi}
\right)}{(2\pi N\Omega^{2N})^{1/4}} 
\exp\left(-\sum_{i<j}\frac{a}{r_{ij}}
e^{-r_{ij}/\xi}
\right),
\label{mbwf}
\end{eqnarray}
where $r_{ij}\equiv|{\bm r}_i-{\bm r}_j|$ and $\xi\equiv(8\pi an)^{-1/2}$ is the correlation (or healing) length.
Thus, the many-body wave function decays exponentially whenever any two of the particles come closer than the scattering length and the correlation length.

The Bogoliubov spectra were first calculated for $f=1$ in Refs.~\cite{Ohmi1998,Ho1998,Murata2007}, and for $f=2$ in Refs.~\cite{Ueda2000,Ueda2002}.
For $f=3$, Barnett {\it et al}. obtained the Bogoliubov spectra
by linearizing the hydrodynamic equations of motion for the vertices which appear in Majorana representation (see Sec.~\ref{sec:Majorana})~\cite{Barnett2009}.

\subsection{Spin-1 BECs}
\label{sec:Bogoliubov(f=1)}

Defining matrices ${\bm \rho}$ and $\tilde{\bm \rho}$ as
\begin{align}
({\bm \rho})_{mm'} \equiv \zeta_m\zeta_{m'}^*,\ \ \ 
(\tilde{\bm \rho})_{mm'} \equiv \zeta_m\zeta_{m'},
\label{eq:rho-tilderho}
\end{align}
and using Eq.~\eqref{eq:C_f=1}, we obtain
\begin{align}
\bm H^{(1)} &= n\left[c_0 ({\bm \rho}+{\bm 1}) + c_1\sum_{\nu=x,y,z}({\rm f}_\nu {\bm \rho} {\rm f}_\nu+f_\nu{\rm f}_\nu)\right],
\label{eq:Bog_def_H1}\\
\bm H^{(2)} &= n\left[c_0\tilde{\bm \rho}+c_1\sum_{\nu=x,y,z}{\rm f}_{\nu}\tilde{\bm \rho}{\rm f}_\nu^{\rm T}\right],
\label{eq:Bog_def_H2}
\end{align}
where $n=N/\Omega$, $f_\nu = \sum_{mm'}({\rm f}_\nu)_{mm'}\zeta_m^*\zeta_{m'}$, and T denotes the matrix transpose.
Note that 
${\rm f}_\nu{\bm \rho}{\rm f}_\nu$ and ${\rm f}_\nu\tilde{\bm \rho}{\rm f}_\nu^{\rm T}$ in Eqs.~\eqref{eq:Bog_def_H1} and \eqref{eq:Bog_def_H2} are products of three matrices, whereas $f_\nu {\rm f}_\nu$ in Eq.~\eqref{eq:Bog_def_H1} is a product of scalar quantity $f_\nu$ and matrix ${\rm f}_\nu$.
We also have
\begin{align}
 E_0&=N\left[-pf_z+q(|\zeta_1|^2+|\zeta_{-1}|^2) +\frac{n}{2}(c_0+c_1|{\bm f}|^2)\right],\label{eq:Bog_spin1E0}\\
 \mu&=-pf_z+q(|\zeta_1|^2+|\zeta_{-1}|^2) + n(c_0+c_1|{\bm f}|^2),\label{eq:Bog_spin1mu}\\
 D^{\rm corr} &= n^2\left[c_0^2+2c_1^2 + c_1(2c_0-c_1) |{\bm f}|^2\right],
\end{align}
where we have neglected the terms on the order of $1/N$.
Here, $E_0/N$ corresponds to the mean-field energy per particle defined by Eq.~\eqref{eq:spin1MFenergy},
while for the stationary solution of GPE $\mu$ agrees with the chemical potential obtained in Sec.~\ref{sec:MFTspin1}.

In the following subsections, we calculate the energy spectrum for each phases in Fig.~\ref{fig:spin-1PD}.
The eigenmodes are calculated under an assumption of ${\rm Im}(E_{\bm k,m})=0$.
The energy spectra and the corresponding eigenmodes are summarized in Table~\ref{table:Bog_spin1}.

\begin{landscape}
\begin{table}[htb]
\begin{center}
{\renewcommand{\arraystretch}{1.2}
\begin{tabular}{llllll}\hline
phase & order parameter $\bm\zeta^{\rm T}$ & energy spectrum $E_{\bm k,m}$ & ${\bm u}^{\rm T}_{\bm k, m}$ & ${\bm v}^{\rm T}_{\bm k, m}$ & quasi-particle $\hat{b}_{\bm k,m}$ \\ \hline \hline
F & $(1,0,0)$ & $\sqrt{\epsilon_{\bm k}[\epsilon_{\bm k}+2(c_0+c_1)n]}$ & $(u,0,0)$& $(v,0,0)$ & Eq.~\eqref{eq:Bog_b_spin1F1}\\
&          & $\epsilon_{\bm k} + p-q$      & $(0,1,0)$& $(0,0,0)$ & Eq.~\eqref{eq:Bog_b_spin1F0} \\
&          & $\epsilon_{\bm k} + 2p-2c_1n$ & $(0,0,1)$& $(0,0,0)$ & Eq.~\eqref{eq:Bog_b_spin1F-1} \\ \hline
P & $(0,1,0)$ & $\sqrt{\epsilon_{\bm k}(\epsilon_{\bm k}+2c_0n)}$ & $(0,u,0)$& $(0,v,0)$ & Eq.~\eqref{eq:Bog_b_spin1P0} \\
&          & $\sqrt{(\epsilon_{\bm k}+q)(\epsilon_{\bm k}+q+2c_1n)}-p$ & $(u,0,0)$& $(0,0,v)$ & Eq.~\eqref{eq:Bog_b_spin1P1} \\
&          & $\sqrt{(\epsilon_{\bm k}+q)(\epsilon_{\bm k}+q+2c_1n)}+p$ & $(0,0,u)$& $(v,0,0)$ & Eq.~\eqref{eq:Bog_b_spin1P1}\\ \hline
AF & $\left(\sqrt{\frac{1+f_z}{2}},0,\sqrt{\frac{1-f_z}{2}}\right)$
          & $\sqrt{(\epsilon_{\bm k}-q)^2+2c_1n(\epsilon_{\bm k}-q)+c_1^2f_z^2}$ & $(0,u,0)$& $(0,v,0)$ & Eq.~\eqref{eq:Bog_b_spin1AF0}\\
&          & $\sqrt{\epsilon_{\bm k}\left[\epsilon_{\bm k}+c_0+c_1 \pm \sqrt{(c_0-c_1)^2+4c_0c_1f_z^2}\right]}$ & $(u,0,u')$& $(v,0,v')$ & Eqs.~\eqref{eq:Bog_b_spin1AF1} and \eqref{eq:Bog_b_spin1AF-1} at $p=0$\\ \hline
BA & $\left(\frac{\sqrt{1-\tilde{q}}}{2},\sqrt{\frac{1+\tilde{q}}{2}},\frac{\sqrt{1-\tilde{q}}}{2}\right)$ 
          & $\sqrt{\epsilon_{\bm k}(\epsilon_{\bm k}+q)}$ & $(u,0,-u)$& $(v,0,-v)$ & Eq.~\eqref{eq:Bog_b_spin1BA1}\\
&          & Eqs.~\eqref{eq:Bog_ene_spin1BA1} and \eqref{eq:Bog_ene_spin1BA2} & $(u,u',u)$& $(v,v',v)$\\ \hline
\end{tabular}}
\caption{Bogoliubov excitation spectra for spin-1 BECs, where F, P, AF, and BA denote ferromagnetic, polar, antiferromagnetic, and broken-axisymmetry phases, respectively.
The mean-field order parameter $\bm \zeta^{\rm T}$, energy spectrum $E_{\bm k, m}$, and eigenmodes ${\bm u}^{\rm T}_{\bm k, m}$ and ${\bm v}_{\bm k, m}^{\rm T}$ are listed.
Here, $f_z$ in the AF phase and $\tilde{q}$ in the BA phase are defined as functions of $p$ and $q$ as $f_z=p/(c_1n)$ and $\tilde{q}=-q/(2c_1n)$.
The order parameter of the BA phase is for $p=0$.
The second and third to last columns show the spin dependence of the eigenmode: for example, $\bm u^{\rm T}_{\bm k,m}=(u,0,0)$ and $\bm v^{\rm T}_{\bm k,m}=(0,0,v)$ show that 
the corresponding quasi-particle is given by $\hat{b}_{\bm k,m}=u \hat{a}_{\bm k,1}+v\hat{a}^\dagger_{-\bm k,-1}$.
}
\label{table:Bog_spin1}
\end{center}
\end{table}
\end{landscape}

\subsubsection{Ferromagnetic phase}
\label{sec:BogFM(f=1)}
The order parameter for the ferromagnetic phase is $\bm\zeta=(1,0,0)^{\rm T}$, which leads to
\begin{align}
 {\bm H}^{(0)}_{\bm k} +
 {\bm H}^{(1)} &= 
\begin{pmatrix}
 \epsilon_{\bm k} + (c_0+c_1)n & 0 & 0 \\ 
 0 & \epsilon_{\bm k}+p-q & 0 \\
 0 & 0 & \epsilon_{\bm k}+2p-2c_1n
\end{pmatrix},
\label{eq:H1_spin1F}\\
 {\bm H}^{(2)} &= 
\begin{pmatrix}
 (c_0+c_1)n & 0 & 0 \\
 0 & 0 & 0 \\
 0 & 0 & 0
\end{pmatrix},
\label{eq:H2_spin1F}
\end{align}
where we have used $\mu=-p+q+(c_0+c_1)n$.
Since both Eqs.~\eqref{eq:H1_spin1F} and \eqref{eq:H2_spin1F} are diagonal, all spin components are decoupled.
For the $m=1$ mode, the matrix ${\bm \sigma}_z{\bm M}^{\rm B}_{\bm k}$ has nonzero off-diagonal parts,
and the energy spectrum and the corresponding eigenmode are given by
\begin{align}
 E_{\bm k,1} &= \sqrt{\epsilon_{\bm k}[\epsilon_{\bm k}+2(c_0+c_1)n]}, \label{eq:Bog_ene_spin1F1}\\
\hat{b}_{\bm k,1} &= {\rm sgn}(c_0+c_1)\sqrt{\frac{\epsilon_{\bm k}+(c_0+c_1)n+E_{{\bm k},1}}{2E_{{\bm k}, 1}}}\hat{a}_{\bm k,1}
 + \sqrt{\frac{\epsilon_{\bm k}+(c_0+c_1)n-E_{{\bm k},1}}{2E_{{\bm k}, 1}}}\hat{a}^\dagger_{-\bm k,1},\label{eq:Bog_b_spin1F1}
\end{align}
which are the same as those of the scalar BEC~\cite{Bogoliubov1947}.
Here, ${\rm sgn}(x)=1$ for $x>0$ and ${\rm sgn}(x)=-1$ for $x<0$.
This mode describes the phonon (density) excitation, which also fluctuates the longitudinal magnetization since the condensate is fully magnetized in the longitudinal direction.
For the system to be mechanically stable, the Bogoliubov excitation spectrum \eqref{eq:Bog_ene_spin1F1} for the phonon mode must be real. 
This implies that the {\it s}-wave scattering length for the total spin-2 channel must be positive [see Eq.~\eqref{c(f=1)}]:
\begin{align}
 c_0+c_1=g_2=\frac{4\pi\hbar^2}{M} a_2>0.
\end{align}
We also note that the Bogoliubov spectrum~\eqref{eq:Bog_ene_spin1F1} is independent of the applied magnetic field and remains gapless in its presence. 
This Nambu-Goldstone mode is a consequence of the global U(1) gauge invariance due to the conservation of the total number of bosons.

On the other hand, the other two modes are single-particle like:
\begin{align}
 E_{\bm k,0} &= \epsilon_{\bm k}+p-q, \label{eq:energy_spec_spin1F0}\\
\hat{b}_{\bm k,0}&= \hat{a}_{\bm k,0}, \label{eq:Bog_b_spin1F0}\\
 E_{\bm k,-1} &= \epsilon_{\bm k}+2p-2c_1n,\\
\hat{b}_{\bm k,-1}&= \hat{a}_{\bm k,-1}. \label{eq:Bog_b_spin1F-1}
\end{align}
For the system to be stable, the single-particle excitation energies must be real and semi-positive definite.
This implies that $p\geq q$ and $p\geq c_1n$; these conditions are satisfied when the system is in the ferromagnetic phase (see Fig.~\ref{fig:spin-1PD}).
The $m=0$ and $-1$ modes describe the fluctuations in the direction and the amplitude of the polarization, respectively.
At $p=q=0$, $E_{{\bm k},0}$ becomes gapless because it is the Nambu-Goldstone mode associated with the spin rotational [SO(3)] symmetry of the Hamiltonian~\cite{Uchino2010a}.
Moreover, the energy spectrum $E_{\bm k,0}$ becomes quadratic, indicating that the critical velocity vanishes in a ferromagnetic BEC.
This is because the spin-gauge symmetry of the ferromagnetic phase allows the decay of superfluid currents through development of spin textures (see Sec.~\ref{sec:vortex_spin1ferro}).

Only the $m=1$ mode contributes to $N^{\rm qntm}$ and $E^{\rm qntm}$, resulting in
\begin{align}
\frac{ N^{\rm qntm}}{N} &= \frac{1}{N}\sum_{\bm k\neq \bm 0} \frac{\epsilon_{\bm k}+g_2n-\sqrt{\epsilon_{\bm k}(\epsilon_{\bm k}+2g_2n)}}{2\sqrt{\epsilon_{\bm k}(\epsilon_{\bm k}+2g_2n)}} = \frac{8}{3}\sqrt{\frac{na_2^3}{\pi}},
\label{eq:Nqntm_spin1F}\\
E^{\rm qntm} &=\frac{1}{2}\sum_{\bm k}\left[
\sqrt{\epsilon_{\bm k}(\epsilon_{\bm k}+2g_2n)}
-\epsilon_{\bm k}-g_2n+\frac{(g_2n)^2}{2\epsilon_{\bm k}}
\right]
= \frac{64}{15}g_2n \sqrt{\frac{na_2^3}{\pi}}N,
\end{align}
where we have used $D^{\rm corr}=(c_0+c_1)^2n^2=(g_2n)^2$.
Thus, the ground-state energy of the system is
\begin{align}
 E^{\rm F}_0 = E_0+E^{\rm qntm} 
=\frac{g_2nN}{2}\left(
1+\frac{128}{15}\sqrt{\frac{na_2^3}{\pi}}
\right)-(p-q)N.
\label{FMGST}
\end{align}
It follows from this expression that the sound velocity is given by
\begin{eqnarray}
c=\sqrt{\frac{\Omega^2}{MN}\frac{\partial^2E_0^{\rm F}}{\partial\Omega^2}}
=\sqrt{\frac{g_2n}{M}}\left(
1+8\sqrt{\frac{na_2^3}{\pi}}
\right),
\end{eqnarray}
where the leading-order term gives the Bogoliubov phonon velocity and the last term is the so-called Lee-Huang-Yang correction to it.

\subsubsection{Polar phase}
\label{sec:BogPolar(f=1)}

The order parameter for the polar phase is $\bm\zeta=(0,1,0)^{\rm T}$, which leads to
\begin{align}
 {\bm H}^{(0)}_{\bm k} +
 {\bm H}^{(1)} &= 
\begin{pmatrix}
 \epsilon_{\bm k} -p+q + c_1n & 0 & 0 \\ 
 0 & \epsilon_{\bm k}+c_0n & 0 \\
 0 & 0 & \epsilon_{\bm k}+p+q+c_1n
\end{pmatrix},\\
 {\bm H}^{(2)} &= 
\begin{pmatrix}
 0 & 0 & c_1n \\
 0 & c_0n & 0 \\
 c_1n & 0 & 0
\end{pmatrix},
\end{align}
where we have used $\mu=c_0n$.
In the matrix ${\bm \sigma}_z{\bm M}^{\rm B}_{\bm k}$, the $m=0$ mode is decoupled from other two modes and describes the phonon mode with its spectrum and eigenmode given by
\begin{align}
 E_{\bm k,0} &= \sqrt{\epsilon_{\bm k}(\epsilon_{\bm k}+2c_0n)}, \label{eq:Bog_ene_spin1P0}\\
 \hat{b}_{\bm k,0} &= {\rm sgn}(c_0)\sqrt{\frac{\epsilon_{\bm k}+c_0n+E_{{\bm k},0}}{2E_{{\bm k}, 0}}}\hat{a}_{\bm k,0}
 + \sqrt{\frac{\epsilon_{\bm k}+c_0n-E_{{\bm k},0}}{2E_{{\bm k}, 0}}}\hat{a}^\dagger_{-\bm k,0}. 
\label{eq:Bog_b_spin1P0}
\end{align}
The Bogoliubov spectrum~\eqref{eq:Bog_ene_spin1P0} is gapless, independent of the applied magnetic field,
because this is the Nambu-Goldstone phonon associated with the global U(1) symmetry breaking.
The other two mode arises from the elementary process of
$({\bm 0},0)+({\bm 0},0)\leftrightarrow({\bm k},\pm 1)+(-{\bm k},\mp 1)$,
with the eigenspectra and the corresponding eigenmodes given by
\begin{align}
 E_{\bm k,\pm1} &= \sqrt{(\epsilon_{\bm k}+q)(\epsilon_{\bm k}+q+2c_1n)}\mp p,\label{eq:energy_spec_spin1P1}\\
\hat{b}_{\bm k,\pm1} &= {\rm sgn}(c_1)\sqrt{\frac{\epsilon_{\bm k}+q+c_1n+(E_{{\bm k},\pm1}\pm p)}{2(E_{{\bm k},\pm1}\pm p)}}\hat{a}_{\bm k,\pm1}
+ \sqrt{\frac{\epsilon_{\bm k}+q+c_1n-(E_{{\bm k},\pm1}\pm p)}{2(E_{{\bm k},\pm1}\pm p)}}\hat{a}^\dagger_{-\bm k,\mp1}. \label{eq:Bog_b_spin1P1}
\end{align}
These two modes are degenerate at $p=0$ and the linear combinations of $\hat{b}_{\bm k,1}+\hat{b}_{\bm k,-1}$ and $\hat{b}_{\bm k,1}-\hat{b}_{\bm k,-1}$ describe
magnetic fluctuations in the $x$ and $y$ directions, respectively.
These modes become gapless at $p=q=0$, because they are the Nambu-Goldstone modes associated with the spin rotational symmetry. 
For the system to be stable, we must have $c_0>0$, $q>{\rm min}(0,-2c_1n)$, and $p^2<q(q+2c_1n)$;
these conditions are satisfied when the system is in the polar phase.

The quantum depletion for the polar phase is given by
\begin{align}
 N^{\rm qntm} &= \sum_{\bm k\neq \bm 0} \left[
 \frac{\epsilon_{\bm k}+c_0n-\sqrt{\epsilon_{\bm k}(\epsilon_{\bm k}+2c_0n)}}{2\sqrt{\epsilon_{\bm k}(\epsilon_{\bm k}+2c_0n)}}
+\frac{\epsilon_{\bm k}+q+c_1n-\sqrt{(\epsilon_{\bm k}+q)(\epsilon_{\bm k}+q+2c_1n)}}{\sqrt{(\epsilon_{\bm k}+q)(\epsilon_{\bm k}+q+2c_1n)}}
\right],
\end{align}
which is analytically calculated only at $q=0$ as
\begin{align}
 \frac{N^{\rm qntm}}{N} = \frac{8}{3}\sqrt{\frac{n}{\pi}}\left(\tilde{a}_0^{3/2} + 2\tilde{a}_1^{3/2}\right).
\label{eq:Nqntm_spin1P}
\end{align}
Here, $\tilde{a}_0$ and $\tilde{a}_1$ are defined so as to be the scattering length corresponding to $c_0$ and $c_1$, respectively:
\begin{align}
 \tilde{a}_0\equiv \frac{M}{4\pi\hbar^2}c_0=\frac{a_0+2a_2}{3},\ \ 
 \tilde{a}_1\equiv \frac{M}{4\pi\hbar^2}c_1=\frac{a_2-a_0}{3}.
\end{align}
From $D^{\rm corr}=(c_0^2+2c_1^2)n^2$, $E^{\rm qntm}$ is calculated as
\begin{align}
 E^{\rm qntm} =&\frac{1}{2}\sum_{\bm k}\left[
\sqrt{\epsilon_{\bm k}(\epsilon_{\bm k}+2c_0n)} - \epsilon_{\bm k}-c_0n + \frac{(c_0n)^2}{2\epsilon_{\bm k}} \right]\nonumber\\
&+
\sum_{\bm k}\left[
\sqrt{(\epsilon_{\bm k}+q)(\epsilon_{\bm k}+q+2c_1n)}
 -(\epsilon_{\bm k}+q+c_1n)
+\frac{(c_1n)^2}{2\epsilon_{\bm k}}\right]\nonumber\\
\to&\ \frac{256\pi\hbar^2}{15M}N\sqrt{\frac{n^3}{\pi}} \left(\tilde{a}_0^{5/2} + 2\tilde{a}_1^{5/2} \right) \ \ \textrm{at}\ \ q=0.
\label{eq:Eqntm_spin1P}
\end{align}
Together with $E^0=c_0nN/2$, the sound velocity at $q=0$ is given by
\begin{align}
c=\sqrt{\frac{c_0n}{M}}\left(
1+8\sqrt{\frac{n\tilde{a}_0^3}{\pi}}+ 16 \frac{\tilde{a}_1}{\tilde{a}_0}\sqrt{\frac{n\tilde{a}_1^3}{\pi}}
\right).
\end{align}
The $q$-dependences of $N^{\rm qntm}$ and $E^{\rm qntum}$ are investigated in Ref.~\cite{Uchino2010a}.

\subsubsection{Antiferromagnetic phase}
\label{sec:BogAF(f=1)}

The order parameter for the antiferromagnetic phase is given by
\begin{align}
 \zeta_{\pm 1} = \sqrt{\frac{1\pm f_z}{2}}, \ \ \zeta_0=0,
\end{align}
where $f_z=p/c_1$. The chemical potential is given by $\mu=q+c_0n$.
The matrix elements of ${\bm \sigma}_z{\bm M}^{\rm B}_{\bm k}$ are calculated as follows:
\begin{align}
 {\bm H}^{(0)}_{\bm k} +
 {\bm H}^{(1)} &= 
\begin{pmatrix}
 \epsilon_{\bm k} +\displaystyle\frac{n}{2}(c_0+c_1)(1+f_z) & 0 & \displaystyle\frac{n}{2}(c_0-c_1)\sqrt{1-f_z^2} \\ 
 0 & \epsilon_{\bm k}-q+c_1n & 0 \\
 \displaystyle\frac{n}{2}(c_0-c_1)\sqrt{1-f_z^2} & 0 & \epsilon_{\bm k}+\displaystyle\frac{n}{2}(c_0+c_1)(1-f_z)
\end{pmatrix},\\
 {\bm H}^{(2)} &= \frac{n}{2}
\begin{pmatrix}
 (c_0+c_1)(1+f_z) & 0 & (c_0-c_1)\sqrt{1-f_z^2} \\
 0 & 2c_1\sqrt{1-f_z^2} & 0 \\
 (c_0-c_1)\sqrt{1-f_z^2} & 0 &  (c_0+c_1)(1-f_z)
\end{pmatrix}.
\end{align}
By diagonalizing the matrix ${\bm \sigma}_z{\bm M}^{\rm B}_{\bm k}$, we obtain the following eigenvalues:
\begin{align}
 E_{\bm k,0} &= \sqrt{(\epsilon_{\bm k}-q)^2+2c_1n(\epsilon_{\bm k}-q)+c_1^2f_z^2}, \label{eq:Bog_ene_spin1AF0}\\
 E_{\bm k,\pm1} &= \sqrt{\epsilon_{\bm k}\left[\epsilon_{\bm k}+c_0+c_1 \pm \sqrt{(c_0-c_1)^2+4c_0c_1f_z^2}\right]}.\label{eq:Bog_AF_magnon-phonon}
\end{align}
The spectrum~\eqref{eq:Bog_ene_spin1AF0} becomes quadratic and gapful in the presence of an external field,
with the corresponding quasi-particle given by
\begin{align}
\hat{b}_{\bm k,0} &= {\rm sgn}(c_1)\sqrt{\frac{\epsilon_{\bm k}-q+c_1n+E_{{\bm k},0}}{2E_{{\bm k}, 0}}}\hat{a}_{\bm k,0}
 + \sqrt{\frac{\epsilon_{\bm k}-q+c_1n-E_{{\bm k},0}}{2E_{{\bm k}, 0}}}\hat{a}^\dagger_{-\bm k,0}. \label{eq:Bog_b_spin1AF0}
\end{align}
On the other hand, the spectrum~\eqref{eq:Bog_AF_magnon-phonon} is always gapless and linear in $k$, independent of the external field.
The corresponding eigenmodes are described with linear combinations of $\hat{a}_{{\bm k},\pm1}$ and $\hat{a}^\dagger_{-{\bm k},\pm1}$.
Phonon and magnon excitations are coupled in these modes.
At $p=0$, Eq.~\eqref{eq:Bog_AF_magnon-phonon} reduces to the excitation spectra of the uncoupled phonon and magnon modes:
\begin{align}
 E_{\bm k,1} &= \sqrt{\epsilon_{\bm k}(\epsilon_{\bm k}+2c_0n)},\\
\hat{b}_{\bm k,1} &= {\rm sgn}(c_0)\sqrt{\frac{\epsilon_{\bm k}+c_0n+E_{{\bm k},1}}{2E_{{\bm k}, 1}}}\frac{\hat{a}_{\bm k,1}+\hat{a}_{\bm k,-1}}{\sqrt{2}}
 + \sqrt{\frac{\epsilon_{\bm k}+c_0n-E_{{\bm k},1}}{2E_{{\bm k}, 1}}}\frac{\hat{a}^\dagger_{\bm k,1}+\hat{a}^\dagger_{-\bm k,-1}}{\sqrt{2}}, \label{eq:Bog_b_spin1AF1}\\
 E_{\bm k,-1} &= \sqrt{\epsilon_{\bm k}(\epsilon_{\bm k}+2c_1n)},\\
\hat{b}_{\bm k,-1} &= {\rm sgn}(c_1)\sqrt{\frac{\epsilon_{\bm k}+c_1n+E_{{\bm k},-1}}{2E_{{\bm k},-1}}}\frac{\hat{a}_{\bm k,1}-\hat{a}_{\bm k,-1}}{\sqrt{2}}
 + \sqrt{\frac{\epsilon_{\bm k}+c_1n-E_{{\bm k},-1}}{2E_{{\bm k},-1}}}\frac{\hat{a}^\dagger_{\bm k,1}-\hat{a}^\dagger_{-\bm k,-1}}{\sqrt{2}}. \label{eq:Bog_b_spin1AF-1}
\end{align}
In the absence of an external field ($p=q=0$), the energy spectra, and hence the quantum depletion and the zero-point fluctuation energy,
coincide with those of the polar phase,
in agreement with the fact that the order parameters $(0,1,0)^{\rm T}$ and $(1,0,1)^{\rm T}/\sqrt{2}$ are transformed to each other by a spin rotation.

\subsubsection{Broken-axisymmetry phase}
\label{sec:BogbA(f=1)}

Here we consider the case of $p=0$. 
The case of $p\neq 0$ is discussed in Ref.~\cite{Murata2007}.
The order parameter for the BA phase with $p=0$ is given by
\begin{align}
 \zeta_{\pm 1} = \frac{\sqrt{1-\tilde{q}}}{2},\ \ \zeta_0=\sqrt{\frac{1+\tilde{q}}{2}},
\end{align}
where $\tilde{q}=-q/(2c_1n)$. The chemical potential is given by $\mu=c_0n+c_1n(1+\tilde{q})$.
In this case, the condensate is polarized along the $x$ direction with the polarization $f_x=\sqrt{1-\tilde{q}^2}$.
Hence, the energy spectra should coincide with those of the ferromagnetic (or polar) phase at $q=0$ (or $q=-2c_1n$).

The matrix elements of ${\bm \sigma}_z{\bm M}^{\rm B}_{\bm k}$ for this case are given by
\begin{align}
& {\bm H}^{(0)}_{\bm k} +
 {\bm H}^{(1)} \nonumber\\
&=\epsilon_{\bm k}{\bm 1}+
\frac{n}{4}\begin{pmatrix}
(c_0-c_1)-(c_0-3c_1)\tilde{q} & (c_0+3c_1)\sqrt{2(1-\tilde{q}^2)} & (c_0-c_1)(1-\tilde{q}) \\ 
 (c_0+3c_1)\sqrt{2(1-\tilde{q}^2)} & 2(c_0-c_1)+2(c_0+c_1)\tilde{q} & (c_0+3c_1)\sqrt{2(1-\tilde{q}^2)} \\
 (c_0-c_1)(1-\tilde{q}) & (c_0+3c_1)\sqrt{2(1-\tilde{q}^2)} & (c_0-c_1)-(c_0-3c_1)\tilde{q} 
\end{pmatrix},\\
& {\bm H}^{(2)} = \frac{n}{4}
\begin{pmatrix}
 (c_0+c_1)(1-\tilde{q}) & (c_0+c_1)\sqrt{2(1-\tilde{q}^2)} & (c_0+c_1)-(c_0-3c_1)\tilde{q} \\
 (c_0+c_1)\sqrt{2(1-\tilde{q}^2)} &  2(c_0+c_1)+2(c_0-c_1)\tilde{q} & (c_0+c_1)\sqrt{2(1-\tilde{q}^2)} \\
 (c_0+c_1)-(c_0-3c_1)\tilde{q} & (c_0+c_1)\sqrt{2(1-\tilde{q}^2)} &  (c_0+c_1)(1-\tilde{q})
\end{pmatrix}.
\end{align}
Since ${\bm H}^{(0)}_{\bm k} + {\bm H}^{(1)}$ and ${\bm H}^{(2)}$ are symmetric under exchange of $m=1$ and $-1$,
the following mode is decoupled from the other two modes:
\begin{align}
 E_{\bm k,0}&=\sqrt{\epsilon_{\bm k}(\epsilon_{\bm k}+q)},\\
\hat{b}_{\bm k,0} &= {\rm sgn}(q)\sqrt{\frac{\epsilon_{\bm k}+q/2+E_{{\bm k},0}}{2E_{{\bm k}, 0}}}\frac{\hat{a}_{\bm k,1}-\hat{a}_{\bm k,-1}}{\sqrt{2}}
 + \sqrt{\frac{\epsilon_{\bm k}+q/2-E_{{\bm k},0}}{2E_{{\bm k}, 0}}}\frac{\hat{a}^\dagger_{\bm k,1}-\hat{a}^\dagger_{-\bm k,-1}}{\sqrt{2}}. \label{eq:Bog_b_spin1BA1}
\end{align}
This energy spectrum coincides with that in Eq.~\eqref{eq:energy_spec_spin1F0} at $p=q=0$ and that in Eq.~\eqref{eq:energy_spec_spin1P1} at $p=0$ and $q=-2c_1n$.
The other two modes are given by
\begin{align}
E_{\bm k,\pm1} &= \sqrt{\epsilon_{\bm k}^2 + (c_0-c_1)n\epsilon_{\bm k} + 2(c_1n)^2(1-\tilde{q}^2)\pm \Lambda_{\bm k}}, \label{eq:Bog_ene_spin1BA1}\\
\Lambda_{\bm k} &= \sqrt{[(c_0+3c_1)n\epsilon_{\bm k}+2(c_1n)^2(1-\tilde{q}^2)]^2-4c_1(c_0+2c_1)n^2\tilde{q}^2\epsilon_{\bm k}^2}.\label{eq:Bog_ene_spin1BA2}
\end{align}
The $q$-dependence of $N^{\rm qntm}$ and $E^{\rm qntm}$ are investigated in Ref.~\cite{Uchino2010a}.

\subsection{Spin-2 BECs}
\label{sec:Bogoliubov(f=2)}

Substituting Eq.~\eqref{eq:C_f=2} in Eqs.~\eqref{eq:def_H^1} and \eqref{eq:def_H^2}, we obtain
\begin{align}
\bm H^{(1)} &= n\left[c_0 ({\bm \rho}+{\bm 1}) + c_1\sum_{\nu=x,y,z}({\rm f}_\nu {\bm \rho} {\rm f}_\nu+f_\nu{\rm f}_\nu) + 2c_2 P_0{\bm \rho} P_0\right], \label{eq:Bog_H1_spin2}\\
\bm H^{(2)} &= n\left[c_0\tilde{\bm \rho}+c_1\sum_{\nu=x,y,z}{\rm f}_{\nu}\tilde{\bm \rho}{\rm f}_\nu^{\rm T}+c_2 a_{00}P_0\right],\label{eq:Bog_H2_spin2}
\end{align}
where $n=N/\Omega$, $f_\nu = \sum_{mm'}({\rm f}_\nu)_{mm'}\zeta_m^*\zeta_{m'}$, $a_{00}=\sum_{mm'}(P_0)_{mm'}\zeta_m\zeta_{m'}$, and $\bm \rho$ and $\tilde{\bm \rho}$ are defined in the same manner as in Eq.~\eqref{eq:rho-tilderho}.
$E_0$, $\mu$, and $D^{\rm corr}$ for a spin-2 system are given by 
\begin{align}
 E_0&=N\left[-pf_z+q(|\zeta_1|^2+|\zeta_{-1}|^2) +\frac{n}{2}(c_0+c_1|{\bm f}|^2+c_2|a_{00}|^2)\right],\label{eq:Bog_spin2E0}\\
 \mu&=-pf_z+q(|\zeta_1|^2+|\zeta_{-1}|^2) + n(c_0+c_1|{\bm f}|^2+c_2|a_{00}|^2),\label{eq:Bog_spin2mu}\\
 D^{\rm corr} &= n^2\left\{c_0^2+12c_1^2 + c_1(2c_0+c_1) |{\bm f}|^2 + [30c_1^2-12c_1c_2+c_2(2c_0+c_2)]|a_{00}|^2\right\},
\end{align}
where we have neglected the terms on the order of $1/N$.
Here, $E_0/N$ agrees with the mean-field energy defined in Eq.~\eqref{eqS2ad6},
while $\mu$ for the stationary solution of the GPE agrees with the chemical potential.

In the following subsections, we calculate the energy spectrum for several stationary states obtained in Sec.~\ref{sec:MFTspin2} (see Table~\ref{table3}).
The eigenmodes are calculated under the assumption of ${\rm Im}(E_{\bm k,m})=0$.
The obtained energy spectra and the corresponding eigenmodes are summarized in Table~\ref{table:Bog_spin2}.

\begin{landscape}
\begin{table}[htb]
\begin{center}
{\renewcommand{\arraystretch}{1.2}
\begin{tabular}{llllll} \hline
phase & order parameter $\bm\zeta^{\rm T}$ & energy spectrum $E_{\bm k,m}$ & ${\bm u}^{\rm T}_{\bm k, m}$ & ${\bm v}^{\rm T}_{\bm k, m}$ & quasi-particle  \\ \hline \hline
F$_2$ & $(1,0,0,0,0)$ & $\sqrt{\epsilon_{\bm k}[\epsilon_{\bm k}+2(c_0+4c_1)n]}$ & $(u,0,0,0,0)$& $(v,0,0,0,0)$ & Eq.~\eqref{eq:Bog_b_spin2F2}\\
&          & $\epsilon_{\bm k} + p-3q$      & $(0,1,0,0,0)$& $(0,0,0,0,0)$ &  \\
&          & $\epsilon_{\bm k} + 2p-4q-4c_1n$      & $(0,0,1,0,0)$& $(0,0,0,0,0)$ &  \\
&          & $\epsilon_{\bm k} + 3p-3q-6c_1n$      & $(0,0,0,1,0)$& $(0,0,0,0,0)$ &  \\
&          & $\epsilon_{\bm k} + 4p-(8c_1-2c_2/5)n$      & $(0,0,0,0,1)$& $(0,0,0,0,0)$ &  \\ \hline
UN & $(0,0,1,0,0)$ & $\sqrt{\epsilon_{\bm k}[\epsilon_{\bm k}+2(c_0+c_2/5)n]}$ & $(0,0,u,0,0)$& $(0,0,v,0,0)$ & Eq.~\eqref{eq:Bog_b_spin2UN0} \\
&          & $\sqrt{(\epsilon_{\bm k}+q)[\epsilon_{\bm k}+q+2(3c_1-c_2/5)n]}-p$ & $(0,u,0,0,0)$& $(0,0,0,v,0)$ & Eq.~\eqref{eq:Bog_b_spin2UN1} \\
&          & $\sqrt{(\epsilon_{\bm k}+q)[\epsilon_{\bm k}+q+2(3c_1-c_2/5)n]}+p$ & $(0,0,0,u,0)$& $(0,v,0,0,0)$ & Eq.~\eqref{eq:Bog_b_spin2UN1} \\
&          & $\sqrt{(\epsilon_{\bm k}+4q)(\epsilon_{\bm k}+4q-2c_2n/5)}-2p$ & $(u,0,0,0,0)$& $(0,0,0,0,v)$ & Eq.~\eqref{eq:Bog_b_spin2UN2}\\ 
&          & $\sqrt{(\epsilon_{\bm k}+4q)(\epsilon_{\bm k}+4q-2c_2n/5)}+2p$ & $(0,0,0,0,u)$& $(v,0,0,0,0)$ & Eq.~\eqref{eq:Bog_b_spin2UN2}\\ \hline
C$_4$ & $\left(\sqrt{\frac{1+f_z/2}{2}},0,0,0,\sqrt{\frac{1-f_z/2}{2}}\right)$
          & $\sqrt{(\epsilon_{\bm k}-4q)^2-\frac{2c_2n}{5}(\epsilon_{\bm k}-4q)+\frac{(f_zc_2n)^2}{100}}$ & $(0,0,u,0,0)$& $(0,0,v,0,0)$ & Eq.~\eqref{eq:Bog_b_spin2C40}\\ 
&          & Eq.~\eqref{eq:Bog_ene_spin2C41} with $+$ sign for the last term & $(0,u,0,0,0)$& $(0,0,0,v,0)$ & Eq.~\eqref{eq:Bog_b_spin2C41} \\
&          & Eq.~\eqref{eq:Bog_ene_spin2C41} with $-$ sign for the last term & $(0,0,0,u,0)$& $(0,v,0,0,0)$ & Eq.~\eqref{eq:Bog_b_spin2C41} \\
&          & Eq.~\eqref{eq:Bog_ene_spin2C42} & $(u,0,0,0,u')$& $(v,0,0,0,v')$ & Eqs.~\eqref{eq:Bog_b_spin2C42} and \eqref{eq:Bog_b_spin2C4-2} at $p=0$ \\\hline
C & $(1,0,\sqrt{2},0,-1)/2$ 
          & $\sqrt{\epsilon_{\bm k}(\epsilon_{\bm k}+2c_0)}$ & $(u,0,\sqrt{2}u,0,-u)$& $(v,0,\sqrt{2}v,0,-v)$ & Eq.~\eqref{eq:Bog_b_spin2C0}\\
&          & $\sqrt{\epsilon_{\bm k}(\epsilon_{\bm k}+4c_1)}$ & $(u,0,0,0,u)$& $(v,0,0,0,v)$ & Eq.~\eqref{eq:Bog_b_spin2C2}\\
&          & $\epsilon_{\bm k}+2c_2n/5$ & $(-1,0,\sqrt{2},0,1)$& $(0,0,0,0,0)$ & Eq.~\eqref{eq:Bog_b_spin2C-2}\\
&          & $\sqrt{\epsilon_{\bm k}(\epsilon_{\bm k}+4c_1)}$ & $(0,\sqrt{3}u,0,u,0)$& $(0,v,0,0,0)$ & Eq.~\eqref{eq:Bog_b_spin2C1}\\
&          & $\sqrt{\epsilon_{\bm k}(\epsilon_{\bm k}+4c_1)}$ & $(0,-u,0,\sqrt{3}u,0)$& $(0,0,0,v,0)$ & Eq.~\eqref{eq:Bog_b_spin2C1}\\ \hline
\end{tabular}}
\caption{Bogoliubov excitation spectra for spin-2 BECs, where F2, UN, and C denote ferromagnetic, uniaxial nematic, and cyclic phases, respectively, where $f_z=p/(c_1n-c_2n/20)$ in the C$_4$ phase.
The C$_4$ state reduces to the biaxial nematic state at $p=0$.
The cyclic order parameter becomes a stationary solution of the GPE only at $p=q=0$.
The second and third to last columns show the spin dependence of the eigenmode as in Table~\ref{table:Bog_spin1}.
}
\label{table:Bog_spin2}
\end{center}
\end{table}
\end{landscape}

\subsubsection{Ferromagnetic state}
\label{sec:BogFM(f=2)}

The order parameter for the ferromagnetic state is given by $\bm\zeta=(1,0,0,0,0)^{\rm T}$, which leads to
\begin{align}
{\bm H}^{(0)}_{\bm k}+{\bm H}^{(1)} 
=& {\rm Diag}[
\epsilon_{\rm k}+(c_0+4c_1)n,\epsilon_{\bm k}+p-3q,\epsilon_{\bm k}+2p-4q-4c_1n,\nonumber\\
&\hspace{15mm}\epsilon_{\bm k}+3p-3q-6c_1n,\epsilon_{\bm k}+4p-\left(8c_1-\frac{2c_2}{5}\right)n],\\
{\bm H}^{(2)}=&{\rm Diag}[(c_0+4c_1)n,0,0,0,0],
\end{align}
where $\mu=-2p+4q+(c_0+4c_1)n$, and ${\rm Diag}$ denotes the diagonal matrix.
Since both ${\bm H}^{(0)}_{\bm k}+{\bm H}^{(1)}$ and ${\bm H}^{(2)}$ are diagonal matrices, 
all spin components are decoupled from each other.
Moreover, only the $m=2$ mode has a linear dispersion:
\begin{align}
E_{{\bm k},2}&=\sqrt{\epsilon_{\bm k}(\epsilon_{\bm k}+2g_4n)}, \label{eq:Bog_ene_spin2F2}\\
\hat{b}_{{\bm k},2}&= {\rm sgn}(g_4)\sqrt{\frac{\epsilon_{\bm k}+g_4n+E_{\bm k,2}}{2E_{\bm k,2}}}\hat{a}_{\bm k,2}+\sqrt{\frac{\epsilon_{\bm k}+g_4n-E_{\bm k,2}}{2E_{\bm k,2}}}\hat{a}^\dagger_{-\bm k,2},
\label{eq:Bog_b_spin2F2}
\end{align}
where we have used $c_0+4c_1=g_4=(4\pi\hbar^2/M)a_4$ [Eq.~\eqref{c(f=2)}].
For the system to be stable, the Bogoliubov excitation energy~\eqref{eq:Bog_ene_spin2F2} must be real, implying that the {\it s}-wave scattering length for the total spin-4 channel must be positive: $a_4>0$.
As in the case of the spin-1 ferromagnetic phase, $E_{\bm k, 2}$ is independent of an applied magnetic field and remains gapless in its presence,
since this is the Nambu-Goldstone mode associated with the global U(1) gauge invariance.

On the other hand, all the other modes give single-particle spectra:
\begin{align}
E_{{\bm k},1}&=\epsilon_{\bm k}+p-3q,\\
E_{{\bm k},0}&=\epsilon_{\bm k}+2p-4q-4c_1n,\\
E_{{\bm k},-1}&=\epsilon_{\bm k}+3p-3q-6c_1n,\\
E_{{\bm k},-2}&=\epsilon_{\bm k}+4p-\left(8c_1-\frac{2c_2}{5}\right)n.
\end{align}
For the system to be stable, the single-particle excitation energies must be positive.

The quantum depletion and the zero-point fluctuation energy for the ferromagnetic phase are calculated as follows:
\begin{align}
\frac{ N^{\rm qntm}}{N} &= \frac{1}{N}\sum_{\bm k\neq \bm 0} \frac{\epsilon_{\bm k}+g_4n-\sqrt{\epsilon_{\bm k}(\epsilon_{\bm k}+2g_4n)}}{2\sqrt{\epsilon_{\bm k}(\epsilon_{\bm k}+2g_4n)}} = \frac{8}{3}\sqrt{\frac{na_4^3}{\pi}},
\\
E^{\rm qntm} &=\frac{1}{2}\sum_{\bm k}\left[
\sqrt{\epsilon_{\bm k}(\epsilon_{\bm k}+2g_4n)}
-\epsilon_{\bm k}-g_4n+\frac{(g_4n)^2}{2\epsilon_{\bm k}}
\right]
= \frac{64}{15}g_4n \sqrt{\frac{na_4^3}{\pi}}N
\end{align}
where we have used $D^{\rm corr}=(c_0+4c_1)^2n^2=(g_4n)^2$.

\subsubsection{Uniaxial-nematic state}
\label{sec:BogAFM(f=2)}

The order parameter for the UN state is given by $\bm \zeta=(0,0,1,0,0)^{\rm T}$ with the chemical potential $\mu=c_2n/5$.
Substituting $\bm\zeta$ and $\mu$ in Eqs.~\eqref{eq:Bog_H1_spin2} and \eqref{eq:Bog_H2_spin2}, we obtain
\begin{align}
{\bm H}^{(0)}_{\bm k}+{\bm H}^{(1)} 
=& {\rm Diag}[
\epsilon_{\rm k}-2p+4q-\frac{c_2n}{5}, \epsilon_{\bm k}-p+q+\left(3c_1-\frac{c_2}{5}\right)n,\epsilon_{\bm k}+\left(c_0+\frac{c_2}{5}\right)n,\nonumber\\
&\hspace{15mm}\epsilon_{\bm k}+p+q+\left(3c_1-\frac{c_2}{5}\right)n,\epsilon_{\bm k}+2p+4q-\frac{c_2n}{5}],\\
{\bm H}^{(2)} =&
\begin{pmatrix}
0 & 0 & 0 & 0 & c_2n/5\\
0 & 0 & 0 & (3c_1-c_2/5)n & 0\\
0 & 0 & (c_0+c_2/5)n & 0 & 0\\
0 & (3c_1-c_2/5)n & 0 & 0 & 0\\
c_2n/5 & 0 & 0 & 0 & 0\\
\end{pmatrix}.
\end{align}
The eigenvalues of the matrix ${\bm \sigma}_z{\bm M}^{\rm B}_{\bm k}$ are calculated to be
\begin{align}
E_{{\bm k},0}    &= \sqrt{\epsilon_{\bm k}[\epsilon_{\bm k}+2(c_0+c_2/5)n]},\label{eq:Bog_ene_spin2UN0}\\
E_{{\bm k},\pm1} &= \sqrt{(\epsilon_{\bm k}+q)[\epsilon_{\bm k}+q+2(3c_1-c_2/5)n]}\mp p,\label{eq:Bog_ene_spin2UN1}\\
E_{{\bm k},\pm2} &= \sqrt{(\epsilon_{\bm k}+4q)(\epsilon_{\bm k}+4q-2c_2n/5)} \mp 2p,\label{eq:Bog_ene_spin2UN2}
\end{align}
with the corresponding eigenmodes given respectively by
\begin{align}
\hat{b}_{{\bm k},0} &=  {\rm sgn}(c_0+c_2/5)\sqrt{\frac{\epsilon_{\bm k}+(c_0+c_2/5)n+E_{{\bm k},0}}{2E_{{\bm k},0}}}\hat{a}_{{\bm k},0} + \sqrt{\frac{\epsilon_{\bm k}+(c_0+c_2/5)n-E_{{\bm k},0}}{2E_{{\bm k},0}}}\hat{a}^\dagger_{-{\bm k},0},
\label{eq:Bog_b_spin2UN0}\\
\hat{b}_{{\bm k},\pm1} &= {\rm sgn}(3c_1-c_2/5)\sqrt{\frac{\epsilon_{\bm k}+q+(3c_1-c_2/5)n+(E_{{\bm k},\pm1}\pm p)}{2(E_{{\bm k},\pm1}\pm p)}}\hat{a}_{{\bm k},\pm1} \nonumber\\
&\hspace{20mm}+ \sqrt{\frac{\epsilon_{\bm k}+q+(3c_1-c_2/5)n-(E_{{\bm k},\pm2}\pm p)}{2(E_{{\bm k},\pm2}\pm p)}}\hat{a}^\dagger_{-{\bm k},\mp1},
\label{eq:Bog_b_spin2UN1}\\
\hat{b}_{{\bm k},\pm2} &=  {\rm sgn}(-c_2)\sqrt{\frac{\epsilon_{\bm k}+4q-c_2n/5+(E_{{\bm k},\pm2}\pm 2p)}{2(E_{{\bm k},\pm1}\pm 2p)}}\hat{a}_{{\bm k},\pm2} \nonumber\\
&\hspace{20mm}+ \sqrt{\frac{\epsilon_{\bm k}+4q-c_2n/5-(E_{{\bm k},\pm2}\pm 2p)}{2(E_{{\bm k},\pm2}\pm 2p)}}\hat{a}^\dagger_{-{\bm k},\mp2}.
\label{eq:Bog_b_spin2UN2}
\end{align}
The spectrum~\eqref{eq:Bog_ene_spin2UN0} is massless (linear dispersion) and gapless regardless of an external field, describing the phonon excitation mode,
whereas other spectra become massive (quadratic in $k$) and gapped in an presence of the external field.
When $p=q=0$, all spectra becomes massless and gapless: $E_{\bm k,\pm1}$ is the Nambu-Goldstone mode associated with the SO(3) spin rotational symmetry, whereas $E_{\bm k,\pm2}$ is the quasi-Nambu-Goldstone mode arising from the hidden SO(5) symmetry (see Sec.~\ref{sec:QNDmode}).

The quantum depletion at $p=q=0$ is calculated as
\begin{align}
 \frac{N^{\rm qntm}}{N} =& \frac{1}{N}\sum_{\bm k\neq \bm 0} \frac{\epsilon_{\bm k}+(c_0+c_2/5)n-\sqrt{\epsilon_{\bm k}[\epsilon_{\bm k}+2(c_0+c_2/5)n]}}{2\sqrt{\epsilon_{\bm k}[\epsilon_{\bm k}+2(c_0+c_2/5)n]}}\nonumber \\
&+ \frac{1}{N}\sum_{\bm k\neq \bm 0} \frac{\epsilon_{\bm k}+(3c_1-c_2/5)n-\sqrt{\epsilon_{\bm k}[\epsilon_{\bm k}+2(3c_1-c_2/5)n]}}{\sqrt{\epsilon_{\bm k}[\epsilon_{\bm k}+2(3c_1-c_2/5)n]}}\nonumber \\
&+ \frac{1}{N}\sum_{\bm k\neq \bm 0} \frac{\epsilon_{\bm k}-c_2n/5-\sqrt{\epsilon_{\bm k}(\epsilon_{\bm k}-2c_2n/5)}}{\sqrt{\epsilon_{\bm k}(\epsilon_{\bm k}-2c_2n/5)}} \nonumber\\
=&  \frac{8}{3}\sqrt{\frac{n}{\pi}}\left[ \left(\tilde{a}_0+\frac{\tilde{a}_2}{5}\right)^{3/2} + 2 \left(3\tilde{a}_1-\frac{\tilde{a}_2}{5}\right)^{3/2} + 2\left(-\frac{\tilde{a}_2}{5}\right)^{3/2}\right],
\end{align}
where $\tilde{a}_{0,1,2}$ are defined as $\tilde{a}_{0,1,2}\equiv M/(4\pi\hbar^2)c_{0,1,2}$ and written in terms of $a_{\mathcal{F}=0,2,4}$ as
\begin{align}
 \tilde{a}_0=\frac{4a_2+3a_4}{7},\ \ 
 \tilde{a}_1=\frac{a_4-a_2}{7},\ \ 
 \tilde{a}_2=\frac{7a_0-10a_2+3a_4}{7}.
\end{align}
For the system to be stable, the energy spectra~\eqref{eq:Bog_ene_spin2UN0}--\eqref{eq:Bog_ene_spin2UN2} should be semi-positive real, implying that $\tilde{a}_0+\tilde{a}_2/5>0$, $3\tilde{a}_1-\tilde{a}_2/5>0$, and $-\tilde{a}_2/5>0$ for $p=q=0$.
In a manner similar to the case of the spin-1 polar phase,
$E^{\rm qntm}$  at $p=q=0$ is calculated as
\begin{align}
 E^{\rm qntm} =\frac{256\pi\hbar^2}{15M}N\sqrt{\frac{n^3}{\pi}} \left[ \left(\tilde{a}_0+\frac{\tilde{a}_2}{5}\right)^{5/2} + 2 \left(3\tilde{a}_1-\frac{\tilde{a}_2}{5}\right)^{5/2} + 2\left(-\frac{\tilde{a}_2}{5}\right)^{5/2}\right]. \label{eq:Bog_Eqntm_UN}
\end{align}

\subsubsection{C$_4$ state}

The order parameter for the C$_4$ state is given by
\begin{align}
 \bm\zeta = \left(\sqrt{\frac{1+f_z/2}{2}},0,0,0,\sqrt{\frac{1-f_z/2}{2}}\right)^{\rm T},
\label{eq:Bog_OP_spin2C4}
\end{align}
where $f_z=p/(c_1n-c_2n/20)$.
Substituting Eq.~\eqref{eq:Bog_OP_spin2C4} and $\mu=4q+(c_0+c_2)n/5$ in Eqs.~\eqref{eq:Bog_H1_spin2} and \eqref{eq:Bog_H2_spin2}, we obtain
\begin{align}
{\bm H}^{(0)}_{\bm k} + {\bm H}^{(1)} 
=& {\rm Diag}\bigg[\epsilon_{\bm k}+ \frac{2+f_z}{4}(c_0+4c_1)n,\epsilon_{\bm k}-3q+\frac{2+f_z}{2}c_1n-\frac{4-f_z}{20}c_2n,\epsilon_{\bm k}-4q-\frac{c_2n}{5},\nonumber\\
&\hspace{15mm}\epsilon_{\bm k}-3q+\frac{2-f_z}{2}c_1n-\frac{4+f_z}{20}c_2n,\epsilon_{\bm k}+\frac{2-f_z}{4}(c_0+4c_1)n\bigg]\nonumber\\
&+\frac{\sqrt{4-f_z^2}}{20}(5c_0-4c_1+2c_2)n
\begin{pmatrix}
 0 & 0 & 0 & 0 & 1 \\
 0 & 0 & 0 & 0 & 0 \\
 0 & 0 & 0 & 0 & 0 \\
 0 & 0 & 0 & 0 & 0 \\
 1 & 0 & 0 & 0 & 0 \\
\end{pmatrix},\\
{\bm H}^{(2)} 
=& {\rm Diag}\left[\frac{2+f_z}{4}(c_0+4c_1)n,0,0,0,\frac{2-f_z}{4}(c_0+4c_1)\right]\nonumber\\
&+\frac{\sqrt{4-f_z^2}}{20}\begin{pmatrix}
 0 & 0 & 0 & 0 & 5c_0-20c_1+2c_2 \\
 0 & 0 & 0 & 10c_1-2c_2 & 0 \\
 0 & 0 & 2c_2 & 0 & 0 \\
 0 & 10c_1-2c_2 & 0 & 0 & 0 \\
 5c_0-20c_1+2c_2 & 0 & 0 & 0 & 0 \\
\end{pmatrix},
\end{align}
from which we can decouple the eigenvalue equation~\eqref{eq:Bog_eigenvalue_equation} for 
the modes of $(u_{\bm k,2},u_{\bm k,-2},v_{\bm k,2},v_{\bm k,-2})^{\rm T}$, $(u_{\bm k,\pm1},v_{\bm k,\mp1})^{\rm T}$, and  $(u_{\bm k,0},v_{\bm k,0})^{\rm T}$.

The energy spectra for the $m=\pm2$ components are given by
\begin{align}
E_{\bm k,\pm2}=&\sqrt{\epsilon_{\bm k}\left[\epsilon_{\bm k}+(c_0+4c_1)n\pm\frac{n}{5}\sqrt{(5c_0-20c_1+2c_2)^2+(20c_1-c_2)(3c_0+c_2)f_z^2}\right]}.
\label{eq:Bog_ene_spin2C42}
\end{align}
The spectra~\eqref{eq:Bog_ene_spin2C42} are massless (linear dispersion) and gapless regardless of the presence of the external field.
This is because they are the Nambu-Goldstone modes associated with the U(1) gauge symmetry and the relative gauge symmetry, which is the rotational symmetry about the direction of the applied field.
When $p=0$ or $f_z=0$, Eq.~\eqref{eq:Bog_ene_spin2C42} reduces to the decoupled density and spin waves:
\begin{align}
 E_{\bm k,2}=&\sqrt{\epsilon_{\bm k}[\epsilon_{\bm k}+2(c_0+c_2/5)n]},\\
 E_{\bm k,-2}=&\sqrt{\epsilon_{\bm k}[\epsilon_{\bm k}+2(4c_1-c_2/5)n]},
\end{align}
whose eigenmodes are given by
\begin{align}
 \hat{b}_{\bm k,2}=& {\rm sgn}(c_0+c_2/5)\sqrt{\frac{\epsilon_{\bm k}+(c_0+c_2/5)n+E_{\bm k,2}}{2E_{\bm k,2}}}\frac{\hat{a}_{\bm k,2}+\hat{a}_{\bm k,-2}}{\sqrt{2}}\nonumber\\
 &\hspace{10mm}+\sqrt{\frac{\epsilon_{\bm k}+(c_0+c_2/5)n-E_{\bm k,2}}{2E_{\bm k,2}}}\frac{\hat{a}^\dagger_{-\bm k,2}+\hat{a}^\dagger_{-\bm k,-2}}{\sqrt{2}},
\label{eq:Bog_b_spin2C42}\\
 \hat{b}_{\bm k,-2}=&  {\rm sgn}(4c_1-c_2/5)\sqrt{\frac{\epsilon_{\bm k}+(4c_1-c_2/5)n+E_{\bm k,-2}}{2E_{\bm k,-2}}}\frac{\hat{a}_{\bm k,2}-\hat{a}_{\bm k,-2}}{\sqrt{2}}\nonumber\\
 &\hspace{10mm}+\sqrt{\frac{\epsilon_{\bm k}+(4c_1-c_2/5)n-E_{\bm k,-2}}{2E_{\bm k,-2}}}\frac{\hat{a}^\dagger_{-\bm k,2}-\hat{a}^\dagger_{-\bm k,-2}}{\sqrt{2}}.
\label{eq:Bog_b_spin2C4-2}
\end{align}

The energy spectra and the corresponding eigenmodes for the $m=\pm1$ modes are given by
\begin{align}
 E_{\bm k,\pm1}=&\sqrt{(\epsilon_{\bm k}-3q)^2+2\left(c_1-\frac{c_2}{5}\right)n(\epsilon_{\bm k}-3q)+\frac{1}{4}\left[f_z\left(c_1-\frac{c_2}{5}\right)n\right]^2} \pm \frac{f_z}{2}\left(c_1+\frac{c_2}{10}\right)n, \label{eq:Bog_ene_spin2C41}\\
 \hat{b}_{\bm k,\pm1} =&  {\rm sgn}(c_1-c_2/5)\sqrt{\frac{\epsilon_{\bm k}-3q+(c_1-c_2/5)n+[E_{\bm k,\pm1}\mp \frac{f_z}{2}(c_1+c_2/10)n]}{2[E_{\bm k,\pm1}\mp \frac{f_z}{2}(c_1+c_2/10)n]}}\hat{a}_{\bm k,\pm1} \nonumber\\
&+\sqrt{\frac{\epsilon_{\bm k}-3q+(c_1-c_2/5)n-[E_{\bm k,\pm1}\mp \frac{f_z}{2}(c_1+c_2/10)n]}{2[E_{\bm k,\pm1}\mp \frac{f_z}{2}(c_1+c_2/10)n]}}\hat{a}^\dagger_{-\bm k,\mp1},
\label{eq:Bog_b_spin2C41}
\end{align}
whereas those for the $m=0$ mode are given by
\begin{align}
 E_{\bm k,0}&=\sqrt{(\epsilon_{\bm k}-4q)^2-\frac{2c_2n}{5}(\epsilon_{\bm k}-4q)+\frac{(f_zc_2n)^2}{100}}, \label{eq:Bog_ene_spin2C40}\\
 \hat{b}_{\bm k,0} &=  {\rm sgn}(-c_2)\sqrt{\frac{\epsilon_{\bm k}-4q-c_2n/5+E_{\bm k,0}}{2E_{\bm k,0}}}\hat{a}_{\bm k0} + \sqrt{\frac{\epsilon_{\bm k}-4q-c_2n/5-E_{\bm k,0}}{2E_{\bm k,0}}}\hat{a}^\dagger_{-\bm k,0}.
\label{eq:Bog_b_spin2C40}
\end{align}
These spectra become massive (quadratic in $k$) and gapful in the presence of an external field.

At $p=q=0$, all five dispersion relations in Eqs.~\eqref{eq:Bog_ene_spin2C42}, \eqref{eq:Bog_ene_spin2C41}, and \eqref{eq:Bog_ene_spin2C40} become gapless and linear, 
reflecting the fact that in the absence of the magnetic field, the ground state is degenerate with respect to five continuous variables (see Sec.~\ref{sec:QNDmode}).
Note, however, that only the first four are the Nambu-Goldstone modes, and the last one is the quasi-Nambu-Goldstone mode arising from the hidden SO(5) symmetry (see Sec.~\ref{sec:QNDmode}).

For the BN state, i.e., C$_4$ state at $p=q=0$, the quantum depletion and the zero-point fluctuation energy are calculated as
\begin{align}
 \frac{N^{\rm qntm}}{N} 
=&  \frac{8}{3}\sqrt{\frac{n}{\pi}}\left[ \left(\tilde{a}_0+\frac{\tilde{a}_2}{5}\right)^{3/2} + 2 \left(\tilde{a}_1-\frac{\tilde{a}_2}{5}\right)^{3/2} + \left(-\frac{\tilde{a}_2}{5}\right)^{3/2}+\left(4\tilde{a}_1-\frac{\tilde{a}_2}{5}\right)^{3/2}\right], \\
 E^{\rm qntm} =&\frac{256\pi\hbar^2}{15M}N\sqrt{\frac{n^3}{\pi}} 
\left[ \left(\tilde{a}_0+\frac{\tilde{a}_2}{5}\right)^{5/2} + 2 \left(\tilde{a}_1-\frac{\tilde{a}_2}{5}\right)^{5/2} + \left(-\frac{\tilde{a}_2}{5}\right)^{5/2}+\left(4\tilde{a}_1-\frac{\tilde{a}_2}{5}\right)^{5/2}\right]. \label{eq:Bog_Eqntm_BN}
\end{align}
Although the mean-field energy for the BN and UN states are degenerate (see Sec.~\ref{sec:MF_spin2_GS}),
the degeneracy is lifted by zero-point fluctuations. From Eqs.~\eqref{eq:Bog_Eqntm_UN} and \eqref{eq:Bog_Eqntm_BN}, the energy difference between the BN and UN states is given by
\begin{align}
 \Delta E=E^{\rm UN}-E^{\rm BN} \propto
 2 \left(3\tilde{a}_1-\frac{\tilde{a}_2}{5}\right)^{5/2} + \left(-\frac{\tilde{a}_2}{5}\right)^{5/2}
-2 \left(\tilde{a}_1-\frac{\tilde{a}_2}{5}\right)^{5/2} - \left(4\tilde{a}_1-\frac{\tilde{a}_2}{5}\right)^{5/2}.
\label{eq:Bog_DeltaE}
\end{align}
In the region of the nematic phase, i.e., when $\tilde{a}_2<0$ and $\tilde{a}_2<20\tilde{a}_1$ (see Fig.~\ref{fig:spin-2PD_B=0}),
$\Delta E$ is a monotonically decreasing function of $\tilde{a}_1$ satisfying $\Delta E < 0\, (\Delta E >0)$ for $\tilde{a}_1 >0\, (\tilde{a}_1<0)$.
Hence, the BN state is the ground state for $\tilde{a}_1<0$ and $\tilde{a}_2<20\tilde{a}_1$, while the UN state is the ground state for $\tilde{a}_1>0$ and $\tilde{a}_2<0$.

\subsubsection{Cyclic state}
\label{sec:BogCyclic(f=2)}

In the absence of an external field (i.e., $p=q=0$) the C3$_{\pm}$ and D2' states reduce to the cyclic state%
\footnote{The spinor in Eq.~\eqref{eq:Bog_OP_spin2C} is related to $\tilde{\bm\zeta}\equiv(1,0,i\sqrt{2},0,1)^{\rm T}/2$ by $\bm\zeta = e^{-i\pi/2}e^{i{\rm f}_z\pi/4}\tilde{\bm \zeta}$.}:
\begin{align}
 \bm\zeta=\left(\frac{1}{2},0,\frac{1}{\sqrt{2}},0,-\frac{1}{2}\right)^{\rm T},
\label{eq:Bog_OP_spin2C}
\end{align}
with the chemical potential given by $\mu=c_0n$.
By solving the eigenvalue equation~\eqref{eq:Bog_eigenvalue_equation}, we obtain the energy spectra and quasi-particles as follows.
\begin{enumerate}
\item Phonon mode describing the density fluctuation:
\begin{align}
E_{\bm k,0}=&\sqrt{\epsilon_{\bm k}(\epsilon_{\bm k}+2c_0n)},\\
 \hat{b}_{\bm k,0} =&  {\rm sgn}(c_0)\sqrt{\frac{\epsilon_{\bm k}+c_0n+E_{\bm k,0}}{2E_{\bm k,0}}}\frac{\hat{a}_{\bm k,2}+\sqrt{2}\hat{a}_{\bm k,0}-\hat{a}_{\bm k,-2}}{2}\nonumber\\
&\hspace{10mm}+\sqrt{\frac{\epsilon_{\bm k}+c_0n-E_{\bm k,0}}{2E_{\bm k,0}}}\frac{\hat{a}^\dagger_{\bm k,2}+\sqrt{2}\hat{a}^\dagger_{\bm k,0}-\hat{a}^\dagger_{\bm k,-2}}{2}.
\label{eq:Bog_b_spin2C0}
\end{align}

\item Spin mode describing the rotation about the $z$ axis:
\begin{align}
E_{\bm k,2}=&\sqrt{\epsilon_{\bm k}(\epsilon_{\bm k}+4c_1n)},
\label{eq:Bog_cyclic_2}\\
 \hat{b}_{\bm k,2} =&  {\rm sgn}(c_1)\sqrt{\frac{\epsilon_{\bm k}+2c_1n+E_{\bm k,2}}{2E_{\bm k,2}}}\frac{\hat{a}_{\bm k,2}+\hat{a}_{\bm k,-2}}{\sqrt{2}}
+\sqrt{\frac{\epsilon_{\bm k}+2c_1n-E_{\bm k,2}}{2E_{\bm k,2}}}\frac{\hat{a}^\dagger_{\bm k,2}+\hat{a}^\dagger_{\bm k,-2}}{\sqrt{2}}.
\label{eq:Bog_b_spin2C2}
\end{align}

\item Fluctuations in the singlet-pair amplitude due to the coupling between the $m=0$ and $m=\pm2$ states via the $c_2$ term:
\begin{align}
E_{\bm k,-2}=&\epsilon_{\bm k} + \frac{2c_2}{5},\\
\hat{b}_{\bm k,-2}=&\frac{-\hat{a}_{\bm k,2}+\sqrt{2}\hat{a}_{\bm k,0}+\hat{a}_{\bm k,-2}}{2}.
\label{eq:Bog_b_spin2C-2}
\end{align}

\item Spin mode describing the rotation about an axis perpendicular to the $z$ axis:
\begin{align}
E_{\bm k,\pm1}=&\sqrt{\epsilon_{\bm k}(\epsilon_{\bm k}+4c_1n)}, \label{eq:Bog_cyclic_pm1}\\
 \hat{b}_{\bm k,\pm1} =&  {\rm sgn}(c_1)\sqrt{\frac{\epsilon_{\bm k}+2c_1n+E_{\bm k,\pm1}}{2E_{\bm k,\pm1}}}\frac{\sqrt{3}\hat{a}_{\bm k,\pm1}\pm \hat{a}_{\bm k,\mp1}}{2}
\pm\sqrt{\frac{\epsilon_{\bm k,\pm1}+2c_1n-E_{\bm k,\pm1}}{2E_{\bm k,\pm1}}}\hat{a}^\dagger_{\bm k,\pm1}.
\label{eq:Bog_b_spin2C1}
\end{align}
\end{enumerate}

Since the first two modes are the Nambu-Goldstone modes associated with the U(1) gauge symmetry and the SO(2) spin rotational symmetry about the $z$ axis,
they remain gapless in the D2' state, which appears continuously from the order parameter~\eqref{eq:Bog_OP_spin2C} in the presence of an external magnetic field (applied in the $z$ direction).
On the other hand, the energy spectrum~\eqref{eq:Bog_cyclic_pm1} describes a gapless Nambu-Goldstone mode associated with the SO(3) spin rotational symmetry in the absence of an external field; however, it becomes gapful in the D2' state.

The quantum depletion and the zero-point fluctuation energy of the cyclic phase (i.e., at $p=q=0$) are given by
\begin{align}
 N^{\rm qntm} &= \frac{8}{3}\sqrt{\frac{n}{\pi}}\left(\tilde{a}_0^{3/2}+3\tilde{a}_1^{3/2}\right),\\
 E^{\rm qntm} &= \frac{256\pi\hbar^2}{15M}N\sqrt{\frac{n^3}{\pi}}\left[\tilde{a}_0^{5/2}+3(2\tilde{a}_1)^{5/2}\right].
\end{align}

\subsection{Quasi Nambu-Goldstone mode}
\label{sec:QNDmode}

As discussed in Sec.~\ref{sec:MF_spin2_GS}, the mean-field ground-state energy is degenerate with respect to the uniaxial nematic (UN) and biaxial nematic (BN) phases at zero magnetic field; in this case, there is a gapless excitation that does not belong to the Nambu-Goldstone mode. In fact, the total number of spontaneously broken symmetry generators is three for the UN phase and four for the BN phase~\cite{Uchino2010a}, whereas five gapless Bogoliubov modes exist in the nematic phase~\cite{Ueda2002}.
The physical origin of this phenomenon may be understood as follows. 
The nematic phase is characterized with a maximum spin-singlet pair amplitude ($|a_{00}|=1/\sqrt{5}$) and zero magnetization ($|\bm f|=0$).
Since the magnetization is kept zero as long as the spin-singlet pair amplitude takes on its maximum [see the discussion below Eq.~\eqref{eq:def_a=A/n}], 
the mean-field energy is degenerate for the states that are obtained from the UN or BN order parameter by SO(5) transformations which  make the spin-singlet pair amplitude invariant~\cite{Uchino2010a}.
Consequently, the order parameter manifold is enlarged with a dimensionality of five, which is equal to the number of gapless modes predicted by the Bogoliubov theory. Therefore, the number of extra gapless modes is equal to the dimensionality of the enlarged manifold minus the dimensionality of the original order-parameter manifold.
The difference is two for the UN phase and one for the BN phases. These extra gapless modes behave as soft modes and correspond to quasi-Nambu-Goldstone modes in elementary particle physics (see, e.g., p.~489 of Ref.~\cite{Weinberg1996});
however, their experimental realization has been elusive. The nematic phase of a spin-2 BEC may provide the first experimental manifestation of a quasi-Nambu-Goldstone mode~\cite{Uchino2010b}.

Concretely speaking, one of the modes in Eq.~\eqref{eq:Bog_ene_spin2UN2} is the quasi Nambu-Goldstone mode for the UN phase, and the modes in Eq.~\eqref{eq:Bog_ene_spin2C40} are the quasi Nambu-Goldstone modes in the BN phase.
The continuous symmetries concerning these modes reflect not the symmetry of the Hamiltonian but the hidden SO(5) symmetry
due to the degeneracy in the mean-field solution $\bm \zeta=(\cos\xi/\sqrt{2},0,\sin\xi,0,\cos\xi/\sqrt{2})^{\rm T}$ for $^\forall\xi\in\mathbb{R}$~\cite{Uchino2010a, Uchino2010b}.
This degeneracy is lifted when we take into account the zero-point energy [see Eq.~\eqref{eq:Bog_DeltaE}].
Since the beyond-mean-field correction lifts the degeneracy between UN and BN states, the excitation spectra~\eqref{eq:Bog_ene_spin2UN2} and \eqref{eq:Bog_ene_spin2C40} should become gapful even at zero magnetic field in such higher-order approximations with which the spin-singlet pair amplitude is quantum-mechanically depleted.

\subsection{Dynamical instability}
\label{sec:dynamicalinstability}

Since the Bogoliubov equation~\eqref{eq:Bog_eigenvalue_equation} is an eigenvalue equation of a non-Hermitian matrix,
it may have a complex eigenvalue.
In such cases, we cannot diagonalize the effective Hamiltonian~\eqref{eq:Bog_Heff} as in the form of Eq.~\eqref{eq:Bog_Heff_diagonalized},
and the number of the quasi-particles increases even when we start from a vacuum of the quasi-particles $|{\rm vac}_{\rm B}\rangle$.
Such an instability is referred to as the dynamical instability.
The dynamical instability may arise for a state which is stationary in the mean-field approximation but not the real ground state of the system, such as 
a multiply quantized vortex state~\cite{Pu1999b, Skryabin2000,Mottonen2003, Kawaguchi2004},
a state with a nonzero current velocity in an optical lattice~\cite{Wu2001},
and spinor BECs in non-ground-state spin configurations~\cite{Robins2001,Saito2005,Sadler2006}.

On the other hand, the Bogoliubov equation may also have a negative real eigenvalue, which is called the Landau instability.
In the presence of the Landau instability,
the system can lower its energy by exciting a negative-eigenvalue mode,
which is possible only when there is an energy dissipation.
In experiments of cold atomic gases, the energy dissipation is small because the system is isolated in vacuum.
Therefore, the dynamical rather than Landau instability is crucial for describing the instability in cold atomic systems.

Mathematically speaking, the Bogoliubov Hamiltonian~\eqref{eq:Bog_Heff} cannot be diagonalized 
because the orthonormal relation~\eqref{eq:norm_uv} is violated for the eigenmode with a complex eigenvalue.
To see this, we rewrite Eq.~\eqref{eq:Bog_eigenvalue_equation} as
\begin{align}
{\bm M}^{\rm B}_{\bm k}\begin{pmatrix} {\bm u}_{\bm k,m} \\ {\bm v}_{\bm k,m}\end{pmatrix}
=E_{\bm k,m}{\bm \sigma}_z\begin{pmatrix} {\bm u}_{\bm k,m} \\ {\bm v}_{\bm k,m}\end{pmatrix}.
\label{eq:Bog_eigeneq1}
\end{align}
Taking the Hermitian conjugate of Eq.~\eqref{eq:Bog_eigeneq1}, we obtain
\begin{align}
\begin{pmatrix} {\bm u}^\dagger_{\bm k,m'}, & {\bm v}^\dagger_{\bm k,m'}\end{pmatrix}
{\bm M}^{\rm B}_{\bm k}
=E_{\bm k,m'}^*\begin{pmatrix} {\bm u}^\dagger_{\bm k,m'}, & {\bm v}^\dagger_{\bm k,m'}\end{pmatrix}{\bm \sigma}_z,
\label{eq:Bog_eigeneq2}
\end{align}
where we have used the fact that ${\bm M}^{\rm B}_{\bm k}$ is Hermitian.
Multiplying $\begin{pmatrix} {\bm u}^\dagger_{\bm k,m'}, & {\bm v}^\dagger_{\bm k,m'}\end{pmatrix}$ to Eq.~\eqref{eq:Bog_eigeneq1} from the left
and $\begin{pmatrix} {\bm u}_{\bm k,m} \\ {\bm v}_{\bm k,m}\end{pmatrix}$ to Eq.~\eqref{eq:Bog_eigeneq2} from the right, and subtracting them,
we obtain
\begin{align}
 (E^*_{\bm k,m'}-E_{\bm k,m})({\bm u}_{\bm k,m'}^\dagger{\bm u}_{\bm k,m}-{\bm v}_{\bm k,m'}^\dagger{\bm v}_{\bm k,m})=0.
\label{eq:orthnorm_uv_complex}
\end{align}
Equation~\eqref{eq:orthnorm_uv_complex} with $m=m'$ is consistent with the orthonormal condition~\eqref{eq:norm_uv}, if the eigenvalue $E_{\bm k,m}$ is a real number.
However, if $E_{\bm k,m}$ has a nonzero imaginary part, Eq.~\eqref{eq:orthnorm_uv_complex} leads to
\begin{align}
 {\bm u}_{\bm k,m}^\dagger{\bm u}_{\bm k,m}-{\bm v}_{\bm k,m}^\dagger{\bm v}_{\bm k,m}=0,
\label{eq:uv_cmp1}
\end{align}  
and Eq.~\eqref{eq:norm_uv} is no longer satisfied.
Instead, for a set of complex conjugate eigenvalues, $E_{\bm k,m'}=E_{\bm k,m}^*$,
we can impose the following normalization condition for the corresponding eigenmodes:
\begin{align}
{\bm u}_{\bm k,m'}^\dagger{\bm u}_{\bm k,m}-{\bm v}_{\bm k,m'}^\dagger{\bm v}_{\bm k,m}=1.
\label{eq:orthnorm_uv_complex2}
\end{align}
In other words, to define the Bogoliubov quasi-particles that satisfy the canonical commutation relations,
the complex eigenvalues appear in complex conjugate pairs.
In such a case, the Bogoliubov quasi-particles can be defined as follows.
Suppose that $\bm \sigma_z {\bm M}^{\rm B}_{\bm k}$ has a pair of complex conjugate eigenvalues $E_{a}$ and $E_{b}=E_{a}^*$ with the respective eigenmodes
$\begin{pmatrix} {\bm u}_{a} \\ {\bm v}_{a}\end{pmatrix}$ and $\begin{pmatrix} {\bm u}_{b} \\ {\bm v}_{b}\end{pmatrix}$.
From Eqs.~\eqref{eq:uv_cmp1} and \eqref{eq:orthnorm_uv_complex2}, these modes satisfy the following equations:
\begin{align}
 {\bm u}_{a}^\dagger {\bm u}_{a} -  {\bm v}_{a}^\dagger {\bm v}_{a} = 
 {\bm u}_{b}^\dagger {\bm u}_{b} -  {\bm v}_{b}^\dagger {\bm v}_{b} = 0, \label{eq:orthonorm_ab1}\\
 {\bm u}_{b}^\dagger {\bm u}_{a} -  {\bm v}_{b}^\dagger {\bm v}_{a} =
 {\bm u}_{a}^\dagger {\bm u}_{b} -  {\bm v}_{a}^\dagger {\bm v}_{b} = 1. \label{eq:orthonorm_ab2}
\end{align}
Because $\begin{pmatrix} {\bm v}^*_{a} \\ {\bm u}^*_{a}\end{pmatrix}$ and $\begin{pmatrix} {\bm v}^*_{b} \\ {\bm u}^*_{b}\end{pmatrix}$ are the eigenmodes of $\bm \sigma_z \bm M^{\rm B}_{-\bm k}$ with 
the eigenvalues $-E^*_{a}$ and $-E^*_{b}=-E_{a}$,
a pair of quasi-particles with momenta $\bm k$ and $-\bm k$ should be constructed from the above two eigenmodes.
Indeed, when we define 
\begin{align}
 \begin{pmatrix} {\bm u}_{{\bm k},m} \\ {\bm v}_{\bm k,m} \end{pmatrix} =& 
 \frac{1}{\sqrt{2}}\begin{pmatrix} {\bm u}_{a}+ {\bm u}_{b} \\ {\bm v}_{a} + {\bm v}_{b} \end{pmatrix},  \label{eq:Bog_compuv1}\\
 \begin{pmatrix} {\bm u}_{-{\bm k},m} \\ {\bm v}_{-\bm k,m} \end{pmatrix} =& 
 \frac{i}{\sqrt{2}}\begin{pmatrix} {\bm v}^*_{a}- {\bm v}^*_{b} \\ {\bm u}^*_{a} - {\bm u}^*_{b} \end{pmatrix}, \label{eq:Bog_compuv2}
\end{align}
we can confirm from Eqs.~\eqref{eq:orthonorm_ab1} and \eqref{eq:orthonorm_ab2} that $\begin{pmatrix} {\bm u}_{\pm \bm k,m} \\ {\bm v}_{\pm \bm k,m}\end{pmatrix}$ satisfy the conventional orthonormal conditions [see Eqs.~\eqref{eq:uv_relation_3} and \eqref{eq:uv_relation_4}]:
\begin{align}
 {\bm u}_{\bm k,m}^\dagger {\bm u}_{\bm k,m} -  {\bm v}_{\bm k,m}^\dagger {\bm v}_{\bm k,m} &= 1,\\
 {\bm u}_{-\bm k,m}^\dagger {\bm u}_{-\bm k,m} -  {\bm v}_{-\bm k,m}^\dagger {\bm v}_{-\bm k,m} &=1, \\
 {\bm u}_{\bm k,m}^{\rm T} {\bm v}_{-\bm k,m} -  {\bm v}_{\bm k,m}^{\rm T} {\bm u}_{-\bm k,m} &= 0.
\end{align}
Hence, the Bogoliubov quasi-particles for the complex modes are defined by
\begin{align}
 \hat{\bm b}_{\bm k,m} &= \frac{{\bm u}_{a}^{\dagger}+{\bm u}_{b}^{\dagger}}{\sqrt{2}}\hat{\bm a}_{\bm k} - \frac{{\bm v}_{a}^{\dagger}+{\bm v}_{b}^{\dagger}}{\sqrt{2}}\hat{\bm a}^\dagger_{-\bm k},\\
 \hat{\bm b}_{-\bm k,m} &= -i\frac{{\bm v}_{a}^{\rm T}-{\bm v}_{b}^{\rm T}}{\sqrt{2}}\hat{\bm a}_{-\bm k} + i\frac{{\bm u}_{a}^{\rm T}-{\bm u}_{b}^{\rm T}}{\sqrt{2}}\hat{\bm a}^\dagger_{\bm k}.
\end{align}
Finally, defining $E_{\bm k,m}=E_a$ and $E_{-\bm k,m}=-E_a^*=-E_{\bm k,m}^*$ for the complex modes,
Eq~\eqref{eq:Bog_Heff_diagonalized} is modified as
\begin{align}
\hat{H}^{\rm B} =& E_0 + \frac{1}{2}\sum_{{\bm k}\ne{\bm 0}}\left\{-{\rm Tr}[\bm H_{\bm k}^{(0)}+\bm H^{(1)}]+\frac{D^{\rm corr}}{2\epsilon_{\bm k}}\right\}
+\sum_{\bm k\neq\bm 0,\, m} {\rm Re}(E_{{\bm k},m})\left(\hat{b}^\dagger_{{\bm k},m}\hat{b}_{{\bm k},m}+\frac{1}{2}\right)\nonumber\\
&+  {\sum_{\bm k\neq \bm0,\,m}}'\,{\rm Im}(E_{{\bm k},m})\left(\hat{b}_{{\bm k},m}\hat{b}_{-{\bm k},m} + \hat{b}^\dagger_{{\bm k},m}\hat{b}^\dagger_{-{\bm k},m}\right),
\label{eq:Bog_Heff_diagonalized2}
\end{align}
where the prime in $\sum'$ indicates that except for $\bm k=\bm0$ we sum over a half momentum space to count each pair $(\bm k, -\bm k)$ only once.
The Hamiltonian~\eqref{eq:Bog_Heff_diagonalized2} clearly shows that a pair of $\hat{b}_{\bm k,m}$ and $\hat{b}_{-{\bm k},m}$ quasi-particles are generated 
even from a vacuum $|{\rm vac}_{\rm B}\rangle$ of the Bogoliubov quasi-particles.
Actually, 
from the Heisenberg equations of operators
$\hat{A}=\hat{b}^\dagger_{\bm k,m}\hat{b}_{\bm k,m} + \hat{b}^\dagger_{-\bm k,m}\hat{b}_{-\bm k,m}$ and
$\hat{B}=-\hat{b}_{\bm k,m}\hat{b}_{-\bm k,m} + \hat{b}^\dagger_{\bm k,m}\hat{b}^\dagger_{-\bm k,m}$,
\begin{align}
i\hbar\frac{d}{dt} \hat{A}
&= 2 {\rm Im}(E_{\bm k,m}) \hat{B}, \\
i\hbar\frac{d}{dt}\hat{B}
&= -2{\rm Im}(E_{\bm k,m}) (\hat{A}+1),
\end{align}
we obtain
\begin{align}
\langle{\rm vac}_{\rm B}| \hat{A}(t)|{\rm vac}_{\rm B}\rangle = \frac{e^{2\gamma t}+e^{-2\gamma t}}{2} -1 
= 2 \sinh^2 (\gamma t),
\end{align}
where $\gamma\equiv {\rm Im}(E_{\bm k,m})/\hbar$
and we have used $\langle{\rm vac}_{\rm B}| \hat{A}(0)|{\rm vac}_{\rm B}\rangle = \langle{\rm vac}_{\rm B}| \hat{B}(0)|{\rm vac}_{\rm B}\rangle =0$.
Since $\hat{b}_{\bm k,m}$ and $\hat{b}_{-\bm k,m}$ particles are created pairwise, the number of these particles increases as
\begin{align}
\langle \hat{b}^\dagger_{\pm\bm k,m}\hat{b}_{\pm\bm k,m}\rangle
=\sinh^2 \left[\frac{{\rm Im}(E_{\bm k,m})}{\hbar}t\right].
\label{eq:Bog_DI_manybody}
\end{align}
The above discussions can straightforwardly be generalized to a nonuniform system by writing down the Bogoliubov eigenvalue equation in the coordinate space.

\subsection{Mean-field description}
In the mean-field theory, the collective motion of the condensate is described with the Bogoliubov spectrum.
The GPE of a spin-$f$ system is generally written in the following form:
\begin{align}
 i\hbar\frac{\partial}{\partial t} \psi_m = \left[-\frac{\hbar^2}{2M}\nabla^2 +U_{\rm trap}(\bm r) + pm + qm^2 \right] \psi_m + \sum_{nm'n'}C^{m_1m_2}_{m_1'm_2'}\psi_n^*\psi_{n'}\psi_{m'},
\label{eq:GPE_general}
\end{align}
where $C^{m_1m_2}_{m_1'm_2'}$ is defined in Eq.~\eqref{eq:def_Cmnm'n'}.
We expand the order parameter as
\begin{align}
\psi_m(\bm r,t) = [\psi_m^{(0)}(\bm r) + \delta\psi_m(\bm r,t)]e^{-i\mu t/\hbar},
\label{eq:Bog_expand_psi}
\end{align}
where $\psi_m^{(0)}$ is a stationary solution of the GPE with the chemical potential $\mu$, i.e.,
\begin{align}
 \mu\psi_m^{(0)} =  \left[-\frac{\hbar^2}{2M}\nabla^2 +U_{\rm trap}(\bm r) + pm + qm^2 \right] \psi^{(0)}_m + \sum_{nm'n'}C^{m_1m_2}_{m_1'm_2'}\psi^{(0)*}_n\psi^{(0)}_{n'}\psi^{(0)}_{m'}.
\end{align}
Substituting Eq.~\eqref{eq:Bog_expand_psi} to the GPE~\eqref{eq:GPE_general} and keeping only the terms linear in $\delta\psi$, we obtain
\begin{align}
 i\hbar\frac{\partial}{\partial t}\delta\psi_m 
=& \left[-\frac{\hbar^2}{2M}\nabla^2 +U_{\rm trap}(\bm r) -\mu - pm + qm^2 \right] \delta\psi_m \nonumber\\
&+ \sum_{nm'n'}C^{m_1m_2}_{m_1'm_2'}\left[\psi_n^{(0)*}\psi^{(0)}_{n'}\delta\psi_{m'}+\psi_n^{(0)*}\delta\psi_{n'}\psi^{(0)}_{m'}+\delta\psi_n^*\psi^{(0)}_{n'}\psi^{(0)}_{m'} \right].
\label{eq:GPE_deltapsi}
\end{align}
We consider a mode with eigen frequency $E/\hbar$, and substitute
\begin{align}
 \delta\psi_m(\bm r,t)=e^{-iE t/\hbar}\tilde{u}_m(\bm r) + e^{iE^* t/\hbar}\tilde{v}_m^*(\bm r)
\end{align}
in Eq.~\eqref{eq:GPE_deltapsi}. Then we obtain the following eigenvalue equation:
\begin{align}
E \begin{pmatrix} \tilde{\bm u}(\bm r) \\ \tilde{\bm v}(\bm r) \end{pmatrix} 
  = \begin{pmatrix} \tilde{\bm H}^{(0)}(\bm r)+\tilde{\bm H}^{(1)}(\bm r) & \tilde{\bm H}^{(2)}(\bm r) \\ -[\tilde{\bm H}^{(2)}(\bm r)]^* & -[\tilde{\bm H}^{(0)}(\bm r)+\tilde{\bm H}^{(1)}(\bm r)]^* \end{pmatrix} 
\begin{pmatrix} \tilde{\bm u}(\bm r) \\ \tilde{\bm v}(\bm r) \end{pmatrix},
\end{align}
where $\tilde{\bm H}^{(0,1,2)}$ are $(2f+1)\times(2f+1)$ matrices defined by
\begin{align}
\tilde{\bm H}^{(0)}(\bm r)  &= \left[-\frac{\hbar^2}{2M}\nabla^2 +U_{\rm trap}(\bm r) -\mu \right] {\bm 1} -p{\rm f}_z + q{\rm f}_z^2,\\
\tilde{H}^{(1)}_{mn}(\bm r)  &= \sum_{m'n'}(C^{mn'}_{nm'}+C^{mn'}_{m'n})\psi^{(0)}_{m'}(\bm r)\psi^{(0)*}_{n'}(\bm r), \label{eq:tildeH1}\\
\tilde{H}^{(2)}_{mn}(\bm r)  &= \sum_{m'n'}C^{m_1m_2}_{m_1'm_2'}\psi^{(0)}_{m'}(\bm r)\psi^{(0)}_{n'}(\bm r). \label{eq:tildeH2}
\end{align}
For the case of a uniform system [$U_{\rm trap}(\bm r)=0$] with $\psi_m^{(0)}=\sqrt{n}\zeta_m$,
$\tilde{\bm H}^{(1)}$ and $\tilde{\bm H}^{(2)}$ are exactly the same as ${\bm H}^{(1)}$ and ${\bm H}^{(2)}$ defined in Eqs.~\eqref{eq:def_H^1} and \eqref{eq:def_H^2}.
Moreover, expanding $\tilde{u}(\bm r)$ and $\tilde{v}(\bm r)$ in terms of plane waves as
\begin{align}
 \tilde{\bm u}(\bm r)&=\frac{1}{\sqrt{\Omega}}\sum_{\bm k} {\bm u}_{\bm k}e^{i\bm k\cdot\bm r},\\
 \tilde{\bm v}(\bm r)&=\frac{1}{\sqrt{\Omega}}\sum_{\bm k} {\bm v}_{\bm k}e^{i\bm k\cdot\bm r},
\end{align}
we obtain Eq.~\eqref{eq:Bog_eigenvalue_equation}.

The above results can be summarized as follows. If $E_{\bm k,m}$ is real for any $\bm k$ and $m$, the time evolution of the order parameter can be approximated as
\begin{align}
 \bm\psi(\bm r,t) =  \sqrt{n}\bm\zeta e^{-i\mu t/\hbar} +\frac{1}{\sqrt{\Omega}}\sum_{\bm k, m} \Lambda_{\bm k,m}\left[{\bm u}_{\bm k, m}e^{-i(E_{\bm k,m}t/\hbar -\bm k\cdot\bm r)} + {\bm v}^*_{\bm k, m}e^{i(E_{\bm k,m}t/\hbar-\bm k\cdot\bm r)} \right]e^{-i\mu t/\hbar}.
\label{eq:Bog_psi_rt}
\end{align}
Here, index $m$ identifies the eigenmode and $\Lambda_{\bm k,m}$ is the amplitude of each eigenmode at $t=0$, which is calculated using the orthonormal condition of $\bm u_{\bm k,m}$ and $\bm v_{\bm k,m}$ as
\begin{align}
 \Lambda_{\bm k, m} \equiv \int d{\bm r}e^{-i\bm k\cdot\bm r}\left[{\bm u}^\dagger_{\bm k,m}\bm\psi(\bm r,0)-{\bm v}^\dagger_{\bm k,m}\bm\psi^*(\bm r,0)\right].
\end{align}
Equation~\eqref{eq:Bog_psi_rt} shows that the Bogoliubov mode with a real eigenvalue $E_{\bm k,m}$ describes
a steady oscillation of the order parameter with a fixed amplitude $\Lambda_{\bm k,m}$.
On the other hand, if there are complex conjugate pairs of eigenmodes, their contributions to Eq.~\eqref{eq:Bog_psi_rt} are given by
\begin{align}
 \Lambda_a \left[{\bm u}_{a}e^{-i(E_{a}t/\hbar -\bm k\cdot\bm r)} + {\bm v}^*_{a}e^{i(E_{a}^*t/\hbar-\bm k\cdot\bm r)} \right]
 +\Lambda_b \left[{\bm u}_{b}e^{-i(E_{b}t/\hbar -\bm k\cdot\bm r)} + {\bm v}^*_{b}e^{i(E_{b}^*t/\hbar-\bm k\cdot\bm r)} \right],
\label{eq:Bog_Lambda_ab}
\end{align}
which indicates an exponential growth or decay of the order parameter, depending on the imaginary part of the eigenvalues,
where $\bm u_{a,b}$, $\bm v_{a,b}$ and $E_a=E_b^*$ are the same as those defined in the previous subsection (Sec.~\ref{sec:dynamicalinstability})
and the amplitudes $\Lambda_{a,b}$ are obtained by using Eqs.~\eqref{eq:orthonorm_ab1} and \eqref{eq:orthonorm_ab2} as
\begin{align}
 \Lambda_{a} \equiv \int d{\bm r}e^{-i\bm k\cdot\bm r}\left[{\bm u}^\dagger_{b}\bm\psi(\bm r,0)-{\bm v}^\dagger_{b}\bm\psi^*(\bm r,0)\right],\\
 \Lambda_{b} \equiv \int d{\bm r}e^{-i\bm k\cdot\bm r}\left[{\bm u}^\dagger_{a}\bm\psi(\bm r,0)-{\bm v}^\dagger_{a}\bm\psi^*(\bm r,0)\right].
\end{align}

\subsection{Domain formation}
\label{sec:Bog_Domainformation}

To see how the system evolves in the presence of dynamical instability, 
let us consider a spin-1 polar state: $\bm\zeta=(0,1,0)$.
The Bogoliubov spectra for the polar state are given by [see Eqs.~\eqref{eq:Bog_ene_spin1P0} and \eqref{eq:energy_spec_spin1P1}]
\begin{align}
E_{{\bm k},{\rm ph}} = \sqrt{\epsilon_{\bm k}(\epsilon_{\bm k}+2c_0n)},\ \ 
E_{{\bm k},{\rm sp}} = \sqrt{(\epsilon_{\bm k}+q)(\epsilon_{\bm k}+q+2c_1n)},
\label{eq:Bog_DM_polar}
\end{align}
where $E_{{\bm k},{\rm ph}}$ and $E_{{\bm k},{\rm sp}}$ correspond to the density (phonon) and spin waves, respectively,
and we have chosen $p$ to be zero since the total longitudinal magnetization is conserved and remains zero in the subsequent dynamics.
When the spin-dependent interaction is ferromagnetic ($c_1<0$), the polar state becomes the ground state at $q>2|c_1|n$, and the Bogoliubov spectra are always real and positive in this region.
Then, consider what happens if we suddenly quench the value of $q$ to below $2|c_1|n$.
This is what Sadler {\it et al}. have done in their experiment~\cite{Sadler2006}.
At $q<2|c_1|n$, the spin mode with a momentum lying in the range
\begin{align}
 {\rm Max}(0,-q) < \epsilon_{\bm k} < 2|c_1|n-q
\end{align}
becomes pure imaginary and
the number of atoms in the $m=\pm 1$ states 
grows according to Eqs.~\eqref{eq:Bog_DI_manybody} or \eqref{eq:Bog_Lambda_ab}.
For $q=0$, the imaginary part of $E_{{\bm k},{\rm sp}}$ becomes maximal at $\epsilon_{\bm k} = |c_1|n$, which means that the spin wave with the wavelength
\begin{align}
\lambda_{\rm c} = \sqrt{\frac{3\pi}{2|a_2-a_0| n}}
\end{align}
grows fastest, resulting in a formation of magnetic patterns with the characteristic wavelength $\lambda_{\rm c}$~\cite{Saito2005,Saito2007a}.
For the parameters of the experiment~\cite{Sadler2006}, $n\sim 2.8\times 10^{20}~{\rm m}^{-3}$ and $|a_2-a_0|=1.07 a_{\rm B}$ with $a_{\rm B}$ the Bohr radius, we obtain $\lambda_{\rm c}\sim 17~\mu$m, in good agreement with the observed domain size (which corresponds to $\lambda_{\rm c}/2$) of about 10~$\mu$m~\cite{Sadler2006}.

Another example of the dynamical instability is the formation of metastable spin domains~\cite{Miesner1999}.
In Ref.~\cite{Miesner1999,Goldstein1997,Mueller1999,Ueda2000}, Miesner {\it et al.} first prepared $^{23}$Na atoms in the $m=1$ state and transferred half of them to the $m=0$ state by irradiating an rf field. 
Then, letting the system evolve freely in time while applying a uniform magnetic field to  
prevent the $m=-1$ component from appearing due to the quadratic Zeeman effect, they found that spin domains develop with the two components being aligned alternatively.
The experimental conditions in effect amount to setting
\begin{eqnarray}
\zeta_1=\zeta_0=\frac{1}{\sqrt{2}},\ \ \zeta_{-1}=0.
\end{eqnarray}
With this initial state, $\mu$ defined in Eq.~\eqref{eq:Bog_defmu} is given by $\mu=(q-p)/2+(c_0+3c_1/4)n$.
Because the $m=-1$ component does not appear due to the large quadratic Zeeman energy,
we can neglect the elements $H^{(1,2)}_{m,-1}$ and $H^{(1,2)}_{-1,m}$ and choose $p=q=0$.
Then, ${\bm M}^{\rm B}_{\bm k}$ in the Bogoliubov equation reduces to a $4\times 4$ matrix
constructed from
\begin{align}
{\bm H}^{(0)}_{\bm k} + {\bm H}^{(1)}&= \epsilon_{\bm k}{\bm 1}+
\frac{n}{4}\begin{pmatrix} 2c_0 + 3c_1 & 2c_0+2c_1 \\
 2c_0+2c_1 & 2c_0-c_1 \end{pmatrix},\\
{\bm H}^{(2)}&=\frac{n}{2}\begin{pmatrix}c_0+c_1 & c_0+c_1 \\ c_0+c_1 & c_0 \end{pmatrix}.
\end{align}
The eigenvalues of ${\bm M}^{\rm B}_{\bm k}$ are given by
\begin{align}
E_{\bm k}&=\sqrt{\epsilon_{\bm k}^2 + \left(c_0+\frac{c_1}{2}\right)n\epsilon_{\bm k}+\frac{3(c_1n)^2}{16} \pm\Lambda_{\bm k}}, \label{eq:Bog_E_domain}\\
\Lambda_{\bm k}&=\sqrt{(c_0^2+2c_0c_1+2c_1^2)n^2\epsilon_{\bm k}^2+\frac{1}{4}(2c_0+c_1)c_1^2n^3\epsilon_{\bm k} -\frac{1}{64}(4c_0+3c_1)c_1^3n^4}.
\end{align}
Since $c_0 \gg c_1 (>0)$ for $^{23}$Na atoms (see Table~\ref{table:scattering_length}),
by ignoring higher-order powers of $c_1/c_0$, we obtain
\begin{eqnarray*}
E_{\bm k}=\sqrt{\epsilon_{\bm k}^2+\left(c_0n+\frac{c_1n}{2}\right)\epsilon_{\bm k}\pm
(c_0n+c_1n)\epsilon_{\bm k}},
\end{eqnarray*}
where the plus (minus) sign
corresponds to the density (spin) wave.
We see that the energy of the spin wave becomes pure imaginary for 
$\epsilon_{\bm k}<c_1n/2$, implying the formation of spin domains.  
The wavelength for the most unstable mode, that is,
the wavelength at which the imaginary part of $E_{\bm k}$ takes its maximum,
is~\cite{Mueller1999,Ueda2000}
\begin{eqnarray}
\lambda_{\rm c}=\sqrt{\frac{6\pi}{(a_2-a_0)n}}.
\end{eqnarray}
For the parameters of the experiment~\cite{Miesner1999}, $n\sim 10^{20}~{\rm m}^{-3}$ and $a_2-a_0=2.47 a_{\rm B}$ (the latest data measured in Ref.~\cite{Black2007}), we obtain $\lambda_{\rm c}\sim 38~\mu$m;
this is in reasonable agreement with the observed value ($\lambda_{\rm c}/2$) of approximately about 40~$\mu$m~\cite{Miesner1999}.

The dynamical instability and pattern-formation dynamics in a spinor BEC have also been discussed in various situations such as 
a non-stationary state~\cite{Zhang2005b}, 
a spin helix~\cite{Cherng2008a},
a system confined in a two-dimensional harmonic trap~\cite{Saito2006,Scherer2010},
and a coreless-vortex state~\cite{Pietila2007,Takahashi2009}.
The effects of the magnetic dipole-dipole interactions for the experiment~\cite{Sadler2006} are discussed in Refs.~\cite{Sau2009,Hoshi2010,Deuretzbacher2010},
and those for different setups are discussed in Refs.~\cite{Cherng2009,Lamacraft2008,Kawaguchi2010,Kudo2010} (see also Sec.~\ref{sec:dipole_Bogoliubov}).
The vortex-formation dynamics via the Kibble-Zurek mechanism is discussed in Refs.~\cite{Lamacraft2007,Uhlmann2007,Saito2007b,Damski2007} (see also Sec.~\ref{sec:KZ}).

\subsection{Parametric amplification}
\label{sec:parametric_amplification}
We see in the previous subsections that when the system is dynamically unstable, the unstable modes, that is, the Bogoliubov modes that have complex eigenvalues, exponentially grow, leading to a formation of magnetic domains.
Now, a question arises as to what is the initial seed that triggers the instability.
In the mean-field description, Eqs.~\eqref{eq:Bog_psi_rt} and \eqref{eq:Bog_Lambda_ab} indicates that the condensate $\bm\zeta$ is stable if there is no fluctuations that triggers the instability in the initial state, i.e., if $\Lambda_{a,b}=0$.
In reality, however, such a state cannot be stable even at absolute zero because quantum fluctuations trigger the instability.
This fact can be understood from Hamiltonian~\eqref{eq:Bog_Heff_diagonalized2};
the number of the unstable quasi-particles increases no matter how they distribute in the initial state.

To quantify the effects of quantum and classical fluctuations, let us consider again 
the first example in the previous subsection~\cite{Sadler2006}.
We consider a ferromagnetically interacting gas ($c_1<0$) and prepare the initial state in the polar phase, $\bm\zeta=(0,1,0)^{\rm T}$.
The Bogoliubov Hamiltonian~\eqref{eq:Bog_Heff} for this state is given by
\begin{align}
\hat{H}^{\rm B} = 
&E_0
+\sum_{\bm k\neq \bm 0}\frac{D^{\rm corr}}{4\epsilon_{\bm k}}\nonumber\\
&+\sum_{{\bm k}\neq{\bm 0}}\left[(\epsilon_{\bm k}+q +c_1n)(\hat{a}^\dagger_{\bm k,1}\hat{a}_{\bm k,1}+\hat{a}^\dagger_{\bm k,-1}\hat{a}_{\bm k,-1})
+(\epsilon_{\bm k}+q + c_0n)\hat{a}^\dagger_{\bm k,0}\hat{a}_{\bm k,0}\right]\nonumber\\
&+\sum_{\bm k\neq \bm 0}\left( \frac{c_0n}{2} \hat{a}_{\bm k,0}\hat{a}_{-\bm k,0} +  c_1n \hat{a}_{\bm k,1}\hat{a}_{-\bm k,-1} + {\rm H. c.}\right).
\end{align}
The Bogoliubov mode that becomes unstable at $p=0$ and $q<2|c_1|n$ excites the atoms in the $m=\pm1$ states [see Eqs.~\eqref{eq:energy_spec_spin1P1} and \eqref{eq:Bog_b_spin1P1}].
The Heisenberg equations of motion for these atoms are~\cite{Lamacraft2007,Saito2007b}
\begin{eqnarray}
i\hbar\frac{d}{dt}\hat{a}_{\pm{\bm k},\pm1}=(\epsilon_{\bm k}+q+c_1n)\hat{a}_{\pm{\bm k},\pm1}+c_1n\hat{a}^\dagger_{\mp{\bm k},\mp 1},
\end{eqnarray}
with the solutions given by
\begin{align}
\hat{a}_{{\bm k},\pm1}(t)
=&\left( \cos\frac{E_{{\bm k},{\rm sp}}t}{\hbar}-i\frac{\epsilon_{\bm k}+q+c_1n}{E_{{\bm k},{\rm sp}}}
\sin\frac{E_{{\bm k},{\rm sp}}t}{\hbar}
\right)\hat{a}_{{\bm k},\pm1}(0) \nonumber\\
&-i\frac{c_1n}{E_{{\bm k},{\rm sp}}}\sin\frac{E_{{\bm k},{\rm sp}}t}{\hbar} \ 
\hat{a}_{-{\bm k},\mp 1}^\dagger(0),
\label{eq:hat_a_t}
\end{align}
where $E_{{\bm k},{\rm sp}}$ is  the Bogoliubov spectrum defined in Eq.~\eqref{eq:Bog_DM_polar}.
When $c_1<0$ and $q<2|c_1|n$, $E_{{\bm k},{\rm sp}}$ becomes pure imaginary.
Rewriting $E_{{\bm k},{\rm sp}}=i\hbar \gamma_{\bm k}$ for the unstable mode,
the number of atoms in these modes increases as
\begin{align}
\langle \hat{n}_{\bm k,\pm1}(t) \rangle
=& \langle \hat{a}^\dagger_{\bm k,\pm1}(t)\hat{a}_{\bm k,\pm1}(t) \rangle \nonumber\\
=& \left[\cosh^2 (\gamma_{\bm k}t) + \left(\frac{\epsilon_{\bm k}+q+c_1n}{\hbar\gamma_{\bm k}}\right)^2\sinh^2(\gamma_{\bm k}t)\right] \langle \hat{n}_{{\bm k},\pm1}(0)\rangle \nonumber\\
&+\left(\frac{c_1n}{\hbar\gamma_{\bm k}}\right)^2 \sinh^2(\gamma_{\bm k}t)
\left(\langle \hat{n}_{-{\bm k},\mp 1}(0)\rangle +1\right),
\label{eq:Bog_nt}
\end{align}
where we have assumed $\langle \hat{a}_{\bm k,\pm1}(0)\hat{a}_{-\bm k,\mp1}(0)\rangle = \langle \hat{a}^\dagger_{\bm k,\pm1}(0)\hat{a}^\dagger_{-\bm k,\mp1}(0)\rangle = 0$.
Equation~\eqref{eq:Bog_nt} clearly shows that even when all atoms are prepared exactly in the $m=0$ state, i.e., $\langle \hat{n}_{\pm\bm k,\pm1}(0)\rangle=0$, 
the number of atoms in the $m=\pm1$ states increases as $[c_1n/(\hbar\gamma_{\bm k})]^2 \sinh^2(\gamma_{\bm k}t)$. 
This is the amplification of quantum fluctuations.
In addition, classical fluctuations, that is, the atoms initially populated in the $m=\pm1$ states, are also amplified.
The amplification of quantum fluctuations is investigated both experimentally and theoretically in Refs.~\cite{Leslie2009,Sau2009} in which the truncated Wigner approximation is used to let classical spin fluctuations numerically evolve in time.
The theoretical results, under the assumption that the observed amplification of spin fluctuations is quantum limited, are consistent with 
the growth rate observed in the experiment within the error bar of the scattering length $a_2-a_0$ (see Table~\ref{table:scattering_length}).
The number fluctuations in each spin component during the parametric amplification are discussed in Ref.~\cite{Mias2008}.
Similar experiments were performed for spin-2 $^{87}$Rb condensates in Refs.~\cite{Klempt2009,Klempt2010,Scherer2010}, where the effects of quantum fluctuations are more prominent due to strong confinement of the trapping potential.

\subsection{Kibble-Zurek mechanism}
\label{sec:KZ}

The dynamical instability and the parametric amplification discussed above predict how the quantum phase transition is triggered.
Actually, the example in the previous subsection describes the phase transition triggered by a sudden change in a magnetic field (magnetic field quench)~\cite{Sadler2006,Saito2007a,Saito2007b,Lamacraft2007,Uhlmann2007,Damski2007,Damski2008}.
The ground-state phase of a spin-1 ferromagnetic BEC is shown in Fig.~\ref{fig:spin-1PD}, where $p$ and $q$ are the coefficients of the linear and quadratic Zeeman effects, respectively.
When we consider a type of magnetic-field quench for which the total magnetization of the system is conserved, the quench occurs along a constant-$p$ line of the phase diagram. 
As in the experiment described in Ref.~\cite{Sadler2006}, we consider the case of $p=0$ which implies that the total longitudinal magnetization of the system is kept to be zero.
Then, when we decrease $q$ across the critical value $q_{\rm c}= 2|c_1|n$, the ground state changes from the polar phase to the broken-axisymmetry (BA) phase:
\begin{align}
\bm\psi^{\rm polar} = \sqrt{n}
\begin{pmatrix}0 \\ 1 \\ 0
\end{pmatrix}\ \ \to \ \ 
\bm\psi^{\rm BA} = \sqrt{n}
\begin{pmatrix}e^{-i\alpha} \sqrt{1-\tilde{q}}/2 \\ \sqrt{(1+\tilde{q})/2} \\ e^{i\alpha}\sqrt{1-\tilde{q}}/2
\end{pmatrix},
\label{eq:BAphase}
\end{align}
where $\tilde{q}\equiv q/q_{\rm c}$ and $\alpha$ is an arbitrary real number.
As we can see from Eq.~\eqref{eq:BAphase}, the populations in the $m=\pm1$ components grow from zero in this phase transition, and
Eq.~\eqref{eq:Bog_nt} describes their time evolution.

Note here that the BA phase has transverse magnetization $F_+\equiv F_x+iF_y = e^{i\alpha}\sqrt{1-\tilde{q}^2}$ which breaks the SO(2) symmetry corresponding to the direction of the transverse magnetization,
whereas the polar phase has no magnetization and preserves the SO(2) symmetry.
Therefore, this is the SO(2)-symmetry-breaking phase transition.
By rapidly quenching the magnetic field from above to below $q_{\rm c}$, 
$\alpha$ in Eq.~\eqref{eq:BAphase} is randomly chosen at each spatial point,
which implies that local transverse magnetizations develop in random directions, and thus, spin vortices are spontaneously created. 

Such a mechanism for topological defect formation associated with a phase transition is known as the Kibble-Zurek (KZ) mechanism.
The basic idea of the KZ mechanism is as follows. After the phase transition domains of the new phase emerge at causally disconnected places, and the order parameters of the new phase at different locations are not correlated. When the order parameters grow to overlap, they may be able to adjust dynamically so that they are connected smoothly; if they cannot, singularities develop in the order parameter space, giving rise to topological excitations. Such a scenario of topological defect formation was first discussed by Kibble~\cite{Kibble1976} in the context of cosmic-string and monopole formation in the early Universe, and an experimental test for the Kibble scenario in condensed matter systems was proposed by Zurek~\cite{Zurek1985} (see Ref.~\cite{Zurek1996} for a review). 

Bose-Einstein condensates of dilute atomic vapor offer an ideal testbed for studying the KZ mechanism because the temperature, strength of interaction, and external parameters such as magnetic field and trapping potentials can be changed in a time shorter than the characteristic time scale of topological defect formation. In fact, scalar vortices~\cite{Scherer2007,Weiler2008} and spin vortices~\cite{Sadler2006,Vengalattore2008} have been observed to emerge spontaneously upon either temperature or magnetic-field quench. 
In this subsection, we discuss the KZ mechanism for a quantum phase transition, which is induced by a magnetic-field quench, from the viewpoint of dynamical instability.

\subsubsection{Instantaneous quench}

We first consider the situation in which the condensate is prepared in the polar phase $\bm\psi^{\rm polar}$ at $q>q_{\rm c}$ and then $q$ is suddenly quenched to below $q_{\rm c}$.
We can use solution~\eqref{eq:hat_a_t} to investigate how the transverse magnetization develops with time. 
Because $\hat{\psi}_0\simeq\sqrt{n}$, the magnetization operator $\hat{F}_+=\hat{F}_-^\dagger=\hat{F}_x+i\hat{F}_y$ is written as
\begin{eqnarray}
\hat{F}_+=\sqrt{2n}[\hat{\psi}_1^\dagger({\bm r})+\hat{\psi}_{-1}({\bm r})].
\label{eq:KZ_F+}
\end{eqnarray}
Using $\hat{\psi}_{\pm 1}(\bm r)=1/\sqrt{\Omega}\sum_{\bm k}\hat{a}_{\bm k,\pm 1}e^{i\bm k\cdot\bm r}$ and Eq.~\eqref{eq:hat_a_t},
the correlation function of the transverse magnetization is given by
\begin{eqnarray}
\langle\hat{F}_+({\bm r},t)\hat{F}_-({\bm r}',t) \rangle
\simeq\frac{n}{2\Omega}\sum_{{\bm k},|\bm k|<k_{\rm c}}\frac{q_{\rm c}}{q_{\rm c}-q-\epsilon_{\bm k}}\exp\left[{\frac{2}{\hbar}|E_{\bm k,{\rm sp}}|t+i{\bm k}\cdot({\bm r}-{\bm r}')}\right],
\label{eq:KZ_corr}
\end{eqnarray}
where $E_{{\bm k},{\rm sp}}$ is given in Eq.~\eqref{eq:Bog_DM_polar} as
\begin{align}
E_{{\bm k},{\rm sp}} = \sqrt{(\epsilon_{\bm k}+q)(\epsilon_{\bm k}+q+2c_1n)},
\label{eq:KZ_Eksp}
\end{align}
which becomes pure imaginary for ${\bm k}$ satisfying $|\bm k|<k_{\rm c}\equiv \sqrt{2M(q_{\rm c}-q)}/\hbar$. 
The exponential terms in Eq.~\eqref{eq:KZ_corr} describe dynamical instabilities, the dominant term of which is the one for which the imaginary part of $E_{\bm k}$ is maximal. 
Let $k_{\rm mu}$ be the ``most unstable'' wave number for which $|{\rm Im}E_{\bm k}|$ is maximal. 
It follows from Eq.~\eqref{eq:KZ_Eksp} that $k_{\rm mu}=0$ for $q_{\rm c}/2<q<q_{\rm c}$ and 
\begin{eqnarray}
k_{\rm mu}=\sqrt{\frac{2M}{\hbar^2}\left(\frac{q_{\rm c}}{2}-q\right)}
\label{k_mu}
\end{eqnarray}
for $q<q_{\rm c}/2$.
We then expand $2|E_{{\bm k},{\rm sp}}|t/\hbar$ in the exponent around $k_{\rm mu}$ as
\begin{align}
 \frac{2|E_{{\bm k},{\rm sp}}|t}{\hbar}=\frac{t}{\tau}\left(1-\frac{1}{4}\xi_{\rm corr}^2\Delta k^2\right) + O(\Delta k^4),
\end{align}
where $\Delta k=k-k_{\rm mu}$.
It is clear that $\tau$ sets the time scale for exponential growth.
Magnetization can be observed when it has grown sufficiently, i.e., after time $t\sim \tau$ has elapsed.
Replacing the summation with the Gaussian integral in Eq.~\eqref{eq:KZ_corr}, we find that $\xi_{\rm corr}$ represents the correlation length.
For $q_{\rm c}/2<q<q_{\rm c}$, $k_{\rm mu}=0$ and
\begin{align}
 \tau = \frac{\hbar}{2\sqrt{q(q_{\rm c}-q)}},\ \ \ \xi_{\rm corr} = \sqrt{\frac{\hbar^2}{M}\frac{2q-q_{\rm c}}{q(q_{\rm c}-q)}}.
\end{align}
In this case, $\xi_{\rm corr}$ gives the typical size of magnetic domains.
For $q<q_{\rm c}/2$, $k_{\rm mu}$ is given by Eq.~\eqref{k_mu} and
\begin{align}
 \tau = \frac{\hbar}{q_{\rm c}},\ \ \ \xi_{\rm corr} = \sqrt{\frac{8\hbar^2}{M}\frac{q_{\rm c}-2q}{q_{\rm c}^2}}.
\end{align}
Since $k_{\rm mu}\neq 0$, the factor $e^{i{\bm k}\cdot({\bm r-\bm r'})}$ in Eq.~\eqref{eq:KZ_corr} creates additional spin winding within the correlation length $\xi_{\rm corr}$,
namely, magnetic domains in the region of $\Delta x<\xi_{\rm corr}$ are correlated.

We consider the Kibble scenario in a two-dimensional (2D) system.
The number of generated spin vortices, or the spin winding number $w$, is obtained from the algebraic relation
\begin{align}
w=\frac{1}{2\pi}\oint_{C(R)}\frac{F_-\bm\nabla F_+ - F_+\bm\nabla F_-}{2i|F_+|^2}\cdot d\bm\ell = \frac{1}{2\pi}\oint \bm\nabla\alpha \cdot d\bm\ell,
\label{eq:KZ_w}
\end{align}
where $C(R)$ is a circle of radius $R$ and $w$ gives the number of rotations of the spin vector in the 2D plane along $C(R)$.
The ensemble average of the winding number, $\langle w \rangle_{\rm avg}$, vanishes due to the random nature of the initial noise (vacuum fluctuations), and the standard deviation, $\langle w^2\rangle^{1/2}_{\rm avg}$, should be regarded as a typical winding number.
For $q_{\rm c}/2<q<q_{\rm c}$, $\langle w^2\rangle_{\rm avg}$ is proportional to the number of random domains $\sim R/\xi_{\rm corr}$ on the boundary $C(R)$.
For $q_{\rm c}/2<q<q_{\rm c}$, an additional spin winding within the correlated region should be multiplied to the winding number.
In particular for $k_{\rm mu}\xi_{\rm corr}\gg 1$, $\langle w^2\rangle_{\rm avg} \propto (k_{\rm mu}\xi_{\rm corr})^2 R/\xi_{\rm corr}$.
In both cases, the number of vortices increases linearly with $\sqrt{R}$ and monotonically decreases with the final value of $q$.

\subsubsection{Finite-time quench}

Next, we consider the case in which $q$ is quenched linearly in a finite time $\tau_{\rm Q}$~\cite{Lamacraft2007,Saito2007b,Damski2007}:
\begin{eqnarray}
q(t)=q_{\rm c}\left(
1-\frac{t}{\tau_{\rm Q}}
\right).
\end{eqnarray}
In this case, the Bogoliubov excitation energy \eqref{eq:KZ_Eksp} changes in time and the spin correlation function is estimated to be
\begin{align}
\langle\hat{F}_+({\bm r},t)\hat{F}_-({\bm r}',t) \rangle
&\propto
\int d{\bm k}
\exp\left[\frac{2}{\hbar}\int^t_0|E_{\bm k}(t')|t'dt'+i{\bm k}\cdot({\bm r}-{\bm r}')\right] \nonumber \\
&\propto \exp\left[
f(t)-|{\bm r}-{\bm r}'|^2/\xi_{\rm Q}^2
\right],
\end{align}
where since we are interested in the vicinity of the critical point, we expanded $|E_{\bm k}(t)|$ around $k_{\rm mu}=0$, and
 $\xi_{\rm Q}$  and $f(t)$ are given by~\cite{Saito2007b}
\begin{align}
\xi_{\rm Q}&=\left[
\frac{16\hbar^2}{M^2}t(\tau_{\rm Q}-t)
\right]^{1/4},
\label{tau_Q}\\
f(t)&=\frac{\tau_{\rm Q}q_{\rm c}}{2\hbar}
\left[\tan^{-1}\sqrt{\frac{t}{\tau_{\rm Q}-t}}
-\sqrt{\frac{t}{\tau_{\rm Q}}\left(1-\frac{t}{\tau_{\rm Q}}\right)}
\left(1-\frac{2t}{\tau_{\rm Q}}\right)\right].
\label{f(t)}
\end{align}
The time scale for magnetization is determined by $f(t)=1$. Solving this equation by assuming $t\ll\tau_{\rm Q}$, we obtain
\begin{eqnarray}
t_{\rm Q}\sim\left(\frac{\hbar^2\tau_{\rm Q}}{q_{\rm c}^2}\right)^{1/3}.
\end{eqnarray}
Substituting this into Eq.~(\ref{tau_Q}), we obtain
\begin{eqnarray}
\xi_{\rm Q}\sim
\left(\frac{\hbar^4}{M^3q_{\rm c}}\right)^{1/6}
\tau_{\rm Q}^{1/3}.
\end{eqnarray}
The same power law was obtained in Refs.~\cite{Lamacraft2007,Damski2007}.
Because $\xi_{\rm Q}$ is the spin correlation length in the present case,
$\langle w^2\rangle$ is proportional to $R/\xi_{\rm Q}\propto R \tau_{\rm Q}^{-1/3}$.

\subsubsection{Numerical simulations}

\begin{figure}[ht]
\begin{center}
\resizebox{\hsize}{!}{
\includegraphics{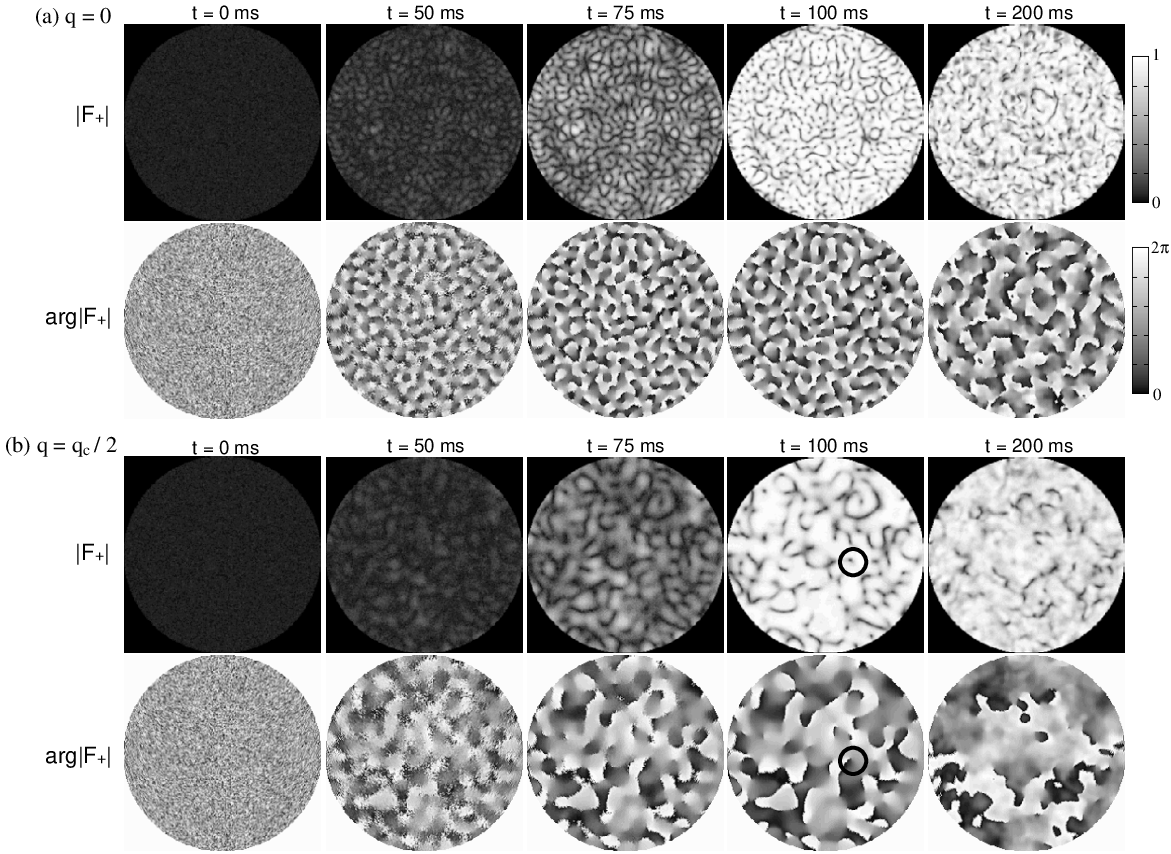}
}
\end{center}
\caption{Spontaneous magnetization following the instantaneous quench from $q>q_{\rm c}$ to $q<q_{\rm c}$. Shown are the time developments of the magnetization $|F_+|$ (upper) and its
direction ${\rm arg}$ $|F_+|$ (lower) for (a) $q = 0$ and (b) $q = q_{\rm c} / 2$.
The black circles in (b) indicate a topological defect.
Reprinted from Ref.~\cite{Saito2007b}.
}
\label{fig:KZmag}
\end{figure}

The topological defect formation caused by a quantum phase transition can be illustrated by numerical simulations. 
Figure~\ref{fig:KZmag} shows the time evolution of transverse magnetization of a two-dimensional spin-1 BEC following an instantaneous quench of the magnetic field from $q>q_{\rm c}$ to (a) $q=0$ and (b) $q=q_{\rm c}/2$. Here, we assume that the system is a 2D disk with a hard wall at radius 100~$\mu$m and the potential is assumed to be flat inside the wall.
The initial state is considered to be a stationary state of the GPEs and the following quench dynamics is obtained by using the time-dependent GPEs. To trigger the dynamical instabilities that lead to defect formation, we introduce small fluctuations in the initial amplitudes of the $m=\pm1$ spin components according to
\begin{eqnarray}
a_{{\bm k},\pm1}(0)=\alpha_{\rm rnd}+i\beta_{\rm rnd},
\end{eqnarray}
where $\alpha_{\rm rnd}$ and $\beta_{\rm rnd}$ are random variables whose amplitudes $x$ follow the normal distribution $p(x)=\sqrt{2/\pi} \exp(-2x^2)$.

\begin{figure}[ht]
\begin{center}
\resizebox{0.6\hsize}{!}{
\includegraphics{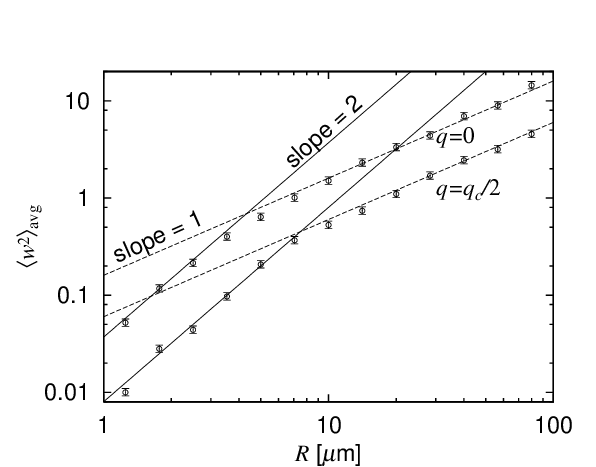}
}
\end{center}
\caption{
$R$ dependence of the variance of the spin winding number $w(R)$ for 
instantaneous quench of the magnetic field to $q = 0$ and to $q = q_{\rm c} / 2$.
The dashed and dotted lines are respectively proportional to $R$ and $R^2$.
Reprinted from Ref.~\cite{Saito2007b}.
}
\label{fig:KZrdep}
\end{figure}

Figure~\ref{fig:KZmag} shows snapshots of the amplitude $|F_+({\bm r})|$ and the phase arg$F_+({\bm r})$ of the transverse magnetization after the instantaneous quench. We observe that many holes emerge spontaneously after 100~ms. These holes represent topological excitations called polar-core vortices in which the $m=\pm1$ components have vortices with the cores filled by the $m=0$ components. 
Figure~\ref{fig:KZrdep} shows the $R$ dependence of the ensemble average of $w^2(R)$.
We note that $\langle w^2(R)\rangle_{\rm avg}$ is proportional to $R$ for large $R$ in agreement with the KZ theory,
whereas it is proportional to $R^2$ for small $R$.

If spin vortices can be generated randomly, $\langle w^2(R)\rangle_{\rm avg}$ should be proportional to the area of the system and hence to $R^2$. The KZ scaling law $\langle w^2(R)\rangle\propto R$ that we find for large $R$ results from the fact that the number of vortices matches that of anti-vortices inside the disk due to spin conservation.
In other words, for small $q$, magnetic domains are aligned in such a manner as to cancel the local spin when averaged over the spin correlation length. Thus, over a larger length scale, the magnetic domains can be created independently. This is the underlying physics that makes the spin conservation compatible with the KZ postulate of independent defect creation at long distance and the KZ scaling law valid for 2D systems.
For the case of slow quench, the $\tau_{\rm Q}$ dependence of the winding number in numerical simulation also shows good agreement with the KZ scaling law as detailed in Ref.~\cite{Saito2007b}.

%% file: dipole.tex
\section{Dipolar Bose-Einstein condensates} 
\label{sec:DipolarBEC}

In this section, we consider the dipole-dipole interaction (DDI) that, unlike the $s$-wave contact interaction, is long-range and anisotropic.
The DDI between alkali atoms is thousand times smaller than the short-range interaction for the background scattering length.
However, recent experimental developments, such as the realization of BECs of $^{52}$Cr atoms~\cite{Griesmaier2005, Beaufils2008}, $^{164}$Dy atoms~\cite{Mingwu2011}, and $^{168}$Er atoms~\cite{Aikawa2012}, creation of ultracold molecules~\cite{Doyle1995,Meerakker2008,Ni2008,Shuman2010},
and precision measurements~\cite{Vengalattore2008, Fattori2008} and control~\cite{Pollack2009}, have enabled us to study several consequences of the DDI.
Here, we introduce the general properties of the DDIs in ultracold gases (Sec.~\ref{sec:dipole_interaction}), and discuss two important cases: spin-polarized dipolar gases (Sec.~\ref{sec:dipole_polarized}) and spinor dipolar gases (Sec.~\ref{sec:dipole_spinor}).
Excellent reviews on dipolar gases have been published by Baranov~\cite{Baranov2002, Baranov2008} and by Lahaye {\it et al}.~\cite{Lahaye2009}.

\subsection{Dipole--dipole interaction}\label{sec:dipole_interaction}
\subsubsection{Scattering properties of dipole interaction}
\begin{figure}[htb]
\begin{center}
\resizebox{0.8\hsize}{!}{\includegraphics{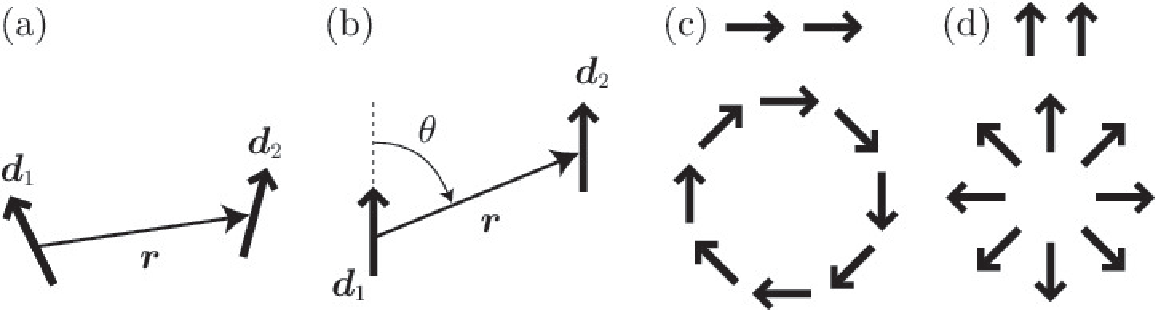}}
\end{center}
\caption{Interaction between two dipole moments for (a) an unpolarized case and (b) a polarized case.
The DDI is most attractive for (c) the head-to-tail configuration and most repulsive for (d) the side-by-side configuration. The bottom figures below (c) and (d) show the associated characteristic spin textures.
}
\label{fig:dipole_config}
\end{figure}

The DDI between two dipole moments
with relative position ${\bm r}$ [see Fig.~\ref{fig:dipole_config} (a)] is given by
\begin{align}
 V_{\rm dd}({\bm r}) &= c_{\rm dd} \frac{\hat{\bm d}_1\cdot\hat{\bm d}_2-3(\hat{\bm d}_1\cdot\hat{\bm r})(\hat{\bm d}_2\cdot \hat{\bm r})}{r^3}
 = c_{\rm dd} \sum_{\nu,\nu'=x,y,z}\hat{d}_{1\nu}Q_{\nu\nu'}({\bm r}) \hat{d}_{2\nu'},
\label{eq:dipole_2body}
\end{align}
where $\hat{\bm d}_{1,2}$ are unit vectors for the directions of dipole moments,
$r=|{\bm r}|$, $\hat{\bm r}={\bm r}/r$, and $Q_{\nu\nu'}({\bm r})\ (\nu,\nu'=x,y,z)$ is a rank-2 traceless symmetric tensor
that can be expressed in terms of rank-2 spherical harmonics $Y_2^m(\hat{\bm r})$ as
\begin{align}
 &Q_{\nu\nu'}({\bm r}) \equiv \frac{\delta_{\nu\nu'}-3\hat{r}_{\nu} \hat{r}_{\nu'}}{r^3}\\
 &= \sqrt{\frac{6\pi}{5}}\frac{1}{r^3}\begin{pmatrix}
\sqrt{\frac{2}{3}}Y_2^{0}(\hat{\bm r}) - Y_2^{2}(\hat{\bm r}) - Y_2^{-2}(\hat{\bm r}) 
& i Y_2^{2}(\hat{\bm r}) - i Y_2^{-2}(\hat{\bm r}) 
&   Y_2^{1}(\hat{\bm r}) -   Y_2^{-1}(\hat{\bm r}) \\
  i Y_2^{2}(\hat{\bm r}) - i Y_2^{-2}(\hat{\bm r}) 
& \sqrt{\frac{2}{3}}Y_2^{0}(\hat{\bm r}) + Y_2^{2}(\hat{\bm r}) + Y_2^{-2}(\hat{\bm r}) 
&-i Y_2^{1}(\hat{\bm r}) - i Y_2^{-1}(\hat{\bm r}) \\
    Y_2^{1}(\hat{\bm r}) -   Y_2^{-1}(\hat{\bm r})
&-i Y_2^{1}(\hat{\bm r}) - i Y_2^{-1}(\hat{\bm r})
& -2\sqrt{\frac{2}{3}}Y_2^{0}(\hat{\bm r})
\end{pmatrix}.
\end{align}
In the special case in which the dipole moments are polarized under an external field, the DDI takes a simple form:
\begin{align}
V_{\rm dd}({\bm r}) 
= c_{\rm dd} Q_{zz}({\bm r}) = c_{\rm dd} \frac{1-3\cos^2\theta}{r^3}
=-\sqrt{\frac{16\pi}{5}}c_{\rm dd} \frac{Y_2^0(\hat{\bm r})}{r^3},
\label{eq:dipole_polarized}
\end{align}
where $\theta$ is the angle between the direction of the polarization and the relative position ${\bm r}$ [see Fig.~\ref{fig:dipole_config}(b)].
Because $V_{\rm dd}$ is negative (positive) for $\theta=0$ ($\pi/2$), the DDI favors a ``head-to-tail'' [Fig.~\ref{fig:dipole_config} (c)] rather than ``side-by-side'' [Fig.~\ref{fig:dipole_config} (d)] configuration.
The long-range character ($\sim 1/r^{3}$) of the DDI becomes prominent in low-energy scattering.
When a potential decreases as $1/r^{n}$ for large $r$,
the phase shift $\delta_l$ behaves in the zero-energy limit as
$k^{2l+1}$ if $l<(n-3)/2$ and as $k^{n-2}$ otherwise (see, e.g., \textsection 132 of Ref.~\cite{LandauLifshitz_QM}).
For a potential with $n>3$, such as van der Waals potential ($n=6$),
the main contribution at $k\to0$ is an {\it s}-wave channel with $\delta_0\sim k$,
and therefore, the potential can be described with a single parameter, namely, the {\it s}-wave scattering length.
On the other hand, when $n=3$, the scattering amplitude logarithmically diverges at the low-energy limit, and therefore, the scattering process is no longer described with a single parameter such as the {\it s}-wave scattering length.
Moreover, the anisotropy of the DDI induces coupling between different partial waves.
Because the dipole interaction is described by rank-2 spherical harmonics and it therefore has {\it d}-wave symmetry, 
a partial wave of angular momentum $(l,m)$ is coupled to $(l,m+\Delta m_f)$ and $(l\pm2,m+\Delta m_f)$, 
where $\Delta m_f$ is the change in the projected total spin of colliding particles and $\Delta m_f=0$ for the polarized case. 
Although there is no diagonal term for the $l=0$ channel, i.e., $\langle l=0,m=0|V_{\rm dd}(\bm r)|l=0,m=0\rangle=0$,
the second-order perturbation induces an effective potential for $l=0$ that behaves as $r^{-6}$ at large distances.
Hence, this part of the dipole interaction can be described with a short-range interaction,
and it is important to keep in mind that the $s$-wave scattering
length may be altered when we change
the strength of the DDI by an applied electric field.

\subsubsection{Dipolar systems}
There are several systems that undergo long-range and anisotropic DDIs.
The coefficient $c_{\rm dd}$ in Eq.~\eqref{eq:dipole_2body} is given by $c_{\rm dd}=d^2/(4\pi\epsilon_0)$ and $c_{\rm dd}=\mu_0d^2/(4\pi)$ for electric and magnetic dipole moments, respectively, where $\epsilon_0$ and $\mu_0$ are the dielectric constant and magnetic permeability of vacuum, respectively,
and $d$ is the magnitude of the dipole moment.
The typical order of magnitude of the electric dipole moment is the product of the elementary charge $e$ and the Bohr radius $a_0$,
whereas the magnetic dipole moment of an alkali atom is of the order of the Bohr magneton $\mu_{\rm B}$.
Therefore, the ratio of the energy of the magnetic DDI to that of the electric DDI is given by
\begin{align}
\frac{\mu_0 \mu_{\rm B}^2}{(ea_0)/\epsilon_0} \sim \alpha^2 \sim 10^{-4},
\end{align}
where $\alpha=e^2/(4\pi\epsilon\hbar c)\simeq 1/137$ is the fine structure constant.

Due to the strong electric dipole moments, 
gases of polar molecules are ideal candidates for dipole-dominant systems.
The permanent dipole moments of the lowest $^{1,3}\Sigma^+$ states of heteronuclear alkali dimers are calculated to be of the order of 1 Debye ($\simeq 3.335\times 10^{-30}$~Cm)~\cite{Aymar2005}.
However, because the ground state of the molecule is rotationally symmetric and the dipole moment averages to zero,
we need to apply a strong electric field, typically of the order of $10^4$ V/cm, to polarize the dipole moments.
Several research groups are now trying to cool molecules using techniques such as buffer-gas cooling~\cite{Doyle1995, Weinstein1998, Egorov2002}
and Stark deceleration (see~\cite{Meerakker2008} for a review).
Direct laser cooling of SrF molecules was experimentally demonstrated~\cite{Shuman2010}.
Another way to achieve quantum degeneracy of heteronuclear molecules is 
to create weakly bound Feshbach molecules or photoassociative molecules from cold atoms and to transfer the molecules to the tightly bound rovibrational ground state.
To our knowledge, heteronuclear Feshbach resonance have been observed in
$^6$Li$^{23}$Na~\cite{Stan2004},
$^6$Li$^{40}$K~\cite{Wille2008},
$^6$Li$^{87}$Rb~\cite{Deh2008},
$^{40}$K$^{85}$Rb~\cite{Hodby2005},
and $^{87}$Rb$^{133}$Cs~\cite{Pilch2009},
while photoassociative molecules have been produced in
$^7$Li$^{133}$Cs~\cite{Deiglmayr2008}, 
$^6$Li$^{40}$K~\cite{Ridinger2011},
$^{23}$Na$^{133}$Cs~\cite{Haimberger2004},
$^{39}$K$^{85}$Rb~\cite{Wang2004},
$^{40}$K$^{87}$Rb~\cite{Ni2008}, 
$^{41}$K$^{87}$Rb~\cite{Aikawa2010}, 
$^{85}$Rb$^{133}$Cs~\cite{Sage2005}, 
$^{174}$Yb$^{87}$Rb~\cite{Nemitz2009},
and $^{176}$Yb$^{87}$Rb~\cite{Nemitz2009}.
Among them, the rovibrational ground state is achieved in
$^7$Li$^{133}$Cs~\cite{Deiglmayr2008},
$^{40}$K$^{87}$Rb~\cite{Ni2008}, 
$^{41}$K$^{87}$Rb~\cite{Aikawa2010},
and $^{85}$Rb$^{133}$Cs~\cite{Sage2005}.
In the quantum degenerate regime, however, 
the lifetime of the trapped $^{40}$K$^{87}$Rb (fermionic) molecules was found to be limited by atom-exchange chemical reactions~\cite{Ni2008}, whose dependence on the magnitude of the electric dipole moment was experimentally investigated in Ref.~\cite{Ni2010}. Heteronuclear molecules involving Li are also unstable against chemical reactions, whereas NaK, NaRb, NaCs, KCs, and RbCs are stable against chemical reactions. 
A strong electric DDI has also been realized in BECs with Rydberg excitations~\cite{Heidemann2008}

The magnetic DDI has been observed directly and indirectly in several systems.
Griesmaier {\it et al}. have achieved a BEC of $^{52}$Cr atoms~\cite{Griesmaier2005}.
The Cr atom has a magnetic dipole moment of $6\mu_{\rm B}$, which is six times as large as that of alkali-metal atoms, and therefore, the DDI is 36 times larger than that of alkali-metal atoms.
Moreover, by quenching the {\it s}-wave scattering length of $^{52}$Cr by means of Feshbach resonance~\cite{Werner2005}, a DDI-dominated BEC was realized in a $^{52}$Cr system in which magnetostriction~\cite{Stuhler2005,Giovanazzi2006,Lahaye2007} and anisotropic collapse~\cite{Koch2008,Lahaye2008} were observed.
On the other hand, Bismut {\it et al}. observed a modification of collective excitation frequencies due to DDIs 
in a regime where DDIs are relatively small compared with contact interactions~\cite{Bismut2010}.
Because of the large dipole moment, Cr atoms are unstable against the dipolar relaxation~\cite{Hensler2003}.
Pasquiou {\it et al}. observed that the dipolar relaxation is resonantly suppressed under a nonzero magnetic field~\cite{Pasquiou2010} and that it is enhanced in a 2D optical lattice~\cite{Pasquiou2010,Pasquiou2011a}.
Subsequently, they created a BEC of $^{52}$Cr atoms in an ambient magnetic field below 1mG,
opening up the possibility to investigate the spinor dipolar nature of the system~\cite{Pasquiou2011b,Pasquiou2012}.
Recently, BECs of $^{164}$Dy atoms~\cite{Mingwu2011} and $^{168}$Er atoms~\cite{Aikawa2012} were realized, where due to its large magnetic dipole moment ($10\mu_{\rm B}$ and $7\mu_{\rm B}$, respectively) combined with a large atomic mass, a very strong dipolar effect was observed.

A dipole-dominant BEC, albeit a very weak one, has also been realized in a BEC of $^7$Li by quenching the $s$-wave contact interaction using Feshbach resonance~\cite{Pollack2009}.
Fattori {\it et al}. observed the decoherence of the interferometer of $^{39}$K due to the dipole interaction~\cite{Fattori2008}.
Vengalattore {\it et al}. observed formation of spin domains in a BEC of $^{87}$Rb~\cite{Vengalattore2008,Vengalattore2010}, the origin of which is still under controversy (see Sec.~\ref{sec:dipole_Bogoliubov}).

\subsubsection{Tuning of dipole-dipole interaction}
\label{sec:dipole_tunability}
\begin{figure}[ht]
\begin{center}
\resizebox{0.4\hsize}{!}{\includegraphics{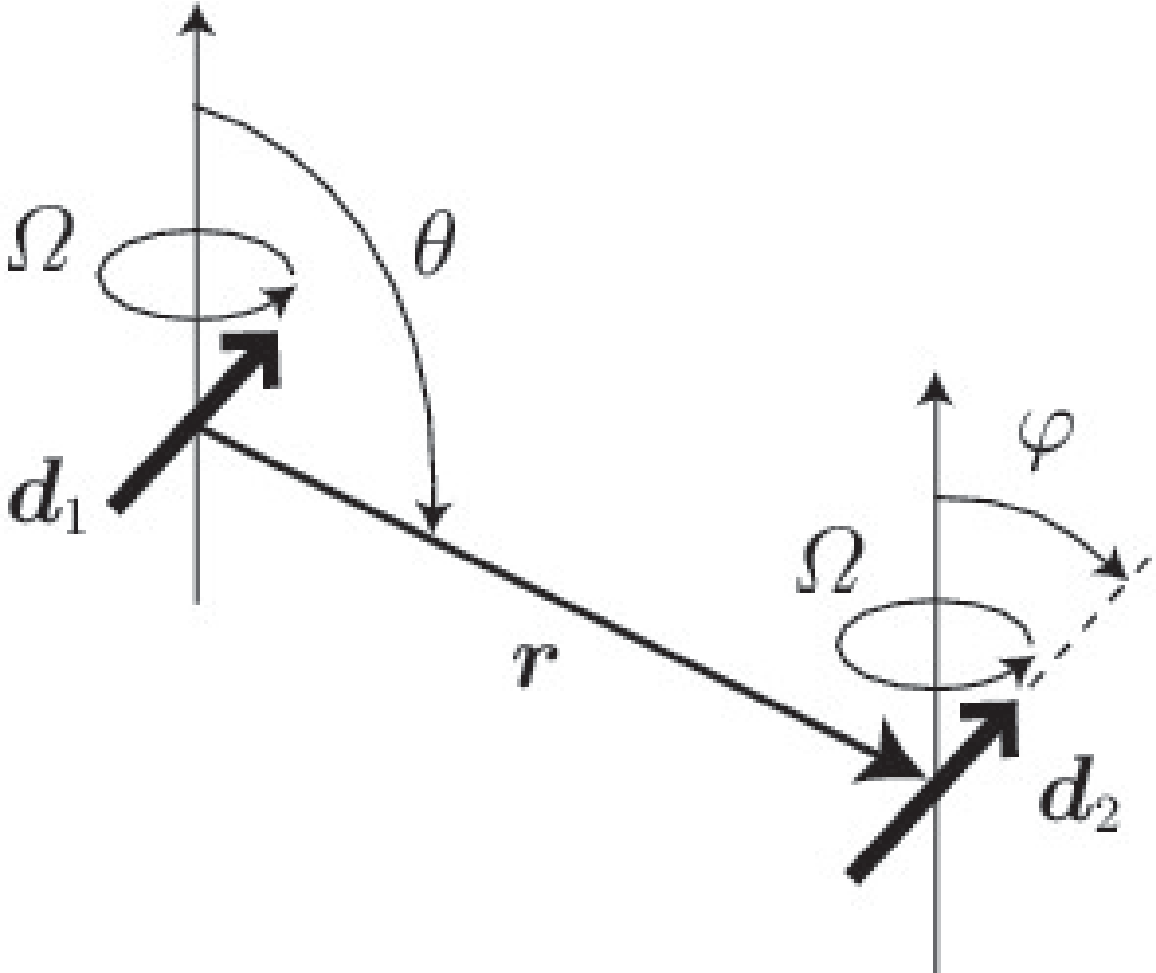}}
\end{center}
\caption{Two parallel dipoles adiabatically following an external magnetic field (\ref{B-field}) that rotates with angular frequency $\Omega$. The sign and strength of the dipole-dipole interaction (\ref{eq:dipole_tuning}) can be controlled by varying the angle $\varphi$.
}
\label{fig:dipole_tuning}
\end{figure}

Tunability of the DDI is crucial for systematically investigating the properties of dipolar BECs.
The strength of the electric dipole moments can be manipulated by an external electric field with the magnetic dipole moment kept constant.
Giovanazzi {\it et al}.~\cite{Giovanazzi2002} proposed a method to control the sign of the dipole interaction as well as its strength by using a rotating field.
This method is applicable to both electric and magnetic dipole interactions.
Consider a magnetic dipole moment under a rotating magnetic field described by
\begin{align}
{\bm B}(t)=B \begin{pmatrix} \sin\varphi\cos\Omega t \\ \sin\varphi\sin\Omega t \\ \cos\varphi\end{pmatrix},
\label{B-field}
\end{align}
where $\varphi$ is the angle between the rotating axis and the dipole moments (see Fig.~\ref{fig:dipole_tuning}).
The angular frequency $\Omega$ is set to be much smaller than the Larmor frequency so that the magnetic dipole moment adiabatically follows the direction of ${\bm B}$.
When $\Omega$ is much larger than the trap frequency, the effective dipole interaction is time-averaged over $2\pi/\Omega$.
Substituting $\hat{\bm d}={\bm B}/|{\bm B}|$ in Eq.~\eqref{eq:dipole_2body} and taking its time-average, we obtain the following effective interaction:
\begin{align}
\langle V_{\rm dd} \rangle = c_{\rm dd} \frac{1-3\cos^2\theta}{r^3}\frac{3\cos^2\varphi -1}{2}.
\label{eq:dipole_tuning}
\end{align}
The last factor on the right-hand side of Eq.~\eqref{eq:dipole_tuning} can take any value between $-1/2$ to $1$.
Hence, one can change the sign of the dipole interaction as well as its strength by varying the angle $\varphi$.

\subsubsection{Numerical method}
In the mean-field treatment,
we need to calculate the effective potential
induced by the distribution of the dipole moments ${\bm d}({\bm r})$,
which includes an integral of the form
\begin{align}
 \sum_{\nu'} \int d^3{\bm r}' Q_{\nu\nu'}({\bm r}-{\bm r}') d_{\nu'}({\bm r}').
 \label{Qint}
\end{align}
While this integral converges,
the $1/r$ divergence of the integrand makes it difficult to numerically carry out the integration in real space.
We therefore use the convolution theorem to rewrite (\ref{Qint}) in momentum space as ~\cite{Goral2002b}
\begin{align}
 \sum_{\nu'} \sum_{\bm k} \tilde{Q}_{\nu\nu'}({\bm k})\tilde{d}_{\nu'}({\bm k})e^{i{\bm k}\cdot{\bm r}},
\label{eq:dipole_convolution}
\end{align}
where $\tilde{Q}_{\nu\nu'}({\bm k})\equiv\int d{\bm r} e^{-i{\bm k}\cdot{\bm r}} Q_{\nu\nu'}({\bm r})$ and
$\tilde{d}_{\nu}({\bm k})\equiv\int d{\bm r} e^{-i{\bm k}\cdot{\bm r}} d_{\nu}({\bm r})$.
Here, the Fourier components of the kernel $Q_{\nu\nu'}$ can be found as follows.
First, we note that
the spherical harmonics is the eigenstate of the three-dimensional Fourier transformation, i.e., 
\begin{align}
 \int e^{i{\bm k}\cdot {\bm r}}Y_l^m(\hat{\bm r}) d\Omega_{\bm r}= 4\pi i^l j_l(kr)Y_l^m(\hat{\bm k}),
\end{align}
where $k=|{\bm k}|, \hat{\bm k}={\bm k}/k$, $j_l(x)$ is the $l$th spherical Bessel function and $d\Omega_{\bm r}$ denotes the integration of the angular part of ${\bm r}$.
Because the radial integration is then carried out as
\begin{align}
\int_{0}^\infty \frac{j_2(kr)}{r^3} r^2 dr = \frac{1}{3},
\label{eq:dipole_uv_cutoff}
\end{align}
we obtain
\begin{align}
\tilde{Q}_{\nu\nu'}({\bm k}) \equiv \int d{\bm r} e^{i{\bm k}\cdot{\bm r}} Q_{\nu\nu'}({\bm r}) 
= -\frac{4\pi}{3}\left(\delta_{\nu\nu'}-3\hat{k}_{\nu}\hat{k}_{\nu'}\right).
\label{eq:dipole_spinor_fourier}
\end{align}
For spin-polarized gases, we only need
\begin{align}
\tilde{Q}_{zz}({\bm k}) = -\frac{4\pi}{3}(1-3\cos^2\theta_k),
\label{eq:dipole_polarized_fourier}
\end{align}
where $\theta_k$ is the angle between the direction of the polarization and the momentum ${\bm k}$.

In the numerical calculation, we evaluate $\tilde{d}_{\nu'}({\bm k})$ from $d_{\nu'}({\bm r})$ by the standard fast Fourier transform algorithm---we 
multiply $\tilde{d}_{\nu'}({\bm k})$ by $\tilde{Q}_{\nu\nu'}({\bm k})$, sum the result over $\nu'$, and perform the inverse Fourier transform.
The result is not sensitive to the grid size $\Delta r$ in the coordinate space as long as $\tilde{d}_{\nu}(1/\Delta r)$ is negligible.
However, a small numerical error arises, depending on the system size $R$,
because the discrete Fourier transform presupposes a 3D periodic lattice with a unit cell of linear dimension $R$.
If the Fourier transform is performed in a cubic region of $R^3$,
the accuracy improves if we introduce an infrared cutoff to calculate
the Fourier transform of the integration kernel as
$\tilde{Q}_{\nu\nu'}^{\rm cut}({\bm k})=\int_{|{\bm r}|<R/2} d{\bm r}
Q_{\nu\nu'}({\bm r})e^{-i{\bm k}\cdot{\bm r}}$ \cite{Ronen2006}.

\subsection{Spin-polarized dipolar BEC}\label{sec:dipole_polarized}

When the dipole moment of every atom is polarized in the direction of an external field, say in the $z$-direction, the condensate is described with a single-component order parameter $\psi({\bm r})$ that obeys the non-local GPE:
\begin{align}
 i\hbar\frac{\partial}{\partial t} \psi({\bm r},t) 
 & =  \left[-\frac{\hbar^2}{2M}\nabla^2 + V_{\rm trap}({\bm r}) + g|\psi({\bm r},t)|^2 + \Phi_{\rm dd}({\bm r},t) \right] \psi({\bm r},t),
\label{eq:dipole_polarized_GP}
\end{align}
where $V_{\rm trap}({\bm r})$ is a trapping potential, $g=4\pi\hbar^2 a/M$ with $a$ being the scattering length for this component,
and 
\begin{align}
 \Phi_{\rm dd}({\bm r},t) = c_{\rm dd} \int d{\bm r}' \frac{1-3\cos^2\theta}{|{\bm r}-{\bm r}'|^3} |\psi({\bm r}',t)|^2
\label{eq:dipole_Phi_dd}
\end{align}
is a mean-field effective potential due to the DDI.
Here, we assume a repulsive short-range interaction $a>0$ so that the condensate is always stable in the absence of the DDI, whereas $c_{\rm dd}$ can be either positive or negative due to the tunability discussed in Sec.~\ref{sec:dipole_tunability}.

\subsubsection{Equilibrium shape and instability}\label{sec:diploe_equilibrium_shape}
Due to the anisotropy of the interaction, the equilibrium shape of a dipolar BEC is highly nontrivial~\cite{Goral2000,Martikainen2001,Ronen2006}.
The dipole interaction works attractively along the direction of dipole moments and repulsively in the perpendicular direction (see Fig.~\ref{fig:dipole_config}).
The attractive part makes a dipole-dominant BEC unstable in a homogeneous system. This instability can be qualitatively understood from the Bogoliubov spectrum in a homogeneous system~\cite{Goral2000}:
\begin{align}
E_{\bm k} = \sqrt{\epsilon_{\bm k}\left[\epsilon_{\bm k} + 2ng\left\{1+\epsilon_{\rm dd}(3\cos^2\theta_k-1) \right\}\right]},
\label{eq:dipole_polarized_bogoliubov}
\end{align}
where $n$ is the number density of atoms and 
\begin{align}
\epsilon_{\rm dd} \equiv \frac{4\pi c_{\rm dd}}{3g}
\end{align}
is a parameter that characterizes the relative strength of the DDI against the short-range interaction.
Equation~\eqref{eq:dipole_polarized_bogoliubov} shows that for $\epsilon_{\rm dd}>1$, the BEC becomes dynamically unstable because $E_{\bm k}$ is imaginary for wave numbers below $k_{\rm c}=\sqrt{4gnM(\epsilon_{\rm dd}-1)}/\hbar$.

When the system is confined in a trapping potential, the quantum pressure due to the confinement can stabilize the condensate against the attractive interaction. Santos {\it et al}.~\cite{Santos2000} numerically investigated the equilibrium shapes of pure dipolar gases with $a=0$.
They considered an axisymmetric trap with trap frequencies $\omega_r$ and $\omega_z$ in the radial and axial directions, respectively.
When the trapping potential is prolate, i.e., when the aspect ratio $l=\sqrt{\omega_r/\omega_z}>1$, the dipole interaction is always attractive.
Therefore, as in the case of attractive {\it s}-wave interaction,
there exists a critical number of atoms above which the BEC becomes unstable.
On the other hand, if the trap potential is sufficiently oblate so that $l<l^*\simeq 0.4$ and if the DDI is not too strong (see the last paragraph of this subsection), the dipole interaction is always positive
and the BEC is stable.
In the region of $l^* < l <1$, the condensate for a small number of atoms is oblate and the DDI is positive.
However, as the number of atoms increases, the aspect ratio increases due to the DDI, which turns to attraction, and finally, the condensate becomes unstable.
Yi and You~\cite{Yi2000,Yi2001} investigated the stability including the effect of repulsive short-range interaction using a Gaussian variational ansatz, 
and obtained results consistent with those of Ref.~\cite{Santos2000}.

In Refs.~\cite{Yi2001,Goral2002b}, the low-energy excitation spectrum is calculated using a time-dependent Gaussian variational method.
It is shown that the instabilities of dipolar BECs in prolate ($l>1.29$) and oblate ($l^*<l<0.75$) traps originate from dynamical instabilities of the breathing and quadrupole modes, respectively.
For an intermediate region, the lowest excitation frequency corresponds to a breathing mode far below the above mentioned critical number of atoms, and to a quadrupolar one near the critical value.

In the Thomas-Fermi limit, $gn\gg 1$, the GPE~\eqref{eq:dipole_polarized_GP} has an exact solution~\cite{ODell2004}.
Using the relationship $(\delta_{\nu\nu'}-3\hat{r}_{\nu}\hat{r}_{\nu'})/r^3 = -\nabla_{\nu}\nabla_{\nu'} (1/r)- 4\pi\delta_{\nu\nu'}\delta({\bm r})/3$,
the mean-field dipolar potential defined in Eq.~\eqref{eq:dipole_Phi_dd} can be rewritten as
\begin{align}
\Phi_{\rm dd}({\bm r}) &= -c_{\rm dd}\left[ \frac{\partial^2}{\partial z^2} \phi({\bm r}) + \frac{4\pi}{3}n({\bm r}) \right],
\end{align}
where $\phi({\bm r}) \equiv \int d{\bm r}' n({\bm r}')/|{\bm r}-{\bm r}'|$, and therefore, $\phi({\bm r})$ obeys Poisson's equation: $\nabla^2\phi = -4\pi n({\bm r})$.
Because $n({\bm r})=|\psi(\bm r)|^2$ is parabolic in the Thomas-Fermi limit, Poisson's equation is satisfied by 
$\phi=a_0 + a_1 x^2 + a_2 y^2 + a_3 z^2 + a_4 x^2 y^2 + a_5 y^2 z^2 + a_6 z^2 x^2 + a_7 x^4 + a_8y^4 + a_9 z^4$, and hence, $\Phi_{\rm dd}$ is also quadratic.
Therefore, for a harmonically trapped BEC, all terms appearing in the GPE~\eqref{eq:dipole_polarized_GP} are quadratic, and as in the simple {\it s}-wave case, an inverted parabola for the density profile gives a self-consistent solution of the dipolar hydrodynamic equations. (For an analytic form of the solution, see Refs.~\cite{ODell2004, Eberlein2005}.)

Eberlein {\it et al}.~\cite{Eberlein2005} pointed out that the potential seen by atoms exhibits a local minimum immediately outside the condensate, i.e., the sum of the trapping and dipole potentials is locally smaller than the chemical potential, causing an instability that brings atoms out from the central part of the condensate to fill the dip in the potential.
The presence of this instability leads to a biconcave shape of a condensate pointed out by Ronen {\it et al}.~\cite{Ronen2007a}.
By developing a high-precision numerical code for dipolar BECs, Ronen {\it et al}.~\cite{Ronen2007a} have shown that the same instability causes a dipolar condensate to become unstable in the limit of strong dipole interaction, even though the trap is strongly oblate, in disagreement with the result of Refs.~\cite{Santos2000, Yi2000,Yi2001}.
They have also shown that the instability of a strongly oblate dipolar BEC is 
analogous to the ``roton-maxon'' instability reported for 2D dipolar gases (see Sec.~\ref{sec:dipole_roton_maxon}).
Interestingly, in a strongly oblate trap, the condensate just before collapse becomes biconcave, with its maximum density away from the center of the gas. 
Such biconcave-shaped condensates were also shown to be stabe at finite temperature~\cite{Ronen2007b}.

\subsubsection{Dipolar collapse}\label{sec:dipole_collapse}
Lahaye {\it et al}. observed magnetostriction of a $^{52}$Cr BEC, where $\epsilon_{\rm dd}=0.16$ for the background scattering length.
By reducing the {\it s}-wave scattering length using a Feshbach resonance,
they observed a change in the aspect ratio of the condensate~\cite{Lahaye2007}.

They also observed a collapse dynamics due to the DDI~\cite{Lahaye2008}.
In the experiment, the $s$-wave scattering length was adiabatically decreased to 30\% of the background value, and  it was then suddenly decreased below the critical value of the collapse. After a certain hold time, the BEC was released from the trap and the TOF image was taken after expansion.
	
The collapse dynamics proceeds as follows.
After a sudden decrease in the scattering length, atoms begin to gather at the center of the trap due to the attractive interaction; 
with an increase in the atomic density at the trap center, the three-body recombination losses become predominant; as a consequence of the atom loss, the attractive interaction weakens, and the atoms are ejected outward due to the quantum pressure. The crucial observation here is that due to the {\it d}-wave nature of the DDI, the atomic ejection is expected to be not spherical but highly anisotropic as experimentally observed and numerically demonstrated~\cite{Lahaye2008}.

\begin{figure}[ht]
\begin{center}
\resizebox{0.7\hsize}{!}{\includegraphics{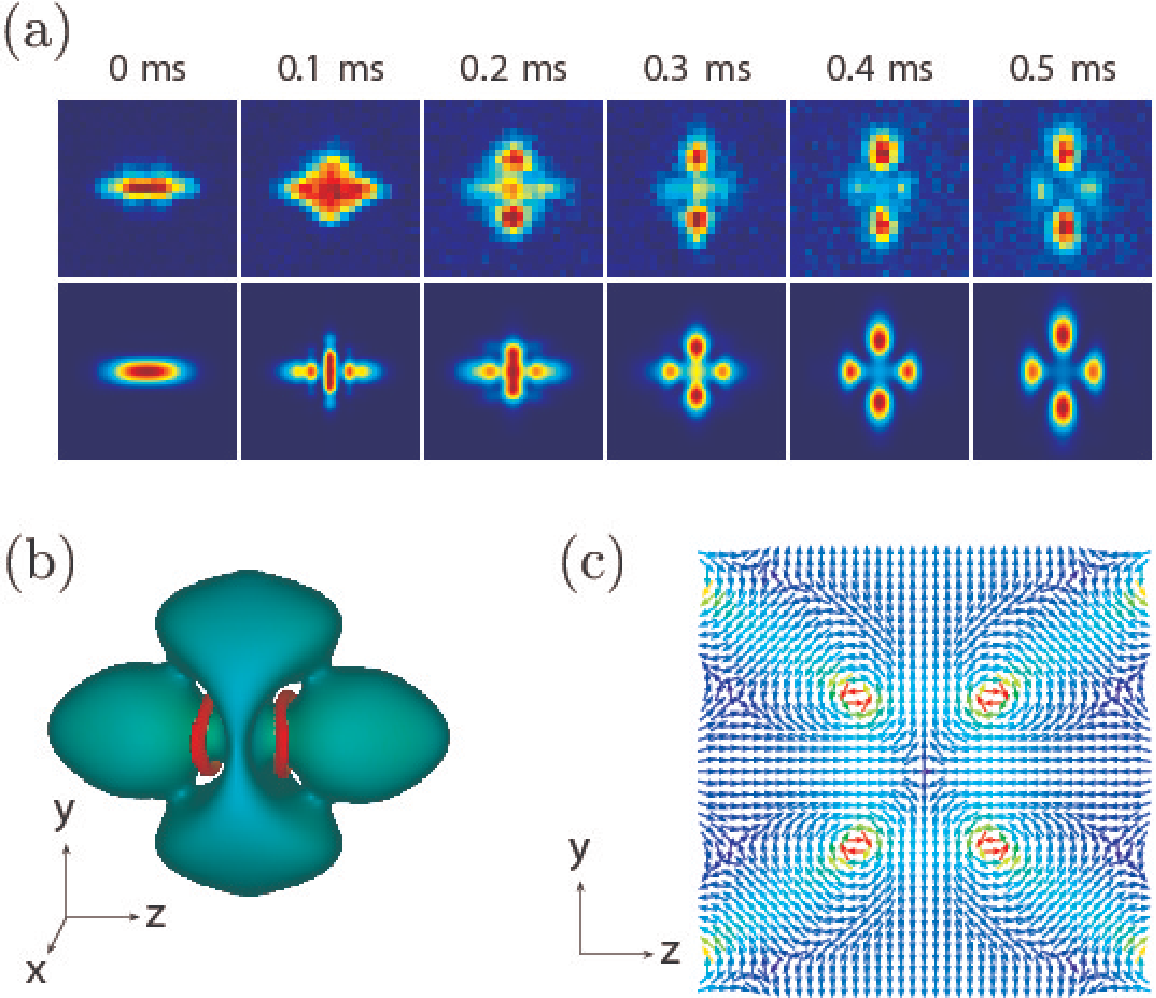}}
\end{center}
\caption{
(a) Snapshots of absorption images of the collapsing condensates for different values of the hold time (top) and the corresponding results of the numerical simulations obtained without adjustable parameters (bottom).
(b) Isodensity surface of an in-trap condensate. The locations of a pair of vortex rings are indicated by the red rings.
(c) Velocity field of the atomic flow in the $x = 0$ plane. The red color shows the region where the velocity field is larger.
Reprinted from Ref.~\cite{Lahaye2008}.
}
\label{fig:dipole_collapse}
\end{figure}

Figure~\ref{fig:dipole_collapse} (a) shows the measured TOF image (upper panels) together with the results of numerical simulations (lower ones).
We have numerically solved the GPE that involves the three-body loss with the loss rate coefficient determined so as to best fit the measured loss curve (Fig.~3 of Ref.~\cite{Lahaye2008}).
The excellent agreement between the experiment and the theory demonstrates the validity of the mean-field description even for dipole-dominant BECs.
Moreover, the numerical simulation revealed that the cloverleaf pattern in the TOF image arises due to the creation of a pair of vortex rings, as indicated in Fig.~\ref{fig:dipole_collapse} (b).
When collapse occurs, the atoms were ejected in the $x$-$y$ plane (vertical direction to the dipole moments), whereas atoms still flow inward along the $z$ direction, giving rise to the circulation.
The velocity field shown in Fig.~\ref{fig:dipole_collapse} (c) clearly shows the {\it d}-wave nature of the collapse dynamics.

\subsubsection{Roton-maxon excitation}
\label{sec:dipole_roton_maxon}
Santos {\it et al}.~\cite{Santos2003} have shown that the Bogoliubov excitation spectrum exhibits roton-maxon behavior, i.e., the excitation energy has a local maximum and minimum as a function of the momentum $k$, in a system that is harmonically confined in the direction of the dipole moments (i.e., the $z$ direction) and free in the $x$ and $y$ directions.
If the in-plane momentum $k$ are much smaller than the inverse size $L^{-1}$ of the condensate in the $z$ direction, excitations have a 2D character.
Because the dipoles are perpendicular to the plane of the trap, particles efficiently repel each other and the in-plane excitations are phonons.
Then, the DDI increases the sound velocity.
For $k\gg 2\pi/L$, excitations acquire a 3D character and the interparticle repulsion is reduced due to the attractive force in the $z$ direction.
This decreases the excitation energy with an increase in $k$.
When $\epsilon_{\rm dd}>1$, the excitation energy reaches a minimum (roton) and then begins increasing, and the nature of the excitations continuously become single-particle like.
As the dipole interaction becomes stronger, the energy at the roton minimum decreases and as it touches zero, an instability arises.
The 3D character is essential for the appearance of the roton minimum, which does not appear in the quasi-2D system where the confining potential in the $z$ direction is strong and the BEC has no degrees of freedom in this direction~\cite{Fischer2006}. The roton-maxon spectrum is also predicted for a 1D system with laser-induced DDI~\cite{ODell2003}.

\subsubsection{Two-dimensional solitons}
It is known that a nonlinear Schr\"{o}dinger equation with short-range interactions admits stable soliton solutions in one dimension, but it does not in higher dimensions.
However, Pedri and Santos ~\cite{Pedri2005} showed that with a nonlocal DDI, a 2D BEC can have a stable soliton.

We consider a dipolar BEC polarized in the $z$ direction, and assume that the system is confined only in the $z$ direction by a harmonic potential $V_{\rm trap}({\bm r})=\frac{1}{2}M\omega_z^2 z^2$, and that it is free in the $x$-$y$ plane.
Then, the DDI is isotropic in the $x$-$y$ plane.
Let us consider a Gaussian variational ansatz
\begin{align}
 \psi(x,y,z) = \frac{N^{1/2}}{\pi^{3/4}l_0^{3/2}L_\rho L_z^{1/2}} \exp\left[-\frac{x^2+y^2}{2l_0^2L_\rho^2}-\frac{z^2}{2l_0^2L_z^2}\right],
\label{eq:dipole_gaussian_ansatz}
\end{align}
where $l_0\equiv\sqrt{\hbar/(M\omega_z)}$ and $L_\rho$ and $L_z$ are dimensionless variational parameters that characterize the widths in the $x$-$y$ plane and the $z$ direction, respectively,
and $N$ is the total number of atoms.
Using this ansatz, the mean-field energy is evaluated to give
\begin{align}
E(L_\rho,L_z) = \frac{N\hbar\omega_z}{2}\left(\frac{1}{L_\rho^2} + \frac{1}{2L_z^2} + \frac{L_z^2}{2}\right)
+ \frac{1}{\sqrt{2\pi}L_\rho^2 L_z l_0^3}\frac{gN^2}{4\pi}\left[1+\epsilon_{\rm dd}f(\kappa)\right],
\end{align}
where $\kappa\equiv L_\rho/L_z$ is the aspect ratio and
\begin{align}
f(\kappa) &\equiv \frac{2\kappa^2+1}{\kappa^2-1} - \frac{3\kappa^2}{(\kappa^2-1)} \frac{\arctan \sqrt{\kappa^2-1}}{\sqrt{\kappa^2-1}}.
\end{align}
For fixed $L_z$ and in the absence of the DDI ($\epsilon_{\rm dd}=0$),
$E(L_\rho,L_z)$ monotonically increases or decreases as a function of $L_\rho$,
resulting in collapse or expansion. (Here, we assume that $g$ can take both positive and negative values.)
However, in the presence of the DDI, $f(\kappa)$ is a monotonically increasing function of $\kappa$ with $f(0)=-1$ and $f(\kappa\to \infty)=2$, and therefore $E(L_\rho, L_z)$ may have a minimum.
When the trapping potential is strong and $L_z=1$, for simplicity, the $L_\rho$-dependence of the mean-field energy is given by
\begin{align}
E(L_\rho,1) = \frac{N\hbar\omega_z}{2} \frac{1}{L_\rho^2}\left[1+\tilde{g}+\tilde{g}\epsilon_{\rm dd}f(L_\rho)\right] + {\rm const.},
\end{align}
where $\tilde{g}\equiv gN/[(\sqrt{2\pi}l_0)^3\hbar\omega_z]=\sqrt{2/\pi}\, aN/l_0$ with $a$ being the {\it s}-wave scattering length.
This function has a minimum when $1+\tilde{g}+\tilde{g}\epsilon_{\rm dd}f(0)>0$ and $1+\tilde{g}+\tilde{g}\epsilon_{\rm dd}f(\infty)<0$, that is,
\begin{align}
\epsilon_{\rm dd}<1+\sqrt{\frac{\pi}{2}}\frac{l_0}{aN}<-2\epsilon_{\rm dd}\ \ \ \textrm{for}\ \ a>0, \label{eq:dipole_2Dsoliton1}\\
\epsilon_{\rm dd}>1+\sqrt{\frac{\pi}{2}}\frac{l_0}{aN}>-2\epsilon_{\rm dd}\ \ \ \textrm{for}\ \ a<0. \label{eq:dipole_2Dsoliton2}
\end{align}
The condition~\eqref{eq:dipole_2Dsoliton1} holds only for $\epsilon_{\rm dd}<0$, which can be achieved using a rotational field.
In this case, the repulsive contact interaction and attractive DDI (in the 2D plane) balance and stabilize 2D solitons.
On the other hand, when $a<0$, the condition~\eqref{eq:dipole_2Dsoliton2} requires $\epsilon_{\rm dd}>0$, which means that the repulsive DDI balances the attractive contact interaction and stabilizes 2D solitons.
The case in which dipole moments are polarized in the 2D plane is discussed in Ref.~\cite{Tikhonenkov2008}.
In this case, stable 2D soliton waves can be stabilized for $a>0$ and $\epsilon_{\rm dd}>0$,
and the solitons are anisotropic and elongated along the direction of polarization.

Because a 2D bright soliton in a dipolar gas is stable, the roton instability discussed in the preceding subsection does not cause a collapse of the BEC but creates 2D solitons~\cite{Nath2009}.
If the dipole moments are polarized perpendicular to the 2D plane, these solitons are stable as long as the gas remains 2D. However, if the dipoles are parallel to the 2D plane, the (anisotropic) solitons may become unstable even in 2D if the number of particles per soliton exceeds a critical value.

\subsubsection{Supersolid}
\label{sec:supersolid}
The Bose-Hubbard model with long-range interactions exhibits a rich variety of phases such as the density wave (DW), supersolid (SS), and superfluid (SF) phases~\cite{Matsuda1970, Liu1973, Bruder1993, Otterlo1994}.
While SF and SS phases have a nonzero superfluid density,
DW and SS phases have a nontrivial crystalline order in which the particle density modulates with a periodicity that is different from that of the external potential. Thus, in the SS phase, both diagonal and off-diagonal long-range orders coexist~\cite{Andreev1969, Chester1970, Legget1970}.

A dipolar gas in an optical lattice is an ideal system to realize such exotic phases, because the interaction parameters can be experimentally controlled.
G\'{o}ral {\it et al}.~\cite{Goral2002a} showed that DW and SS phases, as well as SF and Mott insulator (MI) phases, 
can be accomplished in dipolar gases in a 2D optical lattice,
and Yi {\it et al}.~\cite{Yi2007} discussed detailed phase diagrams in 2D and 3D optical lattices.
A dipolar gas in an optical lattice can be described with an extended Bose-Hubbard model given by
\begin{align}
 H_{\rm EBH} = -t\sum_{\langle i,j\rangle} (\hat{b}_i^+ \hat{b}_j + \hat{b}_j^+\hat{b}_i) + \frac{1}{2} U_0 \sum_i \hat{n}_i(\hat{n}_i-1)
 + \frac{1}{2} \sum_i U^{ii}_{\rm dd} \hat{n}_i(\hat{n}_i-1) + \frac{1}{2}\sum_{i\neq j} U^{ij}_{\rm dd}  \hat{n}_i\hat{n}_j,
\label{eq:dipole_EBH}
\end{align}
where $\hat{b}_i$ is the annihilation operator of a particle at the lattice site $i$, $\hat{n}_i=\hat{b}_i^+ \hat{b}_i$ is the corresponding particle-number operator, $t$ is the hopping matrix element between the nearest neighbors, and $U_0>0$ is the on-site Hubbard repulsion due to the {\it s}-wave scattering.
Here, $t$ and $U_0$ can be expressed in terms of the Wannier function $w({\bm r}-{\bm r}_i)$ of the lowest energy band
as $U_0=4\pi a\hbar^2/M\int d{\bm r} |w({\bm r})|^4$ and $t=\int d{\bm r} w^*({\bm r}-{\bm r}_i)[-\frac{\hbar^2}{2M}\nabla^2+V_0({\bm r})]w({\bm r}-{\bm r}_j)$,
where $V_0$ is the optical lattice potential~\cite{Jaksch1998}. 
The last two terms in Eq.~\eqref{eq:dipole_EBH} describe the on-side and inter-site DDIs with coupling parameters given by
\begin{align}
 U^{ij}_{\rm dd} &= c_{\rm dd}\int d{\bm r} \int d{\bm r}'|w({\bm r}-{\bm r}_i)|^2
 \frac{1-3\cos^2\theta}{|{\bm r}-{\bm r}'|^3}  |w({\bm r}'-{\bm r}_j)|^2,
\end{align}
where $\theta$ is the angle between the dipole moment and vector ${\bm r}-{\bm r}_i$.
Here, $c_{\rm dd}$ can take both positive and negative values by using a fast rotating field (see Sec.~\ref{sec:dipole_tunability}).
Furthermore, $t$, $U_0$, and $\{U_{\rm dd}^{ij}\}$ can be tuned independently by changing the depth of an optical lattice, the longitudinal confinement (for a 2D lattice), and the orientation of the dipoles.

Let us begin by reviewing the case of $U_{\rm dd}^{ij}=0$, i.e., the on-site Bose-Hubbard model.
At $t=0$ and commensurate filling, i.e., the average number of particles per site is an integer $n$, the interaction energy is minimized by populating every lattice site with exactly $n$ atoms; therefore, the Mott insulator (MI) phase is realized. The energy cost required to create a particle-hole excitation in the MI phase is $U_0$ (on-site interaction energy),
and therefore, the MI state is gapped and incompressible.
On the other hand, for nonzero $t$, the kinetic energy favors particle hopping.
When $t$ is sufficiently large to overcome the interaction-energy cost ($\sim U_0$), the system undergoes a phase transition to the superfluid (SF) phase that is gapless and compressible.

In the presence of the long-range interaction,
the ground state of the extended Bose-Hubbard Hamiltonian~\eqref{eq:dipole_EBH} in a 2D optical lattice is investigated in Refs.~\cite{Goral2002a, Yi2007}
using a variational approach based on the Gutzwiller ansatz $|\Psi\rangle = \Pi_i \sum_{n=0}^\infty f^i_n|n\rangle_i$, 
where $|n\rangle_i$ denotes the state with $n$ particles at lattice site $i$.
The coefficients $f^i_n$ can be found by minimizing the expectation value 
$\langle \Psi|H_{\rm EBH}-\mu \sum_i n_i|\Psi\rangle$ under the constraint of a fixed chemical potential $\mu$.

When dipole moments are polarized perpendicular to the 2D lattice plane,
the dipole interaction is repulsive (attractive) for $c_{\rm dd}>0$ ($c_{\rm dd}<0$).
If site $i$ is occupied, it is energetically favorable
that its neighboring sites are equally populated for negative $c_{\rm dd}$ and less populated for positive $c_{\rm dd}$.
Therefore, for negative $c_{\rm dd}$, only MI and SF phases can appear as in the case of the on-site Bose-Hubbard model,
and a local collapse occurs for large $|c_{\rm dd}|$ due to the strong attractive interaction.
On the other hand, for positive $c_{\rm dd}$, the DW phase emerges as an insulating state.
According to the Landau theory of the second-order phase transition, phases with distinct symmetry breaking patterns cannot be continuously connected with each other.
Therefore, as the hopping increases, the DW state first changes to the SS state which then turns to the SF phase.

When dipole moments are polarized along the 2D lattice, say in the $y$-axis, the DDI is anisotropic.
The inter-site interaction for $c_{\rm dd}>0$ ($c_{\rm dd}<0$) is attractive (repulsive) in the $y$ direction and repulsive (attractive) in the $x$ direction.
In this case, the particle density along the repulsive direction becomes periodically modulated, whereas it remains constant along the attractive direction, resulting in striped DW and striped SS states.
A local collapse occurs for large $|c_{\rm dd}|$ for either sign of $c_{\rm dd}$.

In Ref.~\cite{Yi2007}, the phase diagram in a 3D lattice is also investigated, 
where dipole moments are polarized along the $z$-axis.
For $c_{\rm dd}>0$, the DDI is attractive in the $z$ direction and repulsive in the $x$-$y$ direction.
Hence, the DW phase forms a checkerboard-type pattern on the $x$-$y$ plane and it is uniform in the $z$ direction.
The density pattern in the SS phase reflects the same symmetry as the DW phase.
On the other hand, for $c_{\rm dd}<0$, 
layers of high-density sites align along the $z$ direction due to the repulsive interaction.
Hence, layered DW and SS phases emerge perpendicular to the $z$-axis.

Menotti {\it et al}.~\cite{Menotti2007} studied a polarized dipolar gas in a 2D optical lattice beyond the ground state.
They have shown that there exist many metastable states in the insulator region.
The number of metastable states and pattern variations increases rapidly with the number of lattice sites.
Metastable states also appear in magnetic domains in solid-state ferromagnets, classical ferrofulids, and quantum ferrofluids, an analogue of which is discussed in the next subsection.

\subsubsection{Ferrofluid}

\begin{figure}[ht]
\begin{center}
\resizebox{0.5\hsize}{!}{\includegraphics{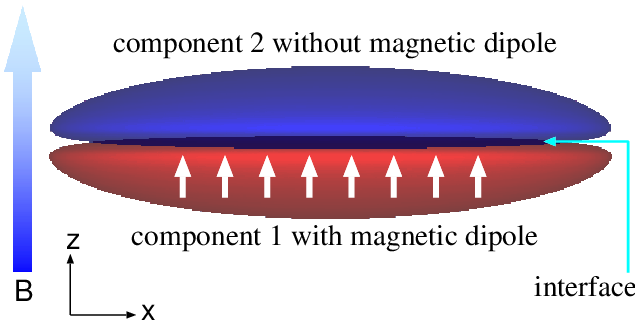}}
\end{center}
\caption{Schematic illustration of a two-component BEC system in which atoms in component 1 have parallel magnetic dipole moments and those in component 2 are nonmagnetic. 
The magnetic dipoles are polarized in the $z$ direction by an external magnetic field.
The two components phase-separate due to a field gradient $B'(z)<0$.
Reprinted from Ref.~\cite{Saito2009}.
}
\label{fig:dipole_ffsystem}
\end{figure}

In classical ferrofluids, that is, colloidal liquids made of nanoscale ferromagnetic particles suspended in a carrier fluid, 
the strong magnetic DDIs modulate the surface of the fluids to form a regular hexagonal pattern of peaks and valleys, which is called a Rosensweig pattern.
A similar phenomenon is expected to occur also in spin-polarized dipolar BECs, i.e., a quantum version of the Rosensweig pattern.
However, as we have discussed in Sec.~\ref{sec:diploe_equilibrium_shape}, 
the density profile of a BEC becomes biconcave just before the collapse,
and the condensate cannot sustain the short-wavelength density modulations.
To prevent the condensate from collapsing, we here consider a repulsively interacting two-component BEC, in which atoms in component 1 have parallel magnetic dipole moments, whereas those in component 2 are nonmagnetic.
The $m=-3$ and 0 states of $^{52}$Cr can be a candidate for such a binary system.
We apply a magnetic-field gradient in the $z$ direction, so that the two components phase-separate as shown in Fig.~\ref{fig:dipole_ffsystem}. 

\begin{figure}[ht]
\begin{center}
\resizebox{0.6\hsize}{!}{\includegraphics{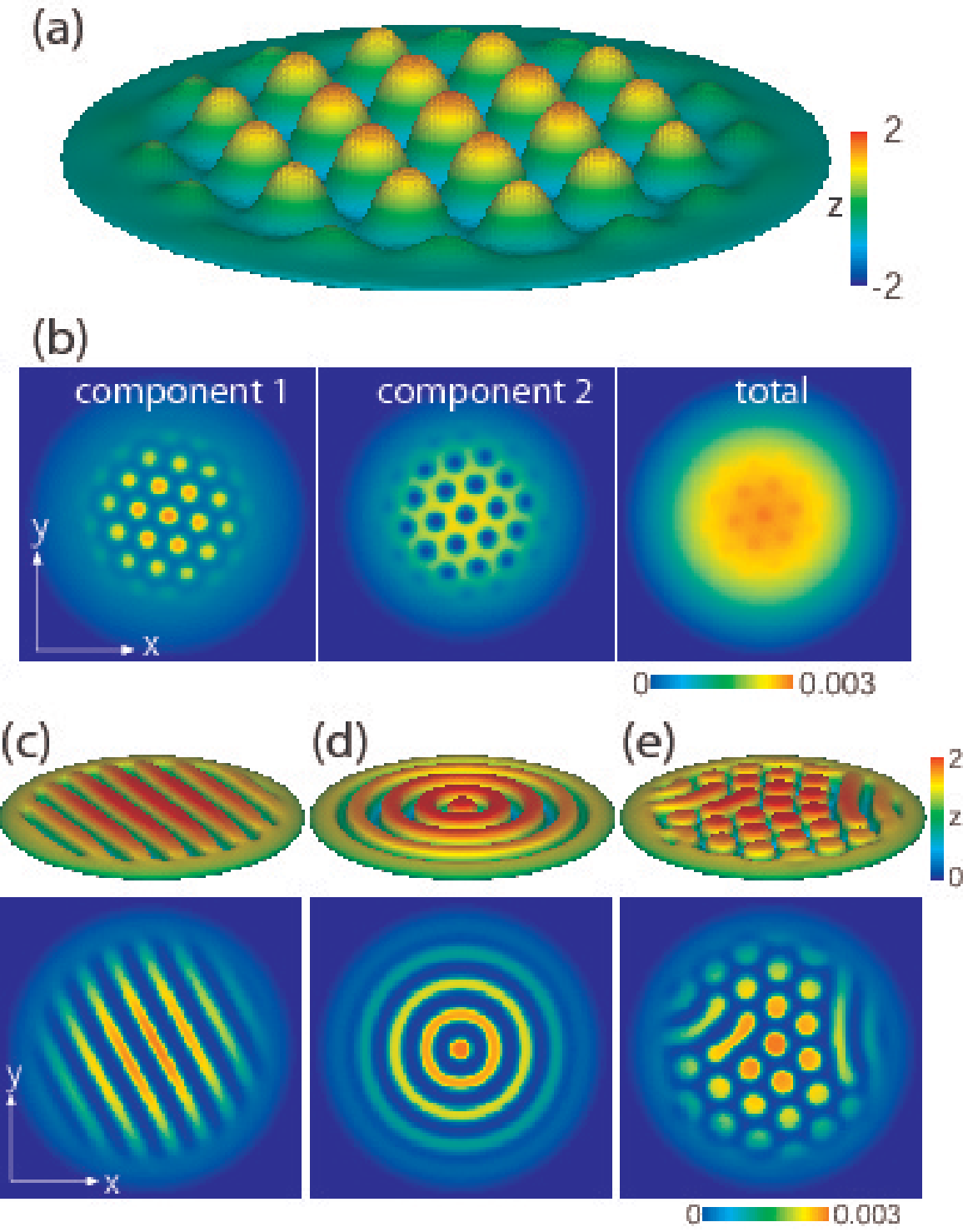}}
\end{center}
\caption{
(a) Isodensity surface of a hexagonal pattern formed in component 1 at $B'=-1$~G/cm.
(b) Column densities for component 1 (left), component 2 (center), and the total system (right).
(c)--(e) Metastable configurations found at $B'=-0.3$~G/cm, where the isodensity surface (top) and column density (bottom) for component 1 are shown.
Reprinted from Ref.~\cite{Saito2009}.
}
\label{fig:dipole_ffpattern}
\end{figure}

The equilibrium configuration of this system can be found by solving the two-component GPE in imaginary time~\cite{Saito2009}.
We find a stable hexagonal pattern, which is called a Rosensweig pattern, of the density distribution of component 1 as shown in Figs.~\ref{fig:dipole_ffpattern} (a) and (b).
We also find several metastable states starting from different initial states, such as stripe, concentric, and deformed hexagonal patterns,
as shown in Figs.~\ref{fig:dipole_ffpattern} (c)--(e).
The energies for these patterns are almost degenerate; however, each pattern is robust against small perturbations.

A striking difference between the present system and magnetic liquids (classical ferrofluids) is that the present system is a superfluid that supports a persistent current.
We also find a metastable state with a nonzero circulation in component 1 and zero circulation in component 2 in a stationary trap.
Remarkably, the Rosensweig pattern emerges even in the presence of a superflow, indicating that the system simultaneously exhibits a diagonal and off-diagonal long-range order, although the latter is embedded as a fundamental postulate of the GP theory.

\subsection{Spinor dipolar BEC}
\label{sec:dipole_spinor}
Let us now consider a dipolar BEC with spin degrees of freedom. By ``spinor dipolar,'' we imply that the direction of the atomic spin or the magnetization of the condensate is not polarized by an external field but can vary in space.
In such a situation, the system develops a nontrivial spin texture due to the DDI and self-organized magnetic patters may emerge.
The length scale of the dipole-induced magnetic structure is on the order of the dipole healing length:
\begin{align}
 \xi_{\rm dd}=\frac{\hbar}{\sqrt{2Mc_{\rm dd}n}}.
\label{eq:Dipole_def_xidd}
\end{align}
Different from spin-polarized BECs, in which DDIs are detectable in experiments only when it is stronger than or comparable to the density-density contact interaction ($c_0$ term),
DDIs in spinor BECs can induce a significant change if only it is comparable with the spin-dependent interactions ($c_1$, $c_2$, and $c_3$ terms), which is usually much weaker than the density-density contact interaction.
In particular, when the contact interaction favors the ferromagnetic state, the DDI is compatible with the ferromagnetic interaction, and the spin texture can be formed even when DDI is weaker than the spin-dependent interactions
if the system size is comparable or larger than the diple healing length $\xi_{\rm dd}$~\cite{Kawaguchi2006b}.

In the second-quantized form, the magnetic DDI for a spinor BEC is given by
\begin{align}
 \hat{V}_{\rm dd} = \frac{c_{\rm dd}}{2} \int d{\bm r}\int d{\bm r}'
\sum_{\nu\nu'} :\hat{F}_{\nu}({\bm r}) Q_{\nu\nu'}({\bm r}-{\bm r}') \hat{F}_{\nu'}({\bm r}'):,
\label{eq:dipole_2quantization}
\end{align}
where $\hat{F}_{\nu}({\bm r})$ is defined by Eq.~\eqref{spindensity} and
$c_{\rm dd} = \mu_0(g \mu_{\rm B})^2/(4\pi)$, with $g$ being the Land\'e g-factor for the atom.
The dipole interaction yields a non-local term in the GPEs. In fact, the GPEs in the presence of the DDIs for a spin-1 BEC [Eq.~\eqref{spin-1GPE}] are given by
\begin{align}
i\hbar \frac{\partial \psi_m}{\partial t}
=& \left[ -\frac{\hbar^2 \nabla^2}{2M} + U_{\rm trap}({\bm r}) - pm + qm^2 \right] \psi_m
\nonumber \\
& + c_0 n \psi_m 
+ c_1 \sum_{m'=-1}^1 {\bm F} \cdot {\bf f}_{mm'} \psi_{m'}
+ c_{\rm dd} \sum_{m'=-1}^1 {\bm b} \cdot {\bf f}_{mm'} \psi_{m'},
\label{spin-1GPE-DD}
\end{align}
and those for spin-2 BEC [Eqs.~\eqref{f=2GPE2}--\eqref{f=2GPE0}] are
given by
\begin{align}
i\hbar \frac{\partial \psi_m}{\partial t}
=& \left[ -\frac{\hbar^2 \nabla^2}{2M} + U_{\rm trap}({\bm r}) - pm + qm^2 \right] \psi_m
\nonumber \\
& + c_0 n \psi_m + c_1 \sum_{m'=-2}^2
{\bm F} \cdot {\bf f}_{mm'} \psi_{m'}  + \frac{c_2}{\sqrt{5}}A \psi_{-m}^\ast
+ c_{\rm dd} \sum_{m'=-2}^2 {\bm b} \cdot {\bf f}_{mm'} \psi_{m'}
\label{f=2GPE-DD}
\end{align}
Here, ${\bm b}$ is the effective dipole field defined by
\begin{align}
b_{\nu} \equiv \int d{\bm r}'
\sum_{\nu\nu'} Q_{\nu\nu'}({\bm r}-{\bm r}') F_{\nu'}({\bm r}'),
\label{eq:effective_dipole_field}
\end{align}
which is nonlocal, and ${\bm B}_{\rm dd}=c_{\rm dd}{\bm b}/(g\mu_B)$
works as an effective magnetic field.
Early studies on spin ordering and spin waves in an array of spinor BECs trapped in deep 1D and 2D optical lattices were made in Refs.~\cite{Pu2001,Gross2002,Zhang2002}.

\subsubsection{Einstein-de Haas effect}

We first consider the dynamics of a spinor dipolar BEC at zero external magnetic field.
Suppose that we prepare a spin polarized BEC in an external field. We consider what happens if we suddenly turn off the external field.
In the absence of the DDI, a spin polarized BEC is stable because the total spin of the system will then be conserved.
However, in the presence of the DDI, the situation drastically changes because the DDI does not conserve the spin angular momentum; it only conserves the total (i.e., spin plus orbital) angular momentum.
This can be understood from the fact that the DDI [see Eq.~\eqref{eq:dipole_2body}] is not invariant under a rotation in spin space due to the spin-orbit coupling term $({\bm d}\cdot{\bm r})$.
On the other hand, because $V_{\rm dd}({\bm r})$ is invariant under simultaneous rotations in spin and coordinate spaces, the DDI conserves the total angular momentum of the system.
Therefore, the DDI causes spin relaxation to occur by transferring the angular momentum from the spin to the orbital part; as a consequence, the BEC starts to rotate as the spin angular momentum relaxes.
This is the Einstein-de Haas (EdH) effect in atomic BECs.

\begin{figure}[ht]
\begin{center}
\resizebox{0.6\hsize}{!}{\includegraphics{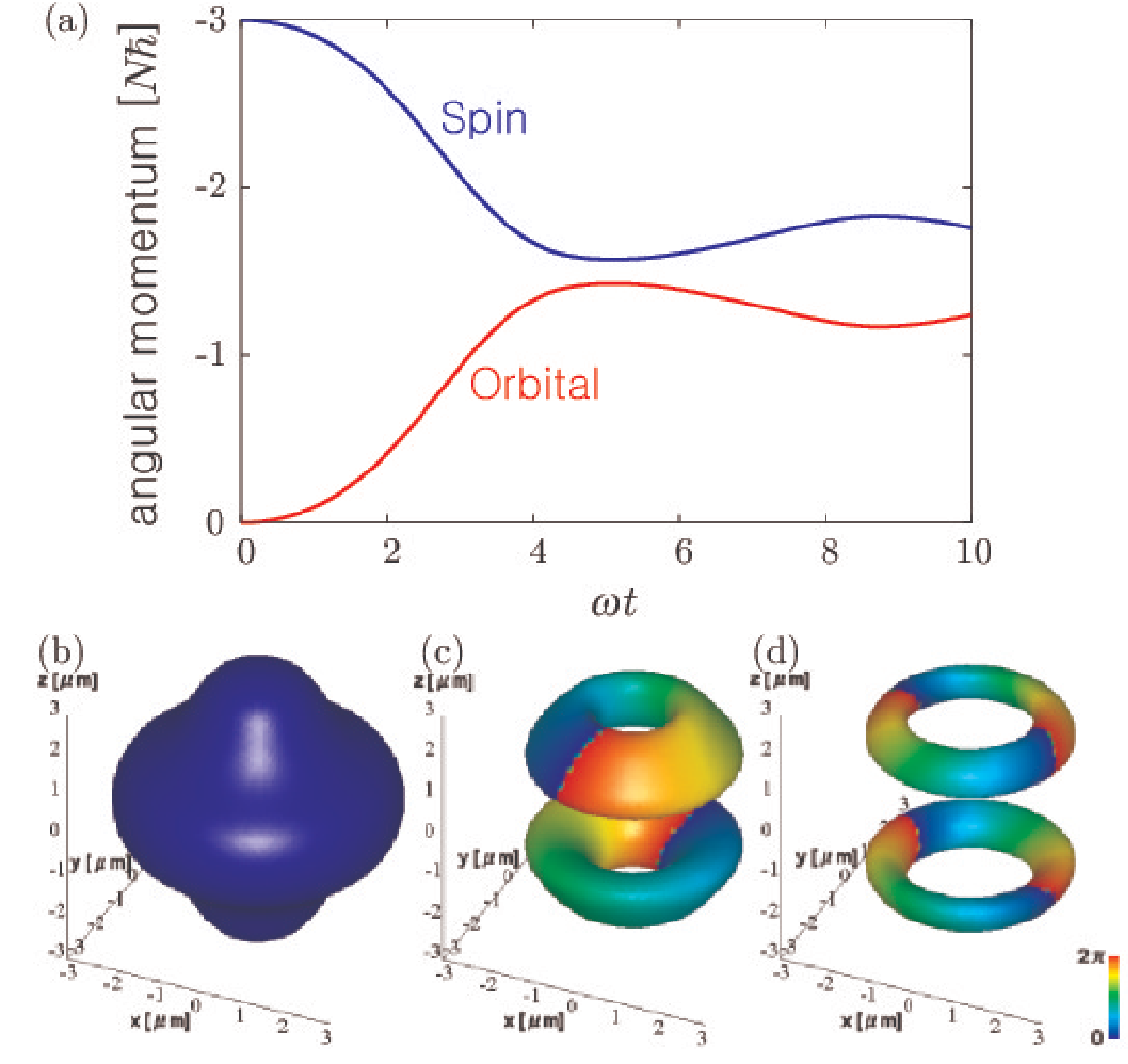}}
\end{center}
\caption{(a) Time evolution of spin and orbital angular momenta due to the EdH effect,
calculated for a $^{52}$Cr BEC in a spherical trap with trap frequency $\omega=2\pi \times 820~{\rm Hz}$.
(b)--(d) Isopycnic surfaces of (b) $\psi_{-3}$, (c) $\psi_{-2}$, and (d) $\psi_{-1}$ at $\omega t=2$,
where the color on the surface represents the phase of the order parameter.
Reprinted from Ref.~\cite{Kawaguchi2006a}.
}
\label{fig:dipole_EdH_AM}
\end{figure}

Such dynamics are discussed in Refs.~\cite{Kawaguchi2006a,Santos2006} for a BEC of $^{52}$Cr atoms.
Figure~\ref{fig:dipole_EdH_AM} (a) shows the time evolution of the spin angular momentum $M_z = \int d {\bm r} \sum_{m=-3}^3 m|\psi_m({\bm r})|^2$
and the orbital angular momentum $L_z = \int d {\bm r} \sum_{m=-3}^3 \psi_m^*({\bm r})\left(-i\frac{\partial}{\partial \varphi}\right) \psi_m({\bm r})$,
where $\varphi$ is the azimuthal angle around the $z$ axis.
The initial state is assumed to be spin-polarized in the magnetic sublevel $m=-3$.
As time progresses, the spin angular momentum decreases with a concomitant increase in the orbital angular momentum, so that the total angular momentum of the system is conserved.

In the dynamics, the order parameter is given by
\begin{align}
\psi_m(r,\varphi,z) = e^{i(J_z-m)\varphi}\eta_m(r,z),
\label{eq:dipole_EdH_OP}
\end{align}
where $(r,\varphi,z)$ are the cylindrical coordinates, $\eta_m$ is a complex function of $r$ and $z$, and
$J_z$ is the projected total angular momentum on the $z$ axis ($J_z=-3$ in the present case).
The order parameter~\eqref{eq:dipole_EdH_OP} is the eigenstate of the projected total angular momentum that satisfies
\begin{align}
({\rm f}_z+\ell_z)\psi_m({\bm r}) =  \left(m-i\frac{\partial}{\partial\varphi}\right)\psi_m({\bm r}) =J_z \psi_m({\bm r}).
\end{align}
The order parameters for magnetic sublevels $m=-3$, $-2$, and $-1$ at $\omega t=2$ are shown in Fig.~\ref{fig:dipole_EdH_AM} (b)--(d).
Each spin component except the $m=-3$ state is a vortex state with the winding number $J_z-m$, and the BEC as a whole forms a coreless vortex.

\begin{figure}[ht]
\begin{center}
\resizebox{0.6\hsize}{!}{\includegraphics{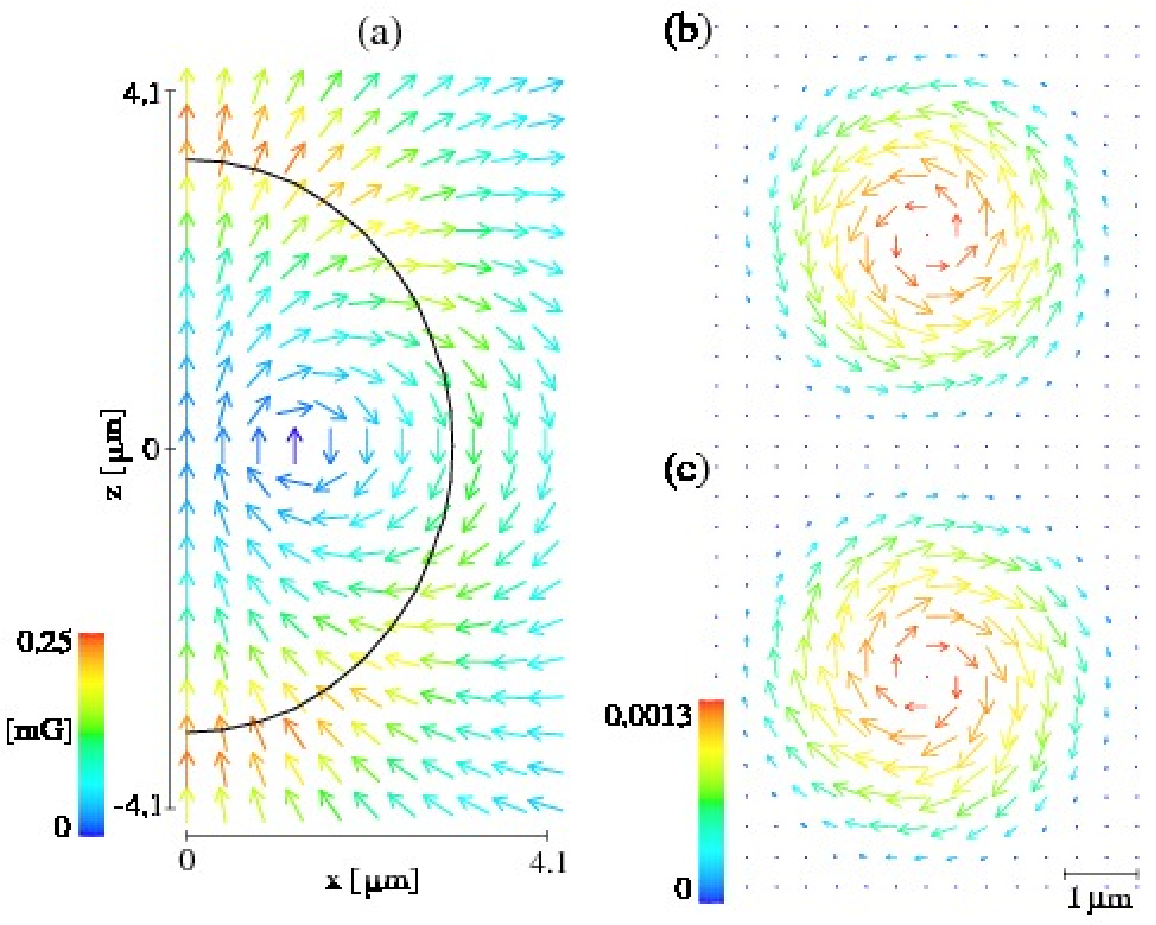}}
\end{center}
\caption{(a) Dipole field induced by a spin-polarized BEC in a spherical trap, where the color of each arrow denotes the field strength according to the left color gauge, and the solid curve indicates the periphery of the condensate.
(b),(c) Spin textures on the (b) $z=2~\mu$m plane and (c) $z=-2~\mu$m plane at $\omega t=2$.
The length of the arrows represents the magnitude of the spin vector projected on the $x$-$y$ plane, 
and the color indicates $|{\bm F}| a_{\rm ho}^3/N$ according to the right color gauge, where $a_{\rm ho}$ is the harmonic oscillator length and $N$ is the total number of atoms.
Reprinted from Ref.~\cite{Kawaguchi2006a}.
}
\label{fig:dipole_dipole_field}
\end{figure}

The physical origin of the spin relaxation is the Larmor precision of atomic spins around the dipole field given by Eq.~\eqref{eq:effective_dipole_field}.
Hence, the time scale for the spin relaxation is determined by the DDI and is on the order of $h/(c_{\rm dd}n)$ with $n$ is the atomic density.
Figure~\ref{fig:dipole_dipole_field} (a) shows the dipole field induced by a spin-polarized BEC in a spherical trap,
and Figs.~\ref{fig:dipole_dipole_field} (b) and (c) show the spin textures at $\omega t=2$.
The whirling patterns in the upper and lower hemispheres exhibit opposite directions, reflecting the fact that the $x$-$y$ components of the dipole field are antisymmetric with respect to the $z=0$ plane.

In the case of a $^{52}$Cr BEC, the energy gain due to the DDI compensates for the kinetic energy cost to form the spin texture.
As discussed in the next section, a spin-1 $^{87}$Rb BEC is ferromagnetic, and therefore a spin-polarized state is rather stable.
In such a case, an external magnetic field applied in the direction opposite to the polarization causes a resonance between the rotational kinetic energy and the linear Zeeman energy via the DDI, leading to the EdH effect~\cite{Gawryluk2007}.
Furthermore, the EdH effect in spin-1 $^{87}$Rb BEC was predicted to be resonantly enhanced using an oscillating magnetic field~\cite{Gawryluk2011}.
In contrast, when a condensate favors a non-magnetic state (e.g., a BEC of spin-1 $^{23}$Na atoms and that of spin-3 $^{52}$Cr atoms),
a spin-polarized state will be intrinsically unstable. Therefore, the EdH effect is expected to occur significantly at low magnetic field.

Spin relaxation due to the DDI was observed to lead to thermalization both in a thermal gas~\cite{Hensler2003} and in a condensate~\cite{Pasquiou2010} of $^{52}$Cr atoms.
The thremalization is suppressed in an optical lattice~\cite{Pasquiou2011a} or in an ultralow magnetic field~\cite{Pasquiou2011b}.

\subsubsection{Ground-state spin textures at zero magnetic field}
\label{sec:dipole_gs}
Next, we consider ground-state spin structures in the absence of an external magnetic field.
The DDI is known to create magnetic domains in solid-state ferromagnets.
Similar structures are expected to appear in ferromagnetic BECs.

The mean-field energy for the DDI can be rewritten as
\begin{align}
\langle \hat{V}_{\rm dd} \rangle
&=\frac{c_{\rm dd}}{2}\int d {\bm r}\int d {\bm r}' \sum_{\nu\nu'}  F_{\nu}({\bm r})  Q_{\nu\nu'}({\bm r}-{\bm r}')  F_{\nu'} ({\bm r}') \nonumber\\
&=\frac{c_{\rm dd}}{2}\sum_{\bm k} \sum_{\nu\nu'}  \tilde{F}_{\nu}({\bm k}) 
 \tilde{Q}_{\nu\nu'}({\bm k})  \tilde{F}_{\nu'}(-{\bm k}) \nonumber\\
&=\frac{c_{\rm dd}}{2}\frac{4\pi}{3}\sum_{\bm k}
\left[3|\hat{\bm k}\cdot\tilde{\bm F}({\bm k})|^2-|\tilde{\bm F}({\bm k})|^2\right],
\label{eq:dipole_energy_in_k}
\end{align}
where $ \tilde{\bm F}({\bm k})$ is the Fourier transformation of $ {\bm F}({\bm r})$  and $\tilde{Q}_{\nu\nu'}({\bm k})$ is given by Eq.~\eqref{eq:dipole_spinor_fourier}.
When the BEC is fully magnetized, i.e., when $|{\bm F}({\bm r})|=fn({\bm r})$ with $f$ being the atomic spin angular momentum and $n({\bm r})$ the
number density, the last term in Eq.~\eqref{eq:dipole_energy_in_k} makes a constant contribution determined by the density distribution.
Then, the DDI favors the configuration that satisfies $\hat{\bm k}\cdot\tilde{\bm F}({\bm k})=0$ for all ${\bm k}$,
or in real space, ${\bm \nabla}\cdot{\bm F}({\bm r})=0$.
This condition is the same as that for solid-state ferromagnets~\cite{LandauLifshitz_ED}, 
and the corresponding stable domain structure is known as a flux-closure structure.
In solid-state ferromagnets, the magnitude of magnetization is constant and the domain structure is constrained by the boundary conditions such that the magnetization is parallel to the surface. In trapped condensates, on the other hand, a spatial variation of $|{\bm F}({\bm r})|$, which is proportional to the particle density, yields a spin texture.

As compared with solid-state ferromagnets, a unique feature of ferromagnetic BECs is that they exhibit the spin-gauge symmetry that generate a supercurrent by developing spin textures (see Sec.~\ref{sec:vortex_spin1ferro}).
The supercurrent, defined by Eq.~\eqref{eq:def_for_mass_supercurrent}, is generated by a spatial variation of the direction of magnetization. Therefore, even in the ground state in a stationary trap, a supercurrent can flow due to a DDI-induced spin texture.

\begin{figure}[ht]
\begin{center}
\resizebox{0.5\hsize}{!}{\includegraphics{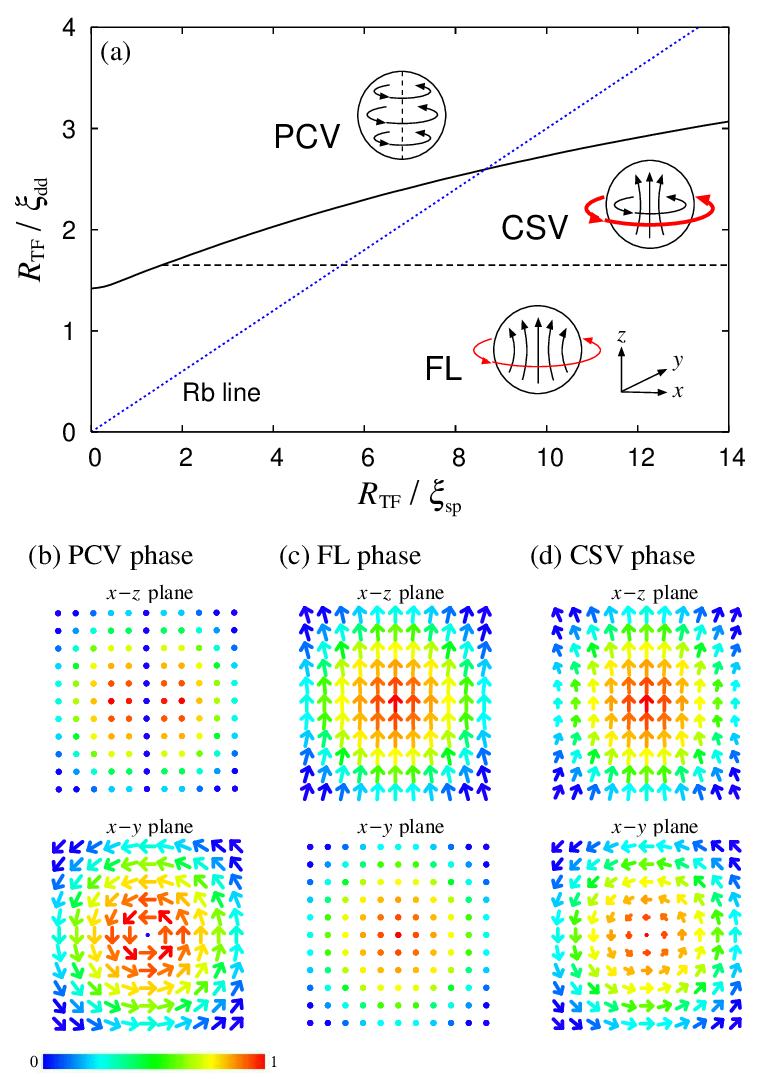}}
\end{center}
\caption{(a) Phase diagram of a ferromagnetic BEC in a spherical trap. The state of a BEC of spin-1 $^{87}$Rb atoms traces along the dotted line.
(b)--(d) Cross sections of spin textures in the $x$-$y$ (upper) and $x$-$z$ (bottom) planes in the (b) flower phase, (c) chiral spin-vortex phase, and (d) polar-core vortex phase, where the color and length of each arrow indicate $|{\bm F}| a_{\rm ho}^3/N$ and the spin projection onto the planes.
Reprinted from Ref.~\cite{Kawaguchi2006b}.
}
\label{fig:dipole_phase_diagram}
\end{figure}

Figure~\ref{fig:dipole_phase_diagram} (a) shows the numerically determined phase diagram for a spin-1 ferromagnetic BEC~\cite{Kawaguchi2006b} as a function of $R_{\rm TF}/\xi_{\rm sp}$ and $R_{\rm TF}/\xi_{\rm dd}$, 
where $R_{\rm TF}$ is the Thomas-Fermi radius, $\xi_{\rm sp}=\hbar/\sqrt{2M|c_1|n}$ is the spin healing length, and $\xi_{\rm dd}$ is the dipole healing length defined in Eq.~\eqref{eq:Dipole_def_xidd}.
There are three phases---flower (FL), chiral spin-vortex (CSV), and polar-core vortex (PCV)---whose spin configurations are 
shown in Fig.~\ref{fig:dipole_phase_diagram} (b), (c), and (d), respectively.
The FL and CSV phases are eigenstates of the projected total angular momentum $J_z=1$,  whose order parameter is given by Eq.~\eqref{eq:dipole_EdH_OP}.
The orbital angular momentum for the order parameter~\eqref{eq:dipole_EdH_OP} is calculated as
\begin{align}
L_z = \int d {\bm r} \sum_{m=0,\pm1} \psi_m^*({\bm r})\left( -i \frac{\partial}{\partial \varphi}\right)\psi_m({\bm r})
= \int d {\bm r} \sum_{m=0,\pm1} (J_z-m)|\psi_m({\bm r})|^2,
\end{align}
which may take a nonzero value depending on the population in each magnetic sublevel.
When the system size is smaller than $\xi_{\rm dd}$, the kinetic energy, which hinders the spin texture from developing, is dominant, and the spin is almost polarized.
Most of the atoms therefore reside in the magnetic sublevel $m=1$ and the orbital angular momentum is quite small.
As the size of the BEC becomes larger, the spin texture develops to form a flux-closure structure.
Consequently, the number of atoms in the $m=0$ and $-1$ magnetic sublevels increases, leading to an increase in the orbital angular momentum.
The difference between FL and CSV phases is the symmetry of spin textures:
The spin texture has chirality in the latter phase.
Furthermore, an increase in the size of the BEC results in the phase transition to the polar-core vortex state with $J_z=0$, in which magnetizations lie in the $x$-$y$ plane to form a whirling pattern and vanish on the $z$ axis.
This is because the $J_z=1$ state costs a large kinetic energy when the population in the $m=-1$ state increases.
In the $J_z=0$ phase, the order parameter becomes polar on the symmetry axis, and hence, that structure is called a polar-core vortex (see Sec.~\ref{sec:Vortices}).

\begin{figure}[ht]
\begin{center}
\resizebox{0.6\hsize}{!}{\includegraphics{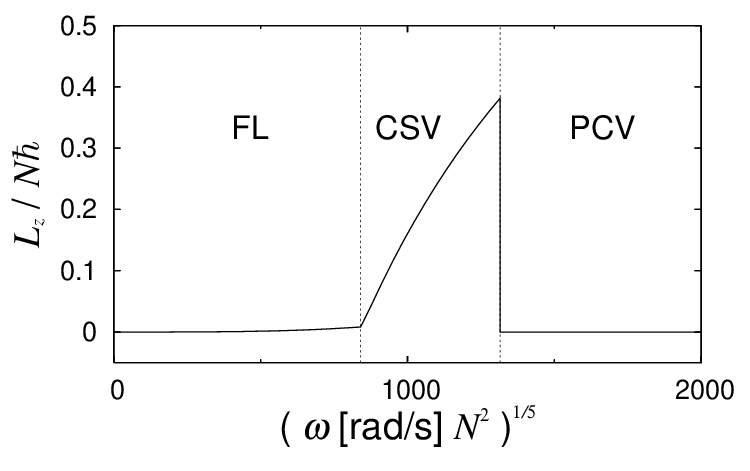}}
\end{center}
\caption{Orbital angular momentum for a spin-1 $^{87}$ Rb BEC in a spherical trap with trap frequency $\omega$.
Reprinted from Ref.~\cite{Kawaguchi2006b}.
}
\label{fig:dipole_Lz}
\end{figure}

All three phases in Fig.~\ref{fig:dipole_phase_diagram} can be realized in spin-1 $^{87}$Rb BECs,
the state of which traces along the dotted line shown in Fig.~\ref{fig:dipole_phase_diagram} (a).
The orbital angular momentum along this line is plotted in Fig.~\ref{fig:dipole_Lz} as a function of the total number of atoms $N$ and trap frequency $\omega$.
The orbital angular momentum in the ground state has a significant value, 
and it increases up to approximately $0.4\hbar$ per atom in the CSV phase.

For higher-spin BECs, similar spin textures are expected to appear in the ferromagnetic phase, although the mass current depends on the value of the atomic spin
and the core structure in the polar-core vortex region is nontrivial.
Spin textures in the limit of a strong ferromagnetic interaction are discussed in Ref.~\cite{Takahashi2007}.
On the other hand, even in the case of a non-ferromagnetic phase such as a spin-1 $^{23}$Na BEC, a spin texture may appear if the dipole interaction is effectively enhanced in a pancake trap with a large aspect ratio~\cite{Yi2006b}.

\subsubsection{Dipole-dipole interaction under a magnetic field}

Suppose that a homogeneous external magnetic field ${\bm B}=B\hat{z}$ is applied in the $z$ direction.
Because the linear Zeeman term $g\mu_{\rm B}B F_z$ rotates the atomic spin
around the $z$ axis at the Larmor frequency $\omega_{\rm L} = g\mu_{\rm B}B/\hbar$,
the DDI is time-averaged over the Larmor precession period when the Larmor precession is much faster than the spin dynamics induced by the DDI, i.e., when $\hbar\omega_{\rm L}\gg c_{\rm dd}n$.

To derive an effective (time-averaged) DDI,
it is convenient to describe the system in the rotating frame of reference with the Larmor frequency in spin space.
The spin vector operators $\hat{\bm F}^{\rm (rot)}$ in the rotating frame are related to those in the laboratory frame by
$\hat{F}_\pm^{\rm (rot)} \equiv \hat{F}_x^{\rm (rot)} \pm i\hat{F}_y^{\rm (rot)}
= e^{\pm i\omega_{\rm L} t} \hat{F}_\pm$ and $\hat{F}_z^{\rm (rot)} = \hat{F}_z$.
Then, $\hat{V}_{\rm dd}$ defined in Eq.~\eqref{eq:dipole_2quantization} can be rewritten in terms of $\hat{\bm F}^{\rm (rot)}$ as
\begin{align}
\hat{V}_{\rm dd} = &-\sqrt{\frac{6\pi}{5}}\frac{c_{\rm dd}}{2}\int d {\bm r} \int d {\bm r}'
\frac{1}{|{\bm r}-{\bm r}'|^3} \nonumber \\
&\times :\bigg[
\frac{Y_2^0({\bm e})}{\sqrt{6}}\left\{
                          4\hat{F}_{z}^{\rm (rot)}({\bm r}) \hat{F}_{z}^{\rm (rot)}({\bm r}') 
                          -\hat{F}_{+}^{\rm (rot)}({\bm r}) \hat{F}_{-}^{\rm (rot)}({\bm r}')
                          -\hat{F}_{-}^{\rm (rot)}({\bm r}) \hat{F}_{+}^{\rm (rot)}({\bm r}') \right\} \nonumber\\
&\ \ +e^{-i\omega_{\rm L}t} Y_2^{-1} ({\bm e})\left\{ \hat{F}_{+}^{\rm (rot)}({\bm r}) \hat{F}_{z}^{\rm (rot)}({\bm r}')
                          +\hat{F}_{z}^{\rm (rot)}({\bm r}) \hat{F}_{+}^{\rm (rot)}({\bm r}') \right\} \nonumber \\
&\ \ -e^{ i\omega_{\rm L}t} Y_2^1    ({\bm e})\left\{ \hat{F}_{-}^{\rm (rot)}({\bm r}) \hat{F}_{z}^{\rm (rot)}({\bm r}')
                          +\hat{F}_{z}^{\rm (rot)}({\bm r}) \hat{F}_{-}^{\rm (rot)}({\bm r}') \right\} \nonumber \\
&\ \ +e^{-2i\omega_{\rm L}t} Y_2^{-2}({\bm e})        \hat{F}_{+}^{\rm (rot)}({\bm r}) \hat{F}_{+}^{\rm (rot)}({\bm r}')         
     +e^{ 2i\omega_{\rm L}t} Y_2^2   ({\bm e})        \hat{F}_{-}^{\rm (rot)}({\bm r}) \hat{F}_{-}^{\rm (rot)}({\bm r}') \bigg]:,
\label{eq:dipole_vdd_rot}
\end{align}
where ${\bm e}\equiv ({\bm r}-{\bm r}')/|{\bm r}-{\bm r}'|$.
In a large magnetic field ($\hbar\omega_{\rm L}\gg c_{\rm dd}n$), we may use an effective DDI that is time-averaged over the Larmor precession period $2\pi/\omega_{\rm L}$:
\begin{align}
\bar{\hat{V}}_{\rm dd} = &-\sqrt{\frac{\pi}{5}}\frac{c_{\rm dd}}{2}\int d {\bm r} \int d {\bm r}'
\frac{Y_2^0({\bm e})}{|{\bm r}-{\bm r}'|^3} \nonumber\\
&\ \ \ \times:\left\{4\hat{F}_{z}^{\rm (rot)}({\bm r}) \hat{F}_{z}^{\rm (rot)}({\bm r}') 
        -\hat{F}_{+}^{\rm (rot)}({\bm r}) \hat{F}_{-}^{\rm (rot)}({\bm r}')
        -\hat{F}_{-}^{\rm (rot)}({\bm r}) \hat{F}_{+}^{\rm (rot)}({\bm r}') \right\}: \nonumber\\
=&-\frac{c_{\rm dd}}{4}\int d {\bm r} \int d {\bm r}'
\frac{|{\bm r}-{\bm r}'|^2-3(z-z')^2}{|{\bm r}-{\bm r}'|^5} \nonumber \\
&\ \ \ \times :\left\{ \hat{\bm F}^{\rm (rot)}({\bm r})\cdot \hat{\bm F}^{\rm (rot)}({\bm r}') - 3 \hat{F}_z^{\rm (rot)}({\bm r})\hat{F}_z^{\rm (rot)}({\bm r}')\right\}:\nonumber\\
= & \frac{c_{\rm dd}}{2}\sum_{\nu\nu'}\int d {\bm r} \int d {\bm r}'
:\hat{\bm F}^{\rm (rot)}({\bm r}) {Q}^{\rm (rot)}_{\nu\nu'}({\bm r}-{\bm r}')\hat{\bm F}^{\rm (rot)}({\bm r}'):,
\label{eq:dipole_vdd_ave}
\end{align}
where ${Q}^{\rm (rot)}_{\nu\nu'}({\bm r})$ is the time-averaged integration kernel in the rotating frame of reference defined by
\begin{align}
{Q}^{\rm (rot)}_{\nu\nu'}({\bm r}) &=-\frac{1-3\hat{r}_z^2}{r^3}\, \frac{\delta_{\nu\nu'}-3\delta_{\nu z}\delta_{\nu' z}}{2},
\label{eq:dipole_vdd_ave_general}
\end{align}
whose Fourier transform is given by
\begin{align}
\tilde{Q}^{\rm (rot)}_{\nu\nu'}({\bm k}) &= \frac{2\pi}{3}(1-3\hat{k}_z^2)(\delta_{\nu\nu'}-3\delta_{\nu z}\delta_{\nu' z}).
\end{align}
In contrast to Eq.~\eqref{eq:dipole_2quantization},
the time-averaged interaction~\eqref{eq:dipole_vdd_ave} separately conserves
the projected total spin angular momentum and the projected relative orbital angular momentum.
However, the long-range and anisotropic nature of the dipolar interaction is still maintained in Eq.~\eqref{eq:dipole_vdd_ave}.
While the total longitudinal magnetization is conserved in the presence of an external magnetic field, the anisotropic dipole interaction can induce spin textures in the transverse magnetization~\cite{Kawaguchi2007}.

\begin{figure}[ht]
\begin{center}
\resizebox{0.7\hsize}{!}{\includegraphics{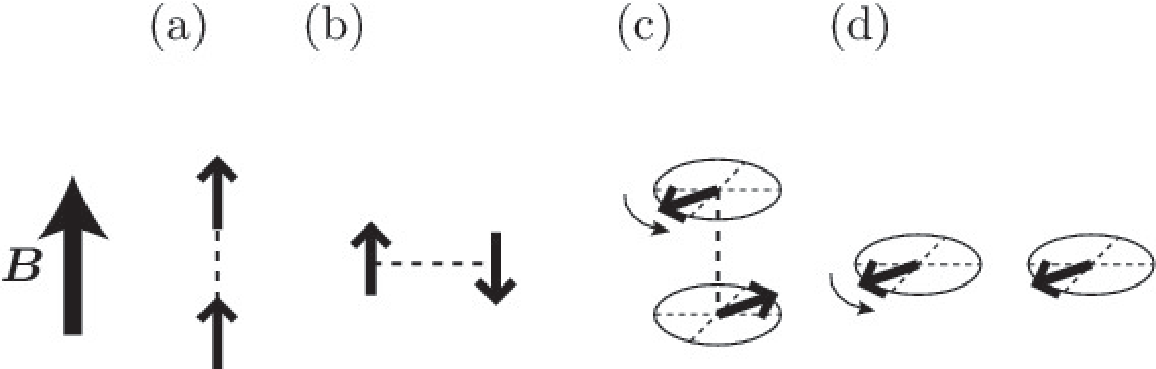}}
\end{center}
\caption{Alignments of (a)(b) longitudinal and (c)(d) transverse magnetizations
that are favored by the time-averaged dipole interaction. 
Due to the Larmor precession, transverse magnetizations tend to be anti-parallel along the magnetic field and parallel perpendicular to the magnetic field.
}
\label{fig:dipole_in_magneticfield}
\end{figure}

Equation~\eqref{eq:dipole_vdd_ave_general} tells us how the dipole moments tend to align in the ground state.
When the condensate magnetizations mostly point in the longitudinal direction, the last factor in Eq.~\eqref{eq:dipole_vdd_ave_general} gives $-1$.
Then, $\hat{\bm F}^{\rm (rot)}(\bm r)$ and $\hat{\bm F}^{\rm (rot)}({\bm r}')$ (assumed to be parallel or antiparallel to $\hat{z}$)
become parallel for $\bm r-\bm r' || {\bm B}$ and antiparallel for $\bm r-\bm r' \perp {\bm B}$ [Figs.~\ref{fig:dipole_in_magneticfield} (a) and (b)].
On the other hand, when the condensate magnetizations lie in the transverse direction, the last factor in Eq.~\eqref{eq:dipole_vdd_ave_general} reduces to $\delta_{\nu\nu'}/2$,
resulting in $\hat{\bm F}^{\rm (rot)}(\bm r) || \hat{\bm F}^{\rm (rot)}({\bm r}')$ for $\bm r-\bm r' \perp {\bm B}$ and
$\hat{\bm F}^{\rm (rot)}(\bm r) || -\hat{\bm F}^{\rm (rot)}({\bm r}')$ for $\bm r-\bm r' || {\bm B}$
[Figs.~\ref{fig:dipole_in_magneticfield} (c) and (d)] (see also Ref.~\cite{Kawaguchi2010}).
The stable configuration shown in Fig.~\ref{fig:dipole_in_magneticfield} (d) is different from that expected for the bare DDI~\eqref{eq:dipole_2quantization} [see Fig.~\ref{fig:dipole_config}].
This is because two transverse spins positioned in the transverse direction ($\bm r-\bm r'\perp {\bm B}$) take both the head-to-tail and side-by-side configurations in the course of the Larmor precession,
resulting in attractive interaction on average.

\subsubsection{Dipole-dipole interaction in quasi-one- and two-dimensional systems}

Because the magnetic DDIs in atomic BECs are, in general, much smaller than the density-density contact interaction,
it is possible to realize a system that is three-dimensional (3D) with respect to density waves and quasi-one-dimensional (quasi-1D) or quasi-two-dimensional (quasi-2D) with respect to spin waves.
Consider, for example, a BEC confined in a strong potential along the $z$ direction.
If the size of the BEC in the $z$ direction is smaller than the dipole healing length $\xi_{\rm dd}$ and the spin healing lengths $\xi_{{\rm sp},i}=\hbar/\sqrt{2M|c_i|n}$ (for $i=1,2,3,\cdots$),
neither the DDI nor the spin-dependent interactions can excite spatial spin modes along the $z$ direction.
This is a quasi 2D system with respect to spin degrees of freedom.
Here, ``quasi'' means that the condensate itself is 3D in the sense that the atoms can occupy the excited states of the 3D trapping potential.
In such a situation, we can describe the spin dynamics with 1D or 2D GPEs by introducing an integrated dipole kernel over the strong confining direction.

Let us first consider a quasi 2D system.
We choose the spin quantization axis in the $z$ direction and apply an external magnetic field in the $z$ direction.
Let $\hat{\bm e}_1, \hat{\bm e}_2$ and $\hat{\bm e}_3$ be the unit vectors along the trap axes.
Assuming that $\hat{\bm e}_3$ is the direction of strong confinement,
the field operator is decomposed as
\begin{align}
 \hat{\psi}(\bm r) = h_1(x_3)\hat{\psi}^{\rm (2D)}({\bm x}_\perp),
\label{eq:dipole_h2D}
\end{align}
where $x_i=\hat{\bm e}_i\cdot \bm r$, ${\bm x}_\perp = (x_1, x_2)$, and $h_1(x_3)$ is assumed to be normalized to unity:
\begin{align}
\int_{-\infty}^\infty |h_1(x_3)|^2dx_3 = 1.
\end{align}
Using Eq.~\eqref{eq:dipole_h2D},
the contact interaction Hamiltonian~\eqref{eq:V_Cmnm'n'} is written as
\begin{align}
 \hat{V}  = \frac{1}{2} \int d\bm x_\perp \sum_{m_1m_2m_1'm_2'} [C^{\rm (2D)}]^{m_1m_2}_{m_1'm_2'} \hat{\psi}_{m_1}^{\rm (2D)*}(\bm x_\perp)\hat{\psi}_{m_2}^{\rm (2D)*}(\bm x_\perp)\hat{\psi}_{m_2'}^{\rm (2D)}(\bm x_\perp)\hat{\psi}_{m_1'}^{\rm (2D)}(\bm x_\perp),
\end{align}
where
\begin{align}
  [C^{\rm (2D)}]^{m_1m_2}_{m_1'm_2'} \equiv  C^{m_1m_2}_{m_1'n_2'}\int_{-\infty}^\infty dx_3 |h_1(x_3)|^4.
\end{align}
It follows that the contact interactions for spin-1, 2, and 3 systems can be written in the same form as Eqs.~\eqref{V(f=1)2}, \eqref{V(f=2)}, and \eqref{V(f=3)}, respectively,
if we replace $c_i$, $\hat{n}(\bm r)$, $\hat{F}_{\nu}(\bm r)$, and $\hat{A}_{\mathcal{F},\mathcal{M}}(\bm r)$ with those defined for a 2D system:
\begin{align}
  c_i^{\rm (2D)} &\equiv  c_i\int_{-\infty}^\infty dx_3 |h_1(x_3)|^4, \label{eq:ci2D}\\
  \hat{n}^{\rm (2D)}(\bm x_\perp) &\equiv \sum_{m=-f}^f\hat{\psi}_m^{\rm (2D)\dagger}(\bm x_\perp)\hat{\psi}_m^{\rm (2D)}(\bm x_\perp),\\
  \hat{F}_{\nu}^{\rm (2D)}(\bm x_\perp) &\equiv \sum_{mm'=-f}^f({\rm f}_{\nu})_{mm'}\hat{\psi}_m^{\rm (2D)\dagger}(\bm x_\perp)\hat{\psi}_{m'}^{\rm (2D)}(\bm x_\perp),\\
  \hat{A}_{\mathcal{F},\mathcal{M}}^{\rm (2D)}(\bm x_\perp) &\equiv \sum_{mm'=-f}^f \langle \mathcal{F},\mathcal{M}|f,m;f,m'\rangle\hat{\psi}_m^{\rm (2D)}(\bm x_\perp)\hat{\psi}_{m'}^{\rm (2D)}(\bm x_\perp). \label{eq:AFM2D}
\end{align}
As for the DDI, Eq.~\eqref{eq:dipole_2quantization} is rewritten as
\begin{align}
 \hat{V}_{\rm dd} = \frac{c^{\rm (2D)}_{\rm dd}}{2} \int d{\bm x}_\perp \int d{\bm x}'_\perp \sum_{\nu\nu'} : \hat{F}^{\rm (2D)}_{\nu}(\bm x_\perp) Q_{\nu\nu'}^{\rm (2D)}({\bm x}_\perp - {\bm x}'_\perp) \hat{F}_{\nu'}^{\rm (2D)}(\bm x'_\perp):,
\end{align}
where $c^{\rm (2D)}_{\rm dd}\equiv c_{\rm dd}\int_{-\infty}^\infty dx_3 |h_1(x_3)|^4$ and $Q_{\nu\nu'}^{\rm (2D)}({\bm x}_\perp-{\bm x}'_\perp)$ is the 2D dipole kernel which is related to the 3D one as
\begin{align}
 Q_{\nu\nu'}^{\rm (2D)}({\bm x}_\perp-{\bm x}'_\perp) = \frac{1}{\int_{-\infty}^\infty dx_3 |h_1(x_3)|^4} \int_{-\infty}^\infty dx_3\int_{-\infty}^\infty dx_3' |h_1(x_3)|^2|h_1(x_3')|^2 Q_{\nu\nu'}({\bm r}-{\bm r}').
\label{eq:dipole_2Dkernel_r}
\end{align}
Equivalently, the Fourier transform is given by
\begin{align}
 \tilde{Q}_{\nu\nu'}^{\rm (2D)}({\bm k}_\perp) = \sum_{k_3}\frac{\left|\int_{-\infty}^\infty dx_3 e^{ik_3 x_3}|h_1(x_3)|^2\right|^2}{\int_{-\infty}^\infty dx_3 |h_1(x_3)|^4} \tilde{Q}_{\nu\nu'}({\bm k}),
\label{eq:dipole_2Dkernel_k}
\end{align}
where $\bm k=(k_x,k_y,k_z)$, $k_i=\hat{\bm e}_i\cdot {\bm k}$, and ${\bm k}_\perp=(k_1,k_2)$.
Equations~\eqref{eq:dipole_2Dkernel_r} and \eqref{eq:dipole_2Dkernel_k} also hold for the time-averaged dipole kernels $Q_{\nu\nu'}^{\rm (rot)}(\bm r)$ and $\tilde{Q}_{\nu\nu'}^{\rm (rot)}(\bm k)$.

\begin{figure}[ht]
\begin{center}
\resizebox{0.8\hsize}{!}{\includegraphics{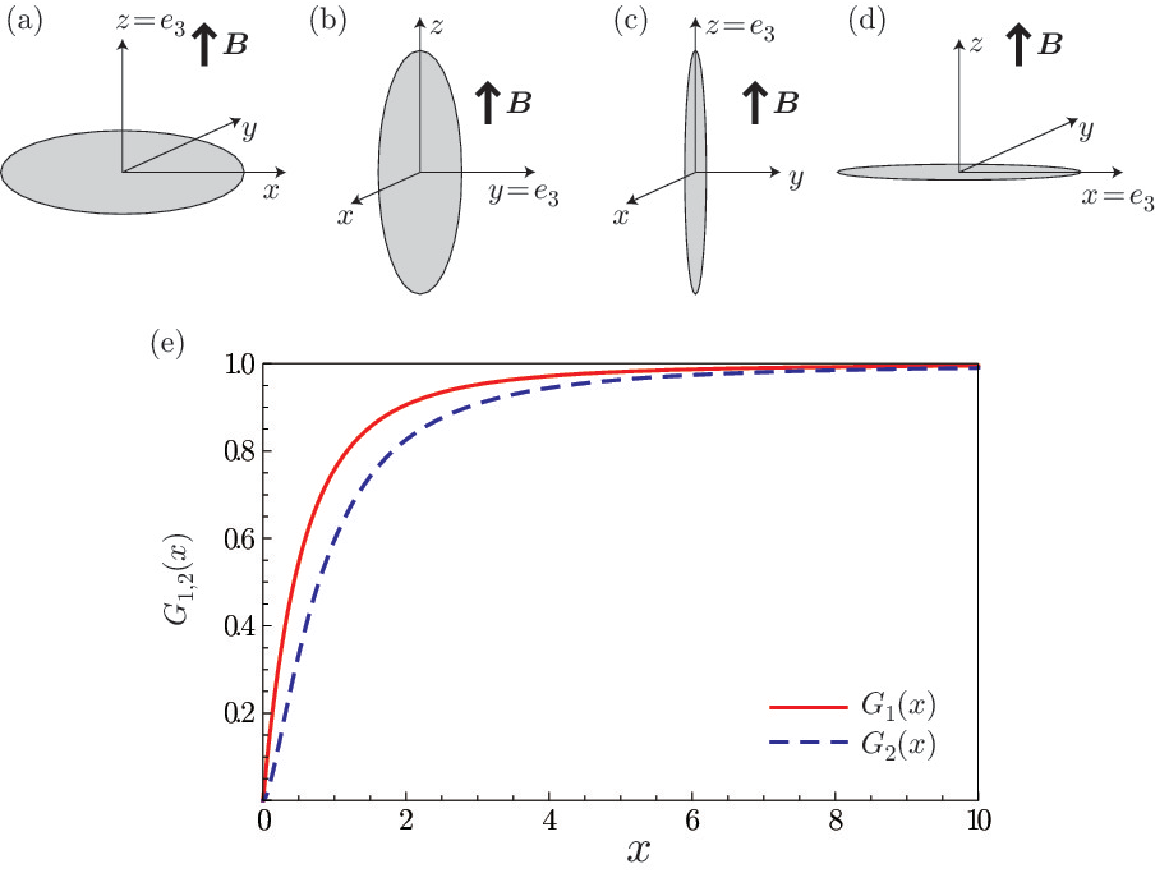}}
\end{center}
\caption{(a)--(d) Configurations of the trap geometry and an external magnetic field for
(a) $\tilde{\bm Q}^{\rm (2D,rot\perp)}$ [Eq.~\eqref{eq:Q2D_rot_perp}],
(b) $\tilde{\bm Q}^{\rm (2D,rot||)}$ [Eq.~\eqref{eq:Q2D_rot_para}],
(c) $\tilde{\bm Q}^{\rm (1D,rot||)}$ [Eq.~\eqref{eq:Q1D_rot_para}],
and (d) $\tilde{\bm Q}^{\rm (1D,rot\perp)}$ [Eq.~\eqref{eq:Q1D_rot_perp}].
(e) Plots of $G_1(x)$ and $G_2(x)$ defined in Eqs.~\eqref{eq:dipole_def_G1} and \eqref{eq:dipole_def_G2}, respectively.
}
\label{fig:dipole_lowD}
\end{figure}

Assuming a Gaussian profile along the $\hat{\bm e}_3$ direction, i.e.,
\begin{align}
 h_1(x_3) = \frac{1}{(2\pi d)^{1/4}}\exp\left(-\frac{x_3^2}{4d^2}\right),
\end{align}
we obtain $c_i^{\rm (2D)} = c_i/\sqrt{4\pi d^2}\, (i=0,1,\cdots)$ and  $c_{\rm dd}^{\rm (2D)} = c_{\rm dd}/\sqrt{4\pi d^2}$, and Eq.~\eqref{eq:dipole_2Dkernel_k} reduces to
\begin{align}
 \tilde{Q}_{\nu\nu'}^{\rm (2D)}({\bm k}_\perp) 
=\frac{d}{\sqrt{\pi}}\int_{-\infty}^\infty d k_3 e^{-d^2 k_3^2}\tilde{Q}_{\nu\nu'}(\bm k).
\end{align}
By expanding $k_{\nu=x,y,z}$ as $k_{\nu}=\sum_{i=1}^3 k_i(\hat{\bm e}_i)_{\nu}$ with $(\hat{\bm e}_i)_{\nu}$ being the $\nu$-component of the unit vector $\hat{\bm e}_i$, and using the integral
\begin{align}
 \frac{d}{\sqrt{\pi}}\int_{-\infty}^\infty d k_3 e^{-d^2k_3^2}\frac{k_ik_j}{k_1^2+k_2^2+k_3^2}
=\left\{\begin{array}{ll}
G_1(k_\perp d)(\hat{\bm k}_\perp)_i(\hat{\bm k}_\perp)_j & (i,j=1,2), \\
1-G_1(k_\perp d) & (i=j=3), \\
0 & (\textrm{otherwise}),
\label{eq:dipole_k_integral}
\end{array}
\right.
\end{align}
with $k_\perp = |{\bm k}_\perp|$, $\hat{\bm k}_\perp = {\bm k}_\perp/k_\perp$, $(\hat{\bm k}_\perp)_i=\hat{\bm e}_i\cdot\hat{\bm k}_\perp$, and
\begin{align}
 G_1(x) \equiv  2x e^{x^2}\int_x^\infty e^{-t^2}dt,
\label{eq:dipole_def_G1}
\end{align}
we obtain
\begin{align}
 \tilde{Q}_{\nu\nu'}^{\rm (2D)}({\bm k}_\perp) 
&= -\frac{4\pi}{3}[\delta_{\nu\nu'}-3(\hat{\bm e}_3)_{\nu}(\hat{\bm e}_3)_{\nu'}] + 4\pi G_1(k_\perp d)[(\hat{k}_\perp)_{\nu}(\hat{k}_\perp)_{\nu'}-(\hat{\bm e}_3)_{\nu}(\hat{\bm e}_3)_{\nu'}], \label{eq:Q2D_lab}\\
 \tilde{Q}_{\nu\nu'}^{\rm (2D,rot)}({\bm k}_\perp) 
&= \frac{2\pi}{3}(\delta_{\nu\nu'}-3\delta_{\nu z}\delta_{\nu' z}) \left\{1-3(\hat{\bm e}_3)_z^2 - 3G_1(k_\perp d)[(\hat{\bm k}_\perp)_z^2 - (\hat{\bm e}_3)_z^2]\right\},\label{eq:Q2D_rot}
\end{align}
where $(\hat{\bm k}_\perp)_{\nu}=\sum_{i=1,2}(\hat{\bm k}_\perp)_i(\hat{\bm e}_i)_{\nu}$
is the $\nu$-component of the unit vector along the wave vector ${\bm k}_\perp$ in the 2D plane.
Substituting $(\hat{\bm e}_1,\hat{\bm e}_2,\hat{\bm e}_3)=(\hat{x},\hat{y},\hat{z})$, Eq.~\eqref{eq:Q2D_lab} reduces to
\begin{align}
 \tilde{\bm Q}^{\rm (2D)}(k_x,k_y) 
&= -\frac{4\pi}{3}\begin{pmatrix} 1 & 0 & 0 \\ 0 & 1 & 0 \\ 0 & 0 & -2 \end{pmatrix}
 + 4\pi \frac{G_1(k_\perp d)}{k_\perp^2} \begin{pmatrix} {k}_x^2 & {k}_x{k}_y & 0 \\ {k}_x{k}_y & {k}_y^2 & 0 \\ 0 & 0 & -k_\perp^2 \end{pmatrix},
\label{eq:Q2D_lab2}
\end{align}
where $k_\perp=\sqrt{k_x^2+k_y^2}$.
Contrary to the 3D dipole kernel~\eqref{eq:dipole_spinor_fourier} which is symmetric in the momentum space, 
an anisotropy arises in $\tilde{\bm Q}^{\rm (2D)}(k_x,k_y)$ due to the confinement.
On the other hand, the dipole kernel for the rotating frame depends on the relative angle between the 2D plane and the external field.
For the case in which an external field is applied perpendicular to the 2D plane [Fig.~\ref{fig:dipole_lowD}(a)],
we substitute $(\hat{\bm e}_1,\hat{\bm e}_2,\hat{\bm e}_3)=(\hat{x},\hat{y},\hat{z})$ in Eq.~\eqref{eq:Q2D_rot}, obtaining
\begin{align}
 \tilde{\bm Q}^{\rm (2D,rot\perp)}(k_x,k_y) 
&= \frac{2\pi}{3}[-2+3G_1(k_\perp d)]\begin{pmatrix} 1 & 0 & 0 \\ 0 & 1 & 0 \\ 0 & 0 & -2 \end{pmatrix},
\label{eq:Q2D_rot_perp}
\end{align}
whereas when an external field is parallel to the 2D plane [Fig.~\ref{fig:dipole_lowD}(b)],
we take $(\hat{\bm e}_1,\hat{\bm e}_2,\hat{\bm e}_3)=(\hat{z},\hat{x},\hat{y})$, obtaining
\begin{align}
 \tilde{\bm Q}^{\rm (2D,rot||)}(k_x,k_z) 
&= \frac{2\pi}{3}\left[1-3\frac{k_z^2}{k_\perp^2}G_1(k_\perp d)\right]\begin{pmatrix} 1 & 0 & 0 \\ 0 & 1 & 0 \\ 0 & 0 & -2 \end{pmatrix},
\label{eq:Q2D_rot_para}
\end{align}
where $k_\perp=\sqrt{k_x^2+k_y^2}$ and $\sqrt{k_x^2+k_z^2}$ for Eqs.~\eqref{eq:Q2D_rot_perp} and \eqref{eq:Q2D_rot_para}, respectively.

The dipole kernel in the quasi-1D system is defined in a similar manner.
Suppose that the potential is tightly confined in the ${\bm e}_1$ and ${\bm e}_2$ directions so that spin structures can develop only along the ${\bm e}_3$ direction.
In this case, we can approximate the field operator as
\begin{align}
\hat{\psi}(\bm r) = h_2({\bm x}_\perp)\hat{\psi}^{\rm (1D)}(x_3),
\end{align}
where $h_2({\bm x}_\perp)$ is assumed to be normalized to unity:
\begin{align}
\int d{\bm x}_\perp |h_2({\bm x}_\perp)|^2 = 1.
\end{align}
The contact interactions for spin-1, 2, and 3 systems are written in the same form as Eqs.~\eqref{V(f=1)2}, \eqref{V(f=2)}, and \eqref{V(f=3)}, respectively,
if we replace $c_i$, $\hat{n}(\bm r)$, $\hat{F}_{\nu}(\bm r)$, and $\hat{A}_{\mathcal{F},\mathcal{M}}(\bm r)$ with those defined for a 1D system:
\begin{align}
  c_i^{\rm (1D)} &\equiv  c_i\int d{\bm x}_\perp |h_2({\bm x}_\perp)|^4, \label{eq:ci1D}\\
  \hat{n}^{\rm (1D)}(x_3) &\equiv \sum_{m=-f}^f\hat{\psi}_m^{\rm (1D)\dagger}(x_3)\hat{\psi}_m^{\rm (1D)}(x_3),\\
  \hat{F}_{\nu}^{\rm (1D)}(x_3) &\equiv \sum_{mm'=-f}^f({\rm f}_{\nu})_{mm'}\hat{\psi}_m^{\rm (1D)\dagger}(x_3)\hat{\psi}_{m'}^{\rm (1D)}(x_3),\\
  \hat{A}_{\mathcal{F},\mathcal{M}}^{\rm (1D)}(x_3) &\equiv \sum_{mm'=-f}^f \langle \mathcal{F},\mathcal{M}|f,m;f,m'\rangle\hat{\psi}_m^{\rm (1D)}(x_3)\hat{\psi}_{m'}^{\rm (1D)}(x_3).\label{eq:AFM1D}
\end{align}
The Hamiltonian for the DDI~\eqref{eq:dipole_2quantization} is rewritten for a quasi 1D system as
\begin{align}
 \hat{V}_{\rm dd} = \frac{c^{\rm (1D)}_{\rm dd}}{2} \int_{-\infty}^\infty dx_3 \int_{-\infty}^\infty dx_3' \sum_{\nu\nu'} : \hat{F}^{\rm (1D)}_{\nu}(x_3) Q_{\nu\nu'}^{\rm (1D)}(x_3-x_3') \hat{F}_{\nu'}^{\rm (1D)}(x_3'):
\end{align}
where $c^{\rm (1D)}_{\rm dd}\equiv c_{\rm dd}\int d{\bm x}_\perp |h_2({\bm x}_\perp)|^4$ and 
\begin{align}
 Q_{\nu\nu'}^{\rm (1D)}(x_3-x_3') = \frac{1}{\int d{\bm x}_\perp |h_2({\bm x}_\perp)|^4} \int d{\bm x}_\perp \int d{\bm x}_\perp' |h_2({\bm x}_\perp)|^2|h_2({\bm x}_\perp')|^2 Q_{\nu\nu'}({\bm r}-{\bm r}')
\label{eq:dipole_1Dkernel_r}
\end{align}
is the 1D dipole kernel.
The Fourier transform of Eq.~\eqref{eq:dipole_1Dkernel_r} is given by
\begin{align}
 \tilde{Q}_{\nu\nu'}^{\rm (1D)}(k_3) = \sum_{{\bm k}_\perp}\frac{\left|\int d{\bm x}_\perp e^{i{\bm k}_\perp\cdot{\bm x}_\perp} |h_2({\bm x}_\perp)|^2\right|^2}{\int d{\bm x}_\perp |h_2({\bm x}_\perp)|^4} \tilde{Q}_{\nu\nu'}({\bm k}).
\label{eq:dipole_1Dkernel_k}
\end{align}

Assuming a Gaussian profile for $h_2({\bm x}_\perp)$:
\begin{align}
 h_2({\bm x}_\perp) = \frac{1}{\sqrt{2\pi d}}\exp\left(-\frac{x_1^2+x_2^2}{4d^2}\right),
\end{align}
we obtain $c_i^{\rm (1D)} = c_i/(4\pi d^2)\, (i=0,1,2,\cdots)$ and  $c_{\rm dd}^{\rm (1D)} = c_{\rm dd}/(4\pi d^2)$, and Eq.~\eqref{eq:dipole_1Dkernel_k} reduces to
\begin{align}
 \tilde{Q}_{\nu\nu'}^{\rm (1D)}(k_3) 
=\frac{d^2}{\pi}\int_{-\infty}^\infty dk_1 \int_{-\infty}^\infty dk_2 e^{-d^2 (k_1^2+k_2^2)}\tilde{Q}_{\nu\nu'}(\bm k).
\end{align}
By using the integral
\begin{align}
 \frac{d^2}{\pi}\int_{-\infty}^\infty d k_1 \int_{-\infty}^\infty d k_2 e^{-d^2(k_1^2+k_2^2)}\frac{k_ik_j}{k_1^2+k_2^2+k_3^2}
=\left\{\begin{array}{ll}
\displaystyle\frac{1}{2}[1-G_2(k_3 d)]& (i=j=1\ {\rm or}\  2), \\
G_2(k_3 d) & (i=j=3), \\
0 & (\textrm{otherwise}),
\end{array}
\right.
\end{align}
with 
\begin{align}
 G_2(x) \equiv  x^2 e^{x^2}\int_{x^2}^\infty \frac{e^{-t}}{t}dt,
\label{eq:dipole_def_G2}
\end{align}
we obtain
\begin{align}
 \tilde{Q}_{\nu\nu'}^{\rm (1D)}(k_3) 
=& -\frac{4\pi}{3} \delta_{\nu\nu'} + 2\pi[(\hat{\bm e}_1)_{\nu}(\hat{\bm e}_1)_{\nu'} + (\hat{\bm e}_2)_{\nu}(\hat{\bm e}_2)_{\nu'}]\nonumber\\
&+2\pi G_2(k_3d)[2(\hat{\bm e}_3)_{\nu}(\hat{\bm e}_3)_{\nu'} - (\hat{\bm e}_1)_{\nu}(\hat{\bm e}_1)_{\nu'} - (\hat{\bm e}_2)_{\nu}(\hat{\bm e}_2)_{\nu'}], \label{eq:Q1D_lab}\\
 \tilde{Q}_{\nu\nu'}^{\rm (1D,rot)}(k_3) 
=& \frac{2\pi}{3}(\delta_{\nu\nu'}-3\delta_{\nu z}\delta_{\nu' z}) \left\{1-\frac{3}{2}[(\hat{\bm e}_1)_z^2+(\hat{\bm e}_2)_z^2] - \frac{3}{2}G_2(k_3 d)[2(\hat{\bm e}_3)_z^2-(\hat{\bm e}_1)_z^2-(\hat{\bm e}_2)_z^2]\right\}.
\label{eq:Q1D_rot}
\end{align}
Taking $(\hat{\bm e}_1,\hat{\bm e}_2,\hat{\bm e}_3)=(\hat{x},\hat{y},\hat{z})$, Eq.~\eqref{eq:Q1D_lab} reduces to
\begin{align}
 \tilde{\bm Q}^{\rm (1D)}(k_z) 
=& \frac{2\pi}{3}[1-3G_2(k_zd)]\begin{pmatrix} 1 & 0 & 0 \\ 0 & 1 & 0 \\ 0 & 0 & -2 \end{pmatrix}.
\label{eq:Q1D_lab2}
\end{align}
The facts that the $k_z$ dependence is factorized in Eq.~\eqref{eq:Q1D_lab2} and that the matrix part is diagonal mean that the spin angular momentum along the 1D direction ($z$ direction) is conserved.
This is because the spin angular momentum cannot be transfered to the orbital one due to the strong confinement.
As for $\tilde{Q}_{\nu\nu'}^{\rm (1D,rot)}$,
when an external field is applied along the 1D condensate [Fig.~\ref{fig:dipole_lowD}(c)],
we substitute $(\hat{\bm e}_1,\hat{\bm e}_2,\hat{\bm e}_3)=(\hat{x},\hat{y},\hat{z})$ to Eq.~\eqref{eq:Q1D_rot} and obtain
\begin{align}
 \tilde{\bm Q}^{\rm (1D,rot||)}(k_z) 
&= \frac{2\pi}{3}[1-3G_2(k_z d)]\begin{pmatrix} 1 & 0 & 0 \\ 0 & 1 & 0 \\ 0 & 0 & -2 \end{pmatrix},
\label{eq:Q1D_rot_para}
\end{align}
whereas when an external field is perpendicular to the 1D axis [Fig.~\ref{fig:dipole_lowD}(d)],
we take $(\hat{\bm e}_1,\hat{\bm e}_2,\hat{\bm e}_3)=(\hat{y},\hat{z},\hat{x})$, obtaining
\begin{align}
 \tilde{\bm Q}^{\rm (1D,rot\perp)}(k_x) 
&= -\frac{\pi}{3}[1-3G_2(k_x d)]\begin{pmatrix} 1 & 0 & 0 \\ 0 & 1 & 0 \\ 0 & 0 & -2 \end{pmatrix}.
\label{eq:Q1D_rot_perp}
\end{align}

To summarize the above results,
the DDI Hamiltonian for the $\ell$-dimensional system is written in a general form:
\begin{align}
 \hat{V}_{\rm dd} = \frac{c_{\rm dd}^{\rm (\ell D)}}{2} \int d^\ell r \int d^\ell r' \sum_{\nu,\nu'=x,y,z}:\hat{\bm F}^{\rm (\ell D)}_{\nu}(\bm r) Q^{\rm (\ell D)}_{\nu\nu'}(\bm r-\bm r')\hat{\bm F}^{\rm (\ell D)}_{\nu'}(\bm r'):,
\label{eq:Vdd_general}
\end{align}
and only the dipole kernel $Q^{\rm (\ell D)}_{\nu\nu'}(\bm r-\bm r')$ changes depending on the dimension and the presence or absence of an external field (i.e., whether we take time average):
Eqs.~\eqref{eq:Q2D_lab} and \eqref{eq:Q2D_lab2} are the bare dipole kernel for a 2D system,
Eqs.~\eqref{eq:Q2D_rot}, \eqref{eq:Q2D_rot_perp}, and \eqref{eq:Q2D_rot_para} are the time-averaged dipole kernel for a 2D system,
Eqs.~\eqref{eq:Q1D_lab} and \eqref{eq:Q1D_lab2} are the bare dipole kernel for a 1D system,
and Eqs.~\eqref{eq:Q1D_rot}, \eqref{eq:Q1D_rot_para}, and \eqref{eq:Q1D_rot_perp} are the time-averaged dipole kernel for a 1D system.
The functions $G_1(x)$ and $G_2(x)$ appearing in the dipole kernels are monotonically increasing functions
satisfying $G_1(0)=G_2(0)=0$ and $G_1(\infty)=G_2(\infty)=1$ [see Fig.~\ref{fig:dipole_lowD}(e)].
The contact interaction Hamiltonian is written in the same form as that for a 3D system 
by replacing the interaction coefficients {\it etc}. according to Eqs.~\eqref{eq:ci2D}--\eqref{eq:AFM2D} for a 2D system, and 
Eqs.~\eqref{eq:ci1D}--\eqref{eq:AFM1D} for a 1D system.

\subsubsection{Bogoliubov analysis}
\label{sec:dipole_Bogoliubov}

In this subsection, we extend the Bogoliubov theory discussed in Sec.~\ref{sec:Bogoliubov} so as to include the DDI.
In the presence of long-range interactions, the Bogoliubov equation becomes a complicated integral equation.
However, when atoms are condensed in the zero-momentum state, which is not always the ground state in the presence of the DDI, a simple extension of the Bogoliubov theory gives an analytic form of the excitation spectrum, from which we can investigate the Landau and dynamical instabilities.

We start from the general form of the DDI given in Eq.~\eqref{eq:Vdd_general} where the dimension $\ell$ takes 1, 2, or 3, and we use either the bare or time-averaged dipole kernel. 
Using the Fourier expansion of the field operator given by Eq.~\eqref{eq:FT_psi_m}, the Hamiltonian for the DDI is written as
\begin{align}
\hat{V}_{\rm dd} = \frac{1}{2\Omega} \sum_{m_1m_2m_1'm_2'}\sum_{\bm k_1 \bm k_2 \bm k_3 \bm k_4} [C_{\rm dd}(\bm k_4-\bm k_1)]^{m_1m_2}_{m_1'm_2'}
\delta_{\bm k_1+\bm k_2, \bm k_3+\bm k_4} \hat{a}^\dagger_{\bm k_1,m_1}\hat{a}^\dagger_{\bm k_2,m_2}\hat{a}_{\bm k_3,m_2'}\hat{a}_{\bm k_4,m_1'},
\label{eq:dipole_Vdd_FT}
\end{align}
where $[C_{\rm dd}(\bm k)]^{m_1m_2}_{m_1'm_2'}$ is defined by
\begin{align}
[C_{\rm dd}(\bm k)]^{m_1m_2}_{m_1'm_2'}
 = c_{\rm dd}\sum_{\nu\nu'} \tilde{Q}_{\nu\nu'}(\bm k) ({\rm f}_{\nu})_{m_1m_1'}({\rm f}_{\nu'})_{m_2m_2'}.
\label{eq:dipole_def_Cmnm'n'}
\end{align}
Here, $[C_{\rm dd}(\bm k)]^{m_1m_2}_{m_1'm_2'}$ satisfies the following symmetry property:
\begin{align}
[C_{\rm dd}(\bm k)]^{m_1m_2}_{m_1'm_2'} = [C_{\rm dd}(-\bm k)]^{m_1m_2}_{m_1'm_2'} = [C_{\rm dd}(\bm k)]^{m_2m_1}_{m_2'm_1'} = [C^*_{\rm dd}(\bm k)]_{m_1m_2}^{m_1'm_2'} = [C^*_{\rm dd}(\bm k)]_{m_2m_1}^{m_2'm_1'} ,
\end{align}
where we have used $\tilde{Q}_{\nu\nu'}(\bm k) = \tilde{Q}_{\nu\nu'}(-\bm k) = \tilde{Q}_{\nu'\nu}(\bm k) $ and ${\rm f}_\nu^{\rm T}={\rm f}_\nu^*$.
As in the conventional Bogoliubov theory, 
when the Bose-Einstein condensation occurs in the $\bm k=\bm 0$ state, we expand Eq.~\eqref{eq:dipole_Vdd_FT} up to the second order of $\hat{a}_{\bm k\neq\bm0,m}$ obtaining
\begin{align}
\hat{V}_{\rm dd}\simeq&\frac{1}{2\Omega}\sum_{m_1m_2m_1'm_2'}[C_{\rm dd}(\bm 0)]^{m_1m_2}_{m_1'm_2'} \hat{a}_{{\bm 0},m_1}^\dagger\hat{a}_{{\bm 0},m_2}^\dagger\hat{a}_{{\bm 0},m_2'}\hat{a}_{{\bm 0},m_1'}\nonumber\\
&+\frac{1}{\Omega}\sum_{m_1m_2m_1'm_2'}\sum_{{\bm k}\neq{\bm 0}}\left\{
[C_{\rm dd}(\bm k)]^{m_1m_2}_{m_1'm_2'} \hat{a}_{{\bm 0},m_1}^\dagger\hat{a}_{{\bm k},m_2}^\dagger\hat{a}_{{\bm 0},m_2'}\hat{a}_{{\bm k},m_1'}+
[C_{\rm dd}(\bm 0)]^{m_1m_2}_{m_1'm_2'} \hat{a}_{{\bm 0},m_1}^\dagger\hat{a}_{{\bm k},m_2}^\dagger\hat{a}_{{\bm k},m_2'}\hat{a}_{{\bm 0},m_1'}
\right\} \nonumber\\
&+\frac{1}{2\Omega}\sum_{m_1m_2m_1'm_2'}\sum_{{\bm k}\neq{\bm 0}}[C_{\rm dd}(\bm k)]^{m_1m_2}_{m_1'm_2'} \left(
\hat{a}_{{\bm 0},m_1}^\dagger\hat{a}_{{\bm 0},m_2}^\dagger\hat{a}_{-{\bm k},m_2'}\hat{a}_{{\bm k},m_1'}
+\hat{a}_{{\bm k},m_1}^\dagger\hat{a}_{-{\bm k},m_2}^\dagger\hat{a}_{{\bm 0},m_2'}\hat{a}_{{\bm 0},m_1'}\right).
\end{align}
By replacing $\hat{a}_{\bm 0,m}$'s according to Eqs.~\eqref{eq:Bog_a2zeta}--\eqref{eq:Bog_norm},
the dipolar interacting part the Bogoliubov Hamiltonian is given by
\begin{align}
\hat{H}^{\rm B}_{\rm dd} = &
\frac{N(N-1)}{2\Omega} \sum_{m_1m_2m_1'm_2'} [C_{\rm dd}({\bm 0})]^{m_1m_2}_{m_1'm_2'}\zeta_{m_1}^*\zeta_{m_2}^*\zeta_{m_2'}\zeta_{m_1'}\nonumber\\
&-\frac{2N-1}{2\Omega} \sum_{m_1m_2m_1'm_2'} [C_{\rm dd}({\bm 0})]^{m_1m_2}_{m_1'm_2'}\zeta_{m_1}^*\zeta_{m_2}^*\zeta_{m_2'}\zeta_{m_1'} \sum_{\bm k\neq\bm 0,m''}\hat{n}_{\bm km''}
\nonumber\\
&+\frac{N}{\Omega}\sum_{m_1m_2m_1'm_2'}\sum_{{\bm k}\neq{\bm 0}}\left\{
[C_{\rm dd}(\bm k)]^{m_1m_2'}_{m_1'm_2} + [C_{\rm dd}(\bm 0)]^{m_1m_2'}_{m_2m_1'}\right\}\zeta_{m_1'}\zeta_{m_2'}^*\hat{a}_{{\bm k},m_1}^\dagger\hat{a}_{{\bm k},m_2}
\nonumber\\
&+\frac{N}{2\Omega}\sum_{m_1m_2m_1'm_2'}\sum_{{\bm k}\neq{\bm 0}}\left\{
[C_{\rm dd}^*(\bm k)]^{m_1m_2}_{m_1'm_2'} \zeta_{m_1'}^*\zeta_{m_2'}^*\hat{a}_{-{\bm k},m_2}\hat{a}_{{\bm k},m_1}
+ [C_{\rm dd}(\bm k)]^{m_1m_2}_{m_1'm_2'}\zeta_{m_2'}\zeta_{m_1'}\hat{a}_{{\bm k},m_1}^\dagger\hat{a}_{-{\bm k},m_2}^\dagger\right\}.
\label{eq:Bog_Heff_dd}
\end{align}
Comparing Eqs.~\eqref{eq:Bog_Heff} and \eqref{eq:Bog_Heff_dd}, we can rewrite the Hamiltonian $\hat{H}^{\rm B}+\hat{H}^{\rm B}_{\rm dd}$ 
in the same form as Eq.~\eqref{eq:Bog_Heff_matrixform}:
\begin{align}
\hat{H}^{\rm B} + \hat{H}^{\rm B}_{\rm dd} =& E_{0} + E_{\rm dd}
-\frac{1}{2}\sum_{\bm k\neq\bm 0} \left\{{\rm Tr}[\bm H_{-\bm k}^{(0)}+ \bm H^{(1)} + \bm H^{(1,{\rm dd})}_{-\bm k}]-\frac{D^{\rm corr}}{2\epsilon_{\bm k}}\right\}\nonumber\\
&+ \frac{1}{2}\sum_{\bm k\neq\bm 0}
\begin{pmatrix} \bar{\hat{\bm a}}^\dagger_{\bm k} & \bar{\hat{\bm a}}_{-\bm k} \end{pmatrix}
\begin{pmatrix} \bm H^{(0)}_{\bm k} + \bm H^{(1)} + \bm H^{(1,{\rm dd})}_{\bm k} & \bm H^{(2)} + \bm H^{(2,{\rm dd})}_{\bm k} \\
[\bm H^{(2)} + \bm H^{(2,{\rm dd})}_{-\bm k}]^* & [\bm H^{(0)}_{-\bm k} + \bm H^{(1)} + \bm H^{(1,{\rm dd})}_{-\bm k}]^*\end{pmatrix}
\begin{pmatrix} \hat{\bm a}_{\bm k} \\ \hat{\bm a}^\dagger_{-\bm k} \end{pmatrix},
\end{align}
where
\begin{align}
 H^{\rm (1,dd)}_{\bm k,m_1m_2} =& \frac{N}{\Omega}\sum_{m_1'm_2'}\left\{[C_{\rm dd}(\bm k)]^{m_1m_2'}_{m_1'm_2} + [C_{\rm dd}(\bm 0)]^{m_1m_2'}_{m_2m_1'}\right\}\zeta_{m_1'}\zeta^*_{m_2'},
\label{eq:def_H^1dd}\\
 H^{\rm (2,dd)}_{\bm k,m_1m_2} =& \frac{N}{\Omega}\sum_{m_1'm_2'}[C_{\rm dd}(\bm k)]^{m_1m_2}_{m_1'm_2'}\zeta_{m_1'}\zeta_{m_2'},
\label{eq:def_H^2dd}\\
E_{\rm dd} =& \frac{N(N-1)}{2\Omega}\sum_{m_1m_2m_1'm_2'}[C_{\rm dd}({\bm 0})]^{m_1m_2}_{m_1'm_2'}\zeta_{m_1}^*\zeta_{m_2}^*\zeta_{m_2'}\zeta_{m_1'},
\end{align}
and $\mu$ in the definition of ${\bm H}^{(0)}_{\bm k}$ [see Eq.~\eqref{eq:def_H^0}] is replaced by $\mu-\mu_{\rm dd}$ with
\begin{align}
\mu_{\rm dd} =& \frac{2N-1}{2\Omega}\sum_{m_1m_2m_1'm_2'}[C_{\rm dd}({\bm 0})]^{m_1m_2}_{m_1'm_2'}\zeta_{m_1}^*\zeta_{m_2}^*\zeta_{m_2'}\zeta_{m_1'}.
\end{align}
Using Eqs.~\eqref{eq:rho-tilderho} and \eqref{eq:dipole_def_Cmnm'n'}, $\bm H^{\rm (1,dd)}_{\bm k}$ and $\bm H^{\rm (2,dd)}_{\bm k}$ are written in matrix forms as
\begin{align}
\bm H^{\rm (1,dd)}_{\bm k} =& c_{\rm dd}n\sum_{\nu\nu'}\left[\tilde{Q}_{\nu\nu'}(\bm k)\,{\rm f}_{\nu}{\bm \rho}{\rm f}_{\nu'} + \tilde{Q}_{\nu\nu'}({\bm 0})\,f_\nu{\rm f}_{\nu'}\right],\\
\bm H^{\rm (2,dd)}_{\bm k} =& c_{\rm dd}n\sum_{\nu\nu'}\tilde{Q}_{\nu\nu'}({\bm k})\,{\rm f}_{\nu}\tilde{\bm \rho}{\rm f}_{\nu'}^{\rm T}.
\end{align}
Then, the Bogoliubov analysis proceeds in the same manner as discussed in Sec.~\ref{sec:Bogoliubov} with ${\bm M}_{\bm k}^{\rm B}$ replaced by
\begin{align}
M^{\rm B,dd}_{\bm k}\equiv\begin{pmatrix} \bm H^{(0)}_{\bm k} + \bm H^{(1)} + \bm H^{(1,{\rm dd})}_{\bm k} & \bm H^{(2)} + \bm H^{(2,{\rm dd})}_{\bm k} \\
[\bm H^{(2)} + \bm H^{(2,{\rm dd})}_{-\bm k}]^* & [\bm H^{(0)}_{-\bm k} + \bm H^{(1)} + \bm H^{(1,{\rm dd})}_{-\bm k}]^*\end{pmatrix}.
\end{align}

An interesting example of the Bogoliubov analysis is the situation of Refs.~\cite{Vengalattore2008,Vengalattore2010}.
In Refs.~\cite{Vengalattore2008,Vengalattore2010}, Vengalattore {\it et al}. observed that a periodic magnetic pattern emerges in an $f=1$ BEC of $^{87}$Rb atoms in a quasi 2D system.
An external magnetic field was applied parallel to the 2D plane, and the longitudinal magnetization of the BEC was zero during the dynamics.
For this system, as we shall show below, one of the Bogoliubov spectra exhibits Landau and dynamical instabilities, which implies that the uniform spin configuration is unstable in this system.

\begin{figure}[ht]
\begin{center}
\resizebox{0.7\hsize}{!}{
\includegraphics{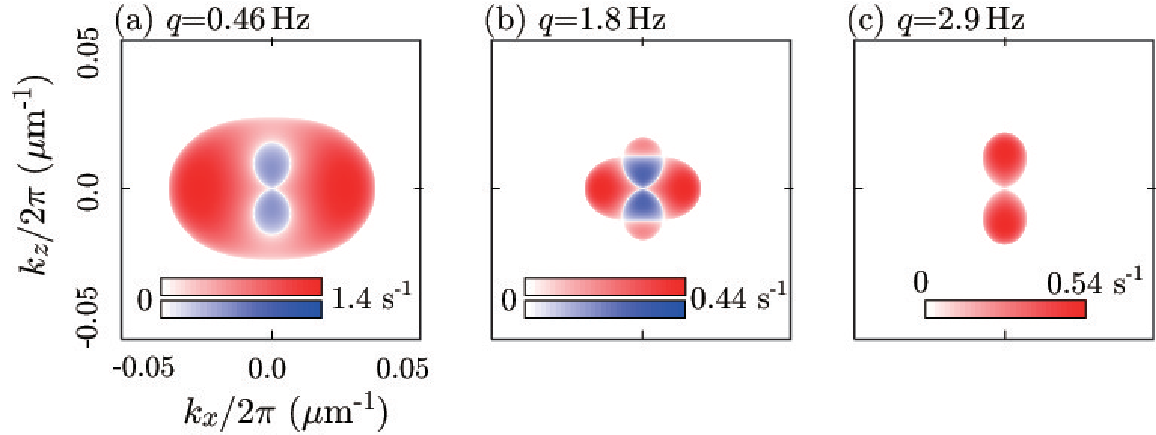}
}
\end{center}
\caption{Real and imaginary parts of $E_{\bm k,x}$ in Eq.~\eqref{dipole_spectrum} for the case of $d=1~\mu$m and $n=2.3\times 10^{14}$~cm$^{-3}$.
Shown are $|{\rm Im}\,E_{\bm k,x}|$ (red) and $-{\rm Re}\,E_{\bm k,x}$ (blue) which correspond to the dynamical and Landau instabilities, respectively.
Reprinted from Ref.~\cite{Kawaguchi2010}.
}
\label{fig:dipole_BdG}
\end{figure}

We take the $y$ and $z$ axes in the direction of the strong confinement and that of an external magnetic field, respectively.
In this situation, the dipole kernel is given by Eq.~\eqref{eq:Q2D_rot_para},
which, for convenience, is rewritten as follows:
\begin{align}
 \tilde{\bm Q}^{\rm (2D,rot||)}(\bm k) &= \mathcal{Q}(\bm k)\begin{pmatrix} 1 & 0 & 0 \\ 0 & 1 & 0 \\ 0 & 0 & -2 \end{pmatrix},\\
 \mathcal{Q}(\bm k) &= \frac{2\pi}{3}\left[1-3\frac{k_z^2}{k^2}G_1(k d)\right].
\end{align}
Here and hereafter we omit the index $\perp$ and define ${\bm k}=(k_x,k_z)$ and $k=\sqrt{k_x^2+k_z^2}$.
For a ferromagnetic interaction ($c_1<0$) and for a positive quadratic Zeeman energy satisfying $0<q<2|c_1|n$,
there is a uniform stationary solution of GPE~\eqref{spin-1GPE-DD} given by~\cite{Cherng2009,Kawaguchi2010}
\begin{align}
\bm \zeta = 
\begin{pmatrix}  
\sqrt{1-\tilde{q}}/2   \\
\sqrt{(1+\tilde{q})/2} \\
\sqrt{1-\tilde{q}}/2
\end{pmatrix},
\label{eq:dipole_zeta_q}
\end{align}
where 
\begin{eqnarray}
\tilde{q}=\frac{q}{2n\left[|c^{\rm (2D)}_1|+c^{\rm (2D)}_{\rm dd}\mathcal{Q}({\bm 0})\right]}.
\end{eqnarray}
The expectation value of the spin vector for this spinor is nonvanishing only for the $x$-component:
\begin{eqnarray}
f_{\nu} = \sum_{mm'} \zeta_m^* ({\rm f}_{\nu})_{mm'} \zeta_{m'}=\sqrt{1-\tilde{q}^2}\,\delta_{\nu x}.
\end{eqnarray}
Although all the matrix elements of ${\bm M}^{\rm B,dd}_{\bm k}$ are nonzero for the order parameter~\eqref{eq:dipole_zeta_q},
it is simplified if we use the Cartesian spin basis (see Sec.~\ref{sec:spin1_Cartesian}),
in which the eigenmatrix of the Bogoliubov equation is given by
\begin{align}
{\bm \sigma}_z
\begin{pmatrix} \mathcal{\bm U}^\dagger[\bm H^{(0)}_{\bm k} + \bm H^{(1)} + \bm H^{(1,{\rm dd})}_{\bm k}]\mathcal{\bm U} & \mathcal{\bm U}^\dagger [\bm H^{(2)} + \bm H^{(2,{\rm dd})}_{\bm k}]\mathcal{U} \\
\mathcal{U}^{\rm T}[\bm H^{(2)} + \bm H^{(2,{\rm dd})}_{-\bm k}]^*\mathcal{U}^* & \mathcal{U}^{\rm T}[\bm H^{(0)}_{-\bm k} + \bm H^{(1)} + \bm H^{(1,{\rm dd})}_{-\bm k}]^*\mathcal{U}^*\end{pmatrix},
\end{align}
where $\mathcal{U}$ is defined in Eq.\eqref{eq:spin1_orthogonal_U}.
In this basis, the $x$ mode is decoupled from the other two ($y$ and $z$) modes, and the spectrum is analytically obtained as~\cite{Kawaguchi2010}
\begin{align}
E_{\bm k,x}^2 
 =&\left\{\epsilon_{\bm k} + q - c^{\rm (2D)}_{\rm dd} n (1-\tilde{q})[2\mathcal{Q}({\bm k}) +\mathcal{Q}({\bm 0}) ] \right\}
\left\{\epsilon_{\bm k} + c^{\rm (2D)}_{\rm dd} n (1+\tilde{q})[\mathcal{Q}({\bm k})-\mathcal{Q}({\bm 0})] \right\}.
\label{dipole_spectrum}
\end{align}
Note here that when $c^{\rm (2D)}_{\rm dd}\neq 0$, the right-hand side of Eq.~\eqref{dipole_spectrum} can be negative, implying that an uniform initial state~\eqref{eq:dipole_zeta_q} is dynamically unstable~\cite{Cherng2009,Kawaguchi2010}.
Figures~\ref{fig:dipole_BdG} (a)--(c) show the distributions of $|{\rm Im}\,E_{\bm k,x}|/h$ (red) and $-{\rm Re}\,E_{\bm k,x}/h$ (blue)
which correspond to the dynamical instability and Landau instability, respectively, calculated for parameters in Ref.~\cite{Vengalattore2008}.
When $E_{\bm k,x}^2>0$, the sign of $E_{\bm k,x}$ is determined so that 
the corresponding eigenmode satisfies the normalization condition~\eqref{eq:norm_uv}.
The anisotropic distribution of the instability suggests the existence of the periodic pattern in real space.
However, the minimum wave length of the unstable modes ($\sim 30~\mu$m) is about 3 times larger than that of the observed magnetic pattern ($\sim 10~\mu$m),
and the origin of the observed magnetic pattern is under controversy~\cite{Cherng2009,Zhang2010,Kjall2009,Kawaguchi2010}.
References~\cite{Kjall2009,Kawaguchi2010} investigate stable spin textures in this system,
showing that the length scale of the stable pattern is larger than the unstable wavelength.
The equilibrium spin configuration achieved after a long-time evolution~\cite{Guzman2011} seems to be consistent with the theoretical calculations.

In the context of the Bogoliubov analysis for spinor dipolar BECs, the effects of the magnetic dipole-dipole interactions for the quench experiment~\cite{Sadler2006} are investigated in Refs.~\cite{Sau2009,Hoshi2010,Deuretzbacher2010}, and the energy spectra for spin textures appearing in the ground states (see Sec.~\ref{sec:dipole_gs}) were numerically calculated~\cite{Huhtamaki2011}.
Other related works include 
the study on possible quantum phases due to the DDI based on the single-mode approximation~\cite{Yi2004,Yi2006c},
the investigation of effects of the DDI on the spin dynamics in a Mott Insulator state (see Sec.~\ref{sec:exp_MI}) by using a variational approach~\cite{Sun2006}, and
the renormalization-group study on the phase transition in a 3D uniform spin-1 dipolar gas~\cite{Pietila2011}.

%% file: hydro.tex
\section{Hydrodynamic equations}
\label{sec:hydro}
In this section, we discuss basic properties of mass current and spin current of spinor condensates by deriving the hydrodynamic equations of motion for supercurrent and magnetization.

\subsection{Equations of motion for the superfluid velocity}

A spinor order parameter can, in general, be decomposed into a scalar part $\tilde{\psi}$ and a normalized spinor $\bm \zeta$ as
\begin{align}
 \bm\psi(\bm r,t) = \tilde{\psi}(\bm r, t)\bm\zeta(\bm r,t).
\label{eq:tildepsi}
\end{align}
We first show that the time and space dependences of $\bm\zeta$ 
act as scalar and vector potentials for the scalar field $\tilde{\psi}$.
We start with the Lagrangian density:
\begin{align}
 \mathcal{L} &= \mathcal{L}_0 - V,\\
 \mathcal{L}_0 &= \sum_m\left[i\hbar \psi_m^* \frac{\partial}{\partial t}\psi_m - \frac{\hbar^2}{2M} \bm\nabla\psi_m^* \cdot\bm \nabla \psi_m\right],\label{eq:hydro_L0}\\
 V &= \sum_m\left[U_{\rm trap}(\bm r) +pm+qm^2\right]|\psi_m|^2 + \frac{1}{2}\sum_{m_1m_2m_1'm_2'} C^{m_1m_2}_{m_1'm_2'}\psi_{m_1}^*\psi_{m_2}^*\psi_{m_2'}\psi_{m_1'}, \label{eq:hydro_V}
\end{align}
where $U_{\rm trap}(\bm r)$ is the trapping potential, $p$ and $q$ are the linear and quadratic Zeeman energies, and $C^{m_1m_2}_{m_1'm_2'}$ is defined in Eq.~\eqref{eq:def_Cmnm'n'}.
The GPEs are obtained from $\delta \mathcal{L}/\delta \psi^*_m=0$.
Substituting Eq.~\eqref{eq:tildepsi} in Eq.~\eqref{eq:hydro_L0}, $\mathcal{L}_0$ is expressed in terms of $\tilde{\psi}$ and $\bm\zeta$ as
\begin{align}
 \mathcal{L}_0 = i\hbar \tilde{\psi}^* \frac{\partial}{\partial t}\tilde{\psi} - \frac{\hbar^2}{2M} (\bm\nabla+i\bm A)\tilde{\psi}^* \cdot (\bm \nabla-i\bm A) \tilde{\psi} - \left(\frac{\hbar^2}{2M}\Lambda + \hbar \Phi\right)|\tilde{\psi}|^2,
\label{eq:hydro_L02}
\end{align}
where
\begin{align}
\Phi &= -i \bm\zeta^\dagger \frac{\partial}{\partial t}  \bm\zeta, \label{eq:hydro_A0}\\
{\bm A} &= i \bm\zeta^\dagger \bm \nabla \bm\zeta, \label{eq:hydro_vectA}\\
\Lambda &=  (\bm \nabla \bm\zeta^\dagger) \cdot (\bm\nabla \bm \zeta) - {\bm A}^2.
\end{align}
Here, $\Phi$ and ${\bm A}$ are real because $\bm\zeta^\dagger\bm\zeta=1$.
The Lagrangian density~\eqref{eq:hydro_L02} shows that $\hbar \Phi$ and $\hbar \bm A$ work as a scalar potential and a vector potential for a scalar field $\tilde{\psi}$ (with charge $1$), respectively.
The additional term $\Lambda$ gives the kinetic energy of $\bm \zeta$.
Defining the electromagnetic four-potential as $\elemag{A}^j = (\hbar\Phi,\hbar{\bm A})\ (j=0,1,2,3)$,
the electromagnetic field tensor is given by
\begin{align}
 \elemag{F}_{jk} &\equiv \partial_j \elemag{A}_k - \partial_k \elemag{A}_j\nonumber\\
 & = i\hbar\left[(\partial_j \bm\zeta^\dagger ) (\partial_k \bm\zeta) - (\partial_k \bm\zeta^\dagger ) (\partial_j \bm\zeta) \right]\ \ \ \ (j,k=0,1,2,3),
\label{eq:elemagF}
\end{align}
where we have used the metric $\eta_{jk}={\rm Diag}[-1,1,1,1]$.

The physical meaning of ${\bm A}$ is the superfluid velocity.
The superfluid velocity $\bm v^{\rm (mass)}$ for a spinor BEC is defined as the sum of currents in all spin components:
\begin{align}
n{\bm v}^{\rm (mass)} &= \frac{\hbar}{2Mi}\sum_{m=-f}^f \left[\psi_m^*(\nabla\psi_m)-(\nabla{\psi_m^*})\psi_m\right].
\label{eq:def_for_mass_supercurrent}
\end{align}
where $n=\sum_m|\psi_m|^2$ is the number density.
Substituting Eq.~\eqref{eq:tildepsi} in Eq.~\eqref{eq:def_for_mass_supercurrent}
and rewriting the scalar wavefunction as
\begin{align}
 \tilde{\psi}(\bm r,t) = \sqrt{n(\bm r,t)}e^{i\phi(\bm r,t)},
\label{eq:tildepsi2}
\end{align}
we obtain
\begin{align}
 \bm v^{\rm (mass)} = \frac{\hbar}{M}\left(\bm\nabla\phi - \bm A\right).
\label{eq:vmass_vs_A}
\end{align}
Hence, different from scalar BECs, the rotation of the superfluid velocity, in general, does not vanish in spinor BECs,
and the vorticity corresponds to the magnetic field $\elemag{\bm B}\equiv \bm \nabla \times (\hbar\bm A)$:
\begin{align}
\bm \omega \equiv \bm\nabla \times \bm v^{\rm (mass)} = -\frac{\elemag{\bm B}}{M}.
\label{eq:vorticity-B}
\end{align}

On the other hand, the scalar potential $\hbar\Phi$ works as an additional (local) chemical potential.
To see this, we rewrite the GPEs as 
\begin{align}
 i\hbar \sum_{m=-f}^f \psi_m^*\frac{\partial}{\partial t}\psi_m = \sum_{m=-f}^f \psi_m^*\left(-\frac{\hbar^2}{2M}\nabla^2 \psi_m + \frac{\delta V}{\delta \psi_m^*}\right),
\label{eq:GPE_hydro}
\end{align}
and substitute Eqs.~\eqref{eq:tildepsi} and \eqref{eq:tildepsi2} in Eq.~\eqref{eq:GPE_hydro}.
Then, the imaginary part reduces to the continuity equation for the particles:
\begin{align}
 \frac{\partial n}{\partial t} + \nabla \cdot [n{\bm v}^{\rm (mass)}]&=0,
\label{eq:continuity_mass}
\end{align}
where we have used the identity $\bm\zeta^\dagger \nabla^2\bm\zeta =  -i\bm\nabla\cdot\bm A - {\bm A}^2 - \Lambda$.
On the other hand, the real part of Eq.~\eqref{eq:GPE_hydro} is rewritten as 
\begin{align}
 \hbar\left(\frac{\partial \phi}{\partial t} + \Phi\right) + \mu_{\rm local}+\frac{1}{2}M[{\bm v}^{\rm (mass)}]^2 = 0,
\label{eq:GPE_hydro_real}
\end{align}
where
\begin{align}
 \mu_{\rm local}(\bm r, t) = -\frac{\hbar^2}{2M}\frac{\nabla^2 \sqrt{n}}{\sqrt{n}} + \frac{\hbar^2}{2M}\Lambda + \frac{1}{n}\sum_{m=-f}^f \psi_m^* \frac{\delta V}{\delta \psi_m^*}
\end{align}
is the local chemical potential.
The gradient of Eq.~\eqref{eq:GPE_hydro_real} gives the equation of motion for ${\bm v}^{\rm (mass)}$:
\begin{align}
 \frac{\partial}{\partial t}[M{\bm v}^{\rm (mass)}] + \bm \nabla \left\{\mu_{\rm local}+\frac{1}{2}M [{\bm v}^{\rm (mass)}]^2 \right\} = \elemag{E},
\label{eq:dvmass_dt}
\end{align}
where $\elemag{E}=\hbar (-\bm\nabla \Phi - \partial {\bm A}/\partial t)$ is the electric field due to ${\bm A}$ and $\Phi$.
Moreover, introducing the material derivative
\begin{align}
\frac{D}{Dt} \equiv \frac{\partial}{\partial t} + [{\bm v}^{\rm (mass)}\cdot \bm\nabla],
\label{eq:def_Dt}
\end{align}
and using the vector calculus formula
\begin{align}
 \bm\nabla ({\bm v}^2) = 2 {\bm v} \times (\bm\nabla \times \bm v) + 2 (\bm v\cdot\bm\nabla)\bm v,
\label{eq:nablav^2}
\end{align}
Eq.~\eqref{eq:dvmass_dt} reduces to the equation of motion which is analogous to that for a charged particle:
\begin{align}
 \frac{D}{Dt}[M{\bm v}^{\rm (mass)}] =  \elemag{E} + {\bm v}^{\rm (mass)}\times \elemag{B} - \bm \nabla \mu_{\rm local}.
\label{eq:dvmass_dt2}
\end{align}

\subsection{Spin supercurrent}
Spinor condensates also support the non-dissipative flow of spins.
As in the case of the superfluid velocity [see Eq.~\eqref{eq:def_for_mass_supercurrent}], 
the velocity field ${\bm v}^{\rm (spin)}_\nu$ for the $\nu$ component of spin is defined as follows:
\begin{align}
n{\bm v}^{\rm (spin)}_\nu &= \frac{\hbar}{2Mi}\sum_{m,m'=-f}^f({\rm f}_\nu)_{mm'}\left[\psi_m^*(\nabla\psi_{m'})-(\nabla{\psi_m^*})\psi_{m'}\right] \ \ \ (\nu=x,y,z).
\label{eq:def_vspin}
\end{align}
From the GPEs at zero magnetic field, we obtain the continuity equation for spin:
\begin{align}
 \frac{\partial }{\partial t} F_\nu + \nabla \cdot [n{\bm v}^{\rm (spin)}_\nu]&=0,
\label{eq:continuity_spin}
\end{align}
where $F_\nu(\bm r)=\sum_{mm'}({\rm f}_\nu)_{mm'}\psi_m^*(\bm r)\psi_{m'}(\bm r)$.
In the presence of an external magnetic field, the spin angular momentum perpendicular to it is not conserved,
and the torque term appears in the equation of motion:
\begin{align}
 \frac{\partial }{\partial t}F_\nu + \nabla \cdot [n{\bm v}^{\rm (spin)}_\nu] = 
\frac{p}{\hbar}\left[\hat{\bm B}\times{\bm F}\right]_\nu + \frac{2q}{\hbar} \left[\hat{\bm B} \times \mathcal{N}\hat{\bm B}\right]_\nu,
\label{eq:equation_of_mortion_for_F}
\end{align}
where $\hat{\bm B}$ is a unit vector in the direction of the magnetic field, $\mathcal{N}$ is the $3\times 3$ symmetric nematic tensor
defined in Eq.~\eqref{eq:def_nematic_tensor}, and $(\hat{\bm B}\times\mathcal{\bm N}\hat{\bm B})_\nu = \sum_{\nu_1\nu_2\nu_3} \epsilon_{\nu\nu_1\nu_2}\hat{B}_{\nu_1}\mathcal{N}_{\nu_2\nu_3}\hat{B}_{\nu_3}$~\cite{Kudo2010}.
The magnetic DDI induces an effective magnetic field and therefore modifies the first term on the right-hand side of Eq.~\eqref{eq:equation_of_mortion_for_F}.

\subsection{Long-wavelength limit}
In the long-wavelength limit, i.e., when the length scale of interest is much larger than the healing lengths $\xi_i\equiv \hbar/\sqrt{2M|c_i|n}$ ($i=0,1,\cdots$),
we can assume that the order parameter at every point belongs to the same phase.
Here we consider the case for $p=q=0$.
Then, the spin-dependent part of the order parameter is written as
\begin{align}
\bm\zeta(\bm r,t) = U(\alpha(\bm r,t),\beta(\bm r,t),\gamma(\bm r,t))\bm\zeta_0,
\label{eq:Uzeta0}
\end{align}
where 
\begin{align}
U(\alpha,\beta,\gamma) \equiv e^{-i{\rm f}_z\alpha}e^{-i{\rm f}_y\beta}e^{-i{\rm f}_z\gamma}
\label{eq:def_EulerU}
\end{align}
is the Euler rotation with $\alpha,\beta$ and $\gamma$ being Euler angles, and $\bm\zeta_0$ is a representative order parameter that minimizes the interaction energy.
Since the interatomic interaction is invariant under SO(3) spin rotations, 
the order parameter~\eqref{eq:Uzeta0} for arbitrary Euler angles minimizes the interaction energy.

Using Eq.~\eqref{eq:Uzeta0}, the electromagnetic four-potential is calculated as follows:
\begin{align}
 \elemag{A}_j \equiv& i \hbar \bm\zeta^\dagger_0 U^\dagger (\partial_jU) \bm\zeta_0\nonumber\\
=& \hbar\bm\zeta_0^\dagger \left[(\partial_j \alpha) e^{i{\rm f}_z\gamma} e^{i{\rm f}_y\beta} {\rm f}_z e^{-i{\rm f}_y\beta}e^{-i{\rm f}_z\gamma} 
+ (\partial_j\beta) e^{i{\rm f}_z\gamma} {\rm f}_y e^{-i{\rm f}_z\gamma}
+ (\partial_j\gamma) {\rm f}_z\right]\bm\zeta_0
\nonumber\\
=& \hbar[-(\partial_j\alpha)\sin\beta\cos\gamma + (\partial_j\beta)\sin\gamma]\langle{\rm f}_x\rangle_0\nonumber\\
& + [(\partial_j\alpha)\sin\beta\sin\gamma + (\partial_j\beta)\cos\gamma]\langle{\rm f}_y\rangle_0 + [(\partial_j\alpha)\cos\beta+\partial_j\gamma] \langle{\rm f}_z\rangle_0,
\end{align}
where $\langle{\rm f}_\nu\rangle_0\equiv \bm\zeta_0^\dagger{\rm f}_\nu\bm\zeta_0$.
Without loss of generality, 
we can choose the representative order parameter $\bm\zeta_0$ so that the spontaneous magnetization, if it exists,  becomes parallel to the $z$ axis.
Then, we obtain
\begin{align}
 \elemag{A}_j =\hbar|{\bm f}| [(\partial_j\alpha)\cos\beta+\partial_j\gamma],
\label{eq:elemagA_gen}
\end{align}
where $|\bm f|$ is the amplitude of the spontaneous magnetization, which is a constant by assumption.
Equation~\eqref{eq:elemagA_gen} implies that the scalar and vector potentials arise only when the condensate has spontaneous magnetization.
Moreover, noting that the Euler angles $\alpha$ and $\beta$ describe the direction of the local magnetization:
\begin{align}
\bm f
= \begin{pmatrix} \bm\zeta_0^\dagger U^\dagger {\rm f}_x U\bm\zeta_0 \\ \bm\zeta^\dagger_0 U^\dagger {\rm f}_y U \bm\zeta_0 \\ \bm\zeta^\dagger_0 U^\dagger {\rm f}_z U\bm\zeta_0  \end{pmatrix}
= |{\bm f}|\begin{pmatrix} \cos\alpha\sin\beta \\ \sin\alpha\sin\beta \\ \cos\beta \end{pmatrix},
\end{align}
the electromagnetic field tensor defined in Eq.~\eqref{eq:elemagF} is written
in terms of a unit vector $\hat{\bm s}\equiv \bm f/|\bm f|$ as
\begin{align}
\elemag{F}_{jk} &= \hbar|{\bm f}| \sin\beta[ (\partial_j\alpha) (\partial_k\beta) - (\partial_k\alpha)(\partial_j\beta)]\nonumber\\
&= -\hbar |{\bm f}| \sum_{\nu_1\nu_2\nu_3} \epsilon_{\nu_1\nu_2\nu_3}\hat{s}_{\nu_1} (\partial_j\hat{s}_{\nu_2}) (\partial_k\hat{s}_{\nu_3}).
\label{eq:elemagF_MFform}
\end{align}
In particular, Eq.~\eqref{eq:elemagF_MFform} for $j,k\neq 0$ gives the vorticity [see Eq.~\eqref{eq:vorticity-B}]:
\begin{align}
\bm\nabla \times \bm v^{\rm (mass)} = \frac{\hbar|\bm f|}{M} \sum_{\nu_1\nu_2\nu_3} \epsilon_{\nu_1\nu_2\nu_3}\hat{s}_{\nu_1} (\bm \nabla\hat{s}_{\nu_2} \times \bm \nabla\hat{s}_{\nu_3}),
\label{eq:MH-relation}
\end{align}
which is non-vanishing for $|\bm f|\neq 0$.
Equation~\eqref{eq:MH-relation} is known as the Mermin-Ho relation~\cite{Mermin1976}.

As a special case of interest,
we consider the ferromagnetic phase in which the condensate is fully polarized at every point and only the direction of the magnetization varies in space and time.
Using Eq.~\eqref{eq:elemagF_MFform}, 
the equation of motion of ${\bm v}^{\rm (mass)}$ [Eq.~\eqref{eq:dvmass_dt}] for a spin-$f$ system is given by~\cite{Barnett2009,Kudo2011}
\begin{align}
 \frac{\partial}{\partial t}[M{\bm v}^{\rm (mass)}] &=  \hbar f \sum_{\nu_1\nu_2\nu_3}\epsilon_{\nu_1\nu_2\nu_3}\hat{s}_{\nu_1}
\frac{\partial \hat{s}_{\nu_2}}{\partial t}\bm \nabla \hat{s}_{\nu_3}
 - \bm \nabla \left\{\mu_{\rm local} +\frac{1}{2}M[{\bm v}^{\rm (mass)}]^2\right\},
\label{eq:dvmass_dt_F}\\
\mu_{\rm local} &= -\frac{\hbar^2}{2M}\frac{\nabla^2\sqrt{n}}{\sqrt{n}} + \frac{\hbar^2 f}{4M}(\bm\nabla \hat{\bm s})^2 + U_{\rm trap}(\bm r) + \frac{4\pi\hbar^2}{M}a_{2f}n,
\end{align}
where $a_{2f}$ is the scattering length for the spin polarized atoms, and we have used $\Lambda=f(\bm \nabla\hat{\bm s})^2/2$.
On the other hand, the spin superfluid velocity~\eqref{eq:def_vspin} for a spin-$f$ system is written in terms of $\hat{\bm s}$ as~\cite{Lamacraft2008,Barnett2009,Kudo2010}
\begin{align}
 {\bm v}^{\rm (spin,F)}_\nu &= f\left(\hat{s}_\nu{\bm v}^{\rm (mass,F)} - \frac{\hbar}{2M}\sum_{\nu_1\nu_2}\epsilon_{\nu\nu_1\nu_2}\hat{s}_{\nu_1}{\bm \nabla}\hat{s}_{\nu_2}\right).
\label{eq:vspin_F_gen}
\end{align}
The first term on the right-hand side of Eq.~\eqref{eq:vspin_F_gen} is the flow of spins carried by the particle,
and the second term originates from the spin flow induced by the gradient of spins that is perpendicular to $\hat{\bm s}$.
Since the energy scale of spin waves is much smaller than that of density waves ($|c_i|n\ll c_0n$),
spin dynamics does not affect the density profile of the condensate, especially in the long-wavelength limit.
We then take the incompressible limit: $\partial n/\partial t=-\bm\nabla\cdot \bm (nv^{\rm (mass)})=0$ [see Eq.~\eqref{eq:continuity_mass}].
Substituting Eq.~\eqref{eq:vspin_F_gen} in Eq.~\eqref{eq:continuity_spin},
we obtain the equation of motion for the magnetization~\cite{Lamacraft2008}
\begin{align}
\frac{D \hat{\bm s}}{D t}  = - \hat{\bm s} \times {\bm B}_{\rm eff},
\label{eq:HDE_spinf}
\end{align}
where 
\begin{align}
 \bm B_{\rm eff} = -\frac{\hbar^2}{2M} \left[\left(\frac{\bm \nabla n}{n}\cdot\bm \nabla\right)\hat{\bm s}+\nabla^2\hat{\bm s}\right].
\label{eq:HDE_Beff}
\end{align}
We note that Eq.~\eqref{eq:HDE_spinf} has the same form as the Landau-Lifshits equation (LLE),
which is widely used to describe the magnetization dynamics in ferromagnets, 
if the partial time derivative in the LLE is replaced by the material derivative~\eqref{eq:def_Dt}.
An external field merely changes the effective field ${\bm B}_{\rm eff}$ to~\cite{Kudo2010}
\begin{align}
 \bm B'_{\rm eff} = -\frac{\hbar^2}{2M} \left[\left(\frac{\bm\nabla n}{n}\cdot\bm \nabla\right)\hat{\bm s}+\nabla^2\hat{\bm s}\right] + \frac{p}{\hbar}\hat{\bm B} + \frac{q}{\hbar}(\hat{\bm B}\cdot\hat{\bm s})\hat{\bm B}.
\end{align}
When we introduce a phenomenological energy dissipation to the GPE by replacing $i(\partial/\partial t)$ with $(i-\Gamma) (\partial/\partial t)$,
the dissipative hydrodynamic equation, which is obtained in a manner similar to Eq.~\eqref{eq:HDE_spinf}, includes the so-called Girbart damping $\Gamma \hat{\bm s}\times \partial \hat{\bm s}/\partial t$
and takes the same form as the Landau-Lifshits-Girbart equation~\cite{Kudo2011}.
Equations~\eqref{eq:dvmass_dt_F} and \eqref{eq:HDE_spinf}, together with the Mermin-Ho relation~\eqref{eq:MH-relation} with $|\bm f|=f$, provide a complete set of equations of motion that describe a general ferromagnetic condensate.
The above set of equations of motion is used to analyze the stable Skyrmion lattice configurations~\cite{Cherng2011a,Cherng2011b}.
The hydrodynamic equation for a general state is discussed in Refs.~\cite{Barnett2009,Lamacraft2010}:
In Ref.~\cite{Barnett2009}, the equations of motion for the vertices which appear in Majorana representation (see Sec.~\ref{sec:Majorana}) are derived,
whereas Ref.~\cite{Lamacraft2010} gives the equations of motion for the SO(3) rotation matrix ($\sim U^\dagger \partial_j U$).

%% file: vortices.tex
\section{Vortices and hydrodynamic properties}
\label{sec:Vortices}

A scalar BEC can host only one type of vortex, that is, a U(1) vortex or a gauge vortex.
However, a spinor BEC can host many different types of vortices.  
The properties of a vortex can be characterized by looking at how the order parameter changes along a loop that encircles the vortex. 
For the case of a scalar BEC,
the order parameter is a single complex function which can be written as
\begin{align}
\psi({\bm r}) = \sqrt{n(\bm r)}e^{i\phi(\bm r)},
\label{eq:Vort_psi_scalar}
\end{align}
where $n({\bm r})$ is the particle number density.
The corresponding superfluid velocity ${\bm v}_{\rm s}$ is defined by
\begin{align}
n{\bm v}_{\rm s}
=\frac{\hbar}{2M i} [ \psi^\ast \boldsymbol{\nabla} \psi - (\boldsymbol{\nabla}\psi^\ast) \psi].
\label{eq:Vort_vmass_scalar}
\end{align}
Substituting Eq.~\eqref{eq:Vort_psi_scalar} in Eq.~\eqref{eq:Vort_vmass_scalar}, we obtain
\begin{align}
{\bm v}_{\rm s}=\frac{\hbar}{M}{\bm\nabla}\phi.
\end{align}
Because the order parameter must be single-valued, the phase $\phi$ can change only by an integer multiple of $2\pi$ when we make a complete circuit of a vortex.
Hence, the mass circulation along a closed path is quantized in units of $\kappa\equiv h/M$:
\begin{align}
 \oint_\mathcal{C} {\bm v}_{\rm s}\cdot d{\bm \ell} = \frac{\hbar}{M}\oint_\mathcal{C} d\phi = n_{\rm w} \kappa,
\label{eq:Vort_scalar_quantization}
\end{align}
where $\oint_{\,\mathcal{C}}d{\bm \ell}$ is the line integral along a closed path $\mathcal{C}$, and $2\pi n_{\rm w}$ is the phase change along the path
with $n_{\rm w}$ being an integer (winding number).
When the vortex lies on the $z$ axis and the system is axisymmetric about the vortex line,
the order parameter is described in the cylindrical coordinate $(r,\varphi, z)$ by
\begin{align}
 \psi(r,\varphi,z) = \sqrt{n(r)}\, e^{in_{\rm w}\varphi}.
\end{align}
If $n_{\rm w}\neq 0$, the density at the vortex core must vanish so as to avoid the phase singularity.

The situation changes drastically for spinor BECs.
To see this, we consider how the spinor order parameter changes along a loop encircling a vortex.
When the distance of the loop from the vortex core is larger than the healing lengths $\hbar/\sqrt{2Mc_i n}$ ($i=0,1,\cdots$),
the order parameter changes slowly along the loop, and therefore, the kinetic energy density associated with the spatial variation of the order parameter is negligible.
In such a case, the order parameter takes the form that minimizes the interaction energy.
Because the spinor system has the U(1) global gauge symmetry and the SO(3) spin rotational symmetry (in the absence of an external field),
the order parameter far from the vortex core is generally
described using gauge-transformation and spin-rotation operators by
\begin{align}
\bm \psi(t) = \sqrt{n(t)}\,e^{i\phi(t)}U(\alpha(t),\beta(t),\gamma(t))\bm \zeta_0,
\label{eq:Vort_genpsi_spinor}
\end{align}
where $0\le t < 1$ is a parameter describing a closed contour,
$\phi$ is an overall gauge, $U(\alpha,\beta,\gamma)$ is an SO(3) rotation defined in Eq.~\eqref{eq:def_EulerU},
and ${\bm \zeta}_0$ is a representative spinor which minimizes the interaction energy.
The single-valuedness condition for the order parameter~\eqref{eq:Vort_genpsi_spinor} is given by
\begin{align}
e^{i\phi(0)}U(\alpha(0),\beta(0),\gamma(0))\bm \zeta_0 = e^{i\phi(1)}U(\alpha(1),\beta(1),\gamma(1))\bm \zeta_0,
\label{eq:Vort_SVC}
\end{align}
whereas the number density $n(t)$ may fluctuate but should satisfy $n(0)=n(1)$.
The condition for $\phi,\alpha,\beta$, and $\gamma$ depends on the symmetry of $\bm \zeta_0$.
For example, as we shall see below, $\phi(1)-\phi(0)$ may take integer multiples of $\pi$ or $2\pi/3$, 
depending on how $\alpha, \beta$ and $\gamma$ vary.
In such cases, the circulation of the velocity field becomes fractional in units of $\kappa$.

As we shrink the loop, 
the kinetic energy density increases, and becomes comparable with the interaction energy density.
In such a situation, the order parameter on the loop is no longer described by Eq.~\eqref{eq:Vort_genpsi_spinor}.
It does not mean, however, that the number density $n(\bm r)$ should vanish at the vortex core: the vortex core can be filled with particles that show different magnetism from those far from the vortex.
This is also a feature of vortices in spinor BECs that is different from scalar ones~\cite{Kobayashi2009b,Kobayashi2012b}.

In the following sections, we first discuss general properties of the mass circulation in Sec.~\ref{sec:Vort_circulation},
and then investigate what types of vortices are allowed in each phase of spin-1 and spin-2 BECs in Secs.~\ref{sec:Vort_spin1} and \ref{sec:Vort_spin2}, respectively,
which are followed by discussions on rotating spinor BECs (Sec.~\ref{sec:Vort_rotating}) and dipolar gases (Sec.~\ref{sec:Vort_dipole}).
In Secs.~\ref{sec:Vort_spin1} and \ref{sec:Vort_spin2}, we consider vortices in a homogeneous system in the absence of an external magnetic field, 
and do not take into account the particle and spin conservations, unless otherwise noted.
We shall discuss vortices in Sec.~\ref{sec:topology} from the viewpoint of topology after reviewing homotopy theory.

\subsection{Mass circulation}
\label{sec:Vort_circulation}

As we have seen in the previous section, the superfluid velocity in a spinor BEC is defined in Eq.~\eqref{eq:def_for_mass_supercurrent} and given by
\begin{align}
 {\bm v}^{\rm (mass)} 
&= \frac{\hbar}{M}\left[\bm\nabla\phi -|{\bm f}|(\cos\beta\bm\nabla\alpha +\bm\nabla\gamma)\right]
\label{eq:Vort_vmass_gen}
\end{align}
for the order parameter~\eqref{eq:Vort_genpsi_spinor} [see Eqs.~\eqref{eq:vmass_vs_A} and \eqref{eq:elemagA_gen}].

Equation \eqref{eq:Vort_vmass_gen} shows that
when $|\bm f|= 0$ the superfluid velocity is, as in the case of a scalar BEC, proportional to the gradient of the gauge angle: ${\bm v}_s = (\hbar/M)\boldsymbol{\nabla}\phi$.
Therefore, spinor BECs with zero magnetization is irrotational ($\bm\nabla \times {\bm  v}_{\rm s} =0$) and the circulation is quantized, though the unit of the quantization is not necessarily equal to $\kappa$ but can be a fraction of it. (See the examples in the following subsections.)

On the other hand, if $|\bm f|\neq 0$, the vorticity $\bm\nabla \times \bm v^{\rm (mass)}$ does not vanish because of the term $\cos\beta\bm\nabla\alpha$,
and hence, the circulation is no longer quantized.
This is due to the contribution from the Berry phase associated with spin textures.
To see this, we use the Mermin-Ho relation~\eqref{eq:MH-relation}.
If there is no singularity inside a loop under consideration, i.e., if the order parameter at every point inside the loop is described by Eq.~\eqref{eq:Vort_genpsi_spinor}, 
the mass circulation along the loop is calculated as follows:
\begin{align}
 \oint_\mathcal{C} {\bm v}^{\rm (mass)}\cdot d{\bm \ell} 
&= \int_S (\bm\nabla \times \bm v^{\rm (mass)})\cdot  d{\bm S} \nonumber\\
&=\frac{\hbar |\bm f|}{M}  \int_S \sin\beta \,(\bm{\nabla}\beta \times \bm{\nabla}\alpha)\cdot d{\bm S}\nonumber \\
&=\frac{\hbar |\bm f|}{M}  \iint\sin\beta d\beta d\alpha\nonumber\\
&=\frac{\hbar|\bm f|}{M} S_{\rm in}(\hat{\bm s}),
\label{eq:Vort_circulation1}
\end{align}
where $\int_S d\bm S$ is the surface integral over a surface $S$ enclosed by $\mathcal{C}$, and
\begin{align}
 S_{\rm in}(\hat{\bm s})\equiv \iint\sin\beta d\beta d\alpha
\label{eq:Berry1}
\end{align}
represents the surface area of the region on a unit sphere on which $\hat{\bm s}({\bm r})\, (\bm r\in S)$ moves.
Here, $S_{\rm in}(\hat{\bm s})$ agrees modulo $4\pi$ with the Berry phase $S(\hat{\bm s})$
that is defined as the surface area of the region on a unit sphere {\it enclosed by the trajectory of} $\hat{\bm s}(\bm r)\, (\bm r\in \mathcal{C})$ [see Eq.~\eqref{eq:Berry2}].
Hence, Eq.~\eqref{eq:Vort_circulation1} shows that the mass circulation of a ferromagnetic condensate is directly related to the Berry phase of the magnetization.
For a general case, we rewrite Eq.~\eqref{eq:Vort_vmass_gen} as
\begin{align}
{\bm v}^{\rm (mass)} -\frac{\hbar}{M}|\bm f| (1-\cos\beta) \bm\nabla\alpha
=\frac{\hbar}{M}\;\boldsymbol{\nabla} [\phi-|\bm f|(\alpha+\gamma)],
\end{align}
and integrate both sides along a closed contour $\mathcal{C}$,
obtaining
\begin{align}
\oint_\mathcal{C} {\bm v}^{\rm (mass)} \cdot d{\bm \ell} - \frac{\hbar}{M}
 |\bm f|\oint_\mathcal{C} (1-\cos\beta)\bm\nabla\alpha \cdot d\bm \ell
=\frac{\hbar}{M}\left[\oint_\mathcal{C} d\phi - |\bm f|\oint_\mathcal{C}(d\alpha+d\gamma)\right].
\label{eq:Vort_qntz_vmass}
\end{align}
Since $\alpha$ and $\beta$ specify the direction of the local magnetization, the single-valuedness condition for the order parameter requires that
\begin{align}
 \oint_\mathcal{C} d\alpha &= 2\pi n_\alpha \ \ \ (n_\alpha:\textrm{integer}),\label{eq:Vort_SVC_alpha}\\
 \oint_\mathcal{C} d\beta &= 0. \label{eq:Vort_SVC_beta}
\end{align}
The same condition also dictates that $\oint_{\,\mathcal{C}}d\phi$ and $\oint_{\,\mathcal{C}} d\gamma$ on the right-hand side of Eq.~\eqref{eq:Vort_qntz_vmass} be integer multiples or fractions of $2\pi$.
On the other hand, the second term on the left-hand side is proportional to the Berry phase:
\begin{align}
 \oint_\mathcal{C} (1-\cos\beta)\bm\nabla\alpha \cdot d\bm \ell = \int_{\alpha(0)}^{\alpha(0)+2\pi n_\alpha} d\alpha \int_{0}^{\beta(\alpha)}  d\beta  \sin\beta \equiv S(\hat{\bm s}).
\label{eq:Berry2}
\end{align}
Equation~\eqref{eq:Vort_qntz_vmass} therefore shows that the difference between the circulation and the Berry phase: $\oint_{\,\mathcal{C}} {\bm v}^{\rm (mass)} \cdot d{\bm \ell} - (\hbar/M) |\bm f| S(\bm f)$ is quantized.

\subsection{Spin-1 BEC}
\label{sec:Vort_spin1}

The rotation matrix for the order parameter of a spin-1 BEC is given by
\begin{align}
U(\alpha,\beta,\gamma) 
= \begin{pmatrix}
e^{-i(\alpha+\gamma)} \cos^2 \frac{\beta}{2} \ & \;
-\frac{e^{-i\alpha}}{\sqrt{2}} \sin \beta \ & \; e^{-i (\alpha-\gamma)}
\sin^2 \frac{\beta}{2} \\[2mm]
\frac{e^{-i\gamma}}{\sqrt{2}} \sin \beta \ & \; \cos \beta \ & \;
-\frac{e^{i\gamma}}{\sqrt{2}} \sin \beta \\[2mm]
e^{i(\alpha-\gamma)} \sin^2 \frac{\beta}{2} \ & \;
\frac{e^{i\alpha}}{\sqrt{2}} \sin \beta \ & \; e^{i(\alpha+\gamma)}
\cos^2 \frac{\beta}{2}
\end{pmatrix}.
\label{spin1-rotm}
\end{align}
The spin-1 BEC in the absence of an external magnetic field has two phases: ferromagnetic and polar. Below, we discuss characteristic vortices in each phase.

\subsubsection{Ferromagnetic phase}
\label{sec:vortex_spin1ferro}

A representative order parameter of a spin-1 ferromagnetic BEC is ${\bm \zeta}^{\rm ferro}_0=(1,0,0)^{\rm T}$. 
Substituting this and Eq.~(\ref{spin1-rotm}) in Eq.~\eqref{eq:Vort_genpsi_spinor}, we obtain a general order parameter of a spin-1 ferromagnetic BEC far away from the vortex core:
\begin{align}
{\bm \psi}^{\rm ferro}
=\sqrt{n}e^{i\phi}U(\alpha,\beta,\gamma)\bm\zeta^{\rm ferro}_0
=\sqrt{n}e^{i(\phi-\gamma)}
\begin{pmatrix}
e^{-i\alpha}  \cos^2\frac{\beta}{2} \\[1mm]
\frac{1}{\sqrt{2}} \sin \beta \\[1mm]
e^{i\alpha} \sin^2 \frac{\beta}{2}
\end{pmatrix}.
\label{spin1ferro}
\end{align}
The linear combination $\phi-\gamma$ in Eq.~\eqref{spin1ferro} reflects the spin-gauge symmetry which implies that the rotation in spin space through angle $\gamma$ is equivalent to the gauge transformation by $\phi=-\gamma$.
For simplicity of notation, we set $\phi'=\phi-\gamma$.

Because $|\bm f|\neq 0$, the mass circulation is not quantized in the ferromagnetic BEC.
Instead, substituting $|\bm f|=1$ in Eq.~\eqref{eq:Vort_qntz_vmass}, we obtain
\begin{align}
\oint_\mathcal{C} {\bm v}^{\rm (mass)} \cdot d{\bm \ell} - \frac{\hbar}{M}S(\hat{\bm s})
=\frac{h}{M} n_{\rm w} \ \ \ (n_{\rm w}:\textrm{integer}),
\label{eq:Vort_spin1F_circ}
\end{align}
where we have used the following condition for $\phi'$:
\begin{align}
 \oint_\mathcal{C} d\phi' = 2\pi n_{\phi'} \ \ \ (n_{\phi'}:\textrm{integer}),
\label{eq:Vort_spin1_qntz}
\end{align}
which is derived from the single-valuedness condition~\eqref{eq:Vort_SVC} together with Eqs.~\eqref{eq:Vort_SVC_alpha}, \eqref{eq:Vort_SVC_beta} and \eqref{spin1ferro}, and 
$n_{\rm w}$ in Eq.~\eqref{eq:Vort_spin1F_circ} is defined by $n_{\rm w}\equiv n_{\phi'}-n_\alpha$.

Equation~\eqref{eq:Vort_spin1F_circ} implies that the mass circulation can change continuously by manipulating spin textures, i.e. spatial spin configurations.
As a special case, let us consider the order parameter~\eqref{spin1ferro} with $\alpha=-n_\alpha\varphi$ and $\phi'=n_\alpha\varphi$ where $\varphi$ denotes an azimuthal angle in the cylindrical coordinate.
Then, Eq.~(\ref{spin1ferro}) changes from ${\bm \zeta}=(e^{i2n_\alpha\varphi},0,0)^{\rm T}$ to $(0,0,1)^{\rm T}$ as $\beta$ changes from 0 to $\pi$.
This implies that the vortex with the winding number $2n_\alpha$ is topologically unstable.
On the other hand, if we take $\alpha=-n_\alpha\varphi$ and $\phi'=(n_\alpha+1)\varphi$,
then Eq.~(\ref{spin1ferro}) changes from ${\bm \zeta}=(e^{i(2n_\alpha+1)\varphi},0,0)^{\rm T}$ to $(0,0,e^{i\varphi})^{\rm T}$ as $\beta$ changes from 0 to $\pi$.
Thus, the vortex with the winding number $2n_\alpha+1$ is unstable against the decay into the singly quantized vortex which is stable~\cite{Ho1998}.

We can utilize these properties to create a doubly quantized vortex from a vortex-free state by changing $\beta$ from $0$ to $\pi$ while keeping $\alpha=\varphi$ and $\phi'=-\varphi$~\cite{Nakahara2000,Isoshima2000,Ogawa2002,Mottonen2002}.
Such a scheme is realized if a spin polarized BEC is prepared in the $m=1$ state under a quadrupole field 
${\bm B}(r,\varphi,z)=(B_\perp(r,z) \cos(-\varphi),B_\perp(r,z)\sin(-\varphi),B_z(r,z))^{\rm T}$, 
where $(r,\varphi,z)$ are the cylindrical coordinates and $B_\perp>0$.
At $t=0$, $B_z$ is set to be positive and much greater than $B_\perp$.
As $B_z$ is changed adiabatically from $B_z\gg B_\perp$ to $B_z\ll -B_\perp$, the atomic spins follow
the direction of the local magnetic field as $\alpha=\varphi$ and $\beta=\arctan(B_\perp/B_z)$,
resulting in the transformation of the order parameter from $(1,0,0)^{\rm T}$ to $(0,0,e^{-2i\varphi})^{\rm T}$; thus, a doubly quantized vortex is imprinted.
Using this method, Leanhardt {\it et al}.~\cite{Leanhardt2002} and Kumakura {\it et al}.~\cite{Kumakura2006} observed multiply quantized vortices.

\begin{figure}[ht]
\begin{center}
\resizebox{0.7\hsize}{!}{
\includegraphics{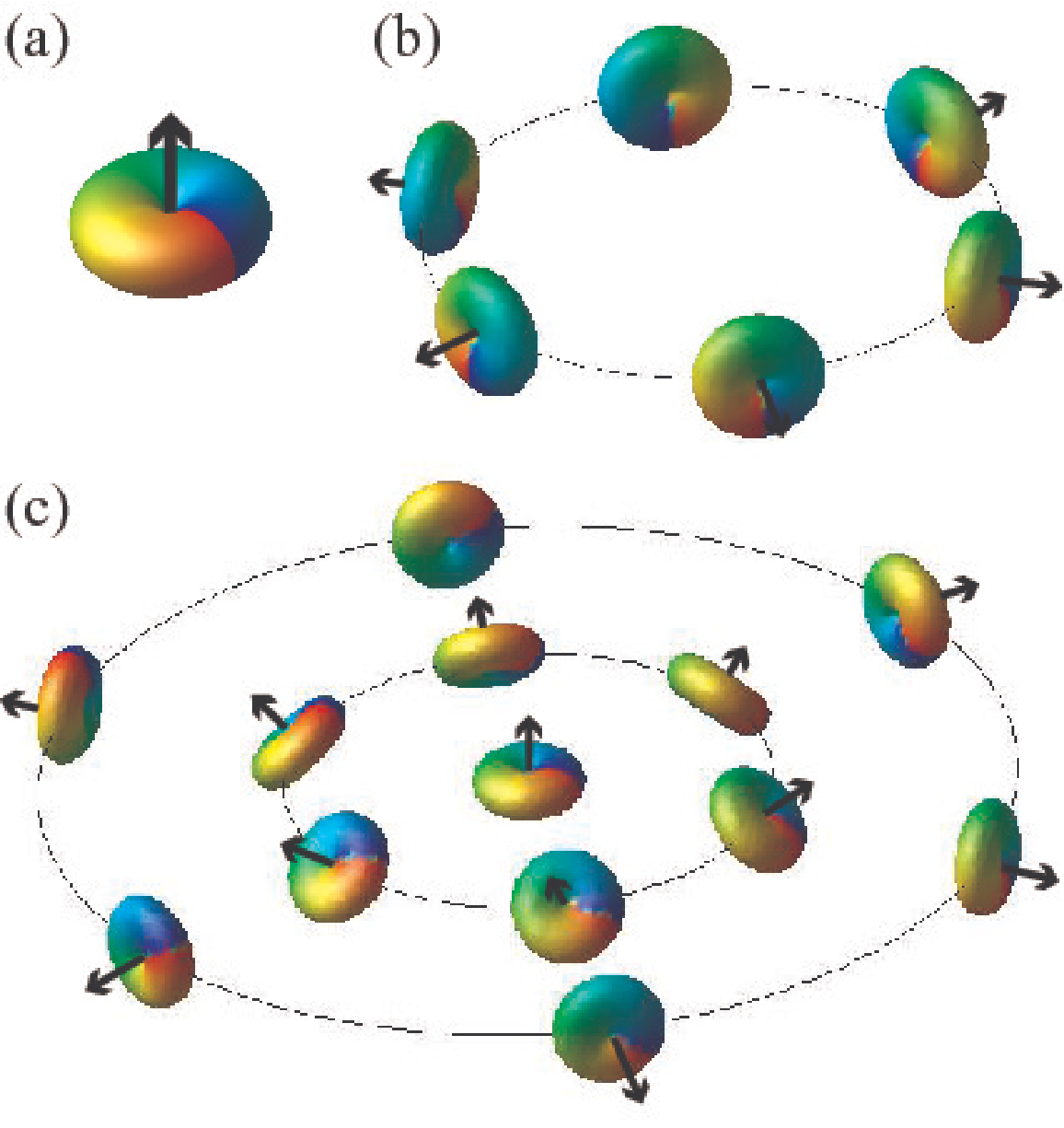}
}
\end{center}
\caption{(a) Spherical-harmonic representation of the order parameter for the spin-1 ferromagnetic phase, where the arrow indicates the direction of the local magnetization.
(b) Polar-core vortex given by Eq.~\eqref{eq:polar_core_vortex}, where the ferromagnetic order parameter has a singularity at the vortex core which is filled by a  polar state.
(c) Coreless vortex given by Eq.~\eqref{eq:coreless_vortex} with $\beta(r=0)=0$ and $\beta(r=r_0)=\pi/2$.
}
\label{fig:spin1ferro}
\end{figure}

From the above argument, any cylindrical configuration of $\bm \psi$ is homotopic to one of the following two types of vortices: 
a coreless vortex [Fig.~\ref{fig:spin1ferro} (c)] and a polar-core vortex [Fig.~\ref{fig:spin1ferro} (b)].
When $\phi'=\pm\alpha$, the vortex is coreless;
in particular, the order parameter for the case of $\phi'=\alpha=\varphi$ is given by
\begin{align}
{\bm \psi}^{\rm coreless}
= \sqrt{n}
\begin{pmatrix}
\cos^2\frac{\beta}{2} \\[1mm]
\frac{e^{i\varphi}}{\sqrt{2}} \sin \beta \\[1mm]
e^{2i\varphi} \sin^2 \frac{\beta}{2}
\end{pmatrix},
\label{eq:coreless_vortex}
\end{align}
in which the singularity can be removed by choosing $\beta=0$ at $r=0$ with the number density $n$ held fixed.
Let us consider the configuration in which $\beta(r=0,\varphi,z)=0$ and $\beta(r=r_0,\varphi,z)=\pi$, where $r_0$ is the radius at the (cylindrical) boundary of the system.
Then, the spinor $(1,0,0)^{\rm T}$ at the origin with no vortex singularity gradually changes to
$(0,0,e^{2i\varphi})^{\rm T}$ at the boundary with a doubly quantized vortex attached to the inverted spin configuration. The corresponding superfluid velocity is nonsingular at the origin; in fact,
\begin{eqnarray}
{\bm v}^{\rm (mass,F)}=\frac{\hbar}{Mr}(1-\cos\beta)\,\hat{\varphi},
\end{eqnarray}
where $\hat{\varphi}$ is the unit vector in the azimuthal direction. 
Mizushima {\it et al}.~\cite{Mizushima2002} investigated the thermodynamical stability of the coreless vortex under rotation with a conserved magnetization.
Based on the Bogoliubov-de Gennes theory,
Ref.~\cite{Pietila2007} discusses the energetical (Landau) and dynamical stabilities of the nonsingular structure of Eq.~\eqref{eq:coreless_vortex} in a condensate subject to a Ioffe-Pritchard field without an external rotation,
whereas the dynamical instability against the splitting of the doubly quantized vortex in the $m=-1$ component of Eq.~\eqref{eq:coreless_vortex} were investigated in Ref.~\cite{Takahashi2009}.
The coreless vortex was observed  by Leanhardt {\it et al}.~\cite{Leanhardt2003}, where a vortex was imprinted using a quadrupole field.

On the other hand, when $\phi'=0$ and $\alpha=\varphi$, the vortex core has a singularity in the ferromagnetic order,
where the order parameter at $r\to\infty$ is given by
\begin{align}
{\bm \psi}^{\rm polar-core}
= \sqrt{n}
\begin{pmatrix}
e^{-i\varphi}\cos^2\frac{\beta}{2} \\[1mm]
\frac{1}{\sqrt{2}} \sin \beta \\[1mm]
e^{i\varphi} \sin^2 \frac{\beta}{2}
\end{pmatrix}.
\label{eq:polar_core_vortex}
\end{align}
Because the $m=\pm1$ components have nonzero vorticities at $r=0$, the populations in these components must vanish at the center of the vortex.
However, the number density $n$ can be non-vanishing even at $r=0$ by increasing the population in the $m=0$ state.
Actually, because $c_0\gg |c_1|$, 
the condensate tends to keep the number density constant and depress only the spin density $\bm f$ in the vortex core [see Eq.~\eqref{eq:spin1MFenergy}].
Since the core is filled by a polar state, this vortex is called a polar-core vortex.
A variational order parameter that describes the system both inside and outside the core is given by
\begin{align}
{\bm \psi}^{\rm polar-core'}
= \sqrt{n}
\begin{pmatrix}
e^{-i\varphi}g_1(r)\cos^2\frac{\beta}{2} \\[1mm]
\sqrt{1-g_1^2(r)\cos^4\frac{\beta}{2}-g_2^2(r)\sin^4\frac{\beta}{2}} \\[1mm]
e^{i\varphi}g_2(r) \sin^2 \frac{\beta}{2}
\end{pmatrix},
\label{eq:polar_core_vortex2}
\end{align}
where $g_1(r)$ and $g_2(r)$ are the variational functions subject to $g_1(0)=g_2(0)=0$ and $g_1(\infty)=g_2(\infty)=1$ and it can be obtained so as to minimize the mean-field energy functional.
The spontaneous formation of the polar-core vortices was observed by Sadler {\it et al}.~\cite{Sadler2006}.

The spin superfluid velocity for the ferromagnetic phase is given by~\eqref{eq:vspin_F_gen} and for $f=1$ we have
\begin{align}
 {\bm v}^{\rm (spin,F)}_\nu &= \hat{s}_\nu{\bm v}^{\rm (mass,F)} - \frac{\hbar}{2M}\sum_{\nu_1\nu_2}\epsilon_{\nu\nu_1\nu_2}\hat{s}_{\nu_1}{\bm \nabla}\hat{s}_{\nu_2},
\label{eq:v_spin_ferro}
\end{align}
which satisfies the following relations:
\begin{align}
{\bm \nabla}\cdot{\bm v}_\nu^{\rm (spin,F)} 
&= \hat{s}_\nu {\bm\nabla}\cdot{\bm v}^{\rm (mass,F)} + {\bm v}^{\rm (mass,F)}\cdot{\bm \nabla}\hat{s}_\nu
-\frac{\hbar}{2M}(\hat{\bm s}\times\nabla^2\hat{\bm s})_\nu,
\label{eq:ferro_div_v_spin}\\
\sum_\nu \hat{s}_\nu {\bm \nabla}\cdot{\bm v}_\nu^{\rm (spin,F)} 
&= {\bm\nabla}\cdot{\bm v}^{\rm (mass,F)},
\label{eq:ferro_div_v_spin_sum}\\
{\bm\nabla}\times {\bm v}^{\rm (spin,F)}_\nu
&=  {\bm\nabla}\hat{s}_\nu\times {\bm v}^{\rm (mass,F)}
+ \frac{\hbar}{2M} \sum_{\nu_1\nu_2\nu_3}(\hat{s}_\nu\hat{s}_{\nu_1}-\delta_{\nu\nu_1})
\epsilon_{\nu_1\nu_2\nu_3}{\bm \nabla}\hat{s}_{\nu_2}\times{\bm\nabla}\hat{s}_{\nu_3},
\label{eq:ferro_rot_v_spin}\\
\sum_\nu \hat{s}_\nu \left[{\bm\nabla}\times {\bm v}^{\rm (spin,F)}_\nu \right] 
&= 0,
\label{eq:ferro_rot_v_spin_sum}
\end{align}
where we have used the Mermin-Ho relation~\eqref{eq:MH-relation}
to derive Eq.~\eqref{eq:ferro_rot_v_spin}.

\subsubsection{Polar phase}
\label{sec:vortex_spin1polar}
We next consider the order parameter of a spin-1 polar BEC for which the representative order parameter is given by ${\bm \zeta}_0=(0,1,0)^{\rm T}$. 
The hydrodynamic properties, vortices, and topological excitations, together with the beyond mean-field properties, in polar BECs are reviewed in Ref.~\cite{Zhou2003}.
A general order parameter of the polar state is given by
\begin{align}
{\bm \psi}^{\rm polar} = \sqrt{n}e^{i\phi}U(\alpha,\beta,\gamma)\begin{pmatrix} 0 \\ 1 \\ 0 \end{pmatrix}
= \sqrt{n} e^{i\phi}
\begin{pmatrix}
-\frac{e^{-i\alpha}}{\sqrt{2}}  \sin\beta \\[1mm]
 \cos \beta \\[1mm]
\frac{e^{i\alpha}}{\sqrt{2}} \sin \beta
\end{pmatrix}.
\label{polar}
\end{align}
The fact that ${\bm \psi}^{\rm polar}$ does not depend on $\gamma$
reflects the SO(2) symmetry around the quantization axis $\hat{\bm d}=(\sin\beta\cos\alpha,\sin\beta\sin\alpha,\cos\beta)$:
$\exp(-i{\bf f}\cdot\hat{\bm d}){\bm \psi}^{\rm polar}={\bm \psi}^{\rm polar}$.
Since magnetization vanishes in the polar phase,
we obtain from Eq.~\eqref{eq:Vort_vmass_gen} that
\begin{align}
{\bm v}^{\rm (mass,P)} 
=\frac{\hbar}{M}\bm\nabla\phi.
\label{polar_s1} 
\end{align}
Thus, unlike the ferromagnetic case, the circulation is quantized. 
However, the unit of quantization is half of $\kappa=h/M$.
To understand this, let us consider a loop that encircles a vortex with a fixed radius.
Then, each point on the loop is specified by the azimuthal angle $\varphi$.
We note that the single-valuedness of the order parameter (\ref{polar}) is met if we take, for example, $\alpha=\phi=n_{\rm w}\varphi/2$ and $\beta=\pi/2$.
Then, the order parameter at $r\to\infty$ is given by
\begin{align}
{\bm\psi}^{\rm half-vortex} = \sqrt{\frac{n}{2}}
\begin{pmatrix}-1 \\ 0 \\ e^{in_{\rm w}\varphi}\end{pmatrix},
\label{eq:polar_half-vortex}
\end{align}
where $n_{\rm w}$ is an integer.
It follows that the circulation is quantized in units of $\kappa/2$ rather than the usual $\kappa=h/M$:
\begin{align}
\oint_\mathcal{C} {\bm v}^{\rm (mass,P)} \cdot d{\bm\ell}
= \frac{\kappa}{2}n_{\rm w}.
\label{polarcorecirculation}
\end{align}
The underlying physics for the half quantum number
is the $\mathbb{Z}_2$ symmetry of the order parameter of the polar phase;
Eq.~\eqref{polar} is invariant under the gauge transformation by $\pi$ ($\phi\to \phi+\pi$) combined with a spin rotation through $\pi$ about an axis perpendicular to $\hat{\bm d}$: $\hat{\bm d}\to-\hat{\bm d}$.
More generally, the single-valuedness of the order parameter is satisfied if $\phi$ changes by $n_{\rm w} \pi$ along the closed loop $\mathcal{C}$, where $n_{\rm w}$ is even 
if $\hat{\bm d}(\varphi=2\pi)=\hat{\bm d}(\varphi=0)$, and odd if
 $\hat{\bm d}(\varphi=2\pi)=-\hat{\bm d}(\varphi=0)$.
Thus, the polar phase of a spin-1 BEC can host a half-quantum vortex~\cite{Zhou2001},
which is also referred to as an Alice vortex or Alice string~\cite{Leonhardt2000}.
The dynamic creation of half-quantum vortices is discussed in Ref.~\cite{Ji2008}.
As in the case of the polar-core vortex in the ferromagnetic phase,
the atomic density can be finite at the vortex core of the half-quantum vortex
by filling the ferromagnetic state [$(1,0,0)^{\rm T}$ for the case of Eq.~\eqref{eq:polar_half-vortex}] in the core.

\begin{figure}[ht]
\begin{center}
\resizebox{0.8\hsize}{!}{
\includegraphics{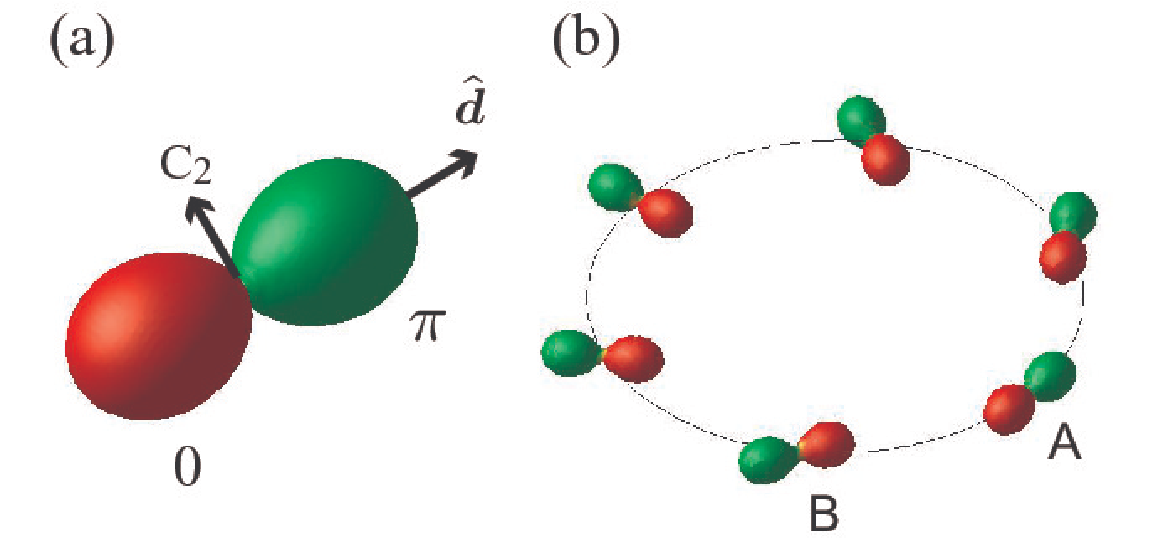}
}
\end{center}
\caption{(a) Spherical-harmonic representation of the order parameter for the polar state,
which is invariant under a $\pi$ rotation about the ${\rm C}_2$ axis together with the $\pi$ gauge transformation.
(b) Half-quantum vortex of the polar phase of a spin-1 BEC. The order parameter rotates through an angle $\pi$ about the ${\rm C}_2$ axis defined in (a) as one makes a complete circuit around the vortex. To satisfy the single-valuedness of the order parameter, this round trip must be accompanied by a phase change by $\pi$ (phase difference between A and B), resulting in a half-quantum vortex.}
\label{fig:spin1polar}
\end{figure}

To calculate ${\bm v}^{\rm (spin)}_\nu$, it is convenient to use the order parameter $\bm \eta = (\eta_x,\eta_y,\eta_z)^{\rm T}$ in the Cartesian basis (see Sec.~\ref{sec:spin1_Cartesian}).
Using Eq.~\eqref{eq:transform_eta_zeta}, the polar order parameter $\bm\eta^{\rm polar}$ that corresponds to Eq.~\eqref{polar} is given by
$\bm\eta^{\rm polar}=e^{i\phi}\hat{\bm d}$.
Then, the spin superfluid velocity is calculated as follows:
\begin{align}
{\bm v}^{\rm (spin,P)}_\nu &= \frac{\hbar}{2Mi}\left[\bm\eta^\dagger \mathcal{U}^{\rm T}{\rm f}_\nu\mathcal{U}^* (\bm\nabla\bm\eta) -  (\bm\nabla\bm\eta^\dagger) \mathcal{U}^{\rm T}{\rm f}_\nu\mathcal{U}^* \bm\eta\right]\nonumber\\
&=-\frac{\hbar}{M} \sum_{\nu_1,\nu_2=x,y,z}\epsilon_{\nu\nu_1\nu_2}\hat{d}_{\nu_1} {\bm \nabla}\hat{d}_{\nu_2},
\label{eq:Vort_vspin_polar}
\end{align}
where $\mathcal{U}$ is defined in Eq.~\eqref{eq:spin1_orthogonal_U}, and we have used $(\mathcal{U}^{\rm T}{\rm f}_\nu\mathcal{U}^*)_{\nu_1\nu_2}=-i\epsilon_{\nu\nu_1\nu_2}$.
If $\hat{\bm d}$ is restricted in a two-dimensional plane perpendicular to a unit vector $\hat{\bm a}$ (in the spin space),
the circulation of $\sum_\nu \hat{a}_\nu {\bm v}^{\rm (spin,P)}_\nu$ is quantized:
\begin{align}
 \oint\sum_\nu \hat{a}_\nu {\bm v}^{\rm (spin,P)}_\nu \cdot d{\bm\ell} = -\frac{\kappa}{2}n_{\rm w}\ \ (n_{\rm w}:\textrm{integer}).
\end{align} 
Here, the unit of quantization is $\kappa/2$ again due to the spin-gauge $\mathbb{Z}_2$ symmetry.
However, in general, the circulation of ${\bm v}^{\rm (spin,P)}_{\nu}$ is not quantized.
The rotation and divergence of ${\bm v}^{\rm (spin,P)}_\nu$ are calculated to be
\begin{align}
{\bm \nabla}\times{\bm v}^{\rm (spin,P)}_\nu &= -\frac{\hbar}{M} \sum_{\nu_1,\nu_2=x,y,z}\epsilon_{\nu\nu_1\nu_2}({\bm \nabla}\hat{d}_{\nu_1}) \times ({\bm \nabla}\hat{d}_{\nu_2}),
\label{eq:polar_spincirculation}
\\
{\bm \nabla}\cdot{\bm v}^{\rm (spin,P)}_\nu &= -\frac{\hbar}{M} (\hat{\bm d} \times \nabla^2\hat{\bm d})_\nu.
\label{eq:polar_div_vspin}
\end{align}
Note that at zero magnetic field, the continuity equation~\eqref{eq:continuity_spin} becomes
\begin{align}
 \frac{\partial F_\nu}{\partial t} &= -{\bm \nabla}\cdot\left(n{\bm v}^{\rm (spin,P)}_\nu\right)\nonumber\\
 &= \frac{\hbar}{M}\sum_{i=x,y,z} \sum_{\nu_1,\nu_2=x,y,z} \epsilon_{\nu\nu_1\nu_2}\left[ (\partial_i n) \hat{d}_{\nu_1} (\partial_i\hat{d}_{\nu_2})
+ n \hat{d}_{\nu_1}(\partial_i\partial_i\hat{d}_{\nu_2}) \right]\nonumber\\
 &= \frac{\hbar}{M}\sum_{i=x,y,z} \sum_{\nu_1,\nu_2=x,y,z} \epsilon_{\nu\nu_1\nu_2} \hat{d}_{\nu_1} \partial_i(n\partial_i \hat{d}_{\nu_2}),
\end{align}
or, in vector notation, it is written as
\begin{align}
 \frac{\partial {\bm F}}{\partial t}
 &= \frac{\hbar}{M}\sum_{i=x,y,z} \hat{\bm d} \times \partial_i(n\partial_i \hat{\bm d}).
\label{eq:dFdt_polar}
\end{align}
This result implies that 
$\sum_i \partial_i (n \partial_i \hat{\bm d})$ has to be either zero or parallel to $\hat{\bm d}$ in a stationary $\hat{\bm d}$ texture.
Otherwise, the local magnetization would develop with time and the order parameter would no longer belong to a manifold for the polar state.
The $\hat{\bm d}$-field around the half-quantum vortex of Eq.~\eqref{eq:polar_half-vortex} is $\hat{\bm d}=(\cos n_{\rm w}\varphi/2, \sin n_{\rm w}\varphi/2, 0)$, which satisfies
$\sum_i \partial_i (n \partial_i \hat{\bm d}) = -n_{\rm w}^2\hat{\bm d}/(4r^2)$ if $n(r,\varphi)$ is axisymmetric, i.e., $\hat{e}_\varphi\cdot{\nabla}n=0$, 
where $\hat{e}_\varphi$ is the unit vector in the azimuthal direction.
Therefore, the half-quantum vortex can be a stationary state.

\subsubsection{Stability of vortex states}

The above discussions can be rephrased in the parlance of homotopy theory.
We first consider an order-parameter manifold, that is, how the order parameter can vary while keeping the nature of the ferromagnetic or polar phases [Eqs.~\eqref{spin1ferro} and \eqref{polar}],
and construct a vortex state so that the order parameter remains in the manifold (for more details, see Sec.~\ref{sec:topology}).
However, such configurations are not always stable in realistic situations because total magnetization are usually conserved in ultracold atomic systems. Here, we discuss the stability of vortex states from the point of view of multi-component order parameter.

Suppose that the order parameter for each spin component is axisymmetric about the vortex.
Then, the single-valuedness condition for the spinor order parameter must be met for each $\psi_m$, and
each spin component accommodates a vortex with an individual quantum number.
However, this leaves relative phases between spinor components arbitrary, and they are determined from the energetics point of view. From Eq.~(\ref{energy_functional(f=1)}), we find that only the term $c_1|{\bm F}|^2/2$ depends explicitly on the relative phases. Expressing the term in terms of components, we have
\begin{align}
\frac{c_1}{2}|{\bm F}|^2=&\, \frac{c_1}{2}\left[
(|\psi_1|^2-|\psi_{-1}|^2)^2+2|\psi_0|^2(|\psi_1|^2+|\psi_{-1}|^2)
\right. \nonumber \\
& \hspace{15mm}\left.
+4|\psi_0^2\psi_1\psi_{-1}|\cos\left(\chi_1+\chi_{-1}-2\chi_0\right)
\right],
\end{align}
where $\chi_m\equiv{\rm arg}(\psi_m)$.
To minimize the energy, $\chi_1+\chi_{-1}-2\chi_0$ must be $\pi n_{\rm w}$, where $n_{\rm w}$ is even if $c_1<0$ and odd if $c_1>0$. Substituting $\chi_m=\chi'_m+\chi''_m\varphi$, where $\varphi$ is the azimuthal angle, we have $\chi'_1+\chi'_{-1}-2\chi'_0=\pi n_{\rm w}$ and $\chi''_1+\chi''_{-1}-2\chi''_0=0$. Assuming that the maximum vorticity is 1, we find that the following three vortex states are allowed~\cite{Isoshima2001}:
\begin{eqnarray}
(\chi''_1,\chi''_0,\chi''_{-1})=(1,1,1),\  (1,0,-1),\  (1,1/2,0),
\end{eqnarray}
where the last state is allowed only if $\psi_0=0$, and therefore, it is described as (1, none, 0).

The $(1,1,1)$ state is the usual singly quantized vortex state with an empty vortex core, where the order parameter is given by
\begin{eqnarray}
e^{i\varphi}(|\psi_1|e^{i\chi'_1},|\psi_0|e^{i\chi'_0},|\psi_{-1}|e^{i\chi'_{-1}})^{\rm T}.
\label{eq:111vortex}
\end{eqnarray} 
The $(1,0,-1)$ state has a vortex in the $m=1$ component and an anti-vortex in the $m=-1$ component with the vortex core filled by the $m=0$ state, where the order parameter of this state is given by
\begin{eqnarray}
(|\psi_1|e^{i(\chi'_1+\varphi)},|\psi_0|e^{i\chi'_0},|\psi_{-1}|e^{i(\chi'_{-1}-\varphi)})^{\rm T}.
\label{eq:11-1vortex}
\end{eqnarray}
The polar-core vortex in the ferromagnetic phase and an integer spin vortex in the polar phase belong to this case.
The (1, none, 0) state corresponds to the Alice vortex [Eq.~\eqref{eq:polar_half-vortex}] whose order parameter is given by
\begin{eqnarray}
(|\psi_1|e^{i(\chi'_1+\varphi)},0,|\psi_{-1}|e^{i\chi'_{-1}})^{\rm T},
\end{eqnarray}
where the vortex core is filled by the $m=-1$ component. Except for the (1,1,1) state, the vortex core is filled by a non-vortex component which plays the role of a pinning potential, stabilizing the vortex state in the absence of an external potential~\cite{Isoshima2001}.

\subsection{Spin-2 BEC}
\label{sec:Vort_spin2}

The rotation matrix of the spin-2 BEC is given by
\begin{eqnarray}
& &  U(\alpha,\beta,\gamma)= \nonumber\\
& & 
\begin{small}
\begin{pmatrix}
e^{-2i(\alpha+\gamma)}C^4  & -2e^{-i(2\alpha+\gamma)}C^3S & \sqrt{6}e^{-2i\alpha}C^2S^2& -2e^{-i(2\alpha-\gamma)}CS^3& e^{-2i(\alpha-\gamma)}S^4\\
2e^{-i(\alpha+2\gamma)}C^3S & e^{-i(\alpha+\gamma)}C^2(C^2-3S^2) & -\sqrt{\frac{3}{8}}e^{-i\alpha}\sin 2\beta& -e^{-i(\alpha-\gamma)}S^2(S^2-3C^2)& -2e^{-i(\alpha-2\gamma)}CS^3\\
\sqrt{6}e^{-2i\gamma}C^2S^2  & \sqrt{\frac{3}{8}}e^{-i\gamma}\sin2\beta& \frac{1}{4}(1+3\cos2\beta)& -\sqrt{\frac{3}{8}}e^{i\gamma}\sin2\beta& \sqrt{6}e^{2i\gamma}C^2S^2\\
2e^{i(\alpha-2\gamma)}CS^3 & -e^{i(\alpha-\gamma)}S^2(S^2-3C^2) & \sqrt{\frac{3}{8}}e^{i\alpha}\sin2\beta& e^{i(\alpha+\gamma)}C^2(C^2-3S^2) & -2e^{i(\alpha+2\gamma)}C^3S\\
e^{2i(\alpha-\gamma)}S^4  & 2e^{i(2\alpha-\gamma)}CS^3& \sqrt{6}e^{2i\alpha}C^2S^2& 2e^{i(2\alpha+\gamma)}C^3S& e^{2i(\alpha+\gamma)}C^4\\
\end{pmatrix}, \end{small}
\nonumber\\
\label{spin2U}
\end{eqnarray}
where $C\equiv\cos\frac{\beta}{2}$ and $S\equiv\sin\frac{\beta}{2}$.
The spin-2 BEC has four mean-field ground-state phases in the absence of the magnetic field: ferromagnetic, uniaxial nematic, biaxial nematic, and cyclic.
The uniaxial and biaxial nematic phases are degenerate at the mean-field level (see Sec.~\ref{sec:MF_spin2_GS}).

\subsubsection{Ferromagnetic phase}

The representative order parameter for the ferromagnetic phase is ${\bm \zeta}^{\rm ferro}_0=(1,0,0,0,0)^{\rm T}$. Substituting this and Eq.~\eqref{spin2U} in Eq.~\eqref{eq:Vort_genpsi_spinor}, we obtain
\begin{align}
{\bm \psi}^{\rm ferro}
=  \sqrt{n}e^{i(\phi-2\gamma)}
\begin{pmatrix}
e^{-2i\alpha}  \cos^4\frac{\beta}{2} \\
2 e^{-i\alpha} \cos^3\frac{\beta}{2}\sin   \frac{\beta}{2} \\
\sqrt{6}       \cos^2\frac{\beta}{2}\sin^2 \frac{\beta}{2} \\
2 e^{i\alpha}  \cos  \frac{\beta}{2}\sin^3 \frac{\beta}{2} \\
e^{2i\alpha}   \sin^4\frac{\beta}{2}
\end{pmatrix}.
\label{spin2ferro}
\end{align}
Similar to the spin-1 case [see Eq.~(\ref{spin1ferro})], the order parameter has the spin-gauge symmetry with $\gamma$ now replaced by $2\gamma$ because the spin is 2 rather than 1. 
From Eq.~\eqref{eq:Vort_vmass_gen},
the superfluid velocity is similarly obtained as
\begin{align}
{\bm v}^{\rm (mass,F)} =\frac{\hbar}{M} [ {\bm\nabla} (\phi - 2\gamma) - 2\cos\beta \bm{\nabla} \alpha ],
\label{ferrov_s2}
\end{align}
whose circulation is not quantized as in the case of the spin-1 ferromagnetic BEC.
Applying the discussions about the stability of the vortex states for the ferromagnetic spin-1 BEC,
it can be shown that the $4n_w+m_w$ vortex ($m_w=0,1,2,3$) is unstable against the decay to the $m_w$ vortex.

As for the spin superfluid velocity, Eq.~\eqref{eq:vspin_F_gen} for $f=2$ reduces to
\begin{align}
 {\bm v}^{\rm (spin,F)}_\nu &= 2\hat{s}_\nu{\bm v}^{\rm (mass,F)} - \frac{\hbar}{M}\sum_{\nu_1\nu_2}\epsilon_{\nu\nu_1\nu_2}\hat{s}_{\nu_1}{\bm \nabla}\hat{s}_{\nu_2}.
\end{align}
Then, the relationships for the spin superfluid velocity corresponding to Eqs.~\eqref{eq:ferro_div_v_spin}--\eqref{eq:ferro_rot_v_spin_sum} are given by
\begin{align}
{\bm \nabla}\cdot{\bm v}_\nu^{\rm (spin,F)} 
&= 2\left[\hat{s}_\nu {\bm\nabla}\cdot{\bm v}^{\rm (mass,F)} + {\bm v}^{\rm (mass,F)}\cdot{\bm \nabla}\hat{s}_\nu
-\frac{\hbar}{2M}(\hat{\bm s}\times\nabla^2\hat{\bm s})_\nu\right],
\\
\sum_\nu \hat{s}_\nu {\bm \nabla}\cdot{\bm v}_\nu^{\rm (spin,F)} 
&= 2{\bm\nabla}\cdot{\bm v}^{\rm (mass,F)},
\\
{\bm\nabla}\times {\bm v}^{\rm (spin,F)}_\nu
&=  2{\bm\nabla}\hat{s}_\nu\times {\bm v}^{\rm (mass,F)}
+ \frac{\hbar}{M} \sum_{\nu_1\nu_2\nu_3}(2\hat{s}_\nu\hat{s}_{\nu_1}-\delta_{\nu\nu_1})
\epsilon_{\nu_1\nu_2\nu_3}({\bm \nabla}\hat{s}_{\nu_2}\times{\bm\nabla}\hat{s}_{\nu_3}),
\label{eq:ferro_rot_v_spin_f=2}\\
\sum_\nu \hat{s}_\nu \left[{\bm\nabla}\times {\bm v}^{\rm (spin,F)}_\nu \right] 
&= \frac{\hbar}{M} \sum_{\nu_1\nu_2\nu_3}
\epsilon_{\nu_1\nu_2\nu_3}\hat{s}_{\nu_1}({\bm \nabla}\hat{s}_{\nu_2}\times{\bm\nabla}\hat{s}_{\nu_3})
= \bm\nabla\times{\bm v}^{\rm (mass, F)},
\label{eq:ferro_rot_v_spin2_f=2}
\end{align}
where we have used the Mermin-Ho relation~\eqref{eq:MH-relation}
to derive Eqs.~\eqref{eq:ferro_rot_v_spin_f=2} and \eqref{eq:ferro_rot_v_spin2_f=2}.

\subsubsection{Uniaxial nematic phase}

The representative order parameter for the spin-2 uniaxial nematic phase is ${\bm \zeta}^{\rm uniax}_0=(0,0,1,0,0)^{\rm T}$.
Substituting this and Eq.~(\ref{spin2U}) in Eq.~\eqref{eq:Vort_genpsi_spinor}, we obtain the general order parameter of this phase:
\begin{align}
{\bm \psi}^{\rm uniax}
= \frac{\sqrt{6n}}{4} e^{i\phi}
\begin{pmatrix}
    e^{-2i\alpha} \sin^2 \beta \\
- 2 e^{- i\alpha} \sin \beta \cos\beta \\
 \sqrt{\frac{2}{3}} (3\cos^2\beta-1) \\
  2 e^{  i\alpha} \sin \beta\cos\beta\\
    e^{ 2i\alpha} \sin^2\beta
\end{pmatrix}.
\label{spin2uniax}
\end{align}

Similar to the case of the spin-1 polar phase, 
the spin degrees of freedom do not contribute to the superfluid velocity that only depends on the spatial variation of the gauge angle:
\begin{align}
{\bm v}^{\rm (mass,uniax)} =\frac{\hbar}{M}{\bm\nabla}\phi.
\end{align}
This is a general property for the non-magnetized phase in spinor BECs [see Eq.~\eqref{eq:Vort_vmass_gen}].

\begin{figure}[ht]
\begin{center}
\resizebox{0.8\hsize}{!}{
\includegraphics{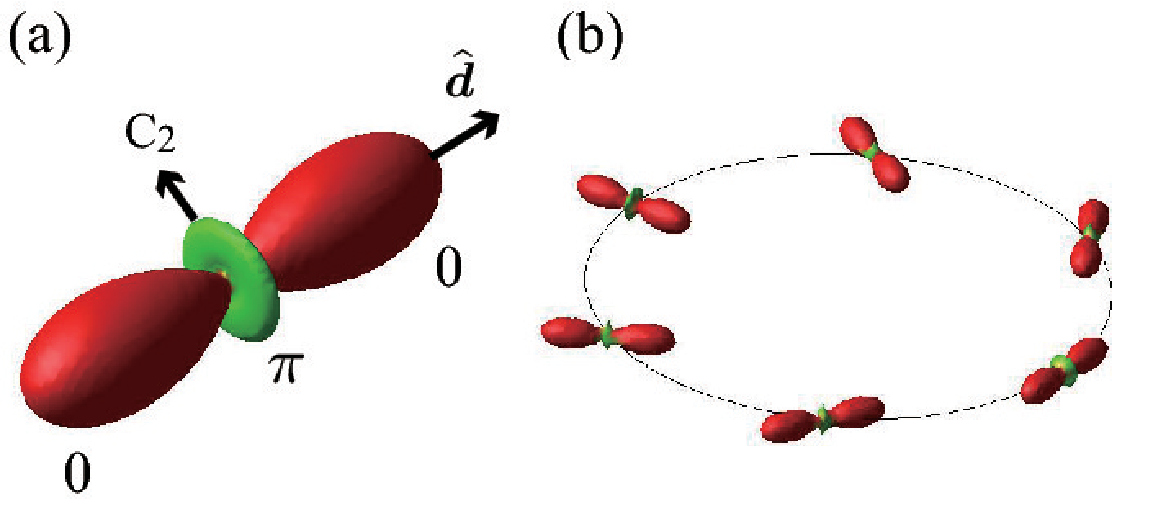}
}
\end{center}
\caption{(a) Order parameter for the spin-2 uniaxial nematic phase, which has an SO(2) symmetry corresponding to rotations about $\hat{\bm d}$.
The uniaxial nematic phase is also invariant under a $\pi$ rotation about the ${\rm C}_2$ axis, i.e., this phase has a $\mathbb{Z}_2$ symmetry.
(b) The order parameter configuration for the 0--1/2 vortex, around which the order parameter rotates by $\pi$ about the ${\rm C}_2$ axis in (a).
}
\label{fig:spin2uniax}
\end{figure}

The shape of the uniaxial order parameter in the spin space is shown in Fig.~\ref{fig:spin2uniax} (a).
As one can see from Fig.~\ref{fig:spin2uniax} (a),
the uniaxial nematic phase has an SO(2) symmetry around the direction of $\hat{\bm d}=(\cos\alpha\sin\beta,\sin\alpha\sin\beta,\cos\beta)$.
The uniaxial nematic phase also has the $\mathbb{Z}_2$ symmetry as in the case of the spin-1 polar phase.
However, in the present case, the $\mathbb{Z}_2$ symmetry is not coupled to the gauge angle $\phi$:
${\bm \psi}^{\rm uniax}$ is invariant under a $\pi$ rotation around an axis perpendicular to $\hat{\bm d}$.
Hence, although the uniaxial nematic phase can accommodate a vortex, as shown in Fig.~\ref{fig:spin2uniax} (b),
the mass circulation for this vortex is zero. It is a spin vortex which may also be called a 0--1/2 vortex, where 0 and 1/2 indicate the gauge transformation and rotation angle in spin space around the vortex, respectively.
An example of the order parameter around the 0--1/2 vortex (far from the vortex core) is given as
\begin{align}
 {\bm \psi}^{\rm spin-vortex} = \frac{\sqrt{6n}}{4}
\begin{pmatrix} e^{i\varphi} \\ 0 \\ \sqrt{2/3} \\ 0 \\ e^{-i\varphi} \end{pmatrix},
\end{align}
where $\varphi$ is an azimuthal angle around the vortex.

The spin superfluid velocity for the uniaxial nematic phase is calculated to be
\begin{align}
 {\bm v}^{\rm (spin,uniax)}_\nu = -\frac{3\hbar}{M} \sum_{\nu_1,\nu_2=x,y,z} \epsilon_{\nu\nu_1\nu_2} \hat{d}_{\nu_1}{\bm\nabla}\hat{d}_{\nu_2}.
\label{eq:Vort_vspin_uniax}
\end{align}
Hence, if $\hat{\bm d}$ is restricted in a plane perpendicular to $\hat{\bm a}$,
the circulation of the spin superfluid velocity $\sum_\nu \hat{a}_\nu {\bm v}^{\rm(spin,uniax)}_\nu$ is quantized in units of $3\kappa$;
however, in general, the circulation of ${\bm v}^{\rm (spin,uniax)}_{\nu}$ is not quantized.
The rotation and divergence of the ${\bm v}^{\rm (spin,uniax)}_{\nu}$ are the same as Eqs.~\eqref{eq:polar_spincirculation} and \eqref{eq:polar_div_vspin}
except for the factor 3 [see Eqs.~\eqref{eq:Vort_vspin_uniax} and \eqref{eq:Vort_vspin_polar}].

Because the spin and gauge is completely decoupled in this phase,
the mass circulation is quantized in terms of units of $\kappa$.
An example of a vortex that has nonzero mass circulation is $(0,0,e^{i\varphi},0,0)^{\rm T}$.

\subsubsection{Biaxial nematic phase}

\begin{figure}[ht]
\begin{center}
\resizebox{0.8\hsize}{!}{
\includegraphics{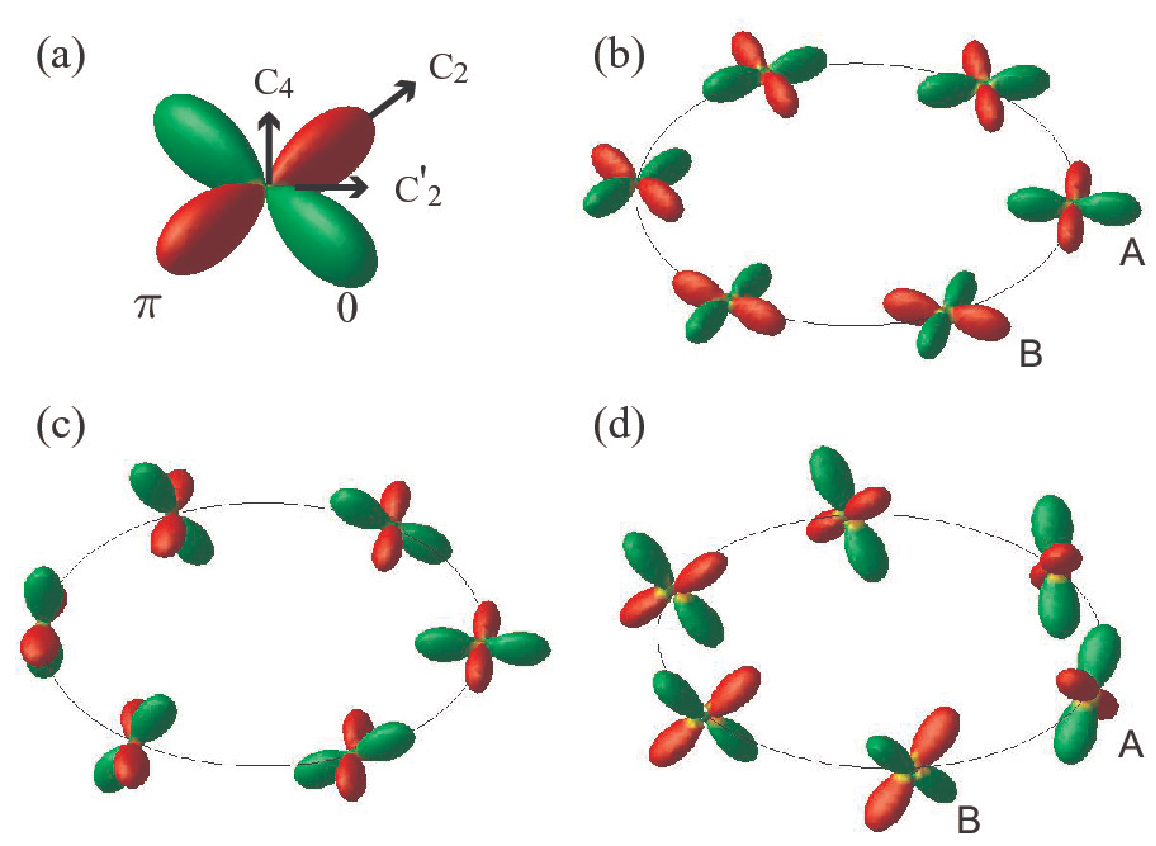}
}
\end{center}
\caption{(a) Order parameter for the spin-2 biaxial nematic phase, 
which has one symmetry axis of the fourth order (${\rm C}_4$) and two axes of the second order (${\rm C}_2$ and ${\rm C}'_2$).
(b)--(d) The configuration of the order parameter around (b) a 1/2--1/4 vortex, (c) a 0--1/2 vortex, and (d) a 1/2--1/2 vortex.}
\label{fig:spin2biax}
\end{figure}

The representative order parameter for the spin-2 biaxial nematic phase is ${\bm \zeta}^{\rm biax}_0=(1/\sqrt{2},0,0,0,1/\sqrt{2})^{\rm T}$ and substituting this and Eq.~(\ref{spin2U}) in Eq.~\eqref{eq:Vort_genpsi_spinor}, 
we obtain the general order parameter of this phase:
\begin{align}
{\bm \psi}^{\rm biax}
= \sqrt{\frac{n}{2}}e^{i\phi}
\begin{pmatrix}
e^{-2i\alpha} \left[\left(1-\frac{1}{2}\sin^2\beta\right)\cos2\gamma-i\cos\beta\sin2\gamma \right]  \\
e^{-i\alpha} \sin\beta(\cos\beta\cos2\gamma-i\sin2\gamma)    \\
\sqrt{\frac{3}{2}}   \sin^2\beta\cos2\gamma \\
-e^{i\alpha} \sin\beta(\cos\beta\cos2\gamma+i\sin2\gamma)  \\
e^{2i\alpha} \left[\left(1-\frac{1}{2}\sin^2\beta\right)\cos2\gamma+i\cos\beta\sin2\gamma \right]
\end{pmatrix}.
\label{spin2biax}
\end{align}
The superfluid velocity only depends on the spatial variation of the gauge angle:
\begin{align}
{\bm v}^{\rm (mass,biax)} 
=\frac{\hbar}{M}\bm\nabla\phi.
\label{biaxial} 
\end{align}
The profile of ${\bm \psi}^{\rm biax}$ is depicted in Fig.~\ref{fig:spin2biax} (a).

When we rotate the order parameter by $\pi/2$ about the ${\rm C}_4$ axis in Fig.~\ref{fig:spin2biax} (a) as we circumnavigate a vortex,
the order parameter changes its sign, as shown in Fig.~\ref{fig:spin2biax} (b) (see, the phase difference between A and B).
Therefore, this vortex has a mass circulation of $\kappa/2$. We call this vortex a 1/2--1/4 vortex,
where 1/2 and 1/4 denote the gauge transformation and the spin-rotation angle, respectively, around the vortex.
If we rotate the order parameter by $\pi$ about the ${\rm C}_2$ axis around a vortex, then it goes back to the initial state, as shown in Fig.~\ref{fig:spin2biax} (c).
In this case, the vortex has no mass circulation and it is referred to as a 0--1/2 vortex.
On the other hand, if we rotate the order parameter by $\pi$ about the ${\rm C}'_2$ axis, the sign of the order parameter is reversed [Fig.~\ref{fig:spin2biax} (d)].
This vortex has a mass circulation of $\kappa/2$, and it is called a 1/2--1/2 vortex.
Because the rotations about ${\rm C}_4$, ${\rm C}_2$, and ${\rm C}'_2$ are non-commutable,
these vortices obey the non-Abelian algebra, which is discussed in Sec.~\ref{sec:Vort_nonAbelian}.

The vortices shown in Fig.~\ref{fig:spin2biax} are some of the simplest examples.
In general, the order parameter can vary in a more complex manner.
However, in order to meet the single-valuedness condition for the order parameter, the mass circulation in the biaxial nematic phase must be quantized in units of $\kappa/2$.
The structure of the order parameter inside the vortex core is discussed in Ref.~\cite{Kobayashi2009b}.

\subsubsection{Cyclic phase}

\begin{figure}[ht]
\begin{center}
\resizebox{0.8\hsize}{!}{
\includegraphics{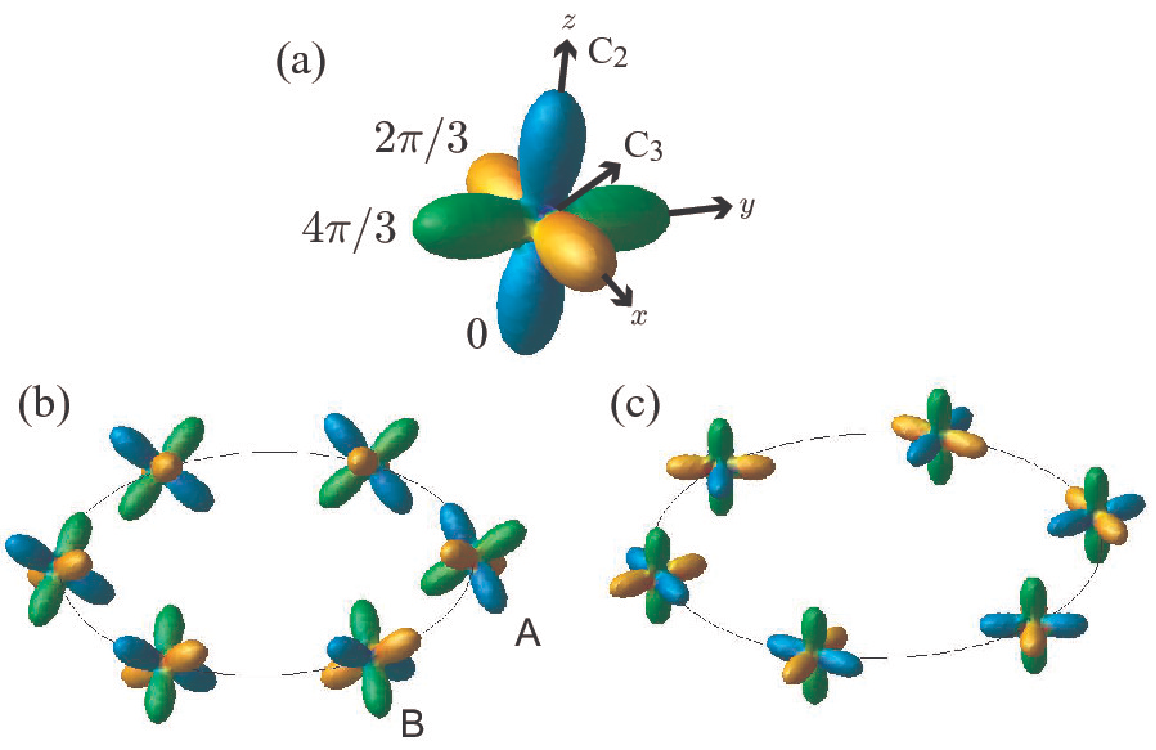}
}
\end{center}
\caption{Spherical-harmonic representation of the order parameter for the cyclic phase $(1,0,i/\sqrt{2},0,1)^T$, where the blue, yellow, and green colors denote ${\rm arg}[\sum_m Y_2^m\psi_m^{\rm cyclic}]=0, 2\pi/3$, and $4\pi/3$, respectively.
The order parameter has a two-fold symmetry about the $x$, $y$, and $z$ axes and a three-fold symmetry about the $(\pm1,\pm1,\pm1)$ axes.
(b) The 1/3--1/3 vortex, where the order parameter is rotated by $-2\pi/3$ about the ${\rm C}_3$ axis from A to B in a clockwise direction, results in the gauge transformation of $-2\pi/3$.
(c) The 0--1/2 vortex, around which the order parameter is rotated by $\pi$ about the $y$-axis.}
\label{fig:spin2cyclic}
\end{figure}

A representative order parameter for the spin-2 cyclic phase is ${\bm \zeta}^{\rm cyclic}_0=(1/2,0,i/\sqrt{2},0,1/2)$. 
Figure~\ref{fig:spin2cyclic}~(a) shows the profile of the order parameter of the cyclic phase.
One can clearly see that it has a three-fold symmetry about the $(1,1,1)$ axis [${\rm C}_3$ axis in Fig.~\ref{fig:spin2cyclic} (a)]. 
This suggests that the mass circulation in a cyclic BEC is quantized in units of $\kappa/3$~\cite{Makela2003,Semenoff2007}.
Under a rotation about the $(1,1,1)$ axis through angle $\theta$ together with the gauge transformation by $\phi$, the oder parameter ${\bm \zeta}^{\rm cyclic}_0$ transforms into
\begin{eqnarray}
e^{i\phi}e^{-i\frac{{\rm f}_x+{\rm f}_y+{\rm f}_z}{\sqrt{3}}\theta}{\bm \zeta}^{\rm cyclic}_0.
\label{(111)}
\end{eqnarray}
For $\theta=-2\pi/3$, Eq.~(\ref{(111)}) reduces to $e^{-2\pi i/3}e^{i\phi}{\bm\zeta}^{\rm cyclic}_0$; the single-valuedness of the order parameter is met by the gauge transformation by $\phi=2\pi/3$. Because this is one-third of the usual $2\pi$, the cyclic phase can possess a one-third vortex. Similarly, for $\theta=-4\pi/3$, Eq.~(\ref{(111)}) reduces to $e^{-4\pi i/3}e^{i\phi}{\bm \zeta}^{\rm cyclic}_0$; the single-valuedness of the order parameter is met by the choice of $\phi=4\pi/3$. Thus, the cyclic phase also has a two-third vortex.
The cyclic phase can also accommodate the 0--1/2 vortex,
around which the order parameter rotates by $\pi$ about the direction of one of the lobes [Fig.~\ref{fig:spin2cyclic} (c)].
The vortices in the cyclic phase obey the non-Abelian algebra (see Sec.~\ref{sec:Vort_nonAbelian}).
The same types of vortices as the cyclic phase were also predicted to exist in {\it d}-wave Fermi condensates~\cite{Adachi2009}.

\subsection{Rotating spinor BEC}
\label{sec:Vort_rotating}

The stationary state of a rotating BEC can be obtained by minimizing $\mathcal{F}=E-\Omega  L_z$, where the rotation axis is assumed to be in the $z$-direction. In the case of a spin-1 BEC, we have
\begin{eqnarray}
\mathcal{F}=\int d{\bm r}\left[\sum_m\psi^*_m\left(-\frac{\hbar^2\nabla^2}{2M}+U({\bm r})-\Omega \ell_z\right)\psi_m
+\frac{c_0}{2}n^2+\frac{c_1}{2}|{\bm F}|^2\right],
\end{eqnarray}
where $\ell_z=-i\hbar\partial/\partial\varphi$.
The stationary state is obtained by requiring that 
\begin{eqnarray}
\frac{\delta (\mathcal{F}-\mu N)}{\delta \psi_m^*}=0,
\label{eqofm}
\end{eqnarray}
where $\mu$ is the chemical potential which is determined so as to conserve the total number of particles $N=\int d{\bm r}\sum_m|\psi_m|^2$ .

The vortex lattice structures in a multi-component BEC was observed for a pseudo-spin-1/2 system~\cite{Schweikhard2004},
where the quantized vortices in each component form a square lattice structure~\cite{Mueller2002,Kasamatsu2003}.
There are a number of theoretical studies on vortex structures in spin-1 and 2 BECs under rotation.
When the total magnetization is conserved in a trapped system, the spin configuration of a single vortex is strongly dependent
on the total magnetization as well as the rotation frequency~\cite{Isoshima2001,Mizushima2002,Pogosov2005}.
For a spin-1 ferromagnetic BEC, a lattice of two-dimensional Skyrmions (Mermin-Ho vortices) was predicted to appear~\cite{Mizushima2004}.
The dependence on the total magnetization and rotation frequency of the vortex lattices was also investigated in Ref.~\cite{Mizushima2004}.
The dynamic creation of half-quantum vortices in a rotating $^{23}$Na BEC was discussed in Ref.~\cite{Ji2008}.
In this system, the dynamical instabilities for creation of integer vortices occur almost at the same frequencies as those for half-quantum vortices.
Ji {\it et al}.~\cite{Ji2008} pointed out that additional pulsed magnetic potentials trigger an instability toward the square lattice of half-quantum vortices,
where the vortices in the $m=1$ and $-1$ are located at the grid point of the nested sublattices.
When spinor gases are rapidly cooled under external rotation,
a two-dimensional Skyrmion lattice and a lattice of half-quantum virtices are predicted to arise in equilibrium states of $^{87}$Rb and $^{23}$Na condensates, respectively~\cite{Su2011}.
For a spin-2 cyclic BEC, Barnett {\it et al.} predicted that the triangular lattice of integer spin vortices dissociates to 
the honeycomb structure of 1/3 and 2/3 vortices when the external magnetic field becomes lower than the critical value~\cite{Barnett2008}.
They also pointed out that the honeycomb lattice structure deforms depending on temperature since the ratio of the stiffnesses of spin and superfluid phase depends on temperature~\cite{Barnett2008}.
The experimental scheme to create a 1/3-vortex is proposed in Ref.~\cite{Huhtamaki2009}:
it is nucleated by simply rotating the cyclic BEC initially prepared in the magnetic sublevels $m=2$ and $-1$.

Since all the above mentioned vortices have nonzero mass circulation, they can be nucleated by an external rotation.
However, the vortices that have no mass circulation, such as a 0--1/2 vortex in the spin-2 uniaxial nematic phase and that in the spin-2 cyclic phase,
do not response to an external rotation.
In such cases, a spin-dependent rotating potential created by means of near-resonant circularly polarized laser beams~\cite{Chiba2008}
may be used to create a vortex state.

In a rapidly rotating system, the lowest-Landau-level approximation becomes relevant.
In this regime, strongly correlated states of a spin-1 Bose gase is discussed in the context of quantum Hall states~\cite{Ardonne1999,Ho2002,Reijnders2002,Paredes2002}.

\subsection{Rotating dipolar BEC}
\label{sec:Vort_dipole}

The dipole-dipole interaction (DDI) is long-range and anisotropic, and it significantly affects the ground states of rotating BECs~\cite{Cooper2005a,Cooper2005b,Zhang2005,Yi2006a,Komineas2007,Simula2011}.
Let us assume that all the dipoles of the atoms are polarized in the same direction. Then, the interaction is described by
\begin{eqnarray}
V({\bm r})=c_0\delta({\bm r})+c_{\rm dd}\frac{1-3\cos^2\theta}{r^3},
\label{contact-dipole}
\end{eqnarray}
where the first and second terms on the right-hand side are the {\it s}-wave contact interaction and the DDI, respectively, and $\theta$ is the angle between the dipole moment and the relative coordinate ${\bm r}$ between two atoms. Thus, depending on the direction of the polarization, the DDI can be either attractive or repulsive. This anisotropy of the interaction leads to the deformation of the vortex core and the structure of the vortex lattice.

Yi and Pu~\cite{Yi2006a} considered a situation in which a system in an axisymmetric harmonic trap is rotated about the $z$-direction, and the dipoles are polarized in the $x$-direction in the frame of reference co-rotating with the system, and found that the vortex core is deformed into an elliptic shape. This is because the dipoles undergo repulsive and attractive interaction in the $x$- and $y$-directions, respectively. 
On the other hand,
Cooper {\it et al}.~\cite{Cooper2005a} and Zhang and Zhai~\cite{Zhang2005} considered a dipolar gas polarized along the rotation axis in which the interparticle interaction is so weak that the lowest Landau level approximation holds, and found that with an increase in the strength of the DDI, the vortex lattice undergoes structural changes from triangular to square and to stripe configurations. 

The rotating spinor dipolar BEC was investigated by Simula {\it et al}.~\cite{Simula2011} who calculated the ground state spin textures for a spin-1 BEC for various rotation frequencies, spin-exchange interaction, and magnetic DDI.
Even for a fixed rotation frequency, the configurational structure of vortices changes depending on the strength of the DDI, as in the case discussed in Sec.~\ref{sec:dipole_spinor}.
In particular, for the polar phase, the magnetization localized at the cores of half quantum vortices aligns due to the DDI.
They also have pointed out that a nontrivial density pattern appears in the limit of strong DDI and rapid rotation: In this limit, the DDI, which effectively works as an attraction, tends to collapse the condensate, whereas the centrifugal potential of the rapid rotation (slower than the trapping frequency) extends the condensate.
The balance between these two forces results in a ring or a figure-of-eight distribution of atoms.
Their result suggests the possibility of entering quantum-Hall like states for which the DDI and centrifugal effect would be counterbalanced.

In the aforementioned work, the dipole interaction is treated as a perturbation to the {\it s}-wave contact interaction. Recently, Pollack {\it et al}.~\cite{Pollack2009} used a shallow zero crossing in the wing of a Feshbach resonance of ${}^7$Li in the $|f=1,m=1\rangle$ state to decrease the scattering length to as low as $0.01a_{\rm B}$. This opens up several new possibilities on the dipole-interaction-dominated BEC. An interesting question is how the degeneracy of a fast rotating BEC is lifted by a weak but genuine dipole interaction and what is the corresponding many-body ground state.

%% file: symmetry.tex
\section{Symmetry classification}
\label{sec:symmetry}

When a system undergoes Bose-Einstein condensation, certain symmetries of the original Hamiltonian are spontaneously broken. 
The classification of symmetry breaking can be carried out systematically using a group-theoretic method, and the resulting symmetry of the order parameter determines the types of possible topological excitations, as discussed in the next section.
In this section, we introduce the concept of the order-parameter manifold $\mani{R}$ and present a brief overview of some basic notions and theoretical tools for investigating symmetries of the order parameter.

\subsection{Order-parameter manifold}
\label{sec:sym_OPM}
The order parameter for a spin-$f$ BEC is described with a set of $2f+1$ complex amplitudes $\bm\psi=(\psi_{-f},\psi_{-f+1},\cdots,\psi_f)^{\rm T}$, that moves in the space of $\mani{M}=\mathbb{C}^{2f+1}=\mathbb{R}^{4f+2}$.
In this section, we assume that the number density is nonzero everywhere
and treat a normalized spinor $\bm\zeta=\bm\psi/\sqrt{\bm\psi^\dagger \bm\psi}$.
The space of $\bm\zeta$ is isomorphic to the $(4f+1)$-dimensional sphere:
\begin{eqnarray}
\mani{M}=S^{4f+1}.
\end{eqnarray}

\begin{figure}[ht]
\begin{center}
\resizebox{0.5\hsize}{!}{
\includegraphics{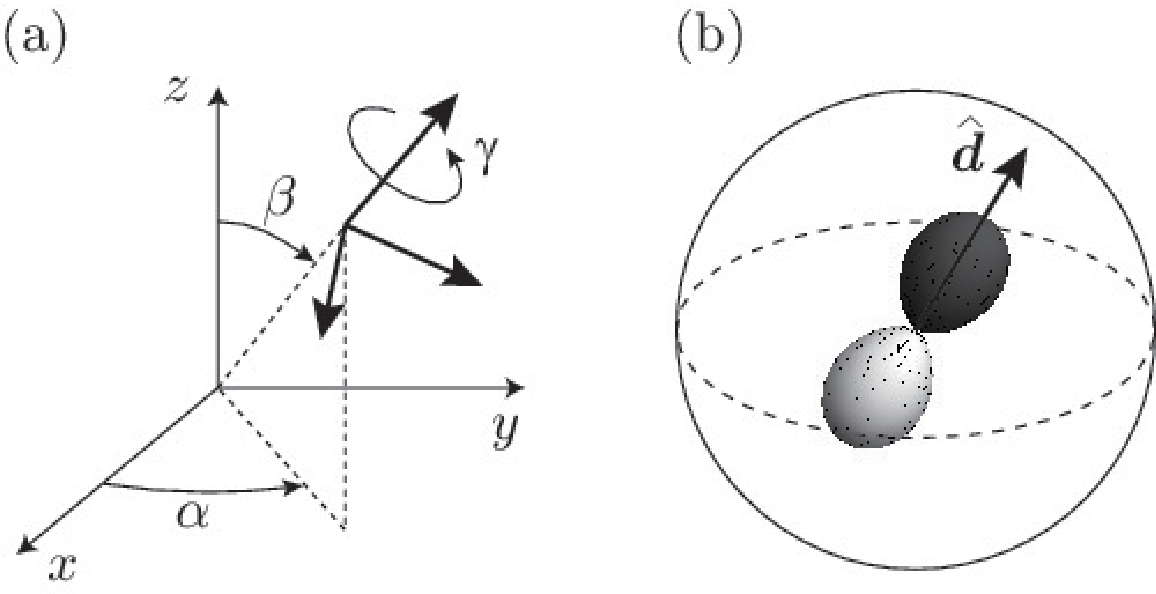}
}
\end{center}
\caption{(a) Euler rotation with Euler angles $\alpha, \beta$, and $\gamma$. (b) Two-sphere $S^2$ which is the manifold of a topological space of a three-dimensional unit vector $\hat{\bm d}$.}
\label{fig:sym_topologicalspace}
\end{figure}

Consider a group $G$ of transformations that act on $\mani{M}$ while leaving the mean-field energy functional invariant.
We assume that the total spin and total number of particles in the system of our concern are conserved.
Then, the energy functional in the absence of an external field is invariant under SO(3) rotations in spin space and U(1) gauge transformations.
The full symmetry group $G$ for a spinor BEC is therefore given by%
\footnote{Strictly speaking, the Hamiltonian is also invariant under the inversion of spins, that is, the system has the time-reversal symmetry. However, we will ignore the time-reversal symmetry because it does not affect the discussions in this section. For more detail, see Ref.~\cite{Kawaguchi2011}.}
\begin{align}
 G = {\rm SO}(3)_{\bf f} \times {\rm U}(1)_\phi,
\label{eq:G_spinor}
\end{align}
where subscripts ${\bf f}$ and $\phi$ denote the spin and gauge degrees of freedom, respectively.
Any element $g$ of $G$ defined in Eq.~\eqref{eq:G_spinor} can be represented as $g=e^{i\phi}e^{-i{\rm f}_z\alpha}e^{-i{\rm f}_y\beta}e^{-i{\rm f}_z\gamma}$,
where $\alpha$, $\beta$, and $\gamma$ are Euler angles [see Fig.~\ref{fig:sym_topologicalspace} (a)] and ${\rm f}_{x,y,z}$ are the $x,y,z$-components of spin-$f$ matrices.
Note that Eq.~\eqref{eq:G_spinor} is the symmetry for a general set of scattering lengths:
For some special cases, the symmetry of the system becomes larger, for example, it is $G={\rm U}(2f+1)$ if all scattering lengths are the same.

In the presence of the linear Zeeman effect (i.e., $p\neq 0$), $G$ reduces to 
\begin{align}
 G_{B1} = {\rm SO}(2)_{{\rm f}_z} \times {\rm U}(1)_\phi,
\label{eq:G_B1_spinor}
\end{align}
because the system will remain invariant only under rotations about the $z$ axis (the direction of the magnetic field) apart from the gauge transformation.
Any element $g'\in G_{B1}$ is represented as  $g'=e^{i\phi}e^{-i{\rm f}_z\alpha}$.
On the other hand, when $p=0$ and $q\neq 0$, the Hamiltonian is invariant under a $\pi$-rotation about an axis in the $x$--$y$ plane,
resulting in
\begin{align}
 G_{B2} = (D_\infty)_{{\rm f}_z} \times {\rm U}(1)_\phi,
\label{eq:G_B2_spinor}
\end{align}
where $D_\infty={\rm SO}(2)\rtimes \mathbb{Z}_2$ with $\mathbb{Z}_2\cong \{1,e^{-i{\rm f}_x \pi}\}$.
Here, the symbol $\rtimes$ is the semi-direct product which implies that the action of elements of SO(2) is affected by elements of $\mathbb{Z}_2$. To see this, let us consider an element of $D_\infty$ which is described by $g_{\alpha,n}=e^{-i{\rm f}_z\alpha}e^{-i{\rm f}_xn\pi}$ with $0\le\alpha < 2\pi$ and $n=0$ or 1.
When we consider a product of two elements $g_{\alpha,n_1}$ and $g_{\beta,n_2}$, we have
\begin{align}
 g_{\alpha,n_1}\circ g_{\beta,n_2} 
&= e^{-i{\rm f}_z\alpha}e^{-i{\rm f}_x n_1\pi}e^{-i{\rm f}_z\beta}e^{-i{\rm f}_x n_2\pi}\nonumber\\
&= e^{-i{\rm f}_z\alpha}e^{-i{\rm f}_x n_1\pi}e^{-i{\rm f}_z\beta}e^{i{\rm f}_x n_1\pi} e^{-i{\rm f}_x (n_1+n_2)\pi}\nonumber\\
&= \left\{ \begin{array}{l}
g_{\alpha+\beta,n_1+n_2} \ \ \textrm{for}\ \  n_1=0, \\
g_{\alpha-\beta,n_1+n_2} \ \ \textrm{for}\ \  n_1=1, \end{array}
\right.
\end{align}
Because the composition rule for the first indices ($\alpha$ and $\beta$) is affected by the second one $(n_1)$,
$D_\infty$ is not a direct product of SO(2) and $\mathbb{Z}_2$ but a semi-direct product of them.

For each $\bm\zeta\in \mani{M}$, the isotropy group (or the little group) 
$H_{\bm\zeta}$ is defined as the subgroup of $G$ that leaves $\bm\zeta$ invariant:
\begin{align}
H_{\bm\zeta} =\{g\in G\,|\,g\bm\zeta=\bm\zeta\} \subset G.
\end{align}
We also define the orbit $\mani{M}_{\rm O}(\bm\zeta)$ of $\bm\zeta$ by 
\begin{align}
\mani{M}_{\rm O}(\bm\zeta)=\{g\bm\zeta\,|\,g\in G\} \subset \mani{M},
\end{align}
i.e.,
$\mani{M}_{\rm O}(\bm\zeta)$ is the set of all points that are obtained by letting all transformations in $G$ act on $\bm\zeta$.
Because elements of $G$ do not change the mean-field energy, the orbit $\mani{M}_{\rm O}(\bm\zeta)$ constitutes a degenerate space in $\mani{M}$.
For the case of a spin-1 polar BEC, a general element of the orbit $\mani{M}_{\rm O}(\bm\zeta^{\rm polar})$ for the ground state is given by~\cite{Ho1998}
\begin{align}
g{\bm\zeta}^{\rm polar} = e^{i\phi}e^{-i{\rm f}_z\alpha}e^{-i{\rm f}_y\beta}e^{-i{\rm f}_z\gamma}\begin{pmatrix} 0\\1\\0 \end{pmatrix}
= e^{i\phi}
\begin{pmatrix}
-\frac{e^{-i\alpha}}{\sqrt{2}} \sin\beta \\
\cos\beta \\
\frac{e^{i\alpha}}{\sqrt{2}} \sin\beta
\end{pmatrix}.
\label{eqSCB95}
\end{align}
The orbit is thus parametrized by $\phi$ and $\hat{\bm d}\equiv(\cos\alpha\sin\beta,\sin\alpha\sin\beta,\cos\beta)$ that specifies the anisotropic direction of the nematic tensor [see Eq.~\eqref{eq:nematictensor_FP}].
The manifold of $\hat{\bm d}$ (a three-dimensional unit vector) is a unit sphere which is referred to as $S^2$, the two-sphere [see Fig.~\ref{fig:sym_topologicalspace} (b)],
whereas the manifold of $\phi$ is ${\rm U}(1)$.
Therefore, we conclude that the order-parameter manifold for the spin-1 polar state is given by
$\mani{R}^{\rm polar} \cong S^2_{\bf f} \times {\rm U(1)}_\phi$.
[Strictly speaking, we need to consider the discrete symmetries, see Eq.~\eqref{eq:R_polar}].

Mathematically, such a manifold is defined as a left coset of $H_{\bm \zeta}$.
For a subgroup $H\subset G$, the left cosets of $H$ are defined as $G/H=\{gH\,|\,g\in G\}$.
The coset space is identical to the orbit of $H$, and it forms a group if $H$ is a normal subgroup of $G$, i.e., if $gHg^{-1}=H$ for $^\forall g\in G$.
The coset space $G/H$ is equivalent to $G/H'$ if and only if $H$ and $H'$ are conjugate: $H'=gHg^{-1}$.
On the other hand, the isotropy groups of all points on the same orbit are conjugate to each other: $H_{g\bm\zeta}=gH_{\bm\zeta}g^{-1}$ for $^\forall g\in G$.
Therefore, the coset space does not depend on the location in the orbit.
We use $H$ as a representation of the conjugate class of the isotropy groups $gH_{\bm\zeta}g^{-1}$,
and define the order-parameter manifold as
\begin{align}
\mani{R}=G/H.
\end{align}
Here, because $H$ represents the remaining symmetry of the ordered state, $\mani{R}$ describes the broken symmetry, or a manifold of the degenerate space.

We also need to take into account the discrete symmetry of the order parameter~\cite{Zhou2001}.
The spin-1 polar phase has a spin-gauge coupled discrete symmetry. In fact, the polar order parameter $\bm\zeta^{\rm polar}=(0,1,0)^{\rm T}$ is invariant under a $\pi$-rotation about an axis
in the $x$--$y$ plane followed by the gauge transformation $e^{i\pi}$.
This can also be understood from the fact that Eq.~\eqref{eqSCB95} is invariant under the transformations of
$(\hat{\bm d},\phi)\to (-\hat{\bm d},\phi+\pi)$.
A group for such an operation $\{1,e^{i\pi}e^{-i{\rm f}_x\pi}\}$ is isomorphic to $\mathbb{Z}_2=\{-1,1\}$ because $(e^{i\pi}e^{-i{\rm f}_x\pi})^2=1$.
Therefore, the correct isotropy group of $\bm\zeta^{\rm polar}$ is the dihedral group $(D_\infty)_{{\rm f}_z,\phi} \cong {\rm SO(2)}_{{\rm f}_z} \rtimes (\mathbb{Z}_2)_{{\bf f},\phi}$,
and the order-parameter manifold is given by
\begin{align}
\mani{R}^{\rm polar} = \frac{{\rm SO(3)}_{\bf f} \times {\rm U(1)}_\phi}{{\rm SO(2)}_{{\rm f}_z}\rtimes(\mathbb{Z}_2)_{{\bf f},\phi}} 
\cong \frac{S^2_{\bf f}\times {\rm U(1)}_\phi }{(\mathbb{Z}_2)_{{\bf f},\phi}}.
\label{eq:R_polar}
\end{align}

For the case of the spin-1 ferromagnetic phase,
the order parameter $\bm\zeta^{\rm ferro}=(1,0,0)^{\rm T}$ is invariant under a spin rotation about the $z$ axis followed by a gauge transformation by the same amount as the rotation angle: $e^{i\phi}e^{-i{\rm f}_z\phi} \bm\zeta^{\rm ferro} = \bm\zeta^{\rm ferro}$.
This is a spin-gauge coupled SO(2) symmetry which is denoted as $H={\rm SO}(2)_{{\rm f}_z+\phi}$, where the subscript ${\rm f}_z+\phi$ is shown to indicate the nature of the coupling.
A general form of the spin-1 ferromagnetic order parameter is given by~\cite{Ho1998}
\begin{align}
g{\bm\zeta}^{\rm ferro} = e^{i\phi}e^{-i{\rm f}_z\alpha}e^{-i{\rm f}_y\beta}e^{-i{\rm f}_z\gamma}\begin{pmatrix} 1\\0\\0\end{pmatrix}
= e^{i(\phi-\gamma)}
\begin{pmatrix}
e^{-i\alpha}  \cos^2 \frac{\beta}{2} \\[1mm]
\frac{1}{\sqrt{2}} \sin \beta \\[1mm]
e^{i\alpha} \sin^2 \frac{\beta}{2}
\end{pmatrix}.
\label{eqSCB88}
\end{align}
Note that the linear combination $\phi-\gamma$ in Eq.~\eqref{eqSCB88} manifestly exhibits the spin-gauge symmetry, i.e., the equivalence between phase change and spin rotation.
Individual configurations of ${\bm\zeta}^{\rm ferro}$ are completely specified by the entire range of Euler angles $(\alpha,\beta, \gamma-\phi)$,
and therefore, the order-parameter manifold is given by
\begin{align}
\mani{R}^{\rm ferro} = \frac{{\rm SO(3)}_{\bf f}\times {\rm U(1)}_\phi}{{\rm SO(2)}_{{\rm f}_z+\phi}} \cong {\rm SO}(3)_{{\bf f},\phi}.
\end{align}

\subsection{Finding ground states from symmetry consideration}
\label{sec:sym_procedure}

The symmetry consideration gives further clues for finding the ground-state order parameters~\cite{Michel1971,Michel1980,Makela2007a,Yip2007,Kawaguchi2011}.
We first note that two order parameters on different orbits may share the same isotropy group.
The orbits of two such points are considered to be of the same type,
and we classify the types of orbits according to the conjugacy classes of subgroups of $G$.
In other words, for each conjugacy class of a subgroup of $G$, we obtain a set of orbits, the union of which is called a {\it stratum} and denoted by $\mani{M}_{\rm S}(\bm \zeta)$ in this review;
$\bm\zeta$ and $\bm\zeta'\in \mani{M}$ belong to the same stratum, if and only if their isotropy groups are conjugate to each other.
Clearly, we have $\mani{M}_{\rm O}(\bm\zeta)\subseteq \mani{M}_{\rm S}(\bm\zeta)\subset \mani{M}$.

It is proved by Michel~\cite{Michel1971,Michel1980} that the gradient of the energy functional with respect to the order parameter is tangent to the stratum $\mani{M}_{\rm S}(\bm\zeta)$.
In other words, the gradient of the energy functional vanishes in the direction along which the order parameter changes its symmetry, or changes the stratum.
Moreover, since the energy functional is a $G$-invariant function, its gradient is perpendicular to the tangential plane of $\mani{M}_{\rm O}(\bm\zeta)$.
Then, we obtain the following theorems known as Michel's theorems:

\vspace{2mm}
\noindent{\it Theorem 1 (inert state).}
If an orbit is isolated in the stratum, the orbit is stationary.

\vspace{2mm}
\noindent{\it Theorem 2 (non-inert state).}
If an orbit is not isolated in the stratum, 
we define a submanifold $\mani{M}_H\subset \mani{M}$ such that
\begin{align}
 \mani{M}_H=\{\bm\zeta\in \mani{M}| h\bm\zeta=\bm\zeta\ \textrm{for}\ ^\forall h\in H\},
\end{align}
where $H$ is a subgroup of $G$ that characterizes the stratum under consideration.
Let $\epsilon_H[\bm\zeta]$ be a real function which is the same as the energy functional $\epsilon[\bm\zeta]$ but the domain of which is restricted to $\mani{M}_H$.
Then, the stationary point of $\epsilon_H[\bm\zeta]$ on $\mani{M}_H$ is always a stationary point of $\epsilon[\bm\zeta]$ on $\mani{M}$.

\vspace{2mm}
\noindent
As an illustration of Michel's theorems,
we consider a smooth real function $f$ on a three-dimensional Euclidean space $(x,y,z)$ and assume that $f$ is invariant under SO(2) rotations about the $z$ axis and under the inversion with respect to the $x$--$y$ plane, namely,
$\mani{M}=\mathbb{R}^3$ and $G={\rm SO(2)}\times \mathbb{Z}_2$.
In this case, there are four strata: 
(i) the origin $x=y=z=0$ with isotropy group ${\rm SO(2)}\times \mathbb{Z}_2$, 
(ii) the $z$ axis with isotropy group SO(2),
(iii) the $x$--$y$ plane with isotropy group $\mathbb{Z}_2$, and 
(iv) all other points, the isotropy group of which has only the identity element.
From the symmetry consideration, it is clear that $\partial f/\partial x=\partial f/\partial y=\partial f/\partial z=0$ on (i), 
$\partial f/\partial x=\partial f/\partial y=0$ on (ii), and $\partial f/\partial z=0$ on (iii).
It follows that the function $f$ always takes its extremum (minimum, maximum, or saddle point) at the origin, that there exists at least one extremum along the $z$ axis (although it might coincide with the origin),
and that there exists at least one extremum on the $x$ axis, the orbit of which (a circle on the $x$--$y$ plane) also gives the extremum.
Michel's theorem is the generalization of this example.

It follows from Theorem 1 that all $G$-invariant functions on a manifold $\mani{M}$ have a common stationary orbit.
The corresponding state is called an inert state.
The order parameter for an inert state is independent of the interaction parameters in the Hamiltonian.
Examples include the polar and ferromagnetic phases in a spin-1 BEC, whose order parameters are given by $(0,1,0)^{\rm T}$ and $(1,0,0)^{\rm T}$, respectively.
Based on Theorem 1, inert states have been obtained for {\it p}- and {\it d}-wave superconductors~\cite{Volovik1985,Ozaki1985}, superfluid helium three~\cite{Bruder1986}, and spinor BECs~\cite{Makela2007a, Yip2007}.
On the other hand, Theorem 2 is instrumental in finding non-inert states, whose order parameter is dependent on the interaction parameters.
An example of a non-inert state is the AF phase in a spin-1 BEC, which appears in the presence of an external magnetic field.
For the case of spin-3 BECs, non-inert states arise even in the absence of an external field (see Table~\ref{table:spin3})~\cite{Kawaguchi2011}.

Based on Michel's theorems, the procedure for finding the ground state 
is summarized as follows:
\begin{enumerate}
\item Classify all subgroups of $G$ according to conjugacy classes.

\item Let $H$ be a subgroup of $G$.
Find such $\bm\zeta \in \mani{M}$ that it is invariant under $H$.

\item (inert state) If $\bm\zeta$ is uniquely determined, then $\bm\zeta$ is a stationary point of the energy functional, and the corresponding orbit $\mani{M}_{\rm O}(\bm\zeta)$ is a stationary orbit.

\item (non-inert state) If $\bm\zeta$ is not uniquely determined,
      then find the minimum of the energy functional in the submanifold $\mani{M}_H\equiv \{ \bm\zeta\in \mani{M}|h\bm\zeta=\bm\zeta\ \textrm{for}\ ^\forall h\in H\}$.
      A stationary point $\bm\zeta\in \mani{M}_H$ is also a stationary point of the energy functional in the whole space of $\mani{M}$.

\item Finally, compare the energies for the obtained stationary states and find the lowest one.
\end{enumerate}

Here, we present a list of the subgroups of SO(3) and show how to carry out step 2 of the above procedure for the case when $G$ is given by Eq.~\eqref{eq:G_spinor}.
The subgroups of ${\rm SO}(3)$ are well known and they are classified into continuous and discrete groups, as listed below.

\begin{itemize}
\item {\bf Continuous symmetries}\\
The only continuous subgroup of SO(3) is SO(2). 
Let us choose the $z$ axis as the symmetry axis.
An infinitesimal transformation $h\in H$ can be expressed as a combination of spin rotation and gauge transformation: 
\begin{align}
h=e^{i\delta_\phi}e^{-i {\rm f}_z \delta_z } \simeq 1+i\delta_\phi-i{\rm f}_z \delta_z,
\label{eq:h_continuous}
\end{align}
where $\delta_z$ and $\delta_\phi$ are infinitesimal real values.
The order parameter that is invariant under $h$ is an eigenstate of ${\rm f}_z$. 
Conversely, the eigenstate of ${\rm f}_z$ with eigenvalue $m$ has SO(2) symmetry and it is invariant under a spin rotation $e^{-i{\rm f}_z \phi}$ accompanied with gauge transformation $e^{im\phi}$.
For a spin-$f$ system, there are $2f+1$ states that have the SO(2) symmetry.
We denote this symmetry as ${\rm SO}(2)_{{\rm f}_z+m\phi}$, 
where the subscript is shown to explicitly indicate the symmetry between the spin and the gauge degrees of freedom.
The combined symmetry for the $m \neq 0$ states is called {\it continuous spin-gauge symmetry}, and it relates spin textures to the superfluid current, as discussed in Sec.~\ref{sec:hydro}.
On the other hand, when $m=0$, the order parameter $\zeta_m=\delta_{m0}$ has an additional $\mathbb{Z}_2$ symmetry: $\exp(-i {\rm f}_x\pi)\bm\zeta=(-1)^f\bm\zeta$,
where the spin and gauge degrees of freedom are coupled for odd $f$ and decoupled for even $f$.
In this case, the isotropy group is given by $D_\infty\cong {\rm SO}(2)\rtimes  \mathbb{Z}_2$.

\item {\bf Discrete symmetries}\\
The elements of the point groups of  ${\rm SO}(3)$ can be described with a combination of the following generators~(see, e.g., \textsection 93 of Ref.~\cite{LandauLifshitz_QM}):
\begin{description}
\item [$C_n$]: the cyclic group of rotations about the $z$ axis through angle $2\pi k/n$ with $k=1,\cdots,n-1$. 
The group is isomorphic to $\mathbb{Z}_n$ and it has generators $\{C_{n,z}\}$.
\item [$D_n$]: the dihedral group generated by $C_{n,z}$ and an additional rotation through $\pi$ about an orthogonal axis.
The group is isomorphic to $\mathbb{Z}_n \rtimes \mathbb{Z}_2$ and it has generators $\{C_{n,z}, C_{2,x}\}$.
\item [\ \ $T$]: the point group of the tetrahedron composed of four
three-fold axes and three two-fold axes. The generators of this group are given by $\{C_{3,z}, C_{2,\sqrt{2}x+z}\}$.
\item [\ \ $O$]: the point group of the octahedron composed of three
four-fold axes, four three-fold axes, and six two-fold axes. The generators of this group are given by $\{C_{4,z}, C_{2,x+z}\}$.
\item [\ \ $Y$]: the point group of the icosahedron composed of six
five-fold axes, ten three-fold axes, and 15 two-fold axes. The generators of this group are given by $\{C_{5,z}, C_{2,2x+(1+\sqrt{5})z}\}$.
\end{description}
Here, for the sake of simplicity of description, we choose the symmetry axis such that the highest symmetry axis is parallel to the $z$ axis.
The generator $C_{n,\Omega_x x+\Omega_y y + \Omega_z z}$ performs an action of a $2\pi/n$ rotation about the direction $\bm \Omega=(\Omega_x,\Omega_y,\Omega_z)$, and can be expressed in terms of the spin operator vector ${\bf f}$ as
\begin{align}
 C_{n,\Omega_x x+\Omega_y y + \Omega_z z} \equiv \exp\left[-i \frac{{\rm f}_x\Omega_x+{\rm f}_y\Omega_y+{\rm f}_z\Omega_z}{|\Omega|}\frac{2\pi}{n}\right],
\end{align}
which is described by a $(2f+1)$ $\times$ $(2f+1)$ matrix when acting on $\bm \zeta$.
Combined with a gauge transformation, 
the state invariant under the transformation $h=C_{n,\Omega_x x+\Omega_y y + \Omega_z z}e^{i\phi}$ 
can be obtained by solving the eigenvalue equation
\begin{align}
 C_{n,\Omega_x x+\Omega_y y + \Omega_z z}{\bm\zeta} = e^{-i\phi}{\bm\zeta}.
\end{align}
If the eigenvalue is 1, 
the invariance of the eigenstate is satisfied by a spin rotation alone, 
and therefore, the gauge symmetry is completely broken.
On the other hand, the eigenstate with an eigenvalue $e^{-i\phi}\neq 1$
is invariant under a simultaneous discrete transformation in spin and gauge.
Such a symmetry is called {\it discrete spin-gauge symmetry}.
Our task here is to find simultaneous eigenstates of
a set of generators for all discrete symmetry groups.
\end{itemize}

It should be noted that although there exist an infinite number of discrete subgroups of SO(3),
only a finite number of subgroups appear as isotropy groups of spin-$f$ system.
This can be understood from the Majorana representation of the order parameter (see Sec.~\ref{sec:Majorana}):
A spin-$f$ order parameter can be described with $2f$ vertices on a unit sphere;
the point group that includes more than $2f$ vertices does not appear as an isotropy group of a spin-$f$ system.
For example, the icosahedral group includes 12 vertices, and hence, it does not appear as an isotropy group for $f<5$~\cite{Makela2007a,Yip2007}.

In the presence of the quadratic Zeeman effect, 
the full symmetry of the system reduces to that in Eq.~\eqref{eq:G_B2_spinor}.
Therefore, the tetrahedron, octahedron, and icosahedron groups cannot be isotropy groups.
In the presence of the linear Zeeman effect, 
the full symmetry further reduces to Eq.~\eqref{eq:G_B1_spinor},
and only the cyclic groups can be the isotropy group.

We emphasize that the above procedure works well for the case in which stationary states have a certain symmetry.
If no symmetry remains (i.e., if $H=1$), the above procedure amounts to solving the GPEs directly.
We therefore do not consider the case of $H=1$.
In the absence of an external field,
all ground states of spinor BECs with spin $f=1,2$ and 3, superfluid $^3$He, and {\it p}- and {\it d}-wave superconductors have nontrivial remaining symmetries, whereas in the presence of an external field,
the symmetry of the system is sometimes completely broken, as in the case of the spin-1 broken-axisymmetry (BA) phase (see next subsection).

\subsection{Symmetry property of spin-1 BECs}
\label{sec:sym_spin1}

We apply the above procedure to spin-1 BECs.
Clearly, there are three inert states whose isotropy groups are continuous:
\begin{align}
{\rm F}_+ &:\ \ \bm\zeta^{\rm (F_+)} =(1,0,0)^{\rm T}, \ \ H^{\rm (F_+)}={\rm SO(2)}_{{\rm f}_z+\phi}=\{e^{i\alpha}e^{-i{\rm f}_z\alpha}|\,0\le\alpha<2\pi\},\label{eq:sym_spin1F+}\\
{\rm F}_- &:\ \ \bm\zeta^{\rm (F_-)} =(0,0,1)^{\rm T}, \ \ H^{\rm (F_-)}={\rm SO(2)}_{{\rm f}_z-\phi}=\{e^{-i\alpha}e^{-i{\rm f}_z\alpha}|\,0\le\alpha<2\pi\},\label{eq:sym_spin1F-}\\
{\rm P}   &:\ \ \bm\zeta^{\rm (P)} =(0,1,0)^{\rm T}, \ \ H^{\rm (P)}=(D_\infty)_{{\rm f}_z,\phi}=\{e^{-i{\rm f}_z\alpha}, e^{i\pi}U_2^\gamma|\,0\le\alpha,\gamma<2\pi\}, \label{eq:sym_spin1P}
\end{align}
where $U_2^\gamma\equiv e^{-i{\rm f}_z\gamma}C_{2,x}e^{i{\rm f}_z\gamma}$ is a $\pi$-rotation about an axis in the $x$--$y$ plane.
The isotropy groups of the above states are the subgroups of all $G$, $G_{B1}$, and $G_{B2}$, and therefore, these states are stationary, independently of the presence of an external magnetic field.
(The isotropy group of the polar state reduces to ${\rm SO(2)}_{{\rm f}_z}$ in the presence of the linear Zeeman effect.)
In the absence of an external field, ${\rm F}_{\pm}$ states are transformed to each other by a spin rotation, and therefore, describe the same phase.

\begin{figure}[ht]
\begin{center}
\resizebox{0.7\hsize}{!}{
\includegraphics{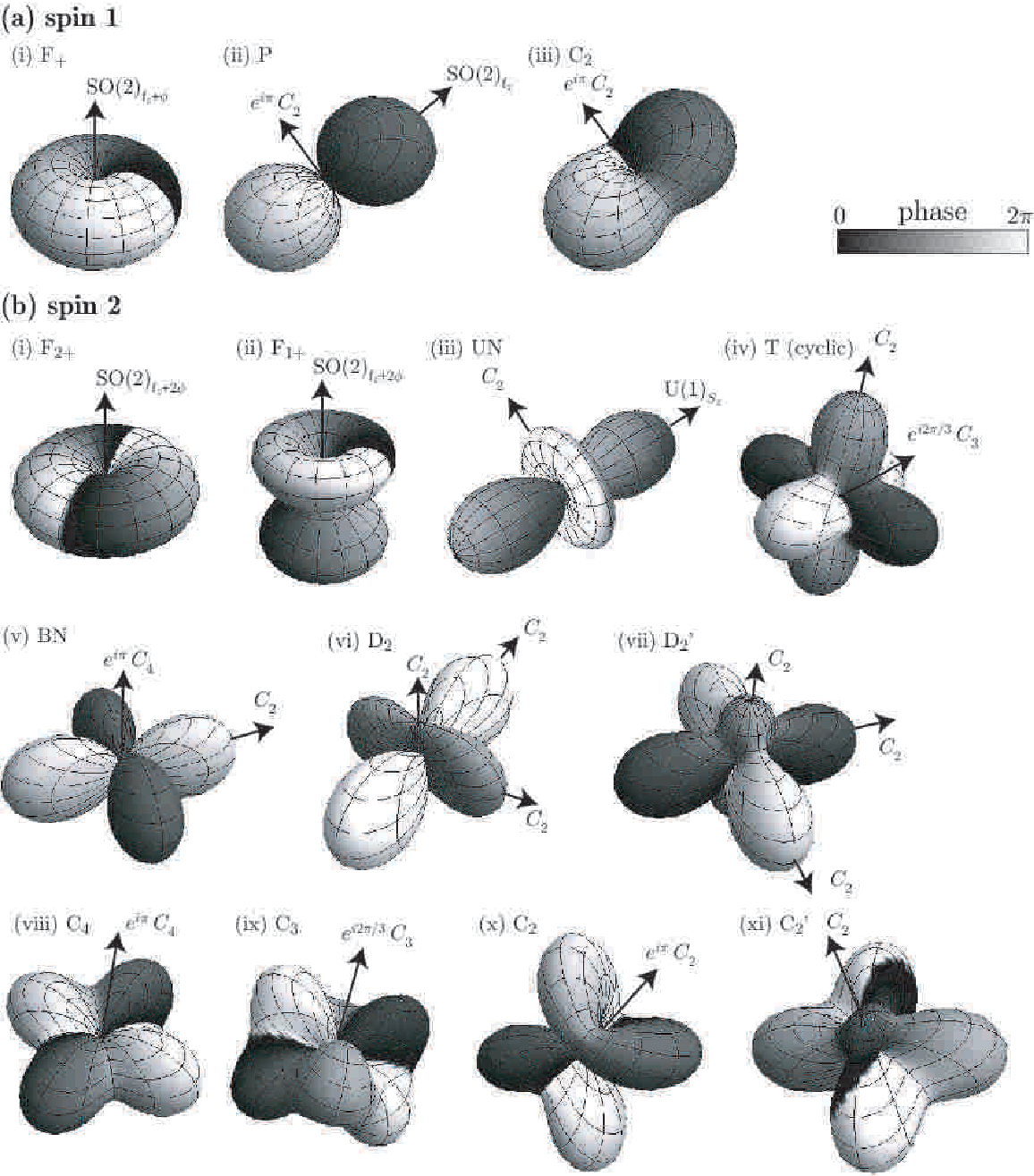}
}
\end{center}
\caption{Spherical-harmonic representation of the order parameters for (a) $f=1$ and (b) $f=2$ BECs.
The continuous and discrete symmetry axes are indicated along with the exponential factors of the gauge transformation.
}
\label{fig:OP}
\end{figure}

As for the discrete symmetry, only the $C_2$ symmetry can be an isotropy group of a spin-1 system,
because the spin-1 order parameter is described by two vertices in the Majorana representation.
The matrix form of $C_{2,z}$ in the spin-1 representation is given by
\begin{align}
 C_{2,z} = {\rm Diag}[-1,1,-1],
\end{align}
where ${\rm Diag}$ denotes the diagonal matrix.
An eigenstate of $C_{2,z}$ that has more than two nonzero components
is given by
\begin{align}
C_2:\ \ \bm\zeta^{(C_2)} = (\sqrt{1-\eta},0,\sqrt{\eta})^{\rm T},\label{eq:sym_spin1C2}
\end{align}
where $0< \eta <1$ and the isotropy group is given by
\begin{align}
H^{(C_2)}=\{1,C_{2,z}e^{i\pi}\}.
\end{align}
Here, different values of $\eta$ lead to different orbits,
whereas the relative phase between $\zeta_{\pm 1}$ and the global phase may vary in each orbit.
According to Theorem 2, we need to minimize the mean-field energy [see Eq.~\eqref{eq:spin1MFenergy}] for $\bm\zeta^{(C_2)}$ with respect to $\eta$.
In the absence of an external field, it ends up with the ferromagnetic ($\eta=0,1$) or polar ($\eta=1/2$) phases,
whereas in the presence of an external field, we obtain a nontrivial solution:
\begin{align}
 \bm\zeta^{\rm (AF)}& = \frac{1}{\sqrt{2}}\left( \sqrt{1-\frac{p}{c_1n}}, 0, \sqrt{1+\frac{p}{c_1n}}\right)^{\rm T},\ \ \ \left|\frac{p}{c_1n}\right|<1, \label{eq:sym_spin1AF}
\end{align}
which is identical to the antiferromagnetic state in Table~\ref{spin1table}.
The morphology of the order parameters in Eqs.~\eqref{eq:sym_spin1F+}, \eqref{eq:sym_spin1P} and \eqref{eq:sym_spin1C2} are depicted in Fig.~\ref{fig:OP} (a), together with the symmetry axes.

Comparing the energies for the order parameters in Eqs.~\eqref{eq:sym_spin1F+}--\eqref{eq:sym_spin1P} and \eqref{eq:sym_spin1AF}, 
we obtain the phase diagram shown in Fig.~\ref{fig:spin-1PD}, except for the region V (the BA phase).
The morphology of the order parameter for the BA phase is the same as the $C_2$ state~\eqref{eq:sym_spin1C2} [see Figs.~\ref{fig:spin1_OP_shape} and \ref{fig:OP} (a)(iii) ].
However, the two-fold symmetry axis is not parallel to the $z$ axis (the symmetry axis of the system) in the BA phase,
and the symmetry $G_{B1}$ or $G_{B2}$ of the system is completely broken.
Such a state cannot be obtained from symmetry considerations.

\subsection{Symmetry property of spin-2 BECs}
\label{sec:sym_spin2}
Since the spin-2 order parameter is described with four vertices,
the possible isotropy groups are SO(2), $T, D_4, C_4, D_3, C_3, D_2$, and $C_2$. 

\begin{itemize}
\item Continuous group SO(2):\\
There are five states that have continuous symmetries:
\begin{align}
{\rm F}_{2+}:\ \ &\bm\zeta^{\rm (F_{2+})}=(1,0,0,0,0)^{\rm T},\ \ H^{\rm (F_{2+})}={\rm SO(2)}_{{\rm f}_z+2\phi}=\{e^{i2\alpha}e^{-i{\rm f}_z\alpha}|0\le\alpha<2\pi\},\\
{\rm F}_{2-}:\ \ &\bm\zeta^{\rm (F_{2-})}=(0,0,0,0,1)^{\rm T},\ \ H^{\rm (F_{2-})}={\rm SO(2)}_{{\rm f}_z-2\phi}=\{e^{-i2\alpha}e^{-i{\rm f}_z\alpha}|0\le\alpha<2\pi\},\\ 
{\rm F}_{1+}:\ \ &\bm\zeta^{\rm (F_{1+})}=(0,1,0,0,0)^{\rm T},\ \ H^{\rm (F_{1+})}={\rm SO(2)}_{{\rm f}_z+\phi}=\{e^{i\alpha}e^{-i{\rm f}_z\alpha}|0\le\alpha<2\pi\},\\
{\rm F}_{1-}:\ \ &\bm\zeta^{\rm (F_{1-})}=(0,0,0,1,0)^{\rm T},\ \ H^{\rm (F_{1-})}={\rm SO(2)}_{{\rm f}_z-\phi}=\{e^{-i\alpha}e^{-i{\rm f}_z\alpha}|0\le\alpha<2\pi\},\\
{\rm UN}  :\ \ &\bm\zeta^{\rm (UN)}  =(0,0,1,0,0)^{\rm T},\ \ H^{\rm (UN)}  =(D_\infty)_{{\rm f}_z}=\{e^{-i{\rm f}_z\alpha}, U_2^\gamma|\,0\le\alpha,\gamma<2\pi\}.
\end{align}
The isotropy groups of the above states are the subgroups of all $G, G_{B1}$ and $G_{B2}$, and therefore, these states are stationary independently of the presence of an external magnetic field.
The isotropy group of the UN state reduces to ${\rm SO(2)}_{{\rm f}_z}$ in the presence of the linear Zeeman effect.
The above all states are inert states.

\item Tetrahedron group $T$:\\
The largest discrete group that can be an isotropy group of the spin-2 system is the symmetry of the tetrahedron.
We solve the simultaneous eigenstate of $C_{3,z}$ and $C_{2,\sqrt{2}x+z}$, 
whose matrix representations are given by
\begin{align}
C_{3,z} &= {\rm Diag}[e^{i2\pi/3},e^{i4\pi/3},1,e^{i2\pi/3},e^{i4\pi/3}],\label{eq:sym_matrixC3z_spin2}
\\
C_{2,\sqrt{2}x+z} &= \frac{1}{9}
\begin{pmatrix}
        1 &  2\sqrt{2} &  2\sqrt{6} &  4\sqrt{2} &          4 \\
2\sqrt{2} &          5 &  2\sqrt{3} &         -2 & -4\sqrt{2} \\
2\sqrt{6} &  2\sqrt{3} &         -3 & -2\sqrt{3} &  2\sqrt{6} \\
4\sqrt{2} &         -2 & -2\sqrt{3} &          5 & -2\sqrt{2} \\
        4 & -4\sqrt{2} &  2\sqrt{6} & -2\sqrt{2} &          1
\end{pmatrix}.
\end{align}
The only solution (up to the overall phase factor) is 
\begin{align}
 T \ ({\rm cyclic}):\ \ \bm\zeta^{(T)} = (1/\sqrt{3},0,0,\sqrt{2/3},0)^{\rm T},\label{eq:sym_spin2T}
\end{align}
or its time reversal $(0,\sqrt{2/3},0,0,1/\sqrt{3})^{\rm T}$.
This is the cyclic phase that arises for $c_1>0$ and $c_2>0$ in the absence of an external magnetic field (see Sec.~\ref{sec:MFTspin2}).
Since $\bm\zeta^{(T)}$ satisfies
$C_{3,z}\bm\zeta^{(T)} = e^{i2\pi/3}\bm\zeta^{(T)}$ and 
$C_{2,\sqrt{2}x+z}\bm\zeta^{(T)} = \bm\zeta^{(T)}$,
the isotropy group is generated by
\begin{align}
\bar{H}^{(T)} = \{e^{-i2\pi/3}C_{3,z}, C_{2,\sqrt{2}x+z}\}.
\label{eq:H_for_cyclic}
\end{align}
Here and henceforth, we denote a set of generators of $H$ by $\bar{H}$.
The cyclic phase is not necessarily stationary in the presence of a magnetic field,
because $T\not\subset G_{B1}, G_{B2}$.
In fact, the tetrahedron symmetry reduces to $C_3$, $D_2$, or $C_2$ in the presence of a magnetic field (see below and Sec.~\ref{sec:MFTspin2}).

\item Dihedral-four group $D_4$:\\
We solve the simultaneous eigenstate of $C_{4,z}$ and $C_{2,x}$, whose matrix forms are given by
\begin{align}
 C_{4,z}&={\rm Diag}[-1,-i,1,i,-1],\\
 C_{2,x}&=\begin{pmatrix} 
0 & 0 & 0 & 0 & 1\\
0 & 0 & 0 & 1 & 0\\
0 & 0 & 1 & 0 & 0\\
0 & 1 & 0 & 0 & 0\\
1 & 0 & 0 & 0 & 0\\
\end{pmatrix},\label{eq:sym_matrixC2x_spin2}
\end{align}
obtaining 
\begin{align}
 D_4\ ({\rm BN}):\ \ \bm\zeta^{(D_4)}=\frac{1}{\sqrt{2}}(1,0,0,0,1)^{\rm T}.
\label{eq:sym_spin2BN}
\end{align}
Since $\bm\zeta^{(D_4)}$ satisfies $C_{4,z}\bm\zeta^{(D_4)}=e^{i\pi}\bm\zeta^{(D_4)}$ and $C_{2,x}\bm\zeta^{(D_4)}=\bm\zeta^{(D_4)}$,
the isotropy group is generated by
\begin{align}
\bar{H}^{(D_4)} = \{ e^{-i\pi}C_{4,z}, C_{2,x}\}.
\end{align}
This is the biaxial nematic (BN) state that appears for $p=0$.

\item Cyclic group $C_4$:\\
In the presence of the linear Zeeman effect, the $D_4$ symmetry of the BN state reduces to $C_4$.
An eigenstate of $C_{4,z}$ with the eigenvalue $e^{i\pi}$ is written as
\begin{align}
 \frac{1}{\sqrt{2}}\left(\sqrt{1+\eta},0,0,0,\sqrt{1-\eta}\right)^{\rm T},\label{eq:sym_spin2C4}
\end{align}
where $-1<\eta< 1$.
We substitute Eq.~\eqref{eq:sym_spin2C4} in the mean-field energy of a spin-2 BEC [Eq.~\eqref{eqS2ad6}] and minimize it with respect to $\eta$.
Then, for $p=0$, we obtain the ${\rm F}_{2\pm}$ states ($\eta=\pm1$) or the BN state ($\eta=0$).
On the other hand, for a nonzero $p$, there is a non-inert state given by
\begin{align}
 C_4:\ \ \bm\zeta^{(C_4)} = \frac{1}{\sqrt{2}}\left(\sqrt{1+\frac{p/2}{(c_1-c_2/20)n}},0,0,0,\sqrt{1-\frac{p/2}{(c_1-c_2/20)n}}\right)^{\rm T}.
\end{align}
The isotropy group of $\bm\zeta^{(C_4)}$ is generated by
\begin{align}
\bar{H}^{(C_4)} = \{e^{-i\pi}C_{4,z}\}.
\end{align}

\item Dihedral-three group $D_3$:\\
There is no state that has the $D_3$ symmetry, because there is no simultaneous eigenstate of 
$C_{3,z}$ and $C_{2,x}$, the matrix forms of which are given by Eqs.~\eqref{eq:sym_matrixC3z_spin2} and \eqref{eq:sym_matrixC2x_spin2}, respectively.

\item Cyclic group $C_3$:\\
The eigenstates of $C_{3,z}$ are given by
\begin{align}
 &\frac{1}{\sqrt{3}}\left(\sqrt{1-\eta},0,0,\sqrt{2+\eta},0\right)^{\rm T},\\
 &\frac{1}{\sqrt{3}}\left(0,\sqrt{2+\eta},0,0,\sqrt{1-\eta}\right)^{\rm T},
\end{align}
where $-2 < \eta < 1$.
Substituting the order parameters in the mean-field energy and minimizing with respect to $\eta$, we obtain the following non-inert state:
\begin{align}
 C_{3+}&:\ \ \bm\zeta^{(C_{3+})}=\frac{1}{\sqrt{3}}\left(\sqrt{1+\frac{p-q}{c_1n}},0,0,\sqrt{2-\frac{p-q}{c_1n}},0\right)^{\rm T},\\
 C_{3-}&:\ \ \bm\zeta^{(C_{3-})}=\frac{1}{\sqrt{3}}\left(0,\sqrt{2+\frac{p+q}{c_1n}},0,0,\sqrt{1-\frac{p+q}{c_1n}}\right)^{\rm T},
\end{align}
whose isotropy groups are respectively generated by
\begin{align}
 \bar{H}^{(C_{3+})} &= \{e^{-i2\pi/3}C_{3,z}\},\\
 \bar{H}^{(C_{3-})} &= \{e^{i2\pi/3}C_{3,z}\}.
\end{align}
For $p=q=0$,
both $\bm\zeta^{(C_{3+})}$ and $\bm\zeta^{(C_{3-})}$ reduce to the cyclic state~\eqref{eq:sym_spin2T}.
In this case, $\bm\zeta^{(C_{3+})}$ and $\bm\zeta^{(C_{3-})}$ are transfered to each other by a rotation in spin space, i.e., they belong to the same orbit.

\item Dihedral-two group $D_2$:\\
The matrix representation of $C_{2,z}$ is given by
\begin{align}
 C_{2,z} = {\rm Diag}[1,-1,1,-1,1].
\end{align}
There are two simultaneous eigenstates of $C_{2,z}$ and $C_{2,x}$.

{\it Case} (i): The order parameter
\begin{align}
\frac{1}{\sqrt{2}}(0,1,0,1,0)^{\rm T}
\label{eq:sym_spin2C20}
\end{align}
is the simultaneous eigenstate of $C_{2,z}$ and $C_{2,x}$ with eigenvalues $-1$ and $1$, respectively.
However, this state has the same symmetry as that of the biaxial nematic ($D_4$) state.
This fact can be understood if we visualize the order parameter using the spherical-harmonic representation [Fig.~\ref{fig:SHR} (g) and (h)] or the Majorana representation [Fig.~\ref{fig:MR} (g) and (h)].
Both states in Eqs.~\eqref{eq:sym_spin2BN} and \eqref{eq:sym_spin2C20}, survive even in the presence of the quadratic Zeeman effect.
The linear Zeeman term breaks the $D_2$ symmetry of the order parameter~\eqref{eq:sym_spin2C20} to $C_2$.

{\it Case} (ii): The other simultaneous eigenstate can be written as
\begin{align}
\frac{1}{2}\left(\sqrt{1-\eta},0,e^{i\delta}\sqrt{2(1+\eta)},0,\sqrt{1-\eta}\right)^{\rm T},\label{eq:sym_spin2D2}
\end{align}
where $-1< \eta< 1$, $0\le \delta< 2\pi$, and the eigenvalues of $C_{2,z}$ and $C_{2,x}$ are both equal to $1$.
(Note that we can always choose the $m=\pm2$ components to be real by applying a spin rotation about the $z$ axis and a gauge transformation.)
Substituting the order parameter~\eqref{eq:sym_spin2D2} in the expression of the mean-field energy,
we find that in the absence of an external field (i.e., $p=q=0$) 
the stationary points of the mean-field energy exist at $\delta=0$ and arbitrary $\eta$, and at $\delta=\pi$ and $\eta=0$: The former is the nematic state, the order parameter of which is given by a superposition of the UN and BN states:
\begin{align}
 D_2\ ({\rm nematic})&:\ \ \bm\zeta^{(D_2)} = \left(\frac{\cos\chi}{\sqrt{2}},0,\sin\chi,0,\frac{\cos\chi}{\sqrt{2}}\right)^{\rm T}\ \ (0 \le \chi \le \pi/2),
\end{align}
whereas the latter describes the cyclic state.
Here, the mean-field energy for $\bm\zeta^{(D_2)}$ does not depend on $\chi$, implying that there is an additional symmetry in the nematic phase, i.e., the SO(5) symmetry (see also Sec.~\ref{sec:QNDmode}).
When we turn on the quadratic Zeeman effect, the degeneracy in the nematic state is lifted and the UN ($\eta=0$) and BN ($\eta=-1$) states arise, which have the symmetry of $D_\infty$ and $D_4$, respectively. Which state appears depends on the sign of $q$. 
On the other hand, the tetrahedral symmetry of the cyclic state is broken into $D_2$ and the order parameter deforms as
\begin{align}
 D_2'&:\ \ \bm\zeta^{(D_2')} = \frac{1}{2}\left(\sqrt{1-\frac{10q}{c_2n}},0,i\sqrt{2\left(1+\frac{10q}{c_2n}\right)},0,\sqrt{1-\frac{10q}{c_2n}}\right)^{\rm T}.
\end{align}
While the isotropy groups for $\bm\zeta^{(D_2)}$ and $\bm\zeta^{(D_2')}$ are the same and is given by
\begin{align}
 \bar{H}^{(D_2, D_2')} = \{C_{2,z},C_{2,x}\},
\end{align}
they differ in the time reversal symmetry: The time reversal symmetry is preserved in $\bm\zeta^{(D_2)}$, and it is broken in $\bm\zeta^{(D_2')}$.

\item Cyclic group $C_2$:\\
In the presence of the linear Zeeman effect, the $D_2$ symmetry of Eq.~\eqref{eq:sym_spin2C20} is broken and the symmetry reduces to $C_2$, where the order parameter is given by
\begin{align}
C_2:\ \ &\bm\zeta^{(C_2)}\frac{1}{\sqrt{2}}\left(0,\sqrt{1+\frac{p}{(c_1-c_2/5)n}},0,\sqrt{1-\frac{p}{(c_1-c_2/5)n}},0\right)^{\rm T},
\end{align}
and the corresponding isotropy group is given by
\begin{align}
 H^{(C_2)} = \{1,e^{i\pi}C_{2,z}\}.
\end{align}
On the other hand, the linear Zeeman deforms the order parameters $\bm\zeta^{(D_2)}$ and $\bm\zeta^{(D_2')}$ to
\begin{align}
C_2':\ \ &\bm\zeta^{(C_2')}=\left(a,0,be^{i\delta},0,c\right)^{\rm T}\ \ \ (a,b,c\in \mathbb{R}, a^2+b^2+c^2=1, \delta=0,\pi/2).
\end{align}
The isotropy group for $\bm\zeta^{(C_2')}$ is given by
\begin{align}
 H^{(C_2')}=\{1,C_{2,z}\}.
\end{align}
Since the time reversal symmetry is broken for both $\delta=0$ and $\pi/2$,
we cannot distinguish the corresponding states based on the symmetry classification alone.

\end{itemize}

The obtained results are the same as those listed in Table~\ref{table3}.
The structure of each order parameter is shown in Fig.~\ref{fig:OP} (b).

\subsection{Symmetry and order parameter structure of spin-3 spinor BECs}
It is straightforward to carry out the procedure for spin-3 BECs but it is rather involved.
For the case of zero external field,
the ground states obtained in the above procedure is summarized in Table~\ref{table:spin3}.
Here, we just show the the isotropy group of each ground-state phase in Table~\ref{table:spin3symmetry}.
The detailed discussion is given in Ref.~\cite{Kawaguchi2011}.

\begin{table}[!h]
\begin{center}
{\renewcommand{\arraystretch}{1.2}
\begin{tabular}{cll}\hline
phase & symmetry & $H, \bar{H}$  \\ \hline\hline
A     & $D_6$    & $\{e^{i\pi}C_{6,z}, e^{i\pi} C_{2,x}\}$ \\
B     & $D_2$    & $\{C_{2,z}, e^{i\pi}C_{2,x}\}$ \\
C     & $C_2$    & $\{e^{i\pi}C_{2,z}\}$ \\
D     & $O$      & $\{e^{i\pi}C_{4,z},e^{i\pi}C_{2,x+z}\}$ \\
E     & $D_3$    & $\{C_{3,z},e^{i\pi}C_{2,x}\}$ \\
FF    & SO(2)    & $\{e^{-i{\rm f}_z\alpha}e^{i3\alpha}|0\le\alpha< 2\pi\}$\\
F     & SO(2)    & $\{e^{-i{\rm f}_z\alpha}e^{i2\alpha}|0\le\alpha< 2\pi\}$\\
G     & $C_2$    & $\{C_{2,z}\}$ \\
H     & $C_5$    & $\{e^{-i4\pi/5}C_{5,z}\}$ \\
I(HH) & $C_3$    & $\{e^{-i2\pi/3}C_{3,z}\}$ \\
J     & $C_4$    & $\{e^{i\pi/2}C_{4,z}\}$ \\ \hline
\end{tabular}}
\caption{Isotropy group for each phase of a spin-3 spinor BEC given in Table~\ref{table:spin3}.
The right column represents the isotropy groups for phase F and FF, and the generators of the isotropy groups for other phases~\cite{Kawaguchi2011}.}
\label{table:spin3symmetry}
\end{center}
\end{table}

%% file: topology.tex
\section{Topological excitations}
\label{sec:topology}

Bose-Einstein condensates can accommodate topological excitations such as vortices, monopoles,
and Skyrmions. 
These topological excitations are diverse in their physical properties but have one thing in common; they can move freely in space and time without changing their characteristics that are distinguished by topological charges. 
The topological charges take on discrete values and have very distinct characteristics independently of the material properties. 
It is these material-independent universal characteristics and robustness to external perturbations that make topological excitations unique.

In the following subsections, we first briefly review the homotopy theory in Sec.~\ref{sec:HomotopyTheory}, which gives mathematical classification of topological excitations,
and then show some examples of topological excitations in spinor BECs in Secs.~\ref{sec:Line defects}--\ref{sec:skyrmions}.

\subsection{Homotopy theory}
\label{sec:HomotopyTheory}

\subsubsection{Classification of topological excitations}
\label{sec:homotopy_classification}

The classification of topological excitations is best made by homotopy theory~\cite{Mermin1979,Mineev1998}. 
This theory describes what types of topological excitations are allowed in what order-parameter manifold. For example, $\pi_2({\rm U}(1))=0$ implies that monopoles, which are characterized by the second homotopy group $\pi_2$, are not topologically stable in systems described by scalar order parameters. This theory also describes what happens if two defects coexist and how they coalesce or disintegrate.

\begin{figure}[ht]
\begin{center}
\resizebox{0.8\hsize}{!}{
\includegraphics{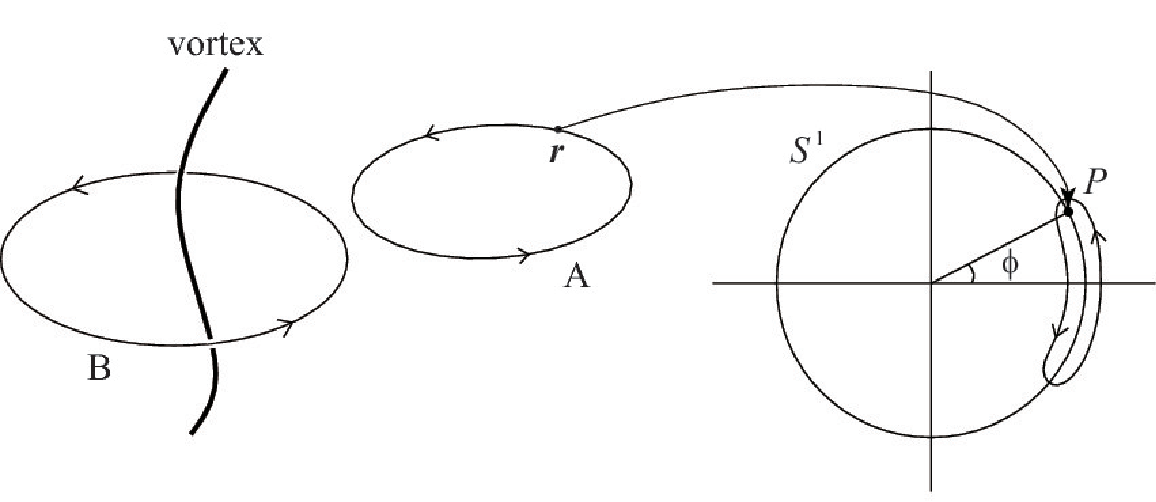}
}
\end{center}
\caption{Mapping from a loop in real space onto the order-parameter manifold $S^1$ (unit circle) according to the correspondence $\phi({\bm r}):  {\bm r}\rightarrow{\rm arg}\psi({\bm r})$. This mapping defines the first homotopy group $\pi_1(S^1)$.}
\label{fig:homotopy2}
\end{figure}

By way of introduction, we first consider a BEC described by a scalar order parameter $\psi({\bm r})=|\psi({\bm r})|e^{i\phi({\bm r})}$. To classify line defects such as vortices, we take a loop in the condensate and consider a map from every point ${\bm r}$ on the loop to the phase $\phi({\bm r})$ of the order parameter $\psi$: 
\begin{eqnarray}
 {\bm r}\mapsto \phi({\bm r})\equiv {\rm arg}\psi({\bm r}).
\label{homotopy1}
\end{eqnarray}
If the loop encircles no vortex (e.g., loop A in Fig.~\ref{fig:homotopy2}), the image of the map covers only a part of the unit circle, as shown on the right-hand side of Fig.~\ref{fig:homotopy2}. If the loop encircles a vortex (e.g., loop B), the image covers the entire unit circle at least once. If it covers the circle $n_{\rm w}$ times, it is said that the winding number of the vortex is $n_{\rm w}$. The crucial observation here is that the winding number does not change if the loop deforms or moves in space as long as it does not cross the vortex line. This property can be used to classify loops according to their winding number.

For general cases, we assume that the kinetic energy density associated with the topological excitation is small enough (at least on the loop under consideration) so that
the order parameter is restricted to the order-parameter manifold $\mani{R}$ of a certain phase. 
Then, a loop $\ell\in\mani{R}$ can be defined as an image of a map from a loop in real space to the order-parameter manifold.
If two loops $\ell_a$ and $\ell_b$ are continuously transformable within the order-parameter manifold $\mani{R}$, they are said to be homotopic to each other and written as $\ell_a\sim\ell_b$. 
The homotopic relationship is an equivalent one in that it is symmetric (i.e., $\ell_a\sim\ell_a$), reflexive (i.e., if $\ell_a\sim\ell_b$, then $\ell_b\sim\ell_a$), and transitive (i.e., if $\ell_a\sim\ell_b$ and $\ell_b\sim\ell_c$, then $\ell_a\sim\ell_c$).
By this equivalence relationship, all loops in $\mani{R}$ are classified into equivalent classes called homotopy classes: 
\begin{eqnarray}
[\ell_1], \ [\ell_2], \ [\ell_3], \cdots, 
\label{homotopyclass}
\end{eqnarray}
where $\ell_i \ (i=1,2,3,\cdots)$ is an arbitrarily chosen loop from the homotopy class $[\ell_i]$ because all loops belonging to the same class are equivalent and continuously deformable to each other.

Mathematically, a loop $\ell$ is defined as a mapping from $I=[0,1]$ to a topological space such that $\ell(0)=\ell(1)=x_0$, where $x_0\in\mani{R}$ is called a base point. If two loops $\ell_a$ and $\ell_b$ share the same base point $x_0$ and they are continuously deformable to each other, they are said to be homotopic at $x_0$. If the base point is not shared, they are said to be freely homotopic. The constant loop $c$ is defined as the one such that $c(t)=x_0$ for ${}^\forall t\in[0,1]$. The inverse of loop $\ell$ is defined as $\ell^{-1}(t)\equiv\ell(1-t)$ for ${}^\forall t\in[0,1]$.

\begin{figure}[ht]
\begin{center}
\resizebox{0.4\hsize}{!}{
\includegraphics{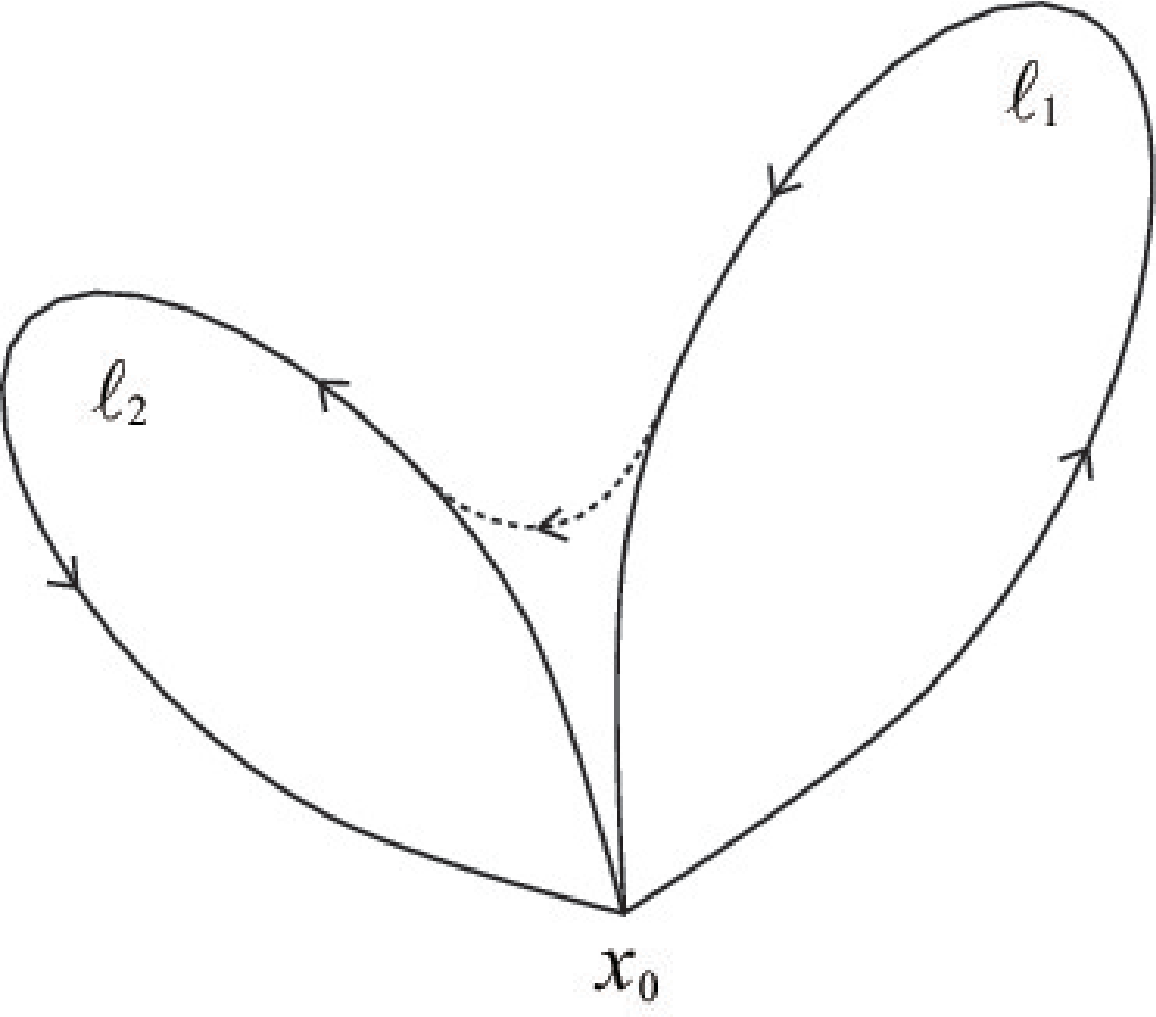}
}
\end{center}
\caption{The product of two loops $\ell_1 \cdot \ell_2$ in which $\ell_1$ is first traversed and then $\ell_2$ is traversed. Loops homotopic to  $\ell_1 \cdot \ell_2$ need not return to $x_0$ en route to the terminal point, as shown in the dashed curve.}
\label{fig:homotopy1}
\end{figure}

The product of two homotopy classes $[\ell_1]$ and $[\ell_2]$ is defined as 
\begin{eqnarray}
[\ell_1] \cdot [\ell_2]
= [\ell_1 \cdot \ell_2],
\label{productofloops}
\end{eqnarray}
where $\ell_1 \cdot \ell_2$ denotes the product of two loops in which $\ell_1$ is first traversed and then $\ell_2$ is traversed. Members of $[\ell_1 \cdot \ell_2]$ that are homotopic to $\ell_1 \cdot \ell_2$ need not return to the base point $x_0$ en route to the terminal point (see the dashed curve in Fig.~\ref{fig:homotopy1}).

With the definition of the product~(\ref{productofloops}), a set of homotopy classes form a group. In fact, they satisfy the associative law
\begin{eqnarray}
([\ell_1] \cdot [\ell_2]) \cdot [\ell_3]
= [\ell_1] \cdot ([\ell_2] \cdot [\ell_3]);
\label{eq10012}
\end{eqnarray}
those loops that are homotopic to the constant loop $c$ constitute the identity element $[c]$ such that
\begin{eqnarray}
[c] \cdot [\ell]
= [\ell] \cdot [c] 
\text{ for any } [\ell];
\label{eq10013}
\end{eqnarray}
finally, the inverse $[\ell^{-1}]$ of $[\ell]$ is the homotopy class that consists of inverse loops of $[\ell]$ so that
\begin{eqnarray}
[\ell^{-1}] \cdot [\ell]
= [\ell] \cdot [\ell^{-1}]
= [c].
\label{eq10014}
\end{eqnarray}
The group defined above is called the fundamental group\index{fundamental group} or the first homotopy group\index{first homotopy group} and it is defined as $\pi_1 (\mani{R}, x_0)$, where $x_0$ is the base point of all loops. Since the choice of the base point is not important in many situations in physics, we shall omit it and write $\pi_1 (\mani{R}, x_0)$ simply as $\pi_1(\mani{R})$.

The fundamental group is said to be Abelian if any two elements commute (i.e., $[\ell]\cdot [\ell']=[\ell']\cdot [\ell]$ for $^\forall [\ell], ^\forall [\ell']\in\mani{R}$) and non-Abelian if some elements do not commute.
For example, for the case of $\mani{R}={\rm U}(1)\cong S^1$, the fundamental group is isomorphic to the additive group of integers
\begin{eqnarray}
\pi_1(S^1)\cong {\mathbb Z},
\label{pi_1S^1}
\end{eqnarray}
and hence, the fundamental group is Abelian.
Here, each integer of $\mathbb{Z}$ describes the number of times the loop $\ell$ encircles the unit circle on the right-hand side of Fig.~\ref{fig:homotopy1},
corresponding to the winding number, or the quantization number of mass circulation.
In this case, the product ``$\,\cdot\,$'' simply means the addition of the winding numbers;
if two singly quantized vortices coalesce, a doubly quantized vortex results.

\begin{figure}[ht]
\begin{center}
\resizebox{0.8\hsize}{!}{
\includegraphics{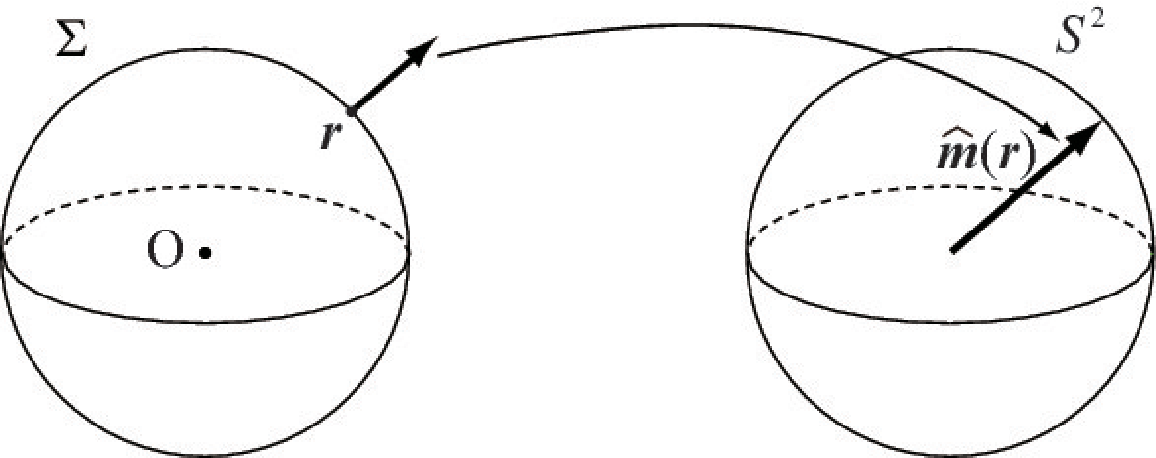}
}
\end{center}
\caption{Mapping from a closed surface in real space onto the order-parameter manifold $\mani{R}=S^2$ according to the correspondence $\hat{\bm m}({\bm r}):  {\bm r}\rightarrow \mani{R}$. 
This mapping defines the second homotopy group $\pi_2(\mani{R})$. }
\label{fig:homotopy3}
\end{figure}

\begin{figure}[!ht]
\begin{center}
\resizebox{0.7\hsize}{!}{
\includegraphics{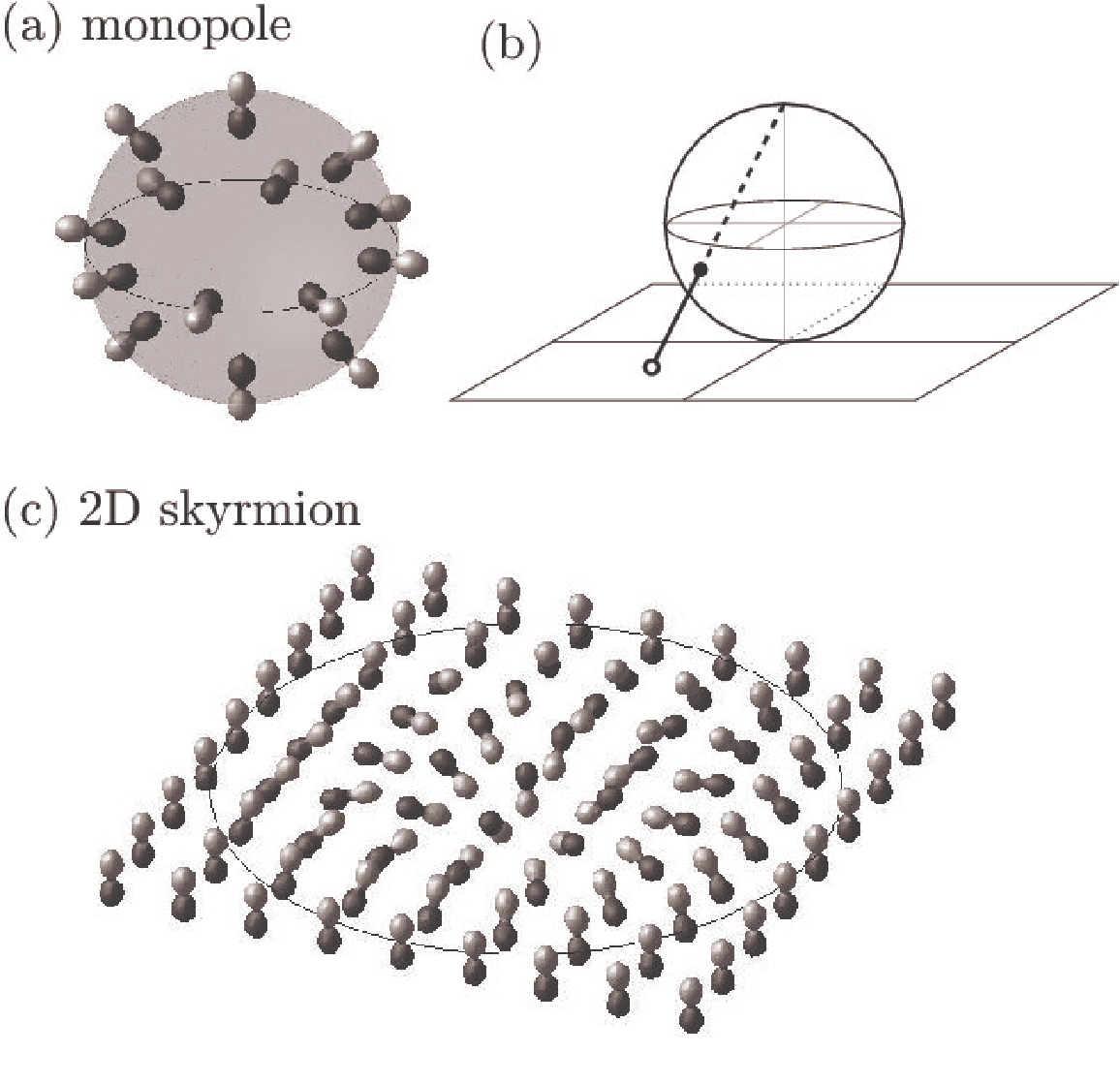}
}
\end{center}
\caption{(a) Monopole with charge 1 in the spin-1 polar phase. (b) 2D plane $I^2$ is compactified into a unit sphere $S^2$ by identifying all infinite points. (c) 2D Skyrmion in the spin-1 polar phase, having the same charge as (a).
}
\label{fig:polar_monopole}
\end{figure}

Point defects such as monopoles can be characterized by considering a closed surface $\Sigma$ that covers the object (see Fig.~\ref{fig:homotopy3}).
We consider a map from each point ${\bm r}$ on $\Sigma$ to a point ${\bm \psi}$ in the order-parameter manifold $\mani{R}$:
\begin{eqnarray}
{\bm \psi}: \Sigma \to \mani{R},
\label{pi_2M}
\end{eqnarray}
and classify elements of such a map by regarding as equivalent any two elements that can transform into each other in a continuous manner. 
For the case when $\mani{R}$ is isomorphic to a two-dimensional sphere or 2-sphere $S^2$, as shown on the right-hand side of Fig.~\ref{fig:homotopy3},
the map (\ref{pi_2M}) is classified by the number of times the image of $\Sigma$ wraps $S^2$.
In this case, we have $\pi_2(S^2)\cong \mathbb{Z}$.
An example of a monopole (with the winding number 1) in the spin-1 polar phase is depicted in Fig.~\ref{fig:polar_monopole} (a).

The mathematical definition for the $n$th homotopy group $\pi_n(\mani{R},x_0)$ 
[which is also written as $\pi_n(\mani{R})$ when the choice of the base point $x_0$ is not important]
is a group of homotopy equivalent classes of maps from an $n$-dimensional cube $I^n =\{ (t_1,t_2,\cdots,t_n)|0\le t_i \le 1 (1\le i \le n)\}$ 
to a manifold $\mani{R}$:
\begin{align}
\alpha : I^n \to \mani{R},
\end{align}
such that the boundary of $I^n$, $\partial I^n =\{(t_1,t_2,\cdots,t_n)\in I^n| \textrm{some }t_i=0 \textrm{  or  } 1\}$, is mapped to the base point $x_0 \in \mani{R}$.
If $I^n/\partial I^n$ denotes the cube $I^n$ whose boundary $\partial I^n$ is shrunk to a point, we have $I^n/\partial I^n\cong S^n$.
An example for $n=2$ is depicted in Fig.~\ref{fig:polar_monopole} (b) where $I^2$ is compactified to $S^2$.
It follows from the above definition that the second homotopy group classifies not only point defects (monopoles) but also two-dimensional (2D) nonsingular structures under a fixed boundary condition.
Figure~\ref{fig:polar_monopole}(c) describes an example of a nonsingular structure on the 2D plane, which has the same charge as Fig.~\ref{fig:polar_monopole}(a).
Such a 2D nonsingular structure is called a 2D Skyrmion.

In a similar manner, we can construct a three-dimensional (3D) nonsingular structure, Skyrmions and knots, which are classified by the third homotopy group. 
In this case, we assume that the order parameter becomes uniform at the boundary of a 3D cube (which is sometimes taken at spatial infinity).
Then, the three-dimensional space is compactified into a three-dimensional sphere $S^3$. 
On the other hand, the zeroth homotopy $\pi_0(\mani{R})$ indicates the number of disconnected pieces in $\mani{R}$, classifying domain walls; 
in particular, $\pi_0(\mani{R})=0$ means that the order-parameter manifold is a connected space, implying no topologically stable domain. 
Table~\ref{tb:defects_solitons} summarizes the topological objects, with and without defects, classified by homotopy groups.
Here defects mean the regions in which the order parameter does not belong to the order-parameter manifold under consideration.
\begin{table}[!ht]
\begin{center}
{\renewcommand{\arraystretch}{1.2}
\begin{tabular}{lll}\hline
$\pi_n$ & \ defects      & \ \ solitons                 \\ \hline\hline
$\pi_0$ & \ domain walls & \ \ dark solitons            \\
$\pi_1$ & \ vortices     & \ \ nonsingular domain walls \\
$\pi_2$ & \ monopoles    & \ \ 2D Skyrmions             \\
$\pi_3$ & \              & \ \ Skyrmions, knots         \\ 
$\pi_4$ & \              & \ \ instantons               \\ \hline
\end{tabular}}
\caption{Topological objects (defects and solitons) described by homotopy groups.}
\label{tb:defects_solitons}
\end{center}
\end{table}

Using homotopy theory, one can elicit insights into the topology of complicated structures, which is difficult to obtain intuitively. Here, we list some useful formulas of homotopy theory~\cite{NakaharaTextbook}:
\begin{align}
\pi_n(S^m) \cong \left\{
     \begin{array}{cl}
       {\mathbb Z}  & \mbox{if\ \ \  $m=n\geq1$;}\\
	 0          & \mbox{if\ \ \  $m\geq n\geq1$;}\\
	 0          & \mbox{if\ \ \  $m=1$ and $n\geq2$;}\\
       {\mathbb Z}  & \mbox{if\ \ \  $n=3$ and $n=2$ (Hopf charge);}\\
       {\mathbb Z}_2& \mbox{if\ \ \  $n=m+1\geq4$ or $n=m+2\geq4$,}
     \end{array}
\right.
\label{general}
\end{align}
where 0 implies that there exist only the trivial configurations (i.e., those contractible to a point in $\mani{R}$) and ${\mathbb Z}_2=\{0,1\}$ is the two-element group.

The following formulas are also useful.
Let $G$ be a Lie group and $H$ be its subgroup. 
If $G$ is connected (i.e., $\pi_0(G)=0$) and simply connected (i.e., $\pi_1(G)=0$), the following isomorphism holds:
\begin{align}
\pi_1(G/H)\cong\pi_0(H).
\label{eq:pi1_G/H}
\end{align}
Similarly, if $\pi_2(G)=\pi_1(G)=0$, then
\begin{align}
\pi_2(G/H)\cong\pi_1(H).
\end{align}
As a corollary, when $G={\rm SU}(2)$ and $H={\rm U}(1)$, we obtain
\begin{align}
\pi_2({\rm SU}(2)/{\rm U}(1))\cong\pi_1({\rm U}(1))\cong {\mathbb Z}.
\end{align}

For the case of spinor BECs, we have $\mani{R}=G/H$ where $G={\rm SO(3)}_{\bf f}\times {\rm U(1)}_\phi$ and $H$ is an isotropy group of the phase under consideration.
Because $G$ is not simply connected, we introduce the universal covering space of $G$, which is denoted by $\tilde{G}$:
\begin{align}
 \tilde{G} = {\rm SU(2)}_{\bf f} \times \mathbb{R}_\phi.
\end{align}
We call $\tilde{G}$ {\it a lift} of $G$.
We can reconstruct an element of $G$ from that of $\tilde{G}$ as follows:
\begin{align}
 (e^{-i{\sigma}_z\alpha/2}e^{-i{\sigma}_y\beta/2}e^{-i{\sigma}_z\gamma/2}, \frac{\phi}{2\pi}) \mapsto e^{i\phi}e^{-i{\rm f}_z\alpha}e^{-i{\rm f}_y\beta}e^{-i{\rm f}_z\gamma}
\label{eq:SU2toSO3}
\end{align}
where $\sigma_{x,y,z}$ are the Pauli matrices.
We define a lift $\tilde{H}$ of $H$ such that $G/H\cong \tilde{G}/\tilde{H}$.
Then, according to Eq.~\eqref{eq:pi1_G/H}, the fundamental group for the order-parameter manifold $\mani{R}=G/H$ is given by
\begin{align}
 \pi_1(\mani{R}) \cong \pi_0(\tilde{H}).
\end{align}
If $\tilde{H}$ is a discrete group, we have $\pi_0(\tilde{H})\cong \tilde{H}$.

Using the above formulas, 
the first, second, and third homotopy groups of some order-parameter manifolds are obtained as summarized in Table~\ref{tb:homotopy}
(for more details, see Refs.~\cite{Mermin1979,NakaharaTextbook}).
For example, in the case of spin-1 ferromagnetic BECs, $\pi_1({\rm SO}(3))\cong{\mathbb Z}_2$ implies that there exist two types of line structures that are singular and nonsingular;
$\pi_2({\rm SO}(3))\cong 0$ implies the absence of point defects and 2D Skyrmions; and
$\pi_3({\rm SO}(3))\cong{\mathbb Z}$ implies the presence of a
nonsingular soliton-like object or a (3D) Skyrmion~\cite{Shankar1977,VolovikMineev1977}.

\begin{table}[!ht]
\begin{center}
{\renewcommand{\arraystretch}{1.2}
\begin{tabular}{llccc}\hline
system & $\mani{R}$ & $\pi_1$ & $\pi_2$ & $\pi_3$ \\ \hline\hline
scalar BEC & ${\rm U}(1)$ & $\mathbb{Z}$ & 0 & 0 \\  
Heisenberg spin & $S^2$ & 0 & $\mathbb{Z}$ & $\mathbb{Z}$ \\ 
uniaxial nematics & ${\mathbb R}P^2\cong {S^2}/{\mathbb Z}_2$ & ${\mathbb Z}_2$ & $\mathbb{Z}$ & $\mathbb{Z}$ \\[2mm] 
biaxial nematics & ${\rm SU}(2)/Q$ & $Q$ & 0 & $\mathbb{Z}$ \\ 
spin-1 ferromagnetic BEC & ${\rm SO}(3)_{{\bf f},\phi}\cong{\mathbb R}P^3_{{\bf f},\phi}$ & ${\mathbb Z}_2$ & 0 & $\mathbb{Z}$ \\ 
spin-1 polar BEC & $[S^2_{{\bf f}}\times {\rm U}(1)_\phi]/({\mathbb Z}_2)_{{\bf f},\phi}$
& $\mathbb{Z}$ & $\mathbb{Z}$ & $\mathbb{Z}$ \\ 
cyclic BEC & $[{\rm SO}(3)_{{\bf f}}\times {\rm U}(1)_\phi]/T_{{\bf f},\phi}$ &
$\mathbb{Z} \times_h T^\ast_{\mathbf{f},\phi} $ & 0 & $\mathbb{Z}$ \\ \hline
\end{tabular}}
\caption{List of homotopy groups. Subscripts ${\bf f}$ and $\phi$ denote the manifolds for the spin and gauge degrees of freedom, respectively. The symbol $\mathbb{R}P^n\cong S^n/\mathbb{Z}_2$ is a projective space, and $Q$ and $T^\ast$ respectively denote the quaternion group and the binary tetrahedral group, the latter being a lift of $T$.
The $h$-product $\times_h$ stipulates a special product rule of the group elements in which spin and gauge degrees of freedom are coupled~\cite{Kobayashi2012} (see the text for detail). 
}
\label{tb:homotopy}
\end{center}
\end{table}

\subsubsection{Relative homotopy groups}
\label{sec:RelativeHomotopy}

To fully understand broken-symmetry systems, it is important to take into account not only the global boundary conditions but also various constraints imposed on the order-parameter field. Relative homotopy groups provide a useful tool for characterizing effects when the order-parameter manifold changes its character on a surface or when it is constrained by boundary conditions. 
Relative homotopy groups can also be used to discuss properties of a defect that is itself regarded as a singularity in the order-parameter field in ordinary homotopy theory. 
The relative homotopy group is also useful to calculate the absolute homotopy group via the exact sequence of homomorphisms.

As an illustrative example, let us consider a spin-1 ferromagnetic BEC. 
Suppose that the system is contained in a cylinder of radius $r_0$ and the order parameter on every point inside the cylinder belongs to the order-parameter manifold $\mani{R}^{\rm ferro}={\rm SO(3)}$ of the ferromagnetic phase.
We then impose a boundary condition that the spin points outward ($\hat{\bm s}=\hat{\bm r}$) or downward ($\hat{\bm s}=-\hat{z}$) on the wall at $r=r_0$, where $r$ is the radial coordinate.
Then, the spin texture would look like 
\begin{align}
{\bm \psi}^{\rm coreless}(r,\varphi,z)
= \sqrt{n}
\begin{pmatrix}
\cos^2 \frac{\beta(r)}{2} \\[1mm]
\frac{e^{i\varphi}}{\sqrt{2}} \sin \beta(r) \\[1mm]
e^{2i\varphi} \sin^2 \frac{\beta(r)}{2}
\end{pmatrix}.
\label{M-H}
\end{align}
where we have chosen $\phi-\gamma=\alpha=\varphi$ in Eq.~\eqref{eqSCB88} so as to eliminate the singularity in the ferromagnetic order.
Because the direction of spin is $\hat{\bm s}(r,\varphi,z)=(\hat{\bm e}_x\cos\varphi+\hat{\bm e}_y\sin\varphi)\sin\beta(r)+\hat{\bm e}_z\cos\beta(r)$,
the above boundary conditions are met if we choose $\beta(0)=0$ and $\beta(r_0)=\pi/2$ or $\beta(r_0)=\pi$.
The topological objects corresponding to $\beta(r_0)=\pi/2$ and $\beta(r_0)=\pi$ are referred to as the Mermin-Ho (MH) vortex and the Anderson-Toulouse (AT) vortex, respectively~\cite{Vollhardt1990}.
It can be shown that the circulation at the wall is 1 for the MH vortex and 2 for the AT vortex.
Both of them are not stable in free space because they belong to the trivial element of the fundamental group, whereas there is no non-trivial 2D structure due to $\pi_2(\mani{R}^{\rm ferro})=0$.
They are classified in terms of the second relative homotopy group $\pi_2(\mani{R}^{\rm ferro}, S^1)$,
and can be stabilized only if the boundary conditions on the wall are imposed by some means.

In general, the $n$th relative homotopy group $\pi_n(\mani{R},\tilde{\mani{R}})$ is constructed out of maps
from $n$-dimensional cube $I^n$ to $\mani{R}$:
\begin{align}
\alpha : I^n\to \mani{R},
\label{eq:def_rel_homo}
\end{align}
under which the boundary $\partial I^n$, which is an $(n-1)$-dimensional sphere or an $(n-1)$-dimensional cube with the boundary condition, is mapped to $\tilde{\mani{R}}\subset \mani{R}$.
If $\tilde{\mani{R}}$ includes only a point $x_0$, the relative homotopy group $\pi_n(\mani{R},\tilde{\mani{R}})$ reduces to the absolute one $\pi_n(\mani{R},x_0)$.
For general cases, structures on the boundary are classified with $\pi_{n-1}(\tilde{\mani{R}})$.
Moreover, since the inside of the $n$ cube should be mapped to $\mani{R}$,
only the elements of $\pi_{n-1}(\tilde{\mani{R}})$ that are mapped to the trivial element of $\pi_{n-1}(\mani{R})$
under an inclusion map $\tilde{\mani{R}}\to\mani{R}$
can appear on the boundary. The set of such elements corresponds to the kernel of the map $\pi_{n-1}(\tilde{\mani{R}})\to\pi_{n-1}(\mani{R})$, which is denoted as
\begin{align}
\mathcal{K}_{n}={\rm Ker}[\pi_{n-1}(\tilde{\mani{R}})\to\pi_{n-1}(\mani{R})].
\end{align}

On the other hand, the structures inside the $n$-dimensional cube is classified by $\pi_n(\mani{R})$.
Note that, however, if the image of $\alpha$ [Eq.~\eqref{eq:def_rel_homo}] can be continuously deformed to a subspace of $\tilde{\mani{R}}$,
it can then be deformed to a point in $\tilde{\mani{R}}$.
Such structures should belong to the trivial element of $\pi_n(\mani{R},\tilde{\mani{R}})$.
Hence, $\pi_n(\mani{R},\tilde{\mani{R}})$ includes the structure of a quotient group given by
\begin{align}
\mathcal{I}_n=\frac{\pi_n(\mani{R})}{{\rm Im}[\pi_{n}(\tilde{\mani{R}})\to\pi_{n}(\mani{R})]},
\end{align}
where the map $\pi_{n}(\tilde{\mani{R}})\to\pi_{n}(\mani{R})$ is a homomorphism induced by an inclusion map $\tilde{\mani{R}}\to\mani{R}$, and ${\rm Im}$ denotes the image.

Mathematically, it can be proved using the exact sequence and the homomorphism theorem that the relative homotopy group satisfies~\cite{Mermin1979,Mineev1998}
\begin{align}
 \pi_n(\mani{R},\tilde{\mani{R}})/\mathcal{I}_n\cong \mathcal{K}_{n},
\end{align}
indicating that an element of $\pi_n(\mani{R},\tilde{\mani{R}})$ is characterized by a set of two elements: one in $\mathcal{I}_n$ and the other in $\mathcal{K}_{n}$.
For the case of above examples on a spin-1 ferromagnetic BEC, we have $\mathcal{I}_2=0$ (where $0$ means a group that includes only the trivial element),
and the MH and AT vortices differ in the elements of $\mathcal{K}_2\cong \mathbb{Z}$~\cite{Mineyev1978}.

An example for $n=1$ is a planar soliton (Maki domain wall) in the $^3$He A phase, where the order-parameter manifold is $\mani{R}=(S^2\times {\rm SO}(3))/\mathbb{Z}_2$ within a dipole healing length from the domain wall and $\tilde{\mani{R}}={\rm SO}(3)$ outside this layer due to dipole locking~\cite{Mineyev1978}. 
We take the axis perpendicular to the domain wall as the $z$-axis. 
Then, the order-parameter field defines a mapping from the $z$-axis into the contour in $\mani{R}$ whose end points belong to $\tilde{\mani{R}}$. 
Because $\tilde{\mani{R}}$ is a connected space, that is $\pi_0(\tilde{\mani{R}})=0$, we have $\mathcal{K}_1=0$.
On the other hand, the image of the map from $\pi_1(\tilde{\mani{R}})=\mathbb{Z}_2$ to $\pi_1(\mani{R})=\mathbb{Z}_4$ is $\mathbb{Z}_2$, resulting in $\pi_1(\mani{R},\tilde{\mani{R}})\cong \mathcal{I}_1=\mathbb{Z}_4/\mathbb{Z}_2=\mathbb{Z}_2$~\cite{Mineyev1978}.
Note that in general $\pi_1(\mani{R},\tilde{\mani{R}})$ need not even be a group;
it becomes a group when $\tilde{\mani{R}}$ is a connected space, or when $\mani{R}$ is a group and $\tilde{\mani{R}}$ is a subgroup of $\mani{R}$.

An example for $n=2$ is a nonsingular vortex state discussed above.
The second relative homotopy group can also classify point defects stuck on the boundary, which is called a boojum;
the relative homotopy group is constructed out of a map from hemisphere that encloses the defect on the boundary to the order-parameter manifold, together with the boundary condition at the edge of the hemisphere~\cite{Mineyev1978}.
Topological structures at the interface of two phases of spinor BECs were discussed in Refs.~\cite{Takeuchi2006, Borgh2012}.
The relative homotopy group is also used to classify vortex core structures~\cite{Mineyev1978},
and deformation of topological excitations under a phase transition~\cite{Trebin1982}.

\subsection{Line defects}
\label{sec:Line defects}

\subsubsection{Homotopy theory analysis of vortex structures}
Line defects or vortices are characterized by the fundamental group $\pi_1(\mani{R})$. 
In Sec.~\ref{sec:Vortices} we have examined vortex structures in spinor BECs.
Here, we analyze these structures in terms of the homotopy theory.
For the case of {\it s}-wave superconductors, liquid $^4$He, and spin-polarized or spinless gaseous BECs,  the mass circulation around a vortex line is quantized and vortices are classified by the quantum number $n_{\rm w}$ as in Eq.~\eqref{eq:Vort_scalar_quantization}.
In this case, the order-parameter manifold for a scalar order parameter is given by $\mani{R}={\rm U(1)}$. 
The fundamental group is therefore the additive group of integers:
\begin{eqnarray}
\pi_1({\rm U(1)}) \cong {\mathbb Z}.
\end{eqnarray}
There is a one-to-one correspondence between an element of $\pi_1({\rm U(1)})$ and a particular value of the winding number $n_{\rm w}$.

For the case of a spin-1 ferromagnetic BEC, 
the order-parameter manifold is SO(3) and the fundamental group is given by $\pi_1({\rm SO(3)})={\mathbb Z}_2=\{0,1\}$ which is the two-element group. 
The spin configuration that corresponds to class 0 is continuously transformable to a uniform configuration [see Eq.~\eqref{eq:coreless_vortex}],
whereas the spin configuration that corresponds to class 1 describes a polar-core vortex that is topologically stable [see Eq.~\eqref{eq:polar_core_vortex}]. 
Because 1 + 1 = 0 in ${\mathbb Z}_2$, the coalescence of two singly quantized vortices is homotopic to a uniform configuration. 
The fact that there are only two topologically distinguishable structures is confirmed by the discussion below Eq.~\eqref{eq:Vort_spin1_qntz}.

For the case of a spin-1 polar BEC, the order-parameter manifold is $\mani{R}^{\rm polar}=[S^2_{\bf f}\times {\rm U(1)}_\phi]/({\mathbb Z}_2)_{{\bf f},\phi}$.
To calculate the fundamental group of this order parameter manifold, we use Eq.~\eqref{eq:pi1_G/H} as follows.
As we have discussed in Sec.~\ref{sec:sym_OPM},
the $\mathbb{Z}_2$ symmetry arises from the fact that the order parameter given by Eq.~\eqref{eqSCB95}
is invariant under the transformation of $(\hat{\bm d},\phi) \to (-\hat{\bm d},\phi+\pi)$.
Because $G=S^2\times {\rm U(1)}$ is not simply connected, 
to calculate $\pi_1(\mani{R}^{\rm polar})$,
we lift $G$ to $\tilde{G}=S^2\times \mathbb{R}$.
With this lift, the points $(\hat{\bm d},\phi)$ and $(-\hat{\bm d},\phi+\pi)$ in $G$ are mapped to
$(\hat{\bm d}, \phi+2n_{\rm w}'\pi)$ and $(-\hat{\bm d}, \phi+(2n_{\rm w}'+1)\pi)$, respectively, in $\tilde{G}$ where $n_{\rm w}'$ is an integer.
These identical points are simply labeled with an integer $n_{\rm w}$ ($=2n_{\rm w}'$ or $2n_{\rm w}'+1$), implying $\tilde{H}= \mathbb{Z}$.
Hence, from Eq.~\eqref{eq:pi1_G/H}, we obtain
\begin{align}
\pi_1(\mani{R}^{\rm polar})  \cong \mathbb{Z}_{{\bf f},\phi}.
\end{align}
As in the case of a scalar BEC, vortices in the polar phase are classified by integers.
However, due to the $\mathbb{Z}_2$ discrete symmetry, 
the minimum unit of circulation is one-half of the usual value of $h/M$, 
as discussed in Sec.~\ref{sec:vortex_spin1polar}.

In a similar manner, we can calculate the fundamental groups of the order-parameter manifolds for the uniaxial nematic, biaxial nematic, and cyclic phases of spin-2 spinor BECs.
The obtained fundamental groups indicate the existence of various types of vortices due to the discrete symmetry~\cite{Makela2003,Makela2006}.
Some of them have fractional circulations, and some others have no mass circulation.
The classification of vortices in spin-3 BECs using the homotopy theory is given in Ref.~\cite{Barnett2007}.
Different from spin-1 BECs, some phases in spin-2 and 3 BECs support non-Abelian vortices as discussed in the next subsection.

\subsubsection{Non-Abelian vortices}
\label{sec:Vort_nonAbelian}

We consider vortices in the spin-2 cyclic phase,
which are non-Abelian in the sense that the fundamental group of the cyclic order-parameter manifold is non-Abelian%
\footnote{In literature, three different definitions of non-Abelian vortices are used: (i) $G$ is non-Abelian, (ii) $H$ is non-Abelian, and (iii) $\pi_1(G/H)$ is non-Abelian,
where $G/H$ is the order parameter manifold for the system under consideration.
Here, we adopt the definition (iii).}.
As shown in Sec.~\ref{sec:sym_spin2}, the cyclic order parameter has the symmetry of tetrahedron, which is a non-Abelian group.
The isotropy group is generated by $\bar{H}^{(T)}$ defined in Eq.~\eqref{eq:H_for_cyclic}.
To make the following discussion simple, we change the symmetry axis and choose
the two-fold symmetry axis in the $z$ direction
and the three-fold symmetry axis in the (1,1,1) direction.
Then, the isotropy group, which is generated by $I_z\equiv C_{2,z}$ and $\bar{C}\equiv e^{-i2\pi/3}C_{3,x+y+z}$, is given by~\cite{Makela2003,Semenoff2007,Kobayashi2009},
\begin{align}
 H^{(T)}=\{1,I_x,I_y,I_z,\bar{C},I_x\bar{C},I_y\bar{C},I_z\bar{C},\bar{C}^2,I_x\bar{C}^2,I_y\bar{C}^2,I_z\bar{C}^2\},
\label{eq:nonAbe_HT}
\end{align}
where $I_x\equiv C_{2,x}$, $I_y\equiv C_{2,y}$, and we have used $\bar{C}I_z\bar{C}^{-1}=I_x$ and $\bar{C}^2 I_z\bar{C}^{-2}=I_y$.
The order parameter that is invariant under all elements of Eq.~\eqref{eq:nonAbe_HT} is $\bm\zeta^{\rm cyclic}_0 = (1/2,0,i/\sqrt{2},0,1/2)$.
Since the gauge transformation is coupled with the spin rotation in $\bar{C}$, this is the spin-gauge coupled tetrahedral symmetry,
and the order-parameter manifold for this phase is written as
\begin{align}
 \mani{R}^{\rm cyclic} = \frac{{\rm SO(3)}_{\bf f}\times {\rm U(1)}_\phi}{T_{{\bf f},\phi}}.
\end{align}

To calculate the fundamental group of $\mani{R}^{\rm cyclic}$,
we first lift $G={\rm SO(3)}_{\bf f}\times {\rm U(1)}_\phi$ to its universal covering space $\tilde{G}={\rm SU(2)}_{\bf f}\times\mathbb{R}_\phi$.
The elements in $\tilde{G}$ are mapped to the ones in $G$ according to Eq.~\eqref{eq:SU2toSO3}.
Under this lift, the elements of $H^{(T)}$ change as follows:
\begin{subequations}
\begin{align}
 1 &\to \{(\pm \bm 1,n_{\rm w})\,|\ n_{\rm w} \in \mathbb{Z}\},\\
 I_{x,y,z} &\to \{(\pm i\sigma_{x,y,z},n_{\rm w})\,|\ n_{\rm w} \in \mathbb{Z}\},\\
 \bar{C} &\to \bigg\{(\pm \tilde{\sigma},n_{\rm w}+\frac{1}{3})\,\bigg|\ n_{\rm w} \in \mathbb{Z}\bigg\},\\
 I_{x,y,z}\bar{C} &\to \bigg\{(\pm i\sigma_{x,y,z}\tilde{\sigma},n_{\rm w}+\frac{1}{3})\,\bigg|\ n_{\rm w} \in \mathbb{Z}\bigg\},\\
 \bar{C}^2 &\to \bigg\{(\pm \tilde{\sigma}^2,n_{\rm w}+\frac{2}{3})\,\bigg|\ n_{\rm w} \in \mathbb{Z}\bigg\},\\
 I_{x,y,z}\bar{C}^2 &\to \bigg\{(\pm i\sigma_{x,y,z}\tilde{\sigma}^2,n_{\rm w}+\frac{2}{3})\,\bigg|\ n_{\rm w} \in \mathbb{Z}\bigg\},
\end{align}
\label{eq:tildeHT}
\end{subequations}
where
\begin{align}
\bm 1\equiv \begin{pmatrix} 1 & 0 \\ 0 & 1 \end{pmatrix},\ \ \tilde{\sigma}\equiv \frac{1}{2}({\bm 1}+i\sigma_x+i\sigma_y+i\sigma_z),
\end{align}
with $\sigma_{x,y,z}$ being the Pauli matrices.
Hence, an element of the lifted isotropy group $\tilde{H}^{(T)}$ is written as
$(\sigma,x)$ where $\sigma \in T^*=\{\pm \bm 1, \pm i\sigma_{x,y,z}, \pm \tilde{\sigma}, \pm i \sigma_{x,y,z}\tilde{\sigma}, \pm \tilde{\sigma}^2, \pm i \sigma_{x,y,z}\tilde{\sigma}^2\}$ and $x=n_{\rm w}$ for $\sigma=\pm \bm 1$ and $\pm i\sigma_{x,y,z}$, $x=n_{\rm w}+1/3$ for $\sigma=\pm \tilde{\sigma}$ and $\pm i \sigma_{x,y,z}\tilde{\sigma}$, 
and $x=n_{\rm w}+2/3$ for $\sigma=\pm \tilde{\sigma}^2$ and $\pm i \sigma_{x,y,z}\tilde{\sigma}^2$ with $n_{\rm w}$ being an arbitrary integer.
These elements represent how the order parameter changes around the vortex under consideration.
The composition of two elements $(\sigma_1,x_1)$ and $(\sigma_2,x_2)$ in $\tilde{H}^{(T)}$ is defined by $(\sigma_1\sigma_2,x_1+x_2)$.
Note, however, that $\tilde{H}^{(T)}$ is not a direct product of $T^*$ and $\mathbb{Z}$ because the choice of $x$ depends on the element of $T^*$ 
\footnote{
Writing the elements of $\tilde{H}^{(T)}$ as
$(e^{i\theta}\sigma, n_{\rm w})$ where
\begin{align*}
\theta=\left\{
\begin{array}{ll}
0 & \textrm{ for } \sigma=\pm{\bm 1}, \pm i\sigma_{x,y,z},\\
2\pi/3 & \textrm{ for } \sigma=\pm\tilde{\sigma}, \pm
i\sigma_{x,y,z}\tilde{\sigma},\\
4\pi/3 & \textrm{ for } \sigma=\pm\tilde{\sigma}, \pm
i\sigma_{x,y,z}\tilde{\sigma}^2,
\end{array}
\right.
\end{align*}
we find that they obey the following composition low:
\begin{align}
(e^{i\theta}_1\sigma_1,n_1)\cdot (e^{i\theta_2}\sigma_2,n_2) =
(e^{i[\theta_1+\theta_2-2\pi h(\sigma_1,\sigma_2)]}\sigma_1\sigma_2,
n_1+n_2+h(\sigma_1,\sigma_2)),
\end{align}
where
\begin{align}
h(\sigma_1,\sigma_2) =
\left\{
\begin{array}{ll}
0 & \textrm{ if} \theta_1+\theta_2<2\pi, \\
1 & \textrm{ if} \theta_1+\theta_2\ge 2\pi, \\
\end{array}
\right.
\end{align}
which we call an $h$ product.
For more details, see Ref.~\cite{Kobayashi2012}.
}.
Since $\pi_0(\tilde{H})=\tilde{H}$ if $\tilde{H}$ is a discrete group~\cite{NakaharaTextbook},
Eq.~\eqref{eq:pi1_G/H} leads to $\pi_1(\mani{R}^{\rm cyclic})=\tilde{H}^{(T)}$,
implying that vortices in the cyclic phase is labeled as $(\sigma, x)$.
These vortices are non-Abelian because $\tilde{H}^{(T)}$ is non-Abelian.
The mass circulation of a vortex labeled with $\sigma=\pm I$ or $\pm i\sigma_{x,y,z}$ is quantized in units of $\kappa=h/M$,
whereas the mass circulation of a vortex labeled with  $\sigma=\pm \tilde{\sigma}$ or $\pm i\sigma_{x,y,z}\tilde{\sigma}$ is quantized in units of $(n_{\rm w}+1/3)\kappa$,
and that with  $\sigma=\pm \tilde{\sigma}^2$ or $\pm i\sigma_{x,y,z}\tilde{\sigma}^2$ is quantized in units of $(n_{\rm w}+2/3)\kappa$.

A single vortex can be classified in terms of conjugacy classes of the fundamental group.
Let us consider a vortex labeled with $(i\sigma_x,n_{\rm w})$.
This vortex can be continuously deformed by globally applying an operator $g\in\tilde{H}^{(T)}$ as $g(i\sigma_x,n_{\rm w})g^{-1}$.
If we choose $g=(\tilde{\sigma},n_{\rm w}')$, we obtain $g(i\sigma_x,n_{\rm w})g^{-1}=(i\sigma_z,n_{\rm w})$ which belongs to a different element of the fundamental group.
The classes of the elements obtained in such a manner are called conjugacy classes. 
For the present case, there are seven classes for each $n_{\rm w}$~\cite{Semenoff2007}:
\begin{subequations}
\begin{align}
(1) &\{(\bm 1,n_{\rm w})\}, \\
(2) &\{(-\bm 1,n_{\rm w})\}, \\
(3) &\{(i\sigma_x,n_{\rm w}),(-i\sigma_x,n_{\rm w}),(i\sigma_y,n_{\rm w}),(-i\sigma_y,n_{\rm w}),(i\sigma_z,n_{\rm w}),(-i\sigma_z,n_{\rm w})\}, \\
(4) &\{(\tilde{\sigma},n_{\rm w}+1/3),(-i\sigma_x\tilde{\sigma},n_{\rm w}+1/3),(-i\sigma_y\tilde{\sigma},n_{\rm w}+1/3),(-i\sigma_z\tilde{\sigma},n_{\rm w}+1/3)\}, \\
(5) &\{(-\tilde{\sigma},n_{\rm w}+1/3),(i\sigma_x\tilde{\sigma},n_{\rm w}+1/3),(i\sigma_y\tilde{\sigma},n_{\rm w}+1/3),(i\sigma_z\tilde{\sigma},n_{\rm w}+1/3)\}, \\
(6) &\{(\tilde{\sigma}^2,n_{\rm w}+2/3),(-i\sigma_x\tilde{\sigma}^2,n_{\rm w}+2/3),(-i\sigma_y\tilde{\sigma}^2,n_{\rm w}+2/3),(-i\sigma_z\tilde{\sigma}^2,n_{\rm w}+2/3)\}, \\
(7) &\{(-\tilde{\sigma}^2,n_{\rm w}+2/3),(i\sigma_x\tilde{\sigma}^2,n_{\rm w}+2/3),(i\sigma_y\tilde{\sigma}^2,n_{\rm w}+2/3),(i\sigma_z\tilde{\sigma}^2,n_{\rm w}+2/3)\}.
\end{align} 
\end{subequations}

\begin{figure}[ht]
\begin{center}
\resizebox{0.9\hsize}{!}{
\includegraphics{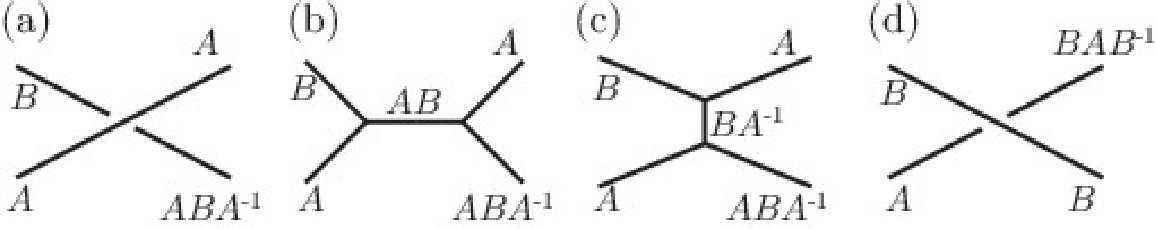}
}
\end{center}
\caption{Collision dynamics of two vortices.
(a) Initial configuration, where 
$A$ or $B$ represents an operation that characterizes the corresponding vortex (a set of spin rotations and gauge transformations for the case of a cyclic BEC).
The vortex on the right end, which is connected to $B$, is identified as $ABA^{-1}$.
The configuration in (a) is topologically equivalent to (b) or (c), where a rung $AB$ or $BA^{-1}$ is formed.
If the vortices are a pair of vortex and anti-vortex, i.e., $A=B^{-1}$, the rung in (b) disappears, giving rise to reconnection,
whereas the rung in (c) corresponds to a doubly quantized vortex that costs a large kinetic energy.
If $A$ and $B$ are commutative, passing through is also possible because the configurations of (a) and (d) will then be topologically equivalent.
However, when $A$ and $B$ are not commutative, the collision always results in the formation of a rung.
}
\label{fig:nonAbelian1}
\end{figure}

\begin{figure}[ht]
\begin{center}
\resizebox{0.8\hsize}{!}{
\includegraphics{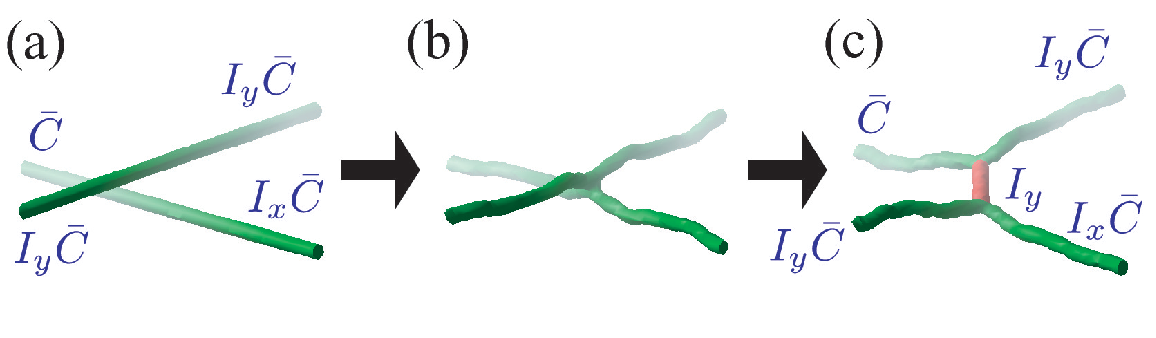}
}
\end{center}
\caption{Numerical simulation for the collision dynamics of non-Abelian vortices,
where $\bar{C}$, $I_y$, $I_{x}\bar{C}$, and $I_y\bar{C}$ represents the elements of $H^{(T)}$ which are related to the elements of the fundamental group according to Eq.~\eqref{eq:tildeHT}.
}
\label{fig:nonAbelian2}
\end{figure}

In the presence of more than two vortices, however, the conjugacy class of the fundamental group cannot completely characterize individual vortices,
because, for example, two vortices described by two different elements of the same conjugacy class cannot be transformed under a global transformation to those described by a single element.
The non-Abelian characteristics of the vortices manifest themselves most dramatically in such a situation, namely, the collision dynamics of two vortices. 
In general, when two vortices collide, they reconnect themselves, pass through [Fig.~\ref{fig:nonAbelian1}(d)], or form a rung that bridges the two vortices [Fig.~\ref{fig:nonAbelian1}(b),(c)]. When two Abelian vortices collide, all these three cases are possible, and one of them occurs depending on the kinematic and initial conditions.
However, when two non-Abelian vortices collide, the only possibility is the formation of a rung. Reconnection and passing through are topologically forbidden because the corresponding generators do not commute with each other.
Figure~\ref{fig:nonAbelian2} illustrates a typical rung formation which is obtained numerically for the cyclic BEC.
When the core of a 1/3 vortex in the cyclic phase is filled with the ferromagnetic state, which is possible under certain conditions, it is possible to observe such dynamics of vortex lines by using a phase-contrast imaging technique that is sensitive to local magnetization.

\subsection{Point defects}
\label{sec:point_defects}

Point defects are characterized by the second homotopy group.
Because $\pi_2({\rm U(1)})=0$, scalar BECs have no topologically stable point defects. 
When the order-parameter manifold is $S^2$, it can host monopoles because $\pi_2(S^2)\cong\mathbb{Z}$.

\subsubsection{Monopole charge}
Here, we consider the situation shown in Fig.~\ref{fig:homotopy3}, where the order parameter $\hat{\bm m}({\bm r})$ is interpreted as a mapping from $S^2$ to $S^2$.
The topological charge $N_2$ of the point defect can be calculated as follows.
By definition, $N_2$ gives the number of times the order parameter $\hat{\bm m}({\bm r})$ shown in Fig.~\ref{fig:homotopy3} wraps its manifold $S^2$. Expressing the components of this vector in spherical coordinate as $m_x=\sin\alpha\cos\beta$, $m_y=\sin\alpha\sin\beta$, and $m_z=\cos\alpha$, we can express $N_2$ as
\begin{eqnarray}
N_2=\frac{1}{4\pi}\int_\Sigma d\Omega
\left|\frac{\partial(\alpha,\beta)}{\partial(\theta,\varphi)}\right|,
\label{N_2}
\end{eqnarray}
where $d\Omega\equiv  \sin\theta d\theta d\varphi$, with $\theta$ and $\varphi$ are the spherical coordinates of ${\bm r}$, and the last term on the right-hand side is the Jacobian of the transformation of the coordinates.
The fact that the integrand is the Jacobian explicitly indicates that $N_2$ gives the degree of mapping from $S^2$ in real space to $S^2$ in order-parameter space.
The right-hand side of Eq.~(\ref{N_2}) can be directly expressed in terms of $\hat{\bm m}$ as
\begin{eqnarray}
N_2=\frac{1}{4\pi}\int_\Sigma d\theta d\phi \hat{\bm m}\cdot
\left(\frac{\partial \hat{\bm m}}{\partial\theta}\times\frac{\partial\hat{\bm m}}{\partial\varphi}\right) =\frac{1}{4\pi}\int_\Sigma d{\bm S} \cdot {\bm j},
\label{N_22}
\end{eqnarray}
where
\begin{align}
{\bm j}=\frac{1}{2}\sum_{\nu_1\nu_2\nu_3=x,y,z}\epsilon_{\nu_1\nu_2\nu_3}m_{\nu_1}(\bm\nabla m_{\nu_2}\times\bm\nabla m_{\nu_3}),
\label{j}
\end{align}
and $d{\bm S}$ represents an areal element perpendicular to $\Sigma$. 
We can use Gauss' law to rewrite the right-hand side of Eq.~(\ref{N_22}) as a volume integral:
\begin{align}
N_2=\int_{V_\Sigma} d \bm r \, n_{\rm m} ,
\label{n_2}
\end{align}
where $V_\Sigma$ is the region enclosed by the surface $\Sigma$, and
\begin{align}
n_{\rm m}({\bm r}) =\frac{1}{4\pi} \bm\nabla\cdot {\bm j}({\bm r})
\end{align}
gives the density distribution of point singularities.

\subsubsection{'t-Hooft-Polyakov monopole (hedgehog)}

As an example of monopoles we consider the spin-1 polar phase.
The second homotopy group of the polar order-parameter manifold is given by $\pi_2(\mani{R}^{\rm polar})={\mathbb Z}$, and therefore, it can accommodate point defects.
To investigate their properties, we write the order parameter in the following form:
\begin{eqnarray}
{\bm \psi}^{\rm polar} 
= \sqrt{\frac{n}{2}}\,e^{i\phi}
\begin{pmatrix}
-d_x+id_y   \\
\sqrt{2}d_z \\
 d_x+id_y
\end{pmatrix}.
\label{P2}
\end{eqnarray}
When we write $\hat{\bm d}\equiv (d_x,d_y,d_z)=(\sin\beta\cos\alpha,\sin\beta\sin\alpha,\cos\beta)$, Eq.~\eqref{P2} reduces to Eq.~\eqref{eqSCB95}.
By setting $\phi=0$, we obtain the spherical monopole which is called a 't-Hooft-Polyakov monopole or a hedgehog:
\begin{eqnarray}
\hat{\bm d}({\bm r})=\frac{{\bm r}}{|{\bm r}|}.
\label{hedgehog}
\end{eqnarray}
It follows from Eq.~\eqref{N_22} that the topological charge of the hedgehog is $N_2=1$.
Moreover, ${\bm j}$ defined in Eq.~\eqref{j} is related to the circulation of spin currents given by Eq.~\eqref{eq:polar_spincirculation} as
\begin{align}
\sum_{\nu=x,y,z} \hat{d}_\nu \left({\bm \nabla} \times {\bm v}_\nu^{(\rm spin,P)}\right) = - \frac{2\hbar}{M} {\bm j},
\end{align}
which implies that the surface integral of the spin circulation is quantized in terms of $4h/M$;
\begin{align}
\int \sum_{\nu=x,y,z} \hat{d}_\nu \left({\bm \nabla} \times {\bm v}_\nu^{(\rm spin,P)}\right)\cdot d{\bm S} = - \frac{4h}{M} N_2.
\end{align}

\begin{figure}[ht]
\begin{center}
\resizebox{0.6\hsize}{!}{
\includegraphics{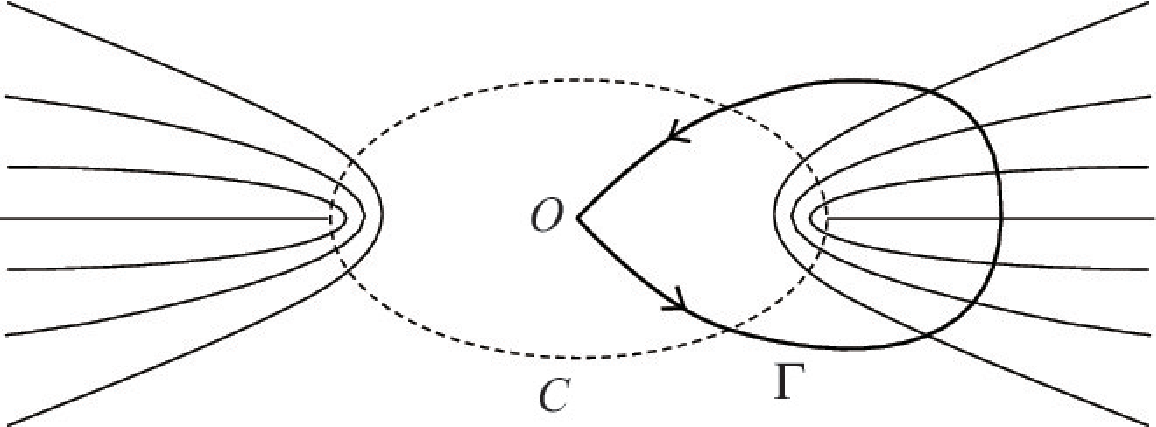}
}
\end{center}
\caption{Alice ring which is a vortex ring of a half-quantum vortex located on $C$. Far from the origin, it appears to be a monopole.}
\label{fig:homotopy7}
\end{figure}

Although a monopole is topologically stable in the spin-1 polar phase,
it is energetically unstable against deformation into a ring of a half-quantum vortex, which is called an Alice ring~\cite{Ruostekoski2003}.
The Alice ring is a combined object that possesses two topological charges $\pi_1$ and $\pi_2$ (see Fig.~\ref{fig:homotopy7}). 
Viewed far from the origin, it appears to be a point defect ($N_2=1$); however, this monopole field is created by a ring of the Alice vortex ($N_1=1$),
which is located on $C$ in Fig.~\ref{fig:homotopy7}.
Since the monopole and the Alice ring have the same $\pi_2$ charge, they can be continuously deformed to each other.
While the monopole is always accompanied with the density singularity,
the core of the Alice vortex can be filled with a ferromagnetic state.
As a result, the Alice ring is energetically favored because $|c_1|\ll c_0$ for the $^{23}$Na BEC~\cite{Ruostekoski2003}.

\subsubsection{Dirac monopole}

The magnetic monopole was originally envisaged as a magnetic analogue of the quantized electric charge:
\begin{eqnarray}
\bm\nabla\cdot{\bm B}=4\pi g\delta({\bm r}),
\label{Bfield}
\end{eqnarray}
where $g$ denotes the strength of the magnetic monopole. The solution to this equation is found to be ${\bm B}=g{\bm r}/r^3$ and the corresponding vector potential is given by
\begin{eqnarray}
{\bm A}
=-\frac{g}{r(r-z)}(-y,x,0)
=-\frac{g(1+\cos\theta)}{r\sin\theta}\hat{\bm e}_\varphi,
\label{vectorA}
\end{eqnarray}
where $(r,\theta,\varphi)$ are the polar coordinates and $\hat{\bm e}_\varphi$ is the unit vector in the $\varphi$-direction.
The vector potential (\ref{vectorA}) reproduces the magnetic field (\ref{Bfield}), except on the positive $z$-axis along which the magnetic field exhibits a singularity called the Dirac string. In fact, taking the rotation of (\ref{vectorA}), we obtain
\begin{eqnarray}
{\rm rot}{\bm A}=g\frac{\bm r}{r^3}-4\pi g\delta(x)\delta(y)\Theta(z)\hat{\bm e}_z,
\end{eqnarray}
where $\hat{\bm e}_z$ is the unit vector along the $z$-direction, and $\Theta(z)$ is the Heaviside unit step function. 

The Dirac monopole can be created in the ferromagnetic phase of a spin-1 BEC~\cite{Savage2003}.
Substituting $\phi= -\varphi$, $\alpha=\varphi$, and $\beta=\theta$ in Eq.~(\ref{eqSCB88}), we obtain
\begin{eqnarray}
{\bm \zeta}^{\rm Dirac}
= 
\begin{pmatrix}
e^{-2i\varphi}  \cos^2 \frac{\theta}{2} \\[1mm]
\frac{e^{-i\varphi}}{\sqrt{2}} \sin\theta \\[1mm]
             \sin^2 \frac{\theta}{2}
\end{pmatrix}.
\label{Dirac}
\end{eqnarray}
The corresponding superfluid velocity ${\bm v}^{\rm (mass)}=(\hbar/M){\rm Im}({\bm \zeta}^\dagger\nabla{\bm \zeta})$ [defined in Eq.~\eqref{eq:def_for_mass_supercurrent}]
is the same as the vector potential (\ref{vectorA}) of the Dirac monopole with identification $g=\hbar/M$ [see also Eq.~\eqref{eq:vorticity-B}]. 

We note that the order parameter (\ref{Dirac}) exhibits a doubly quantized vortex only along the positive $z$-axis ($\theta=0$) and that it has no singularity along the negative $z$-axis ($\theta=\pi$).  We also note that since the fundamental group of SO(3) is $\mathbb{Z}_2$, the only topologically stable vortex is a singly quantized vortex which is called a polar-core vortex. In fact, the singular Dirac monopole can deform continuously to a nonsingular spin texture if we take the following parametrization:
\begin{eqnarray}
{\bm \zeta}^{\rm Dirac}
= \begin{pmatrix}
e^{-2i\varphi}  \cos^2 \frac{\beta}{2} \\[1mm]
\frac{e^{-i\varphi}}{\sqrt{2}} \sin\beta \\[1mm]
             \sin^2 \frac{\beta}{2}
\end{pmatrix}, \ \ \
\beta(r,\theta,\varphi,t)=\Theta (1-t)+\pi t.
\label{Dirac2}
\end{eqnarray}
As $t$ increases from 0 to 1, the order parameter continuously changes from the Dirac monopole at $t=0$ to a nonsingular texture ${\bm \zeta}=(0,0,1)^{\rm T}$ at $t=1$.
An experimental method for the creation of the Dirac monopole was discussed in Ref.~\cite{Pietila2009a}. Topologically stable monopoles under a non-Abelian gauge field are discussed in Refs.~\cite{Ruseckas2005,Pietila2009b}.

\subsection{Skyrmions}
\label{sec:skyrmions}
\subsubsection{Shankar Skyrmion}

The third homotopy group characterizes topological objects called Skyrmions that extend over the entire three-dimensional space. A prime example of this is the so-called Shankar monopole in the $^3$He-A phase~\cite{Shankar1977,VolovikMineev1977}, 
although it is actually not a monopole but a Skyrmion.
The topology of the Shankar Skyrmion is $\pi_3({\rm SO}(3))\cong{\mathbb Z}$, which is also supported by the ferromagnetic phase of a spin-1 BEC. 
The order parameter of the ferromagnetic BEC is characterized
by the direction of the local spin $\hat{\bm s}$ and the superfluid phase

\begin{figure}[ht]
\begin{center}
\resizebox{0.6\hsize}{!}{
\includegraphics{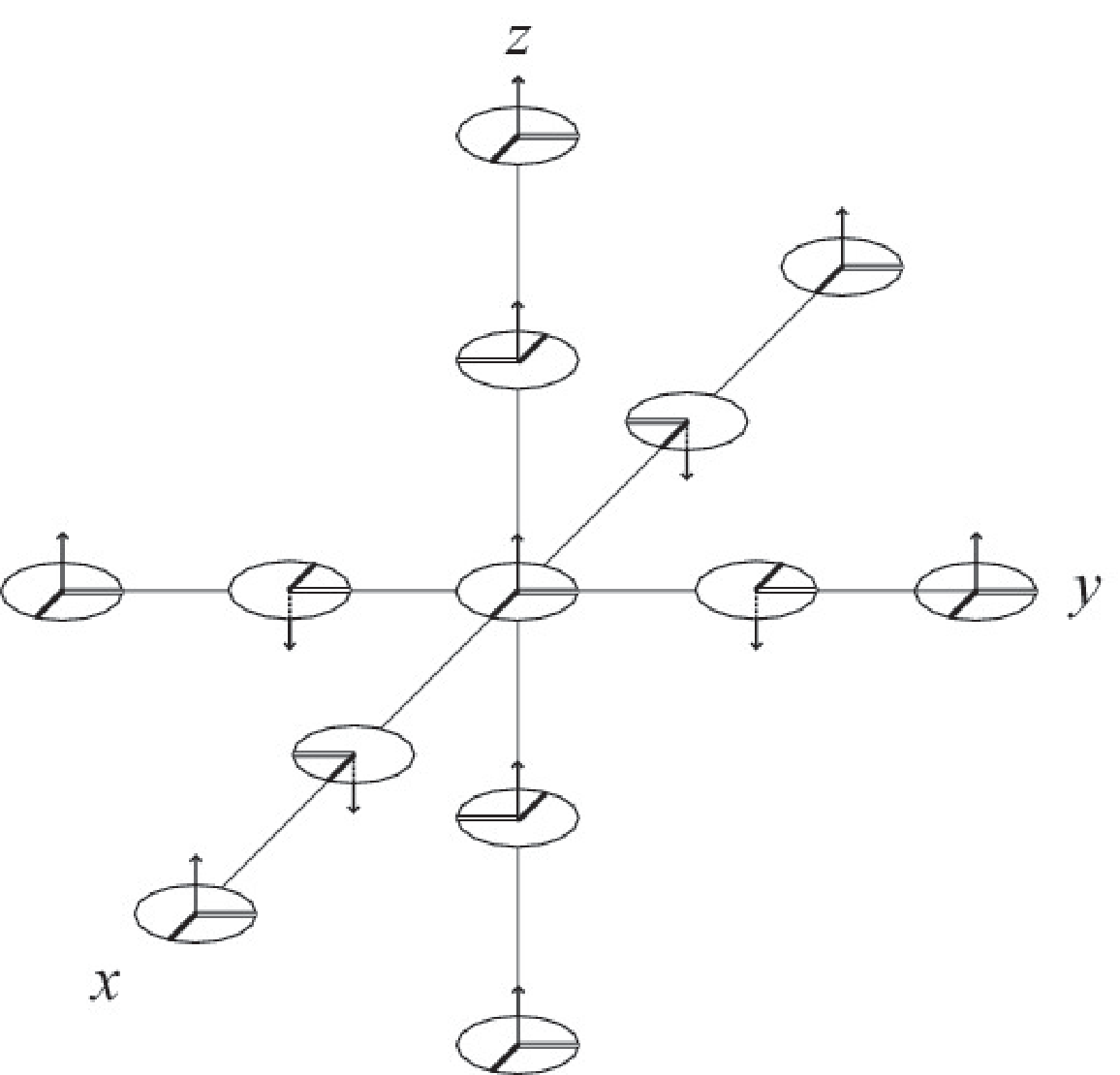}
}
\end{center}
\caption{Shankar Skyrmion with charge one, 
where the arrows represent the direction of the local magnetization $\hat{\bm s}$, 
whereas the sold and double lines on each disk indicate the direction of phase $0$ and $\pi/2$ in the spherical-harmonic representation, respectively.}
\label{fig:homotopy8}
\end{figure}

An example of a topological object corresponding to an element of $\pi_3({\rm SO}(3))\cong{\mathbb Z}$ may be realized by rotating the order parameter at position ${\bm r}$ about the direction $\bm\Omega={\bm r}/r$ through an angle $f(r)N_3$, where $f(0)=2\pi$, $f(\infty)=0$, and $N_3\in\mathbb{Z}$ is the Skyrmion charge~\cite{VolovikMineev1977,Shankar1977,Khawaja2001Nature,Khawaja2001}. 
It follows from the condition $f(\infty)=0$ that the order parameter is uniform at spatial infinity, and thus, the three-dimensional space is compactified to $S^3$. The case of $N_3=1$ is shown in Fig.~\ref{fig:homotopy8}.

The topological charge of the third homotopy group can be introduced in a manner similar to that of the second homotopy group. Now, the mapped vector has a four-component vector that is normalized to unity. One such representation is 
\begin{eqnarray}
\hat{\bm m}=\left(\frac{x}{r}\sin\frac{f(r)}{2},\frac{y}{r}\sin\frac{f(r)}{2},\frac{z}{r}\sin\frac{f(r)}{2},\cos\frac{f(r)}{2} \right).
\end{eqnarray}
The invariant of such a mapping is again given by the Jacobian of the transformations:
\begin{eqnarray}
N_3=\frac{1}{12\pi^2}\int d{\bm r}\sum_{\alpha,\beta,\gamma,\delta=1,2,3,4}\sum_{i,j,k=x,y,z}\epsilon_{\alpha\beta\gamma\delta}
\epsilon_{ijk}m_\alpha\partial_i m_\beta\partial_j m_\gamma \partial_k m_\delta,\label{N_3}
\end{eqnarray}
where $\epsilon_{ijk}$ and $\epsilon_{\alpha\beta\gamma\delta}$ are the completely antisymmetric tensors of rank 3 and 4, respectively.

\subsubsection{Knot soliton}

The third homotopy group also classifies knot solitons.
Knots are characterized by the linking number and differ from other topological excitations such as vortices, monopoles, and Skyrmions which are classified by winding numbers.
Knot solitons have attracted considerable interest since Faddeev and Niemi suggested that knots might exist as stable solitons in a three-dimensional classical field theory~\cite{Faddeev1997}. 

\begin{figure}[!ht]
\begin{center}
\resizebox{\hsize}{!}{
\includegraphics{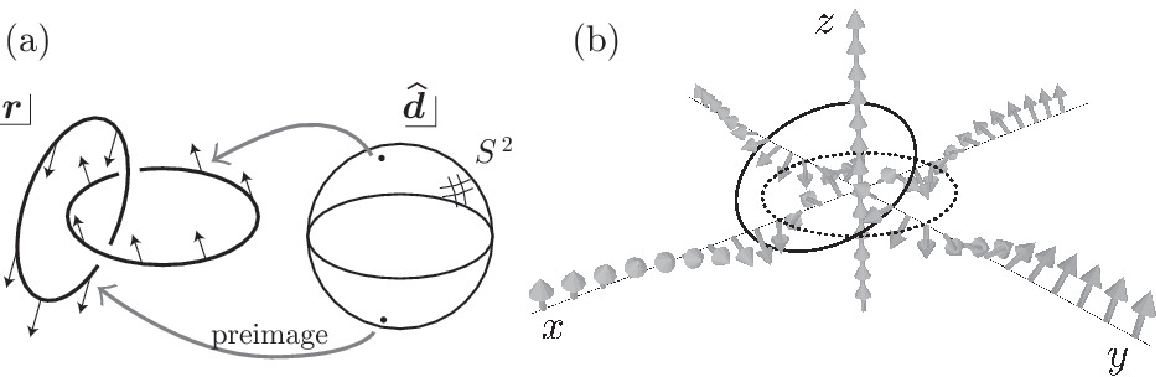}
}
\end{center}
\caption{
(a) Preimages of two distinct points on $S^2$ form a link.
(b) Spin configuration of a knot with Hopf charge 1 in a polar BEC,
where arrows indicate the $\hat{\bm d}$ field of the polar phase.
The solid and dashed loops trace the point where $\hat{\bm d}$ points to $x$ and $-z$, respectively, and form a link.
Reprinted from Ref.~\cite{Kawaguchi2008}.
}
\label{fig:knot_preimage}
\end{figure}

Knots are characterized by mappings from a three-dimensional sphere $S^3$ to $S^2$.
As in the case of Skyrmions, the $S^3$ domain is prepared by imposing a boundary condition that the order parameter takes on the same value 
in every direction at spatial infinity.
Here, we consider the spin-1 polar phase~\cite{Kawaguchi2008}.
The order parameter manifold of the polar phase is given by $\mani{R}^{\rm polar}\cong (S^2_{\bf f}\times {\rm U}(1)_\phi)/(\mathbb{Z}_2)_{{\bf f},\phi}$, of which
neither U(1) nor $\mathbb{Z}_2$ symmetry contributes to the homotopy groups in spaces higher than one dimension; hence, we find that $\pi_3(\mani{R}^{\rm polar})\cong \pi_3(S^2)\cong \mathbb{Z}$.
The corresponding topological charge $Q\in \mathbb{Z}$ is known as the Hopf charge.
Note that the domain (${\bm r}$) is three-dimensional, while the target space ($\hat{\bm d}$) is two-dimensional%
\footnote{The unit vector $\hat{\bm d}$ is defined as ${\bm n}$ in Ref.~\cite{Kawaguchi2008}}.
Consequently, the preimage of a point on the target $S^2$ constitutes a closed loop in $S^3$,
and the Hopf charge is interpreted as the linking number of these loops:
if the $\hat{\bm d}$ field has Hopf charge $Q$,
two loops corresponding to the preimages of any two distinct points on the target $S^2$ will be linked $Q$ times [see Fig.~\ref{fig:knot_preimage} (a)].
Figure~\ref{fig:knot_preimage} (b) shows an example of the $\hat{\bm d}$ field of a polar BEC with Hopf charge 1%
\footnote{Technically speaking, the configuration in Fig.~\ref{fig:knot_preimage} (b) is an {\it unknot} of a pair of rings with linking number 1,
because the preimage of one point on $S^2$ forms a simple unknotted ring.}.

The Hopf charge is defined as follows~\cite{Faddeev1997,Kawaguchi2008}.
For an areal element in the order-parameter manifold
\begin{align}
\mathcal{F}_{ij} = \hat{\bm d}\cdot(\partial_i \hat{\bm d} \times \partial_j \hat{\bm d}),
\end{align}
we introduce $\mathcal{A}_i$ such that
\begin{align}
\mathcal{F}_{ij}=\partial_i\mathcal{A}_j-\partial_j\mathcal{A}_i.
\label{eq:def_Fknot}
\end{align}
Then, the charge $Q$ is defined as
\begin{align}
Q=\frac{1}{4\pi^2} \int d^3x\ \epsilon_{ijk}\mathcal{F}_{ij} \mathcal{A}_k.
\label{eq:H-charge}
\end{align}
In the practical calculation of $Q$, we first consider 
a four-dimensional unit vector $\hat{\bm m}(\bm r)$ on $\mathbb{R}^3$ which can be written as a two-dimensional complex vector
\begin{align}
Z\equiv \begin{pmatrix} Z_1(\bm r) \\ Z_2(\bm r) \end{pmatrix}  = \begin{pmatrix} m_1(\bm r)+im_2(\bm r) \\ m_3(\bm r)+im_4(\bm r) \end{pmatrix},
\end{align}
where $|Z_1|^2+|Z_2|^2=\sum_{j=1,2,3,4}m_j^2 = 1$.
We then introduce the Hopf map, which is a map from $S^3$ to $S^2$, as
\begin{align}
\hat{\bm d} = Z^\dagger {\bm \sigma} Z,
\label{eq:Hopfmap}
\end{align}
where ${\bm \sigma}=(\sigma_x,\sigma_y,\sigma_z)$ is the vector of Pauli matrices.
The components of $\hat{\bm d}$ are given in terms of $Z_1$ and $Z_2$ as follows:
\begin{align}
d_x&=Z_1^*Z_2+Z_2^*Z_1,\\
d_y&=-i(Z_1^*Z_2-Z_2^*Z_1),\\
d_z&=|Z_1|^2+|Z_2|^2.
\end{align}
Substituting Eq.~\eqref{eq:Hopfmap} to Eq.~\eqref{eq:def_Fknot}, we obtain
\begin{align}
 \mathcal{F}_{jk} = i\left[(\partial_j Z^\dagger)(\partial_k Z) - (\partial_k Z^\dagger)(\partial_j Z)\right],
\end{align}
from which $\mathcal{A}_j$ is obtained as
\begin{align}
 \mathcal{A}_j = \frac{i}{2}\left[Z^\dagger(\partial_j Z) - (\partial_j Z^\dagger)Z\right].
\end{align}
Note that $\hat{\bm d}$ is invariant under the U(1) gauge transformation of $Z$, i.e., $Z\to e^{i\phi}Z$.
Under this transformation $\mathcal{F}_{jk}$ is also invariant,
whereas $\mathcal{A}_j$ is gauge dependent: $\mathcal{A}_j\to \mathcal{A}_j+\partial_j \phi$.
However, by taking the summation in terms of $i,j$ and $k$,
the charge $Q$ defined in Eq.~\eqref{eq:H-charge} becomes independent on $\phi$.
Hence, $Q$ is determined only by the configuration of $\hat{\bm d}(\bm r)$, and classifies the $\hat{\bm d}$ textures.
Writing down Eq.~\eqref{eq:H-charge} in terms of $m_j(\bm r)$, we have the same form as Eq.~\eqref{N_3}, which defines the charge for the mapping $S^3\to S^3$.

Knots can be created in a spin-1 polar BEC by manipulating an external magnetic field~\cite{Kawaguchi2008}.
In the presence of an external magnetic field,
the linear Zeeman effect causes the Larmor precession of $\hat{\bm d}$,
whereas $\hat{\bm d}$ tends to become parallel to the magnetic field because of the quadratic Zeeman effect.
Suppose that we prepare an optically trapped BEC in the magnetic sublevel $m=0$, i.e., $\hat{\bm d}=(0,0,1)^{\rm T}$,
by applying a uniform magnetic field in the $z$-direction.
Then, we suddenly turn off the uniform field and switch on a quadrupole field.
Due to the linear Zeeman effect, $\hat{\bm d}$ starts rotating around the local magnetic field,
and therefore, the $\hat{\bm d}$ field winds as a function of time, resulting in the formation of knots.

\begin{figure}[!ht]
\begin{center}
\resizebox{0.9\hsize}{!}{
\includegraphics{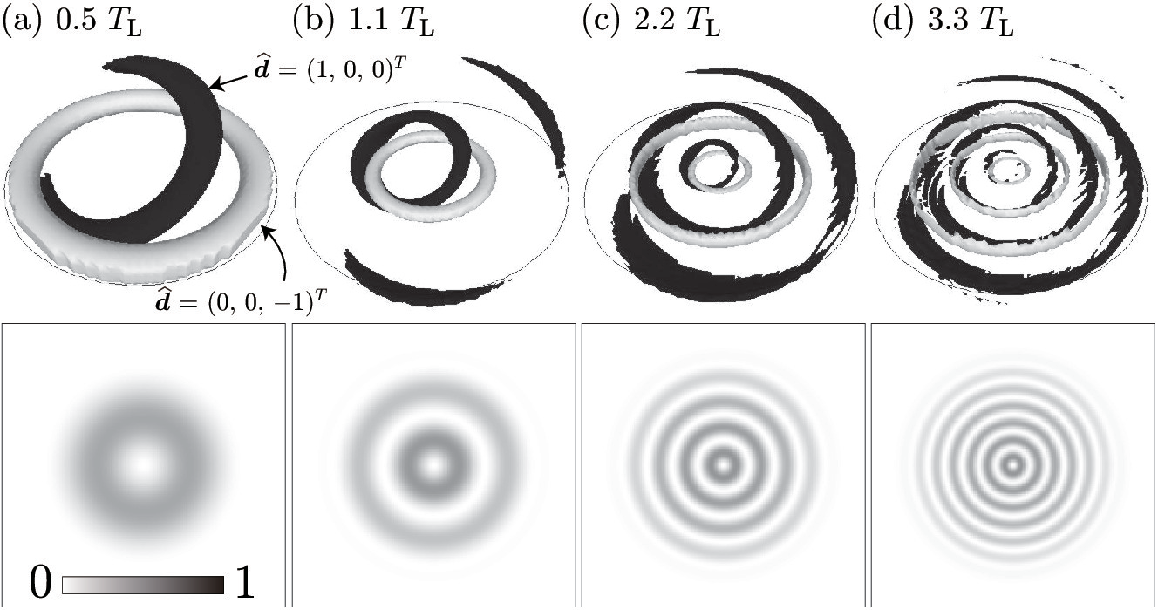}
}
\end{center}
\caption{Dynamics of the creation of knots in a spherical optical trap under a quadrupole magnetic field.
Snapshots of the preimages of $\hat{\bm d}=(0,0,-1)^{\rm T}$ and $\hat{\bm d}=(1,0,0)^{\rm T}$ (top), and the cross sections of the density for the $m=-1$ components on the $x--y$ plane (bottom).
Reprinted from Ref.~\cite{Kawaguchi2008}.
}
\label{fig:knottime}
\end{figure}

Figure~\ref{fig:knottime} shows the dynamics of the creation of knots in an optical trap subject to a quadrupole field,
where the upper panels show the snapshots of the preimages of $\hat{\bm d}=-\hat{z}$ and $\hat{\bm d}=\hat{z}$
and the bottom panels show the cross sections of the density for $m=-1$ (bottom) components on the $x$--$y$ plane.
The density pattern in the $m=-1$ components is characteristic of knots;
a double-ring pattern appears corresponding to each knot.
As the $\hat{\bm d}$ field winds in the dynamics, the number of rings increases.
This prediction can be tested by the Stern-Gerlach experiment.
However, the knot created in the above method is unstable
because the right-hand side of Eq.~\eqref{eq:dFdt_polar} is non-vanishing for the $\hat{\bm d}$ texture,
and therefore, the order parameter goes out of $\mani{R}^{\rm polar}$~\cite{Kawaguchi2008}.

\subsection{Action of vortices on another topological excitation}

\begin{figure}[ht]
\begin{center}
\resizebox{0.6\hsize}{!}{
\includegraphics{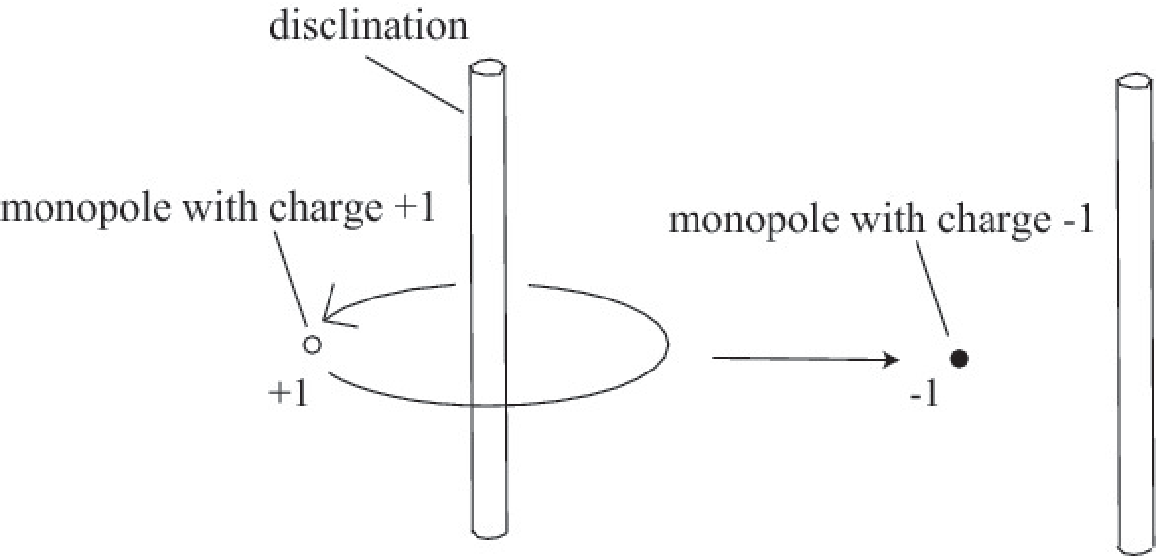}
}
\end{center}
\caption{Influence of a vortex on a monopole. As a monopole with charge $+1$ makes a complete circuit around the vortex, it changes to a monopole with charge $-1$.
Reprinted from Ref.~\cite{Kobayashi2012}.
}
\label{fig:homotopy9}
\end{figure}

As we have seen above, topological excitations are classified by the $n$th homotopy group $\pi_n$, if they are isolated.
However, for topological excitations that coexist with vortices, there are cases in which an element of $\pi_n$ alone cannot properly describe the charge of a topological excitation due to the influence of the vortices. 
This is because an element of $\pi_n$ corresponding to the charge of a topological excitation may change when the topological excitation circumnavigates a vortex.
This phenomenon is referred to as the action of $\pi_1$ on $\pi_n$. 
As an example, let us consider a spin-1 polar phase in which a monopole and a half-quantum vortex (Alice vortex) coexist.
As a monopole makes a complete circuit around the half-quantum vortex, the configuration of the $\hat{\bm d}$-vector undergoes the transformation of $\hat{\bm d}\to -\hat{\bm d}$,
giving rise to the sign change of the monopole charge defined in Eq.~\eqref{N_22} (see Fig.~\ref{fig:homotopy9}).
In such a case, the homotopy groups no longer classify the topological excitations.
Instead, topological excitations coexisting with vortices are classified by the Abe homotopy group $\kappa_n$ which is defined as a semi-direct product of $\pi_1$ and $\pi_n$~\cite{Abe1940}. See Ref.~\cite{Kobayashi2012} for detail.

%% file: manybody.tex
\section{Many-Body Theory}
\label{sec:ManyBodyTheory}

In this section, we examine the many-body spin states of spin-1 and 2 BECs by assuming that a single spatial wave function is shared by all spin states, namely, with the single-mode approximation (SMA, see Sec.~\ref{sec:SMA}).
According to the discussion in Sec.~\ref{sec:SMA},
the field operator of a spin-$f$ BEC in the SMA can be expressed as
\begin{align}
\hat{\psi}_m({\bm r}) = \hat{a}_m\psi_{\rm SMA}({\bm r}) \ \ (m=f,f-1,\cdots,-f).
\label{MB3}
\end{align}
We describe the Hamiltonian in terms of $\hat{a}_m$ and investigate the many-body states.

\subsection{Many-body states of spin-1 BECs} 
\label{sec:spin-1MBS}

\subsubsection{Eigenspectrum and eigenstates} 
\label{sec:spin-1eigenspectrum}

For the sake of simplicity, we consider the case of zero magnetic field (i.e., $p=q=0$).  
Substituting Eq.~(\ref{MB3}) in Eqs.~\eqref{H_0} and \eqref{V(f=1)2} 
and replacing $\sum_m \hat{a}_m^\dagger \hat{a}_m$ and $\sum_{m,m'} \hat{a}_m^\dagger \hat{a}_{m'}^\dagger \hat{a}_{m'} \hat{a}_m $ with $\hat{N}$ and $\hat{N}(\hat{N}-1)$, respectively, 
we obtain~\cite{Law1998,Koashi2000,Ho2000}
\begin{eqnarray}
\hat{H}
= \hat{N} \! \int \! d{\bm r} \; \psi_{\rm SMA}^\ast
\left[ -\frac{\hbar^2 \nabla^2}{2M} + U_{\rm trap}({\bm r}) + \frac{c_0}{2} (\hat{N}-1) |\psi_{\rm SMA}|^2 \right] \psi_{\rm SMA} + \frac{c_1}{2V^{\rm eff}} :\hat{{\bm F}}^2 \! : ,
\label{MB4}
\end{eqnarray}
where
\begin{eqnarray}
\hat{{\bm F}}
=\sum_{m,m'=-1}^1 {\bf f}_{mm'} \hat{a}_m^\dagger \hat{a}_{m'},
\label{MB5}
\end{eqnarray}
$V^{\rm eff}= \left(\int d{\bm r} | \psi_{\rm SMA} |^4\right)^{-1} $, and $\psi_{\rm SMA}$ is determined by 
\begin{align}
\left[-\frac{\hbar^2}{2M}\nabla^2 + U_{\rm trap}(\bm r) + c_0 (N-1) |\psi_{\rm SMA}(\bm r)|^2\right]\psi_{\rm SMA}(\bm r) = \mu\psi_{\rm SMA}(\bm r)
\label{eq:SMA_GP2}
\end{align}
subject to the normalization condition
\begin{eqnarray}
\int | \psi_{\rm SMA} |^2 d{\bm r} =1.
\label{MB6-2}
\end{eqnarray}

Substituting the solution of Eq.~\eqref{eq:SMA_GP2} in Eq.~(\ref{MB4}), we obtain
\begin{eqnarray}
\hat{H} = \mu\hat{N}-\frac{c_0}{2V^{\rm eff}}\hat{N}(\hat{N}-1)+\frac{c_1}{2V^{\rm eff}} : \hat{{\bm F}}^2 \! : .
\label{MB7}
\end{eqnarray}
From Eq.~\eqref{eq:spin1identity3}, we find that the total spin operator $\hat{{\bm F}}$ satisfies
\begin{eqnarray}
 : \hat{{\bm F}}^2 \! :
=\hat{N}(\hat{N}-1)-3 \; \hat{S}^\dagger \hat{S},
\label{MB8}
\end{eqnarray}
where  
\begin{eqnarray}
\hat{S}
\equiv \frac{1}{\sqrt{3}} ( \hat{a}_0^2 - 2 \hat{a}_1 \hat{a}_{-1} )
\label{MB9}
\end{eqnarray}
is the spin-singlet pair operator. We use Eq.~(\ref{MB8}) to rewrite Eq.~(\ref{MB7}) as
\begin{eqnarray}
\hat{H} = \mu\hat{N} - \frac{c_0}{2V^{\rm eff}} \hat{N} (\hat{N}-1) + \frac{c_1}{2V^{\rm eff}}
\left[ \hat{N} (\hat{N}-1) -3 \hat{S}^\dagger \hat{S} \right].
\label{MB10}
\end{eqnarray}
Because $\hat{N}$ and $\hat{S}^\dagger \hat{S}$ commute with each other, the eigenvalue problem reduces to finding their simultaneous eigenstates. 

The simultaneous eigenstate of $\hat{N}$ and $\hat{S}^\dagger\hat{S}$ is characterized with the number of particles and the number of spin-singlet pairs.
Let $|N,0\rangle$ be the $N$-particle state in which there is no spin-singlet pair, i.e., $\hat{N}|N,0\rangle=N|N,0\rangle$ and $\hat{S} |N, 0 \rangle = 0$. 
Using the commutation relation
\begin{align}
\left[ \hat{S}, (\hat{S}^\dagger)^k \right]
= \frac{2}{3}k \, (2\hat{N}-2k+5)(\hat{S}^\dagger)^{k-1},
\label{MB11}
\end{align}
we obtain
\begin{eqnarray}
\hat{S}^\dagger \hat{S} (\hat{S}^\dagger )^k |N,0\rangle
=\frac{2}{3}k\, (2\hat{N}-2k+1) (\hat{S}^\dagger)^k |N,0\rangle .
\label{MB12}
\end{eqnarray}
Noting that $(\hat{S}^\dagger)^k|N,0\rangle$ describes the $N+2k$ particle state, namely,
\begin{align}
\hat{N}(\hat{S}^\dagger)^k|N,0\rangle &= \left\{[\hat{N},(\hat{S}^\dagger)^k]+(\hat{S}^\dagger)^k \hat{N}\right\}|N,0\rangle \nonumber\\
&= 2k(\hat{S}^\dagger)^k|N,0\rangle + (\hat{S}^\dagger)^k N |N,0\rangle\nonumber\\
&= (N+2k) (\hat{S}^\dagger)^k|N,0\rangle,
\label{MB12-2}
\end{align}
we define the $N$-particle $k$-singlet-pair state $|N,k\rangle$ as
\begin{eqnarray}
|N, k \rangle
\equiv \frac{(\hat{S}^\dagger)^k |N-2k, 0\rangle}{\sqrt{\langle N-2k, 0 | \hat{S}^k (\hat{S}^\dagger)^k |N-2k, 0 \rangle}}.
\label{MB13}
\end{eqnarray}
It follows from Eqs.~\eqref{MB12} and \eqref{MB12-2} that $|N,k\rangle$ is the simultaneous eigenstate of $\hat{S}^\dagger\hat{S}$ and $\hat{N}$ satisfying
\begin{align}
\hat{S}^\dagger \hat{S} |N, k \rangle &=\frac{2}{3}k\, (2N-2k+1) |N, k \rangle,\label{MB14}\\
\hat{N}|N,k\rangle  &=  N|N,k\rangle.\label{MB14-2}
\end{align}

Now the question arises: what is the spin state of the remaining $N-2k$ atoms that do not form spin-singlet pairs? 
By definition in Eq.~\eqref{MB5}, $\hat{\bm F}$ satisfies 
\begin{align}
 :\hat{\bm F}^2: = \hat{\bm F}^2 -2 \hat{N},
\label{eq:MB-F^2}
\end{align}
from which Eq.~\eqref{MB7} is rewritten as~\cite{Law1998}
\begin{align}
\hat{H} = \mu\hat{N} - \frac{c_0}{2V^{\rm eff}} \hat{N} (\hat{N}-1) + \frac{c_1}{2V^{\rm eff}} (\hat{\bm F}^2 -2\hat{N}).
\label{eq:MB-H_FN}
\end{align}
Since $\hat{F}_{x,y,z}$ obey the angular momentum commutation relations,
the eigenstate of the system is also specified with the eigenvalues of $\hat{N}$, $\hat{\bm F}^2$, and $\hat{F}_z$.
Moreover, because $\hat{\bm F}^2$ and $\hat{S}^\dagger\hat{S}$ commute [see Eqs.~\eqref{MB8} and \eqref{eq:MB-F^2}],
$|N,k\rangle$ defined in Eq.~\eqref{MB13} is also the simultaneous eigenstate of $\hat{\bm F}^2$ and $\hat{F}_z$.
In fact, using Eqs.~\eqref{MB8} and \eqref{MB14}--\eqref{eq:MB-F^2}, we obtain
\begin{align}
\hat{\bm F}^2 |N,k\rangle &= [\hat{N}(\hat{N}+1)-3\hat{S}^\dagger\hat{S}]|N,k\rangle, \nonumber\\
&=(N-2k)(N-2k+1)|N,k\rangle.
\label{eq:MB-F^2eigenstate}
\end{align}
indicating that the residual $N-2k$ atoms form a total spin $F=N-2k$ state.
The eigenvalue $F_z$ of $\hat{F}_z$ takes $F_z=N-2k, N-2k-1,\cdots, -(N-2k)$.
Hence, specifying the eigenstate with $F=N-2k$ and $F_z$ instead of $N$,
the eigenstate and eigenenergy of the Hamiltonian~\eqref{MB7} are given by
\begin{align}
|k,F,F_z\rangle &= Z^{-1/2} (\hat{S}^\dagger)^k (\hat{F}_-)^{F-F_z} (\hat{a}_1^\dagger)^F |{\rm vac}\rangle,\label{MB18}\\
E_k &= \mu N - \frac{c_0}{2V^{\rm eff}} N (N \! - \!1) + \frac{c_1}{2V^{\rm eff}} \left[ N (N \! - \!1) \! - \! 2k (2N \! - \! 2k+1) \right],\label{eq:E_k_spin1}\\
&=\mu N - \frac{c_0}{2V^{\rm eff}} N (N \! - \!1) + \frac{c_1}{2V^{\rm eff}} \left[ F(F+1) -2N \right],\label{eq:E_k_spin1-2}
\end{align}
where $|{\rm vac}\rangle$ is the vacuum of atoms, $Z$ is the normalization constant, and
\begin{align}
\hat{F}_-\equiv \hat{F}_x-i\hat{F}_y =\sqrt{2} (\hat{a}_0^\dagger\hat{a}_1+\hat{a}_{-1}^\dagger\hat{a}_0)
\end{align}
is the lowering operator of the magnetic quantum number ($F_z$).
It follows from Eq.~\eqref{eq:E_k_spin1} or Eq.~\eqref{eq:E_k_spin1-2} that
the many-body ground state for $c_1>0$ is $|k=N/2,F=0,F_z=0\rangle$ for even $N$ and $|k=(N-1)/2,F=1,F_z=1,0,-1\rangle$ for odd $N$, 
whereas the many-body ground state for $c_1<0$ is $|k=0,F=N,F_z=N,N-1,\cdots,-N\rangle$.

\subsubsection{Fragmentation} 
\label{sec:fragmentation}

We consider the state given in Eq.~(\ref{MB18}) with $F_z=F$, and calculate the number of the atoms in the $m=0$ sublevel, i.e., $\langle \hat{a}_0^\dagger \hat{a}_0 \rangle$. We introduce the basis state $|n_1, n_0, n_{-1} \rangle$ in which the magnetic sublevels $m=1, 0, -1$ are occupied by $n_1, n_0,$ and $n_{-1}$ atoms, respectively. Then, $| {\rm vac} \rangle = | 0,0,0 \rangle$ and
\begin{align}
\langle \hat{a}_0^\dagger \hat{a}_0 \rangle
&=\frac{\langle {\rm vac} | \hat{a}_1^F \hat{S}^k \hat{a}_0^\dagger \hat{a}_0 (\hat{S}^\dagger)^k (\hat{a}_1^\dagger)^F | {\rm vac} \rangle}
{\langle {\rm vac} | \hat{a}_1^F \hat{S}^k  (\hat{S}^\dagger)^k (\hat{a}_1^\dagger)^F| {\rm vac} \rangle } 
\nonumber\\
&=\frac{\langle F,0,0 | \hat{S}^k \hat{a}_0^\dagger \hat{a}_0 (\hat{S}^\dagger)^k |F,0,0 \rangle}
{\langle F,0,0| \hat{S}^k  (\hat{S}^\dagger)^k |F,0,0 \rangle }.
\label{MB20}
\end{align}
Substituting $\hat{a}_0^\dagger \hat{a}_0 = \hat{a}_0 \hat{a}_0^\dagger -1$ and using the fact that $\hat{a}_0^\dagger$ commutes with $\hat{S}^\dagger$, we have
\begin{eqnarray}
\langle \hat{a}_0^\dagger \hat{a}_0 \rangle
=\frac{\langle F, 1, 0 | \hat{S}^k (\hat{S}^\dagger)^k | F, 1, 0 \rangle}
{\langle F, 0, 0 | \hat{S}^k (\hat{S}^\dagger)^k | F, 0, 0 \rangle}-1.
\label{MB21}
\end{eqnarray}
We use Eq.~(\ref{MB11}) to rewrite
\begin{align}
\hat{S} (\hat{S}^\dagger)^k
&= [ \hat{S}, (\hat{S}^\dagger)^k ] + (\hat{S}^\dagger)^k \hat{S}
\nonumber \\
&= \frac{2}{3}k \; (2\hat{N} - 2k +5)(\hat{S}^\dagger)^{k-1} + (\hat{S}^\dagger)^k \hat{S}.
\label{MB22}
\end{align}
Noting that $\hat{S} |F, 1, 0 \rangle = \hat{S} | F, 0, 0 \rangle = 0$ and $\hat{N} | F, 1, 0 \rangle = (F+1)|F,1,0 \rangle$, we have
\begin{align}
\langle \hat{a}_0^\dagger \hat{a}_0 \rangle
&= \frac{(2F+2k+3) \langle F,1,0|\hat{S}^{k-1}(\hat{S}^\dagger)^{k-1} |F,1,0\rangle}
{(2F+2k+1) \langle F,0,0|\hat{S}^{k-1}(\hat{S}^\dagger)^{k-1} |F,0,0\rangle}-1
\\[2mm]
&= \frac{(2F+2k+3) (2F+2k+1) \cdots\cdot (2F+5)}
{(2F+2k+1) (2F+2k-1) \cdots\cdot (2F+3)} -1
\\
&= \frac{2k}{2F+3}
= \frac{N-F}{2F+3}.
\label{eq:MB_n0}
\end{align}
Because $ \langle \hat{F}_z\rangle =\langle \hat{a}_1^\dagger \hat{a}_1 \rangle - \langle \hat{a}_{-1}^\dagger \hat{a}_{-1} \rangle =F $ and $\sum_{m=-1}^1 \langle \hat{a}_m^\dagger \hat{a}_m \rangle =N$, we obtain
\begin{align}
\bar{n}_1
&\equiv \langle \hat{a}_1^\dagger \hat{a}_1 \rangle
= \frac{NF+F^2+N+2F}{2F+3},
\label{MB23}
\\
\bar{n}_0
&\equiv \langle \hat{a}_0^\dagger \hat{a}_0 \rangle
= \frac{N-F}{2F+3},
\label{MB24}
\\
\bar{n}_{-1}
&\equiv \langle \hat{a}_{-1}^\dagger \hat{a}_{-1} \rangle
= \frac{(N-F)(F+1)}{2F+3}.
\label{MB25}
\end{align}
On the other hand, the off-diagonal elements of the one-body density matrix $\langle \hat{a}_m^\dagger \hat{a}_{m'}\rangle$ can be shown to vanish for $F_z=F$ with arbitrary $k$.

When all atoms form spin-singlet pairs, i.e., $F=0$, Eqs.~(\ref{MB23})--(\ref{MB25}) show that all magnetic sublevels are equally populated
\begin{eqnarray}
\langle \hat{a}_1^\dagger \hat{a}_1 \rangle
= \langle \hat{a}_0^\dagger \hat{a}_0 \rangle
= \langle \hat{a}_{-1}^\dagger \hat{a}_{-1} \rangle
= \frac{N}{3},
\label{MB26}
\end{eqnarray}
and that there is no spin coherence between different components: $\langle \hat{a}_1^\dagger \hat{a}_0\rangle = \langle \hat{a}_0^\dagger \hat{a}_{-1}\rangle = \langle \hat{a}_{-1}^\dagger \hat{a}_1\rangle = 0$.
That is, the condensate is fragmented in the sense that more than one single-particle state is macroscopically occupied~\cite{Nozieres1982,Law1998,Koashi2000,Ho2000,Mueller2006}. 
This is in sharp contrast with the mean-field result which dictates by assumption that there exists one and only one BEC, that is, the one-body density matrix $\langle \hat{a}_m^\dagger \hat{a}_{m'}\rangle$ is always diagonalized to ${\rm Diag}[N,0,0]$.
However, such an assumption cannot be justified when the system possesses certain exact symmetries such as spin rotational symmetry in the present case.
Equation~(\ref{MB24}) shows that $\bar{n}_0$ decreases rapidly with an increase in $F$ (see also Ref.~\cite{Ho2000}). In fact, when $F=O(\sqrt{N})$, we have $\bar{n}_0=O(\sqrt{N})$. This implies that although mean-field results break down at zero magnetic field, the validity of mean-field theory quickly recovers as the magnetic field increases. This affords an example of why fragmented BECs are fragile against symmetry-breaking perturbations. 

We also note an interesting relationship between a mean-field state and the many-body spin-singlet state. For $c_1>0$, a mean-field ground state is polar in which all particles occupy the same single-particle state:
\begin{eqnarray}
|\hat{\bm d}\rangle = \frac{\left(\hat{a}_{\hat{\bm d}}^\dagger\right)^N}{\sqrt{N!}} \; |{\rm vac}\rangle ,
\label{MB27}
\end{eqnarray}
where 
\begin{align}
\hat{a}_{\hat{\bm d}} &\equiv d_x\hat{a}_x + d_y\hat{a}_y + d_z\hat{a}_z\nonumber\\
&= -\frac{d_x+id_y}{\sqrt{2}}\hat{a}_1 + d_z\hat{a}_0+ \frac{d_x-id_y}{\sqrt{2}}\hat{a}_{-1}
\label{MB15}
\end{align}
annihilates a particle in the $m=0$ state along the $\hat{\bm d}$ direction ($|\hat{\bm d}|=1$).
Substituting $d_x = \sin\theta\cos\phi , \ d_y=\sin\theta\sin\phi$, and $d_z=\cos\theta$ in Eq.~(\ref{MB15}), we obtain
\begin{eqnarray}
|\hat{\bm d}\rangle
=\! \frac{1}{\sqrt{N!}}
\left( -\frac{\hat{a}_1^\dagger}{\sqrt{2}}  \sin\theta e^{i\phi} + \hat{a}_0^\dagger \cos\theta + \frac{\hat{a}_{-1}^\dagger}{\sqrt{2}}  \sin\theta e^{-i\phi} \right)^N \!\!\!
|{\rm vac}\rangle.
\label{MB28}
\end{eqnarray}
We calculate the equal-weighted superposition state of $|\hat{\bm d}\rangle$ over all solid angles $ d\Omega_{\hat{\bm d}} = \sin\theta d\theta d\phi$:
\begin{align}
|{\rm sym}\rangle
&=
\frac{1}{4\pi} \int_0^\pi \sin\theta \; d\theta \int_0^{2\pi} d\phi \; |\theta, \phi \rangle
\nonumber \\
&=
\left\{
	\begin{array}{ll}
	\ 0 \ \ \ \ &\textrm{if $N$ is odd;} \\[3mm]
	\displaystyle{\frac{1}{(N+1)\sqrt{N!}}} (a_0^{\dagger 2}-2a_1^\dagger a_{-1}^\dagger)^{N/2} |{\rm vac} \rangle \ \ &\textrm{if $N$ is even.}
	\end{array}
\right.
\end{align}
This result shows that the spin-singlet state is the superposition state of a polar state $ |\hat{\bm d}\rangle $ over all directions of $\hat{\bm d}$. Conversely, the mean-field state $|\hat{\bm d}\rangle$ may be interpreted as a broken symmetry state of the spin-singlet state with respect to the direction of quantization $\hat{\bm d}$.

It is natural to ask how the system evolves when all atoms are prepared in the $m=0$ state, i.e., $|\hat{\bm d}=\hat{z}\rangle$.
Although such an initial state is stationary in the mean-field theory, 
the population in the $m=0$ state decreases, due to the many-body effect,
with the characteristic time scale of $\sqrt{N}\tau_{1}$ where $\tau_1=\hbar/(c_1N/V^{\rm eff})$ is the time scale for the (mean-field) spin-mixing process~\cite{Law1998,Pu1999,Zhou2001}.
Hence, such many-body spin-mixing dynamics becomes relevant for small $N$.
The many-body spin-mixing dynamics beyond the SMA is discussed in Ref.~\cite{Pu1999},
and the effect the quadratic Zeeman shift is examined in Refs.~\cite{Cui2008,Barnett2010,Barnett2011}.
The correspondence between spin-1 polar condensates and quantum rotors is studied in Refs.~\cite{Barnett2010,Barnett2011}.

As for related works,
it is pointed out in Refs.~\cite{Pu2000, Duan2000,Duan2002} that macroscopic entangled states can be generated from a BEC of $m=0$ atoms
via the spin-exchanging collisions (i.e., $|m=0\rangle+|m=0\rangle \to |m=1\rangle+|m=-1\rangle$).
Quantum phase diffusions in a ferromagnetic BEC is discussed in Refs.~\cite{Yi2003a,Yi2003b}.
The effects of the magnetic dipole-dipole interaction (DDI) on the many-body ground state are discussed in Refs.~\cite{Yi2004,Yi2006c}:
The DDI with the single-mode approximation breaks the spin-rotational symmetry of the Hamiltonian, and therefore, the spin-singlet-pair condensation is destabilized.
The many-body correction for the spin-mixing dynamics (see Sec.~\ref{sec:SMD}) is discussed in Refs.~\cite{Chang2007,Heinze2010}.
The number fluctuation in each spin component during the parametric amplification (see Sec.~\ref{sec:parametric_amplification}) is discussed in Ref.~\cite{Mias2008}.
The linear and quadratic Zeeman-energy dependences on the energy spectra for a ferromagnetic BEC are investigated in Ref.~\cite{Lamacraft2011}.
Strongly correlated states in a rotating spin-1 Bose gase is discussed in the context of quantum Hall states~\cite{Ardonne1999,Ho2002,Reijnders2002,Paredes2002}.

\subsection{Many-body states of spin-2 BECs} 
\label{sec:spin-2MBS}

Next, we discuss the many-body states of spin-2 BECs in the single-mode approximation. 
Here, we consider the case of $q=0$ and $p\neq 0$.
Substituting $\hat{\psi}_m({\bm r})=\hat{a}_m\psi_{\rm SMA}({\bm r})$ in Eq.~(\ref{V(f=2)}), we obtain the spin-dependent part of the Hamiltonian as
\begin{eqnarray}
\hat{H}_{\rm sp}=
\frac{c_1}{2V^{\rm eff}}:\hat{{\bm F}}^2:
+\frac{2c_2}{5V^{\rm eff}}\hat{\cal S}_+\hat{\cal S}_-
-p\hat{F}_z,
\label{h2}
\end{eqnarray}
where 
\begin{eqnarray}
\hat{{\bm F}}\equiv\sum_{mm'=-2}^2{\bf f}_{mm'}\hat{a}_m^\dagger\hat{a}_{m'}
\end{eqnarray}
is the spin vector operator,
\begin{eqnarray}
\hat{F}_z\equiv\sum_{m}m\hat{a}_m^\dagger\hat{a}_m
\end{eqnarray}
is its $z$-component, and
\begin{eqnarray}
\hat{\cal S}_-=(\hat{\cal S}_+)^\dagger\equiv
\frac{1}{2}\sum_m(-1)^m\hat{a}_m\hat{a}_{-m}
\label{FSF}
\end{eqnarray}
are the creation ($\hat{\cal S}_+$) and annihilation ($\hat{\cal S}_-$) operators of a spin-singlet pair.

While the operator $\hat{\cal S}_+$, when applied to a vacuum, creates
a pair of bosons in the spin-singlet state, the pair should not be regarded as a single composite boson because $\hat{\cal S}_+$ does not satisfy the
commutation relations of bosons.
In fact, the operators $\hat{\cal S}_\pm$ together with
$\hat{\cal S}_z\equiv (2\hat{N}+5)/4$ satisfy the $SU(1,1)$ 
commutation relations:
\begin{eqnarray}
[\hat{\cal S}_z,\hat{\cal S}_\pm]=\pm\hat{\cal S}_\pm, \ \
[\hat{\cal S}_+,\hat{\cal S}_-]=-2\hat{\cal S}_z.
\label{SU(1,1)}
\end{eqnarray}
Here, the minus sign in the last equation is the only difference
from the usual spin commutation relations. This difference, however,
leads to some important consequences. In particular, the
Casimir operator $\hat{\cal S}^{2}$, which commutes with
$\hat{\cal S}_\pm$ and $\hat{\cal S}_z$, does not take the 
usual form of the squared sum of spin components but instead takes the form
\begin{eqnarray}
\hat{\cal S}^{2}\equiv -\hat{\cal S}_+\hat{\cal S}_-
+\hat{\cal S}_z^2-\hat{\cal S}_z.
\label{Casimir}
\end{eqnarray}

\subsubsection{Eigenspectrum and eigenstates}
\label{sec:Eigenspectrum}

The eigenstates are classified according to the eigenvalues of the Casimir operator (\ref{Casimir}). Because 
$\hat{\cal S}_+\hat{\cal S}_-=\hat{\cal S}_z^2-\hat{\cal S}_z
-\hat{\cal S}^{2}$ is positive semidefinite,
$\hat{\cal S}_z$ must have a minimum value ${\cal S}^{\rm min}_z$. 
Recalling that $\hat{\cal S}_z\equiv(2\hat{N}+5)/4$,  
${\cal S}^{\rm min}_z$ can be expressed in terms of a non-negative
integer $N_0$ as ${\cal S}^{\rm min}_z=(2N_0+5)/4$.
Let $|\phi\rangle$ be the eigenvector corresponding to this minimum eigenvalue; then, $\hat{\cal S}_-|\phi\rangle=0$. It follows that
\begin{eqnarray}
\hat{\cal S}^{2}|\phi\rangle=
{\cal S}^{\rm min}_z({\cal S}^{\rm min}_z-1)|\phi\rangle;
\end{eqnarray}
hence, the eigenvalue of $\hat{\cal S}^{2}$ is given by 
$\nu=S(S-1)$ with $S=(2N_0+5)/4$.

The operation of $\hat{\cal S}_+$ on $|\phi\rangle$ increases the eigenvalue of $\hat{\cal S}_z$ by 1 and that of $\hat{N}$ increases by 2, as observed from the commutation relations (\ref{SU(1,1)}). 
Therefore, the allowed combinations of eigenvalues 
$S(S-1)$ and $S_z$ for $\hat{\cal S}^{2}$ and $\hat{\cal S}_z$, respectively, are given by
\begin{equation}
S=(2N_0+5)/4 \;\;
(N_0=0,1,2,\ldots)
\label{eigen_S}
\end{equation}
and
\begin{equation}
S_z=S+N_{\rm S} \;\;
(N_{\rm S}=0,1,2,\ldots),
\label{def_N2}
\end{equation}
where $N_0$ and $N_{\rm S}$  satisfy
\begin{equation}
N=2N_{\rm S}+N_0,
\end{equation}
and we may therefore interpret $N_{\rm S}$ as the number of
spin-singlet ``pairs'' and $N_0$ the number of remaining bosons.

To find the exact energy eigenvalues of Hamiltonian (\ref{h2}), we first note that $\hat{\cal S}_\pm$ and $\hat{\cal S}_z$ commute with $\hat{\bm F}$:
\begin{eqnarray}
[\hat{\cal S}_\pm,\hat{\bm F}]=0, \ \ 
[\hat{\cal S}_z,\hat{\bm F}]=0.
\end{eqnarray}
The energy eigenstates can be classified
according to quantum numbers
$N_0$ and
$N_{\rm S}$,
total spin $F$,
and magnetic quantum number $F_z$.
We denote the eigenstates as
$|N_0,N_{\rm S},F,F_z;\lambda\rangle$, where
$\lambda=1,2,\ldots,g_{N_0,F}$ labels orthonormal
degenerate states with $g_{N_0,F}$ provided in Ref.~\cite{Ueda2002}.

The first commutator in Eq.~(\ref{SU(1,1)}) implies that $\hat{\cal S}_\pm$ plays the role of changing the eigenvalue of $\hat{\cal S}_z$ by $\pm1$ by creating or annihilating a spin-singlet pair. The eigenstate can be constructed by first preparing the eigenstates that do not involve spin-singlet pairs,
\begin{eqnarray}
|N_0,0,F,F_z;\lambda\rangle,
\end{eqnarray}
and then, by generating new eigenstates through successive operations of $\hat{\cal S}_+^{N_{\rm S}}$ on it:
\begin{eqnarray}
|N_0,N_{\rm S},F,F_z;\lambda\rangle
=
\frac{\hat{\cal S}_+^{N_{\rm S}}|N_0,0,F,F_z;\lambda\rangle}{\parallel\hat{\cal S}_+^{N_{\rm S}}|N_0,0,F,F_z;\lambda\rangle\parallel},
\label{eigenstate}
\end{eqnarray}
where $\parallel |\psi\rangle\parallel\equiv
\sqrt{\langle\psi|\psi\rangle}$, and
the orthnormality condition is
\begin{eqnarray}
\langle N_0,N_{\rm S},F,F_z;\lambda^\prime|
N_0,N_{\rm S},F,F_z;\lambda\rangle=
\delta_{\lambda^\prime,\lambda}.
\end{eqnarray}

To write down the eigenstates (\ref{eigenstate}) explicitly, we need to find the matrix elements of $\hat{\cal S}_+$, of which only the nonvanishing ones are
\begin{eqnarray}
\langle N_0,N_{\rm S}+1,F,F_z;\lambda|\hat{\cal S}_+|
N_0,N_{\rm S},F,F_z;\lambda\rangle=
\sqrt{(N_{\rm S}+1)(N_{\rm S}+N_0+5/2)}.
\end{eqnarray}
Finally, the energy eigenvalue for the state $|N_0,N_{\rm S},F,F_z;\lambda\rangle$
is given by
\begin{equation}
E=\frac{c_1}{2V^{\rm eff}}[F(F+1)-6N]
+\frac{c_2}{5V^{\rm eff}}N_{\rm S}(N+N_0+3)
-pF_z,
\label{E2}
\end{equation}
where the relationship $2N_{\rm S}+N_0=N$ is used.

The total spin $F$ can, in general, take integer values in the range
$0\le F \le 2N_0$. However,  there
are some forbidden values~\cite{Ueda2002}. That is, $F=1,2,5,2N_0-1$ are not allowed when $N_0=3k \ (k\in \mathbb{Z})$, and  $F=0,1,3,2N_0-1$ are forbidden when
$N_0=3k\pm1$. We prove this at the end of this subsection.

To gain a physical insight into the nature of the energy eigenstates $|N_0,N_{\rm S},F,F_z;\lambda\rangle$, it is useful to express them in terms of some building blocks. In fact, we can express them in terms of one-, two-, and three-boson creation operators.
Let $\hat{A}^{(n)}_f{}^\dagger$ be the operator such that when applied
to the vacuum state, it creates $n$ bosons that have a total spin $F=f$ and magnetic quantum number $F_z=f$.
While the choice of a complete set of operators for the building blocks is not unique, 
we choose the following five operators for constructing the eigenstates:
\begin{eqnarray}
\hat{A}^{(1)\dagger}_2&=&\hat{a}_2^\dagger
\label{def_A12}
\\
\hat{A}^{(2)\dagger}_0&=&
\frac{1}{\sqrt{10}}
[(\hat{a}_0^\dagger)^2-2\hat{a}_1^\dagger\hat{a}_{-1}^\dagger
+2\hat{a}_2^\dagger\hat{a}_{-2}^\dagger]
=\sqrt{\frac{2}{5}}\hat{\cal S}_+
\\
\hat{A}^{(2)\dagger}_2&=&
\frac{1}{\sqrt{14}}
[2\sqrt{2}\hat{a}_2^\dagger\hat{a}_0^\dagger
-\sqrt{3}(\hat{a}_1^\dagger)^2]
\\
\hat{A}^{(3)\dagger}_0&=&
\frac{1}{\sqrt{210}}
[\sqrt{2}(\hat{a}_0^\dagger)^3-3\sqrt{2}\hat{a}_1^\dagger
\hat{a}_0^\dagger\hat{a}_{-1}^\dagger
+3\sqrt{3}(\hat{a}_1^\dagger)^2\hat{a}_{-2}^\dagger
\nonumber \\
& & 
+3\sqrt{3}\hat{a}_2^\dagger(\hat{a}_{-1}^\dagger)^2
-6\sqrt{2}\hat{a}_2^\dagger\hat{a}_0^\dagger
\hat{a}_{-2}^\dagger]
\label{def_A30}
\\
\hat{A}^{(3)\dagger}_3&=&
\frac{1}{\sqrt{20}}
[(\hat{a}_1^\dagger)^3-\sqrt{6}\hat{a}_2^\dagger\hat{a}_1^\dagger
\hat{a}_0^\dagger
+2(\hat{a}_2^\dagger)^2\hat{a}_{-1}^\dagger].
\end{eqnarray}
Note that $\hat{A}^{(2)\dagger}_1$ and $\hat{A}^{(3)\dagger}_1$ vanish identically due to Bose symmetry, and that the operators $\hat{A}^{(n)\dagger}_f$ commute with $\hat{F}_+$.

We next consider a set ${\cal B}$ of unnormalized states:
\begin{eqnarray}
|n_{12},n_{20},n_{22},n_{30},n_{33}\rangle
\equiv
(\hat{a}^\dagger_2)^{n_{12}}
(\hat{A}^{(2)\dagger}_0)^{n_{20}}
(\hat{A}^{(2)\dagger}_2)^{n_{22}}
(\hat{A}^{(3)\dagger}_0)^{n_{30}}
(\hat{A}^{(3)\dagger}_3)^{n_{33}}
|{\rm vac}\rangle,
\label{def_state_unit}
\end{eqnarray}
with $n_{12},n_{20},n_{22},n_{30}=0,1,2,\ldots,\infty$ and $n_{33}=0,1$. 
Note that the value of $n_{33}$ can restricted to 0 or 1 because $(\hat{A}^{(3)\dagger}_3)^2$ is decomposed as follows:
\begin{align}
 (\hat{A}^{(3)\dagger}_3)^2 = \sqrt{\frac{7}{15}} (\hat{a}_2^\dagger)^2 \hat{A}^{(2)\dagger}_2 \hat{A}^{(2)\dagger}_0 + \sqrt{\frac{14}{45}} (\hat{a}_2^\dagger)^3 \hat{A}^{(3)\dagger}_0
-\frac{1}{20}\left(\sqrt{\frac{14}{3}}\hat{A}^{(2)\dagger}_2\right)^3.
\end{align}
The total number of bosons and the total spin of the state (\ref{def_state_unit}) are
$N=n_{12}+2(n_{20}+n_{22})+3(n_{30}+n_{33})$
and 
$F=F_z=2(n_{12}+n_{22})+3n_{33}$, respectively.
For given $N$ and $F$, $n_{33}$ is uniquely
determined by the parity of $F$:
\begin{eqnarray}
n_{33}=F\;{\rm mod}\; 2.
\end{eqnarray}
It can be shown~\cite{Ueda2002} that ${\cal B}$ forms a nonorthogonal 
complete basis set of the subspace ${\cal H}_{(F_z=F)}$
in which the magnetic quantum number $F_z$
is equal to the total spin $F$.
It can also be shown~\cite{Ueda2002} that 
the energy eigenstates can be represented as
\begin{equation}
(\hat{F}_-)^{\Delta F}(\hat{A}^{(2)\dagger}_0)^{n_{20}}
\hat{P}_{(N_{\rm S}=0)}
(\hat{a}^\dagger_2)^{n_{12}}
(\hat{A}^{(2)\dagger}_2)^{n_{22}}
(\hat{A}^{(3)\dagger}_0)^{n_{30}}
(\hat{A}^{(3)\dagger}_3)^{n_{33}}
|{\rm vac}\rangle,
\label{eigenstates}
\end{equation}
where $\hat{P}_{(N_{\rm S}=0)}$ is the projection onto the subspace with $N_{\rm S}=0$ (the kernel of $\hat{\cal S}_-$),
$n_{12},n_{20},n_{22},n_{30}=0,1,2,\ldots,\infty$,
$n_{33}=0,1$, and $\Delta F=0,1,\ldots, 2F$.
These parameters are related to $\{N_0,N_{\rm S},F,F_z\}$ as
\begin{eqnarray}
N_0&=&n_{12}+2n_{22}+3n_{30}+3n_{33},
\\
N_{\rm S}&=&n_{20},
\\
F&=&2n_{12}+2n_{22}+3n_{33},
\\
F_z&=&F-\Delta F,
\end{eqnarray}
and the corresponding eigenenergy is given by Eq.~(\ref{E2}).
Note that the states defined in (\ref{eigenstates})
are unnormalized, and the states having the same energy 
(i.e., those belonging to the same set of parameter values 
$\{N_0,N_{\rm S},F,F_z\}$) are nonorthogonal.

It might be tempting to envisage a physical picture that the system, as in the case of $^4$He, to be made up of $n_{nf}$ composite bosons whose creation operator 
is given by $\hat{A}^{(n)\dagger}_f$. 
However, this is an oversimplification because the operator $\hat{A}^{(n)}_f$ does not obey the boson commutation relation.
Moreover, the projection operator $\hat{P}_{(N_{\rm S}=0)}$ in
(\ref{eigenstates}) imposes many-body spin correlations such that
the spin correlation between {\it any} two bosons must have a 
vanishing spin-singlet component. 
Note that two bosons with independently fluctuating spins generally have a nonzero overlap with the spin-singlet state.
The many-body spin correlations of the energy eigenstates are 
thus far more complicated than what a naive picture of composite 
bosons suggests. 
On the other hand, as long as quantities such as the number of bosons, 
magnetization, and energy are concerned, 
the above simplified picture is quite helpful. 
As an illustration, we provide an explanation for 
the existence of forbidden values for the total spin $F$.
For example, to construct a state with $F=0$ or $F=3$,
composite particles with total spin 2 must be avoided, namely,
$n_{12}=n_{22}=0$. Then, we have $N_0=3(n_{30}+n_{33})$, implying that
$F=0$ or $F=3$ is only possible when $N_0=3k\,(k\in \mathbb{Z})$.
For a state with $F=2$ or $F=5$, we have $n_{12}+n_{22}=1$ and
$N_0=1+n_{22}+3(n_{30}+n_{33})$, implying that $N_0\neq 3k\,(k\in \mathbb{Z})$.
The above simplified picture is also helpful when we consider the magnetic
response discussed below.

\subsubsection{Magnetic response}
\label{sec:Magnetic response}

Here, we consider how the ground state and the magnetization $F_z$ 
respond to an applied magnetic field $p$.
From Eq.~(\ref{E2}), we see that the minimum energy states always 
satisfy $F_z=F$ when $p>0$. The problem thus reduces to minimizing
the function
\begin{equation}
E(F_z,N_s)=\frac{c_1}{2V^{\rm eff}}[F_z(F_z+1)-6N]
+\frac{c_2}{5V^{\rm eff}}N_{\rm S}(N+N_0+3)
-pF_z.
\label{E3}
\end{equation}

\begin{figure}[ht]
\begin{center}
\resizebox{0.8\hsize}{!}{
\includegraphics{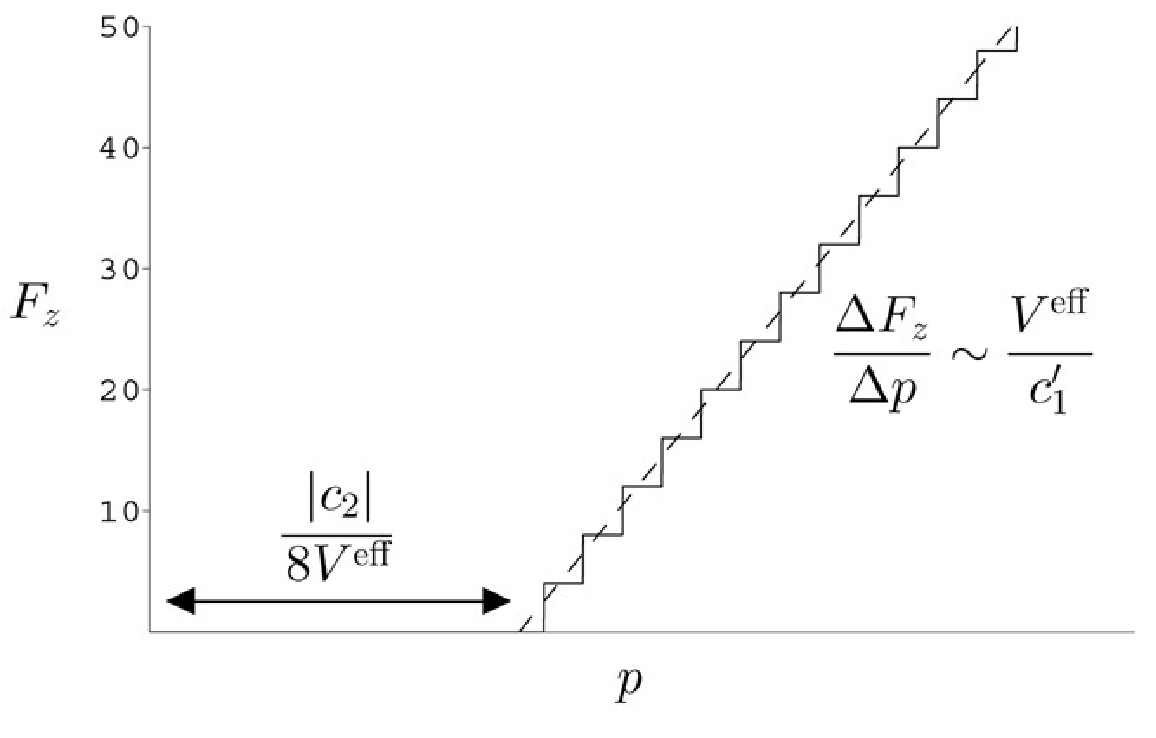}
}
\end{center}
\caption{Typical dependence of the ground-state magnetization
on the applied magnetic-field strength, for $c_2<0$, $c_1'>0$ and $|c_2|\gg c_1'$.
}
\label{fig:spin2Meissner}
\end{figure}

Here, we only consider the case of $c_2<0$. Detailed investigations of this problem can be found in Ref.~\cite{Ueda2002}.
In this case, it is convenient to introduce a new parameter
\begin{equation}
c_1'\equiv c_1-\frac{c_2}{20},
\end{equation}
and consider the energy as a function of $F_z$ and $l\equiv 2N_0-F_z$:
\begin{align}
E(F_z,l)=&\frac{c_1'}{2V^{\rm
eff}}\left[F_z-\frac{V^{\rm eff}}{c_1'}\left(p+\frac{c_2}{8V^{\rm eff}}
\right)+\frac{1}{2}\right]^2
\nonumber \\
 & -\frac{c_2}{40V^{\rm eff}}l(l+2F_z+6)+{\rm const}.
\label{e_fz_l}
\end{align}
For $c_2<0$, the second term in the right-hand side of Eq.~\eqref{e_fz_l} is minimized at $l=0$. Then, the ground-state magnetization is given by
\begin{align}
 F_z = \left\{
\begin{array}{ll}
0 & \textrm{ for \ \ } 0<p\le \displaystyle{\frac{|c_2|+4c_1'}{8V^{\rm eff}}}; \\[3mm]
\displaystyle{\frac{V^{\rm eff}}{c_1'}\left(p-\frac{|c_2|+4c_1'}{8V^{\rm eff}}\right)} & \textrm{ for \ \ } p>\displaystyle{\frac{|c_2|+4c_1'}{8V^{\rm eff}}}.
\end{array}
\right.
\label{eq:MB_spin2Fz}
\end{align}
While the averaged slope $\Delta F_z/\Delta p\sim V^{\rm eff}/c_1'$
coincides with the mean-field result [see below Eq.~\eqref{eq:MFTspin2C4}], the offset term $(|c_2|+4c_1')/(8V^{\rm eff})$ in Eq.~(\ref{eq:MB_spin2Fz}) 
makes a qualitative distinction from the mean-field theory, that is, the onset of the
magnetization displaces from $p=0$ to $p=(|c_2|+4c_1')/(8V^{\rm eff})$. 
A typical behavior of the magnetic response for $|c_2|\gg c_1'$ is shown in Fig.~\ref{fig:spin2Meissner}.

We now calculate the Zeeman-level populations of the ground states for $c_2<0$. 
In the mean-field theory, the lowest-energy states for $c_1>0$, $c_2<0$, $q=0$ and small positive $p$ have vanishing population in the $m=0,\pm 1$ levels ($C_4$ state in Table~\ref{table3}).
In contrast, the exact ground states derived in the preceding
subsection, $(\hat{A}^{(2)\dagger}_0)^{N_{\rm S}}
(\hat{a}^\dagger_2)^{n_{12}}
(\hat{A}^{(2)\dagger}_2)^{n_{22}}
(\hat{A}^{(3)\dagger}_3)^{n_{33}}|{\rm vac}\rangle$
with $n_{22}=0,1$ and $n_{33}=0,1$ have nonzero populations in the $m=0,\pm 1$
levels.  The exact forms for the averaged population
$\langle\hat{a}_m^\dagger\hat{a}_m\rangle$ are calculated as follows.
The above ground states have the form $(\hat{A}_0^{(2)\dagger})^{N_{\rm
S}} |\phi\rangle\propto(\hat{\cal S}_+)^{N_{\rm S}} |\phi\rangle$, with
$|\phi\rangle$ being a state with a fixed number
   ($s\equiv n_{12}+2n_{22}+3n_{33}$) of bosons satisfying
$\hat{\cal S}_-|\phi\rangle=0$.
The average Zeeman population for the ground states,
\begin{equation}
\langle\hat{a}_m^\dagger\hat{a}_m\rangle
\equiv \frac{\langle \phi|(\hat{\cal S}_-)^{N_{\rm S}}
\hat{a}_m^\dagger\hat{a}_m
(\hat{\cal S}_+)^{N_{\rm S}} |\phi\rangle}
{\langle \phi|(\hat{\cal S}_-)^{N_{\rm S}}
(\hat{\cal S}_+)^{N_{\rm S}} |\phi\rangle},
\end{equation}
is then simply related to the average Zeeman populations for
the state $|\phi\rangle$ as
\begin{equation}
\langle\hat{a}_m^\dagger\hat{a}_m\rangle
=
\langle\hat{a}_m^\dagger\hat{a}_m\rangle_0
+\frac{N_{\rm S}}{s+5/2}
(\langle\hat{a}_m^\dagger\hat{a}_m\rangle_0+
\langle\hat{a}_{-m}^\dagger\hat{a}_{-m}\rangle_0+1),
\label{formula_zeeman1}
\end{equation}
where $\langle\hat{a}_m^\dagger\hat{a}_m\rangle_0\equiv
\langle \phi|\hat{a}_m^\dagger\hat{a}_m|\phi\rangle/\langle
\phi|\phi\rangle$.
This formula implies that when $N_{\rm S}\gg s$, the Zeeman populations of the ground states are sensitive to the form of $|\phi\rangle$.

With this formula, it is a straightforward task to calculate the average Zeeman-level populations for the state
\begin{eqnarray}
(\hat{A}^{(2)\dagger}_0)^{N_{\rm S}}
(\hat{a}^\dagger_2)^{n_{12}}
(\hat{A}^{(2)\dagger}_2)^{n_{22}}
(\hat{A}^{(3)\dagger}_3)^{n_{33}}|{\rm vac}\rangle
\end{eqnarray}
with $n_{22}=0,1$ and $n_{33}=0,1$.
A striking feature appears in the leading terms under
the condition $1\ll n_{12} \ll N_{\rm S}$.
The results are summarized as
\begin{equation}
\langle\hat{a}^\dagger_1\hat{a}_1\rangle
\sim
\langle\hat{a}^\dagger_{-1}\hat{a}_{-1}\rangle
\sim
N_{\rm S}(1+n_{33})/n_{12}
\end{equation}
and
\begin{equation}
\langle\hat{a}^\dagger_0\hat{a}_0\rangle
\sim
N_{\rm S}(1+2n_{22})/n_{12}.
\label{population_0}
\end{equation}
These results show that the population of each magnetic sublevel depends very sensitively on the spin correlations. As a consequence, a minor change in the magnetization might lead to a major change in the population. Such dramatic changes originate from bosonic stimulations caused by the term $\hat{a}_2^\dagger\hat{a}_0^\dagger$ in $\hat{A}_2^{(2)\dagger}$ or the term $(\hat{a}_2^\dagger)^2\hat{a}_{-1}^\dagger$ in $\hat{A}^{(3)\dagger}_3$.

\subsubsection{Symmetry considerations on possible phases}

The possible ground states of spinor BECs can be inferred based on symmetry arguments without solving the GPEs. 
Let us consider the eigenvalues of the order parameter matrix ${\bm \chi}$ introduced in Sec.~\ref{sec:spin2_Cartesian}:
\begin{eqnarray}
{\rm det}\left(E\hat{1}-\sqrt{3}{\bm \chi}\right)
=E^3-3S E+\frac{T}{2\sqrt{3}}=0,
\label{determinant}
\end{eqnarray}
where $\sqrt{3}$ is added in front of the matrix for the sake of convenience,
and $S$ and $T$ are the amplitudes of the spin-singlet pair and the spin-singlet trio, respectively:
\begin{align}
S&=\frac{1}{2}\sum_{i,j}\chi_{ij}^2
=\frac{1}{2}\psi_0^2-\psi_1\psi_{-1}+\psi_2\psi_{-2}
,
\label{pair_singlet} \\
T&=-18\sum_{i,j,k}\epsilon_{ijk}\chi_{xi}\chi_{yj}\chi_{zk}
\nonumber \\
&=-9(\psi_2\psi_{-1}^2+\psi_1^2\psi_{-2})+\sqrt{6}\psi_0(-\psi_0^2+3\psi_1\psi_{-1}+6\psi_2\psi_{-2}).
\label{trio_singlet}
\end{align}
Substituting $S=xy$ and $T=2\sqrt{3}(x^3+y^3)$, Eq.~(\ref{determinant}) reduces to
\begin{eqnarray}
E^3+x^3+y^3-3Exy=0,
\end{eqnarray}
and hence, the solutions are 
\begin{eqnarray}
E=-(x+y), -(\omega x+\omega^2 y),-(\omega^2 x+\omega y),
\end{eqnarray}
where $\omega=e^{2\pi i/3}$ and
\begin{eqnarray}
x=\frac{S}{y}=\frac{1}{48^\frac{1}{6}}\left(T+\sqrt{T^2-48S^3}\right)^\frac{1}{3}.
\end{eqnarray}
We can infer from this that the spin-singlet pair and the spin-singlet trio of atoms form building blocks of the ground state of the spin-2 BEC. It is noteworthy that this fact arises solely from the fact that the order parameter matrix of the spin-2 BEC is traceless symmetric.

%% file: summary.tex
\section{Summary and Future Prospects}
\label{sec:Summary}

In the present paper, we have reviewed the basic knowledge concerning spinor Bose-Einstein condensates (BECs) that has been accumulated thus far. The fundamental characteristics of spinor BECs are rotational invariance, coupling between the spin and the gauge degrees of freedom, and magnetism arising from the magnetic moment of the spin.

The rotational invariance and gauge invariance alone uniquely determine the microscopic Hamiltonian of the spinor BEC, as discussed in Sec.~\ref{sec:GeneralTheory}. The mean-field theory introduced in Sec.~\ref{sec:meanfield} describes the ground-state phases and spin dynamics in spinor BECs. A number of experiments have been performed to examine spinor physics as explained in Sec.~\ref{sec:experiments}. In a scalar BEC, the measurements of the collective modes are found to agree with the theoretical values with an accuracy better than 1\% (see Ref.~\cite{Dalfovo1999} for a comprehensive review). 
For a spinor BEC, experiments on parametric amplification via dynamical instability show qualitative agreement with the Bogoliubov theory described in Sec.~\ref{sec:Bogoliubov}.
In the near future, experimental tasks to be carried out include investigating the extent to which the predictions of the Bogoliubov theory can be verified experimentally.

The magnetic dipole-dipole interaction of alkali atoms is so weak that it is negligible in a scalar BEC.
However, in a spinor BEC, it makes a significant contribution to creating spatial spin structures as discussed in Sec.~\ref{sec:DipolarBEC}.
Recently, spin polarized BECs of dysprosium~\cite{Mingwu2011} and erbium~\cite{Aikawa2012}, both having large magnetic moments, were realized,
and a spinor BEC of chromium atoms were created in an ultralow magnetic field~\cite{Pasquiou2011b,Pasquiou2012},
opening up a new paradigm of spinor dipolar physics.

In the low-energy limit, the hydrodynamic equations derived in Sec.~\ref{sec:hydro} give a simple description of the spin dynamics in terms of physical quantities such as superfluid velocity, magnetization, and nematic director.
As discussed in Sec.~\ref{sec:Vortices}, there are many different types of vortices depending on the symmetry of the ground-state order parameter.
In particular, the circulation of the superfluid velocity is not always quantized in units of $h/M$ but can take its rational fraction or even continuous values in spinor condensates.
Properties of vortices and other topological excitations are categorized based on the symmetry of each phase (Sec.~\ref{sec:symmetry}) and by application of homotopy theory to the order parameter manifold (Sec.~\ref{sec:topology}).

When we take into account many-body quantum correlations in a spinor BEC,
some of mean-field ground states are no longer stable:
Rather than condensation into a single-particle state, spin-singlet pairs (spin-1 and 2) and spin-singlet trios (spin-2) undergo Bose-Einstein condensation, as discussed in Sec.~\ref{sec:ManyBodyTheory}. Such states are fragmented in the sense that atoms are condensed in several single-particle states. Fragmented condensates are fragile against external perturbations such as an external magnetic field. The many-body quantum correlations are expected to be prominent in systems of small number of atoms.

While we have discussed spinor BECs at absolute zero in this review, finite-temperature effects were investigated in a number of articles: 
Isoshima {\it et al.}~\cite{Isoshima2000b} investigated the finite-temperature phase diagram for a spin-1 BEC at a fixed magnetization by means of the first-order perturbation method (so called ``Hartree-Fock-Bogoliubov-Popov'' approximation~\cite{Griffin1996}).
They predicted that a Bose gas with antiferromagnetic interaction ($c_1>0$) undergoes two-step phase transitions from a single component BEC to a multiple component BEC,
whereas for ferromagnetic interactions  ($c_1<0$) phase separation occurs at low temperature~\cite{Isoshima2000b}.
Zhang {\it et al.} investigated the same system and argued that three-step phase transitions occur in a ferromagnetic gas, where the lowest temperature phase is not phase-separated but uniform with nonzero transverse magnetization~\cite{Zhang2004}.
However, the latter paper ignores coherence between thermal atoms in different magnetic sublevels,
which is not an adequate approximation in the present situation.
Actually, when the condensate has nonzero magnetization in the transverse direction, 
it create spin coherence in thermal components via collisions with the condensate~\cite{Phuc2011,Kawaguchi2012}.
This property is crucial in determining the phase diagram, in particular, in the presence of the quadratic Zeeman effect~\cite{Kawaguchi2012}.

The BEC transition temperature, $T_{\rm BEC}$, of a spinor gas is suppressed compared with a scalar gas with the same number of atoms.
This is because atoms distribute in several internal states and the phase space density for each spin state is always smaller than that of a scalar gas~\cite{Isoshima2000}.
The mean-field shift of the BEC transition temperature in a harmonic trap was investigated in Ref.~\cite{Huang2002}.
Several papers discuss magnetic ordering above $T_{\rm BEC}$:
Gu and Klemm predicted from a simple mean-field calculation (Hartree approximation) that when the spin exchange interaction is ferromagnetic ($c_1<0$),
a ferromagnetic phase transition always occurs above $T_{\rm BEC}$ no matter how small the interaction is~\cite{Gu2003}.
However, when we take into account the Fock terms of both spin-dependent and spin-independent interactions, which are neglected in Ref.~\cite{Gu2003},
the condition (within the Hartree-Fock approximation) for the appearance of the ferromagnetic order above $T_{\rm BEC}$ is given by $c_1 < -c_0/3 <0$.
Natu and Mueller calculated the spin susceptibility using a random phase approximation,
and obtained the interaction dependence of the ferromagnetic transition temperature~\cite{Natu2011}. 
They also pointed out that for an antiferromagnetic interaction ($c_1>0$), the instability against a formation of spin-singlet pairs occurs above $T_{\rm BEC}$ if the spin-dependent interaction is strong enough ($c_1>c_0/2$).

The finite-temperature effect on the spin dynamics is also important to understand experiments.
For example, in the quench dynamics discussed in Sec.~\ref{sec:Bog_Domainformation}~\cite{Sadler2006,Guzman2011},
the excess energy of an initial non-ground state is expected to thermalize the system so as to increase the entropy~\cite{Mur-Petit2006,Moreno-Cardoner2007,Gawryluk2007b}.
However, the investigation of the spin correlation function revealed that this system does not reach thermal equilibrium within a realistic time scale~\cite{Barnett2011b};
the authors of Ref.~\cite{Barnett2011b} implied that this is due to the low Landau-damping rate.
On the other hand, for a pseudo-spin-1/2 system, a collective spin dynamics is observed even well above $T_{\rm BEC}$~\cite{Lewandowski2002,Oktel2002,Fuchs2002,Williams2002,McGuirk2002}.
The condensate spin dynamics was observed to be enhanced in the presence of a thermal gas in a spin-1/2 BEC~\cite{McGuirk2003} and a spin-1 BEC~\cite{Kronjager2005}.
Motivated by these experiments, properties of spin waves such as sound velocity and damping above and around $T_{\rm BEC}$ were discussed in spin-1 gases~\cite{Gu2004,Endo2008,Natu2010,Endo2011,Szirmai2012}.
However, the magnetic property of the normal component and its effect on the condensate are yet to be understood. 
In particular, to fully understand the spin dynamics at finite temperature, the spin conservation law for the entire system must be taken into account.

Beyond the mean-field treatment, 
the diagrammatic method was applied for spin-1 BECs both for absolute zero and finite temperatures~\cite{Szepfalusy2002,Ohtsuka2003,KisSzabo2007,Phuc2012}.
On the other hand, a perturbative method in terms of noncondensate fraction reveals that the quadratic Zeeman energy dependence of the ground-state phase diagram 
significantly deviates from that in the mean-field theory due to the existence of quantum-depleted atoms~\cite{Phuc2011}.
The true phase diagram even at zero temperature is still yet to be understood.
Finite-temperature phase diagrams were obtained phenomenologically by means of low-energy effective field theories~\cite{Yang2009}.
The renormalization-group approach was applied for low-dimensional systems~\cite{Kolezhuk2010}, $4-\epsilon (\epsilon<1)$ dimensions~\cite{Szirmai2006}, and a three-dimensional system in the presence of a magnetic dipole-dipole interaction~\cite{Pietila2011}.
It is suggested in Ref.~\cite{Pietila2011} that polar and ferromagnetic condensates are unstable when thermal fluctuations dominate quantum fluctuations,
and the possibility of a first-order phase transition in the unstable regime was discussed.

Two-dimensional (2D) spinor BECs exhibit a rich variety of Berezinskii-Kosterlitz-Thouless (BKT) transitions~\cite{Kosterlitz1973}, 
because spinor BECs can accommodate various types of vortices.
Mukerjee {\it et al.} pointed out that the universal jump in the renormalized stiffness for a polar gas in a 2D system is four times larger than that in a scalar 2D Bose gas~\cite{Mukerjee2006}.
The underlying physics is as follows.
In the absence of an external magnetic field, the nematic director can point in an arbitrary direction in the three-dimensional spin space, resulting in no spin (nematic) ordering in 2D~\cite{Kosterlitz1973}.
Hence, only the superfluid phase exhibits quasi-long-range order in a 2D polar gas.
However, since the smallest unit of vortices in the polar phase is a half-quantum vortex,
around which both the superfluid phase and the direction of nematic director $\hat{\bm d}$ change by $\pi$ (see Sec.~\ref{sec:vortex_spin1polar}),
the vortices in the superfluid phase carry half the normal quantum of vorticity,
resulting in four-times-larger universal jump.
This result also indicates that the quasi-long-range order arises not in the phase of each spin component $\psi_m \,(m=0,\pm1)$, but in the phase of a spin-singlet pair $\psi_0^2-2\psi_1\psi_{-1}$.
Song and Zhou numerically showed that even though there is no quasi-long-range order in nematic directors, the $\pi$-disclination of nematic directors is correlated with a half winding of superfluid phase, implying that fluctuations of disclination-antidisclination pairs are strongly suppressed~\cite{Song2009}.
The thermal activation of half-quantum vortices and the BKT-type of crossover to the superfluid state in a quasi-2D condensate is discussed in Ref.~\cite{Pietila2010} by means of the c-field methods~\cite{Blakie2008}.

The effect of the quadratic Zeeman shift on the BKT transition is also investigated~\cite{Podolsky2009,James2011}.
For $q<0$, the nematic director is confined in a 2D plane perpendicular to the external magnetic field, and exhibits a quasi-long-range order.
Hence, there are three distinct types of vortices which undergo binding-unbinding transition:
phase vortices, integer nematic vortices, and half-quantum vortices. However, the conditions for defect unbinding cannot always be treated independently of one another.
Podolsky {\it et al.}~\cite{Podolsky2009} showed that in a certain parameter region, 
unbinding of phase or nematic vortices leads to unbinding of half-quantum vortices.
This phase transition is distinct from the conventional BKT transition because it involves two types of vortices~\cite{Podolsky2009}.
On the other hand, for $q>0$, the Ising-type transition is predicted to occur in addition to the BKT transition~\cite{James2011}.
For $q>0$, the direction of the nematic director $\hat{\bm d}$ tends to be parallel or antiparallel to the external magnetic field.
Then, the $\pi$-disclination in the nematic directors around a half-quantum vortex is deformed to a sine-Gordon domain wall attached to the half-quantum vortex, which terminates at another half-quantum vortex.
At the onset of the BKT transition temperature, the domain wall, starting from and ending up with a bound vortex pair, is still fluctuating.
This situation is equivalent to the 2D Ising model and undergoes phase transition at a lower temperature.
As $q$ increases, the tension of the domain wall increases, and the Ising and BKT transitions meet for sufficiently large $q$~\cite{James2011}.

Spinor BECs in a double well potential exhibit the {\it external} Josephson effect.
In contrast, the spin exchange dynamics in a single well is regarded as an {\it internal} Josephson effect,
where atoms undergo periodic oscillations between internal states (see Sec.~\ref{sec:SMD}).
An interesting result for a spinor Josephson junction is the macroscopic oscillation and self-trapping of the condensate magnetization without any net change of total atom numbers within each well~\cite{Ashhab2002a,Mustecaplioglu2005}.
Moreover, in such dynamics, atoms in different internal states can tunnel as pairs through the potential barrier between the two wells in opposite directions,
leading to the generation of quantum entanglement between condensates in the two wells~\cite{Ng2003,Mustecaplioglu2007}.
The Josephson effect in spin-2 BECs was investigated in Ref.~\cite{Qi2009}.

A spinor gas loaded in an optical lattice can be described with the spinor Bose-Hubbard Hamiltonian in which the spinor interactions are possible at each lattice site. The new control parameters introduced in this system are the ratio of the transfer amplitude $t$ to the on-site spin-independentinteraction $U_0$ and filling fraction. 
For the ferromagnetic case, the system is expected to show the Mott-insulator-superfluid (MI-SF) transition similar to the scalar case.
For the antiferromagnetic case, Mott lobes are expected to be a nematic insulator or a spin-singlet insulator~\cite{deml02break,imam03spinexchange} and feature an even-odd parity effect in which even fillings stabilize the MI state due to the spin gap associated with spin-singlet pair formation~\cite{deml02break,tsuchiya04}.

In one dimension, antiferromagnetic spin-1 systems with odd fillings and small $t/U_0$ are predicted to be dimerized with an excitation gap~\cite{deml02break,yip03,Zhou2003,Zhou2003b}. Since each atom can be dimerized with one of the two nearest neighbors, the dimer phase has two-fold degeneracy and breaks lattice translational symmetry.
A renormalization-group method confirms the dimerized insulating phase for odd fillings~\cite{rizzi05}, while quantum Monte Carlo calculations show that the entire first Mott lobe has the dimerized ground state up to a relatively large $t/U_0$~\cite{apaja06}. A Monte Carlo study also shows that the SF-MI transition is first order for even fillings due to the spin gap of spin-singlet pairs, whereas it is second order for odd fillings as in the ferromagnetic case~\cite{batrouni09}. 
In higher dimensions, insulating states with an odd number of atoms per site are predicted to be nematic, whereas those with an even number are either nematic or singlet with a first-order transition between them~\cite{imam03spinexchange}. A further survey of this subject is found in Ref.~\cite{Stamper-KurnRMP}.

As discussed in Sec.~\ref{sec:supersolid}, dipolar gases in optical lattices exhibit supersolid behavior, where we assumed that dipole moments are polarized.
For the case of polar molecules in a Mott state, if we liberate the polarization direction and take into account the molecular rotational degrees of freedom as well as the electric spin degrees of freedom, the system can be mapped onto a lattice spin model~\cite{Micheli2006,Barnett2006c}.
The mixing of different rotational states is tunable by means of a microwave field, indicating that a system of polar molecules serves as an excellent playground for simulating lattice-spin models.
More recently, the same system proves to be a highly tunable generalization of the $t$-$J$ model~\cite{Gorshkov2011a,Gorshkov2011b}.

From the viewpoint of AMO (atomic, molecular, and optical) physics, 
spinor BECs coupled with a single-mode cavity have received much attention,
motivated by the realization of cavity optomechanics with a scalar BEC~\cite{Brennecke2008}.
The bistability of spinor BECs in a single-mode cavity was discussed in Refs.~\cite{Zhou2009,Zhou2010}.
It has been pointed out that cavity transmission spectra can serve as a probe to detect quantum spin dynamics~\cite{Cui2008} and quantum ground states~\cite{Zhang2009}.
As a direct analog of cavity optomechanics, Brahms and Stamper-Kurn proposed to utilize a spinor BEC as a macroscopic counterpart in optomechanics, showing spin cooling and amplification and spin optodynamic squeezing of light~\cite{Brahms2010}.
Moreover, by utilizing a dissipative process in a cavity, polar condensates can be cooled down to a many-body ground state~\cite{Jing2011}.
In such a quantum region, the measurement backaction should also be took into consideration~\cite{Ashhab2002b,Zhang2011}.

Finally, we make a brief overview of the subjects related to the spin-orbit (SO) interaction in ultracold atomic systems.
The implementation of the synthetic gauge field~\cite{Lin2009b} and the SO coupling interaction~\cite{Lin2011} are among the most important experimental achievements in recent years.
Rotating atomic gases expand due to the centrifugal force, which makes it difficult to reach the fast-rotating regime of the quantum Hall effect.
As a breakthrough to such a situation, Lin {\it et al.} realized a vector potential, corresponding to a uniform magnetic field by utilizing a pair of counter-propagating Raman lasers and a magnetic field gradient~\cite{Lin2009b}.
Here, a pair of Raman lasers couple atoms in magnetic sublevels $\Delta m=\pm 1$ and momentum difference $\Delta k=\mp 2k_{\rm L}$, where $2k_{\rm L}$ is the momentum difference between two lasers~\cite{Spielman2009}.
In the absence of a magnetic field gradient, atoms in the dressed states that diagonalize the single-particle Hamiltonian experience a uniform vector gauge field~\cite{Lin2009a},
whereas a magnetic field gradient creates a position-dependent effective vector potential that mimics the one for a charged particle in a uniform magnetic field~\cite{Spielman2009,Lin2009b}.
Note that the gauge field created in Ref.~\cite{Lin2009b} is completely determined by the external parameters of Raman lasers and field gradient, and does not possess dynamical degrees of freedom unlike the ordinary gauge fields.
However, this synthetic gauge field is not accompanied by the centrifugal potential, and moreover,
the high tunability of this technique could greatly expand the scope in ultracold atomic physics, enabling the investigation of the quantum-Hall and SO-coupling phenomena, both of which are hot topics also in condensed matter physics.

By applying this technique to multicomponent systems, we can make the gauge potential spin-dependent, thereby engineering a non-Abelian gauge field.
The SO-coupling interaction is a special type of the non-Abelian gauge field:
When the spin-dependent vector potential is written in terms of spin matrices, say ${\bm A}\propto ({\rm f}_y,0,0)$, the single-particle Hamiltonian includes the SO-coupling term $k_x F_y$.
Such a SO-coupling interaction was realized experimentally using a pseudo-spin-1/2 BEC~\cite{Lin2011}.
Several theoretical papers discuss how to create a Rashba ($\propto {\rm f}_x k_y-{\rm f}_y k_x$) and Dresselhaus ($\propto {\rm f}_x k_y+{\rm f}_y k_x$) type of SO couplings, three-dimensional SO couplings ($\propto {\rm f}_x k_x + {\rm f}_y k_y + {\rm f}_z k_z$), and SO couplings in integer-spin systems~\cite{Juzeliunas2010,Campbell2011,Sau2011,Xu2012,Anderson2012}.
As in the case of the dipolar interaction, the SO coupling is expected to generate a spatial spin structures in the ground state.
As an example, we here consider a 2D spinor BEC with SO-coupling interaction $V_{\rm SO}={\rm f}_x k_x+{\rm f}_y k_y$, which is equivalent to the pure Rashba-type interaction if we rotate the spin by $\pi/2$ about the $z$ axis.
When the SO-coupling interaction is sufficiently strong, the spin of an atom moving with momentum ${\bm k}=(k_x,k_y)$ should be antiparallel to ${\bm k}$ so as to minimize $V_{\rm SO}$.
For the case of spin-1 atoms, such an order parameter is given by ${\bm \psi}_{\bm k}(\bm \rho) = \frac{1}{2}e^{i {\bm k}\cdot{\bm \rho}} (-e^{-i\varphi_k},\sqrt{2},-e^{i\varphi_k})^{\rm T}$,
where ${\bm \rho}\equiv(x,y)$ and $\varphi_k\equiv{\rm arg}(k_x+ik_y)$~\cite{Wang2010}.
The interatomic interaction determines in which plane-wave state the condensation occurs:
The ferromagnetic interaction favors a single-plane-wave state whose momentum is spontaneously chosen;
the polar and nematic interactions in spin-1 and 2 systems, respectively, favor stripe structures which are superposition states of a pair of counter-propagating plane waves so as to minimize the spontaneous magnetization;
the cyclic interaction in spin-2 systems leads to triangular and square lattice structures composed of three and four plane waves~\cite{Wang2010,Xu2011,Kawakami2011}.
Inclusion of the trapping potential changes the energy balance among SO-coupling interactions and interatomic interactions, and more exotic lattice structures are predicted to appear~\cite{Sinha2011,Hu2012,Xu2011b}.
For a 3D SO-coupling, a Skyrmion structure is predicted to appear~\cite{Kawakami2012}.
For a review on SO-coupled systems, we suggest Ref.~\cite{Zhai2012}.